\documentclass[12pt, a4paper]{book_modified}

\usepackage{fordthesis}

\begin{document}
\begin{fmffile}{feyndiags}

\renewcommand{\thepage}{\roman{page}}

%

\thispagestyle{empty}
$\phantom{A}$ \vspace{-1.5cm} \\
\begin{center}
\scalebox{1.2}{
\begin{minipage}{11cm}
\begin{center}
\Huge 
{Studies of event shape\\observables with the\\OPAL detector at LEP}
\end{center}
\end{minipage}}
\vspace{6.5cm}

{\Large By}\vspace{0.5cm}

{\LARGE Matthew Thomas Ford} \vspace{0.5cm}

{\Large of}\vspace{0.5cm}

{\LARGE Trinity College} \vspace{6.5cm}

{\Large A dissertation submitted to the University of Cambridge}

{\Large for the degree of Doctor of Philosophy}\vspace{0.5cm}

{\Large February 2004}

\end{center}

\newpage
$\phantom{A}$
\vspace{6cm}
\begin{center}
\textbf{\large Declaration}
\end{center}

This dissertation is the result of my own work, and includes nothing which
is the outcome of work done in collaboration except where specifically
indicated in the~text.\vspace{2cm}

\hspace{10cm} Matthew Ford

\cleardoublepage
\begin{center}
$\phantom{DUMMY}$\vspace{-1.9cm} \\
\textbf{\Large Studies of event shape observables\\with the OPAL detector at LEP} \vspace{0.5cm}

\emph{\large Matthew~T.~Ford}\vspace{0.8cm}

\textbf{\large Abstract}
\end{center}

During the years 1989--2000, the LEP experiments at CERN studied
electron-positron annihilation in the energy range
$\sqrt{s}=91$--209~GeV. Data from the four detectors have been used to
test the Standard Model of particle physics, to measure its
parameters, and to constrain the possibilities for new phenomena.

In quark-antiquark pair production at LEP, many features of the
hadronic final state can be predicted by quantum
chromodynamics~(QCD). Using data collected by the OPAL experiment, we
present the statistical distributions of fourteen ``event shape
observables,'' which describe the inclusive kinematic properties of
events producing three or more jets. For six of these observables, we
compare the measured distributions with those calculated in
perturbative QCD. By optimising the agreement between theory and data,
we measure the strong coupling~$\alpha_\mathrm{S}$ at a
range of energy scales. We also test the predictions of three Monte
Carlo event generators, for all fourteen observables.

Over the years since the LEP experiments began operating, many similar
analyses have been published, and have contributed to the world
average measurements of $\alpha_\mathrm{S}$. However, several
improvements have now been made, both in the theoretical
calculations and in the experimental analysis techniques. We therefore
present a complete reanalysis of the OPAL data, over the full
range of LEP collision energies. Particular attention is given to the
estimation of uncertainties, including the large contribution due to
uncalculated higher-order terms of the theory predictions.

In collaboration with the LEP QCD Working Group, we have combined the
$\alpha_\mathrm{S}$ measurements obtained from event shape observables
by all four experiments. Detailed investigations were undertaken to
ensure consistent implementation of the theoretical predictions and
uncertainty estimates, and to take account of correlations between
measurements. Our combined preliminary value for the strong coupling
at the Z$^0$ mass scale, including contributions from measurements at
higher energies, is\vspace{-0.15cm}
\[
\alpha_\mathrm{S}(M_\mathrm{Z})=0.1201 \pm 0.0003\;\mathrm{(stat.)}\pm
0.0048\;\mathrm{(syst.)} \;\;\;. \vspace{-0.2cm}
\]
Our result is in good agreement with the current world average.

\cleardoublepage
\subsection*{Acknowledgements}

First and foremost I would like to thank my supervisor, David Ward,
for his dedicated support and encouragement throughout my time working on
OPAL, and for his thorough reading of this document. His advice has
always been dependable, and much appreciated.

I also owe considerable gratitude to my collaborators in the OPAL and
LEP QCD Working Groups. Within OPAL my work has profited from the
experience of
Stefan~Kluth\footnotemark[1]\footnotetext[1]{Max-Planck-Institut f\"ur
Physik, Munich, Germany}, who developed many of our current techniques
and software tools during his PhD research with David Ward in the
early 1990s. In the latter stages of my OPAL analysis work,
correspondence with
Mike~Donkers\footnotemark[2]\footnotetext[2]{Carleton University,
Ottawa, Canada} and Christoph~Pahl\footnotemark[1] led to some
invaluable discussion and cross-checking of our results. Throughout my
time on OPAL, our quarterly plenary meetings have provided a
friendly environment in which to present new results and ideas, and to
receive constructive comments and questions; I would like to thank our
Physics Coordinators and plenary organisers for making this possible.

My involvement with the LEP QCD Working Group has enabled me to
participate in a unique and rewarding project, combining results from
the four experiments. I would like to thank all members of the group
for entrusting a major part of this work of this work to me. In
particular, our group convener,
Roger~Jones\footnotemark[3]\footnotetext[3]{Lancaster University,
U.K.}, has been responsible for the efficient organisation and
documentation of countless meetings and telephone conferences, which
have maintained the focus and momentum of our work.  I am also
indebted to
Hasko~Stenzel\footnotemark[4]\footnotetext[4]{Universit\"at Giessen,
Germany}, Daniel~Wicke\footnotemark[5]\footnotetext[5]{Bergische
Universit\"at, Wuppertal, Germany} and
Gavin~Salam\footnotemark[6]\footnotetext[6]{LPTHE, Universit\'es P.~et
M.~Curie (Paris~VI) et D.~Diderot (Paris VII), Paris, France} for many
enthusiastic discussions, and for the work they have contributed to
our LEP \as\ measurements.

For financial support I am obliged to PPARC, who provided three
years of generous funding: this included travel expenses, allowing me
to attend many productive meetings at CERN. In addition, I would like to
thank the Cavendish Laboratory in Cambridge for awarding me a
J.J.~Thomson studentship, which provided a full extra term of
funding. I am also grateful to Christ's College for employing me as a
undergraduate supervisor in the Natural Sciences Tripos for three
years.

Last but not least, I would like to thank my own College, Trinity,
which has been my home for the majority of my eight-and-a-half years
as a student. During my postgraduate studies, Trinity has not only
provided excellent accommodation, but has also funded my attendance at
the CERN Summer School in Beatenberg, Switzerland, and at the QCD~'02
Conference in Montpellier, France.

\begin{flushright}
\emph{MTF}
\end{flushright}
\cleardoublepage
\tableofcontents

\cleardoublepage
\renewcommand{\thepage}{\arabic{page}}
\setcounter{page}{1}

\chapter*{Introduction}
\addcontentsline{toc}{chapter}{\numberline { }Introduction}
{\pagestyle{myheadings}
\markboth{\hfill INTRODUCTION}{INTRODUCTION \hfill}

Since the end of the nineteenth century, several revolutions have
occurred in our understanding of fundamental physics. Firstly, we
have progressed from a world in which the atom was considered
indivisible, to a scenario where both the nucleus and the nucleon are
composite objects. Secondly, the discoveries of relativity and quantum
mechanics have redefined our interpretation of space, time, causality
and observation. The formulation of relativistic quantum mechanics led
to a remarkable explanation for the origin of `spin', and for the
Exclusion Principle, which plays a central r\^ole in atomic physics
and chemistry; it also predicted the existence of an antiparticle for
every particle in nature. By quantising the wavefunctions representing
elementary particles, and by applying the principle of \emph{local
gauge invariance} motivated by classical electromagnetism, a model
emerged to explain three of the four known forces of nature. The
electromagnetic and weak interactions were unified to form a
single electroweak field theory, in which the symmetry of the two
forces is spontaneously broken; the strong interaction, which accounts
for the internal binding of atomic nuclei, was described by a separate
theory called quantum chromodynamics (QCD).

The existence of a `hidden' quantum degree of freedom, which we now
recognise as colour, was initially proposed in 1964 on the basis of
baryon spectroscopy~\cite{greenberg1964}: baryons were suspected to
contain three identical `quarks' of half-integral spin, but in order
to antisymmetrise the baryon wavefunction, one would have to introduce
a new quantum number. In 1972 a non-Abelian SU(3) gauge theory was
developed to describe the dynamics of colour~\cite{fritzsch_gellmann};
this theory, in which the strong interaction of quarks is mediated by
eight gauge bosons called gluons, became known as QCD. As in any
quantum field theory, the interaction vertices in QCD are subject to
an infinite series of perturbative corrections: one is therefore
forced to \emph{renormalise} the theory, resulting in a coupling
strength \as\ which varies as a function of the interaction energy. It
was proven in 1973 that the strength of QCD interactions must decrease
with increasing energy scales~\cite{asymptotic_freedom}; this is in
contrast to quantum electrodymanics, in which the electromagnetic
coupling~$\alpha$ is largest in high-energy interactions. The
phenomenology predicted in QCD was therefore very different from that
of electromagnetism.

Experiments have established conclusively the need for exactly three
quark colours~\cite{webber_qcdbook}. The decay rate of the $\pi^0$
meson into two photons is predicted to scale quadratically with the
number of colours~$N_\text{c}$; measurements as early as
1963~\cite{pi0decay} were sensitive to this effect. More recently, the
LEP Collaborations have measured the partial width for hadronic decays
of the Z$^0$ boson~\cite{lep_electroweak_comb}, which varies linearly
with $N_\text{c}$. Compared with the electroweak theory, however,
which has now been tested and verified to a high degree of
precision,\footnote{The Higgs mechanism is the only aspect of the
electroweak theory \emph{not} to have been confirmed so far. According
to the Standard Model, this is responsible for symmetry-breaking
between the weak and electromagnetic gauge fields.}  the detailed
dynamics of colour presents more difficulties. The established
technique of perturbation theory is not applicable to low-energy
processes in QCD, such as those responsible for the binding of
hadrons. Furthermore, the complicated gauge structure and large
coupling strength have limited the precision of most high-energy
predictions to the level of a few percent. Nonetheless, experiments
have made significant progress to affirm the place of QCD within the
Standard Model. Since 1979, when the first gluon jets were observed in
\epem\ collisions at PETRA~\cite{gluon_discovery}, the large samples
of high-energy collision data at LEP and HERA have confirmed a wide
variety of QCD predictions.

In experimental work, we cannot directly observe the quarks and gluons
participating in hard interactions. The particles seen in our
detectors are hadrons, produced by low-energy fragmentation processes
which cannot be predicted in perturbation theory. One must therefore
define physical observables, based on these final-state hadrons, to
probe the `perturbative' stage of the event. In this work, we will use
data from the OPAL detector to study the distributions of fourteen
\emph{event shape observables}, which offer sensitivity to the QCD
interactions in \epem\ annihilation events. By fitting experimental
data to the theoretical predictions, we will then measure the strong
coupling constant \as.

The analysis presented here represents a continuation of previous OPAL
studies~\cite{OPAL_as_91,OPAL_as_133,OPAL_as_161,OPAL_as_189}, which
have measured eleven of the event shape observables at centre-of-mass
collision energies in the range 91--189~GeV. Several improvements have
been introduced since the original results were published, however,
both in the theoretical predictions and in our experimental
methods. We have therefore performed a complete re-analysis of the
OPAL data, in the energy range 91--207~GeV. These results will
supersede the published measurements where applicable.\footnote{It is
possible that further small modifications will occur before final
publication. The results given in this work should therefore be
regarded as preliminary.}

As the LEP experiments have now ceased operating, we must consider
how to extract global results from the complete dataset. By combining
measurements of \as\ obtained at different centre-of-mass energies, we
can not only test the predicted energy-dependence of the strong
interaction, but can also calculate a single result for \as\ at a
fixed energy scale. The uncertainties is this measurement can be
further reduced by combining fits from different experiments, and from
different event shape observables. Our final result will therefore be
a LEP average value for \as\ obtained from \epem\ event shape
measurements.

The dissertation is organised as follows:
\begin{description}
\item[Chapter \ref{chapter:theory}] introduces the Standard Model of
particle physics, with a particular emphasis on the theory of QCD. We
define the event shape observables to be measured in our analysis, and
describe the techniques used in perturbative QCD to predict their
distributions. The theoretical uncertainties of our \as\ measurements
are then discussed, using a new approach developed in collaboration
with the LEP QCD Working Group~\cite{uncertaintyband}. We also
describe the Monte Carlo simulation models used to account for
non-perturbative effects in our analysis. Finally we list a few of the
other methods by which QCD can be studied experimentally.
\item[Chapter \ref{detectorchapter}] gives an historical and technical
overview of the LEP collider, and of the OPAL detector.
\item[Chapter \ref{opalchapter}] describes our measurements of event
shape observables using multihadronic events at OPAL, and our fits to
the theoretical predictions. We also list the areas in which our new
analysis differs from the published measurements, and compare the old
and new values of \as\ where appropriate. We finally present a
combined OPAL measurement of \as, using the methods to be discussed in
Chapter~\ref{lepcombinationchapter}.
\item[Chapter \ref{lepcombinationchapter}] explains the procedure for
combining \as\ values obtained at different energy scales, with
different observables and different experiments. We first describe an
investigation into the consistency of the analysis procedures used by
the four LEP Collaborations. We then discuss the estimation of
uncertainties and correlations in the \as\ measurements, and the use
of the covariance matrix to extract a weighted mean. We explain in
general terms why the most na\"{\i}ve choice of correlations leads to
an unreliable fit; a more suitable covariance matrix is then proposed
and tested. Finally we present the combined \as\ values from all LEP
event shape measurements, and from various subsets.
\item[Chapter \ref{conclusionchapter}] summarises our final
conclusions, and the outlook for future extensions to the work.
\end{description}

}
\chapter{Theoretical background}
\label{chapter:theory}

We begin this chapter by presenting a brief overview of quantum
chromodynamics~(QCD), and its place within the Standard Model. In
Section~\ref{qcdpert} we discuss the specific application of QCD to
the study of hadron production in \epem\ annihilation. We define in
Section~\ref{evsh_defs} a set of event shape observables, which will
form the basis for our experimental work;
Sections~\ref{pert_predictions}--\ref{evsh_prediction_errors} will
discuss the theoretical predictions for some of these observables, and
their associated uncertainties. In Section~\ref{mcmodels}, we
introduce the models implemented in three Monte Carlo programs to
simulate the non-perturbative aspects of QCD. Finally,
Section~\ref{otherqcd} will give an overview of other techniques used
to measure the coupling~\as\ and to test the validity of QCD.

\section{The elements of the Standard Model}

In our present understanding of particle physics, the elementary
building-blocks of matter comprise twelve spin-$\frac{1}{2}$ fermions
and their twelve antiparticles. The particles can be grouped into
three families, each containing two quarks, one charged lepton and one
neutrino, as shown in Table~\ref{quarksandleptons}. In the absence of
interactions, each fermion is represented by the quanta of a
field~$\psi$, satisfying the Dirac Equation
\begin{equation}
\left(i\gamma^\mu\partial_\mu-m\right)\psi\,=\,0\;\;\;,
\end{equation}
which corresponds to the Lagrangian density
\begin{equation}
\mathcal{L}_\text{D}\,=\,\bar{\psi}\left(i\gamma^\mu\partial_\mu-m\right)\psi\;\;\;.
\end{equation}
The spinor $\psi$ has four components, which collectively correspond
to the helicity states of the particle and antiparticle.

\begin{table}
\begin{center}
\scalebox{0.84}{
\begin{tabular}{|c|c||c|c|}
\hline
\multicolumn{2}{|c||}{Quarks} &
\multicolumn{2}{|c|}{Leptons} \bigstrut \\
\hline
\hline
Down (d) & Up (u) & Electron (e$^-$) & \parbox{1.6cm}{Electron\\[-1.5mm]neutrino} ($\nu_\text{e}$)\rule{0pt}{0.7cm} \bigstrut \\[3.5mm]
$Q=+2/3$ & $Q=-1/3$ &$Q=-1$ & $Q=0$ \\
$m=5.0$--8.5~MeV & $m=1.5$--4.5~MeV & $m=0.511$~MeV & $m<3$~eV~{\footnotesize (95\%~CL)}
\bigstrut \\ \hline
Strange (s) & Charm (c) & Muon ($\mu^-$) & \parbox{1.6cm}{Muon\\[-1.5mm]neutrino} ($\nu_\mu$)\rule{0pt}{0.7cm} \bigstrut \\[3.5mm]
$Q=+2/3$ & $Q=-1/3$ &$Q=-1$ & $Q=0$ \\
$m=80$--155~MeV & $m=1.0$--1.4~GeV & $m=106$~MeV & $m<0.19$~MeV~{\footnotesize (90\%~CL)}
\bigstrut \\ \hline
Bottom (b) & Top (t) & Tau ($\tau^-$) & \parbox{1.6cm}{Tau\\[-1.5mm]neutrino} ($\nu_\tau$)\rule{0pt}{0.7cm} \bigstrut \\[3.5mm]
$Q=+2/3$ & $Q=-1/3$ &$Q=-1$ & $Q=0$ \\
$m=4.0$--4.5~GeV & $m=174\pm 5$~GeV & $m=1.78$~GeV & $m<18$~MeV~{\footnotesize (95\%~CL)}
\bigstrut \\
\hline
\end{tabular}}
\end{center}
\caption{The elementary fermions of the Standard Model. The charge~$Q$
for each particle is expressed in units of the proton charge. The
masses~$m$ are taken from Ref.~\cite{PDbook}.}
\label{quarksandleptons}
\end{table}

Interactions are explained in the
Standard Model by imposing \emph{local gauge symmetries} on the
fields: an example will be described in the next section. These
symmetries require the existence of four vector fields, whose quanta
are the spin-1 gauge bosons listed in
Table~\ref{gaugebosons}. Additional terms are introduced into the
Dirac Lagrangian, leading to interactions between the fermions and
gauge fields, and between the gauge fields themselves. The photon and
the W$^\pm$ and Z$^0$ bosons are responsible for the electroweak
interaction, and the gluon is the carrier of the strong
interaction. Although a further gauge boson, the `graviton' has been
postulated, no complete theory of gravity currently exists within the
Standard Model.

\begin{table}
\begin{center}
\scalebox{0.84}{
\begin{tabular}{|c||c|c|c|c|c|c|c|c|c|}
\hline & \multirow{2}[4]{*}{Charge} &
\multirow{2}[4]{*}{\parbox{1.2cm}{\centering Mass\\(GeV)}} &
\multicolumn{7}{|c|}{Direct couplings to other particles} \bigstrut \\
\cline{4-10} & & & Quarks & e$^\pm$, $\mu^\pm$, $\tau^\pm$ &
$\nu_\text{e}$, $\nu_\mu$, $\nu_\tau$ & $\phantom{A}\gamma\phantom{A}$
& \rule{0pt}{0.5cm}$\phantom{A}\text{W}^\pm\phantom{A}$ &
$\phantom{A}\text{Z}^0\phantom{A}$ & $\phantom{A}$g$\phantom{A}$
\bigstrut \\ \hline \hline Photon ($\gamma$) & 0 & 0 & \textbullet &
\textbullet & & & \textbullet~\textdagger & $\phantom{\text{\textbullet}}$~\textdagger & \bigstrut \\
W$^\pm$ bosons & $\pm
1$ & 80.4 & \textbullet & \textbullet & \textbullet & \textbullet~\textdagger &
$\phantom{\text{\textbullet}}$~\textdagger & \textbullet~\textdagger & \bigstrut[b]\\
Z$^0$ boson & 0 & 91.2 &
\textbullet & \textbullet & \textbullet & $\phantom{\text{\textbullet}}$~\textdagger & \textbullet~\textdagger & &
\bigstrut[b]\\ Gluon (g) & 0 & 0 & \textbullet & & & & & & \textbullet~\textdagger
\bigstrut[b] \\ \hline
\end{tabular}}
\end{center}
\caption{The gauge bosons of the Standard Model, and a summary of
their interactions. Couplings indicated with bullets~({\small
\textbullet}) arise from trilinear terms in the Lagrangian, while
those with daggers~({\small \textdagger}) are quartic; when expressed
in Feynman diagram notation, these correspond respectively to
`three-point' and `four-point' vertices. The direct interaction of the
photon and Z$^0$ boson is possible only via the quartic WWZ$\gamma$
vertex.}
\label{gaugebosons}
\end{table}

One additional particle, the scalar Higgs boson~(H$^0$), is predicted
to exist, but has not yet been observed conclusively. The Higgs field
introduces a \emph{spontaneously broken symmetry} into the Standard
Model, thereby offering an explanation for the non-zero masses of the
W$^\pm$ and Z$^0$ bosons. It also accounts for the mass terms in the
Dirac Lagrangian, but does not predict the masses of the individual
fermions.

Comprehensive discussions of the Standard Model can be found in many
textbooks, such as Refs.~\cite{martin_shaw,aitchison_hey}. A more
detailed treatment of QCD is given in Ref.~\cite{webber_qcdbook}.

\section{The Lagrangian of QCD}

In QCD the six quark flavours are represented by quantum fields
$q=\{u,d,s,c,b,t\}$, which behave identically, apart from their differing masses,
and do not directly interact with one another. The quark fields have
an extra degree of freedom known as colour; each of the three
components~\mbox{$q_a~(a=1,2,3)$} is a Dirac spinor. Treating them as
non-interacting fermion fields, the Dirac Lagrangian would therefore
become
\begin{equation}
\mathcal{L}\,=\,\sum_{a}\bar{q}_a\left(i\gamma^\mu\partial_\mu-m\right)q_a\;\;\;.
\label{non_interacting_colour}
\end{equation}
We now consider the effect of a unitary ``phase transformation''
applied to the three-component colour vector $q$
\begin{equation}
q_a \,\to\, q'_a\,=\,\sum_b \Omega_{ab}q_b \,\equiv\, \sum_b \exp\bigg[i\sum_A
\alpha^A\lambda^A_{ab}\bigg]\,q_b \;\;,
\end{equation}
where the $3\times 3$ Hermitian matrices $\lambda^A~(A=1,2,\ldots 8)$
are the generators of the Lie group SU(3), and $\alpha^A$ are eight
arbitrary constants. The Lagrangian given in
Equation~(\ref{non_interacting_colour}) is invariant under this global
transformation, due to the unitary property of the matrices; this is
analogous to the invariance of the Dirac Lagrangian under the phase
transformation~\mbox{$\psi\to\psi'=\psi e^{i\phi}$}.

Our global colour transformation demonstrates the conservation of
colour in a non-interacting theory, but does not introduce any
physical dynamics. The theory of QCD is derived by requiring the
invariance of the Lagrangian under \mbox{\emph{local}} SU(3) colour
transformations: instead of choosing the same unitary matrix,
$\Omega=\exp [i\sum_A \alpha^A\lambda^A]$, at all points in space and
time, we allow the coefficients $\alpha^A$ to vary, giving\vspace{-0.1cm}
\begin{equation}
q_a \,\to\, q'_a\,=\,\sum_b\Omega_{ab}(x)q_b \,\equiv\, \sum_b \exp\bigg[i\sum_A \alpha^A(x)\lambda^A_{ab}\bigg]\,q_b \;\;.\vspace{-0.05cm}
\label{localgauge}
\end{equation}
Substituting this transformed quark field into
Equation~(\ref{non_interacting_colour}), we find that the Lagrangian
is no longer invariant, because the space-time derivatives act on the
coefficients~$\alpha^A(x)$. To restore the invariance of the
Lagrangian, we must first replace the partial
derivative~$\partial_\mu$ with a \emph{covariant derivative}\vspace{-0.05cm}
\begin{equation}
(D_\mu)_{ab}=\partial_\mu\delta_{ab}+\frac{ig}{2}\sum_A\mathcal{A}_\mu^A\lambda^A_{ab} \;\;\;,\vspace{-0.05cm}
\end{equation}
where we have introduced eight \emph{gauge fields} $\mathcal{A}^A$,
each with four space-time components~$\mu$; the free parameter $g$ is
a universal coupling constant. The Lagrangian now becomes\vspace{-0.05cm}
\begin{eqnarray}
\mathcal{L} & = & \sum_{a,b}\bar{q}_a\left(i\gamma^\mu
D_\mu-m\right)_{ab}q_b \\
& \equiv & \sum_a \bar{q}_a\left(i\gamma^\mu\partial_\mu-m\right)q_a\,+\,
\frac{ig}{2}\sum_{a,b}\sum_A \bar{q}_a \,(\gamma^\mu \mathcal{A}_\mu^A) \,\lambda^A_{ab} \,q_b\;\;\;.\vspace{-0.05cm}
\end{eqnarray}
In the last line, we have decomposed $\mathcal{L}$ into two
contributions: the first is the Dirac Lagrangian for three
non-interacting components of a fermion field, and the second
introduces interactions between the gauge fields and the quarks. The
quanta of the eight fields $\mathcal{A}^A$ are called gluons, and are
responsible for the observed strong interactions of quarks. To
complete the process of establishing local gauge invariance, the
transformation properties of the gluon fields must be chosen such that
the covariant derivative~$\sum_b (D_\mu)_{ab}\,q_b$ transforms in the
same way as the quark field itself,\vspace{-0.05cm}
\begin{equation}
\sum_b(D'_\mu)_{ab} \,q'_b=\sum_{b,c} \Omega_{ab}(x) \,(D_\mu)_{bc}\, q_c \;\;\;.
\end{equation}
\enlargethispage{1.5\baselineskip}This is achieved with the relationship
\begin{equation}\vspace{-0.1cm}
\sum_A {\mathcal{A}'_\mu}^A \lambda^A \,=\, \Omega(x)\bigg[\sum_A
\mathcal{A}_\mu^A
\lambda^A\bigg]\Omega^{-1}(x)\,+\,\frac{2i}{g}(\partial_\mu\Omega(x))\,\Omega^{-1}(x)\;\;\;,\vspace{-0.05cm}
\end{equation}
where we have suppressed the colour indices of the $\lambda^A$ and
$\Omega(x)$ matrices.\footnote{A simpler transformation law, of the
form ${\mathcal{A}'}^A=\mathcal{A}^A+\delta\mathcal{A}^A$, exists when
the gauge transformation $\Omega(x)$ differs only infinitesimally from
from the identity matrix.}

One further contribution must be inserted in the Lagrangian, to
specify the equations of motion for the gluon fields. In quantum
electrodynamics, the Lagrangian for the photon field $A$ is given
by
\begin{equation}
\mathcal{L}_\text{photon}=-\frac{1}{4}F_{\mu\nu}F^{\mu\nu}\;\;\;,
\end{equation}
where $F$ is simply a quantised form of Maxwell's electromagnetic
field strength tensor
\begin{equation}
F_{\mu\nu}=\partial_\mu A_\nu-\partial_\nu A_\mu \;\;\;.
\end{equation}
Applying the Euler-Lagrange Equations to $\mathcal{L}_\text{photon}$
gives the familiar Maxwell Equations, governing the internal dynamics
of the field. An analogous term appears in the Lagrangian of QCD,
\begin{equation}
\mathcal{L}_\text{gluon}=-\frac{1}{4}\sum_A F^A_{\mu\nu}F_A^{\mu\nu}\;\;\;,
\label{lgluon}
\end{equation}
but here the eight field strength tensors for the gluons are
\begin{equation}
F^A_{\mu\nu}=\partial_\mu \mathcal{A}^A_\nu-\partial_\nu \mathcal{A}^A_\mu
-g\sum_{B,C}f^{ABC}\mathcal{A}^B_\mu\mathcal{A}^C_\nu \;\;\;,
\label{gluonfieldstrength}
\end{equation}
where the structure constants $f^{ABC}$ are defined by the commutation
relations of the SU(3) generators,
$\left[\lambda^A,\,\lambda^B\right]=2if^{ABC}\lambda^C$. The last
term of Equation~(\ref{gluonfieldstrength}), which is derived by imposing local SU(3) gauge symmetry on the
octet of gluon fields, arises because the gauge transformations of QCD
do not commute. When we expand out the product
$F^A_{\mu\nu}F_A^{\mu\nu}$ in Equation~(\ref{lgluon}), we find an
array of terms containing products of two, three and four gluon
fields. The three- and four-gluon terms in the Lagrangian give rise to
the self-interaction of the gluon field, which has no analogue in QED.

\begin{figure}[h!]
\begin{center}
\scalebox{0.8}{
\begin{fmfframe}(0.5,0.5)(0.5,0){
\begin{fmfgraph*}(3.5,3.5)
\fmfpen{thin}
\fmfsurroundn{i}{3}
\fmf{fermion}{i2,c,i3}
\fmf{gluon}{c,i1}
\fmflabel{\boldmath $\bar{\mathbf{q}}$}{i3}
\fmflabel{\boldmath $\mathbf{q}$}{i2}
\fmflabel{\boldmath $\mathbf{g}$}{i1}
\end{fmfgraph*}}
\end{fmfframe}
\begin{fmfframe}(0.5,0.5)(0.5,0){
\begin{fmfgraph*}(3.5,3.5)
\fmfpen{thin}
\fmfsurroundn{i}{3}
\fmf{gluon}{c,i1}
\fmf{gluon}{c,i2}
\fmf{gluon}{c,i3}
\fmflabel{\boldmath $\mathbf{g}$}{i3}
\fmflabel{\boldmath $\mathbf{g}$}{i2}
\fmflabel{\boldmath $\mathbf{g}$}{i1}
\end{fmfgraph*}}
\end{fmfframe}
\begin{fmfframe}(0.5,0.5)(0.5,0){
\begin{fmfgraph*}(3.5,3.5)
\fmfpen{thin}
\fmfsurroundn{i}{8}
\fmf{gluon}{c,i2}
\fmf{gluon}{c,i4}
\fmf{gluon}{c,i6}
\fmf{gluon}{c,i8}
\fmflabel{\boldmath $\mathbf{g}$}{i8}
\fmflabel{\boldmath $\mathbf{g}$}{i6}
\fmflabel{\boldmath $\mathbf{g}$}{i4}
\fmflabel{\boldmath $\mathbf{g}$}{i2}
\end{fmfgraph*}}
\end{fmfframe}}
\end{center}
\caption{Feynman vertices in QCD}
\label{qcdrules}
\end{figure}
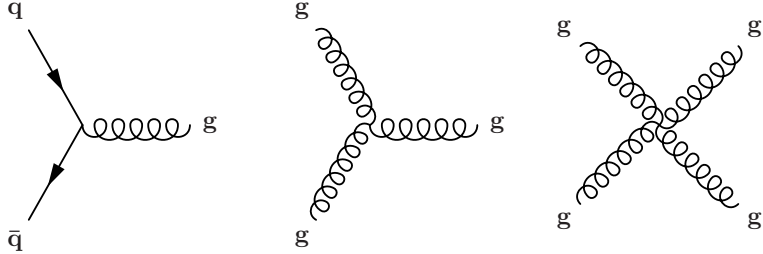

Collecting all terms together, we arrive at the complete Lagrangian
density of QCD: \footnote{When performing practical calculations, some
further terms need to be inserted to fix the gauge. These are
discussed in Ref.~\cite{webber_qcdbook}.}
\begin{equation}
\mathcal{L_\text{QCD}} = 
\sum_a \bar{q}_a\left(i\gamma^\mu\partial_\mu-m\right)q_a\,+\,
\frac{ig}{2}\sum_{a,b}\sum_A \bar{q}_a \,(\gamma^\mu
\mathcal{A}_\mu^A) \,\lambda^A_{ab} \,q_b\,-\,\frac{1}{4}
\sum_AF^A_{\mu\nu}F_A^{\mu\nu} \;\;\;.
\end{equation}
\enlargethispage{1.5\baselineskip}\newpage It is beyond the scope of this work to derive the Feynman rules
associated with the QCD Lagrangian, or indeed to discuss the formal
interpretation of Feynman diagrams; a list of the QCD Feynman rules
can be found in Ref.~\cite{webber_qcdbook}. However, a quick
examination of the terms in the Lagrangian shows that the permitted
vertices are as shown in Figure~\ref{qcdrules}.

Up to this point, the coupling constant of QCD has been denoted~$g$.
From now on, however, we will use the related quantity~$\as=g^2/4\pi$.

The idea of local gauge invariance under
non-Abelian transformations was first proposed by Yang and Mills in
1954~\cite{yang_mills}, and has subsequently provided the foundation
for both the strong and electroweak field theories of the Standard
Model; the simpler Abelian case gave us quantum electrodynamics.

\section{Renormalisation and the running coupling}
\label{renormalisation}

\begin{figure}
\begin{center}
\scalebox{0.8}{
\begin{fmfframe}(1.5,0.5)(1.5,0){
\begin{fmfgraph*}(3,6)
\fmfpen{thin}
\fmftopn{i}{2}
\fmfbottomn{o}{2}
\fmf{fermion}{i1,v1,i2}
\fmf{fermion}{o1,v2,o2}
\fmf{phantom, tension=0.3}{v1,v2}
\fmffreeze
\fmf{photon, label=\boldmath $Q^2$}{v1,v2}
\fmfv{label=\boldmath $\mathbf{e}^-$,label.angle=170}{i1}
\fmfv{label=\boldmath $\mathbf{e}^-$,label.angle=10}{i2}
\fmfv{label=\boldmath $\mathbf{e}^-$,label.angle=-170}{o1}
\fmfv{label=\boldmath $\mathbf{e}^-$,label.angle=-10}{o2}
\fmfv{label=(\emph{a}),label.angle=-90,label.dist=0.35w}{v2}
\end{fmfgraph*}}
\end{fmfframe}
\begin{fmfframe}(1.5,0.5)(1.5,0){
\begin{fmfgraph*}(3,6)
\fmfpen{thin}
\fmftopn{i}{2}
\fmfbottomn{o}{2}
\fmf{fermion}{i1,v1,i2}
\fmf{fermion}{o1,v2,o2}
\fmf{phantom, tension=0.3}{v1,v2}
\fmffreeze
\fmf{photon, label=\boldmath $Q^2$}{v1,v3}
\fmf{photon, label=\boldmath $Q^2$}{v4,v2}
\fmf{fermion, left, tension=0.5}{v3,v4,v3}
\fmfv{label=\boldmath $\mathbf{e}^-$,label.angle=170}{i1}
\fmfv{label=\boldmath $\mathbf{e}^-$,label.angle=10}{i2}
\fmfv{label=\boldmath $\mathbf{e}^-$,label.angle=-170}{o1}
\fmfv{label=\boldmath $\mathbf{e}^-$,label.angle=-10}{o2}
\fmfv{label=(\emph{b}),label.angle=-90,label.dist=0.35w}{v2}
\end{fmfgraph*}}
\end{fmfframe}}
\end{center}
\caption{Leading QED contributions to the scattering of two electrons}
\label{electronscatter}
\end{figure}
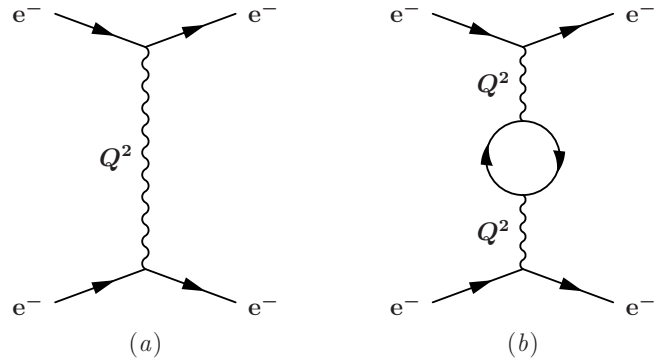

When using perturbation theory to calculate predictions for a physical
observable in a quantum field theory, one must typically sum a series
of Feynman diagrams corresponding to the appropriate initial and final
states. In Figure~\ref{electronscatter}, for example, diagram
(\emph{b}) is a higher-order contribution to the electromagnetic
scattering process shown in diagram~(\emph{a}). Unfortunately,
however, most Feynman diagrams with loops lead to divergent integrals;
whereas diagram (\emph{b}) should introduce a small correction to the
cross section, it appears that it will instead cause a finite cross
section to become infinite! This problem was first identified in QED,
and was solved by the principle of renormalisation.

The apparent paradox of the infinite loop integrals can addressed in
QED by a redefinition of the fine-structure
constant,~$\alpha_\text{em}$. One must remember that no laboratory
experiment can distinguish between the two processes shown in
Figure~\ref{electronscatter}, and therefore any observations of
electromagnetic scattering will already `include' the higher-order
contributions. Remarkably enough, it can be shown that the effect of
these corrections is to modify the effective \emph{coupling strength}
of the photon (albeit by an infinite factor), without altering its
detailed properties as a photon. Hence we should not try to build
Feynman diagrams out of the original `bare' photons, but instead use
the `ready-assembled' photons provided by nature, which include all of
the loop diagrams such as Figure~\ref{electronscatter}(\emph{b}).

We cannot, however, ignore the loop integrals altogether. Even though
the higher-order diagrams contain divergent integrals, they remarkably
have a \emph{finite} dependence on the squared four-momentum~$Q^2$ of
the exchanged photon. Consequently, if two experiments study photons
of different $Q^2$, they will observe different effective coupling
strengths: photons of higher virtuality interact more strongly.

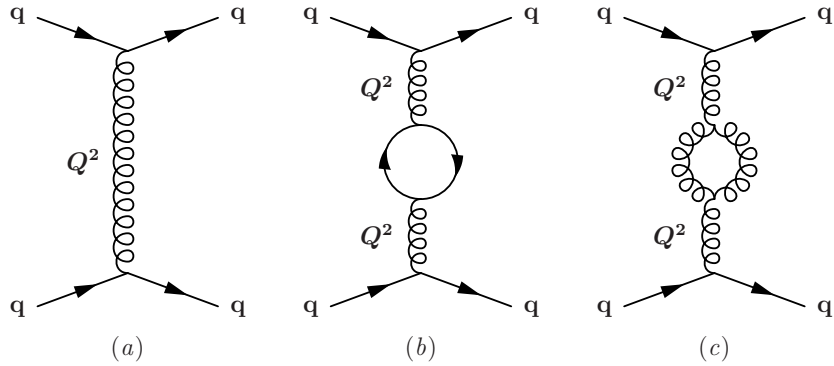
\begin{figure}
\begin{center}
\scalebox{0.8}{
\begin{fmfframe}(0.8,0.5)(0.8,0){
\begin{fmfgraph*}(3,6)
\fmfpen{thin}
\fmftopn{i}{2}
\fmfbottomn{o}{2}
\fmf{fermion}{i1,v1,i2}
\fmf{fermion}{o1,v2,o2}
\fmf{phantom, tension=0.3}{v1,v2}
\fmffreeze
\fmf{gluon, label=\boldmath $Q^2\phantom{I}$}{v1,v2}
\fmfv{label=\boldmath $\mathbf{q}$,label.angle=170}{i1}
\fmfv{label=\boldmath $\mathbf{q}$,label.angle=10}{i2}
\fmfv{label=\boldmath $\mathbf{q}$,label.angle=-170}{o1}
\fmfv{label=\boldmath $\mathbf{q}$,label.angle=-10}{o2}
\fmfv{label=(\emph{a}),label.angle=-90,label.dist=0.35w}{v2}
\end{fmfgraph*}}
\end{fmfframe}
\begin{fmfframe}(0.8,0.5)(0.8,0){
\begin{fmfgraph*}(3,6)
\fmfpen{thin}
\fmftopn{i}{2}
\fmfbottomn{o}{2}
\fmf{fermion}{i1,v1,i2}
\fmf{fermion}{o1,v2,o2}
\fmf{phantom, tension=0.3}{v1,v2}
\fmffreeze
\fmf{gluon, label=\boldmath $Q^2\phantom{I}$}{v1,v3}
\fmf{gluon, label=\boldmath $Q^2\phantom{I}$}{v4,v2}
\fmf{fermion, left, tension=0.5}{v3,v4,v3}
\fmfv{label=\boldmath $\mathbf{q}$,label.angle=170}{i1}
\fmfv{label=\boldmath $\mathbf{q}$,label.angle=10}{i2}
\fmfv{label=\boldmath $\mathbf{q}$,label.angle=-170}{o1}
\fmfv{label=\boldmath $\mathbf{q}$,label.angle=-10}{o2}
\fmfv{label=(\emph{b}),label.angle=-90,label.dist=0.35w}{v2}
\end{fmfgraph*}}
\end{fmfframe}
\begin{fmfframe}(0.8,0.5)(0.8,0){
\begin{fmfgraph*}(3,6)
\fmfpen{thin}
\fmftopn{i}{2}
\fmfbottomn{o}{2}
\fmf{fermion}{i1,v1,i2}
\fmf{fermion}{o1,v2,o2}
\fmf{phantom, tension=0.3}{v1,v2}
\fmffreeze
\fmf{gluon, label=\boldmath $Q^2\phantom{I}$}{v1,v3}
\fmf{gluon, label=\boldmath $Q^2\phantom{I}$}{v4,v2}
\fmf{gluon, left, tension=0.5}{v3,v4,v3}
\fmfv{label=\boldmath $\mathbf{q}$,label.angle=170}{i1}
\fmfv{label=\boldmath $\mathbf{q}$,label.angle=10}{i2}
\fmfv{label=\boldmath $\mathbf{q}$,label.angle=-170}{o1}
\fmfv{label=\boldmath $\mathbf{q}$,label.angle=-10}{o2}
\fmfv{label=(\emph{c}),label.angle=-90,label.dist=0.35w}{v2}
\end{fmfgraph*}}
\end{fmfframe}}
\end{center}
\caption{Leading QCD contributions to the scattering of two quarks}
\label{quarkscatter}
\end{figure}

An analogous situation arises in QCD: the two loop diagrams in
Figure~\ref{quarkscatter} contain divergent integrals, which can be
absorbed into a redefinition of the gluon coupling strength~\as. There
is a very significant difference between the QCD and QED cases,
however. The sign of the $Q^2$-dependence for the gluon
self-interaction loop in Figure~\ref{quarkscatter}(\emph{c}) is
opposite to that for the fermion loops in
Figures~\ref{electronscatter}(\emph{b})
and~\ref{quarkscatter}(\emph{b}). As a consequence of this extra
diagram, which is not present in QED, the strong coupling \as\
\emph{decreases} at high virtualities. Conversely, \as\ grows without
limit as the virtuality of the gluon approaches zero. Therefore quarks
will appear to be ``asymptotically free'' in very high energy
interactions, while at low energy scales their interactions are so
strong that free quarks are never observed.  This low-energy behaviour
of QCD prevents the reliable use of perturbation theory at scales
below a few~GeV; in particular, the process of \emph{hadronisation},
by which free quarks are converted into observable hadrons, is not
well understood.

The scale-dependence of the ``running~coupling'' \asq\ is given by the
Renormalisation Group Equation (RGE):
\begin{equation}
Q\frac{\partial \as}{\partial Q}=2\beta(\as) \;\;\;,
\label{rge}
\end{equation}
where the $\beta$-function is of the form
\begin{equation}
\beta(\as)\,=\,-\frac{\beta_0}{4\pi}\as^2
-\frac{\beta_1}{8\pi^2}\as^3
-\frac{\beta_2}{128\pi^3}\as^4+\mathcal{O}(\as^5)\;\;\;.
\end{equation}
The coefficients $\beta_n$ are functions of the number of
kinematically accessible quark flavours, and are listed in
Ref.~\cite{PDbook}. To first order, the solution of the RGE for
$N_\text{f}$~flavours\footnote{All of the QCD calculations applied in
our analysis will use $N_\text{f}=5$.} can be written as
\begin{equation}
\as(Q)=\as(Q_0){\left[1-\frac{\as(Q_0)}{12\pi}(33-2N_\text{f})\ln\left(\frac{Q^2}{Q_0^2}\right)\right]}^{-1} \;\;\;,
\label{asrun_1storder}
\end{equation}
where \asq\ and $\as(Q_0)$ are the values of \as\ at two different
scales; for the purposes of our analysis, however, we will use
numerical solutions of the RGE with coefficients up to $\beta_2$
included.

When calculating the effects of divergent loop diagrams in a physical
process, one must choose an energy scale~$\mu$ at which to
`renormalise' the diagrams. In QED, for example, we can calculate the
$Q^2$-dependence of a divergent diagram by comparing it with $Q=0$
case. For the strong interaction, however, the scale $Q=0$ would be an
inappropriate reference point. Instead we choose a scale $Q=\mu$,
close to the characteristic energy scale of the physical process. This
``renormalisation scale'' is an unphysical parameter, and plays no part
in the QCD Lagrangian; hence we should not expect physical observables
to depend on it. However, as we shall see later in this chapter, the
cancellation of $\mu$ from physical predictions will be incomplete
unless the calculation itself is complete. Schematically, we can write
the prediction for an observable~$R$ as
\begin{equation}
R(\as)=R_\text{known}(\as,\mu)+R_\text{unknown}(\as,\mu)
\end{equation}
where 
\begin{equation}
\frac{\text{d}R}{\text{d}\mu}=0 \;\;,\;\;\;\;\text{but}\;\;\;
\frac{\text{d}R_\text{known}}{\text{d}\mu}\neq 0
\end{equation}
The magnitude of the derivative $\text{d}R_\text{known}/\text{d}\mu$
is often used to estimate the size of the unknown term, since it
should approach zero as $R_\text{known}\to R$.

\section[QCD perturbation theory in \lowercase{e}$^+$\lowercase{e}$^- \to 
\mathrm{hadrons}$] {QCD perturbation theory in \boldmath e$^+$e$^- \to
\mathrm{hadrons}$}

\label{qcdpert}

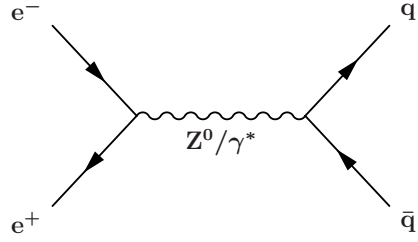
\begin{figure}
\begin{center}
\scalebox{0.8}{
\begin{fmfframe}(1,0.5)(1,0){
\begin{fmfgraph*}(7,3)
\fmfpen{thin}
\fmfleftn{i}{2}
\fmfrightn{o}{2}
\fmf{fermion}{i2,v1,i1}
\fmf{photon, label=\boldmath $\mathbf{Z}^0/\gamma^*$}{v1,v2}
\fmf{fermion}{o1,v2,o2}
\fmflabel{\boldmath $\textbf{e}^+$}{i1}
\fmflabel{\boldmath $\textbf{e}^-$}{i2}
\fmflabel{\boldmath $\bar{\mathbf{q}}$}{o1}
\fmflabel{\boldmath $\mathbf{q}$}{o2}
\end{fmfgraph*}}
\end{fmfframe}}
\end{center}
\caption{Simple quark pair production diagram in e$^+$e$^-$ annihilation}
\label{epemtohadrons}
\end{figure}

In e$^+$e$^-$ annihilation, the simplest process yielding hadrons in
the final state is shown in Figure~\ref{epemtohadrons}. The quark and
antiquark carry both colour and electric charge, however, so they may
interact further before fragmenting into hadrons. Although both strong
and electroweak processes are possible, we will concern ourselves only
with the strong interactions,\footnote{Electroweak processes in which
further Z$^0$ or W$^\pm$ bosons are produced, or in which photons are
radiated from the initial state, will be considered as a background in
our experimental analysis.} which lead to further quark or gluon jets
in the final state.

Provided the running coupling \asq~is suitably small, perturbation
theory should allow us to calculate the matrix element and cross
section for any configuration of partons. So far, however, the
technicalities of handling loops in QCD diagrams have limited the
precision of such predictions to second order in \as.

\subsection[Cross sections at $\mathcal{O}(\as^2)$]
{Cross sections at \boldmath $\mathcal{O}(\as^2)$}

According to Fermi's Golden Rule, the transition rate from an initial
state~`i' to a final state~`f' is given by
\begin{equation}
\Gamma\;=\;\frac{2\pi}{\hbar}\,{\left|\mathcal{M}_{(\text{i}\to\text{f})}\right|}^2 \; \rho\left(p_1, p_2,
\ldots p_{\scriptscriptstyle N}\right) \;\;\;,
\end{equation}
where $\mathcal{M}_{(\text{i}\to\text{f})}$ is the matrix element
connecting states `i' and `f' in the perturbed
Hamiltonian,\footnote{In the annihilation of unpolarised particle
beams, one actually takes the average squared matrix element
$\left\langle{\left|\mathcal{M}_{(\text{i}\to\text{f})}\right|}^2\right\rangle$
for the possible initial states.} and $\rho$ is the density of
possible momentum states in phase space for the final state `f'.

In a quantum field theory such as QCD, the matrix element
$\mathcal{M}_{(\text{i}\to\text{f})}$ is a sum of transition
amplitudes~$\mathcal{M}_i$ represented by Feynman diagrams. The
diagrams may be grouped conveniently by the number of strong
interaction vertices, each of which contributes a factor~$\as^{1/2}$
to the corresponding amplitude. In Figures~\ref{feyn_1storder},
\ref{feyn_2ndorder} and~\ref{feyn_3rdorder}, we list the diagrams
containing one, two and three strong vertices respectively (the
initial e$^+$e$^-$ state is not shown).

\begin{figure}
\begin{center}
\scalebox{0.8}{
\begin{fmfframe}(1,0.5)(1,0){
\begin{fmfgraph*}(5,4)
\fmfpen{thin}
\fmfleft{i}
\fmfrightn{o}{3}
\fmf{photon, tension=2}{i,v1}
\fmf{fermion}{o1,v1}
\fmf{plain, tension=2}{v1,v2}
\fmf{fermion, tension=2}{v2,o3}
\fmffreeze
\fmf{gluon}{v2,o2}
\fmflabel{\boldmath $\bar{\mathbf{q}}$}{o1}
\fmflabel{\textbf{g}}{o2}
\fmflabel{\boldmath $\mathbf{q}$}{o3}
\end{fmfgraph*}
\hspace{1cm}
\begin{fmfgraph*}(5,4)
\fmfpen{thin}
\fmfleft{i}
\fmfrightn{o}{3}
\fmf{photon, tension=2}{i,v1}
\fmf{fermion, tension=2}{o1,v2}
\fmf{plain, tension=2}{v2,v1}
\fmf{fermion}{v1,o3}
\fmffreeze
\fmf{gluon}{v2,o2}
\fmflabel{\boldmath $\bar{\mathbf{q}}$}{o1}
\fmflabel{\textbf{g}}{o2}
\fmflabel{\boldmath $\mathbf{q}$}{o3}
\end{fmfgraph*}}
\end{fmfframe}
}
\end{center}
\caption{First order QCD matrix element contributions, $\mathcal{M}_i
\propto \as^{1/2}$}
\label{feyn_1storder}
\end{figure}
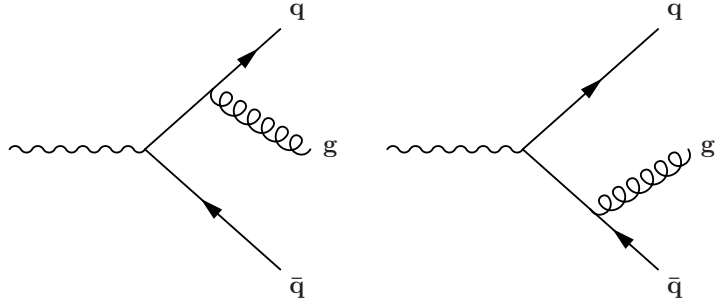

\begin{figure}
\vspace{1cm}
\begin{center}
\emph{(a) q$\bar{\mathit{q}}$ diagram with one loop} \vspace{1cm}\\
\scalebox{0.7}{
\begin{fmfgraph*}(5,4)
\fmfpen{thin}
\fmfleft{i}
\fmfrightn{o}{2}
\fmf{photon, tension=2}{i,v1}
\fmf{fermion, tension=2}{o1,v2}
\fmf{plain, tension=2}{v2,v1,v3}
\fmf{fermion, tension=2}{v3,o2}
\fmffreeze
\fmf{gluon}{v2,v3}
\end{fmfgraph*}}
\vspace{1cm}\\
\emph{(b) q$\bar{\mathit{q}}\mathit{q}\bar{\mathit{q}}$ tree level diagrams} \vspace{1cm}\\
\scalebox{0.7}{
\begin{fmfgraph*}(5,4)
\fmfpen{thin}
\fmfleft{i}
\fmfrightn{o}{4}
\fmf{photon, tension=2}{i,v1}
\fmf{fermion, tension=1}{o1,v2}
\fmf{plain, tension=3}{v2,v1}
\fmf{plain, tension=3}{v1,v3}
\fmf{fermion, tension=1}{v3,o4}
\fmffreeze
\fmf{gluon, tension=2}{v3,v4}
\fmf{phantom, tension=0.5}{v2,v4}
\fmf{fermion}{o2,v4}
\fmf{fermion}{v4,o3}
\end{fmfgraph*}}
\hspace{1cm}
\scalebox{0.7}{
\begin{fmfgraph*}(5,4)
\fmfpen{thin}
\fmfleft{i}
\fmfrightn{o}{4}
\fmf{photon, tension=2}{i,v1}
\fmf{fermion, tension=1}{o1,v2}
\fmf{plain, tension=3}{v2,v1}
\fmf{plain, tension=3}{v1,v3}
\fmf{fermion, tension=1}{v3,o4}
\fmffreeze
\fmf{gluon, tension=2}{v4,v2}
\fmf{phantom, tension=0.5}{v3,v4}
\fmf{fermion}{o2,v4}
\fmf{fermion}{v4,o3}
\end{fmfgraph*}}
\vspace{1cm}\\
\emph{(c) q$\bar{\mathit{q}}\mathit{gg}$ tree level diagrams} \vspace{1cm}\\
\scalebox{0.7}{
\begin{fmfgraph*}(5,4)
\fmfpen{thin}
\fmfleft{i}
\fmfrightn{o}{4}
\fmf{photon, tension=2}{i,v1}
\fmf{fermion, tension=1.5}{o1,v2}
\fmf{plain, tension=2.5}{v2,v1,v3}
\fmf{fermion, tension=1.5}{v3,o4}
\fmffreeze
\fmf{gluon}{o2,v2}
\fmf{gluon}{v3,o3}
\end{fmfgraph*}}
\hspace{0.3cm}
\scalebox{0.7}{
\begin{fmfgraph*}(5,4)
\fmfpen{thin}
\fmfleft{i}
\fmfrightn{o}{4}
\fmf{photon, tension=1.5}{i,v1}
\fmf{fermion, tension=1}{o1,v5}
\fmf{plain, tension=2}{v1,v2,v3}
\fmf{plain, tension=2}{v1,v4,v5}
\fmf{fermion, tension=1}{v3,o4}
\fmffreeze
\fmf{gluon}{v2,o2}
\fmf{gluon}{v3,o3}
\end{fmfgraph*}}
\hspace{0.3cm}
\scalebox{0.7}{
\begin{fmfgraph*}(5,4)
\fmfpen{thin}
\fmfleft{i}
\fmfrightn{o}{4}
\fmf{photon, tension=1.5}{i,v1}
\fmf{fermion, tension=1}{o1,v5}
\fmf{plain, tension=2}{v1,v2,v3}
\fmf{plain, tension=2}{v1,v4,v5}
\fmf{fermion, tension=1}{v3,o4}
\fmffreeze
\fmf{gluon}{o3,v4}
\fmf{gluon}{o2,v5}
\end{fmfgraph*}}
\vspace{0.5cm}\\
\scalebox{0.7}{
\begin{fmfgraph*}(5,4)
\fmfpen{thin}
\fmfleft{i}
\fmfrightn{o}{4}
\fmf{photon, tension=2}{i,v1}
\fmf{fermion, tension=2}{o1,v2}
\fmf{plain, tension=2}{v2,v1,v3}
\fmf{fermion, tension=2}{v3,o4}
\fmffreeze
\fmf{gluon}{o3,v2}
\fmf{gluon, rubout}{v3,o2}
\end{fmfgraph*}}
\hspace{0.3cm}
\scalebox{0.7}{
\begin{fmfgraph*}(5,4)
\fmfpen{thin}
\fmfleft{i}
\fmfrightn{o}{4}
\fmf{photon, tension=1.3}{i,v1}
\fmf{fermion, tension=1}{o1,v5}
\fmf{plain, tension=2}{v1,v2}
\fmf{plain, tension=1}{v2,v3}
\fmf{plain, tension=2}{v1,v4}
\fmf{plain, tension=1}{v4,v5}
\fmf{fermion, tension=1}{v3,o4}
\fmffreeze
\fmf{gluon}{v2,o3}
\fmf{gluon, rubout}{v3,o2}
\end{fmfgraph*}}
\hspace{0.3cm}
\scalebox{0.7}{
\begin{fmfgraph*}(5,4)
\fmfpen{thin}
\fmfleft{i}
\fmfrightn{o}{4}
\fmf{photon, tension=1.3}{i,v1}
\fmf{fermion, tension=1}{o1,v5}
\fmf{plain, tension=2}{v1,v2}
\fmf{plain, tension=1}{v2,v3}
\fmf{plain, tension=2}{v1,v4}
\fmf{plain, tension=1}{v4,v5}
\fmf{fermion, tension=1}{v3,o4}
\fmffreeze
\fmf{gluon}{o2,v4}
\fmf{gluon, rubout}{o3,v5}
\end{fmfgraph*}}
\vspace{0.5cm} \\
\scalebox{0.7}{
\begin{fmfgraph*}(5,4)
\fmfpen{thin}
\fmfleft{i}
\fmfrightn{o}{4}
\fmf{photon, tension=2}{i,v1}
\fmf{fermion, tension=1}{o1,v2}
\fmf{plain, tension=3}{v2,v1}
\fmf{plain, tension=3}{v1,v3}
\fmf{fermion, tension=1}{v3,o4}
\fmffreeze
\fmf{gluon, tension=2}{v3,v4}
\fmf{phantom, tension=0.5}{v2,v4}
\fmf{gluon}{v4,o2}
\fmf{gluon}{v4,o3}
\end{fmfgraph*}}
\hspace{1cm}
\scalebox{0.7}{
\begin{fmfgraph*}(5,4)
\fmfpen{thin}
\fmfleft{i}
\fmfrightn{o}{4}
\fmf{photon, tension=2}{i,v1}
\fmf{fermion, tension=1}{o1,v2}
\fmf{plain, tension=3}{v2,v1}
\fmf{plain, tension=3}{v1,v3}
\fmf{fermion, tension=1}{v3,o4}
\fmffreeze
\fmf{gluon, tension=2}{v4,v2}
\fmf{phantom, tension=0.5}{v3,v4}
\fmf{gluon}{v4,o2}
\fmf{gluon}{v4,o3}
\end{fmfgraph*}}
\end{center}
\caption{Second order QCD matrix element contributions, $\mathcal{M}_i
\propto \as$}
\label{feyn_2ndorder}
\end{figure}

\begin{figure}
\begin{center}
\emph{(a) q$\bar{\mathit{q}}\mathit{g}$ diagrams with one loop} \vspace{1cm}\\
\scalebox{0.7}{
\begin{fmfgraph*}(5,4)
\fmfpen{thin}
\fmfleft{i}
\fmfrightn{o}{3}
\fmf{photon, tension=1.5}{i,v1}
\fmf{fermion, tension=1}{o1,v7}
\fmf{plain, tension=3}{v1,v2,v3,v4}
\fmf{plain, tension=3}{v1,v5,v6,v7}
\fmf{fermion, tension=1}{v4,o3}
\fmffreeze
\fmf{gluon, tension=1}{v3,o2}
\fmf{gluon, left, tension=1}{v2,v4}
\end{fmfgraph*}}
\hspace{0.3cm}
\scalebox{0.7}{
\begin{fmfgraph*}(5,4)
\fmfpen{thin}
\fmfleft{i}
\fmfrightn{o}{3}
\fmf{photon, tension=1.5}{i,v1}
\fmf{fermion, tension=1}{o1,v7}
\fmf{plain, tension=3}{v5,v1,v2}
\fmf{plain, tension=1.5}{v2,v3}
\fmf{plain, tension=1.5}{v5,v6}
\fmf{plain, tension=3}{v3,v4}
\fmf{plain, tension=3}{v6,v7}
\fmf{fermion, tension=1}{v4,o3}
\fmffreeze
\fmf{gluon, tension=1}{v4,o2}
\fmf{gluon, left, tension=1}{v2,v3}
\end{fmfgraph*}}
\hspace{0.3cm}
\scalebox{0.7}{
\begin{fmfgraph*}(5,4)
\fmfpen{thin}
\fmfleft{i}
\fmfrightn{o}{3}
\fmf{photon, tension=1.5}{i,v1}
\fmf{fermion, tension=1}{o1,v5}
\fmf{plain, tension=1.5}{v4,v1,v2}
\fmf{plain, tension=3}{v2,v3}
\fmf{plain, tension=3}{v4,v5}
\fmf{fermion, tension=1}{v3,o3}
\fmffreeze
\fmf{gluon, tension=1}{v3,o2}
\fmf{gluon, tension=1}{v4,v2}
\end{fmfgraph*}}
\vspace{0.5cm} \\
\scalebox{0.7}{
\begin{fmfgraph*}(5,4)
\fmfpen{thin}
\fmfleft{i}
\fmfrightn{o}{3}
\fmf{photon, tension=1.5}{i,v1}
\fmf{fermion, tension=1}{o1,v7}
\fmf{plain, tension=3}{v1,v2,v3,v4}
\fmf{plain, tension=3}{v1,v5,v6,v7}
\fmf{fermion, tension=1}{v4,o3}
\fmffreeze
\fmf{gluon, tension=1}{o2,v6}
\fmf{gluon, left, tension=1}{v7,v5}
\end{fmfgraph*}}
\hspace{0.3cm}
\scalebox{0.7}{
\begin{fmfgraph*}(5,4)
\fmfpen{thin}
\fmfleft{i}
\fmfrightn{o}{3}
\fmf{photon, tension=1.5}{i,v1}
\fmf{fermion, tension=1}{o1,v7}
\fmf{plain, tension=3}{v5,v1,v2}
\fmf{plain, tension=1.5}{v2,v3}
\fmf{plain, tension=1.5}{v5,v6}
\fmf{plain, tension=3}{v3,v4}
\fmf{plain, tension=3}{v6,v7}
\fmf{fermion, tension=1}{v4,o3}
\fmffreeze
\fmf{gluon, tension=1}{o2,v7}
\fmf{gluon, left, tension=1}{v6,v5}
\end{fmfgraph*}}
\hspace{0.3cm}
\scalebox{0.7}{
\begin{fmfgraph*}(5,4)
\fmfpen{thin}
\fmfleft{i}
\fmfrightn{o}{3}
\fmf{photon, tension=1.5}{i,v1}
\fmf{fermion, tension=1}{o1,v5}
\fmf{plain, tension=1.5}{v4,v1,v2}
\fmf{plain, tension=3}{v2,v3}
\fmf{plain, tension=3}{v4,v5}
\fmf{fermion, tension=1}{v3,o3}
\fmffreeze
\fmf{gluon, tension=1}{o2,v5}
\fmf{gluon, tension=1}{v4,v2}
\end{fmfgraph*}}
\vspace{0.5cm} \\
\scalebox{0.7}{
\begin{fmfgraph*}(5,4)
\fmfpen{thin}
\fmfleft{i}
\fmfrightn{o}{3}
\fmf{photon, tension=1.5}{i,v1}
\fmf{fermion, tension=1}{o1,v5}
\fmf{plain, tension=3}{v4,v1,v2}
\fmf{plain, tension=1.5}{v2,v3}
\fmf{plain, tension=1.5}{v4,v5}
\fmf{fermion, tension=1}{v3,o3}
\fmffreeze
\fmf{gluon, tension=1}{v2,o2}
\fmf{gluon, tension=1, rubout}{v5,v3}
\end{fmfgraph*}}
\hspace{0.3cm}
\scalebox{0.7}{
\begin{fmfgraph*}(5,4)
\fmfpen{thin}
\fmfleft{i}
\fmfrightn{o}{3}
\fmf{photon, tension=1.5}{i,v1}
\fmf{fermion, tension=1}{o1,v5}
\fmf{plain, tension=3}{v4,v1,v2}
\fmf{plain, tension=1.5}{v2,v3}
\fmf{plain, tension=1.5}{v4,v5}
\fmf{fermion, tension=1}{v3,o3}
\fmffreeze
\fmf{gluon, tension=1}{v2,v6,o2}
\fmf{gluon, tension=0}{v3,v6}
\end{fmfgraph*}}
\hspace{0.3cm}
\scalebox{0.7}{
\begin{fmfgraph*}(5,4)
\fmfpen{thin}
\fmfleft{i}
\fmfrightn{o}{3}
\fmf{photon, tension=1.5}{i,v1}
\fmf{fermion, tension=1}{o1,v3}
\fmf{plain, tension=2}{v3,v1,v2}
\fmf{fermion, tension=1}{v2,o3}
\fmffreeze
\fmf{gluon, tension=1}{v2,v4,v3}
\fmf{gluon, tension=1}{v4,o2}
\end{fmfgraph*}}
\vspace{0.5cm} \\
\scalebox{0.7}{
\begin{fmfgraph*}(5,4)
\fmfpen{thin}
\fmfleft{i}
\fmfrightn{o}{3}
\fmf{photon, tension=1.5}{i,v1}
\fmf{fermion, tension=1}{o1,v5}
\fmf{plain, tension=3}{v4,v1,v2}
\fmf{plain, tension=1.5}{v2,v3}
\fmf{plain, tension=1.5}{v4,v5}
\fmf{fermion, tension=1}{v3,o3}
\fmffreeze
\fmf{gluon, tension=1}{o2,v4}
\fmf{gluon, tension=1, rubout}{v5,v3}
\end{fmfgraph*}}
\hspace{0.3cm}
\scalebox{0.7}{
\begin{fmfgraph*}(5,4)
\fmfpen{thin}
\fmfleft{i}
\fmfrightn{o}{3}
\fmf{photon, tension=1.5}{i,v1}
\fmf{fermion, tension=1}{o1,v5}
\fmf{plain, tension=3}{v4,v1,v2}
\fmf{plain, tension=1.5}{v2,v3}
\fmf{plain, tension=1.5}{v4,v5}
\fmf{fermion, tension=1}{v3,o3}
\fmffreeze
\fmf{gluon, tension=1}{o2,v6,v4}
\fmf{gluon, tension=0}{v6,v5}
\end{fmfgraph*}}
\hspace{0.3cm}
\scalebox{0.7}{
\begin{fmfgraph*}(5,4)
\fmfpen{thin}
\fmfleft{i}
\fmfrightn{o}{3}
\fmf{phantom, tension=1.5}{i,v1}
\fmf{phantom, tension=1}{o1,v5}
\fmf{phantom, tension=1.5}{v4,v1,v2}
\fmf{phantom, tension=3}{v2,v3}
\fmf{phantom, tension=3}{v4,v5}
\fmf{phantom, tension=1}{v3,o3}
\fmffreeze
\fmf{phantom, tension=1}{o2,v5}
\fmf{phantom, tension=1}{v2,v4}
\end{fmfgraph*}}
\vspace{1cm} \\
\emph{(b)
$\mathit{q}\bar{\mathit{q}}\mathit{q}\bar{\mathit{q}}\mathit{g}$ and
$\mathit{q}\bar{\mathit{q}}\mathit{ggg}$ tree level diagrams}
\vspace{1cm}\\
\scalebox{0.7}{
\begin{fmfgraph*}(5,4)
\fmfpen{thin}
\fmfleft{i}
\fmfrightn{o}{5}
\fmf{photon, tension=2}{i,v1}
\fmf{fermion, tension=1}{o1,v2}
\fmf{plain, tension=3}{v2,v1}
\fmf{plain, tension=3}{v1,v3}
\fmf{fermion, tension=1}{v3,o5}
\fmffreeze
\fmf{gluon, tension=2}{o2,v2}
\fmf{gluon, tension=2}{v3,v4}
\fmf{phantom, tension=0.5}{v2,v4}
\fmf{fermion}{o3,v4}
\fmf{fermion}{v4,o4}
\end{fmfgraph*}}
\hspace{0.3cm}
\scalebox{0.7}{
\begin{fmfgraph*}(5,4)
\fmfpen{thin}
\fmfleft{i}
\fmfrightn{o}{5}
\fmf{photon, tension=1.5}{i,v1}
\fmf{fermion, tension=1}{o1,v5}
\fmf{plain, tension=2}{v1,v2,v3}
\fmf{plain, tension=2}{v1,v4,v5}
\fmf{fermion, tension=1}{v3,o5}
\fmffreeze
\fmf{gluon}{v2,o3}
\fmf{gluon}{v3,o4}
\fmf{gluon}{o2,v4}
\end{fmfgraph*}}
\hspace{0.3cm}
\raisebox{1.3cm}{$+$ many others \ldots}
\end{center}
\caption{Third order QCD matrix element contributions, $\mathcal{M}_i
\propto \as^{3/2}$}
\label{feyn_3rdorder}
\end{figure}

For any given final state, an infinite number of diagrams exist with
differing numbers of loops. For example the process
$\mathrm{e}^+\mathrm{e}^-\to\mathrm{q}\bar{\mathrm{q}}\mathrm{g}$ has
two tree-level diagrams with $\mathcal{M}_i\propto\as^{1/2}$, shown in
Figure~\ref{feyn_1storder}, and eleven one-loop diagrams with
\mbox{$\mathcal{M}_i\propto\as^{3/2}$}, shown in
Figure~\ref{feyn_3rdorder}(\emph{a}).\footnote{We do not include
diagrams containing loops on the external `legs', as these are already
taken into account by the renormalised particle masses.} For a general
$N$-jet final state, the total matrix element is given by
\begin{equation}
\mathcal{M}_{(\text{i}\to\text{f})}\;=\;
\underbrace{\phantom{0}\sum_{\parbox{0.7cm}{\centering \scriptsize $i$\vspace{-0.1cm}\\(tree)}} \mathcal{M}_i\phantom{0}}_{\displaystyle \propto \as^{(N-2)/2}}
+\underbrace{\phantom{0}\sum_{\parbox{1.0cm}{\centering \scriptsize $i$\vspace{-0.1cm}\\(1~loop)}} \mathcal{M}_i\phantom{0}}_{\displaystyle \propto \as^{N/2}}
+\underbrace{\phantom{0}\sum_{\parbox{1.0cm}{\centering \scriptsize $i$\vspace{-0.1cm}\\(2~loop)}} \mathcal{M}_i\phantom{0}}_{\displaystyle \propto \as^{(N+2)/2}} + \ldots \;\;\;.
\end{equation}
To calculate the transition rate, and hence the differential cross
section $\mathrm{d}\sigma/\mathrm{d}\Omega$, we multiply this total
matrix element by its complex conjugate~$\mathcal{M}_{(\text{i}\to\text{f})}^*$:
\begin{eqnarray}
{\left|\mathcal{M}_{(\text{i}\to\text{f})}\right|}^2 & = &
\underbrace{\phantom{0}
\sum_{\parbox{0.7cm}{\centering \scriptsize $i$\vspace{-0.1cm}\\(tree)}}
\sum_{\parbox{0.7cm}{\centering \scriptsize $j$\vspace{-0.1cm}\\(tree)}}
\mathcal{M}^*_i\mathcal{M}_j\phantom{0}}_{\displaystyle \propto \as^{N-2}}
\;+\;\underbrace{\phantom{0}
\sum_{\parbox{0.7cm}{\centering \scriptsize $i$\vspace{-0.1cm}\\(tree)}}
\sum_{\parbox{1.0cm}{\centering \scriptsize $j$\vspace{-0.1cm}\\(1~loop)}}
\!\!\mathcal{M}^*_i\mathcal{M}_j\phantom{0}}_{\displaystyle \propto \as^{N-1}}
\nonumber \\
& & \rule[1.0cm]{0pt}{0pt}
+\underbrace{\phantom{0}
\sum_{\parbox{0.7cm}{\centering \scriptsize $i$\vspace{-0.1cm}\\(tree)}}
\sum_{\parbox{1.0cm}{\centering \scriptsize $j$\vspace{-0.1cm}\\(2~loop)}}
\!\!\mathcal{M}^*_i\mathcal{M}_j \;+\;
\sum_{\parbox{1.0cm}{\centering \scriptsize $i$\vspace{-0.1cm}\\(1~loop)}}
\sum_{\parbox{1.0cm}{\centering \scriptsize $j$\vspace{-0.1cm}\\(1~loop)}}
\!\!\mathcal{M}^*_i\mathcal{M}_j
\phantom{0}}_{\displaystyle \propto \as^N} + \ldots
\label{matsquared}
\end{eqnarray}
Thus for a full description of multihadronic final states
at~$\mathcal{O}(\as^2)$, one must calculate all two-parton diagrams
with up to two loops, all three-parton diagrams with up to one loop,
and all four-parton diagrams at tree level.\footnote{In the
calculation of event shape distributions at $\mathcal{O}(\as^2)$,
however, we are not directly concerned with two-parton events;
these appear only in the total hadronic cross section used to
normalise the distributions.} Many of the higher-order diagrams
contribute only through interference with those of lower order. The
explicit evaluation of these matrix elements is discussed in
Ref.~\cite{cpar_ellis}.

\section{Event shape observables}
\label{evsh_defs}

In order to make experimental tests of perturbative QCD, and to
measure its free parameter \as, one must first define some physical
observables. These should be as sensitive as possible to the
high-energy perturbative process, and as insensitive as possible to
the subsequent non-perturbative effects of hadronisation and
decays. Theorists have proposed many standard observables, called
event shapes, each of which probes slightly different aspects of the
final state. Event shapes are functions of the \mbox{3-momenta}
\mbox{\boldmath $p$}$_i$, and energies $E_i$ of all particles $i$
detected in the final state, and do not require identification of
particle types.\footnote{In principle, one needs to know the mass of
the particle to determine its energy, given a precise measurement of
the momentum. In our analysis, however, we will assume the masses of
all particles to be that of the pion; any bias introduced by this
assumption will be corrected by Monte Carlo simulations.} It is
therefore straightforward to make comparisons between data and theory,
and between different experiments, without resorting to complicated
and inefficient selection criteria. Many other tests of QCD require,
for example, some discrimination between quark and gluon jets; in
event shape measurements, we do not even need to assign the
final-state particles to jets.

\subsection{Thrust, thrust major, thrust minor and oblateness}
\label{thrust_def}

\begin{description}
\item[Thrust, \boldmath $T$] is defined by
\begin{equation}
T= \max_{\mbox{\boldmath $\scriptstyle
                    \hat{n}$}}\left(\frac{\sum_i|\mbox{\boldmath
                    $p$}_i\cdot \naxis |}
                    {\sum_i|\mbox{\boldmath $p$}_i|}\right)\;\;,
\end{equation}
where the thrust axis \nth~is defined as the unit
\mbox{3-vector} \naxis~which maximises the
expression. For a perfectly `pencil-like' two-jet event, the thrust
axis lies parallel to the jets, so $|\mbox{\boldmath $p$}_i \cdot
\nth |=|\mbox{\boldmath $p$}_i|$, yielding
$T=1$. In the case of a `spherical' event, with an
infinite number of particles distributed isotropically in the final
state, the thrust becomes a ratio of solid angle integrals:
\begin{equation}
T=\frac{\int \, |\cos \theta \, | \; \mathrm{d}\Omega}{\int \mathrm{d}\Omega}\;=\;
\frac{2\pi}{4\pi}\;=\;\frac{1}{2}\;\;\;.
\end{equation}
It can be shown that all events satisfy $\frac{1}{2}<T<1$. All the
other event shapes considered here approach zero in the two-jet limit;
for consistency, we will therefore define the observable $y=1-T$ which
shares this property.

The concept of thrust was already in use before the advent of QCD. In
1964~\cite{thrust1964}, a ``principal axis'' equivalent to \nth~was
proposed for the analysis of jets observed in hadron collisions,
though the origin of the jets was hitherto unexplained. Later, in
1977~\cite{thrust1977}, it was recognised that this ``maximum directed
momentum'' represented a calculable quantity in perturbative QCD.

\item[Thrust major, \boldmath $T_{\mathrm{maj.}}$] is defined in the
  same way as thrust, except that the axis \naxis~is
  constrained to be orthogonal to the thrust axis:
\begin{equation}
T_{\mathrm{maj.}} = \max_{\scriptstyle \mbox{\boldmath $\scriptstyle
\hat{n}$} \bot \mbox{\boldmath $\scriptstyle \hat{n}$}_T}
\left(\frac{\sum_i|\mbox{\boldmath $p$}_i\cdot \naxis |} {\sum_i|\mbox{\boldmath $p$}_i|}\right)\;\;.
\end{equation}
The axis which maximises the quantity in parentheses is \ntmaj.
\item[Thrust minor, \boldmath $T_{\mathrm{min.}}$] is analogous to $T$
  and $T_{\mathrm{maj.}}$ except that \naxis~is
  orthogonal to both \nth~and \ntmaj:
\begin{equation}
T_{\mathrm{min.}} = \frac{\sum_i|\,\mbox{\boldmath
                    $p$}_i\cdot \ntmin |}
                    {\sum_i|\mbox{\boldmath $p$}_i|}
                    \;\;\;,\;\;\;\;\mathrm{where}\;\; \ntmin \, = \,
                    \frac{ \nth \times \ntmaj }
                    {\left| \nth \times \ntmaj \right|}
\end{equation}
\item[Oblateness, \boldmath $O$] is simply the difference between thrust
major and thrust minor:
\begin{equation}
O\;=\;T_\mathrm{maj.}-T_\mathrm{min.}
\end{equation}
\end{description}

The thrust major and thrust minor are both zero for a perfect two-jet
event, since all particle momenta are parallel to the thrust axis, and
hence orthogonal to \ntmaj~and~\ntmin. Furthermore, the thrust minor
is zero for a three-jet event,\footnote{These statements are only
valid for perfectly narrow jets, or partons. In reality, hadronisation
and particle decays introduce some transverse momentum within the
jet.}  since momentum conservation dictates that all particles must
lie in the plane orthogonal to~\ntmin. For a spherical event, both
$T_\mathrm{maj.}$ and~$T_\mathrm{min.}$ approach a maximum value of
$1/2$. The oblateness is unusual, in that it vanishes for both two-jet
events and spherical events.

\begin{figure}
\begin{center}
\psfragscanon
$\phantom{AAAA}$\includegraphics[width=0.7\textwidth]{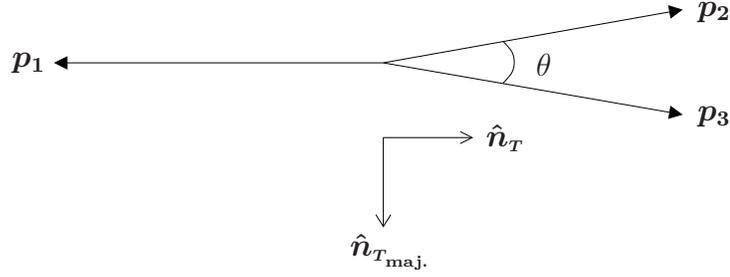}
\psfragscanoff
\end{center}
\caption{A three-jet event approaching the two-jet limit}
\label{threetotwo}
\end{figure}

\begin{figure}
\begin{center}
\psfragscanon
$\phantom{AAAA}$\vspace{0.5cm}\\
$\phantom{AAAA}$\includegraphics[width=0.7\textwidth]{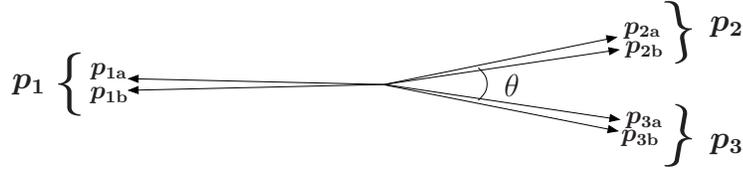}
\psfragscanoff
\end{center}
\caption{Collinear branching of partons within a three-jet event}
\label{threejetsplit}
\end{figure}

\newpage
When a three-jet event approaches the two-jet limit, as depicted in
Figure~\ref{threetotwo}, the four observables have differing
sensitivities to the opening angle~$\theta$:
\begin{center}
\begin{tabular}{r @{\hspace{0.2cm}} c @{\hspace{0.2cm}} l @{\hspace{0.2cm}} c @{\hspace{0.2cm}} l}
$1-T$ & $\sim$ & $1-\cos \theta$ & $\sim$ & $\theta^2$ \\
$T_\mathrm{maj.}$ & $\sim$ & $\sin \theta$ & $\sim$ & $\theta$ \\
$T_\mathrm{min.}$ & = & 0 \\
$O$ & = & $T_\mathrm{maj.}$  & $\sim$ & $\theta$ \\
\end{tabular}
\end{center} \vspace{0.1cm}
Hence we expect, for example, that the thrust major should be
affected to a greater extent than the thrust by the effects of
hadronisation and particle decays.

All four observables are linear with respect to the momenta
$\mbox{\boldmath $p$}_i$. Hence if one particle splits into two, such
that both final state particles continue on the same trajectory as
shown in Figure~\ref{threejetsplit}, the observables do not change. In
perturbative QCD, the matrix element for a process involving soft or
collinear gluon emission becomes infinite in the limit of small
momentum transfers. This leads to ``infrared divergences'' in
quantities which are sensitive to such processes. We can therefore
only calculate distributions for observables such as those considered
above, which are ``infrared safe''.

\subsection[Sphericity, aplanarity and the $C$- and $D$-parameters]
{\boldmath Sphericity, aplanarity and the $C$- and $D$-parameters}
\label{sph_apl_c_d}

The sphericity tensor $S^{\alpha \beta}$ is defined~\cite{sphericity} as
\begin{equation}
S^{\alpha \beta}=\frac{\sum_i p^\alpha_i \, p^\beta_i}{\sum_i
  |\mbox{\boldmath $p$}_i|^2}\;\;\;,
\end{equation}
where $\alpha,\beta=1,2,3$ correspond to the three spatial components.
The three eigenvalues $\lambda_i$ are ordered such that $\lambda_1 <
\lambda_2 < \lambda_3$. Then the \mbox{\boldmath\textbf{sphericity},
$S$} and \mbox{\boldmath\textbf{aplanarity}, $A$} are defined by
\begin{equation}
S=\frac{3}{2}\,(\lambda_1+\lambda_2)\;\;\;\;\;\mathrm{and}\;\;\;\;\;A=\frac{3}{2}\,\lambda_1 \;\;\;.
\end{equation}
Sphericity is a `three-jet' observable, approaching zero in the
two-jet limit, while aplanarity is a `four-jet' observable (like
thrust minor), which vanishes in the three-jet planar limit. Since the
sphericity tensor is quadratic in the momentum components
$p_i^{\alpha,\beta}$, it is not infrared safe; the sphericity will
differ between the events shown in Figures~\ref{threetotwo}
and~\ref{threejetsplit}.\footnote{The aplanarity will also differ, in
general, though it is zero for both of the planar events in
Figures~\ref{threetotwo} and~\ref{threejetsplit}.} While the
sphericity and aplanarity provide an interesting test of the Monte
Carlo models for non-perturbative physics, they cannot be predicted
reliably in perturbation theory.

As a generalisation of $S^{\alpha \beta}$, the power of the momentum
dependence may be modified:
\begin{equation}
S^{(r)\alpha \beta}=\frac{\sum_i {|\mbox{\boldmath $p$}_i|}^{r-2}\,
    p^\alpha_i \, p^\beta_i} {\sum_i {|\mbox{\boldmath
    $p$}_i|}^r}\;\;\;.
\end{equation}
The sphericity tensor corresponds to the case $r=2$. It can be
argued~\cite{pythiamanual} that higher values of $r$ should give better
experimental sensitivity to the parton structure of the event. This is
because transverse momentum fluctuations during fragmentation have a
proportionally smaller effect on particles with large momenta. However,
to avoid infrared divergences in perturbation theory, we must choose
the linearised form $r=1$~\cite{cpar_donoghue}:
\begin{equation}
M^{\alpha \beta}=S^{(1)\alpha \beta}=\frac{\sum_i
p^\alpha_i \, p^\beta_i / |\mbox{\boldmath $p$}_i|}
{\sum_i |\mbox{\boldmath $p$}_i|}\;\;\;.
\end{equation}
Once again, this tensor is conveniently parameterised by its
eigenvalues $\lambda_1$, $\lambda_2$ and $\lambda_3$. Since $M$ is
constructed with unit trace, the eigenvalues are bound by the
constraint \mbox{$\lambda_1+\lambda_2+\lambda_3=1$}. We can thus form
two independent combinations called the \textbf{\boldmath $C$-} and
\textbf{\boldmath $D$-parameters}~\cite{cpar_ellis}, which are
symmetric in the three eigenvalues:
\begin{equation}
C=3\left(\lambda_1\lambda_2 + \lambda_2\lambda_3
+\lambda_3\lambda_1 \right)
\;\;\;\;\textrm{and}\;\;\;\;
D=27\,\lambda_1\lambda_2\lambda_3
\end{equation}
The factors of 3 and 27 ensure that both parameters are in the range
0--1, reaching their maximum value for an isotropic event with
$\lambda_1=\lambda_2=\lambda_3=1/3$. In the case of a planar event,
one of the eigenvalues will be zero; the $D$-parameter is therefore a
four-jet observable, while $C$ is three-jet. It is worth noting that
the $C$-parameter can be expressed explicitly in terms of the particle
momenta $\mbox{\boldmath $p$}_i$, since it is related to the second
Fox-Wolfram moment~\cite{foxwolfram1979} by
\begin{equation}
C=1-H_2\;\;\;,\;\;\;\;\;\textrm{where}\;\;\;
H_2=\sum_{i,j} \frac{|\mbox{\boldmath $p$}_i| \, |\mbox{\boldmath $p$}_j|}
{E_\mathrm{vis.}^2} \; P_2\left(\cos \theta_{ij}\right)\;\;\;.
\end{equation}
Here $E_\mathrm{vis.}$ is the total visible energy of the event,
$\theta_{ij}$ is the opening angle between particles $i$ and $j$,
and $P_2(x) \equiv \frac{1}{2}\left(3x^2-1\right)$ is the second Legendre
polynomial.

\subsection{The heavy and light jet masses}

Unlike the observables discussed so far, the heavy and light jet
masses do not involve a global sum over all particles in the
event. Instead, we split the event into two `hemispheres', divided
by a plane orthogonal to the thrust axis~\nth. Denoting the
hemispheres by $H_1$ and $H_2$, the two corresponding jet masses
$M_{1,2}$ are defined by the invariant masses in each half of the event:
\begin{equation}
M^2_{1,2} = 
{\; \bigg(\sum_{i \in H_{1,2}} E_i \; \bigg)}^2 -\;
{\bigg| \sum_{i \in H_{1,2}} \mbox{\boldmath $p$}_i\; \bigg|}^2 \;\;\;.
\end{equation}
The \textbf{heavy jet mass, \boldmath{\MH}}, and \textbf{light jet
mass, \boldmath{\ML}} are the larger and smaller of these masses
respectively. The dimensionless quantities $\MH/\sqrt{s}$ and
$\ML/\sqrt{s}$ are infrared-safe event shape
observables,\footnote{Often in the literature the factors
$\frac{1}{\sqrt{s}}$ are included in the definitions of \MH~and \ML.}
whose distributions have been predicted by perturbative QCD.
Referring to Figure~\ref{threetotwo}, we see that \ML~vanishes in a
three-particle system, since one of the hemispheres contains only a
single particle.\footnote{In reality the heavy and light jet masses
are never zero, due to the finite masses of the individual
particles. This effect becomes increasingly less significant in high
energy collisions, however, and will be neglected in the theoretical
predictions used for our analysis.} Like all four-jet observables,
however, it will acquire a small positive value due to
hadronisation. The heavy jet mass \MH~is three-jet observable,
sensitive to the momentum transfer of single hard gluon emissions.

\subsection{The wide, narrow and total jet broadenings}

Again the event is divided into two hemispheres $H_1$ and~$H_2$ by a
plane orthogonal to the thrust axis. In each half of the event, we
define the jet broadening~\cite{broadenings_catani}
\begin{equation}
B_{1,2} = \frac{\sum_{i \in H_{1,2}} \left| \mbox{\boldmath $p$}_i
\times \nth \right|}{2\sum_i \left| \mbox{\boldmath $p$}_i \right|}\;\;\;.
\end{equation}
In terms of $B_1$ and $B_2$, the wide, narrow and total broadenings
are given by:
\begin{eqnarray}
\textbf{Wide jet broadening} & & \BW=\max\left(B_1,B_2\right)\nonumber \\
\textbf{Narrow jet broadening} & & \BN=\min\left(B_1,B_2\right)\nonumber \\
\textbf{Total jet broadening} & & \BT=\BW+\BN=B_1+B_2
\end{eqnarray}
The total broadening \BT~is similar to the thrust major,
in that it measures momentum components orthogonal to the thrust axis
of a three-jet event. In fact to $\mathcal{O}(\as)$ in perturbation
theory, $\BT=\BW=\frac{1}{2}T_\mathrm{maj.}=\frac{1}{2}O$. The
broadenings all vanish in the two-jet limit, while for a spherical
event $\BT\!=\!\pi/8$ and \mbox{$\BW\!=\!\BN\!=\!\pi/16$.} The
theoretical maxima, which exceed the values for the spherical case,
are unknown.  The total and wide jet broadenings are three-jet
observables, while \BN~is four-jet.

\subsection[The transition parameter, $y_{23}$, between two and~three jets 
in the Durham algorithm]
{\setstretch{1}The transition parameter, \boldmath $y_{23}$, between two 
and~three jets in the Durham algorithm}
\label{durham_algorithm}

Various algorithms exist to determine the number of jets in an
hadronic final state. Generally the aim is to group particles together
such that we reconstruct the directions and momenta of the partons
produced in the hard interaction. However, information about this
partonic state is inevitably lost during hadronisation and decays. The
jet-finding algorithms typically include at least one free ``resolution
parameter,'' and the number of reconstructed jets will depend on its
chosen value.

In the Durham algorithm\footnote{The Durham algorithm (also known as
the $k_{\scriptscriptstyle \perp\!}$-algorithm) derives its name from
the Durham Workshop on Jet Studies at LEP and HERA, December~1990,
where it was first discussed.}~\cite{durham_catani}, a ``scaled
transverse momentum'' $y_{ij}$ is defined for every pair of particles
$i,j$ in the final state:
\begin{equation}
y_{ij}=\frac{2\,
\min\!\left(E_i^2,E_j^2\right)
\left(1-\cos \theta_{ij}\right)}
{E_\mathrm{vis.}^2}\;\;\;,
\label{y_ij_def}
\end{equation}
where $E_{i,j}$ are the energies of the particles, $\theta_{ij}$ is
the angle between them, and $E_\mathrm{vis.}$ is the total visible
energy in the event. The pair with the smallest $y_{ij}$ is then
replaced by a single `pseudoparticle' with momentum $\mbox{\boldmath
$p$}_{(ij)}=\mbox{\boldmath $p$}_i+\mbox{\boldmath $p$}_j$ and energy
$E_{(ij)}=E_1+E_2$.\footnote{The algorithm has been constructed such
that the combined pseudoparticle has a 4-momentum equal to the sum of
its constituent particles' 4-momenta. This choice is not unique. One
could instead define the energy $E_{(ij)}=|\mbox{\boldmath
$p$}_i+\mbox{\boldmath $p$}_j|$, such that the resulting jets are
massless. Alternatively, the momentum could be scaled such that
$|\mbox{\boldmath $p$}_{(ij)}|=E_i+E_j$. These three ``recombination
schemes'' are known as the E-, P- and E0-schemes respectively; they
have been studied experimentally~\cite{pr054} by OPAL using the
JADE algorithm~\cite{jade}, which defines $y_{ij}=2E_iE_j(1-\cos
\theta_{ij})$ in place of Equation~(\ref{y_ij_def}). Here we will only
consider the E-scheme, as described above. Often the Durham algorithm
itself is confusingly referred to as the D-scheme, when in fact the
same choice of recombination schemes exists in both the Durham and
JADE algorithms.} The process is repeated until all pairs of particles
have $y_{ij}>y^\mathrm{cut}$, for some fixed resolution parameter
$y^\mathrm{cut}$. The remaining pseudoparticles are then regarded as
jets.

For small values of $y^\mathrm{cut}$ the algorithm will terminate at
an early stage, yielding a large number of jets, whereas for values
approaching unity the whole event will be combined into a single jet.
As a measure of how `$N$-jetlike' an event is, we define the
transition parameter $y^\mathrm{cut}_{\scriptscriptstyle N-1,N}$ as
the highest the value of $y^\mathrm{cut}$ for which the event is
resolved into $N$~jets. We consider here the case
$y^\mathrm{cut}_{\scriptscriptstyle 2,3}$, at which the number of
resolved jets changes from two to three. For events which genuinely
contain three or more jets, we expect
$y^\mathrm{cut}_{\scriptscriptstyle 2,3}$ to be large, while for
two-jet events at LEP collision energies
$y^\mathrm{cut}_{\scriptscriptstyle 2,3} \lesssim 10^{-3}$. If an
event is resolved into three identical jets $120^\circ$ apart,
$y^\mathrm{cut}_{\scriptscriptstyle 2,3}$~takes its maximum possible
value of 1/3.

Hereafter we will follow the convention of the
literature\footnote{Some authors use the symbols $D_2$ or $y_3$ for
this same observable.} by denoting this observable \ytwothree; this
should not be confused with our earlier notation $y_{ij}$, which
refers to an individual pair of particles or pseudoparticles within
the event.

\section{Perturbative predictions for the event shapes}
\label{pert_predictions}

In an experiment, one measures the value of each event shape
observable for every event which has been selected as
`multihadronic'.\footnote{Our precise signal definition and
selection criteria will be discussed in Chapter~\ref{opalchapter}.}
Being a quantum theory, QCD makes physical predictions for these event
shapes in the form of frequentist probabilities. We therefore seek to
measure the probability density function for each observable, and
compare it directly with the best available prediction.

For a generic observable $y$, we can statistically estimate the form of the
differential cross section~$\mathrm{d}\sigma/\mathrm{d}y$. This tells
us the expected density of events, per unit luminosity, at a given
value of the event shape~$y$. Dividing this by the total cross section
for multihadron production,~$\sigma_\mathrm{tot.}$, gives the corresponding
probability density function. In theoretical work, authors more often
refer to the \emph{cumulative} probability function~$R(y)$, defined by
\begin{equation}
R(y_0) = \frac{1}{\sigma_\mathrm{tot.}} \int_0^{y_0} \frac{\mathrm{d}\sigma}{\mathrm{d}y} \;\mathrm{d}y \;\;\;.
\end{equation}
The probability density may then be written as the derivative of this
function,~$R'(y)$. The total multihadronic cross
section~$\sigma_\mathrm{tot.}$ is largely determined by electroweak
physics; however it differs slightly from the cross section for the
`bare' process \mbox{$\mathrm{e}^+\mathrm{e}^-\!\to
{(\mathrm{Z}^0/\gamma)}^*\!\to \mathrm{q}\bar{\mathrm{q}}$}, due to QCD
corrections. These have been evaluated as
follows:~\cite{sigmatot_gorishnii,sigmatot_surguladze}
\begin{equation}
\sigma_\mathrm{tot.}\;=\;\sigma_0\left[\;1\;+\;\frac{\as}{\pi}\;+\;1.411{\left(\frac{\as}{\pi}\right)}^2-\;12.8{\left(\frac{\as}{\pi}\right)}^3
+\;\ldots\;\right]\;\;\;,
\end{equation}
where $\sigma_0$ is the Born cross section for the bare electroweak
process.

Before proceeding further, we will list the six event shapes whose
perturbative predictions will be studied. For various reasons, the
variables~$y$ used in theoretical expressions are not always
identical to the event shape itself:
\begin{eqnarray}
\textrm{Thrust:}                & & y=1-T               \nonumber \\
\textrm{Heavy jet mass:}        & & y=M^2_\mathrm{H}/s  \nonumber \\
\textrm{$C$-parameter:}         & & y=C/6               \nonumber \\
\textrm{Total jet broadening:}  & & y=\BT               \nonumber \\
\textrm{Wide jet broadening:}   & & y=\BW               \nonumber \\
\textrm{Durham \ytwothree:}     & & y=\ytwothree
\label{ydef}
\end{eqnarray}

\subsection[$\mathcal{O}(\as^2)$ predictions]
{\boldmath $\mathcal{O}(\as^2)$ predictions}
\label{fixedorder}
In Section~\ref{qcdpert}, we discussed the calculation of cross
sections for two-, three- and four-parton final states at
$\mathcal{O}(\as^2)$ in perturbative QCD.  As we saw in
Equation~(\ref{matsquared}), the squared matrix element
${\left|\mathcal{M}\right|}^2$ for any final state can be expressed as
a power series in~\as:
\begin{equation}
{\left|\mathcal{M}\right|}^2 \; = \;
{\left[{\left|\mathcal{M}\right|}^2\right]}_0 \; + \; \as
{\left[{\left|\mathcal{M}\right|}^2\right]}_1 \; + \; \as^2
{\left[{\left|\mathcal{M}\right|}^2\right]}_2 \;\;\;.
\end{equation}
A Monte Carlo program called EVENT2~\cite{event2} has been
written, which generates parton-level `events' using the
probabilities associated with the squared matrix elements
${\left[{\left|\mathcal{M}\right|}^2\right]}_i$ at each order. By
generating large samples covering the whole of phase space, one can
then infer the cumulative probability function for any event shape~$y$:
\begin{equation}
R(y)\; = \; 1 \; + \; \mathcal{A}(y)\as\, +\, \mathcal{B}(y)\as^2 \;\;\;.
\end{equation}
In the limit of small \as, the effects of QCD will be
``switched~off,'' and all events will be two-jet. We have assumed
$y\to 0$ for such events, so that the cumulative distribution becomes
a step function $R(y)=\Theta(y)$, and the corresponding probability
density becomes a delta function $R'(y)=\delta(y)$. This is true of
all the variables~$y$ listed in Equations~(\ref{ydef}).

In this work, we use numerical values of $\mathcal{A}(y)$ and
$\mathcal{B}(y)$ that were generated for other recent OPAL analyses.

\subsection{NLLA resummations}
\label{nlla}

A perturbation expansion up to $\mathcal{O}(\as^2)$ will only be
useful if it converges rapidly. One would na\"{\i}vely expect the
convergence of $R(y)$ to be satisfactory at LEP energy scales, since
$\as\sim 0.1$. However, this will only be true if the coefficients
$\mathcal{A}(y)$, $\mathcal{B}(y)$, $\mathcal{C}(y)$,~\ldots~do not
grow faster than $\as^{-n}$ ($n=1,2,3,\ldots$)

\begin{figure}[h!]
\begin{center}
\scalebox{0.8}{
\begin{fmfframe}(1,0.5)(1,0){
\begin{fmfgraph*}(10,6)
\fmfpen{thin}
\fmfleftn{i}{11}
\fmfrightn{o}{11}
\fmftopn{t}{5}
\fmfbottomn{b}{5}
\fmf{plain}{l5,l4,l3,l2,l1,c,r1,r2,r3,r4,r5}
\fmf{fermion, tension=2}{l5,i6}
\fmf{fermion, tension=2}{r5,o6}
\fmf{plain}{t2,x}
\fmf{plain}{b4,y}
\fmf{fermion, tension=4}{x,c}
\fmf{fermion, tension=4}{y,c}
\fmffreeze
\fmf{gluon}{i5,l4}
\fmf{gluon}{l3,i8}
\fmf{gluon}{v1,l1}
\fmf{phantom}{i4,v1}
\fmf{gluon}{v2,r2}
\fmf{phantom}{v2,o9}
\fmf{gluon}{o7,r3}
\fmfdot{c}
\fmflabel{\textbf{q}}{i6}
\fmflabel{\boldmath$\bar{\mathbf{q}}$}{o6}
\fmflabel{\boldmath$\mathbf{e}^+$}{t2}
\fmflabel{\boldmath$\mathbf{e}^-$}{b4}
\end{fmfgraph*}}
\end{fmfframe}}
\end{center}
\caption{Multiple low-angle gluon emission near the two jet limit}
\label{lowanglegluons}
\end{figure}
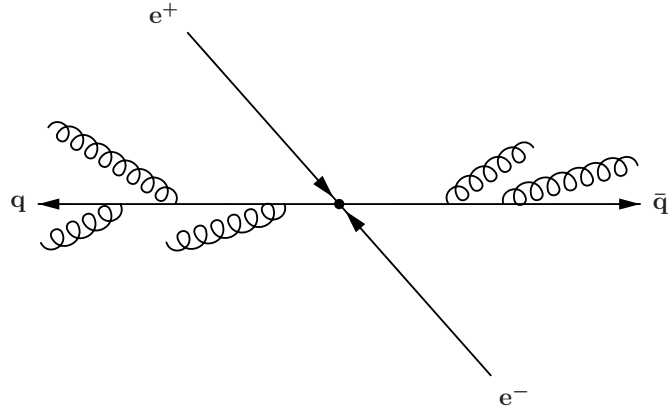

As we have seen, there is a high probability of soft or collinear
gluons being emitted in a quark or gluon jet. To avoid singularities
in the distributions, we defined a set of infrared-safe observables,
which are invariant under any perfectly collinear splitting. However,
if gluons are emitted at a small non-zero angle, there will still be a
corresponding small change in the observables $y$. Since the number of
gluons may be quite large, as depicted in Figure~\ref{lowanglegluons},
this effect will enhance the coefficients of high powers of $\as$ in
the distribution $R(y)$. We therefore expect the convergence of the
series to break down near the two-jet limit, when $y$ is small. This
divergence is manifested in large factors ${\left(-\ln y\right)}^n$,
which appear in the coefficients $\mathcal{A}(y)$, $\mathcal{B}(y)$,
$\mathcal{C}(y)$, \ldots with various powers $n$.

Fortunately help is at hand, as an alternative formalism exists for
dealing with multiple soft and collinear emissions. Rather than
computing matrix elements for individual diagrams such as
Figure~\ref{lowanglegluons}, a system of differential equations is
devised to describe the sequence of emissions from a parton as it
evolves from the hard interaction down to lower mass scales. These
DGLAP\footnote{Dokshitzer-Gribov-Lipatov-Altarelli-Parisi}
equations~\cite{dglap} are based on a set of ``splitting kernels''
$P_{a\to bc}(\as,z)$, which define the probability of a parton $a$
splitting to produce partons $b$ and $c$ with momentum fractions $z$
and $(1-z)$ respectively. One can then derive a Sudakov form factor
$\Delta(t,t_0)$, describing the probability for a parton to evolve
from a mass scale $Q_0$ to a lower scale $Q$ with no detectable
emissions:\footnote{Since there is an infinite probability of emitting
arbitrarily soft or collinear gluons, an ``infrared cut-off'' must be
applied when defining an emission.}
\begin{equation}
\mathcal{P}_\mathrm{no~emission}=\Delta(t,t_0)=\exp\left[-\int_{t_0}^t \mathrm{d}t' \, \sum_{b,c}\int \mathrm{d}z \, \frac{\as}{2\pi} P_{a \to bc}(\as,z)\right]\;\;\;,
\end{equation}
where $t \equiv \ln(Q^2/\Lambda^2)$ for some arbitrary scale~$\Lambda$,
and the sum $\sum_{b,c}$ runs over all possible branchings for a given
parton $a$. The running coupling \as~is evaluated at a scale which
depends on~$t'$. This Sudakov form factor is analogous to the
well-known exponential probability distribution for an unstable
particle to exist for a time $t$; in this case, however, the ``decay
rate'' is dependent on the scale parameter $t$.

Using these form factors, theorists have systematically collected
together the terms containing factors~${(-\ln y)}^n$ in the event shape
distributions $R(y)$. The result is a new series expansion, in which
each term is characterised by the power of \as~minus the power
of~$L\equiv-\ln y$:
\begin{displaymath}
R(y)\;=\;C(\as)\Sigma(\as,y)\;+\;D(\as,y) \;\;\;,
\end{displaymath}
\begin{equation}
\mathrm{where}\;\;\;\ln\Sigma\;=\;Lg_1(\as L) \,+\, g_2(\as L) \,+\,
\as g_3(\as L) \,+\, \ldots \;\;\;.
\label{nllaeqn}
\end{equation}
The functions $g_1(x)$, $g_2(x)$, \ldots~are properties of the event
shape~$y$, which ``resum'' the large logarithms~$L$ to all orders
in~$\as$. The term $Lg_1(\as L)$ resums the leading logarithms~(LL),
while $g_2(\as L)$ resums next-to-leading logarithms~(NLL). Together,
they give a prediction for $R(y)$ in the ``Next-to-Leading Logarithmic
Approximation''~(NLLA). Although the function $D(\as,y)$ also contains
some dependence on the event shape $y$, it is constrained to vanish at
every order in \as~as $y\to 0$.

Expressions for $g_1(x)$ and $g_2(x)$ have been calculated for the
thrust and heavy jet mass~\cite{resum_catani}, the wide and total jet
broadenings~\cite{broadenings_catani,newbroadenings}, the
\mbox{$C$-parameter \cite{cpar_resum}} and the Durham
\ytwothree~parameter~\cite{durham_catani,durham_dissertori,numres}.
Predictions for the function $C(\as)$ are also available, in the form
\begin{equation}
C(\as)\;=\;1\;+\;C_1\as\;+\;C_2\as^2 \;\;\;,
\end{equation}
where the coefficients $C_1$ and $C_2$ are constants for each
observable. The next-to-next-to-leading resummations $g_3(x)$ are
not yet available.

Quarks are assumed to be massless in both the $\mathcal{O}(\as^2)$ and
NLLA predictions used in this work. Although perturbative predictions
do exist for heavy quark production at
$\mathcal{O}(\as^2)$~\cite{quarkmass_oasq}, there are no corresponding
NLLA calculations.\footnote{Recently an NLLA prediction has become
available for jet rate observables, including the Durham \ytwothree\
parameter, in the presence of quark masses~\cite{quarkmass_resum}.
However, it does not have the same form as the massless calculations,
and has not yet been combined with the $\mathcal{O}(\as^2)$
prediction.} In the interests of consistency, we will therefore use only
the massless predictions.

\subsection[Combining $\mathcal{O}(\as^2)$ and NLLA predictions]
{Combining \boldmath $\mathcal{O}(\as^2)$ and NLLA predictions}

So far, we have described two separate predictions for each event
shape distribution: an order-by-order perturbation expansion in the
strong coupling, valid in the 3- and 4-jet regime, and a resummation
of large logarithms, valid in the limit $y\to 0$. In a study of
experimental data, however, we need a prediction covering the complete
spectrum of multihadronic events. In this section we discuss the
methods used to combine the $\mathcal{O}(\as^2)$ and NLLA predictions.

\subsubsection[Log$(R)$ matching scheme]
{\boldmath Log$(R)$ matching scheme}
\label{logrscheme}

We begin by writing the logarithm of the NLLA distribution,
$R_{\scriptscriptstyle \mathrm{NLLA}}(\as,y)$, as a two-variable power
series in \as~and $L\;(\equiv -\ln y)$:\footnote{The ``remainder
function'' $D(\as,y)$ is not included explicitly here. However, the
matching of $\mathcal{O}(\as^2)$ and NLLA calculations can be viewed
as a way of estimating it. Similarly the term $\ln C(\as)$ is dropped,
since it is formally of the same order as the next term $\as g_3(\as
L)$ in the resummation.}
\begin{eqnarray}
\ln R_{\scriptscriptstyle \mathrm{NLLA}}(\as,y) & = &
\ln\Sigma(\as,y)\;=\;Lg_1(\as L) \,+\, g_2(\as L)
\nonumber \\ & = & \ln \sum_{n=1}^\infty \left(G_{n,n+1} \as^n
L^{n+1}\,+\,G_{n,n}\as^n L^n\right)\;\;.
\end{eqnarray}
The coefficients $G_{nm}$ are easily calculated from series expansions
of the functions $g_n(x)$. The above expression can be compared with the
corresponding logarithm of the $\mathcal{O}(\as^2)$ prediction,
$R_{\mathcal{O}(\as^2)}(\as,y)$:
\begin{eqnarray}
\ln R_{\mathcal{O}(\as^2)}(\as,y) &=& \ln \left[\,1\;+\;\mathcal{A}(y)\as\;+\;\mathcal{B}(y)\as^2\;+\;\mathcal{O}(\as^3)\,\right] \nonumber \\
&=& \mathcal{A}(y)\as\;+\;\left[\mathcal{B}(y)-{\textstyle \frac{1}{2}}\mathcal{A}(y)\right]\as^2\;+\;\mathcal{O}(\as^3) \nonumber \\
&\equiv & \widehat{\mathcal{A}}(y)\as\;+\;\widehat{\mathcal{B}}(y)\as^2+\;\mathcal{O}(\as^3)\;\;\;.
\end{eqnarray}

\begin{table}
\begin{center}
\begin{tabular}{r|cc|cc}
\bigstrut[b]
 & $\widehat{\mathcal{A}}(y)\as$ & $\widehat{\mathcal{B}}(y)\as^2$ & $\widehat{\mathcal{C}}(y)\as^3$ & \ldots \\[0.2cm]
\hline
\raisebox{-0.05cm}{$L\,g_1(\as L)$}    & \raisebox{-0.05cm}{\color{red}$G_{12} \as L^2$} &  \raisebox{-0.05cm}{\color{red}$G_{23} \as^2 L^3$} & \raisebox{-0.05cm}{$G_{34} \as^3 L^4$} & \raisebox{-0.05cm}{\ldots} \bigstrut[t]\\[0.2cm]
$g_2(\as L)$     &  \color{red}$G_{11} \as L  $ &  \color{red}$G_{22} \as^2 L^2$ & $G_{33} \as^3 L^3$ & \ldots \bigstrut[b]\\[0.2cm]
\hline
\raisebox{-0.05cm}{$\as \,g_3(\as L)$} & \multirow{3}{*}{\framebox{\rule[-1cm]{0pt}{2.02cm}\raisebox{-0.05cm}{$f_1(y)\,\as$}}}
                                    & \raisebox{-0.05cm}{$G_{21} \as^2 L  $} & \raisebox{-0.05cm}{$G_{32} \as^3 L^2$} & \raisebox{-0.05cm}{\ldots} \bigstrut[t]\\[0.2cm]
$\as^2 \,g_4(\as L)$ &                & \multirow{2}{*}{\framebox{\rule[-0.6cm]{0pt}{1.3cm}$f_2(y)\,\as^2$}} 
                                                         & $G_{31} \as^3 L $ & \ldots \\[0.2cm]
\multicolumn{1}{c|}{\vdots} & & & \framebox{\rule[-0.15cm]{0pt}{0.58cm}$f_3(y)\,\as^3$} & $\ddots$
\end{tabular}
\end{center}
\caption{An expansion of $\ln R$ in powers of \as\ and $L$. The sum of
the first two rows corresponds to the NLLA resummation, while the
first two columns together represent the $\mathcal{O}(\as^2)$
perturbative prediction. The functions $f_1(y)$, $f_2(y)$,\ldots\ have
been introduced merely for consistency, to represent non-logarithmic
contributions to the distribution.}
\label{matchingtable}
\end{table}

In Table~\ref{matchingtable}, we have arranged the terms $G_{nm} \as^n
L^m$ in an array, such that sum of the first two rows is equivalent to
$\ln \left[R_{\scriptscriptstyle \mathrm{NLLA}}(\as,y)\right]$, while
the first two columns give~$\ln
[R_{\mathcal{O}(\as^2)}(\as,y)]$.  To combine the
$\mathcal{O}(\as^2)$ and NLLA predictions, we add them together and
subtract the four double-counted terms indicated in red:
\begin{eqnarray}
\ln R_\mathrm{matched}(\as,y) &=& Lg_1(\as L)\,+\,g_2(\as L)\,+\,\mathcal{A}(y)\as\,+\,
\left[\mathcal{B}(y)-{\textstyle\frac{1}{2}}\mathcal{A}(y)\right]\as^2 \nonumber \\
& & -\, \left(G_{12}L^2+G_{11}L\right)\as \,-\, \left(G_{23}L^3+G_{22}L^2\right)\as^2 \;\;\;.
\end{eqnarray}
After exponentiating the entire expression, this gives the
$\mathcal{O}(\as^2)+\mathrm{NLLA}$ prediction for $R(y)$ in the
Log$(R)$ matching scheme.

\subsubsection[$R$ matching scheme]
{\boldmath $R$ matching scheme}
\label{rscheme}

\enlargethispage{\baselineskip}In the derivation above, we expanded the expressions for
$\ln\left[R(\as,y)\right]$ in powers of \as~and $L$, and identified
the four terms which were common to both expressions. Unfortunately,
however, this procedure has an ambiguity. We could instead have
expanded the two expressions for $R(\as,y)$ itself, or indeed any
other function of $R$; in each case, the double-counted terms would be
slightly different.

In the $R$ matching scheme, we have
\begin{eqnarray}
R_\mathrm{matched}(\as,y) & = & R_{\scriptscriptstyle \mathrm{NLLA}}(\as,y)\;+\;
R_{\mathcal{O}(\as^2)}(\as,y) \nonumber \\
&& -\;\;\Big[\as~\mathrm{and}~\as^2~
\mathrm{terms~in}~R_{\scriptscriptstyle \mathrm{NLLA}}(\as,y)\Big] \;\;\;.
\end{eqnarray}
Expanding $R_{\scriptscriptstyle \mathrm{NLLA}}(\as,y)$ in powers of
\as~eventually leads to:
\begin{eqnarray}
R_\mathrm{matched}(\as,y) & = & \Big[\,1\,+\,C_1
\as\,+\,C_2\as^2\,\Big] \, \exp\Big(\,Lg_1(\as L)\,+\,g_2(\as
L)\,+\,G_{21}\as^2 L\,\Big) \nonumber \\ & & +\;\;
\Big[\mathcal{A}(y)-C_1-G_{11}L-G_{12}L^2\Big]\,\as \nonumber
\\ & & +\;\;
\Big[\mathcal{B}(y)-C_2-C_1(G_{11}L+G_{12}L^2)-{\textstyle
\frac{1}{2}}(G_{11}L+G_{12}L^2)^2 \nonumber \\
& & \phantom{+}\;\; \phantom{\Big[}-G_{21}L-G_{22}L^2-G_{23}L^3\Big]\,\as^2\;\;\;.
\label{rmatching}
\end{eqnarray}
This expression explicitly includes the coefficients $C_1$ and $C_2$,
and also $G_{21}$ which formally belongs to the sub-leading
function~$g_3(\as L)$ of the resummation. The coefficients $C_2$ and
$G_{21}$ cannot be derived analytically, and have been estimated
numerically by comparing the NLLA and $\mathcal{O}(\as^2)$
expressions. Theorists generally advocate the Log$(R)$ matching scheme
in preference to the $R$~scheme, because the former does not depend on
these numerical coefficients; however, it should be emphasised that
both schemes are formally valid at both $\mathcal{O}(\as^2)$ and NLLA
accuracy. We will use the difference between the two schemes as part
of our estimate for the theoretical uncertainty in~$R(y)$.

\subsection{Kinematic constraints}
\label{kinconstraints}

\enlargethispage{\baselineskip}Each of the event shapes $y$ is constrained on kinematic grounds to
lie within a certain range; the thrust, for example, must satisfy
$\frac{1}{2} \le T \le 1$. Outside this range, the perturbative
expansion for $\mathrm{d}R/\mathrm{d}y$ must vanish at every order in
\as. In the NLLA prediction, however, there is nothing to prevent the
missing sub-leading terms from contributing a finite cross section
outside the allowed range. To enforce the kinematic constraint
$y \le y_\mathrm{max}$, the logarithm~$L=-\ln y$ can be replaced
with
\begin{equation}
\widetilde{L}\;=\;\frac{1}{p}\ln\left[\,\frac{1}{y^p}\,-\,\frac{1}{y_\mathrm{max}^p}\,+\,1\,\right]
\;\;\;.
\end{equation}
This substitution does not alter the formal NLLA accuracy of the
predictions, and is valid for any $p\geq 1$. In the Log$(R)$ matching
scheme, we then have
\begin{equation}
\ln R(y_\mathrm{max})=0\;\;, \hspace{0.5cm}\mathrm{and}\hspace{0.5cm}
\left.\frac{\mathrm{d}R}{\mathrm{d}y}\right|_{y=y_\mathrm{max}}\!\!\!\! \equiv
\,R\,\left.\frac{\mathrm{d}L}{\mathrm{d}y}\,\frac{\mathrm{d}(\ln R)}{\mathrm{d}L}\right|_{y=y_\mathrm{max}}\!\!\!\! = 0\;\;\;,
\end{equation}
so the cumulative distribution $R(y)$ smoothly approaches unity at the
kinematic boundary $y_\mathrm{max}$. The parameter~$p$ determines the
`sharpness' with which the constraint is applied. For large values
of~$p$, the distribution is modified drastically in the region
immediately below the cutoff, and almost unchanged elsewhere. In the
past, however, all experimental analyses have implicitly taken $p=1$,
which gives a more smooth modification.\footnote{Some analyses,
including previous OPAL publications, have omitted the kinematic
constraint altogether. This corresponds to the limit $p\to\infty$, in
the region~$y<y_\mathrm{max}$.} In this work we will continue to use
$p=1$, but will consider this ambiguity in our estimate of the
theoretical uncertainty.

In the $R$~matching scheme, a similar approach can be taken: the same
substitution $L\to \widetilde{L}$ is valid, but the condition
$\mathrm{d}R/\mathrm{d}y\to 0$ is not automatically satisfied at
$y=y_\mathrm{max}$. Starting from Equation~(\ref{rmatching}), and
expanding $R_\mathrm{matched}(y)$ in powers of the logarithm~$L$,
one eventually obtains
\begin{equation}
R(y) \; = \; 1\;+\; \Big[\,C_2 \as^3 \,G_{11}\, +\, \left(C_1 \as^3 + C_2
\as^4\right) \,G_{21}\,\Big]\,L\;+\;\mathcal{O}(L^2)\;\;\;.
\end{equation}
After substituting $L\to \widetilde{L}$, we immediately have
$R(y_\mathrm{max})=1$, but we must also impose the smoothness
criterion $\mathrm{d}R/\mathrm{d}\widetilde{L}=0$ at the
boundary. This is achieved by modifying the the coefficients $G_{11}$
and $G_{21}$, and also the function $g_2(\as \widetilde{L})$ which
implicitly contains the term $G_{11}\as\widetilde{L}$:
\begin{eqnarray}
G_{11} & \to & \widetilde{G}_{11}(y) \; = \; \left[\,1-\left(\frac{y}{y_\mathrm{max}}\right)^p\,\right]\,G_{11} \nonumber \\
G_{21} & \to & \widetilde{G}_{21}(y) \; = \; \left[\,1-\left(\frac{y}{y_\mathrm{max}}\right)^p\,\right]\,G_{21} \nonumber \\
g_2(\as L) & \to & \tilde{g}_2(\as \widetilde{L}) \; = \; g_2(\as \widetilde{L}) \; - \;\left(\frac{y}{y_\mathrm{max}}\right)^pG_{11}\as \widetilde{L} \;\;\;.
\end{eqnarray}

Values for the kinematic boundaries~$y_\text{max}$, and also for the
matching coefficients $G_{nm}$ and $C_n$, can be found in
Ref.~\cite{uncertaintyband}.

\section{Recent advances in the NLLA predictions}
\label{nlla_advances}

Over the time since the first NLLA resummations were published,
several improvements and corrections have followed:

\begin{itemize}
\item In Ref.~\cite{durham_dissertori}, a more complete NLLA
resummation was found for \ytwothree, replacing the earlier
predictions in Ref.~\cite{durham_catani}. These updated calculations
were applied to LEP2 data in recent OPAL
publications~\cite{OPAL_as_189}, but have now been superseded by a
further improvement (see~below).
\item In Ref.~\cite{newbroadenings}, a problem was identified in the
resummations of the total and wide jet broadening distributions,
originally published in Ref.~\cite{broadenings_catani}. When the
recoil effects in quark-gluon splittings were correctly taken into
account, the distribution became `shifted' slightly:
$R_\mathrm{new}(B)=R_\mathrm{old}(\lambda B)$. The shift has not yet
been implemented in OPAL publications, but in this work the new
calculations are used throughout.
\item In Ref.~\cite{numres}, a new approach to resummation
was developed; this led to a further improvement in the
\ytwothree~distribution, and also some new NLLA predictions for
observables which had not been resummed previously.
Given the probability density $R'_\mathrm{s}(y_\mathrm{s})$ for a
`simple' observable~$y_s$, one can write down the distribution
$R'(y)$ for any other observable in terms of the conditional
probability density $P(y|y_s)$:
\begin{equation}
R'(y)\;=\;\int\,R'_\mathrm{s}(y_\mathrm{s})\,P(y|y_s)\;\mathrm{d}y_s\;\;\;.
\end{equation}
Provided the observables $y$ and $y_\mathrm{s}$ are suitably similar,
it has been shown that the conditional probability $P(y|y_s)$ can be
calculated quite straightforwardly. This leads to a relationship
between the two NLLA resummed distributions, of the form
\begin{equation}
R'(y)\;=\;R'_\mathrm{s}(y)\;\times\;\mathcal{F}\left(-\frac{\mathrm{d}(\ln
\Sigma)}{\mathrm{d}L}\right) \;\;\;,
\end{equation}
where $\Sigma(\as,L)$ is the exponentiated part of the NLLA
expression, defined in Equation~(\ref{nllaeqn}). The numerical values
of the functions~$\mathcal{F(\lambda)}$ have been tabulated for three
observables $y$, given in each case a corresponding observable
$y_\mathrm{s}$ whose resummation is already known:
  \begin{itemize}
    \item The Durham $\ytwothree$ parameter
    \item The thrust major, $T_\mathrm{maj.}$
    \item The oblateness, $O$
  \end{itemize}
It was shown that certain next-to-leading logarithmic contributions
had been omitted from the previous \ytwothree~calculations in
Ref.~\cite{durham_dissertori}, and are now included in the new
`semi-numerical' result. In this work, we have implemented the
improved calculations for the first time in the analysis of OPAL data.

The NLLA predictions for the thrust major and oblateness are
completely new, and have not yet been compared with OPAL measurements.

\item Resummed distributions have recently become available for some
`four-jet' observables, which provide sensitivity to the three-gluon
vertex. In Ref.~\cite{fourjetresum}, a full NLLA calculation was
presented for the light hemisphere mass~\ML, and the narrow jet
broadening~\BN. The $D$-parameter was similarly treated in
Ref.~\cite{dparresum}. Since these observables vanish for all planar
events, their perturbative distributions are of the form
\begin{equation}
R(y)\;=\;1\;+\;\mathcal{B}(y)\as^2\;+\;\mathcal{C}(y)\as^3\;+\;\ldots \;\;\;.
\end{equation}
Monte Carlo programs analogous to EVENT2 are now
available~\cite{menlo_parc_mc,debrecen_mc,mercutio_mc,eerad2_mc} to
calculate the coefficients $\mathcal{B}(y)$ and $\mathcal{C}(y)$,
using QCD matrix elements with up to one loop. Matching schemes have
also been defined~\cite{fourjetresum}, to combine the
$\mathcal{O}(\as^3)$ and NLLA predictions.

Although it has not been possible to implement these calculations in
the work presented here, we considered it worthwhile to measure the
distributions themselves, at all LEP centre-of-mass energies. Future
measurements using these NLLA predictions should yield
substantial improvements over existing studies of four-jet events based
on matrix elements alone.

\end{itemize}

\section{Theoretical uncertainties}
\label{evsh_prediction_errors}

We consider here the sources of uncertainty in our perturbative
predictions for the event shape distributions, and their effect on the
values of \as\ obtained in experimental measurements. The new methods
presented here have been developed in collaboration with the LEP QCD
Working Group, and are documented in Ref.~\cite{uncertaintyband}.

\subsection{Uncertainties in the event shape distributions}
\label{distribution_theory_errors}

Owing to the truncation of both the $\mathcal{O}(\as^2)$ perturbation
series and the NLLA resummed expressions, our predicted event shape
distributions are not exact. Many techniques have been proposed to
estimate the effect of the missing higher-order terms: one of the
simplest would be to use the last known term of the perturbation
expansion, proportional to $\as^2$, as an estimate of remaining
uncalculated terms. This approach is likely to overestimate the
uncertainty, however, since the sum of the unknown terms is expected
to be of order~$\as^3$. Instead we will identify several arbitrary
assumptions which are made in our $\mathcal{O}(\as^2)+\text{NLLA}$
predictions; the variation of our results with respect to changes in
these assumptions will give a measure of the theoretical uncertainty.

\subsubsection[The renormalisation scale parameter, $x_\mu$]
{The renormalisation scale parameter, \boldmath$x_\mu$}

As we explained in Section~\ref{renormalisation}, a characteristic
energy scale~$\mu$ must be chosen at which to renormalise the QCD
couplings.\footnote{In practice one must also choose a
``renormalisation scheme,'' which specifies the detailed method of
renormalisation; all of the predictions used in this work have been
calculated in the modified minimal subtraction
($\overline{\textsc{ms}}$) scheme. However, the choice of scheme does
not affect the first-order scale dependence of the predictions, and
will therefore be neglected.} It is generally agreed that the most
appropriate choice for $\mu$ is the characteristic energy scale of the
hard interaction (in our case, the \epem\ centre-of-mass energy), but a
review of other possible choices is given in Ref.~\cite{PDbook}.

In general, our perturbation expansion for the cumulative distribution
of an event shape $y$ will be
\begin{equation}
R\left(y, \asq, x_\mu\right)\;=\;1\;+\;\mathcal{A}\left(y\right)\as(\mu)\,+\,\mathcal{B}\left(y,x_\mu\right)\as^2(\mu)\,+\,\mathcal{C}\left(y,x_\mu\right)\as^3(\mu)\,+\,\ldots\;\;\;,
\label{R_mudependence}
\end{equation}
where we have defined $x_\mu=\mu/Q$. The coupling $\as(\mu)$ is a
function of \asq, as described by the Renormalisation Group
Equation~(RGE) given in Equation~(\ref{rge}). Since the
renormalisation scale is an unphysical parameter, the \emph{complete}
prediction for the observable~$R$ cannot depend on $\mu$; therefore
the $x_\mu$-dependence of the coefficients
$\mathcal{B},\mathcal{C},\ldots$ must exactly cancel the variation of
$\as(\mu)$. However, as we discussed briefly in
Section~\ref{renormalisation}, the QCD prediction for $R$ will no
longer be independent of~$\mu$ when truncated to $\mathcal{O}(\as^2)$.
This dependence, which should decrease as more terms are added to the
series, can be used to estimate the effect of the higher-order terms.

Using the RGE, we find that the variation of $\as(\mu)$
with respect to small deviations from our chosen scale $\mu=Q$ is given to first order by
\footnote{This formula, together with certain others in this chapter,
is often expressed in the literature in terms of $\overline{\as}\equiv \as/2\pi$.}
\begin{equation}
\as(\mu)\,=\,\as(Q)\,-\,\frac{\beta_0 \ln x_\mu}{\pi}\,\as^2(Q)\,+\,\ldots\;\;\;.
\end{equation}
Inserting this into Equation~(\ref{R_mudependence}), and requiring
$\text{d}R/\text{d}\mu=0$ when summed to all orders, we find that
\begin{equation}
\mathcal{B}(y,x_\mu)\;=\;\mathcal{B}(y,1)\;+\;\frac{\beta_0 \ln x_\mu}{\pi}\,\mathcal{A}(y)\;\;\;.
\end{equation}
Conventionally, and somewhat
arbitrarily, the theoretical uncertainty of the distribution is
defined by varying $x_\mu$ in the range $1/2<x_\mu<2$.

\subsubsection[The resummation parameter, xl]
{The resummation parameter, \boldmath\xl}

For each event shape observable~$y$, we
have defined a resummation which collects the logarithms $L=-\ln y$
into a series of terms with the form~$Lg_1(\as L)$, $g_2(\as L)$,
\mbox{$\as g_3(\as L)$, \ldots}, where only the first two functions
$g_1$ and $g_2$ are currently known. However, it has been noted
recently~\cite{xscale} that the choice of logarithm~$L$ is not
unique. Suppose, for example, that we defined a new event shape
observable \mbox{$y=\xl\BT$}, where \BT\ is the total jet broadening
and \xl\ is some arbitrary constant. Our resummed distribution for
this observable would be expressed in terms of the `shifted' logarithm
$L=-\ln \BT-\ln \xl$.

The new parameter \xl\ plays a r\^ole analogous to that of the
renormalisation scale parameter $x_\mu$. If the complete resummation
were known, the overall dependence on \xl\ would vanish; however,
while we only have the first two terms, some residual ambiguity will
remain. Just as the coefficient functions of the perturbation series
acquire a dependence on~$x_\mu$ to cancel the variation of $\as(\mu)$,
the functions~$g_n(\as L)$ become dependent on \xl. Writing the
modified logarithm as $\widehat{L}=L-\ln\xl$, our old NLLA
resummation is given by
\begin{eqnarray}
\ln R_{\scriptscriptstyle\text{NLLA}} & \!\!=\!\! & Lg_1(\as L)+g_2(\as L) \nonumber \\
& \!\!\approx\!\! & \Big[\widehat{L}+\ln\xl\Big]\Big[g_1(\as \widehat{L})+g'_1(\as \widehat{L})\as\ln\xl \Big] \nonumber \\
&  & +\,g_2(\as \widehat{L})\,+\,g'_2(\as \widehat{L})\as\ln\xl \;\;\;,
\end{eqnarray}
where $g'_n(\lambda)\equiv \text{d}g_n/\text{d}\lambda$. Regrouping
the terms of this expression, we can now write
\begin{equation}
\ln R_{\scriptscriptstyle\text{NLLA}} \,=\, \widehat{L}\hat{g}_1(\as \widehat{L}) \,+\,
\hat{g}_2(\as \widehat{L}) \,+\, \text{subleading~terms}\;\;\;,
\end{equation}
where
\begin{eqnarray}
\hat{g}_1(\as \widehat{L}) & \!\!=\!\! & g_1(\as \widehat{L}) \nonumber \\
\text{and}\;\;\;\hat{g}_2(\as \widehat{L}) & \!\!=\!\! & g_2(\as \widehat{L})+\Big[g_1(\as \widehat{L})+g'_1(\as \widehat{L})\as \widehat{L}\Big]\ln\xl\;\;\;.
\end{eqnarray}
The ``subleading terms'' here represent an estimate of the theoretical
uncertainty in the resummation. Similar transformation laws can be
derived for the matching coefficients~$G_{nm}$ and $C_n$; these are
listed in Ref.~\cite{uncertaintyband}.

\enlargethispage{\baselineskip}We must now choose a nominal value and a range of variation for the
parameter~\xl, to define our resummed prediction and its associated
uncertainty. In the past, event shape distributions have always been
calculated using the implicit assumption~$\xl=1$. This is, in fact,
the most natural choice. The six event shape observables $y$ have been
defined such that they approach the form
\begin{equation}
\ln y=a\ln \left(\frac{k_{\scriptscriptstyle{\text{T}}}}{Q}\right)-b\eta\;\;\;\,
\end{equation}
in the case of a single soft and collinear gluon emission;
$k_{\scriptscriptstyle\text{T}}$ and $\eta$ are respectively the
transverse momentum and rapidity of the emitted gluon with respect to
the q$\bar{\text{q}}$ axis, and $a$ and $b$ are integers. When we
perform the transformation $y\to \xl y$ an additional constant is
introduced in the above expression. We will therefore continue to use
$\xl=1$ as our nominal choice, to preserve this simple relationship
between the logarithm~$L$ and the physical properties of the
event. The range of variation in \xl\ used to define our theoretical
uncertainties, however, is a more complicated and subjective issue: we
discuss this point in Ref.~\cite{uncertaintyband}. We will use the
range $4/9<\xl<9/4$ for the Durham~$y_{23}$ parameter, and
$2/3<\xl<3/2$ for all other observables.

\subsubsection[The kinematic constraint parameters, $p$ and $y_\text{max}$]
{The kinematic constraint parameters, \boldmath$p$ and $y_\text{max}$}

In Section~\ref{kinconstraints} we
introduced a modified form for the logarithm~$L$, to impose the
kinematic constraint $R(y_\text{max})=1$ on our NLLA
predictions. Including both the \xl\ parameter and the kinematic
constraint, the logarithm is now given by
\begin{equation}
\widetilde{L}=\frac{1}{p}\ln\left[\frac{1}{{(\xl y)}^p}-\frac{1}{{(\xl y_\text{max})}^p}+1\right]
\end{equation}
The parameter~$p$, which determines the sharpness with which the
constraint is applied, will be varied over the range $1<p<2$ to
determine the theoretical uncertainty; for our nominal distribution,
we will continue to use $p=1$.

An ambiguity also exists in the choice of kinematic boundaries
$y_\text{max}$. Although a well-defined limit may exist for each
observable---for example $T>0.5$ in the case of thrust---this may
correspond to a very improbable state containing a large number of
partons. Also, no analytical expressions exist for $y_\text{max}$ in
some cases, so they must be estimated using Monte Carlo
simulations. In Ref.~\cite{uncertaintyband} we define ranges of
variation for $y_\text{max}$, to be used as contributions to our
theoretical uncertainty. This is, however, a negligible effect in
comparison to the $x_\mu$ and \xl\ variations.

\subsubsection{Matching scheme dependence}

The Log($R$) and $R$ matching schemes, defined in
Sections~\ref{logrscheme} and~\ref{rscheme}, provide two alternative
methods for combining the $\mathcal{O}(\as^2)$ and NLLA
predictions. We have chosen to use the Log($R$) scheme as our nominal
theory prediction, and the difference between the two schemes as an
estimate of the theoretical uncertainty. Again, this is usually a
negligible contribution when compared with the $x_\mu$ and \xl\
variations.

\subsubsection{Quark mass effects}

As we noted in Section~\ref{nlla}, the perturbative predictions used
in this work do not take account of quark masses. After averaging over
all accessible quark flavours, the influence of heavy quark effects on
the fitted value of \as\ is expected to be $\mathord{\sim}1\%$ at
$\sqrt{s}=91$~GeV~\cite{uncertaintyband}. When new NLLA predictions
become available, a consistent $\mathcal{O}(\as^2)+\text{NLLA}$ mass
correction can be applied to our results; meanwhile the size of the
uncertainty is much smaller than other contributions such as the
$x_\mu$ dependence, and will be neglected.

\subsection[Uncertainties in \as]
{Uncertainties in \boldmath\as}
\label{as_theoryerrors}

The aim of our experimental analysis
will be to measure the strong coupling \as\ by fitting theoretical
predictions to the observed event shape distributions. We now
investigate the effects of our theoretical uncertainties on the
extracted values of \as; at this stage, we will consider separately
the effects of the $x_\mu$, \xl\ and $p$ variations.

\begin{figure}
\begin{center}
\includegraphics[width=0.9\textwidth]{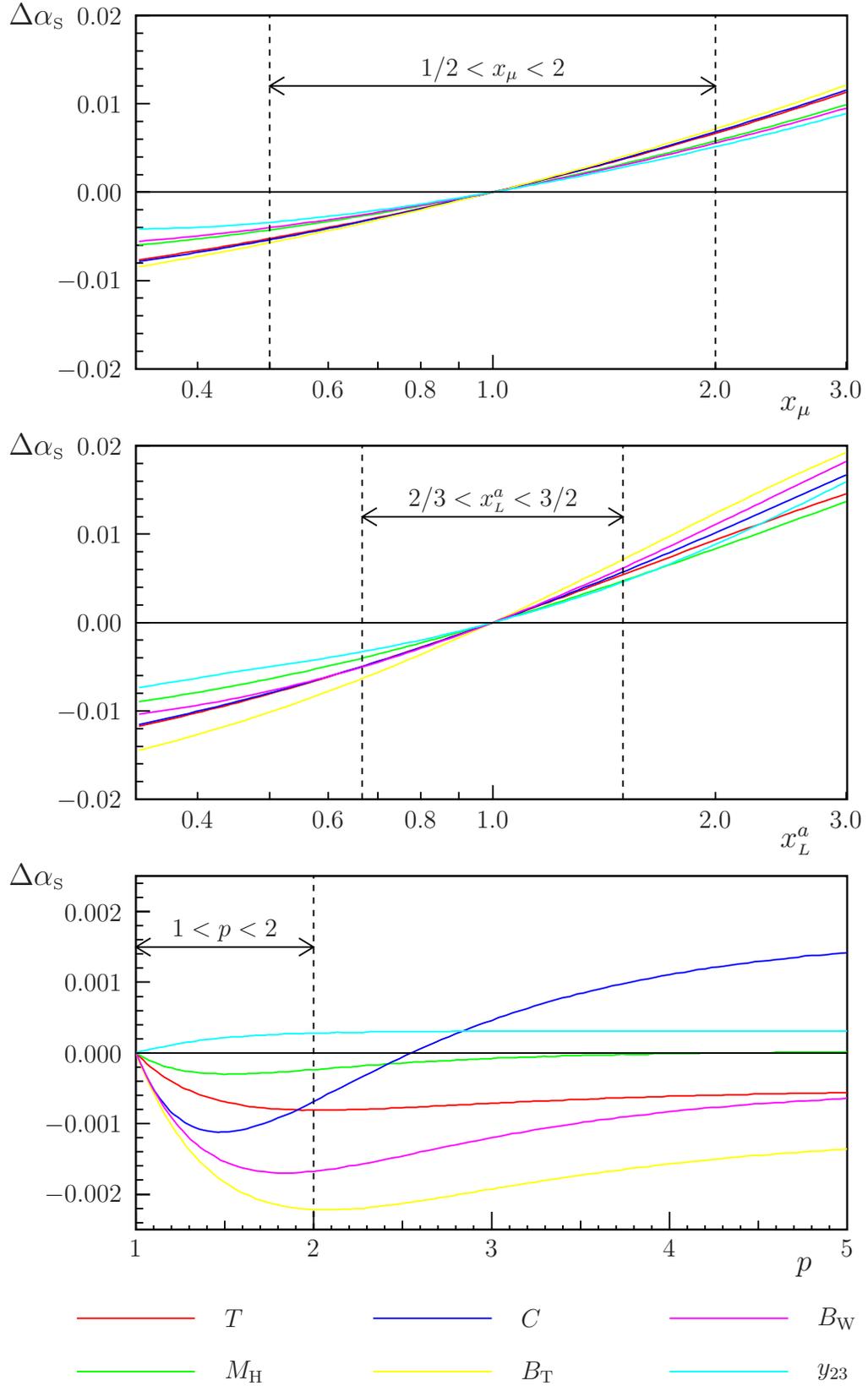}
\end{center}
\vspace{-0.25cm}
\caption{Deviations in \as\ corresponding to variation of the
parameters $x_\mu$, \xl\ and $p$, in the perturbative predictions for
the thrust, heavy jet mass, $C$-parameter, total jet broadening, wide
jet broadening and Durham $y_{23}$ parameter. The power~$a$ is equal
to 1/2 in the case of \ytwothree, and 1 for all other observables.}
\label{xmu_xl_p_variation}
\end{figure}

For each variant of the perturbative prediction, we define two
distributions:
\begin{itemize}
\item The variant distribution, calculated at a fixed value of \as.\\
Example: $R'(\BW)$ with $x_\mu=1$, $\xl=3/2$, $p=1$, $\as=0.12$.
\item The nominal distribution, calculated at an alternative value of \as.\\
Example: $R'(\BW)$ with $x_\mu=1$, $\xl=1$, $p=1$, $\as=0.12+\Delta\as$.
\end{itemize}
We then fit the nominal distribution to the variant distribution, with
\as\ as a free parameter; the deviation~$\Delta\as$ estimates the
theoretical uncertainty in an experimental determination of \as.  Our
fits are calculated using the method of least squares, with the
statistical `uncertainty' of each bin proportional to the square root
of its contents.\footnote{We do not consider the effects of background
and detector biases, which will in practice alter the statistical
weights slightly from a simple Poisson distribution.}  The ranges of
the fits for each observable are the same as those used in the OPAL
analysis, as listed in Section~\ref{opalfitmethod}.  Our results are
shown in Figure~\ref{xmu_xl_p_variation}, for a nominal value of
\as=0.12. The total jet broadening consistently gives the largest
theoretical uncertainties, while the Durham~\ytwothree\ parameter and
heavy jet mass give comparatively small uncertainties. The deviations
in \as\ due to variation of $p$ are much smaller than those due to
$x_\mu$ and~\xl.

\subsection{Combining the theoretical uncertainty estimates}

We would now like to calculate an
overall theoretical uncertainty, combining each of the contributions
listed in Section~\ref{distribution_theory_errors}. One must remember,
however, that the various contributions do not represent distinct
\emph{sources} of uncertainty, in the conventional sense: the sources
here are simply the uncalculated terms of the order-by-order and
resummed predictions. Our calculated uncertainties represent the
sensitivity of the distributions to various subsets of the missing
terms. Since those subsets are likely to overlap, we should not merely
add the contributions in quadrature. Furthermore, different regions of
the event shape distributions will be dominated by different types of
uncertainty; generally the $x_\mu$ variation gives the largest
uncertainty at the two-jet end of the fit range (small~$y$), and the
\xl\ variation dominates the region of multiple gluon emissions
(large~$y$). It is therefore appropriate to combine our theoretical
uncertainty estimates at the distribution level, rather than combining
deviations in the \as\ fits.

\begin{figure}
\begin{center}
\includegraphics[width=0.9\textwidth]{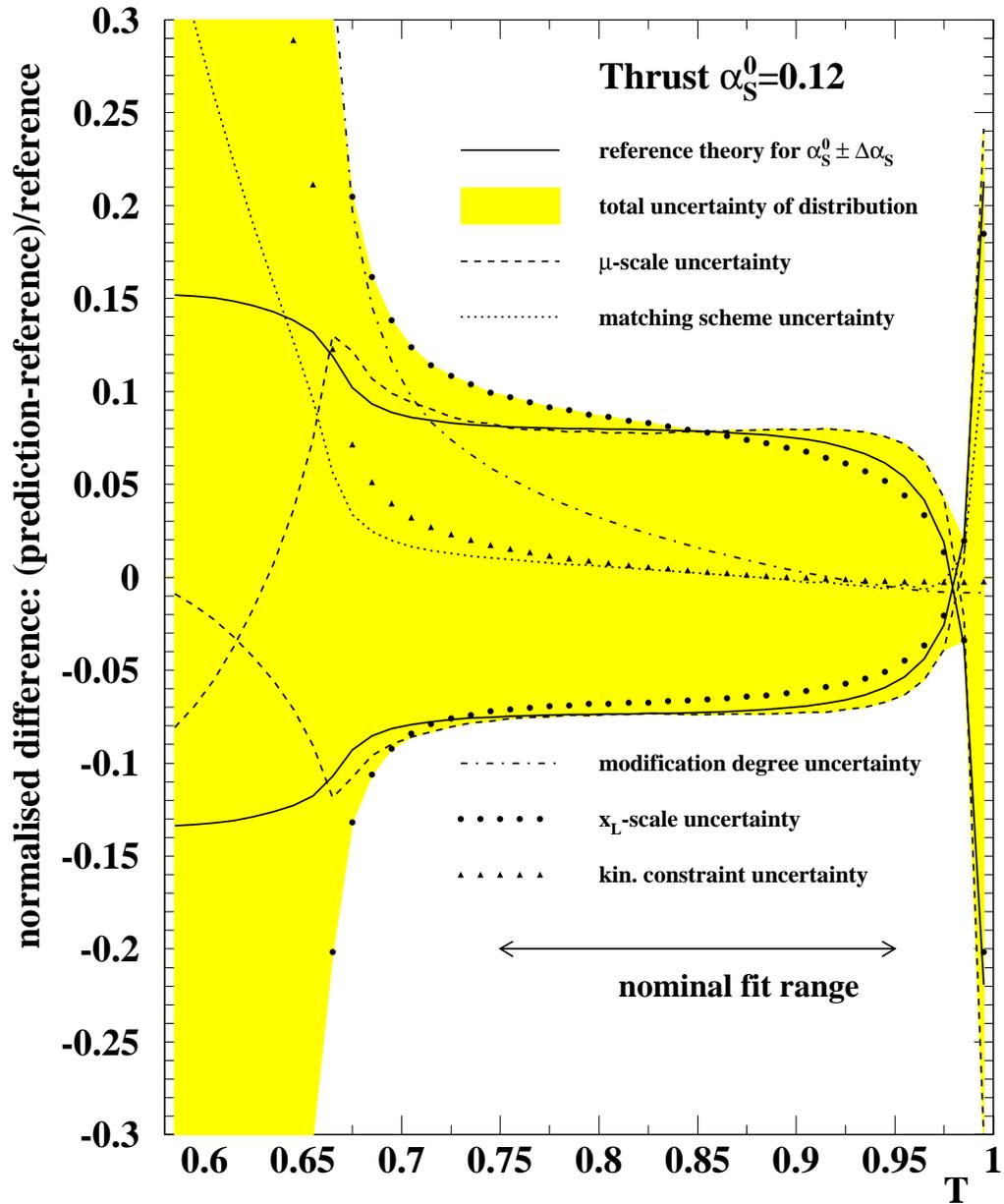}
\end{center}
\vspace{-0.5cm}
\caption{The uncertainty band for the thrust distribution \emph{(Figure prepared by H.~Stenzel, and published in Ref.~\cite{uncertaintyband})}}
\label{thrustband}
\end{figure}

Figure~\ref{thrustband} shows the deviations in the thrust
distribution corresponding to each of the uncertainty contributions
(excluding quark mass effects). In the case of the upper \xl\
variation, for example, we plot the fractional difference between the
distributions predicted for $\xl=3/2$ and $\xl=1$. The outer envelope
of all these variations defines an ``uncertainty~band,'' indicated by
yellow shading.

Just as we have done for the individual variations in the previous
section, this combined uncertainty band can be translated into a
deviation in \as. We once again define a variant distribution,
corresponding to the edge of the band, and a nominal
distribution. While fixing $\as=\as^0$ in the variant distributions,
we change \as\ in the nominal distribution; the upper and lower
uncertainties $\sigma^\pm$ are defined by the range of values
\mbox{$\as^0-\sigma^-<\as<\as^0+\sigma^+$} for which the nominal
distribution lies completely inside the uncertainty band, within the
standard fit range.\footnote{Alternatively, one could simply fit the
nominal distributions to the variant distributions, with \as\ as a
free parameter, as we have done in Section~\ref{as_theoryerrors}. This
approach would give slightly larger uncertainty estimates (an increase
of $\mathord{\sim}15\%$ in the theoretical uncertainty of the combined LEP 
\as\ measurement presented in Chapter~\ref{lepcombinationchapter}).}

\newpage
\section{Monte Carlo models}
\label{mcmodels}

So far, the QCD predictions we have discussed have been concerned with
free quarks and gluons. Before reaching our detector, however, these
partons must `fragment' into bound colourless states. This final phase
of the QCD interaction, which occurs at a low characteristic energy
scale, cannot be predicted by perturbation theory. Instead, we must
use numerical simulations based on semi-empirical models. The Monte
Carlo programs used in our analysis to simulate multihadronic events
are $\mathcal{KK}$2f~4.01/4.13~\cite{kk2f},
PYTHIA~6.125~\cite{pythia6.1}, HERWIG~6.2~\cite{herwig6} and
\mbox{ARIADNE~4.11~\cite{ariadne4}}. 

\begin{figure}[t!]
\begin{center}
\includegraphics[width=0.85\textwidth]{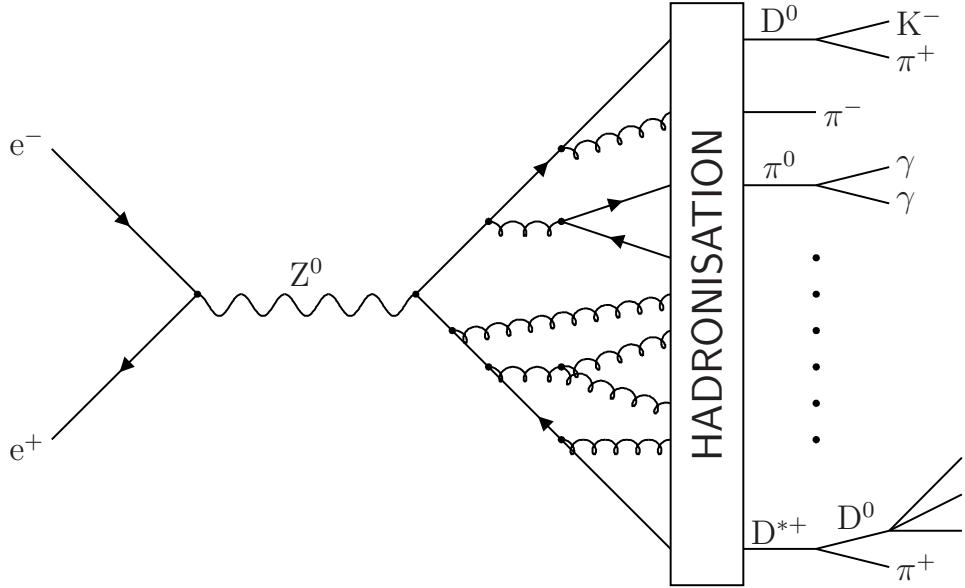}
\end{center}
\caption{A schematic view of the stages to be simulated in an hadronic
decay of the Z$^0$ boson at LEP}\vspace{0.3cm}
\label{eventflowfigure}
\end{figure}

The simulation of each event proceeds in four distinct stages, which
are illustrated schematically in Figure~\ref{eventflowfigure}:
\begin{itemize}
\item Generation of a $\mathrm{q}\bar{\mathrm{q}}$ system from the
initial \epem\ state, possibly with initial-state radiation
\item A ``parton shower,'' in which gluons are radiated from the
$\mathrm{q}\bar{\mathrm{q}}$ pair; the gluons may then radiate other
gluons, or split to form new $\mathrm{q}\bar{\mathrm{q}}$ pairs. This
stage should reproduce as closely as possible the predictions of
perturbative QCD.
\item Fragmentation of the parton system into hadrons
\item Decays of short-lived hadrons such as the $\pi^0$ and
$\text{K}^0_\text{S}$ mesons
\end{itemize}
We will briefly describe the implementation of these stages in the
Monte Carlo programs; not all stages are implemented by all programs.

\subsection[Generation of the $\mathrm{q}\bar{\mathrm{q}}$ system]
{Generation of the \boldmath$\mathrm{q}\bar{\mathrm{q}}$ system}

The first stage of the event is simulated most accurately by
$\mathcal{KK}$2f, which provides a detailed treatment of multiple
photon emission from the initial \epem\ state. The resulting
$\mathrm{q}\bar{\mathrm{q}}$ system can then be passed as an input to
the parton shower models in PYTHIA, HERWIG or ARIADNE. When simulating
events without initial-state radiation, however, we can use PYTHIA or
HERWIG in place of $\mathcal{KK}$2f.

\subsection{The parton shower}

In the PYTHIA~\cite{pythia6.1} and HERWIG~\cite{herwig6} programs,
the perturbative stage of the event is simulated using a numerical
implementation of the DGLAP equations~\cite{dglap}, which were also
used in the derivation of the NLLA analytical predictions. The
progress of each parton in the cascade is parameterised by an
``evolution~variable,'' $t=\ln(Q^2/\Lambda^2)$, where $Q^2$ is related
to the virtuality of the parton, and $\Lambda$ is some arbitrary fixed
scale. The probability~$\mathcal{P}$ for a parton~$a$ of virtuality~$Q^2$ to
split into two partons $b$ and $c$, carrying momentum fractions $z$
and $(1-z)$ respectively, can then be written in the form
\begin{equation}
\frac{\text{d}\mathcal{P}}{\text{d}t} \;\propto\;
\int_0^1 \text{d}z \,\frac{\as}{2\pi} \,P_{a\to bc}(z) \;\;\;\;\;\;\;\;[0<t<t_0]\;\;\;,
\end{equation}
where $t_0$ and $t$ are the evolution variables before and after the
splitting, and $P_{\text{a}\to \text{bc}}$ are the Altarelli-Parisi
splitting kernels for the processes \mbox{$\text{q}\to\text{qg}$},
\mbox{$\text{g}\to\text{gg}$},
\mbox{$\text{g}\to\text{q}\bar{\text{q}}$} and
\mbox{$\text{q}\to\text{q}\gamma$}. The precise definition of the
scale $Q^2$ varies between models: PYTHIA uses $Q^2=m_a^2$, while
HERWIG uses $Q^2\approx m_a^2/2z(1-z)$, where $m_a^2$ is the invariant
mass-squared of the parton~$a$.

The models include various features to incorporate coherence
effects,\footnote{The acronym `HERWIG' stands for ``Hadron Emission
Reactions With \emph{Interfering Gluons}''.} the simplest of which is
the property of \emph{angular~ordering}: the opening
angle~$\theta_{bc}$ of each branching \mbox{$a\to bc$} is required to
be less than that of the previous branch in the cascade. This result
can be explained~\cite{webber_qcdbook} in terms of the Uncertainty
Principle. Other effects include correlations between the azimuthal
branching angles, due to gluon polarisation.

The shower is continued until the virtuality of the partons reaches
some lower limit $Q_0\sim 1$~GeV, which is a tunable parameter of the
model. The final parton configuration is then passed to the
non-perturbative hadronisation stage. More detailed discussions of
parton shower physics are given in Refs.~\cite{webber_qcdbook}
and~\cite{pythiamanual}.

The ARIADNE Monte Carlo~\cite{ariadne4} uses an alternative formalism:
the Colour Dipole Model~\cite{colour_dipole,colour_dipole2}. In this
paradigm, new partons are radiated by the colour fields between the
existing quarks and gluons, and not by the partons themselves. The
q$\bar{\text{q}}$ system represents a single dipole, capable of
radiating a gluon; after the first branching has occurred, the
q$\bar{\text{q}}$g system is described by two independent q--g and
g--$\bar{\text{q}}$ dipoles, and so forth.

Since ARIADNE simulates only the parton shower stage, the quarks and
gluons are then passed to PYTHIA for hadronisation.

\subsection{Fragmentation}

Hadronisation is simulated in PYTHIA using the \emph{Lund string
model}~\cite{stringfrag}. Unlike the electromagnetic field patterns
formed by distributions of charges and currents, the
corresponding fields in QCD are expected to be confined in narrow
regions stretched between the colour charges; this is a result of the
gluon's self-coupling property. According to the string model, the
field lines will eventually `break' at several points to form new
q$\bar{\text{q}}$ or diquark-antidiquark pairs which lead to meson and
baryon production. The model has many tunable parameters, which have
been chosen to give optimal agreement with the LEP data.

HERWIG uses the alternative \emph{cluster model}~\cite{clustermodel},
motivated by the property of preconfinement~\cite{preconfinement} in
perturbative QCD. It has been shown that partons naturally tend to
group themselves into colourless low-mass clusters of quarks and
gluons, during the perturbative shower evolution. Herwig simulates the
production of hadrons through the decay of these `preconfined'
clusters. Any gluons remaining at the end of the parton shower are
first divided non-perturbatively into q$\bar{\text{q}}$ pairs, which
are then paired with neighbouring quarks and antiquarks to produce
colourless clusters.

\subsection{Decays}

Some of the hadrons produced in \epem\ collisions at LEP are expected
to decay very close the interaction point; the experiments will then
measure event shape observables based on the daughter particles. For
convenience, we consider all particles with lifetimes less than
$3\times 10^{-10}$~s to be unstable. PYTHIA and HERWIG simulate the decays of
these particles based on standard branching ratios supplied by the
Particle Data Group~\cite{PDbook}.

\section{Other experimental studies of QCD}
\label{otherqcd}

Our measurements of event shape observables in \epem\ collisions
represent just one of the experimental methods used to study
QCD. Reviews of other tests and measurements can be found in
Refs.~\cite{PDbook} and~\cite{alphas_bethke}; only a very brief
overview will be given here, focusing mainly on OPAL
measurements. The results can be divided broadly into two categories:
those which measure the free parameter~\as, and those which test the
structure of the theory. Although the value of \asmz\ is not predicted
by the Standard Model, the consistency of measurements obtained from
different physical processes at a variety of energy scales is itself a
further success for perturbative QCD.

\subsection[Measurements of \as]
{Measurements of \boldmath\as}

\begin{figure}[t]
\begin{center}
\includegraphics[width=0.66\textwidth]{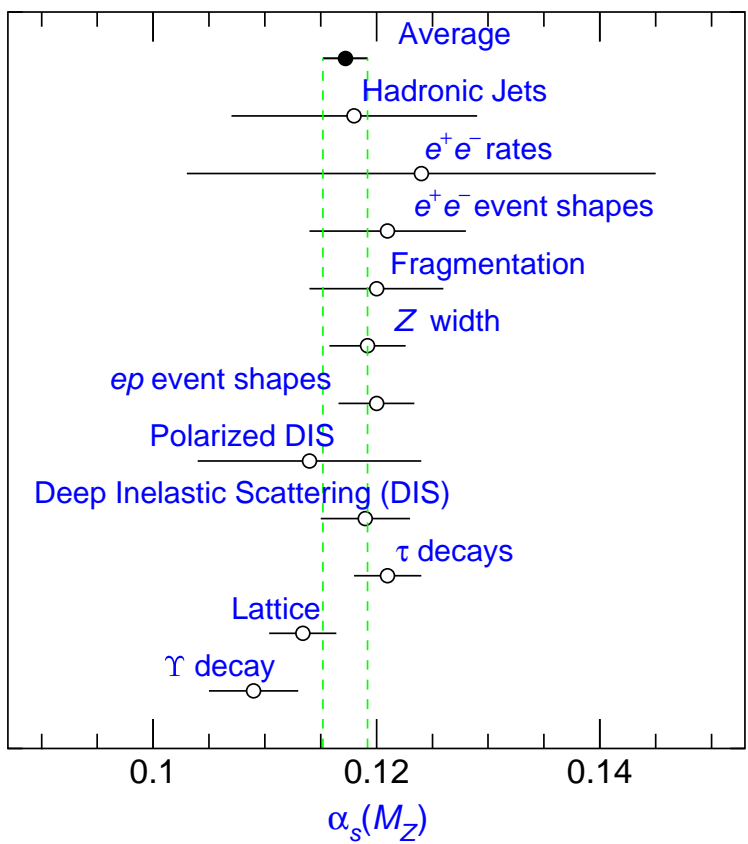}
\end{center}
\vspace{-0.5cm}
\caption{Contributions to the world average measurement of \asmz \newline \emph{(Figure taken from Ref.~\cite{PDbook})}}
\label{pdg_asmz}
\end{figure}

The current world average of \as\ at the Z$^0$ mass scale is
\mbox{$\asmz=0.1172\pm 0.0020$} \cite{PDbook}; a breakdown of the
contributions from different physics processes is shown in
Figure~\ref{pdg_asmz}. The most precise determinations of \asmz\ using
LEP data are those from the hadronic decays of the $\tau$~lepton (for
example, in Ref.~\cite{alphas_taudecay}), and those from the ratio of
hadronic and leptonic partial decay widths of the
Z$^0$~boson~\cite{lep_electroweak_comb}. These methods have two
distinct advantages over the event shape fits: firstly they have
negligible sensitivity to non-perturbative effects, and secondly the
theory predictions are available to three orders of perturbation
theory. Since the measurements themselves are single numbers rather
than a fits to a distribution, however, they do not additionally
provide a test of the QCD gauge structure.

The strong coupling has also been determined at LEP using a variety of
jet multiplicity observables~\cite{alphas_jets}, similar to the
\ytwothree\ variable used in our own measurements; some of these
distributions have only $\mathcal{O}(\as^2)$ predictions, while others
also have NLLA resummations. 

A new determination of \as\ has recently been performed using the
photon structure function $F_2^\gamma$ in \epem\ data from PETRA,
TRISTAN and LEP~\cite{alphas_f2gamma}. By combining data at different
centre-of-mass energies, one can also measure \as\ from the
energy-dependence of the parton fragmentation
functions~\cite{alphas_fragfun}.

In the last ten years, a new approach has been developed to account
for non-perturbative effects in event shape distributions. In these
``power~correction''
models~\cite{powercorr_Dokshitzer,powercorr_Manohar,powercorr_Akhoury,powercorr_gardi1,powercorr_gardi2,powercorr_Korchemsky},
the distributions and their moments are predicted to be shifted by
some factor proportional to $1/Q^n$, where $Q$~is the energy scale of
the hard interaction and $n$~is an integer. The use of power
corrections in experimental analyses (for example,
Ref.~\cite{DELPHI_as_1}) is supposed to remove the need for a Monte
Carlo model to simulate hadronisation corrections,\footnote{This point
is perhaps debatable, since power corrections do not contain detailed
information about resonances, hadron masses and decays.} but also
introduces a new free parameter~$\alpha_0$ which must be determined
from the data.

Outside of \epem\ physics, one of the most precise measurements of
\as\ has been obtained as part of a global determination of parton
distribution functions in electron-proton and proton-antiproton
collisions~\cite{alphas_mrst}.

\subsection{Tests of the QCD gauge structure}

In addition to measuring the parameter \as, experimental data can be
used to test the validity of the SU(3) gauge theory as a description
of the strong interaction. Predictions in perturbative QCD are often
expressed in terms of three ``colour~factors,'' $C_\text{A}=3$,
$C_\text{F}=4/3$ and $T_\text{F}=1/2$, which roughly correspond to the
relative strengths of the $\text{g}\to\text{gg}$,
$\text{q}\to\text{qg}$ and $\text{g}\to\text{q}\bar{\text{q}}$
vertices respectively. These factors are fundamental properties of the
symmetry group, and can be compared against the corresponding factors
in other groups. One can determine $C_\text{A}$, $C_\text{F}$ and
$T_\text{F}$ in \epem\ annihilation~\cite{opal_colourfactors1}, using
four-jet events of the types
$\epem\to\text{Z}^0/\gamma\to\text{q}\bar{\text{q}}\text{q}\bar{\text{q}}$
and $\epem\to\text{Z}^0/\gamma\to\text{q}\bar{\text{q}}\text{gg}$.
The measurements use next-to-leading order QCD
predictions, for the four-jet rate~$R_4$ and several angular
correlation variables. The results are in good agreement with the
standard QCD values.

The QCD colour factors can also be measured using the three-jet \epem\
event shape observables from which we determine~\as\ in this work. The
results~\cite{opal_colourfactors2}, although rather imprecise, are in
agreement with the SU(3) gauge structure.

\subsection{Flavour independence}

QCD makes no distinction between the couplings of the six
flavours. The quarks do have different masses and electric charges,
but these are expected to have a negligible effect, when the energy
scale probed is much higher than the quark masses.
The proton and neutron masses, for example, which are
dominated by QCD interaction potentials, differ by only~0.1\%. At the
energy scales of LEP, the perturbative behaviour of b~quarks should
differ slightly from that of the lighter flavours, but we do not
expect to observe differences in the gluon radiation from u,~d, s
and~c quarks. By tagging hadronic Z$^0$ decays as either u/d/s, c or b
flavoured, based on the hadrons identified in the final state, one can
demonstrate using OPAL data that the corresponding values of \as\ do
not differ by more than a few
\mbox{percent \cite{alphas_flavour1,alphas_flavour2}}. Measurements of
charged particle multiplicities in Z$^0$ decays have also been
consistent with the flavour-independence of
QCD~\cite{multiplicity1,multiplicity2}.

\chapter{The OPAL detector at LEP}
\label{detectorchapter}

The work described in this dissertation is based on data from the
Large Electron-Positron Collider (LEP), which operated at CERN between
the years 1989 and 2000. Interactions were studied at centre-of-mass
energies in the range 91--209~GeV, using four detectors spaced around
the circular collider. Our detailed event shape measurements,
presented in Chapter~\ref{opalchapter}, are derived from collisions
observed in the OPAL detector. In Chapter~\ref{lepcombinationchapter}
we combine our measurements of the strong coupling~\as\ with those
from the other three LEP experiments: ALEPH, DELPHI and L3. This
chapter outlines the main features of the LEP collider and the OPAL
detector, with a focus on the components most relevant to our
measurements. Details of the ALEPH, DELPHI and L3 experiments may be
found in Refs.~\cite{alephdetector},\cite{delphidetector}
and~\cite{l3detector} respectively; further information on the OPAL
detector is given in Ref.~\cite{nimpaper}.

\section{The LEP collider}
\label{lepcollider}

\subsection{Historical background}

In 1973 the Gargamelle bubble chamber at CERN delivered the world's
first experimental evidence for ``neutral~currents,''
mediated by the Z$^0$ boson. A beam of muon neutrinos was observed to
scatter from heavy nuclei, without any production of charged
muons~\cite{gargamelle}. This confirmed the predictions of the
Glashow-Weinberg-Salam electroweak gauge theory, and provoked further
experiments to investigate the properties of the Z$^0$ and W$^\pm$
bosons. Ten years later, in 1983, the UA1 and UA2 experiments both
reported direct evidence for the production of
Z$^0$~\cite{ua1_zdiscovery,ua2_zdiscovery} and
W$^\pm$~\cite{ua1_wdiscovery,ua2_wdiscovery} bosons in
p$\bar{\mathrm{p}}$ interactions at the CERN SPS collider.

Planning had already commenced in 1976 for the building a circular
high energy e$^+$e$^-$ collider, which would act as a `factory' for
Z$^0$ and W$^\pm$ bosons. Inelastic collisions between leptons
are phenomenologically much simpler than those between composite
hadrons, so extremely precise measurements could then be made at
well-defined energy scales. Due to its low mass, an electron emits
many times more energy in the form of synchrotron radiation than a
proton would when travelling with the same energy.\footnote{The energy
loss per turn of a circular accelerator of radius~$R$, for a particle
of mass~$m$, charge~$q$, speed~$\beta c$ and energy~$E=\gamma mc^2$,
is\vspace{-0.2cm}
\[
\Delta E\;=\;\frac{q^2 \beta^3 \gamma^4}{3 \epsilon_0 R}\;=\;\frac{q^2
\beta^3 E^4}{3 \epsilon_0 c^8 m^4 R}\;\;\;,\vspace{-0.1cm}
\]
where $c$ is the speed of light and $\epsilon_0$ is the permittivity
of free space.} It would therefore be necessary to build an
accelerator with the largest feasible radius, in order to minimise the
centripetal acceleration of the beam particles.

\begin{figure}[tb]
\begin{center}
\includegraphics[clip=true,width=\textwidth]{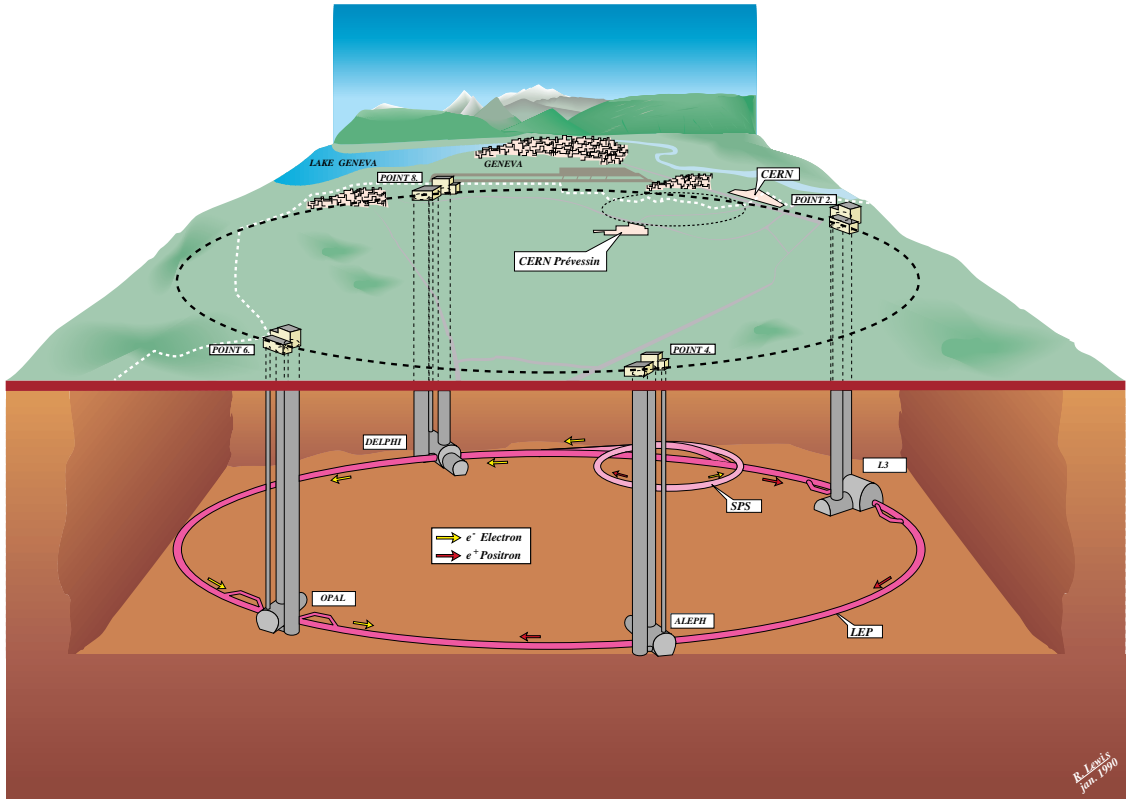}
\end{center}
\vspace{-0.3cm}
\caption[A schematic cut-away impression of the LEP collider and its
geographical location] {A schematic cut-away impression of the LEP
collider and its geographical location; the Swiss-French border is
indicated in white, with the city of Geneva and its airport in the
background. The horizontal and vertical scales are not equal.}
\label{lep_3dview}
\end{figure}

A LEP Study Group was formed, and published an initial design report
in August~1977~\cite{firstlepproposal} for a collider with eight
experimental interaction points and a circumference of 51~km; after
several revisions, the CERN Council in December~1981 approved a
proposal for a 26.7~km accelerator with four experiments. The machine
was designed to run initially with an energy of up to 50~GeV per beam,
which was expected to be sufficient for the production of real Z$^0$
bosons. A later phase (LEP2) was also anticipated, with energies of up to
100~GeV per beam. LEP was to be installed in a tunnel approximately
100~m below the surface of the plain lying between Lake Geneva and the
Jura mountains: a schematic view of the tunnel is shown in
Figure~\ref{lep_3dview}. Construction began in 1983, and the first
beams were injected in 1989. The final designs were published
in Refs.~\cite{lepdesign1,lepdesign2,lepmachine}.

LEP operated successfully with a beam energy of 45.6~GeV, exactly half
of the Z$^0$ mass, from 1989 to 1995. Millions of Z$^0$ decays were
observed by each of the four experiments, yielding precision tests of
the Standard Model. During the winter shutdown of 1995--6, various
hardware upgrades took place~\cite{lepdesign3} including the
installation of new superconducting cavities; these enabled LEP to
start its second phase of running, with energies above the threshold
for W$^+$W$^-$ pair production. The beam energies were gradually
increased over following five years, reaching a maximum of 104.5~GeV
per beam during 2000. LEP was finally closed to make way for
construction of the Large Hadron Collider (LHC), in November~2000.

\subsection{The LEP injector chain}

\begin{figure}
\begin{center}
\includegraphics[clip=true,width=\textwidth]{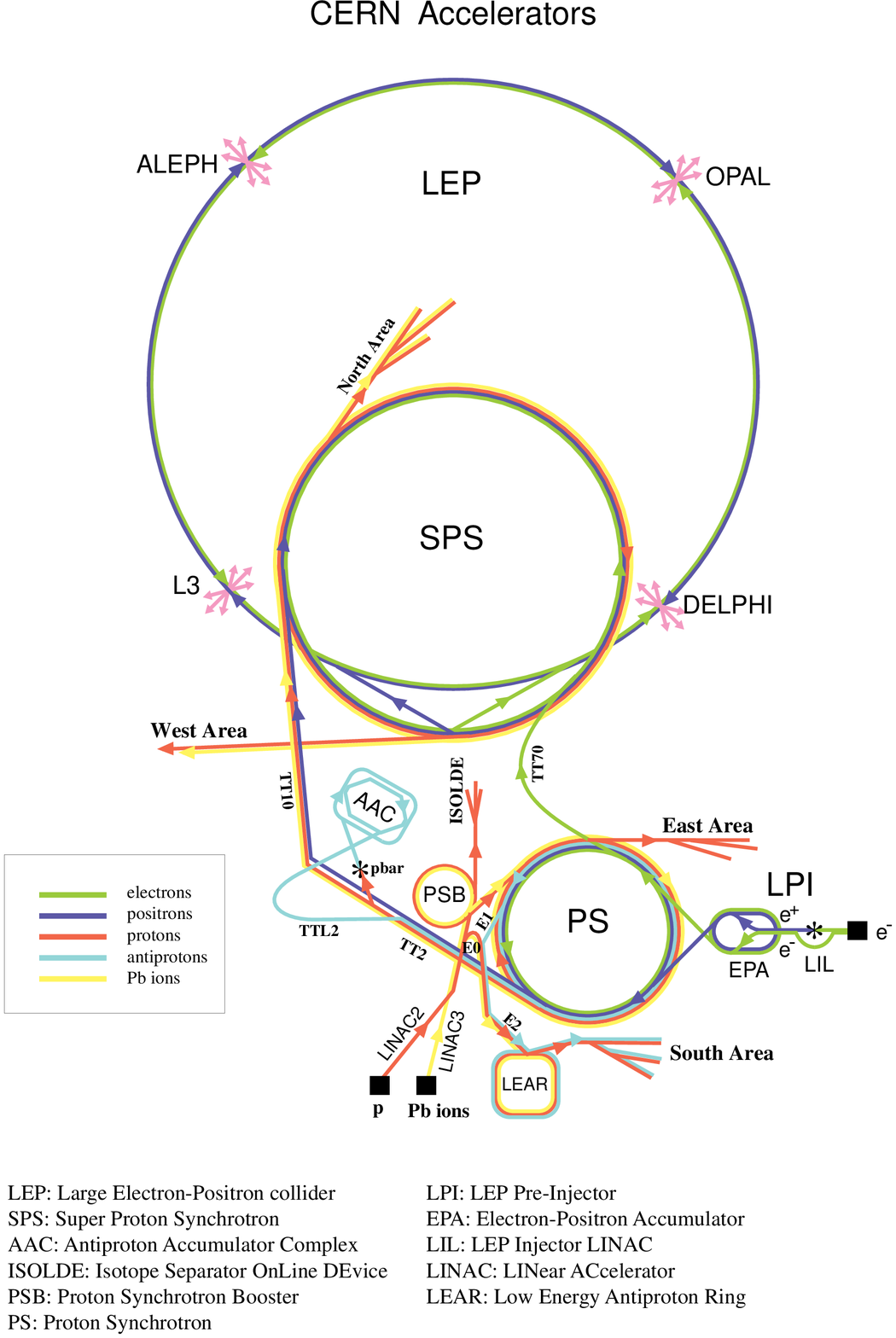}
\end{center}
\caption{A plan of the CERN accelerator complex during the operation of LEP}
\label{lep_planview}
\end{figure}

A complex sequence of accelerators and storage rings were used to
supply electrons and positrons to the LEP beampipe;
Figure~\ref{lep_planview} shows a diagram of the complete CERN
accelerator complex (some of which does not relate to LEP). The
electrons and positrons were produced in pairs from a fixed tungsten
target, using a beam of 200~MeV electrons from the LEP Injector
Linac~(LIL). The beams were separated, and accelerated to 600~MeV,
before injection into the Electron-Positron Accumulator~(EPA) and
thence into the Proton Synchrotron~(PS) ring. At an energy of 3.5~GeV,
the particle bunches were transferred to the Super Proton
Synchrotron~(SPS), which finally accelerated them to 20~GeV for
injection into LEP.

\subsection{The main ring}

The LEP main ring comprised eight circular arcs joined by straight
sections of length 119~m; its total circumference was~26659~m, exactly
27/7 times larger that of the SPS. For geological reasons the tunnel
was constructed with a 1.4\% tilt, resulting in an altitude difference
of 120~m between the highest and lowest points.

Dipole magnets were installed at approximately 50~m intervals around
the ring, providing a field of up to 0.135~T to bend the beams. These
were interspersed with quadrupole and sextupole magnets, which focused
the beam using linear and quadratic field gradients. Other magnets were
used to correct the beam orbit in the vicinity of the straight
sections.

Physics experiments were installed in four of the straight sections,
at 90$^\circ$ intervals around the ring. In the same sections, either
side of each experiment, a system of radio-frequency~(RF) cavities
provided the acceleration needed to reach and maintain the final beam
energy. Initially only copper cavities were used, but these were
gradually replaced by superconducting niobium cavities, cooled by
liquid helium. All of the cavities operated at a frequency of
352.21~MHz, corresponding to the 31320$^\mathrm{th}$ harmonic of the
LEP revolution frequency.

LEP could be operated with either 4+4 or 8+8 bunches of
electrons and positrons; the timing of the bunches was such that they
crossed one another at the centres of the straight sections. A system
of electrostatic separators was used to prevent the beams from
interacting before reaching a stable orbit at their final energy; the
bunches could then be brought into collision at the centre of each
experiment, with a vertical precision of about~4~$\mu$m. When
8+8 bunches were used, the beams were kept separated at the
non-experimental crossing points.

An extremely high vacuum was needed, to keep the beams circulating
close to the speed of light for several hours. Synchrotron radiation
from the beam stimulated the production of electron-positron pairs in
the wall of the aluminium beampipe, and hence the desorption of gas
molecules; continuous pumping was therefore required to maintain a
pressure below 10$^{-9}$~torr. Low pressures were especially important
in the vicinity of the interaction points, where unwanted beam
scattering could contribute to the experimental background. LEP was
the first accelerator to use non-evaporable getter (NEG) strips for
its main pumping system: these act by chemically adsorbing the
majority of gases found inside the beampipe. Conventional rotary vane
pumps were also used, to establish a pressure low enough for the NEG
strips to become effective.

\section{The OPAL detector}
\label{opaldetector}

OPAL\footnote{\textbf{O}mni-\textbf{P}urpose \textbf{A}pparatus at
\textbf{L}EP} was a multi-purpose experiment, designed
to detect and identify nearly all of the possible processes occurring
in e$^+$e$^-$ collisions at LEP. The detector provided a high
acceptance and accurate reconstruction for every type of event, with
the exception of very low angle scattering processes ($\theta <
40$~mrad) and those producing only neutrinos.

\begin{figure}[tb]
\begin{center}
\includegraphics[width=\textwidth]{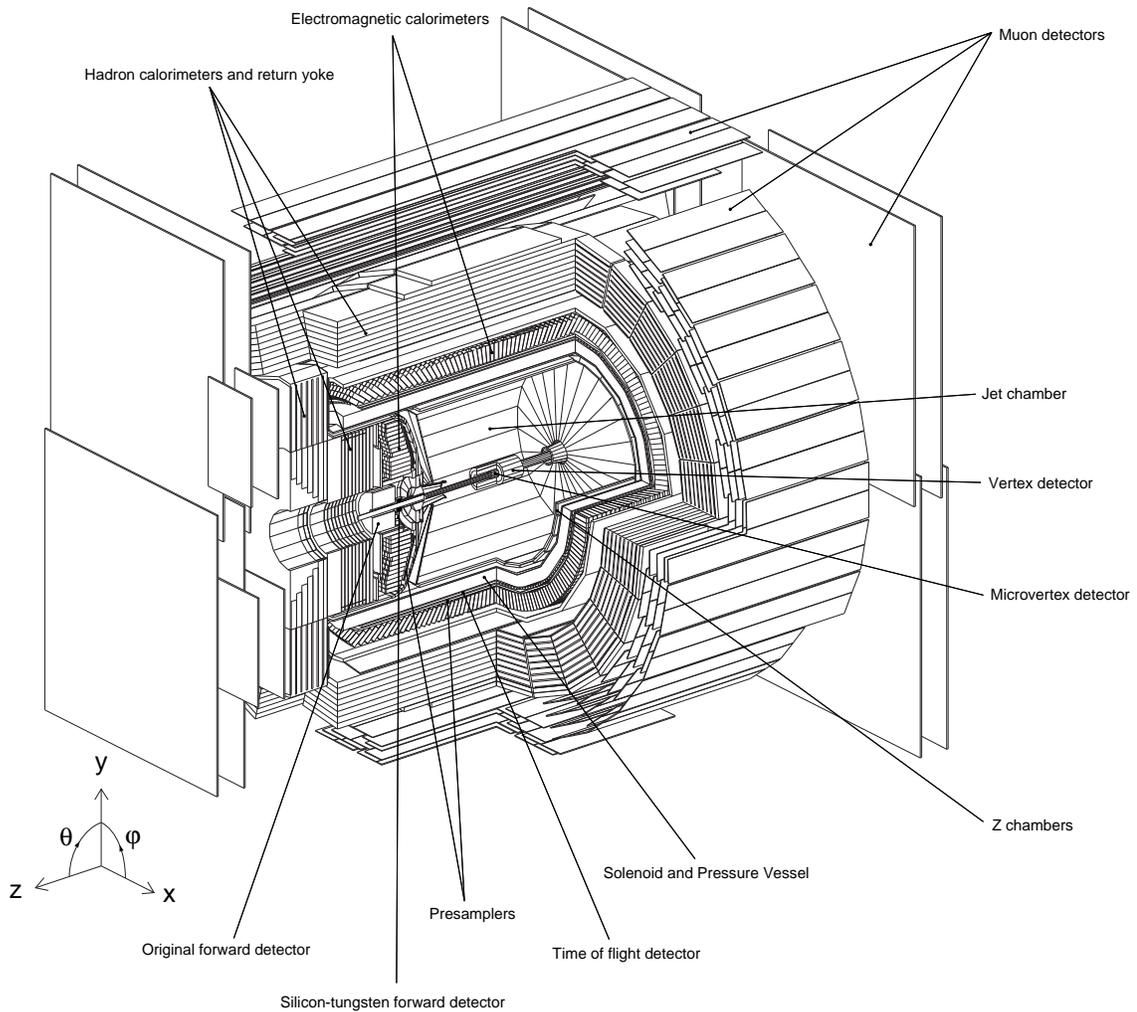}
\end{center}
\caption{A cut-away drawing of the OPAL detector}
\label{opalpic_3d}
\end{figure}

\begin{figure}
\begin{center}
\includegraphics[width=\textwidth]{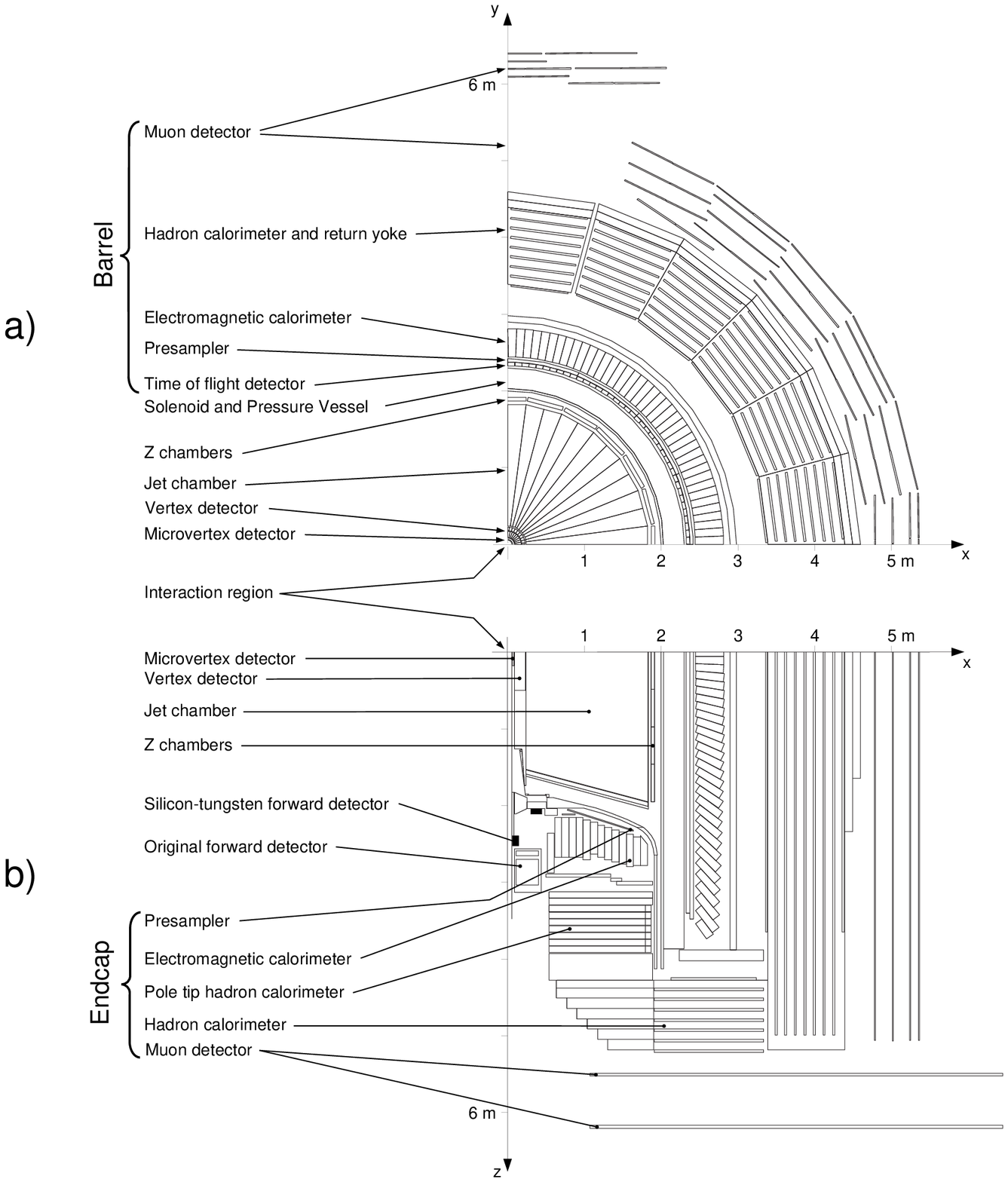}
\end{center}
\caption[Section views of the OPAL detector]{Section views of the OPAL
detector, in planes (a) perpendicular and (b) parallel to the beam
axis.}
\label{opalpic_xsect}
\end{figure}

The overall structure of the OPAL detector was similar to that of the
other three LEP experiments. Each of the major subdetectors formed a
cylindrical layer around the interaction point, offering the most
complete feasible solid angle coverage in each case; a
three-dimensional drawing and two cross sections of these layers are
shown in Figures~\ref{opalpic_3d} and~\ref{opalpic_xsect}. At the
centre, a silicon microvertex detector provided excellent spatial
resolution for charged particles passing through the wall of the
beampipe. This was surrounded by a system of pressurised drift chambers
in a strong magnetic field, which measured the directions, momenta and
energy losses of the charged particles. Outside the drift chambers and
the magnet, a system of scintillation counters was used to determine
the flight times of particles from the interaction region.  An
electromagnetic calorimeter then measured the energies of photons,
electrons and positrons; hadrons were subsequently absorbed by the
return yoke of the magnet, which functioned as an hadronic
calorimeter. Finally, a layer of thin drift chambers identified muons,
which were usually the only detectable particles to escape the
calorimeters. A system of ``forward~detectors'' was used to detect
particles travelling nearly parallel to the beampipe, such as the
electrons and positrons in Bhabha or two-photon events.

OPAL literature conventionally uses a Cartesian coordinate system,
centred on the nominal interaction point. The $z$-axis points along
the e$^-$ beam direction, which is inclined above the horizontal at an
angle of 13.9~mrad. The $x$-axis is precisely horizontal, and points
approximately towards the centre of the LEP ring, while the $y$-axis
is approximately vertical. The polar angle~$\theta$ and azimuthal
angle~$\phi$ are defined in the usual way with respect to the
Cartesian axes.

A more detailed description of the detector components will be given
in the sections that follow; we focus particularly on those
subdetectors most relevant to our analysis. Further details may be
found in Ref.~\cite{nimpaper}, which describes the original detector
as it was in 1989. The silicon microvertex
detector~\cite{opalsilicon1,opalsilicon2,opalsilicon3}, tile
endcaps~\cite{opal_te}, silicon-tungsten luminometers~\cite{opal_sw}
and pretrigger system~\cite{opal_pretrigger} were added later.

\subsection{Central tracking and microvertex detectors}
The colliding beams at the centre of the OPAL detector were contained
in an airtight beryllium pipe with a thickness of~1.1~mm and a minimum
inner radius of 53~mm. Outside the beampipe, a second cylindrical tube
of inner radius~80~mm formed the inner wall of a 4~bar pressure vessel,
which contained the drift chambers of the central tracking system. A
silicon microvertex detector was located in the narrow annular region
between these two pipes.

\subsubsection{Silicon microvertex detector}

The OPAL microvertex detector was first installed in
1991~\cite{opalsilicon1}, and was subsequently upgraded in
1993~\cite{opalsilicon2} and 1995~\cite{opalsilicon3}. The beampipe was
surrounded by two layers of slightly overlapping silicon
`ladders'. The two faces of each ladder were divided perpendicularly
into narrow strips, to give a two-dimensional readout when a charged
particle passed through the ladder. By combining spatial position data
from the two layers of the microvertex detector, one could extrapolate
tracks back to the interaction point with high precision. It was
therefore possible, in many cases, to resolve the secondary decay
vertices associated with $\tau$~leptons and b~flavoured hadrons. One
could also distinguish pairs of high energy particles produced almost
parallel to one another, which would appear as a single track in the
drift chambers.

The resolution of the vertex position was approximately 20--50~$\mu$m
in the $z$~direction and 15~$\mu$m in the $r$-$\phi$~plane; the
azimuthal coverage of the detector was made almost complete in the
most recent upgrade, while the angular coverage in the polar direction
was increased to~$\left|\cos\theta\right|<0.89$.

\subsubsection[Central vertex detector, jet chamber and $z$-chambers]{\boldmath Central vertex detector, jet chamber and $z$-chambers}
\label{cv_cj_cz}

A system of drift chambers provided the principal momentum measurement
for charged particles. The chambers were contained in a sealed
cylindrical vessel around the beam axis, filled with 88.2\%~argon,
9.8\%~methane and 2.0\%~isobutane at 4~bar pressure. A solenoidal
magnet surrounded the curved outer surface of the pressure vessel,
providing a field of 0.435~T; the field was uniform at the 0.5\% level, and
was parallel to the beam axis, so that the beam itself was unperturbed.

\begin{figure}[tb]
\begin{center}
\includegraphics[width=0.7\textwidth]{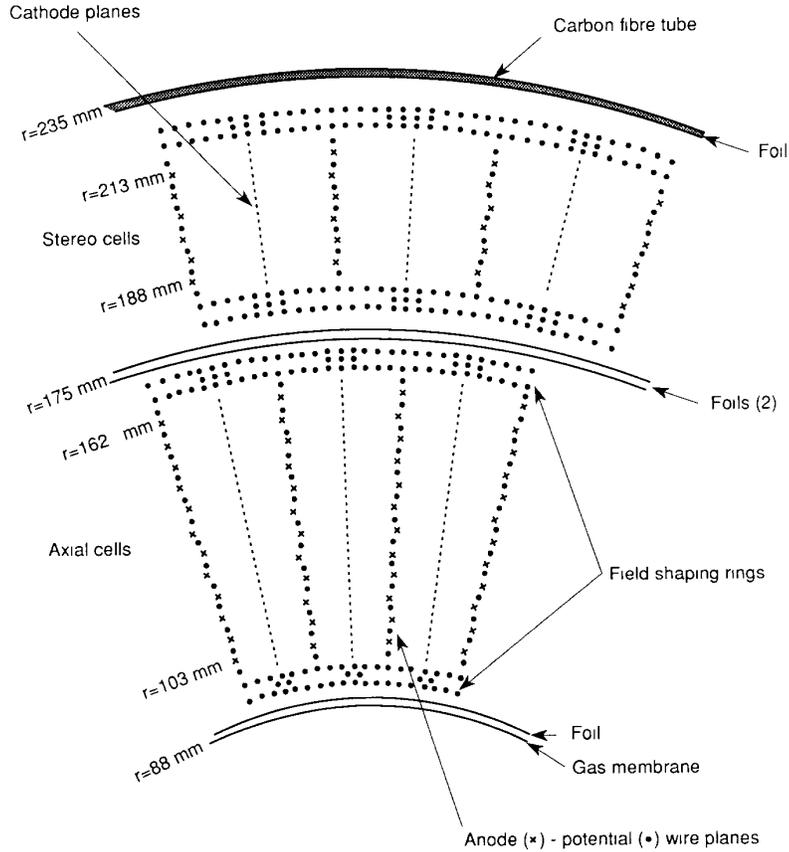}
\end{center}
\caption{A cross section through the OPAL central vertex detector}
\label{opalcv_xsect}
\end{figure}

The \textbf{central vertex detector}~\cite{opalcv} formed the inner
part of the tracking system, and originally fulfilled the same purpose
as the new silicon microvertex detector by extrapolating tracks back
to the interaction point. The chamber was 1~m long with a 470~mm
diameter, and was divided into two layers, each containing 36~identical
sectors, as shown in Figure~\ref{opalcv_xsect}. The inner `axial'
layer contained radial planes of high-voltage wires aligned parallel to
the beam axis; the electric field was directed perpendicular to the
anode planes. By measuring the drift times of electrons released by
the passage of a charged particle, one could determine the position of
the ionised gas molecule with a precision of about~50~$\mu$m in the
$r$-$\phi$~plane. A crude measurement of the $z$~coordinate was also
possible, by comparing the arrival times of the signal at each end of
the wire. This method was used for making fast real-time trigger
decisions, but was inadequate for offline analysis purposes. A second
`stereo' layer was therefore constructed, in which the anode and
cathode wires were not parallel to the beam; instead one of the
endplates was rotated by approximately~4$^\circ$ about the beam
axis. By combining the drift times from the axial and stereo layers,
it was possible to reconstruct the trajectory of a particle in three
dimensions.

The \textbf{jet chamber}~\cite{opalcj} was similar to the axial part of
the central vertex detector, but was 4~m long with an outer diameter of
3.7~m, and was divided into 24~sectors. By determining a large number
of points on the curved trajectory of a charged particle in a known
magnetic field, one could make a precise measurement of its transverse
momentum. The fractional resolution was estimated to be
\mbox{$\sigma_{p_\text{\tiny{T}}}/p_\text{\tiny{T}}=0.018\oplus
0.0016\,p_\text{\tiny{T}}$}, where the transverse
momentum~$p_\text{\tiny{T}}$ is in units of~GeV. The first term of the
uncertainty is due to multiple scattering, and the second is due to
the spatial resolution. The rate of energy loss due to ionisation,
d$E$/d$x$, could also be calculated, from the charge depositions
recorded on the wires. According to the Bethe-Bloch formula, d$E$/d$x$
is directly related to the speed $v/c$ of the particle; by combining
the momentum and speed, one obtains an estimate of the mass, and hence
the identity of the particle. Such identification was only possible for
particles of relatively low energy ($p \lesssim 10$~GeV), and is not
directly relevant to our event shape measurements, which are fully
inclusive quantities.

Of the three drift chamber systems, the jet chamber provided the
most complete angular coverage, with all polar angles in the range
$\left|\cos\theta\right|<0.98$ being covered by at least eight sense
wires.

Around the jet chamber was a cylindrical layer of 24~\textbf{\boldmath
$z$-chambers}~\cite{opalcz}, designed specifically to measure the
$z$~coordinates of the tracks. Each chamber was 50~cm wide and 59~mm
thick, and covered the full 4~m length of the jet chamber. The maximum
drift distance was about 25~cm in the $z$~direction, corresponding to a
resolution of approximately 200~$\mu$m.

In both the jet chamber and the $z$-chambers, it was possible to obtain
track coordinates in the direction parallel to the wires by measuring
the ratio of charges received at the two ends. However, far more
precise reconstruction was possible by combining the transverse
measurements from the jet chamber and $z$-chambers.

\subsection{Electromagnetic calorimeters}

For the purposes of our analysis, the data provided by the central
tracking chambers are supplemented most importantly by energy
measurements from the electromagnetic calorimeters.

The solenoid was surrounded by a cylindrical array of 9440~lead glass
blocks, which formed the barrel section of the electromagnetic
calorimeter system. Each block was $\mathord{\sim} 10 \mathord{\times} 
10$~cm in cross
section and 37~cm in depth, providing 24.6~radiation lengths for
photons travelling from the interaction point; the blocks were oriented
approximately towards the centre of the detector, with a slight tilt
to prevent particles escaping through the gaps. Two further arrays,
each comprising 1132~lead glass blocks, formed the endcap
electromagnetic calorimeters~\cite{opal_ee}: these blocks varied in
depth, but provided a minimum of 20.5~radiation lengths. Together, the
barrel and endcap sections offered complete azimuthal coverage over the
polar range~$\left|\cos\theta\right|<0.98$.

When a high-energy electron or positron entered the lead glass, it
would radiate bremsstrahlung photons due to the nuclear Coulomb fields
of the material; its energy would be reduced by a factor $e^{-1}$
for each radiation length of material. Some of the emitted photons would
then be converted to electron-positron pairs by further interaction with
the material: the mean free path for this process is 9/7 of the
radiation length. The result was an electromagnetic shower comprising
electrons, positrons and photons. The electrons and positrons would
also lose energy due to ionisation and \v{C}erenkov radiation, until
eventually the entire shower had been converted to low-energy
radiation and heat. Each block was wrapped with a reflective aluminium
layer, and was connected via a lightguide to a photomultiplier tube,
which detected the \v{C}erenkov photons. The number of observed photons
was proportional to the total track length contained in the
calorimeter, which in turn was proportional to the energy deposited.

Although the depth of the lead glass blocks ensured total absorption
of electrons, positrons and photons, a considerable fraction of the
initial energy carried by these particles was lost before they entered
the electromagnetic calorimeter. Approximately two radiation lengths
of material lay between the interaction point and the calorimeter,
mostly due to the solenoid and pressure vessel. A system of
\mbox{\textbf{presamplers}}~\cite{opal_ecal_presamplers} was therefore
installed in front of the calorimeters, to measure the numbers and
positions of particles produced through showering in this intervening
material; the number of particles present in the shower at this stage
provided a rough estimate of the energy already deposited. The
presamplers in the barrel region comprised two layers of streamer
tubes, while those in the endcaps consisted of thin high-gain multiwire
chambers.

The energy resolution of the electromagnetic calorimeters was estimated
to be $\sigma_{\scriptscriptstyle E}/E \,\approx\, 0.2\% \,\oplus\,
6.3\%/\sqrt{E[\text{GeV}]}\;$ in the barrel region, and
$\;\sigma_{\scriptscriptstyle E}/E \,\approx\,
5\%/\sqrt{E[\text{GeV}]}\;$ in the endcaps; the $E^{-1/2}$ factors
arise from statistical fluctuations in the number of \v{C}erenkov
photons, whose expected value is proportional to the incident
energy. The resolution for shower positions measured in the
presamplers was approximately 5~mm in the barrel and 2--5~mm in the
endcaps.

\subsection{Hadron calorimeters}

Outside the electromagnetic calorimeters, the return yoke of the
magnet absorbed the vast majority of hadrons emerging from the lead
glass, through nuclear interactions. In the barrel region, the yoke was
divided into eight concentric iron layers of thickness 100~mm, with
25~mm~spaces: these gaps contained streamer
tubes~\cite{opal_hcal_streamers}, which sampled the energy of the
shower. The toroidal endcaps of the magnet were similarly divided into
seven layers of iron, with 35~mm spaces containing streamer tubes. In
the forward regions, the magnet yoke was implemented as a poletip
hadron calorimeter, which extended angular coverage to
$\left|\cos\theta\right|<0.99$.

Since approximately two interaction lengths of material lay in front
of the hadron calorimeters, most hadronic showers were initiated in
the electromagnetic calorimeters. In order to measure the energy of an
incident hadron, one therefore needed to combine the energies recorded
by both calorimeter systems. The intrinsic energy resolution of the
hadron calorimeters was considerably lower than that of the
electromagnetic calorimeters: in all three sections (barrel, endcap
and poletip), the fractional uncertainty was
$\;\sigma_{\scriptscriptstyle E}/E \,\approx\,
120\%/\sqrt{E[\text{GeV}]}\;$.

\subsection{Muon detectors}

Muons were not absorbed significantly by the electromagnetic
calorimeter, due to their mass being far greater than that of the
electron. They could also penetrate the hadron calorimeters, since muons
do not couple to the strong interaction in nuclei. A further layer of
detectors was therefore constructed outside the return yoke of the
magnet, to measure the positions of muons and to distinguish them from
other particles.

Isolated high-energy muons, such as those produced in
\mbox{$\epem\to\mu^+\mu^-$} events, could be identified without the aid
of dedicated muon detectors: the signature consisted of a continuous
charged track in each subdetector, including the hadron
calorimeters. Such events were useful for studying the response of the
muon detector. However, the muons produced indirectly through decays
of heavy hadrons or $\tau$-leptons are perhaps of greater interest;
these had to be distinguished from a background of other
particles. Although the hadron calorimeter could still assist in the
identification of these muons, the principal signal was based on
extrapolation of tracks from the central drift chambers to the muon
detectors.

The muon detectors comprised a cylindrical arrangement of 110~drift
chambers surrounding the barrel~\cite{opal_mb}, and two orthogonal
layers of streamer tubes on each endcap~\cite{opal_me}. The combined
barrel and endcap muon detectors covered 93\% of $4\pi$~solid angle,
with gaps for the beampipe, cables and support structures. Within
this region of coverage, the acceptance for isolated muons with an
energy greater than 3~GeV was essentially~100\%. The probability of
misidentifying a 5~GeV pion as a muon was estimated to be less than
1\%.

\subsection{Time-of-flight detector}

\enlargethispage{-1\baselineskip}Between the jet chamber and the electromagnetic presamplers, a further
subdetector was installed to measure the arrival times of charged
particles. By comparing these times against the LEP bunch crossings,
which occurred every 22~$\mu$s,\footnote{LEP was designed to operate
with either 4+4 or 8+8 bunches of electrons and positrons,
corresponding respectively to periods of 22~$\mu$s and 11~$\mu$s
between bunch crossings. In practice, however, the 8+8
bunch mode was not used extensively.} it was possible for the trigger
system to reject cosmic ray backgrounds. To some extent one could also
perform particle identification, in the energy range 0.6--2.5~GeV, by
measuring deviations from the speed of light.

The barrel time-of-flight system, which was installed before OPAL
began taking data, consisted of 160~scintillation counters. Each of
these counters formed a strip of length 6.8~m, parallel to the beam
axis. Light was collected by phototubes at both ends; a positive
trigger required the two signals to arrive within 50~ns of each other,
and within 50~ns of the expected arrival time of a relativistic
particle from the beam interaction. The time resolution was estimated
to be 280~ps at the centre of the counters, and 350~ps at the ends.

When searching for rare signatures of new particles at LEP2, it was
especially important to eliminate spurious triggers caused by cosmic
rays. An endcap time-of-flight system~\cite{opal_te} was therefore
added in 1996, between the end of the pressure vessel and the
electromagnetic presamplers, to improve rejection of particles
arriving out-of-time with the beam crossing. This comprised an array
of thin scintillating tiles connected to remote photo-transducers via
optical fibres. A further set of tiles, closer to the beampipe, formed
a ``minimum ionising particle (MIP) plug'': this complemented the
poletip hadron calorimeter, by extending the acceptance region for
particles such as muons down to a polar angle of 43~mrad.

\subsection{Forward detectors and silicon-tungsten luminometers}

When measuring the cross section for a process such as Z$^0$
production, it is essential to determine the luminosity of the
machine. In \epem\ collisions, this is typically achieved by observing
low-angle Bhabha scattering of the beam particles; the differential
cross section for this process has been calculated with high precision
in QED, so it can be used as a calibration reference when studying
other processes. A pair of ``forward~detectors'' was therefore placed
in the regions close to the beampipe at either end of the
detector. Each forward detector comprised several components,
including calorimeters, drift chambers and tube chambers: these are
described in Ref.~\cite{nimpaper}.

A more precise silicon-tungsten calorimeter~\cite{opal_sw} was added
in 1993, consisting of 18~tungsten plates interleaved with 19~silicon
sampling wafers. While LEP was operating at 45~GeV per beam, this
detector could determine the positions of Bhabha electrons to a
precision of 0.2~mm, and their energies to $\pm 4\%$. The precision of
the resulting luminosity measurements was limited by a theoretical
systematic uncertainty of 0.05\%~\cite{sw_lumi_pr289}.

\subsection{Trigger system}

After each bunch crossing, a real-time trigger
system~\cite{opal_trigger} was needed to decide within 22~$\mu$s whether
an `interesting' event had occurred. If so, a signal was sent to each
subdetector requesting a full read-out of data collected from the
event. No further triggers could then be received until the read-out
had been completed, approximately 20~ms or 1000~bunch crossings
later. It was therefore essential to keep the trigger rate below about
5~Hz, so that the detector would be responsive more than 90\% of the time;
when running at the Z$^0$ energy with a typical luminosity of about
\mbox{$1\times10^{31}~\text{cm}^{-2}\text{s}^{-1}$}, the production
rate of Z$^0$ bosons was $\mathord{\sim} 0.4$~Hz.

The first stage of triggering was performed by the electronics of the
individual subdetectors. Each subdetector, or group of subdetectors,
computed a set of binary trigger outputs based on various
characteristics of the event. These signals were then combined in a
programmable central trigger logic, which decided whether the event was
to be accepted.

\newpage
The trigger signals provided by the subdetectors fell into two
categories, as follows:
\begin{description}
\item[The \boldmath $\theta$--$\phi$ matrix] The detector was divided
into a grid of overlapping bins of solid angle, in the~$\theta$
and~$\phi$ directions; the number of bins varied between subdetectors,
up to a maximum of 6~bins in~$\theta$ and 24~bins in~$\phi$. Every
subdetector made a trigger decision for each individual bin, with a
relatively low threshold. The central trigger logic could then test for
spatial correlations between layers of the detector. The
$\theta$--$\phi$ matrix is shown schematically in
Figure~\ref{trigger_matrix}.
\begin{figure}
\begin{center}
\includegraphics[width=0.95\textwidth]{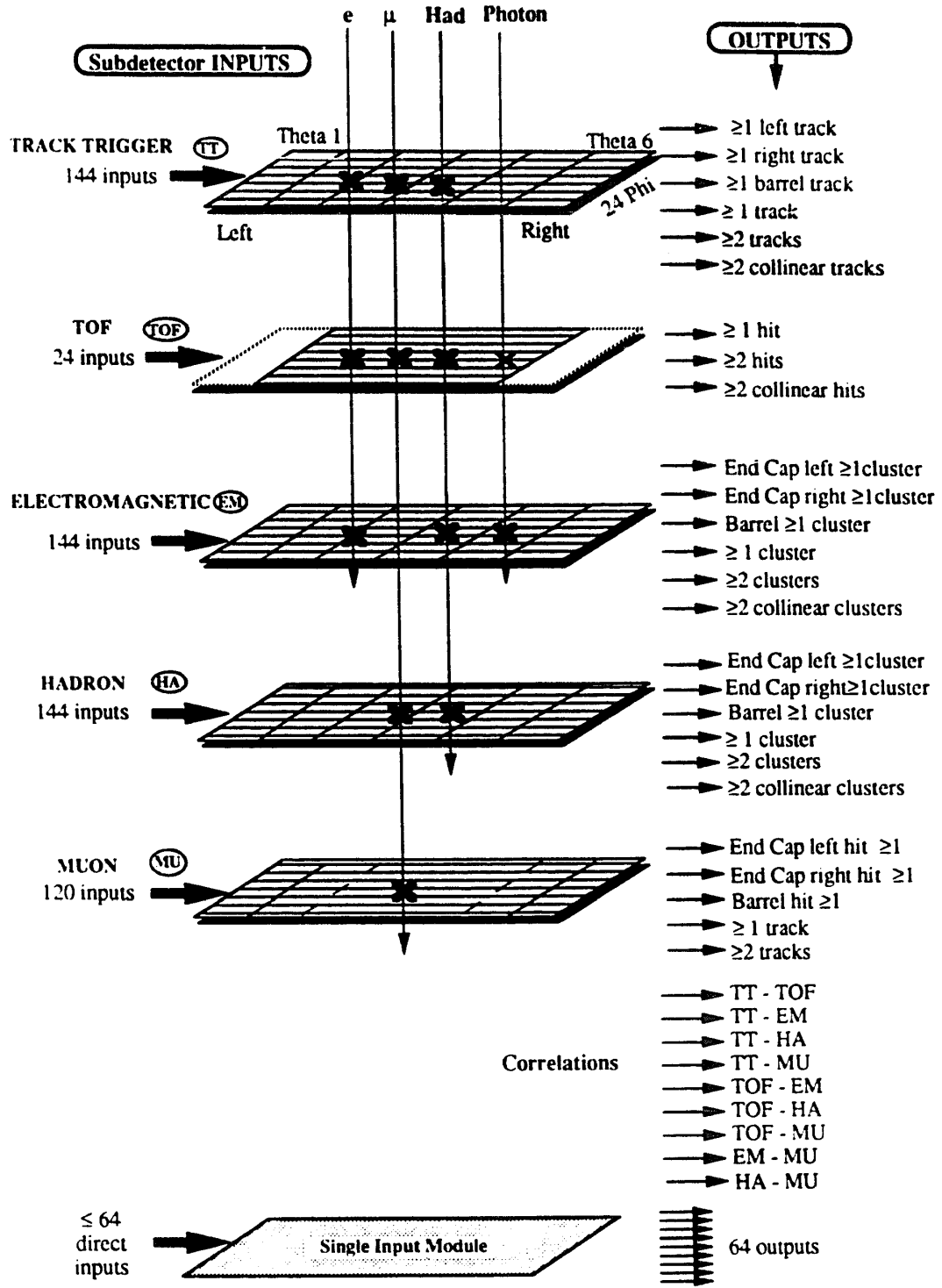}
\end{center}
\caption[The $\theta$--$\phi$ matrix used by the OPAL trigger
system]{A schematic view of the $\theta$--$\phi$ matrix used by the
OPAL trigger system. Crosses on the vertical lines represent
sensitivity of each subdetector to the various types of particle. The
matrix input bits were combined by the central trigger logic to give
the binary output signals listed in the right-hand column.}
\label{trigger_matrix}
\end{figure}
\item[Direct trigger signals] These were based on global properties of
the event, such as total energies or track multiplicities; they
generally had higher thresholds than the $\theta$--$\phi$ matrix
inputs. A full list of the direct signals from each subdetector is
given in Ref.~\cite{opal_trigger}.
\end{description}
The trigger signals from the central vertex and jet chambers were
generated by a sophisticated ``track~trigger''~\cite{opal_tt}. The
$z$-coordinate of each hit was measured rapidly by comparing the
charges or times of signals received at the two ends of the sense
wire.\footnote{As described in Section~\ref{cv_cj_cz}, a slower but
more precise measurement of the $z$-coordinate was obtained for
offline analysis purposes by combining data from the jet chamber and
$z$-chambers.} The ratio $z/r\equiv \cot\theta$ was then calculated to
determine the polar angle of the hit. A genuine track originating at
the interaction point would be expected to have the same $z/r$ value
for all hits, since the force due to the magnetic field acted in the
$\phi$~direction.

An additional pretrigger system~\cite{opal_pretrigger} was introduced
in 1992: this was primarily intended to speed up the decisions of the
central trigger logic, since it was planned that LEP should begin
operating in the 8+8 bunch configuration.

\subsection{Data acquisition, event builder and filter}

After a positive trigger decision had been made, the data from the
subdetectors were collected and merged by a central computer called the
``event~builder''; the hardware and software required to store,
process and transmit data from the digitisers of the subdetectors to
the event builder are described in Ref.~\cite{opal_daq}.

Each complete event was then processed by an online software
filter~\cite{opal_filter}, which acted as a second-level trigger. The
events were partially reconstructed, and classified into several
categories such as multihadron events or lepton pairs; events
classified as background were rejected. The filter software also
provided an online event display, and enabled real-time monitoring of
the detector performance. Events passed by the filter were stored on
disk to await offline reconstruction.

\section{Reconstruction of OPAL events}
\label{rope_section}

Before an event could be used for physics analysis, the raw data from
the subdetectors had to be converted into useful quantities such as
energies, momenta and particle identities. This process was performed
by a program called ROPE (``Reconstruction of OPAL Physics Events''),
which also provides the standard framework for accessing stored
data. The software contains a kernel and a collection of modules
associated with individual tasks and subdetectors; the modules were
developed, maintained and documented independently by their individual
authors. ROPE uses a database of calibration constants, which was
regularly updated to maintain accurate reconstruction; the raw data
were generally reprocessed several times to incorporate improvements in
the software and calibration.

The final reconstructed events were stored on data summary tapes (DST),
where they could be accessed for further offline analysis. Lists were
compiled of the events satisfying various criteria, to enable faster
access to specific classes of event. For some analyses, such as the
event shape measurements presented in this work, the relevant
properties of each event are stored in databases called `ntuples':
these are conveniently accessed using the \texttt{HBOOK} package of
the CERN program library, and can be shared between users with similar
data requirements.

\chapter[Measurements of \as\ at OPAL]{Measurements of \boldmath{\as} 
using event shape distributions at OPAL}
\label{opalchapter}

In Chapter~\ref{chapter:theory}, we defined a set of ``event shape
observables,'' which provide sensitivity to QCD interactions in
e$^+$e$^-$ annihilation. As discussed in
Section~\ref{pert_predictions}, the distributions of certain
observables have been predicted in perturbation theory at
\mbox{$\mathcal{O}(\as^2)+\mathrm{NLLA}$} precision. Our aim is to
measure these distributions experimentally, thereby testing the
validity of the predictions and permitting a measurement of the
coupling parameter~\as.

The OPAL Collaboration has previously published event shape
distributions and \as~measurements using data collected at
centre-of-mass energies in the range
\mbox{91--189~GeV~\cite{OPAL_as_91,OPAL_as_133,OPAL_as_161,OPAL_as_189}}. In
this work, we have extended this set of measurements to include the
full range of LEP collision energies up to 209~GeV. We have also
reprocessed the data recorded at lower energies, to incorporate
various experimental and theoretical developments which have taken
place during the lifetime of LEP. This unification permits a more
robust and consistent comparison between the results presented at
different energies.  The main improvements with respect to the
published OPAL results can be summarised as follows:
\begin{itemize}
\item Improvements have been made in the NLLA theory predictions
for the total jet broadening~\BT, the wide jet broadening~\BW, and the
Durham \ytwothree\ parameter. These were discussed in
Section~\ref{nlla_advances}, and are used throughout this work.
\item No measurements of the $C$-parameter were presented in the
original analysis of LEP1 data. Distributions were published at
$\sqrt{s}=$130--136~GeV~\cite{OPAL_as_133} and
161~GeV~\cite{OPAL_as_161}, but \as~measurements were not performed
using this observable until the NLLA resummation became available in
1998~\cite{cpar_resum}. Here we present distributions and \as~fits for
the $C$-parameter at all LEP energy scales.
\item The NLLA resummations do not automatically force each event
shape distribution to vanish at the edge of phase space; missing
subleading terms can result in a non-zero prediction outside the
kinematically allowed range of the
observable. In~Section~\ref{kinconstraints} we discussed a remedy
for this situation, involving the substitution
\[
L=\ln\left(\frac{1}{y}\right)\;\to\;\widetilde{L}=
\ln \left(\frac{1}{y}-\frac{1}{y_\mathrm{max}}+1\right)\;\;\;.
\]
This method was known~\cite{resum_catani} at the time of the original
LEP1 analysis~\cite{OPAL_as_91}, and was investigated as an
alternative to the unmodified NLLA prediction. However, it was never
adopted as the standard for \as\ measurements by the OPAL
Collaboration. In this work, we introduce the above substitution in
our fits to all OPAL data; this approach is now advocated by most
theorists, and has been adopted by the other three LEP experiments.
\item In 1996, the EVENT2 Monte Carlo program~\cite{event2}
became available, for the computation of event shape distributions at
$\mathcal{O}(\as^2)$ in perturbation theory, as described in
Section~\ref{fixedorder}. This superseded an earlier program,
EVENT~\cite{cpar_ellis}, which was used to generate the
coefficient functions $\mathcal{A}(y)$ and
$\mathcal{B}(y)$~\cite{zphyslep1} for previous LEP analyses. The two
algorithms are ultimately equivalent, but EVENT2 is more
efficient; coupled with the availability of much faster computers,
this has led to a more precise determination of the coefficients. We
therefore use these new $\mathcal{O}(\as^2)$ predictions throughout
this work.
\item The PYTHIA, HERWIG and ARIADNE Monte Carlo programs have evolved
considerably since the first OPAL event shape analyses were
published. In this work we use PYTHIA~6.1 for our central analysis,
and HERWIG~6.2 and ARIADNE~4.11 as alternatives in the estimation of
our systematic uncertainties. In each case we use the most recent
parameter set tuned to OPAL data.
\item A new event selection has been introduced to exclude the
background of four-fermion events from our analysis. This replaces a
cut used by OPAL in previous LEP2 publications, and significantly
increases the purity of our sample. We present an analysis and
justification of this selection in Section~\ref{fourfermioncuts}.
\item Before 1996, no satisfactory algorithm existed for relating the
charged tracks observed in the central detector with the energy
depositions recorded in the electromagnetic calorimeters. The
\texttt{MT}~package (``Matching Tracks and clusters,'' \cite{mt}) was
then developed to combine tracks and clusters, and reduce
double-counted energy.
\item The handling of statistical uncertainties in both the event
shape distributions and the \as\ measurements has changed. In the
past, the uncertainties of the distributions were estimated directly
from the data, and no statistical correlations were calculated between
different bins. We now compute a full covariance matrix for each
distribution, using Monte Carlo simulations, and then use this matrix
when fitting for~\as.
\end{itemize}

The remainder of this chapter is organised as follows: in
Section~\ref{sigdef} we define our signal, and list the dominant
background processes.  Section~\ref{datasamples} lists the OPAL data
and Monte Carlo samples to be used in the analysis. In
Section~\ref{selection} we describe our event selection criteria, and
list the numbers of events selected at each centre-of-mass energy.
Section~\ref{evshmeasure} discusses our event shape measurements, and
compares the resulting distributions with those predicted by Monte
Carlo models; the complete results are in
Appendix~\ref{evshdistappendix}. In Section~\ref{asfits}, we fit
theoretical predictions to our data, and discuss the resulting
measurements of~\as, which are tabulated in
Appendix~\ref{asfitappendix}. Finally, in
Section~\ref{opal_as_combinations}, we discuss the combination of OPAL
results obtained at different energy scales and from different
observables.

\section{Signal definition}
\label{sigdef}

The aim of this analysis is to compare measured event shape
distributions with those predicted by perturbative QCD. Our
experimental signal definition must therefore match, as closely as
possible, the class of processes included in the theoretical
calculations.

In Section~\ref{qcdpert}, we described the basic
$\mathrm{e}^+\mathrm{e}^-\to\left(\mathrm{Z}^0/\gamma\right)^*\to\mathrm{q}\bar{\mathrm{q}}$
process, and an associated class of higher-order diagrams involving
gluon radiation from the $\mathrm{q}\bar{\mathrm{q}}$ pair. The set of
resulting final states defines a statistical population for the event
shape distributions discussed in Section~\ref{pert_predictions}. All
electroweak physics is ``factored out'' from these predictions,
through normalisation of the distributions.

\begin{figure}
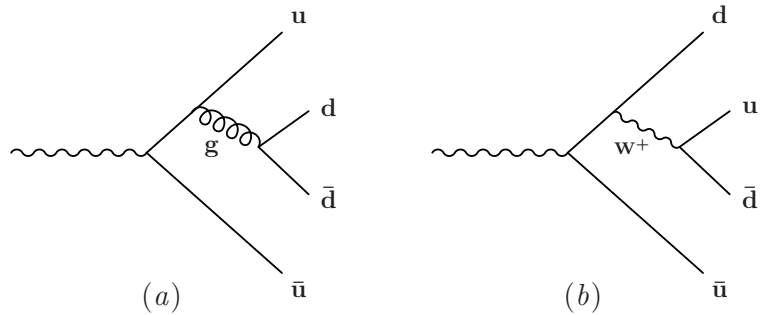

\begin{center}
\scalebox{0.8}{
\begin{tabular}{c @{\hspace{2cm}} c}
$\phantom{DUMMY}$ \\
\end{tabular}}
\end{center}
\caption{Two Feynman diagrams contributing to the
$\mathrm{u}\bar{\mathrm{u}}\mathrm{d}\bar{\mathrm{d}}$ final state}
\label{uudd}
\end{figure}

In reality, of course, one can never define processes or observables
that depend \emph{only} on strong interactions. Quarks are charged
particles, which can radiate real or virtual photons, and also W$^\pm$
and Z$^0$ bosons.\footnote{In principle, H$^0$ bosons may also be
exchanged.  However, direct searches have found no conclusive evidence
for Higgs production at LEP~\cite{lep_higgs}.} Diagrams involving
electroweak gauge bosons may in principle interfere with the
``pure~QCD'' diagrams, so even statistical separation is not strictly
possible. Figure~\ref{uudd}, for example, shows two of the diagrams
contributing to the
$\mathrm{u}\bar{\mathrm{u}}\mathrm{d}\bar{\mathrm{d}}$ final state:
diagram~(\emph{a}) is included in the QCD event shape predictions,
while diagram~(\emph{b}) is not.  Fortunately, however, there is
negligible interference between these processes. The W$^\pm$ and Z$^0$
bosons are narrow high-energy resonances, with total widths of 2.1~GeV
and 2.5~GeV respectively~\cite{PDbook}; their decay products tend to
be well-separated, with large invariant masses. Gluons, in contrast,
are usually emitted at a low angle with respect to the parent quark,
and carry a small momentum transfer. Hence there is very little
overlap in phase space between the two processes shown in
Figure~\ref{uudd}. Furthermore, the vast majority of four-jet final
states arising from QCD processes are q$\bar{\mathrm{q}}$gg, rather
than q$\bar{\mathrm{q}}$q$\bar{\mathrm{q}}$; there is no electroweak
correction to these processes at tree level.  We therefore claim that
the set of QCD `multihadron' processes described in
Section~\ref{qcdpert} constitutes a valid signal definition, and that
electroweak four-fermion processes such as Figure~\ref{uudd}(\emph{b})
form a separable background.

\begin{figure}
\begin{center}
\scalebox{0.8}{
\begin{fmfframe}(1,1)(1,0){
\begin{fmfgraph*}(7,3)
\fmfpen{thin}
\fmfleftn{i}{4}
\fmfrightn{o}{4}
\fmf{plain, tension=1}{i1,v3}
\fmf{plain, tension=1}{i4,v4}
\fmf{plain, tension=2}{v3,v1}
\fmf{plain, tension=2}{v4,v1}
\fmf{photon, label=\boldmath $\mathbf{Z}^0/\gamma^*$}{v1,v2}
\fmf{plain, tension=1}{o1,v5}
\fmf{plain, tension=1}{o4,v6}
\fmf{plain, tension=2}{v5,v2}
\fmf{plain, tension=2}{v6,v2}
\fmffreeze
\fmf{gluon, tension=1}{v6,v8}
\fmf{phantom, tension=3}{v8,o3}
\fmf{photon, tension=1}{v5,v7}
\fmf{phantom, tension=3}{v7,o2}
\fmflabel{\boldmath $\textbf{e}^+$}{i1}
\fmflabel{\boldmath $\textbf{e}^-$}{i4}
\fmflabel{\boldmath $\bar{\mathbf{q}}$}{o1}
\fmflabel{\boldmath $\mathbf{q}$}{o4}
\fmflabel{\boldmath $\gamma$}{v7}
\fmflabel{\boldmath $\mathbf{g}$}{v8}
\end{fmfgraph*}}
\end{fmfframe}}
\end{center}
\caption{Multihadronic event with final-state photon radiation (FSR)}
\label{fsr}
\end{figure}
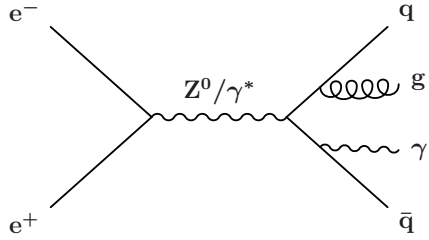

For experimental reasons, we will also include events with
final-state photon radiation in our signal. An example is shown in
Figure~\ref{fsr}.
Such photons tend to be radiated at low angles, and often cannot be
distinguished from the decay products of the corresponding quark jet.
We therefore make no attempt to identify these events as background.
Their influence is expected to be small, since the electromagnetic
coupling $\alpha_\mathrm{em}$ is much smaller than \as, and the
angular distribution of FSR photons is similar to that of gluons at
lowest order.

\section{Event samples}
\label{datasamples}

\subsection{OPAL data}

Our analysis will use all available data recorded in OPAL physics runs
at centre-of-mass energies $\sqrt{s}\geq 130$~GeV, during the years
1995--2000. In most cases, the data are concentrated densely around
well-defined energy points, as reflected by the ranges listed in
Table~\ref{datasummary}. It is natural to treat each of these ranges
separately, so that the running coupling~\asq\ can be regarded as
constant within each data sample. However, the data at 130.1~and
136.1~GeV will be merged into a single sample with mean
$\sqrt{s}=133.3$~GeV, to reduce statistical uncertainties.

\begin{table}
\begin{center}
\begin{tabular}{| c | r@{$\;$--$\;$}l | c | c |}
\hline
Year &
\multicolumn{2}{|c|}{\parbox{2.5cm}{\centering Range of $\sqrt{s}$\\(GeV)}} &
\multicolumn{1}{|c|}{\parbox{2.5cm}{\centering Mean $\sqrt{s}$\\(GeV)}} &
\multicolumn{1}{|c|}{\parbox{2.1cm}{\centering \rule{0pt}{0.4cm}Integrated\\luminosity (pb$^{-1}$)\rule[-0.2cm]{0pt}{0pt}}}
\bigstrut \\ \hline
1996--2000 &  91.0 &  91.5 &  91.3 &  14.7 \bigstrut[t] \\
1995, 1997 & 129.9 & 130.3 & 130.1 &  5.31 \\
1995, 1997 & 135.7 & 136.3 & 136.1 &  5.95 \\
1996       & 161.2 & 161.6 & 161.3 & 10.06 \\
1996       & 170.2 & 172.5 & 172.1 & 10.38 \\
1997       & 180.8 & 184.2 & 182.7 & 57.72 \\
1998       & 188.3 & 189.1 & 188.6 & 185.2 \\
1999       & 191.4 & 192.1 & 191.6 & 29.53 \\
1999       & 195.4 & 196.1 & 195.5 & 76.67 \\
1999, 2000 & 199.1 & 200.2 & 199.5 & 79.27 \\
1999, 2000 & 201.3 & 202.1 & 201.6 & 37.75 \\
2000       & 202.5 & 205.5 & 204.9 & 82.01 \\
2000       & 205.5 & 208.9 & 206.6 & 138.8 \bigstrut[b] \\
\hline
\end{tabular}
\end{center}
\caption{The OPAL data samples used for our analysis}
\label{datasummary}
\end{table}

In addition, several ``calibration runs'' were made during the same
period, at the Z$^0$ energy $\sqrt{s}=M_\mathrm{Z}=91$~GeV. These
primarily served to provide calibration constants for the experiments'
reconstruction software, and to test the response of the detectors in
a known physics environment with a large annihilation cross
section. However, these new data at 91~GeV can also be used for
physics analysis. Since the detector configuration was unchanged
between the calibration runs and the high-energy runs, we expect a
large degree of correlation between the experimental systematic
uncertainties; this is extremely beneficial when measuring the
energy-dependence of a quantity such as~\asq. Our analysis of the
calibration data will therefore supersede previous measurements
performed at LEP1.

\subsection{Monte Carlo events}
\label{opalmcsamples}

\enlargethispage{-1\baselineskip}In order to optimise our selection cuts, and to correct for biases in
our measurements, we require a faithful simulation of the entire
experiment. Standard event generators are used to simulate the various
types of signal and background processes, using parameters tuned to
OPAL data at LEP1. Where necessary, the full response of the OPAL
detector is estimated for each event using the GOPAL simulation
program~\cite{gopal}, and the simulated events are then reconstructed
using ROPE, as described in Section~\ref{rope_section}. The event
generators to be used in this analysis are as follows:
\begin{itemize}
\item {\boldmath $\mathrm{q}\bar{\mathrm{q}}$} events are generated
using $\mathcal{KK}$2f linked with PYTHIA~6.125, as described in
Section~\ref{mcmodels}. Events involving photon radiation from the
initial or final states are included. Some of the default PYTHIA
parameters are replaced with optimised values, based on OPAL data at
the Z$^0$ peak~\cite{pr141}. When testing the model-dependence of our
analysis, and estimating systematic uncertainties, we use HERWIG~6.2
and ARIADNE~4.11 as alternative models to simulate parton showering;
the HERWIG cluster model also provides an alternative to the PYTHIA
string model for fragmentation. These two programs were also described
in Section~\ref{mcmodels}.  The corresponding OPAL parameter sets were
listed in Ref.~\cite{pr379_tunings}.  The same initial
$\mathrm{q}\bar{\mathrm{q}}$ states, simulated by $\mathcal{KK}$2f,
are used in the PYTHIA, HERWIG and ARIADNE event samples.
\item For certain purposes, such as the prediction of hadronisation
effects, we do not require simulation of the OPAL detector, nor the
inclusion of initial-state radiation. In these cases we generate much
larger samples of non-radiative $\mathrm{q}\bar{\mathrm{q}}$ events,
using the OPAL-tuned versions of PYTHIA~6.158
\footnote{The OPAL parameter set for PYTHIA version~6.158 is the same
as that for version~6.125, except that the parameter \texttt{PARJ(55)}
describing b-quark fragmentation has been changed from $-0.0038$ to
$-0.0020$.}, HERWIG~6.2 and ARIADNE~4.11. We do not use
$\mathcal{KK}$2f to generate the initial
$\mathrm{q}\bar{\mathrm{q}}$ systems, since this is only beneficial for
radiative events.
\item As we shall see in Section~\ref{fourfermioncuts}, a significant
background arises from W$^+$W$^-$ and ZZ pair production, at energies
$\sqrt{s} \geq 161$~GeV: the two on-shell gauge bosons decay to
produce a four-fermion final state.
\begin{itemize}
\item For all {\boldmath
$\mathrm{q}\bar{\mathrm{q}}\mathrm{q}\bar{\mathrm{q}}$,
$\mathrm{q}\bar{\mathrm{q}}\ell^\pm\nu_\ell$,
$\mathrm{q}\bar{\mathrm{q}}\ell^+\ell^-$ and
$\mathrm{q}\bar{\mathrm{q}}\nu\bar{\nu}$} background final states at
\mbox{$\sqrt{s}\geq 183$~GeV} (except
$\mathrm{q}\bar{\mathrm{q}}\mathrm{e}^+\mathrm{e}^-$), we use the
KoralW~\cite{koralw} generator, version~1.42~\cite{koralw1.42}. This
uses four-fermion matrix elements calculated with grc4f~\cite{grc4f}
version~2.1, including interference between W$^+$W$^-$ and ZZ
diagrams. KoralW also features accurate predictions for initial- and
final-state photon radiation, which may interfere with one another.
\item For {\boldmath
$\mathrm{q}\bar{\mathrm{q}}\mathrm{e}^+\mathrm{e}^-$} final states,
and for all four-fermion processes at $\sqrt{s}=161$~GeV and 172~GeV,
no KoralW samples are available. We instead use grc4f~2.1. The
simulation of $\mathrm{q}\bar{\mathrm{q}}\mathrm{e}^+\mathrm{e}^-$
processes excludes `multiperipheral' diagrams such as two-photon
processes, which will be rejected efficiently by the selection cuts.
\end{itemize}
\enlargethispage{-1\baselineskip}Hadronisation of the four-fermion
final states is simulated using JETSET~7.4, which contains a slightly
older version of the string fragmentation model used by PYTHIA~6.1.
\end{itemize}

\section{Event selection}
\label{selection}

We now describe in turn each of the selection criteria
applied to our data. For each case, we indicate the class of
background events to be removed.

\subsection{Detector status}
\label{detstatuscut}

To ensure the best possible precision, we require that all relevant
components of the detector are fully operational:
\begin{itemize}
\item The status flags for the central jet chamber, and for both the
barrel and endcap electromagnetic calorimeter systems, must indicate `OK'.
\item At least two out of the following three inputs to the trigger
system must be working, to ensure near-perfect efficiency for
detecting multihadronic events:
\begin{itemize}
\item[(i)] the track trigger
\item[(ii)] the time-of-flight detector
\item[(iii)] both the barrel and endcap electromagnetic calorimeter
systems.
\end{itemize}
\end{itemize}

\subsection[LEP2 multihadron selection (L2MH)]{LEP2 multihadron selection (L2MH)\footnote{For our analysis of calibration data at $\sqrt{s}=91$~GeV, we use the Tokyo multihadron selection~(TKMH) instead of~L2MH. TKMH~is defined in exactly the same way as L2MH, except that the cut on visible energy is slightly looser ($R_\mathrm{vis} \ge 0.10$), and the cut on energy balance is slightly tighter ($\left|R_\mathrm{bal}\right| \le 0.65$). These values are chosen to optimise acceptance of multihadron events, in a regime where background processes and initial-state radiation can be neglected.}}

This standard selection is used for many OPAL analyses. It efficiently
eliminates background events with low multiplicities, such as electron
or muon pair production, and also those with missing energy or
momentum. Untagged and single-tagged two-photon events are rejected,
because the electron or positron is usually not deflected sufficiently
to be observed in the calorimeter.

\begin{itemize}
\item The event must include at least five good tracks in the central
detector. A ``good track'' is defined as follows:
\begin {itemize}
\item Momentum transverse to the beam axis: $p_\mathrm{T} \ge
50~\mathrm{MeV}/c$
\item Number of hits in the jet chamber: $N_\mathrm{hits} \ge 20$
\item The track is extrapolated to find the point $P$ where it passes
closest to the beam axis. The displacement of $P$ from the nominal
interaction point must satisfy the following conditions, when
expressed in cylindrical polar coordinates:\vspace{-0.0cm}
\begin{center}
\begin {tabular}{rl}
$r-\phi$ component (normal to the beam axis): & $d_0 \le 2$~cm \\
$z$ component (parallel to the beam axis): & $|z_0| \le 25$~cm
\end {tabular}
\end{center}
\end {itemize}
\item At least seven good energy clusters are required in the
electromagnetic calorimeters. ``Good~clusters'' are defined as those
having raw energies of at least 100~MeV in the barrel region or
200~MeV in the endcap region.
\item The visible energy ratio, $R_{\mathrm{vis}}$, must satisfy
\[
R_{\mathrm{vis}} \; \equiv \; \frac{\sum E_{\mathrm{raw}}}{2E_{\mathrm{be
am}}}\; \ge\; 0.14
\]
\item The energy balance ratio, $R_{\mathrm{bal}}$, must satisfy
\[
R_{\mathrm{bal}} \; \equiv \; \left| \frac{\sum E_{\mathrm{raw}} \cos \theta }
{\sum E_{\mathrm{raw}}} \right| \; \le \; 0.75
\]
\end {itemize}

\subsection{Number of accepted tracks}

As described above, the L2MH selection requires at least five
``good~tracks'' to be observed in the central detector. For this
analysis we increase the minimum number of tracks to seven, in order
to reduce the remaining background from two-photon and $\tau^+\tau^-$
events to a negligible level. According to Monte Carlo simulations at
$\sqrt{s}=189$~GeV, we expect 98.5\% of signal events passing the L2MH
selection to satisfy this additional criterion.

\subsection{Containment in the detector}

For each event, we find the direction of the thrust axis \nth\ as
defined in Section~\ref{thrust_def}. We require the angle between
\nth\ and the LEP beam axis, $\theta_\mathrm{T}$, to satisfy
\mbox{$|\,\mathrm{cos}~\theta_T\, | \le 0.9$}. This reduces the
probability of final-state hadrons being lost in the beampipe, which
would bias the event shape observables.

\subsection[Rejection of radiative events ($\sqrt{s'}$ cut)]{Rejection of radiative events (\boldmath $\sqrt{s'}$ cut)}
\label{isrcut}

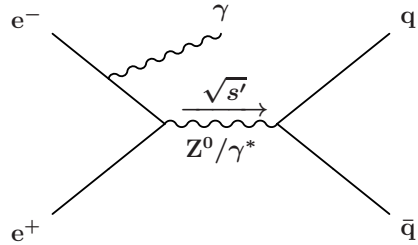
\begin{figure}
\begin{center}
\scalebox{0.8}{
\begin{fmfframe}(1,0.5)(1,0){
\begin{fmfgraph*}(7,3)
\fmfpen{thin}
\fmfleftn{i}{2}
\fmfrightn{o}{2}
\fmftopn{t}{1}
\fmf{plain, tension=1}{i1,v3}
\fmf{plain, tension=1}{i2,v4}
\fmf{plain, tension=1}{v3,v1}
\fmf{plain, tension=1}{v4,v1}
\fmf{photon, label=\boldmath $\mathbf{Z}^0/\gamma^*$}{v1,v2}
\fmf{plain, tension=1}{o1,v5}
\fmf{plain, tension=1}{o2,v6}
\fmf{plain, tension=1}{v5,v2}
\fmf{plain, tension=1}{v6,v2}
\fmffreeze
\fmf{photon, tension=1}{v4,t1}
\fmflabel{\boldmath $\textbf{e}^+$}{i1}
\fmflabel{\boldmath $\textbf{e}^-$}{i2}
\fmflabel{\boldmath $\bar{\mathbf{q}}$}{o1}
\fmflabel{\boldmath $\mathbf{q}$}{o2}
\fmflabel{\boldmath $\gamma$}{t1}
\fmfv{label.dist=0.04w, label.angle=28, label=\boldmath $\underrightarrow{\phantom{A}\textstyle \sqrt{s'}\phantom{A}}$}{v1}
\end{fmfgraph*}}
\end{fmfframe}}
\end{center}
\caption{Multihadronic event with initial-state photon radiation (ISR)}
\label{isr}
\end{figure}

At LEP2 collision energies, the majority of $\text{q}\bar{\text{q}}$
events include at least one photon radiated from the electron or
positron in the initial state, as shown in Figure~\ref{isr}. In most
cases, the photon carries an energy
\begin{equation}
E_\gamma \; \approx \; \frac{1}{2}\left(\sqrt{s}-\frac{M_{\text{Z}^0}^2}{\sqrt{s}}\right) \;\;\;,
\end{equation}
leaving an electron-positron pair with exactly the energy needed to
create a Z$^0$~boson on the mass shell.
Such events are regarded as background, because the quark-antiquark
pairs are not created with the desired centre-of-mass energy, and
would lead to a measurement of the running coupling \asq\ at the wrong
energy scale.

An algorithm has been developed by the OPAL
Collaboration~\cite{sprime} to determine whether an event contains
initial-state radiation. Firstly, any isolated photons seen in the
electromagnetic calorimeter are immediately identified as ISR. The
remaining charged tracks and calorimeter clusters are then formed into
jets using the Durham algorithm. A kinematic fit is performed,
allowing for up to two unseen ISR photons, and imposing
energy-momentum conservation for the whole event. All ISR photons are
then discarded from the event, and the invariant mass $\sqrt{s'}$ of
the remaining hadronic system computed. Events are selected only if
\begin{equation}
\sqrt{s}-\sqrt{s'}\;\le\;10~\mathrm{GeV} \;\;\;.
\end{equation}
This algorithm is compared in Ref.~\cite{sprime} with an older method,
which is still used in the present analysis to estimate systematic
uncertainties. In addition to eliminating radiative multihadronic
events, the above cut removes the vast majority of the remaining
two-photon and $\mathrm{q}\bar{\mathrm{q}}\mathrm{e}^+\mathrm{e}^-$
events. The isolated electrons and positrons in these events are
identified as converted ISR photons by our $\sqrt{s'}$ algorithm.

In principle, the cross section for the process
$\mathrm{e}^+\mathrm{e}^- \to \mathrm{q}\bar{\mathrm{q}}\gamma$ will
include contributions from both initial- and final-state
radiation. ISR diagrams such as Figure~\ref{isr} will interfere with
FSR diagrams such as Figure~\ref{fsr}. Fortunately, however,
final-state radiation is much rarer than ISR at LEP2 energies, and
tends to occupy a different region of phase space; most FSR photons
have low momenta, and are detected close to a quark jet. A detailed
investigation~\cite{ainsley} has shown that ignoring ISR/FSR
interference leads to an error of order 0.1\% in measurements of the
non-radiative multihadron cross section at LEP2 energies.

\subsection{Four-fermion rejection cuts}
\label{fourfermioncuts}

After imposing the criteria described in
Sections~\ref{detstatuscut}--\ref{isrcut}, Monte Carlo simulations
show that about 65--70\% of selected events at the highest LEP2
energies fall within our signal definition.  Nearly all of the
remaining events are due to `four-fermion' processes involving the
production of an on-shell W$^+$W$^-$ or ZZ pair. Such events can
closely resemble our signal, especially when both of the W$^\pm$ or
Z$^0$ bosons decay hadronically, as occurs in about 50\% of
cases. Fortunately however, as we have discussed in
Section~\ref{sigdef}, the $\mathrm{q}\bar{\mathrm{q}}$ and
four-fermion processes tend to populate different regions of phase
space. Hence there is little interference, and we have some
possibility of separating them on an event-by-event basis.

\begin{figure}[p]
\begin{center}
\resizebox{0.8\textwidth}{!}{
\begin{tabular}{cc}
\resizebox{0.4\textwidth}{!}{\begin{fmfframe}(0.5,0.5)(0.5,0.5){
\begin{fmfgraph*}(6,7.5)
\fmfpen{thin}
\fmfstraight
\fmfleftn{i}{5}
\fmfrightn{o}{5}
\fmftopn{t}{4}
\fmf{plain, tension=2}{i2,v1,v2,v3,o2}
\fmf{plain, tension=2}{i4,v4,v5,v6,o4}
\fmf{photon, tension=2, label=\boldmath $\mathbf{Z}^0/\gamma$}{v2,v5}
\fmffreeze
\fmf{photon, label.side=left, label=\boldmath $\mathbf{Z}^0/\gamma$, tension=2}{v4,v7}
\fmf{plain, tension=3}{v7,v9}
\fmf{plain, tension=2}{v7,v8}
\fmf{phantom}{v8,o4}
\fmf{phantom, tension=3.5}{v9,t3}
\fmflabel{\boldmath $\textbf{e}^+$}{i2}
\fmflabel{\boldmath $\textbf{e}^-$}{i4}
\fmflabel{\boldmath $\textbf{e}^+$}{o2}
\fmflabel{\boldmath $\textbf{e}^-$}{o4}
\fmflabel{\boldmath $\mathrm{q}$}{v9}
\fmfv{label=\boldmath $\bar{\mathrm{q}}$, label.angle=0}{v8}
\end{fmfgraph*}}\end{fmfframe}}
\hspace{0.3cm} &
\resizebox{0.4\textwidth}{!}{\begin{fmfframe}(0.5,0.5)(0.5,0.5){
\begin{fmfgraph*}(6,7.5)
\fmfpen{thin}
\fmfstraight
\fmfleftn{i}{5}
\fmfrightn{o}{5}
\fmfbottomn{b}{4}
\fmf{plain, tension=2}{i2,v1,v2,v3,o2}
\fmf{plain, tension=2}{i4,v4,v5,v6,o4}
\fmf{photon, tension=2, label=\boldmath $\mathbf{Z}^0/\gamma$}{v2,v5}
\fmffreeze
\fmf{photon, label.side=right, label=\boldmath $\mathbf{Z}^0/\gamma$, tension=2}{v1,v7}
\fmf{plain, tension=3}{v7,v9}
\fmf{plain, tension=2}{v7,v8}
\fmf{phantom}{v8,o2}
\fmf{phantom, tension=3.5}{v9,b3}
\fmflabel{\boldmath $\textbf{e}^+$}{i2}
\fmflabel{\boldmath $\textbf{e}^-$}{i4}
\fmflabel{\boldmath $\textbf{e}^+$}{o2}
\fmflabel{\boldmath $\textbf{e}^-$}{o4}
\fmflabel{\boldmath $\bar{\mathrm{q}}$}{v9}
\fmfv{label=\boldmath $\mathrm{q}$, label.angle=0}{v8}
\end{fmfgraph*}}\end{fmfframe}}
\\
\end{tabular}}
\end{center}
\caption{$t$-channel exchange diagrams leading to
$\mathrm{q}\bar{\mathrm{q}}\mathrm{e}^+\mathrm{e}^-$ final states}
\label{qqeediagrams}
\end{figure}
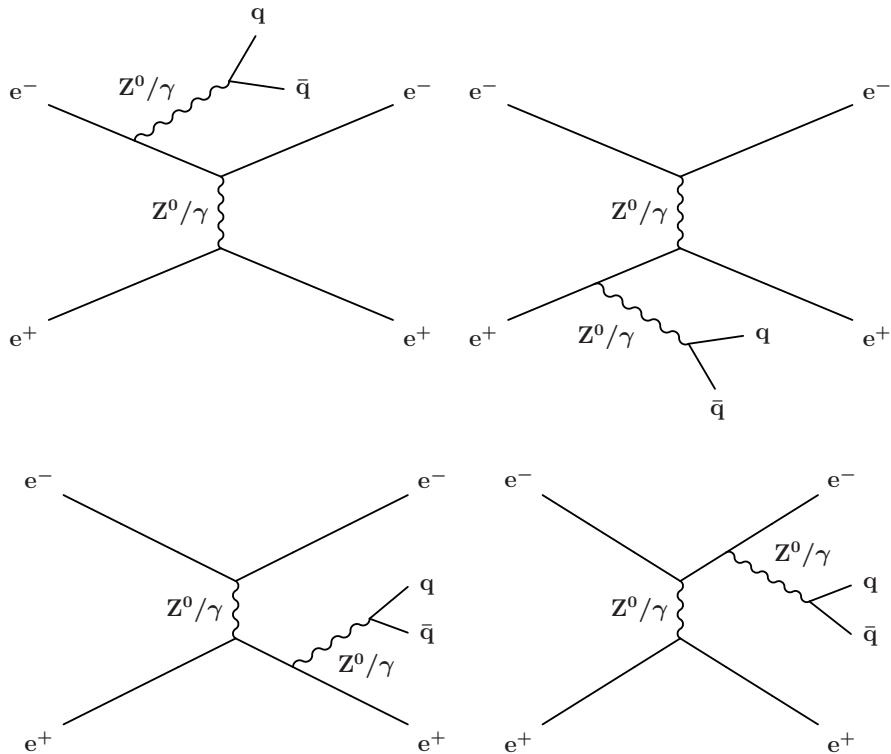

The main four-fermion backgrounds are
$\mathrm{q}\bar{\mathrm{q}}\mathrm{q}\bar{\mathrm{q}}$,
$\mathrm{q}\bar{\mathrm{q}}\ell^\pm\nu_\ell$ and
$\mathrm{q}\bar{\mathrm{q}}\ell^+\ell^-$ final states, produced via
the processes shown in Figure~\ref{4fbackgrounddiagrams}. In addition,
there are some extra $t$-channel exchange diagrams, shown in
Figure~\ref{qqeediagrams}, which lead to
$\mathrm{q}\bar{\mathrm{q}}\mathrm{e}^+\mathrm{e}^-$ final states. The
latter are already excluded to a satisfactory level by our $\sqrt{s'}$
cut, along with two-photon events, because the final-state electron or
positron tends to pass undetected into the beampipe. For convenience,
we will use the term ``qq$\ell\nu$~background'' to imply both
$\mathrm{q}\bar{\mathrm{q}}\ell^\pm\nu_\ell$ and
$\mathrm{q}\bar{\mathrm{q}}\ell^+\ell^-$ states, of which the former
is a more significant contribution.

\begin{figure}[p]
\begin{center}
\vspace{0.5cm}
\hspace{-1.5cm}
\resizebox{0.35\textwidth}{!}{\begin{fmfgraph*}(7,4)
\fmfpen{thin}
\fmfleftn{i}{2}
\fmfrightn{o}{4}
\fmf{plain, tension=1}{i1,v1}
\fmf{plain, tension=1}{i2,v1}
\fmf{photon, tension=2, label=\boldmath $\mathbf{Z}^0/\gamma$}{v1,v2}
\fmf{photon, tension=1, label=\boldmath $\mathrm{W}^-$}{v2,v4}
\fmf{photon, tension=1, label=\boldmath $\mathrm{W}^+$}{v3,v2}
\fmf{plain, tension=1}{v3,o1}
\fmf{plain, tension=1}{v3,o2}
\fmf{plain, tension=1}{v4,o3}
\fmf{plain, tension=1}{v4,o4}
\fmflabel{\boldmath $\textbf{e}^+$}{i1}
\fmflabel{\boldmath $\textbf{e}^-$}{i2}
\fmflabel{\boldmath $\bar{\mathrm{d}},\bar{\mathrm{s}},\bar{\mathrm{b}},\nu_\ell$}{o1}
\fmflabel{\boldmath $\mathrm{u},\mathrm{c},\ell^+$}{o2}
\fmflabel{\boldmath $\mathrm{d},\mathrm{s},\mathrm{b},\bar{\nu}_\ell$}{o3}
\fmflabel{\boldmath $\bar{\mathrm{u}},\bar{\mathrm{c}},\ell^-$}{o4}
\end{fmfgraph*}}
\hspace{2.0cm}
\resizebox{0.30\textwidth}{!}{
\begin{fmfgraph*}(6,4)
\fmfpen{thin}
\fmfstraight
\fmfleftn{i}{2}
\fmfrightn{o}{4}
\fmf{plain, tension=1}{i1,v1}
\fmf{plain, tension=1}{i2,v2}
\fmf{plain, tension=1, label=\boldmath $\nu_\mathrm{e}$}{v1,v2}
\fmf{photon, tension=1, label=\boldmath $\mathrm{W}^+$}{v3,v1}
\fmf{photon, tension=1, label=\boldmath $\mathrm{W}^-$}{v2,v4}
\fmf{plain, tension=1}{v3,o1}
\fmf{plain, tension=1}{v3,o2}
\fmf{plain, tension=1}{v4,o3}
\fmf{plain, tension=1}{v4,o4}
\fmflabel{\boldmath $\textbf{e}^+$}{i1}
\fmflabel{\boldmath $\textbf{e}^-$}{i2}
\fmflabel{\boldmath $\bar{\mathrm{d}},\bar{\mathrm{s}},\bar{\mathrm{b}},\nu_\ell$}{o1}
\fmflabel{\boldmath $\mathrm{u},\mathrm{c},\ell^+$}{o2}
\fmflabel{\boldmath $\mathrm{d},\mathrm{s},\mathrm{b},\bar{\nu}_\ell$}{o3}
\fmflabel{\boldmath $\bar{\mathrm{u}},\bar{\mathrm{c}},\ell^-$}{o4}
\end{fmfgraph*}}
\vspace{1.5cm}\\
\resizebox{0.30\textwidth}{!}{
\begin{fmfgraph*}(6,4)
\fmfpen{thin}
\fmfstraight
\fmfleftn{i}{2}
\fmfrightn{o}{4}
\fmf{plain, tension=1}{i1,v1}
\fmf{plain, tension=1}{i2,v2}
\fmf{plain, tension=1, label=\boldmath $\mathrm{e}^-$}{v1,v2}
\fmf{photon, tension=1, label=\boldmath $\mathrm{Z}^0$}{v3,v1}
\fmf{photon, tension=1, label=\boldmath $\mathrm{Z}^0$}{v2,v4}
\fmf{plain, tension=1}{v3,o1}
\fmf{plain, tension=1}{v3,o2}
\fmf{plain, tension=1}{v4,o3}
\fmf{plain, tension=1}{v4,o4}
\fmflabel{\boldmath $\textbf{e}^+$}{i1}
\fmflabel{\boldmath $\textbf{e}^-$}{i2}
\fmflabel{\boldmath $\bar{\mathrm{q}},\ell^+$}{o1}
\fmflabel{\boldmath $\mathrm{q},\ell^-$}{o2}
\fmflabel{\boldmath $\bar{\mathrm{q}},\ell^+$}{o3}
\fmflabel{\boldmath $\mathrm{q},\ell^-$}{o4}
\end{fmfgraph*}}
\end{center}
\caption{Principal four-fermion background diagrams}
\label{4fbackgrounddiagrams}
\end{figure}
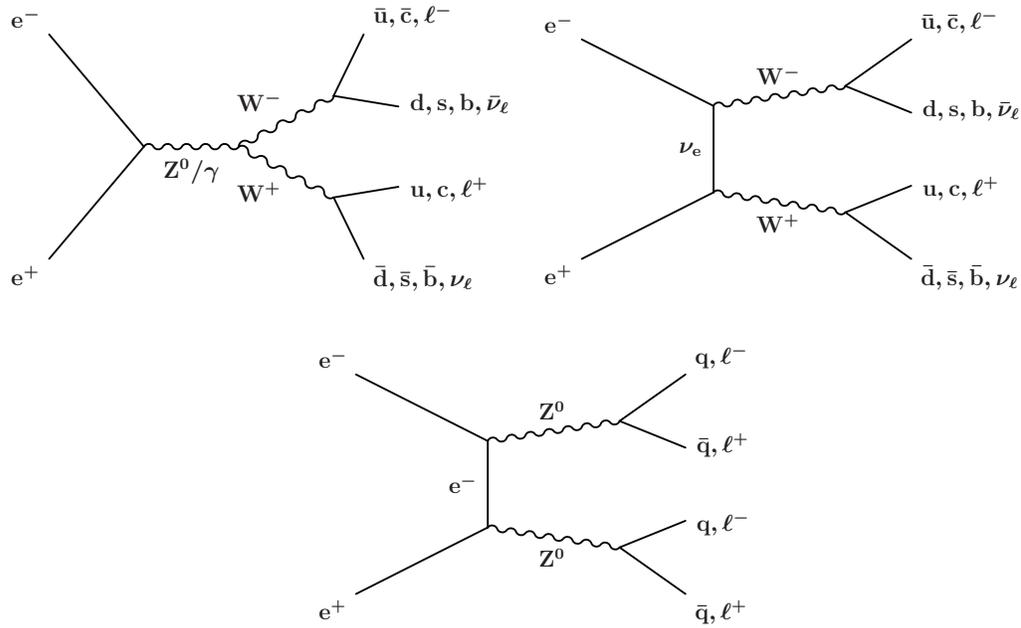

\begin{figure}
\begin{center}
\includegraphics[width=0.9\textwidth]{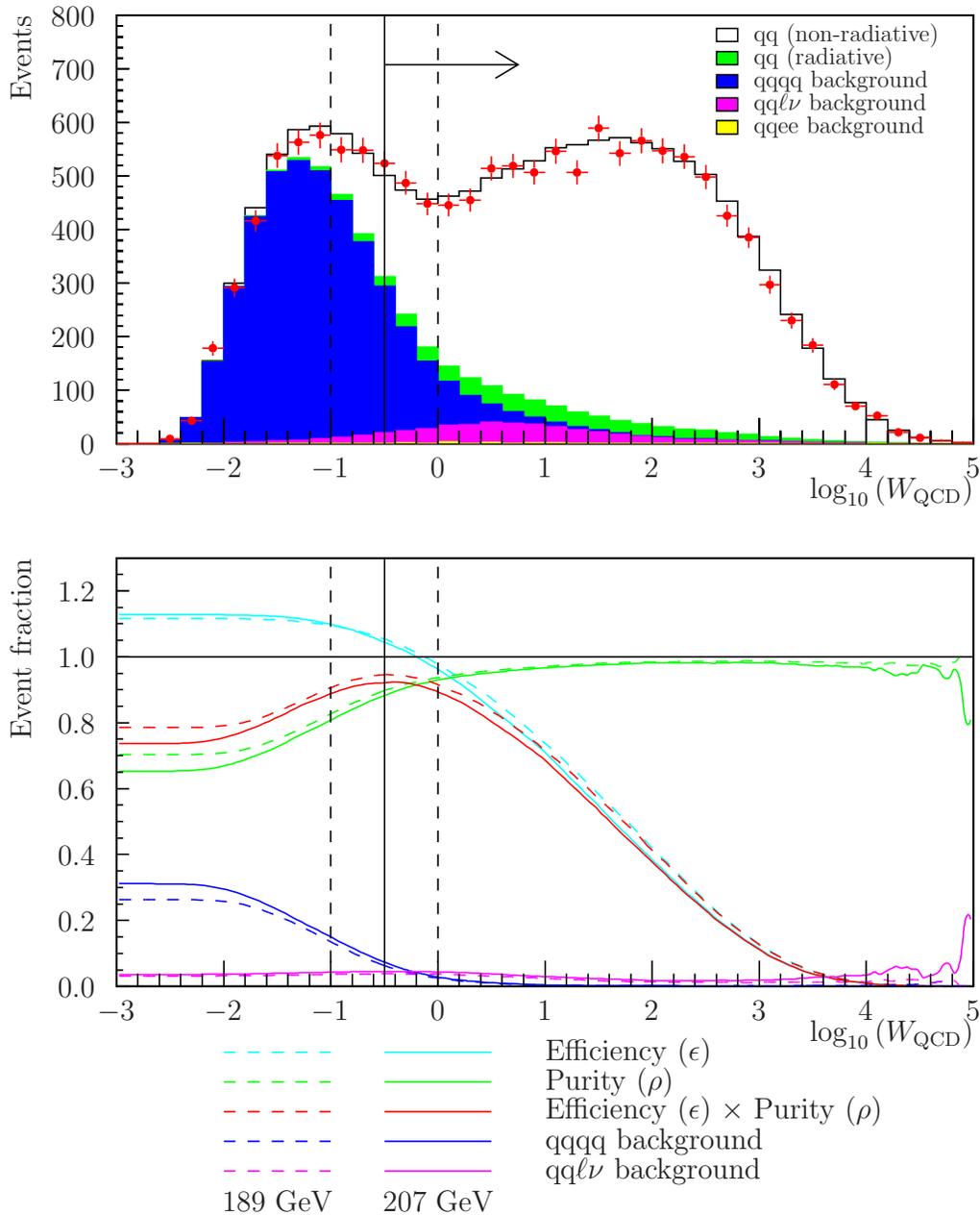}
\end{center}
\vspace{-0.5cm}
\caption[Distribution and event selection properties for the
$W_\mathrm{QCD}$ observable used in previous OPAL event shape
analyses]{Distribution and event selection properties for the
$W_\mathrm{QCD}$ observable used in previous OPAL event shape
analyses. In the upper plot, OPAL data are compared with MC
predictions for the distribution; the data points represent all
selected OPAL events at energies $\sqrt{s}\ge 189~\mathrm{GeV}$, after
imposing the cuts defined in
Sections~\ref{detstatuscut}--\ref{isrcut}. The solid histograms
represent Monte Carlo predictions, using KK2f/PYTHIA~6.15 for the
$\text{q}\bar{\text{q}}$~events, KoralW~for the qqqq and qq$\ell\nu$
backgrounds, and grc4f 2.1 for the qqee background. The
``hadron~level'' events predicted by each Monte Carlo program have
been subjected to a full simulation of the OPAL detector. In the lower
plot, we show the efficiency and purity of the selection as a function
of the cut position, calculated using Monte Carlo simulations at
$\sqrt{s}=189$~GeV and 207~GeV. Also shown are the numbers of qqqq and
qq$\ell\nu$ events expected to pass the selection, as fractions of the
total selected events. The chosen position for the cut is indicated by
a solid vertical line; the dashed lines indicate alternative cuts,
which are used to estimate systematic uncertainties.}
\label{w420sel}
\end{figure}

In previous studies of QCD event shapes at LEP2~\cite{OPAL_as_189},
the following procedure has been used to estimate the compatibility of
each event with our signal definition. The event is first forced into
a four-jet configuration using the Durham algorithm, as described in
Section~\ref{durham_algorithm}. The EVENT2~\cite{event2} program is
then used to compute the matrix element
\mbox{$\mathcal{M}(p_1,p_2,p_3,p_4)$} for each of the possible
processes
$\mathrm{e}^+\mathrm{e}^-\to\mathrm{q}\bar{\mathrm{q}}\mathrm{q}\bar{\mathrm{q}},\mathrm{q}\bar{\mathrm{q}}\mathrm{gg}$. Since
there is no identification of quark and gluon jets, the matrix
elements for all permutations of jet momenta
\mbox{$\left\{p_1,p_2,p_3,p_4\right\}$} are considered. We define the
discriminator variable $W_\mathrm{QCD}$ as the largest of these:
\footnote{$W_\mathrm{QCD}$ is sometimes called $W_{420}$, for example
in Ref.~\cite{pr321}.}
\begin{equation}
W_\mathrm{QCD}=\max_{\left\{p_1,p_2,p_3,p_4\right\}}\Big[\,{\big|\mathcal{M}(p_1,p_2,p_3,p_4)\big|}^2\,\Big]\;\;\;.
\end{equation}
The expected and measured distributions of this observable are shown
in Figure~\ref{w420sel}, for events which have passed the selection
criteria defined in Sections~\ref{detstatuscut}--\ref{isrcut}. The
$W_\mathrm{QCD}$ values for signal events are generally higher than
those for four-fermion events. Although most events to be selected are
not four-jet processes of the type used to compute $W_\mathrm{QCD}$,
the final state particles can nonetheless be resolved into four jets,
of which some pairs are nearly parallel. Since the probability of
collinear gluon emission is high, the matrix element for this four-jet
state will be closely related to that for the `true' two- or
three-jet state. The cut
\begin{equation}
\log_{10}(W_\mathrm{QCD})>-0.5
\end{equation}
is imposed to complete the event selection. After all cuts have been
applied we would then expect 77\% of genuine signal events to be
selected, and 88\% of selected events to be
$\text{q}\bar{\text{q}}$~processes (including those with
ISR).\footnote{These statistics are calculated at $\sqrt{s}=207$~GeV
using the PYTHIA, KoralW and grc4f event generators, and a full
simulation of the OPAL detector.}

In a recent OPAL publication~\cite{pr362}, however, it has been
suggested that the above selection can be improved. While the
$W_\mathrm{QCD}$ variable is an effective measure of compatibility
with our signal definition, it does not contain any information about
an event's consistency with the four-fermion background processes. In
an OPAL measurement of the W$^+$W$^-$ production cross section and
branching fractions~\cite{pr321}, two likelihood variables were used
to select
\mbox{$\mathrm{e}^+\mathrm{e}^-\to\mathrm{W}^+\mathrm{W}^-\!\!/\mathrm{ZZ}\to\mathrm{q}\bar{\mathrm{q}}\mathrm{q}\bar{\mathrm{q}}$}
and
\mbox{$\mathrm{e}^+\mathrm{e}^-\to\mathrm{W}^+\mathrm{W}^-\to\mathrm{q}\bar{\mathrm{q}}\ell^\pm\nu_\ell$}
events. These discriminators, called \Lqqqq\ and \Lqqln\ respectively,
are based on several input variables. \Lqqqq, which is defined in
Ref.~\cite{pr321}, uses both $W_\mathrm{QCD}$ and the corresponding
matrix elements for four-fermion processes; \Lqqln\ is based primarily
on the identification of isolated leptons~\cite{lqqln}. The following
cuts were adopted in Ref.~\cite{pr362} to reject four-fermion
background:
\begin{equation}
\Lqqqq<0.25 \;,\hspace{1cm}
\Lqqln<0.5 \;,\hspace{1cm}
\log_{10}(W_\mathrm{QCD})>-0.5 \;\;\;.
\end{equation}
In the remainder of this section, we will discuss the implementation
of these new cuts. First, however, we will describe in general terms a
criterion for optimising selection algorithms.

\subsubsection{Minimisation of the statistical uncertainty}
\label{effpur_max}

We wish to choose our selection cuts such that the uncertainty in our
final measurements will be minimised. Unfortunately it is difficult to
calculate reliably the dependence of our systematic
uncertainties on the cut parameters; however, the treatment of
statistical uncertainties is straightforward.

Suppose we expect the following numbers of events to be produced at
our interaction point, in a given time interval:
\begin{itemize}
\item $N_\mathrm{nr}$ non-radiative $\text{q}\bar{\text{q}}$ events
($\sqrt{s}-\sqrt{s'}\leq 1$~GeV)
\item $N_\mathrm{r}$ radiative $\text{q}\bar{\text{q}}$ events
($\sqrt{s}-\sqrt{s'}>1$~GeV)
\item $N_\mathrm{b}$ background events
\end{itemize}
The efficiencies for observing these events in our detector, and
selecting them as signal, are $\epsilon_\mathrm{nr}$,
$\epsilon_\mathrm{r}$ and~$\epsilon_\mathrm{b}$ respectively. The
total number of selected events $N_\mathrm{data}$, and its associated
standard deviation, is therefore \footnote{$N_\mathrm{nr}$,
$N_\mathrm{r}$ and $N_\mathrm{b}$ are \emph{constants}, representing
the mean number of events in each category, while $N_\mathrm{data}$ is
a random variable with a statistical distribution.}
\begin{equation}
N_\mathrm{data}\,=\,\epsilon_\mathrm{nr}N_\mathrm{nr}+
\epsilon_\mathrm{r}N_\mathrm{r}+ \epsilon_\mathrm{b}N_\mathrm{b}
\;\pm\sqrt{ \epsilon_\mathrm{nr}N_\mathrm{nr}+
\epsilon_\mathrm{r}N_\mathrm{r}+ \epsilon_\mathrm{b}N_\mathrm{b} }
\;\;\;.
\end{equation}
After subtracting the expected background,\footnote{The background
subtraction and detector correction will be discussed in the next
section.} we have
\begin{equation}
N_\mathrm{data}-\epsilon_\mathrm{b}N_\mathrm{b}
\,=\,\epsilon_\mathrm{nr}N_\mathrm{nr}+
\epsilon_\mathrm{r}N_\mathrm{r}
\;\pm\sqrt{
\epsilon_\mathrm{nr}N_\mathrm{nr}+
\epsilon_\mathrm{r}N_\mathrm{r}+
\epsilon_\mathrm{b}N_\mathrm{b}
} \;\;\;.
\end{equation}
We now multiply this result by a ``detector correction,'' such that
the expected value is equal to the original number of non-radiative
signal events,~$N_\mathrm{nr}$:
\begin{eqnarray}
N_\mathrm{corr.}
& = & \frac{N_\mathrm{nr}
\left(N_\mathrm{data}-\epsilon_\mathrm{b}N_\mathrm{b}\right)}
{\epsilon_\mathrm{nr}N_\mathrm{nr}+\epsilon_\mathrm{r}N_\mathrm{r}}
\;=\;
N_\mathrm{nr}\pm
\frac{N_\mathrm{nr}\sqrt{
\epsilon_\mathrm{nr}N_\mathrm{nr}+
\epsilon_\mathrm{r}N_\mathrm{r}+
\epsilon_\mathrm{b}N_\mathrm{b}
}}{\epsilon_\mathrm{nr}N_\mathrm{nr}+\epsilon_\mathrm{r}N_\mathrm{r}}
\nonumber \\
\rule{0pt}{1cm} & = &
N_\mathrm{nr}\pm
\sqrt{N_\mathrm{nr}\left(\frac{N_\mathrm{nr}}
{\epsilon_\mathrm{nr}N_\mathrm{nr}+\epsilon_\mathrm{r}N_\mathrm{r}}\right)
\left(\frac{
\epsilon_\mathrm{nr}N_\mathrm{nr}+
\epsilon_\mathrm{r}N_\mathrm{r}+
\epsilon_\mathrm{b}N_\mathrm{b}
}{\epsilon_\mathrm{nr}N_\mathrm{nr}+\epsilon_\mathrm{r}N_\mathrm{r}}\right)}
\nonumber \\
\rule{0pt}{1cm} & \equiv &
N_\mathrm{nr}\pm
\sqrt{\frac{N_\mathrm{nr}}{\epsilon \rho}}\;\;\;,
\end{eqnarray}where we have defined
\begin{eqnarray}
\textrm{Efficiency:} & & \epsilon =
\frac{\epsilon_\mathrm{nr}N_\mathrm{nr}+
\epsilon_\mathrm{r}N_\mathrm{r}}{N_\mathrm{nr}}
\rule[-0.8cm]{0pt}{0pt} \label{effdef} \\ \textrm{Purity:} & & \rho =
\frac{\epsilon_\mathrm{nr}N_\mathrm{nr}+
\epsilon_\mathrm{r}N_\mathrm{r}}{ \epsilon_\mathrm{nr}N_\mathrm{nr}+
\epsilon_\mathrm{r}N_\mathrm{r}+
\epsilon_\mathrm{b}N_\mathrm{b}}\;\;\;.
\label{purdef}
\end{eqnarray}
To minimise the statistical uncertainty, we must therefore maximise
the product of efficiency and purity,~$\epsilon\rho$.

In the absence of radiative events ($N_\mathrm{r}=0$), our definitions
for $\epsilon$ and~$\rho$ reduce to their familiar form: the
efficiency~$\epsilon$ is the probability for a given signal event to
be selected, and the purity~$\rho$ is the probability for a given
selected event to be signal. In this analysis, however, the
`efficiency' is not constrained to satisfy \mbox{$0\leq\epsilon\leq1$}, and
cannot be interpreted as a probability.\footnote{If the denominator in
Equation~(\ref{effdef}) were changed from $N_\mathrm{nr}$ to
\mbox{$N_\mathrm{nr}+N_\mathrm{r}$}, we could interpret $\epsilon$ as
the average selection probability of a $\text{q}\bar{\text{q}}$ event,
including those with initial-state radiation. With this definition,
the statistical uncertainty of~$N_\mathrm{corr.}$ would still be
minimised by maximising $\epsilon\rho$. However, the
`efficiency'~$\epsilon$ would then be very small, because most
radiative events are deliberately rejected by our cut on~$\sqrt{s'}$.}
This anomaly arises because the residual contribution from radiative events is
not removed from our measurements by subtraction; instead it is
included in a multiplicative ``detector~correction'' to be described
in Section~\ref{evshmeasure}. When interpreting our results, we will
use $\epsilon_\textrm{nr}$, which is a more conventional measure of
efficiency, involving only non-radiative $\text{q}\bar{\text{q}}$ events:
\begin{equation}
\epsilon_\textrm{nr}\;=\;\lim_{\delta\to 0}\left[\;
P\,\big(\;\textrm{event is selected} \;\; \big| \;\; \sqrt{s}-\sqrt{s'}<\delta\;\big)
\;\right] \;\;\;.
\end{equation}

\subsubsection[The $W_\mathrm{QCD}$, \Lqqqq\ and \Lqqln\ selection cuts]
{\boldmath The $W_\mathrm{QCD}$, \Lqqqq\ and \Lqqln\ selection cuts}
\label{4flikelihood}

In Figure~\ref{w420sel}, we have shown the distribution of the QCD
event weight $W_\mathrm{QCD}$, for events passing the cuts defined
in Sections~\ref{detstatuscut}--\ref{isrcut}. Also shown are the
efficiency~$\epsilon$, the purity~$\rho$, and their
product~$\epsilon\rho$, as a function of the cut position. The
left-hand side of the plot corresponds to a very `loose' cut passed by
every event, while the right-hand side represents a `tight' cut which
rejects all events. As the cut is tightened, the efficiency~$\epsilon$
decreases and the purity~$\rho$ rises. The product~$\epsilon\rho$
reaches a maximum when $W_\mathrm{QCD}^\mathrm{cut}\approx -0.5$, as
indicated by the solid vertical line. This optimised cut value is in
agreement with the choice made in previous OPAL
analyses~\cite{OPAL_as_189}.

\begin{figure}
\begin{center}
\includegraphics[width=\textwidth]{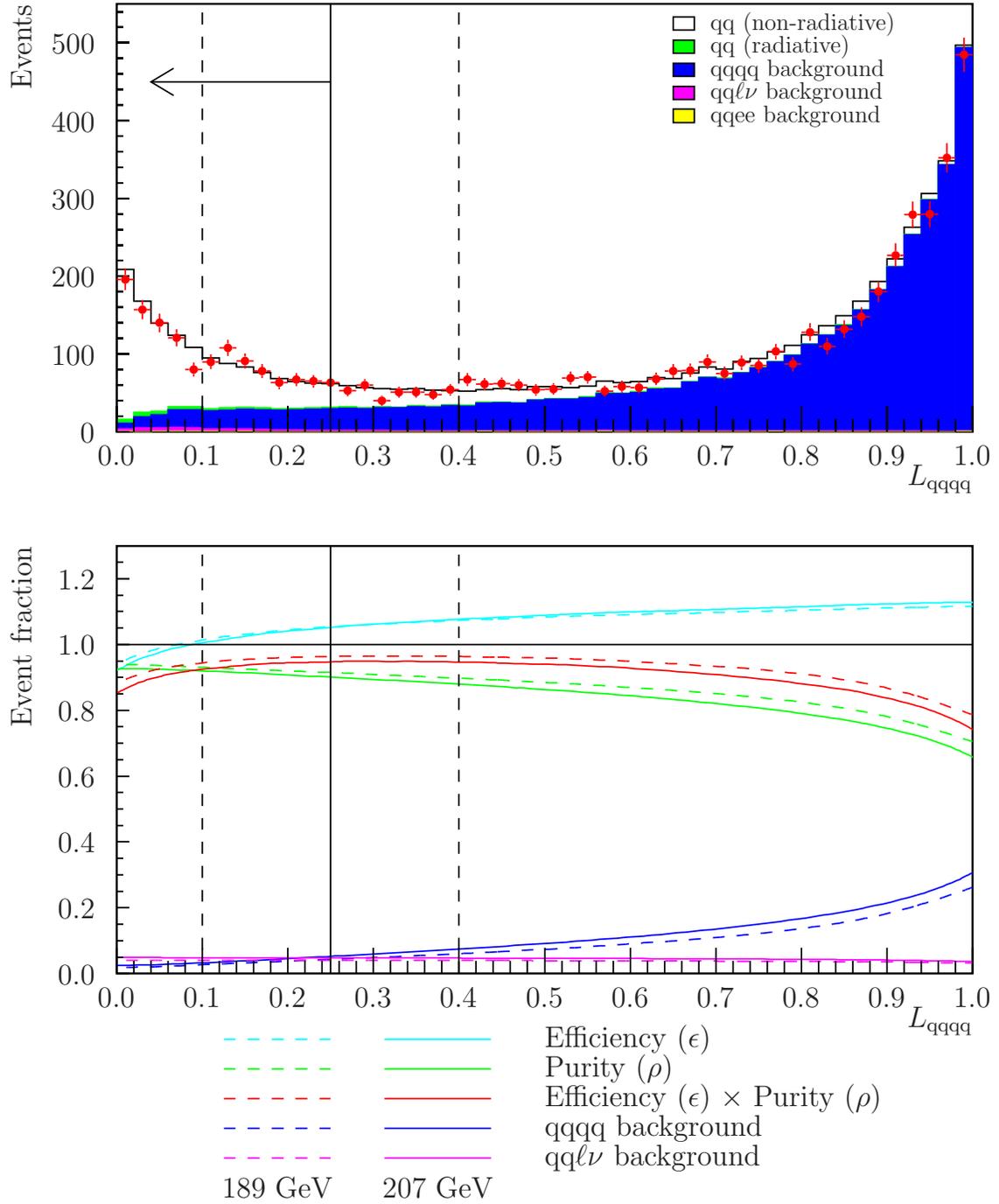}
\end{center}
\caption{Distribution and event selection properties for the \Lqqqq\
likelihood observable. Events with $\Lqqqq=-1$ have been excluded from
the upper plot. See the caption of Figure~\ref{w420sel} for further
details.}
\label{lqqqqsel}
\end{figure}

Treating the $\mathrm{q}\bar{\mathrm{q}}\mathrm{q}\bar{\mathrm{q}}$
likelihood variable \Lqqqq\ in the same way, we obtain the
distribution shown in Figure~\ref{lqqqqsel}. Some events, which fail a
`preselection' in the calculation of \Lqqqq, are given the value
$\Lqqqq=-1$. After applying the cuts listed in
Sections~\ref{detstatuscut}--\ref{isrcut}, we estimate that the events
in this category amount to 83\% of remaining
$\mathrm{q}\bar{\mathrm{q}}$ events, and 81\% of remaining qq$\ell\nu$
background events, but only 4\% of remaining
$\mathrm{q}\bar{\mathrm{q}}\mathrm{q}\bar{\mathrm{q}}$ events. These
events are omitted from Figure~\ref{lqqqqsel}.  The peak
of~$\epsilon\rho$ with respect to \Lqqqq\ is much broader than in the
$W_\mathrm{QCD}$ case; the selection would be well optimised by any
cut in the range \mbox{$\Lqqqq<0.1$} to \mbox{$\Lqqqq<0.5$}. For
consistency with previous analyses, we choose the cut
\mbox{$\Lqqqq<0.25$}.

\begin{figure}
\begin{center}
\includegraphics[width=\textwidth]{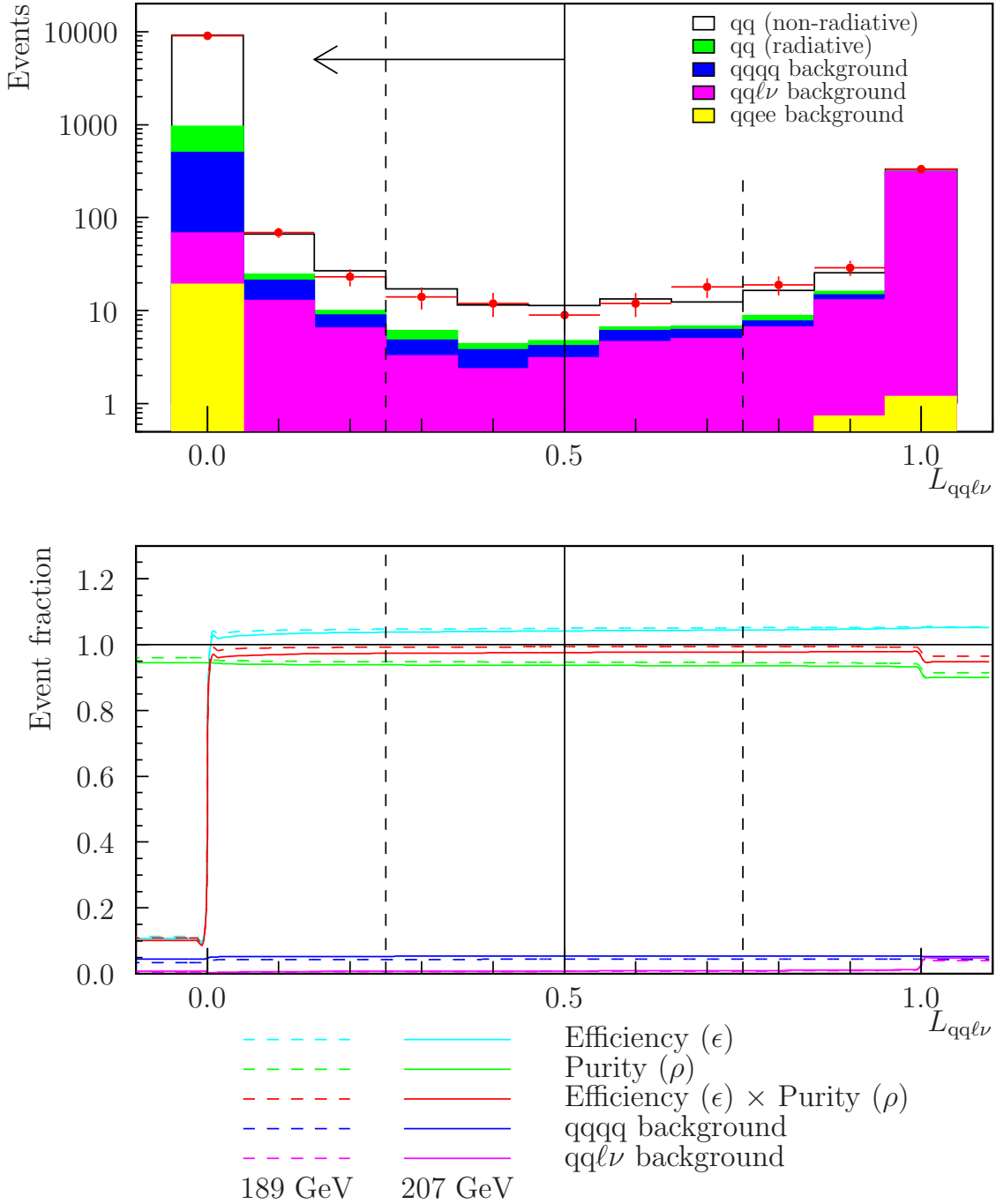}
\end{center}
\caption{Distribution and event selection properties for the \Lqqln\
likelihood observable. In the upper plot, events with $\Lqqln=-1$ have
been redefined to have $\Lqqln=0$. See the caption of
Figure~\ref{w420sel} for further details.}
\label{lqqlnsel}
\end{figure}

We now consider a selection based on the
$\mathrm{q}\bar{\mathrm{q}}\ell^\pm\nu_\ell$ likelihood
variable,~\Lqqln. After imposing the \Lqqqq\ cut, the distribution of
\Lqqln\ is as shown in Figure~\ref{lqqlnsel}. Once again, some events
which fail a preselection are given the value $\Lqqln=-1$; in this case,
however, such events are included in the $\Lqqln=0$ bin of the
histogram. For most events, \Lqqln\ is then very close to zero or one,
while a small minority are scattered evenly in the central range. Any
cut in the range \mbox{$\Lqqln<0.01$} to \mbox{$\Lqqln<0.99$} would
give an almost optimal selection; we again follow the convention of
Ref.~\cite{pr362}, and select events with \mbox{$\Lqqln<0.5$}.

When these new likelihood cuts are used at $\sqrt{s}=207$~GeV, the
predicted efficiency $\epsilon_\mathrm{nr}$ for selecting
non-radiative signal events is~76\%, and the purity~$\rho$
is~94\%. The four-fermion background has been halved when compared
with the original $W_\mathrm{QCD}$ selection, while leaving the
efficiency almost unchanged.

\begin{figure}
\begin{center}
\includegraphics[width=\textwidth]{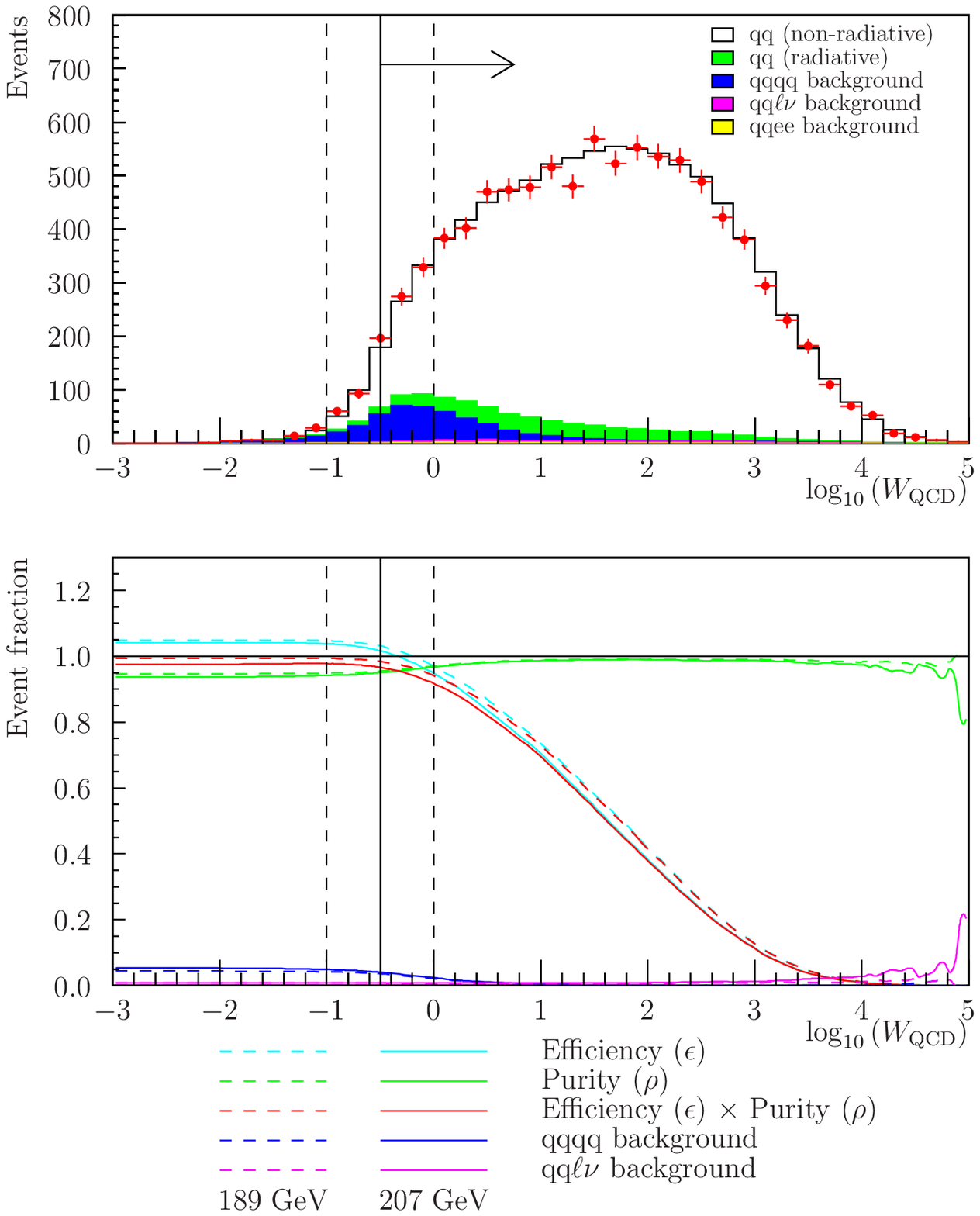}
\end{center}
\caption[Distribution and event selection properties for the
$W_\mathrm{QCD}$ observable, after imposing the cuts
$\Lqqqq<0.25$ and $\Lqqln<0.5$]
{Distribution and event selection properties for the $W_\mathrm{QCD}$
observable, after imposing the cuts $\Lqqqq<0.25$ and
$\Lqqln<0.5$. The vertical lines indicate the original
position of the $W_\mathrm{QCD}$ cut, and its alternative variations.
See the caption of Figure~\ref{w420sel} for a full explanation.}
\label{w420sel+defcuts}
\end{figure}

Finally, we investigate whether any improvement can be achieved by
applying a cut on $W_\mathrm{QCD}$ in addition to the likelihood
selection. Figure~\ref{w420sel+defcuts} shows the distribution of
$W_\mathrm{QCD}$ for events satisfying the cuts defined thus far. We
find that the product of efficiency and purity, $\epsilon\rho$, cannot
be increased significantly by excluding further events. This result is
to be expected, since the information contained in $W_\mathrm{QCD}$
has been used in the calculation of~\Lqqqq. We therefore place no
explicit cut on $W_\mathrm{QCD}$.

In summary, we will use the following two selection criteria to reject
four-fermion events:
\begin{itemize}
\item $\Lqqqq<0.25$
\item $\Lqqln<0.5$\;\;\;.
\end{itemize}
For our analysis of data at $\sqrt{s}=91$~GeV and 130--136~GeV we
do not apply these cuts, because the background from four-fermion
processes is negligible below the W$^+$W$^-$ pair production threshold.

\subsubsection{Aside: a useful cross-check for the position of a cut}

To choose the cut position for a continuous variable such as
$W_\mathrm{QCD}$ or \Lqqqq, we have maximised the product of
efficiency~($\epsilon$) and purity~($\rho$). Our selection criterion
takes the form
\begin{equation}
\lambda\;<\;\lambda_\mathrm{cut} \;\;\;,
\end{equation}
for some some cut variable $\lambda$,\footnote{For cuts of the form
$\lambda>\lambda_\mathrm{cut}$, we can transform to a new cut variable
$\tilde\lambda=-\lambda$.} and the probabilities
$\epsilon_\mathrm{nr}$, $\epsilon_\mathrm{r}$
and~$\epsilon_\mathrm{b}$ defined in Section~\ref{effpur_max} are
functions of the cut position $\lambda_\mathrm{cut}$. Intuitively, one
expects that the optimal value of $\lambda_\mathrm{cut}$ will be at a
position where the differential cross sections
$\mathrm{d}\sigma/\mathrm{d}\lambda$ for signal and background
processes are roughly equal; examples of this can be seen in
Figures~\ref{w420sel} and~\ref{lqqqqsel}. It turns out that a general
theorem exists for the proportions of signal and background events at
the cut threshold.

The product of efficiency and purity can be written as
\begin{equation}
\epsilon\rho\;=\;\frac{1}{N}\;\frac{s^2}{s+b} \;\;\;,
\end{equation}
where $s$ and $b$ are the expected numbers of signal and background
events passing the selection, and $N$ is the expected number of signal
events before any cuts are applied. In the presence of initial-state
radiation, the selected radiative events are included as signal. Using
the notation of Section~\ref{effpur_max}, the numbers $s$, $b$ and $N$
correspond to\vspace{-0.3cm}
\begin{eqnarray} s & \!=\! &
\epsilon_\mathrm{nr}N_\mathrm{nr}+\epsilon_\mathrm{r}N_\mathrm{r}
\nonumber \\ b & \!=\! & \epsilon_\mathrm{b}N_\mathrm{b} \nonumber \\ N &
\!=\! & N_\mathrm{nr} \hspace{2cm} .\\[-1.1cm] \nonumber
\end{eqnarray}
At the maximum of $\epsilon\rho$, we have
\begin{equation}
\frac{\partial(\epsilon\rho)}{\partial s}
{\frac{\mathrm{d}s}{\mathrm{d}\lambda}}_{\rule{0pt}{0.5cm}\mathrm{cut}} +\;
\frac{\partial(\epsilon\rho)}{\partial b}
{\frac{\mathrm{d}b}{\mathrm{d}\lambda}}_{\rule{0pt}{0.5cm}\mathrm{cut}} =\;0 \;\;\;.
\end{equation}
Evaluating the partial derivatives and multiplying the above
equation by \mbox{$N(s+b)^2/s$}, we find
\begin{equation}
(s+2b)\;
{\frac{\mathrm{d}s}{\mathrm{d}\lambda}}_{\rule{0pt}{0.5cm}\mathrm{cut}} -\;
s\;
{\frac{\mathrm{d}b}{\mathrm{d}\lambda}}_{\rule{0pt}{0.5cm}\mathrm{cut}} =\;0 \;\;\;.\vspace{0.2cm}
\end{equation}
We now determine the fraction of events at the cut threshold
$\lambda=\lambda_\mathrm{cut}$ that are signal:
\begin{equation}
\frac{\mathrm{d}s/\mathrm{d}\lambda_\mathrm{cut}}
{\mathrm{d}s/\mathrm{d}\lambda_\mathrm{cut} +\;
 \mathrm{d}b/\mathrm{d}\lambda_\mathrm{cut} }
\;=\;
\frac{s}{2(s+b)}
\;\equiv\;
\frac{\rho}{2}\;\;\;.\vspace{0.2cm}
\label{rhoover2}
\end{equation}
Thus for a reasonably pure selection ($\rho\approx 1$), we expect
signal events to contribute just under half the combined differential
cross section at the optimal cut value. This result is confirmed by
the $W_\mathrm{QCD}$ and \Lqqqq\ distributions, shown in
Figures~\ref{w420sel} and~\ref{lqqqqsel}
respectively. Equation~(\ref{rhoover2}) cannot be applied to
the~\Lqqln\ cut, because the distribution for this
observable is not sufficiently smooth; the vast majority of events
fall into one of three bins, at \mbox{$\Lqqln=-1$, 0
or 1}.

\subsection{Results of the event selection}

In Tables~\ref{cutflow0}--\ref{cutflow2}, we list the numbers of
events passing each stage of the selection procedure, at each
centre-of-mass energy. To confirm that the final quantities of events
selected from OPAL data are consistent with Monte Carlo predictions,
we perform a $\chi^2$ test: the results are given in
Table~\ref{chisquareselection}. We find $\chi^2=17.2$ for 11~degrees
of freedom, corresponding to a $P$-value of 0.90. At
$\sqrt{s}=91$~GeV, the predicted and observed numbers of selected
events differ by 8~standard deviations, and are not included in the
$\chi^2$~test; the luminosity of our data sample has not been
calculated with sufficient precision to make a meaningful comparison
at this energy.\pagebreak[4]

\begin{table}[tbp]
\begin{center}
\scalebox{0.80}{
\begin{tabular}{|r|r|rr|r|c|}
\hline
\multicolumn{1}{|c|}{\multirow{2}[3]{*}[-0.1cm]{\parbox{2cm}{\centering Energy\\and cut}}} &
\multicolumn{1}{|c|}{\multirow{2}[3]{*}[-0.1cm]{\parbox{1.2cm}{\centering Data\\events}}} &
\multicolumn{3}{|c|}{Events predicted by MC sims.} &
\multicolumn{1}{|c|}{\multirow{2}[3]{*}[-0.1cm]{\parbox{1.7cm}{\centering Efficiency\\($\epsilon_\textrm{nr}$)}}} \bigstrut \\
\cline{3-5}
& &
\multicolumn{1}{|c}{\parbox{1.5cm}{\centering \rule{0pt}{0.4cm}q$\bar{\mathrm{q}}$ \vspace{-0.1cm}\\\mbox{\footnotesize \rule[-0.2cm]{0pt}{0.1cm}(non-rad.)}}} &
\multicolumn{1}{c|}{\parbox{1.6cm}{\centering \rule{0pt}{0.4cm}q$\bar{\mathrm{q}}$ \vspace{-0.1cm}\\{\footnotesize \rule[-0.2cm]{0pt}{0.1cm}(radiative)}}} &
\multicolumn{1}{|c|}{Total} &
\bigstrut \\

\hline \hline \multicolumn{1}{|l|}{\textbf{91 GeV}} & & & & & \bigstrut \\

 All MC events                                          &
 \multicolumn{1}{|c|}{--} &   441590$\phantom{0}$ &  \multicolumn{1}{c|}{--} & 441590 &  1.000 \\
 TKMH                                               &
 426193 &   420817$\phantom{0}$ & \multicolumn{1}{c|}{--}  &  420817 &  0.953 \\
 $N_\mathrm{tracks}\geq 7$                          &
 419922 &   414362$\phantom{0}$ & \multicolumn{1}{c|}{--} &  414362 &  0.938 \\
 $\left|\cos\theta_\mathrm{T}\right|<0.9$           &
  \textbf{395695} &   \textbf{390438$\phantom{0}$} &  \multicolumn{1}{c|}{--} &   \textbf{390438} &  \textbf{0.884}

\bigstrut[b] \\
\hline \multicolumn{1}{|l|}{\textbf{133 GeV}} & & & & & \bigstrut \\
\hline
\end{tabular}}
\end{center}
\caption[Numbers of selected events at 91 GeV and 133 GeV] {Numbers of
selected events at $\sqrt{s}=91$~GeV and 133~GeV. Each selection cut
is applied to events passing all previous cuts, and the lines printed
in bold indicate the final selection. The `TKMH' and `L2MH' lines
include an extra cut on the thrust
axis:~$\left|\cos\theta_\mathrm{T}\right|<0.95$. At 133~GeV, the Monte
Carlo predictions have been re-scaled from samples much larger than
the data.  The efficiency $\epsilon_\mathrm{nr}$ is calculated using
only non-radiative $\text{q}\bar{\text{q}}$ events, which are defined
in this table by \mbox{$\sqrt{s}-\sqrt{s'}<1$~GeV}.}
\label{cutflow0}
\end{table}

\begin{table}[p]
\begin{leftfullpage}
\begin{center}
\resizebox{0.95\textwidth}{!}{
\begin{tabular}{|r|r|rrrrr|r|c|c|}
\hline
\multicolumn{1}{|c|}{\multirow{2}[3]{*}[-0.1cm]{\parbox{2cm}{\centering Energy\\and cut}}} &
\multicolumn{1}{|c|}{\multirow{2}[3]{*}[-0.1cm]{\parbox{1.2cm}{\centering Data\\events}}} &
\multicolumn{6}{|c|}{Events predicted by MC simulations} &
\multicolumn{1}{|c|}{\multirow{2}[3]{*}[-0.1cm]{\parbox{1.7cm}{\centering Efficiency\\($\epsilon_\textrm{nr}$)}}} &
\multicolumn{1}{|c|}{\multirow{2}[3]{*}[-0.1cm]{\parbox{1.1cm}{\centering Purity\\($\rho$)}}} \bigstrut \\
\cline{3-8}
& &
\multicolumn{1}{|c}{\parbox{1.5cm}{\centering \rule{0pt}{0.4cm}q$\bar{\mathrm{q}}$ \vspace{-0.1cm}\\\mbox{\footnotesize \rule[-0.2cm]{0pt}{0.1cm}(non-rad.)}}} &
\multicolumn{1}{c}{\parbox{1.6cm}{\centering \rule{0pt}{0.4cm}q$\bar{\mathrm{q}}$ \vspace{-0.1cm}\\{\footnotesize \rule[-0.2cm]{0pt}{0.1cm}(radiative)}}} &
\multicolumn{1}{c}{qqqq} &
\multicolumn{1}{c}{qq$\ell\nu$} &
\multicolumn{1}{c|}{qqee} &
\multicolumn{1}{|c|}{Total} &
& \bigstrut \\

\hline \hline \multicolumn{1}{|l|}{\textbf{161 GeV}} & & & & & & & & & \bigstrut \\

 All MC events                                          &
 \multicolumn{1}{|c|}{--} &   244$\phantom{0}$ &  1327$\phantom{0}$ &    21 &    27 &    52 &  1671 &  1.000 &  0.941 \\
 L2MH                                               &
 1395 &   231$\phantom{0}$ &  1141$\phantom{0}$ &    20 &    21 &     6 &  1419 &  0.945 &  0.967 \\
 $N_\mathrm{tracks}\geq 7$                          &
 1369 &   229$\phantom{0}$ &  1123$\phantom{0}$ &    20 &    21 &     5 &  1398 &  0.938 &  0.967 \\
 $\left|\cos\theta_\mathrm{T}\right|<0.9$           &
 1287 &   215$\phantom{0}$ &  1059$\phantom{0}$ &    19 &    20 &     4 &  1317 &  0.879 &  0.967 \\
 ISR cut                                            &
  304 &   204$\phantom{0}$ &    77$\phantom{0}$ &    15 &     3 &     1 &   300 &  0.834 &  0.938 \\
\hline
\end{tabular}}
\end{center}
\caption[Numbers of selected events at 161--192 GeV] {Numbers of
selected events at $\sqrt{s}=161$--192~GeV. Each selection cut is
applied to events passing all previous cuts, and the lines printed in
bold indicate the final selection. The `L2MH' line includes an extra
cut on the thrust
axis:~\mbox{$\left|\cos\theta_\mathrm{T}\right|<0.95$}. The Monte
Carlo predictions have been re-scaled from samples much larger than
the data, and rounded to the nearest integer. The efficiency
$\epsilon_\mathrm{nr}$ is calculated using only non-radiative
$\text{q}\bar{\text{q}}$ events, which are defined in this table by
\mbox{$\sqrt{s}-\sqrt{s'}<1$~GeV}. The purity $\rho$ includes all
two-fermion multihadron events as signal.}
\label{cutflow1}
\end{leftfullpage}
\end{table}

\begin{table}[p]
\begin{fullpage}
\begin{center}
\resizebox{0.95\textwidth}{!}{
\begin{tabular}{|r|r|rrrrr|r|c|c|}
\hline
\multicolumn{1}{|c|}{\multirow{2}[3]{*}[-0.1cm]{\parbox{2cm}{\centering Energy\\and cut}}} &
\multicolumn{1}{|c|}{\multirow{2}[3]{*}[-0.1cm]{\parbox{1.2cm}{\centering Data\\events}}} &
\multicolumn{6}{|c|}{Events predicted by MC simulations} &
\multicolumn{1}{|c|}{\multirow{2}[3]{*}[-0.1cm]{\parbox{1.7cm}{\centering Efficiency\\($\epsilon_\textrm{nr}$)}}} &
\multicolumn{1}{|c|}{\multirow{2}[3]{*}[-0.1cm]{\parbox{1.1cm}{\centering Purity\\($\rho$)}}} \bigstrut \\
\cline{3-8}
& &
\multicolumn{1}{|c}{\parbox{1.5cm}{\centering \rule{0pt}{0.4cm}q$\bar{\mathrm{q}}$ \vspace{-0.1cm}\\\mbox{\footnotesize \rule[-0.2cm]{0pt}{0.1cm}(non-rad.)}}} &
\multicolumn{1}{c}{\parbox{1.6cm}{\centering \rule{0pt}{0.4cm}q$\bar{\mathrm{q}}$ \vspace{-0.1cm}\\{\footnotesize \rule[-0.2cm]{0pt}{0.1cm}(radiative)}}} &
\multicolumn{1}{c}{qqqq} &
\multicolumn{1}{c}{qq$\ell\nu$} &
\multicolumn{1}{c|}{qqee} &
\multicolumn{1}{|c|}{Total} &
& \bigstrut \\

\hline \hline \multicolumn{1}{|l|}{\textbf{196 GeV}} & & & & & & & & & \bigstrut \\

 All MC events                                          &
\multicolumn{1}{|c|}{--} &  1044$\phantom{0}$ &  5863$\phantom{0}$ &   684 &   704 &  3105 & 11400 &  1.000 &  0.606 \\
 L2MH                                               &
 7213 &   991$\phantom{0}$ &  4546$\phantom{0}$ &   662 &   625 &   171 &  6995 &  0.949 &  0.792 \\
 $N_\mathrm{tracks}\geq 7$                          &
 7078 &   985$\phantom{0}$ &  4454$\phantom{0}$ &   662 &   620 &   161 &  6882 &  0.943 &  0.790 \\
 $\left|\cos\theta_\mathrm{T}\right|<0.9$           &
 6622 &   924$\phantom{0}$ &  4171$\phantom{0}$ &   637 &   597 &   151 &  6480 &  0.884 &  0.786 \\
 ISR cut                                            &
 1732 &   863$\phantom{0}$ &   313$\phantom{0}$ &   510 &    57 &     3 &  1747 &  0.826 &  0.674 \\
\hline
\end{tabular}}
\end{center}
\caption{Numbers of selected events at $\sqrt{s}=196$--207 GeV. For
details see the caption of Figure~\ref{cutflow1}.\textcolor{white}{
FILL FILL FILL FILL FILL FILL FILL FILL FILL FILL FILL FILL FILL FILL
FILL FILL FILL FILL FILL FILL FILL FILL FILL FILL FILL FILL FILL FILL
FILL FILL FILL FILL FILL FILL FILL FILL FILL FILL FILL FILL FILL FILL
FILL FILL FILL FILL FILL FILL FILL FILL FILL FILL FILL FILL FILL FILL
FILL FILL FILL FILL FILL FILL FILL FILL FILL FILL FILL FILL FILL FILL
FILL FILL }}
\label{cutflow2}
\end{fullpage}
\end{table}

\begin{table}[h!]
\begin{center}
\begin{tabular}{|c| c @{\hspace{0.35cm}} r@{.}l| c @{\hspace{0.5cm}} r @{\hspace{0.5cm}} c | c @{\hspace{0.25cm}} r @{$\;\pm\;$} l|c|}
\hline
\multicolumn{1}{|c|}{\parbox{2cm}{\centering $\sqrt{s}$~(GeV)}} &
\multicolumn{3}{|c|}{\parbox{2cm}{\centering Integrated\vspace{-0.08cm}\\luminosity\vspace{-0.08cm}\\(pb$^{-1}$)}} &
\multicolumn{3}{|c|}{\parbox{1.7cm}{\centering Observed\vspace{-0.08cm}\\ events}} &
\multicolumn{3}{|c|}{\parbox{2.3cm}{\centering Predicted\vspace{-0.08cm}\\ events}} &
\multicolumn{1}{|c|}{\parbox{2.5cm}{\centering \rule{0pt}{0.4cm}Difference\vspace{-0.05cm}\\(standard\vspace{-0.08cm}\\deviations)\rule[-0.2cm]{0pt}{0pt}}} \bigstrut \\
\hline
\hline
133.3 &&  11 & 26  &&  630 &&&  684 & 26 & $-$2.1 \bigstrut[t] \\
161.3 &&  10 & 06  &&  281 &&&  275 & 17 & $+$0.4 \\
172.1 &&  10 & 38  &&  218 &&&  232 & 15 & $-$0.9 \\
182.7 &&  57 & 72  && 1077 &&& 1084 & 33 & $-$0.2 \\
188.6 &&  185 & 2  && 3086 &&& 3130 & 56 & $-$0.8 \\
191.6 &&  29 & 53  &&  514 &&&  473 & 22 & $+$1.9 \\
195.5 &&  76 & 67  && 1137 &&& 1161 & 34 & $-$0.7 \\
199.5 &&  79 & 27  && 1090 &&& 1131 & 34 & $-$1.2 \\
201.6 &&  37 & 75  &&  519 &&&  527 & 23 & $-$0.3 \\
204.8 &&  82 & 01  && 1130 &&& 1090 & 33 & $+$1.2 \\
206.6 &&  138 & 8  && 1717 &&& 1804 & 42 & $-$2.0 \bigstrut[b] \\
\hline
\end{tabular}
\end{center}
\caption{Differences between observed and predicted numbers of
selected events at each centre-of-mass energy, excluding 91~GeV. The
uncertainties indicated are statistical, and are based on a Poisson
distribution (they do not represent uncertainties in the predicted mean
of the distribution). Combining the squared deviations gives
$\chi^2=17.2$ with 11~degrees of freedom. The data are therefore
consistent with Monte Carlo expectations at the 90\% confidence
level.}
\label{chisquareselection}
\end{table}

\begin{figure}
\begin{center}
\includegraphics[width=\linewidth,bb=98 290 490 551]{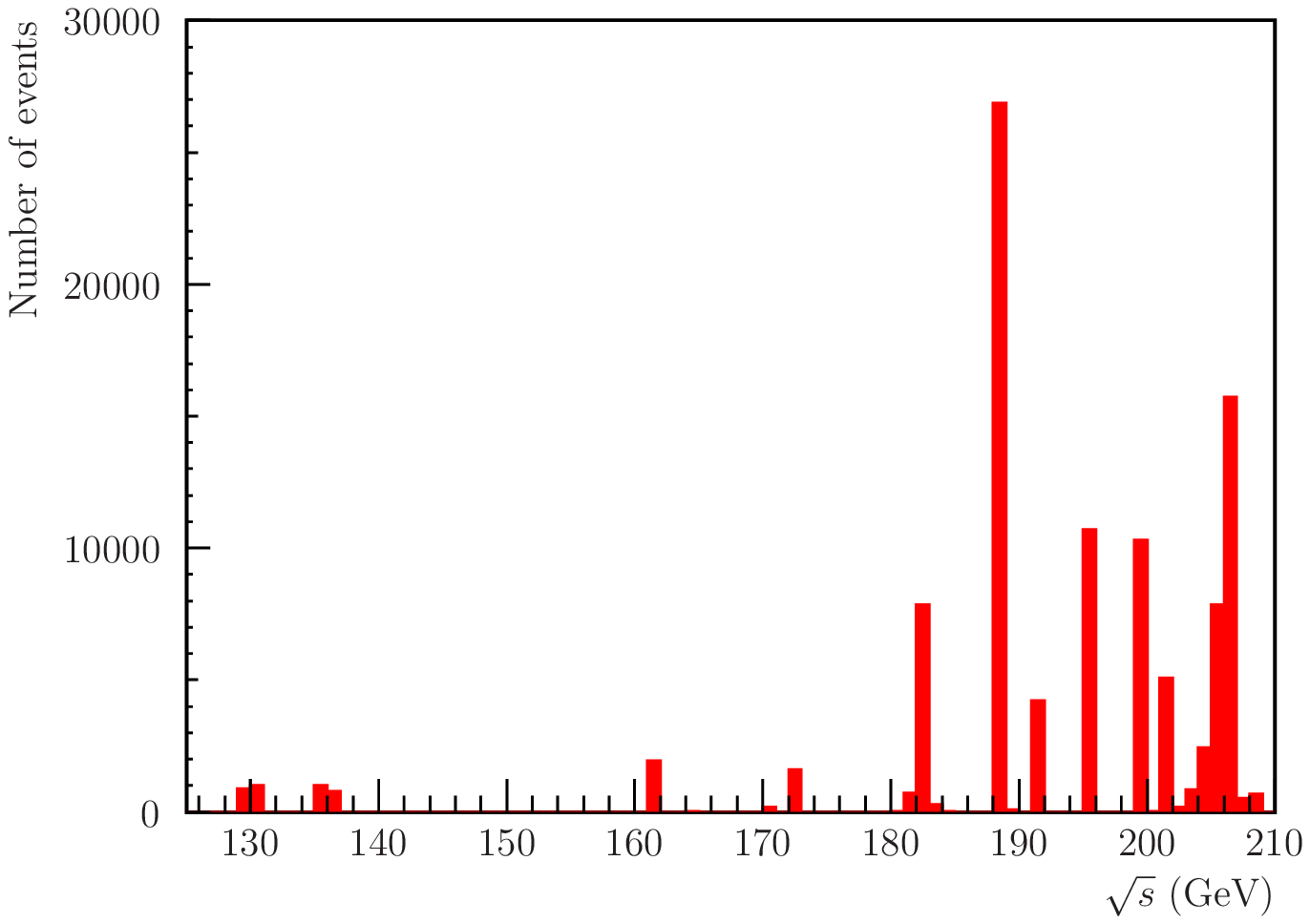}
\end{center}
\caption{Distribution of
centre-of-mass energies for selected non-radiative
$\mathrm{q}\bar{\mathrm{q}}$ events (calibration data at
$\sqrt{s}=91$~GeV is not shown).}
\label{ebeam}
\end{figure}

Finally, in Figure~\ref{ebeam}, we present the distribution of
centre-of-mass energies for all selected events at $\sqrt{s}\geq
130$~GeV.

\section{Measurement of the event shape distributions}
\label{evshmeasure}

Using data from the tracking chambers and the electromagnetic
calorimeters, we now compute the event shape observables defined in
Section~\ref{evsh_defs}, for each selected event.

Charged particles normally produce both a curved track in the central
detector and a cluster of energy deposition in the calorimeters,
while neutral particles are seen only in the calorimeters. To avoid
double-counting of energy, an association must be made between the
charged tracks and the energy clusters; this is not trivial, however,
because the clusters from two neighbouring particles can often overlap
in a high-multiplicity event. To solve this problem a matching
algorithm, \texttt{MT}~\cite{mt}, has been developed. For each charged
track of momentum~{\boldmath $p$}, the expected energy response
$\widehat{E}(\mbox{\boldmath $p$})$ of the calorimeters is
estimated. If the measured energy $E$ of the matched cluster is less
than~\mbox{$\widehat{E}(\mbox{\boldmath $p$})+\Delta E$}, where
$\Delta E$ is the assumed resolution of the calorimeter, then the
cluster is not used. If \mbox{$E>\widehat{E}(\mbox{\boldmath
$p$})+\Delta E$}, however, then two particles are deemed to have
contributed to the cluster: a charged particle of momentum~{\boldmath
$p$}, and a neutral particle of
energy~\mbox{$E-\widehat{E}(\mbox{\boldmath $p$})$}. This algorithm
ensures that high-resolution data from the tracking chambers are given
preference over less precise information from the
calorimeters.\footnote{As noted in Ref.~\cite{mt}, the resolution of
the electromagnetic calorimeters is usually less than that of the
tracking chambers. The only common exception is for very high energy
electrons, which are not generally produced in multihadronic events.}
The four-momentum of each particle is then computed, by assuming
masses equal to the pion mass $m_{\pi^\pm}$ for charged particles, and
zero for neutral electromagnetic clusters.

For each event shape observable, we construct a histogram of about ten
bins. The bin boundaries, listed in Appendix~\ref{evshdistappendix},
have been taken from previous OPAL publications where possible. For
the narrow jet broadening~(\BN), light jet mass~(\ML) and
$D$-parameter, however, the choice of binning is new. At the peak of
each distribution the bin width is determined by the resolution of the
detector, while in the tail we use wider bins to reduce the
statistical uncertainty.

\subsection{Estimation of background}
\label{backgroundsubtraction}

The selection algorithm discussed in Section~\ref{selection} is not
perfectly efficient, and does not remove all of the four-fermion
background events present. Although it was claimed in
Tables~\ref{cutflow1} and \ref{cutflow2} that only 6\% of selected
events at the highest energies are background, this statistic does not
apply to individual bins of the event shape distributions. Among
signal events, it is inevitable that three- and four-jet QCD processes
will be the least distinguishable from four-fermion background
events. It is therefore essential to estimate the expected background
selected within each bin of the distributions, and
subtract it from the number of events selected in the data.

Our predictions for four-fermion background processes are based on the
standard OPAL Monte Carlo runs described in
Section~\ref{opalmcsamples}. A full simulation of the detector
response has been performed for each event. After applying our
selection algorithm to the simulated events, which have been
reconstructed with ROPE in the same way as for real data, we compute
the event shape distributions of the selected background events.
These are then subtracted from the corresponding data histograms.

In this analysis, we have not considered every possible source of
background; other classes of event, such as two-photon processes and
$\tau^+\tau^-$ production, will very occasionally pass our
selection. These rare backgrounds are beyond the scope of this work,
but a detailed study~\cite{ainsley} has shown their contribution to be
negligible in a measurement of the non-radiative
$\mathrm{e}^+\mathrm{e}^-\to \mathrm{q}\bar{\mathrm{q}}$ cross
section.

\subsection{Detector correction}
\label{detectorcorrection}

After the subtraction of four-fermion background, our distributions
must be `unfolded' to correct for three remaining sources of bias:
\begin{itemize}
\item Only 75--80\% of non-radiative q$\bar{\mathrm{q}}$ events pass
the event selection at energies above the W$^+$W$^-$ production
threshold.
\item Some of the selected q$\bar{\mathrm{q}}$ events include initial-state radiation. Their contribution cannot be subtracted with other
backgrounds, since our measurements would then rely on accurate
simulation of the QCD processes which we are aiming to measure;
instead we will include them in a multiplicative ``detector
correction,'' which is less sensitive to the Monte Carlo model.
\item Biases will be present in our measurement of the actual event
shape observables, due to the limitations of the detector. For
example, the spreading of energy clusters within the calorimeters will
tend to increase the apparent width of jets.
\end{itemize}
To accomplish this correction, we use the simulated
q$\bar{\mathrm{q}}$ event samples described in
Section~\ref{opalmcsamples}.  Two sets of normalised event shape
distributions are computed, using the same set of Monte Carlo events
in each case:
\begin{description}
\item[Detector-level distributions,] which include a full simulation
of the OPAL detector, and are subjected to the same selection criteria
as the real data.
\item[Hadron-level distributions,] computed using the true particle
momenta predicted by PYTHIA, without simulation of the
detector. Particles with mean lifetimes longer than $3\times
10^{-10}$~s are treated as stable, while others are forced to
decay. Radiative events are removed by applying the cut
\begin{equation}
\left[\sqrt{s}-\sqrt{s'}\,\right]_\mathrm{true}<\;1~\mathrm{GeV}\;\;\;,
\end{equation}
but no further selection criteria are applied.
\end{description}
The ratio of these two distributions gives a numerical correction
factor for each bin. We apply this to our measured distributions using
a simple bin-by-bin multiplication, and finally normalise the result.

\subsection{Results from the detector simulation}
\label{detsimresults}

We have already listed, in Tables~\ref{cutflow1} and~\ref{cutflow2},
the expected numbers of signal and background events passing our selection.
In this section, we present a more detailed analysis of the predicted
signal and background contributions to each event shape distribution.

\begin{figure}
\begin{leftfullpage}
\begin{center}
\includegraphics[width=\linewidth]{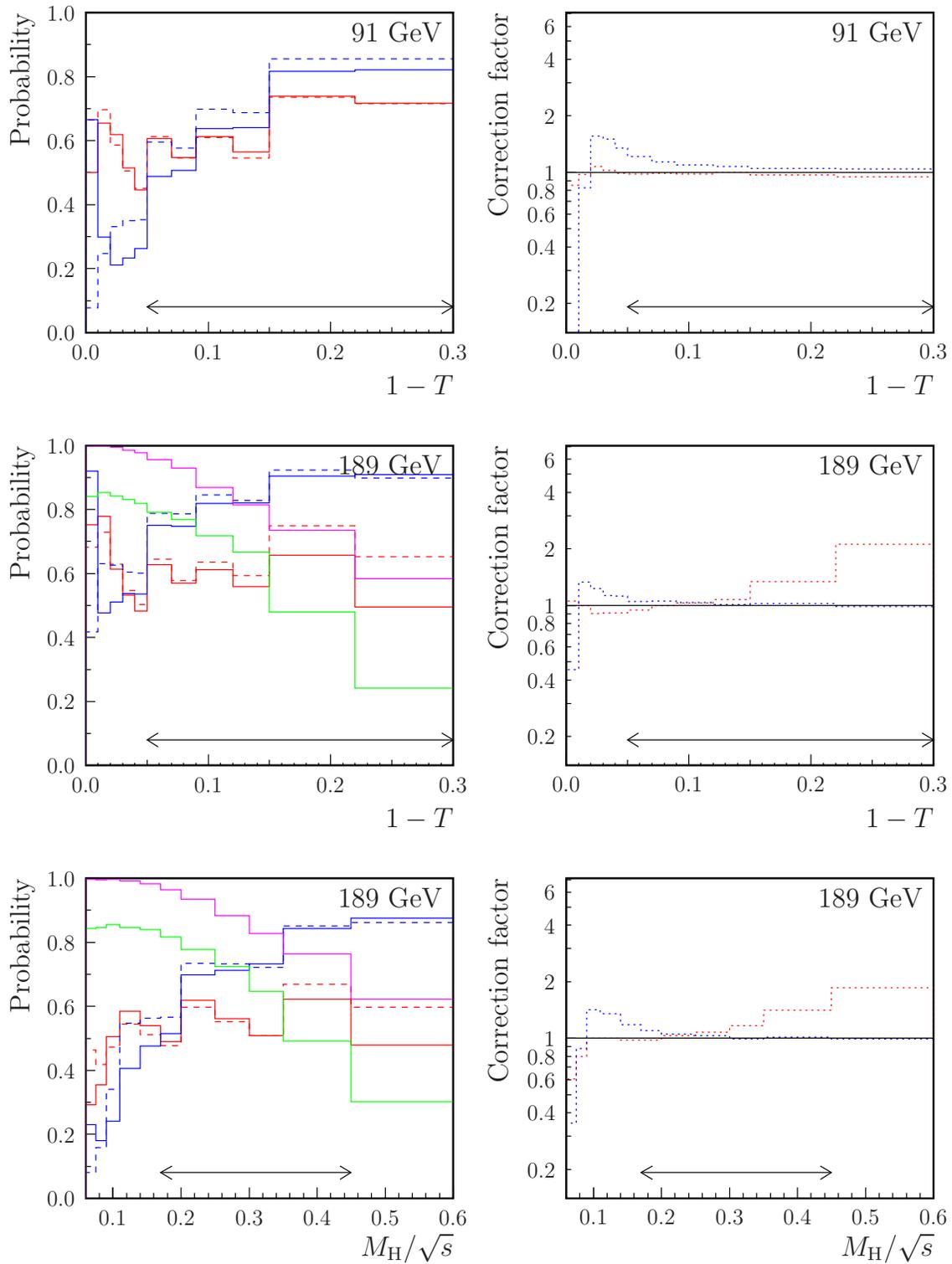}
\end{center}
\caption{Efficiencies, purities, correction factors and bin migration
probabilities for the thrust distribution at $\sqrt{s}=91$~GeV and
189~GeV, and for the heavy jet mass distribution at 189~GeV. See
Figure~\ref{dethad_key} for a full explanation.}
\label{dethad1}
\end{leftfullpage}
\end{figure}

\begin{figure}
\begin{fullpage}
\begin{center}
\includegraphics[width=\linewidth]{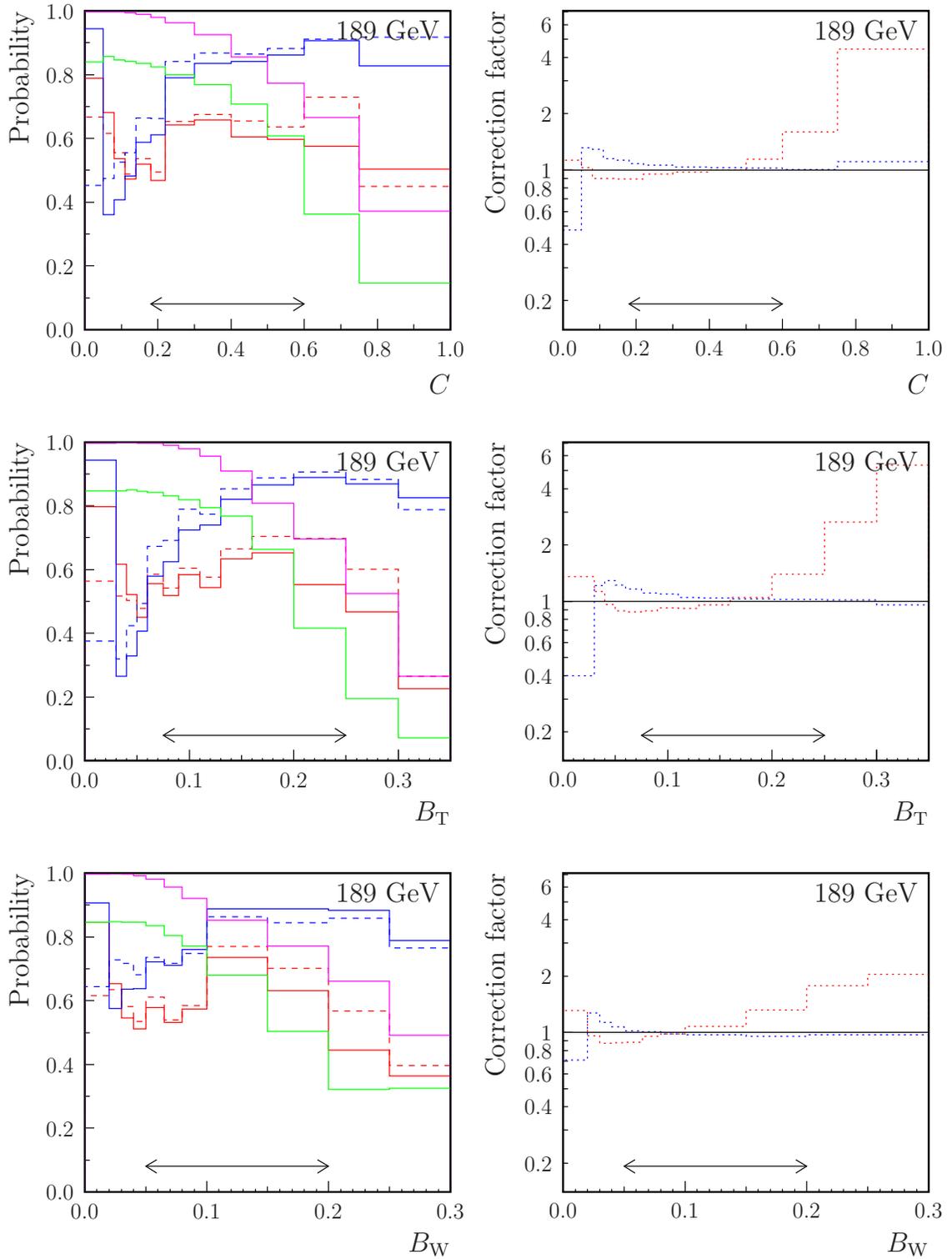}
\end{center}
\caption{Efficiencies, purities, correction factors
and bin migration probabilities for the $C$-parameter and for the
total and wide jet broadenings at $\sqrt{s}=189$~GeV. See Figure~\ref{dethad_key}
for a full explanation.}
\label{dethad2}
\end{fullpage}
\end{figure}

\begin{figure}
\begin{leftfullpage}
\begin{center}
\includegraphics[width=\linewidth]{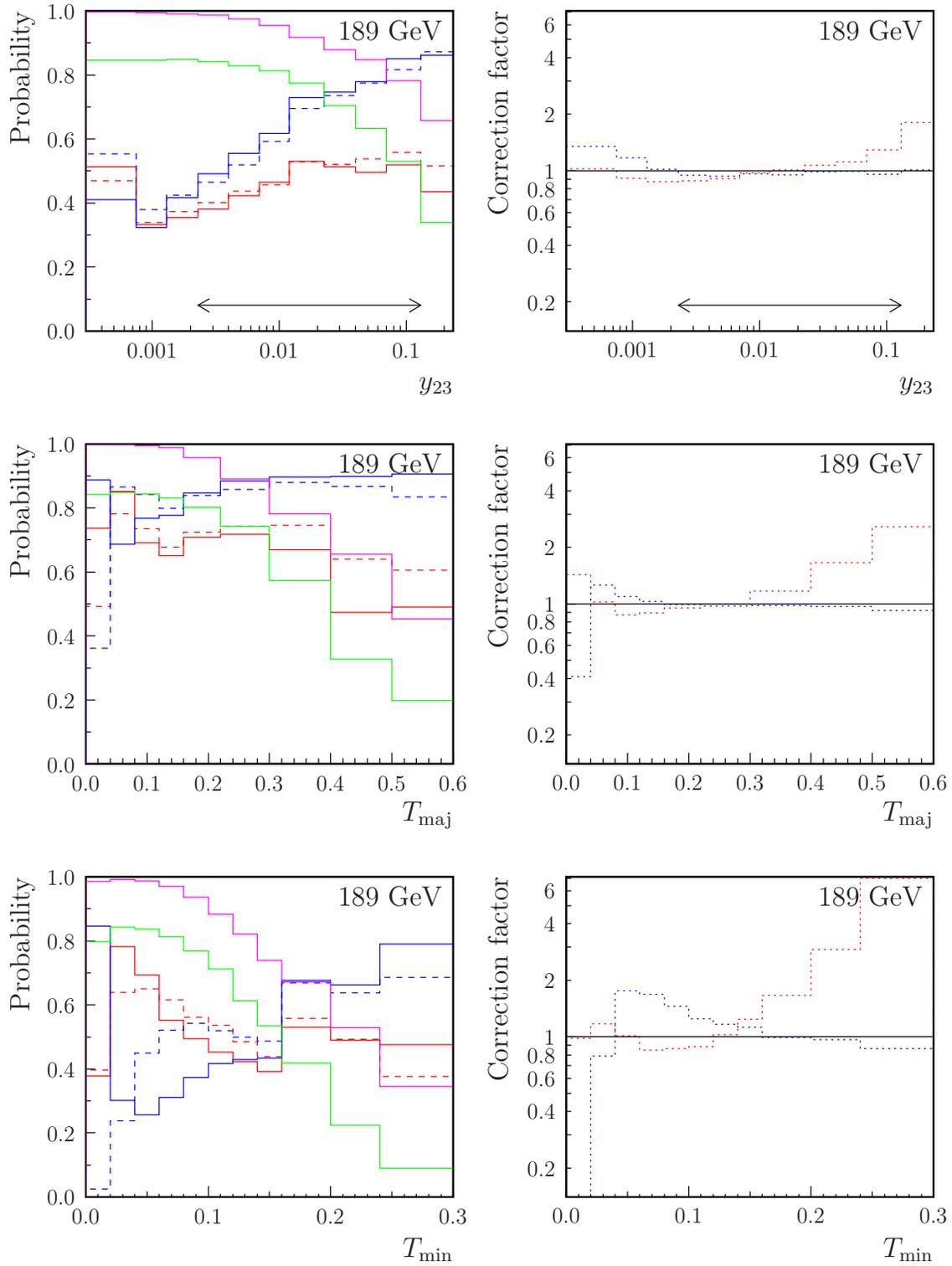}
\end{center}
\caption{Efficiencies, purities, correction factors
and bin migration probabilities for the Durham \ytwothree parameter,
thrust major and thrust minor at $\sqrt{s}=189$~GeV. See Figure~\ref{dethad_key}
for a full explanation.}
\label{dethad3}
\end{leftfullpage}
\end{figure}

\begin{figure}
\begin{fullpage}
\begin{center}
\includegraphics[width=\linewidth]{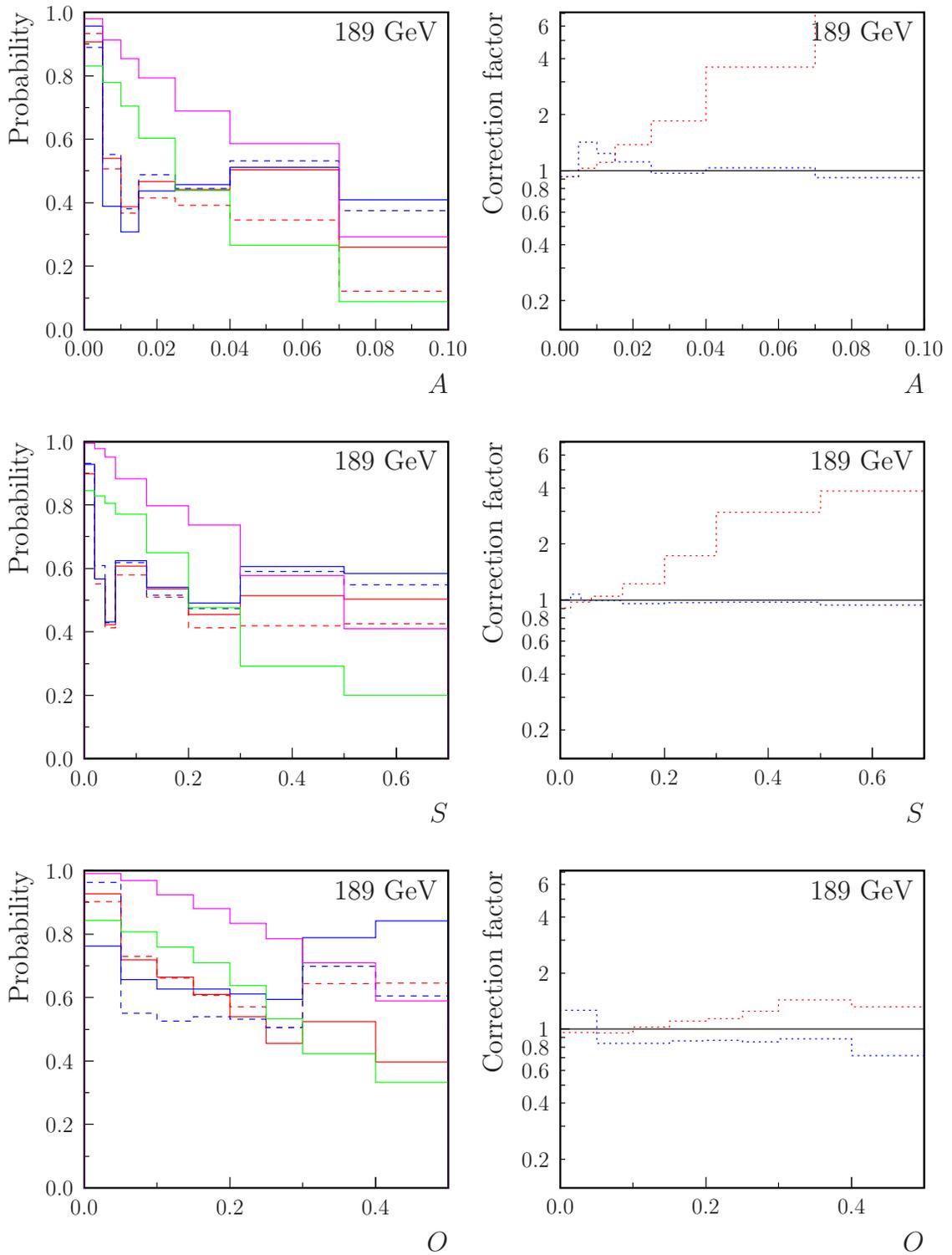}
\end{center}
\caption{Efficiencies, purities, correction factors and bin migration
probabilities for the aplanarity, sphericity and oblateness at
$\sqrt{s}=189$~GeV. See Figure~\ref{dethad_key} for a full explanation.}
\label{dethad4}
\end{fullpage}
\end{figure}

\begin{figure}
\begin{leftfullpage}
\begin{center}
\includegraphics[width=\linewidth]{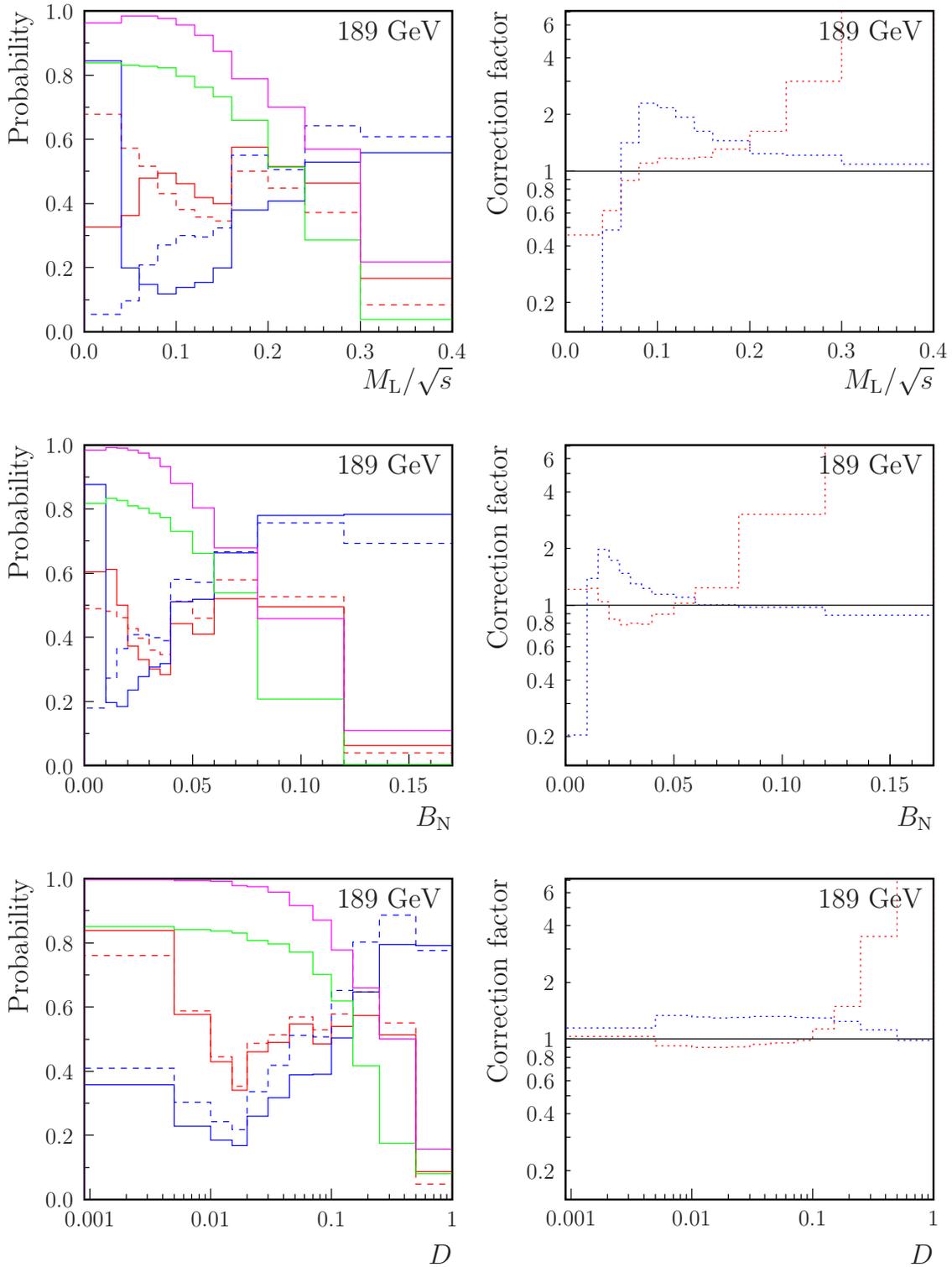}
\end{center}
\caption{Efficiencies, purities, correction factors
and bin migration probabilities for the light jet mass, narrow jet
broadening and $D$-parameter at $\sqrt{s}=189$~GeV. See Figure~\ref{dethad_key}
for a full explanation.}
\label{dethad5}
\end{leftfullpage}
\end{figure}

\begin{figure}
\begin{center}
\setlength\fboxsep{0.5cm}
\framebox{\includegraphics[width=0.7\linewidth]{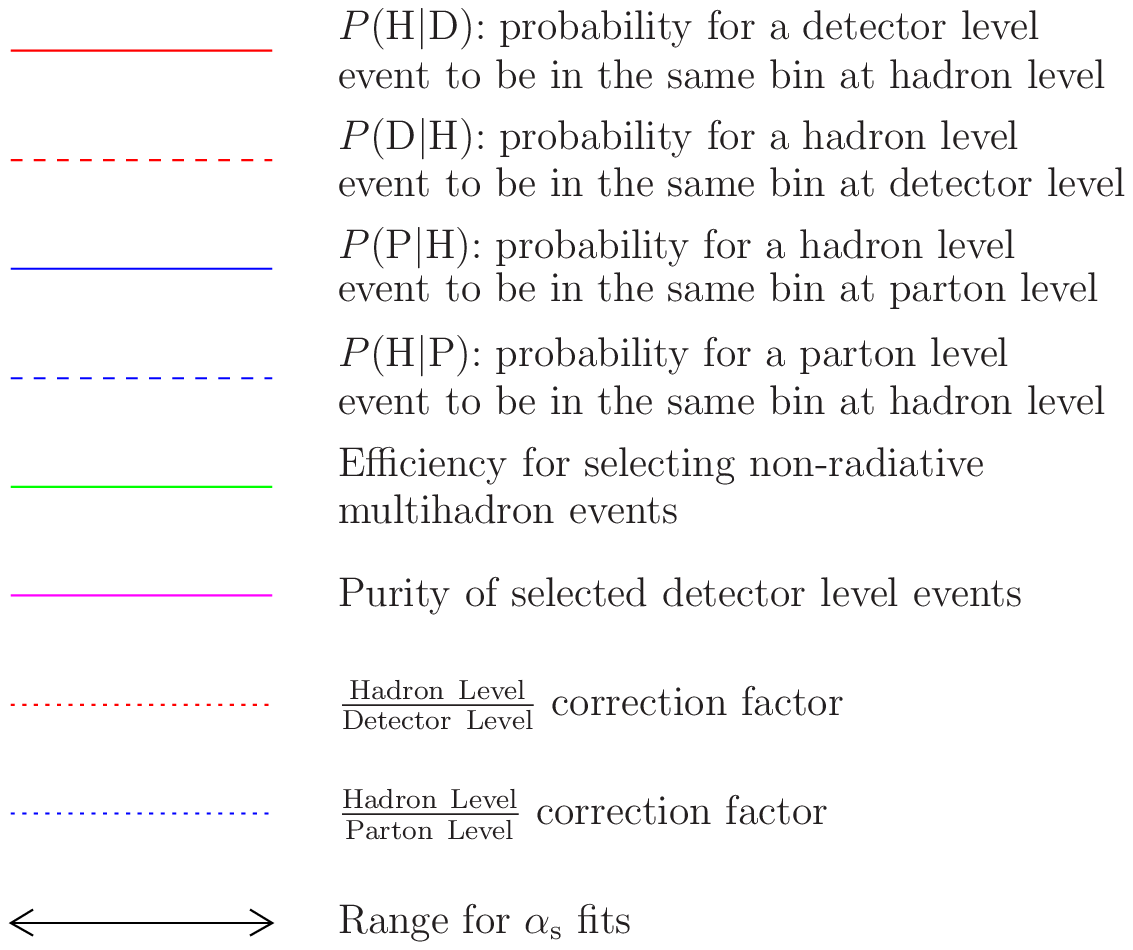}}
\end{center}
\caption{Legend for Figures~\ref{dethad1}--\ref{dethad5}}
\label{dethad_key}
\end{figure}

In Figures~\ref{dethad1}--\ref{dethad5}, the selection purity $\rho$
and the non-radiative efficiency $\epsilon_\mathrm{nr}$ are given as
functions of each event shape observable $y$,\footnote{Recall that the
dimensionless variable $y$ represents a generic event shape
observable. In all cases, a final state comprising two back-to-back
particles is assigned the value $y=0$; in the case of thrust, we must
therefore define $y=1-T$. Further examples were listed in
Section~\ref{pert_predictions}.} at $\sqrt{s}=189$~GeV. The legend for
these plots can be found in Figure~\ref{dethad_key}. In most cases we
find that the selection is extremely pure for events at low values of
$y$, which represent the two-jet regime. As $y$ increases towards the
region of multiple hard gluon emissions, the purity $\rho$ decreases
monotonically in every case. This effect is especially pronounced in
the distributions of the `four-jet' variables \ML, \BN\ and $D$, for
which only \mbox{10--20\%} of selected events in the last bin are
signal. The efficiency $\epsilon_\mathrm{nr}$ also drops with
increasing $y$, as QCD processes become more compatible with
four-fermion background. However, the efficiency does not approach
100\% in the two-jet region, since events are rejected if the thrust
axis lies within 26$^\circ$ of the beam axis; the largest values of
$\epsilon_\mathrm{nr}$ are approximately 0.85.

In the same figures, we have shown the probabilities
$P(\mathrm{D}|\mathrm{H})$ for hadron-level events to be observed in
the correct bin by the detector, and $P(\mathrm{H}|\mathrm{D})$ for
detector-level events to have originated at hadron-level in the
observed bin. These probabilities usually lie in the range
\mbox{$0.4<P<0.8$}, with no common trend across the distributions; we
can conclude that the bin widths are of the same order of magnitude as
the detector resolution. The exception, once again, is in the bins of
largest \ML, \BN\ and $D$, for which the probability of observing an
event in the correct bin is below 10\%. We suggest that this is due to
the limited volume of phase space within the bin.

In the right-hand columns of Figures~\ref{dethad1}--\ref{dethad5}, we
present the bin-by-bin ``detector corrections'' defined in
Section~\ref{detectorcorrection}.\footnote{The hadronisation
corrections, plotted in blue, will be discussed in
Section~\ref{hadronlevelprediction}.} These compensate for selection
inefficiencies, initial-state radiation and bin migrations. For the
observables $T$, \MH, $C$, \BT, \BW, \ytwothree, $T_\mathrm{maj.}$,
$O$ and $D$, we find a central range of several bins for which the
correction is uniform and close to unity. For $T_\mathrm{min.}$, \ML,
\BN, and most notably $A$ and $S$, the detector correction is larger
and varies significantly across the distribution. In the case of the
aplanarity and sphericity, the poor resolution of the detector may be
attributed to the sphericity tensor being non-infrared-safe, as
discussed in Section~\ref{sph_apl_c_d}; its quadratic dependence on
the particle momenta implies that an error will be incurred if two
overlapping calorimeter clusters are assigned to a single particle.

In Figure~\ref{dethad1}, we have compared the detector corrections for
the thrust distribution at $\sqrt{s}=91$~GeV and 189~GeV. We find that the
response of the detector does not differ significantly between these
two energies, for our purposes.

In many similar analyses, such as the OPAL measurement of event shapes
at LEP1~\cite{OPAL_as_91}, detector corrections have been performed
using a complete response matrix. This takes account of the migration
probabilities between bins, and attempts to `invert' the detector
simulation as accurately as possible.  Although mathematically
elegant, however, the inversion of the response matrix can lead to an
unacceptable amplification of statistical
fluctuations~\cite{cowanbook}; it also transfers systematic
uncertainties between bins, by mixing data from regions of different
purity. For these reasons, we have instead adopted the simpler
bin-by-bin approach. As noted in Ref.~\cite{cowanbook}, there may be a
bias associated with the correction, due to imperfect modelling of the
underlying hadron-level distribution.\footnote{This error would still
be present, though perhaps smaller, in the case of the matrix
method. The response matrix itself depends on the complete structure
of the hadron-level event population, and not simply on the event
shape distribution. It would be meaningless to refer to a ``migration
probability'' from one bin to another without specifying the full
kinematics of the events involved.} We will estimate the size of this
systematic error by calculating an alternative set of detector
corrections based on events generated with HERWIG.

\begin{figure}
\begin{leftfullpage}
\begin{center}
\vspace{1cm}
\includegraphics[width=0.75\linewidth]{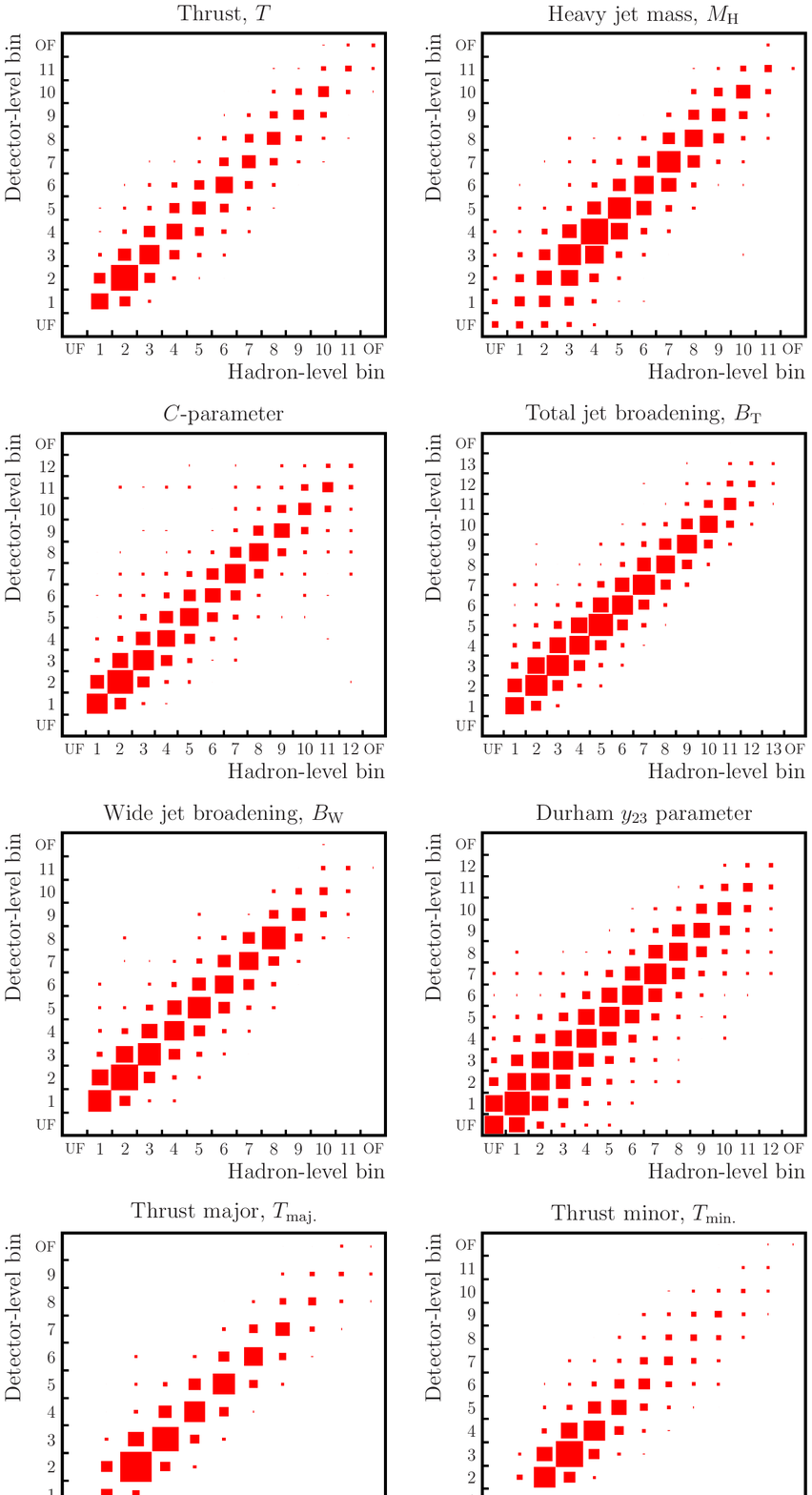}
\end{center}
\vspace{1.0cm}
\caption[Joint hadron- and detector-level distributions predicted for
$T$, \MH, $C$, \BT, \BW, \ytwothree, $T_\mathrm{maj.}$ and
$T_\mathrm{min.}$] {Joint hadron- and detector-level distributions
predicted for $T$, \MH, $C$, \BT, \BW, \ytwothree, $T_\mathrm{maj.}$
and $T_\mathrm{min.}$ at $\sqrt{s}=189$~GeV. The area of each box
represents the expected number of events within the given pair of
bins. The underflow and overflow bins are denoted by UF and OF
respectively.}
\label{migration1}
\end{leftfullpage}
\end{figure}

\begin{figure}
\begin{fullpage}
\begin{center}
\includegraphics[width=0.75\linewidth]{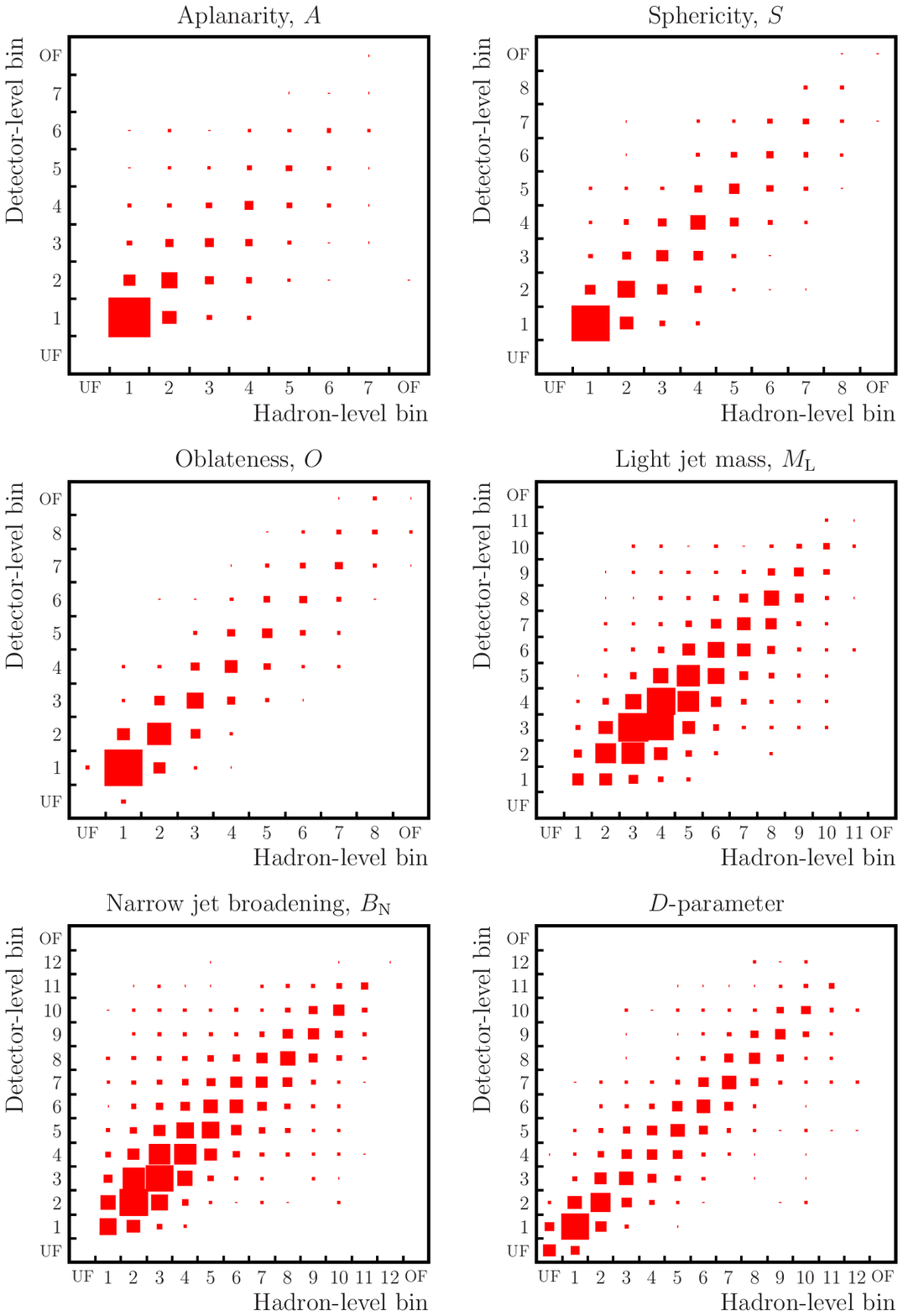}
\end{center}
\caption[Joint hadron- and detector-level distributions predicted for
$A$, $S$, $O$, \ML, \BN\ and $D$.]  {Joint hadron- and detector-level
distributions predicted for $A$, $S$, $O$, \ML, \BN\ and $D$ at
$\sqrt{s}=189$~GeV. The area of each box represents the expected
number of events within the given pair of bins. The underflow and
overflow bins are denoted by UF and OF respectively.}
\label{migration2}
\end{fullpage}
\end{figure}

To give an indication of the off-diagonal elements in the response
matrix, we have calculated joint distributions for the hadron- and
detector-level observables at \mbox{$\sqrt{s}=189$~GeV;} these are presented
in Figures~\ref{migration1} and~\ref{migration2}. The distributions
are not perfectly diagonal, but the vast majority of events do not
migrate by more than one bin from their true hadron-level position. In
most, but not all, cases the probabilities of upward and downward
migrations are roughly equal; this implies that we do not waste a
significant amount of information by using bin-by-bin correction
factors.

\subsection{Estimation of uncertainties}
\label{evsh_errors}

The statistical uncertainties of the distributions are obtained using
the method described in Appendix~\ref{multinomial}. In previous OPAL
publications, this uncertainty has been estimated by treating the raw
number of events in each bin as an independent Poisson variable; the
corresponding uncertainty was then rescaled according to the detector
correction and normalisation. This is a very good approximation in the
case of a narrow bin, but it fails when the relative uncertainty of
the total number of events is comparable with that of the individual
bin contents. Other authors, such as the ALEPH
Collaboration~\cite{ALEPH_as_1}, have assumed a multinomial
distribution which takes into account correlations between the bin
contents and the normalisation. When the bin contents are corrected
for background and detector effects before normalisation, however, the
multinomial distribution is no longer a perfect description; the
precise covariance matrix for a corrected and normalised histogram is
derived in Appendix~\ref{multinomial}.

To estimate the systematic uncertainty, we repeat the complete
measurements using several variants of our standard method. In each
case the difference with respect our central result is taken as a
contribution to the uncertainty:
\begin{itemize}
\item To determine the uncertainty due to modelling of the event
selection, we vary some of the cuts for both data and Monte Carlo
samples:
\begin{itemize}
\item The containment cut is tightened to
$\left|\cos\theta_\mathrm{T}\right|<0.7$.
\item The q$\bar{\mathrm{q}}$q$\bar{\mathrm{q}}$ rejection cut is
tightened to $\Lqqqq<0.1$ and loosened to\\
$\Lqqqq<0.4$. The larger of the two absolute differences is
taken as a systematic uncertainty.
\item The q$\bar{\mathrm{q}}\ell^\pm\nu_\ell$ rejection cut is
tightened to $\Lqqln<0.25$ and loosened to
$\Lqqln<0.75$. The larger of the two absolute differences is
taken as a systematic uncertainty.
\item An simpler algorithm is used to identify initial-state
radiation, allowing only one ISR photon per event~\cite{sprime}.  This
gives an alternative value of $\sqrt{s'}$, which we use for the ISR
rejection cut.
\end{itemize}
\item The uncertainty due to background modelling is estimated by
varying the subtracted background in each bin by $\pm 5\%$. This range
of variation has been investigated by fitting Monte Carlo predictions
for the $W_\text{QCD}$ distribution to the observed data, with the
total background cross section as a free parameter~\cite{ainsley}.
Our systematic uncertainty is taken as the larger of the upper and
lower deviations in the event shape distribution.
\item To estimate the bias due to the bin-by-bin detector correction,
we re-calculate the correction factors using events simulated with
HERWIG~6.2 instead of PYTHIA~6.1. As discussed in
Section~\ref{opalmcsamples}, the PYTHIA and HERWIG samples both use
the same initial q$\bar{\mathrm{q}}$ pairs, generated by
$\mathcal{KK}$2f.
\item Instead of using the \texttt{MT} algorithm to match charged
tracks and electromagnetic energy clusters, we simply add all tracks
and clusters without compensation for double-counting.
\end{itemize}
The ranges of variation for the cuts are essentially arbitrary, and
have been chosen for consistency with other OPAL analyses.

In addition to these systematic variations, we estimate the
statistical uncertainty due to the finite numbers of Monte Carlo
events used to calculate the background subtraction and detector
correction. A complete analysis would involve determining the
statistical correlation between the numbers of hadron- and
detector-level Monte Carlo events in each bin; for our purposes,
however, we have conservatively assumed that the numerator and
denominator of the detector correction have uncorrelated statistical
uncertainties. We include the result as a contribution to the
\emph{systematic} uncertainty of our event shape distributions, though
this classification is debatable.

The total systematic uncertainty within each bin is obtained by
summing all of the above contributions in quadrature.

\subsection{Results of the event shape measurements}
\label{evshresultssection}

\enlargethispage{\baselineskip}Our complete event shape measurements
at each centre-of-mass energy are presented in
Appendix~\ref{evshdistappendix}. Results are quoted as estimates of
the probability density $R'(y)$, which satisfies the normalisation
condition $\int_0^{y_\mathrm{max}}R'(y)\,\mathrm{d}y=1$.

The distributions are shown graphically in
Figures~\ref{figdist10}--\ref{figdist213}, and are compared against
predictions from three hadron-level Monte Carlo generators: PYTHIA
6.1, HERWIG~6.2 and ARIADNE~4.11. In each case, we have generated
finely-binned histograms using a sample of five million
events\footnote{These hadron-level Monte Carlo samples are much larger
than the standard OPAL runs used to compute the detector
corrections. The size of the OPAL samples is limited by the speed of
the GOPAL detector simulation.} with the standard OPAL choice of
parameters.

As an example of the background subtraction, detector correction and
systematic uncertainty composition, we will give detailed breakdowns of
four distributions: the thrust and light jet mass measurements are
presented for $\sqrt{s}=91$~GeV and 189~GeV in
Tables~\ref{thrust91syst}--\ref{ml189syst}. The relative systematic
uncertainties are generally found to be larger at high energies. This
difference is mainly due to the presence of four-fermion background,
but is also caused by statistical fluctuations in the ``systematic
deviations'' used to estimate the uncertainty.

Our detector correction factors are particularly large at the
extremities of the light jet mass distribution, at $\sqrt{s}=189$~GeV. The
corrections are estimated from Monte Carlo models, which do not always
provide an accurate description of four-jet observables such as~\ML;
the relative systematic uncertainties in these bins are therefore
considerably larger than those in the extremities of the thrust
distribution.


 \begin{table}[p]
 \begin{leftfullpage}
 \footnotesize{
 \begin{center} 

 \end{tabular}\end{center}}
 \caption{The thrust distribution measured at $\sqrt{s}=91$~GeV. In
 the first table, the ``detector correction'' represents the ratio of
 two \emph{normalised} distributions of Monte Carlo events, at hadron
 level and detector level; the number of events selected in each bin
 is multiplied by this detector correction, then normalised, to give
 the measured distribution d$R$/d$(1-T)$. In the second table, the
 total systematic uncertainty for each bin is obtained by summing in
 quadrature the four contributions shown.}
 \label{thrust91syst}
 \end{leftfullpage}
 \end{table}


 \begin{table}[p]
 \begin{fullpage}
 \footnotesize{
 \begin{center} 

 \end{tabular}\end{center}}
 \caption{The thrust distribution measured at $\sqrt{s}=189$~GeV. In
 the first table, the measured distribution d$R$/d$(1-T)$ is obtained
 by subtracting the expected background from the number of selected
 events in each bin, then applying the detector correction and
 normalisation as described in Table~\ref{thrust91syst}. In the second
 and third tables, the total systematic uncertainty for each bin is
 calculated by summing in quadrature the eight contributions shown. For
 the background variation and the \Lqqqq\ and \Lqqln\ cut variations,
 the larger of the two absolute deviations is used; for the
 alternative Monte Carlo generators, only HERWIG is used.}
 \label{thrust189syst}
 \end{fullpage}
 \end{table}


 \begin{table}[p]
 \begin{leftfullpage}
 \footnotesize{
 \begin{center} 
\end{center}}
 \caption{The light jet mass distribution measured at $\sqrt{s}=91$~GeV.
 See Table~\ref{thrust91syst} for further explanation.}
 \label{ml91syst}
 \end{leftfullpage}
 \end{table}


 \begin{table}[p]
 \begin{fullpage}
 \footnotesize{
 \begin{center} %
 \end{tabular}\end{center}}
 \caption{The light jet mass distribution measured at
 $\sqrt{s}=189$~GeV. See Table~\ref{thrust189syst} for further
 explanation.}
 \label{ml189syst}
 \end{fullpage}
 \end{table}

\subsection{Comparison with Monte Carlo models}

\begin{figure} \begin{center}
\includegraphics[width=\linewidth]{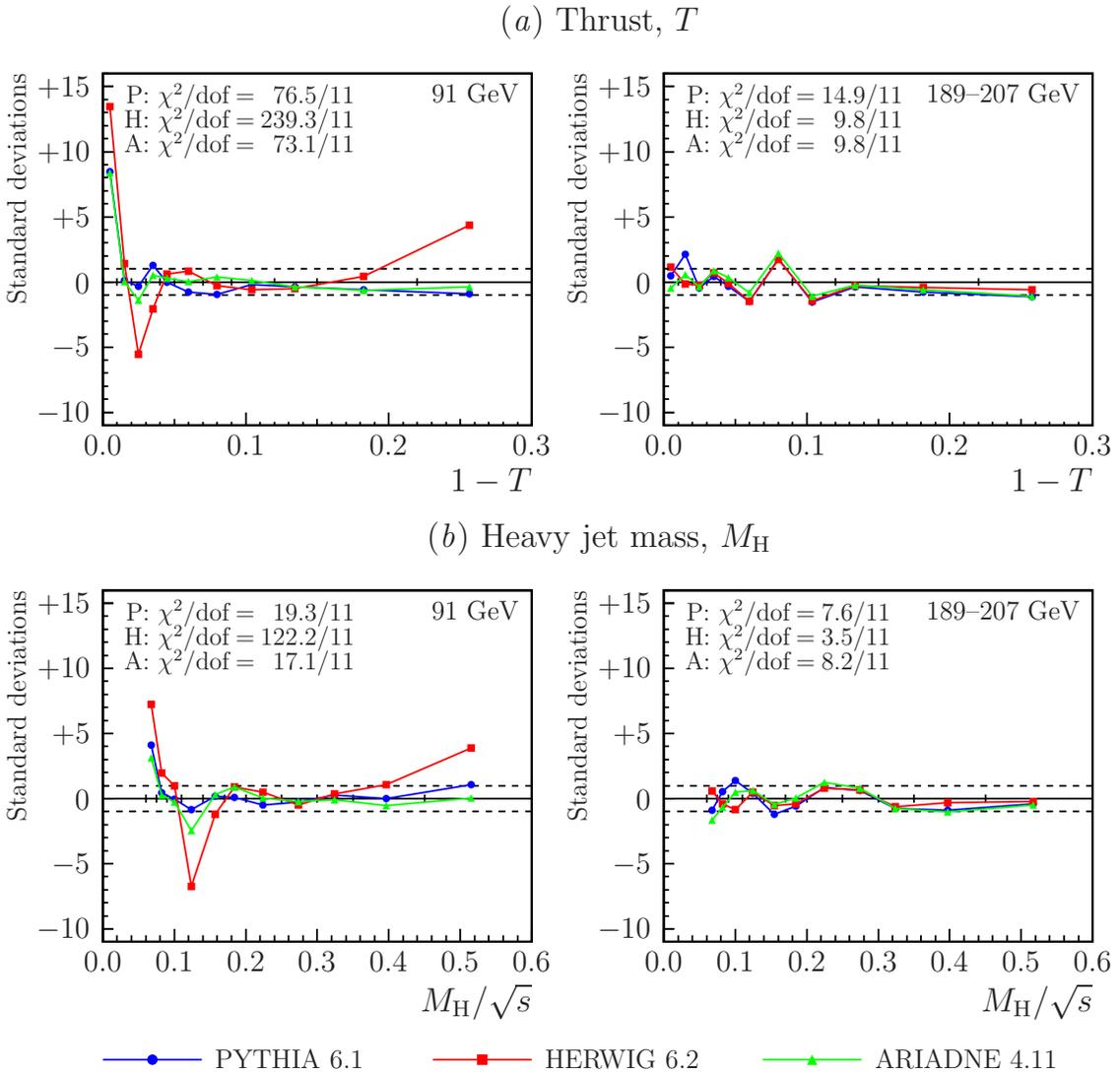} \end{center}
\caption [Deviations of Monte Carlo predictions from OPAL data, for
the thrust and heavy jet mass] {Deviations between Monte Carlo
predictions and OPAL data, for the thrust and heavy jet mass
distributions. In each bin, we show the Monte Carlo prediction minus
the OPAL measurement, expressed in units of the total uncertainty in the
data; the pairs of dashed horizontal lines represent a difference of
one standard deviation.  Bin edges are indicated by small tick-marks
on the central axis. The $\chi^2$ values for PYTHIA~(P), HERWIG~(H)
and ARIADNE~(A) are based on the total uncertainty of the data,
assuming no correlation between bins. Points for each generator are
connected by coloured lines for clarity, though no
interpolation is implied.}
\label{evsh_mcratio1}
\end{figure}

\begin{figure}
\begin{center}
\includegraphics[width=\linewidth]{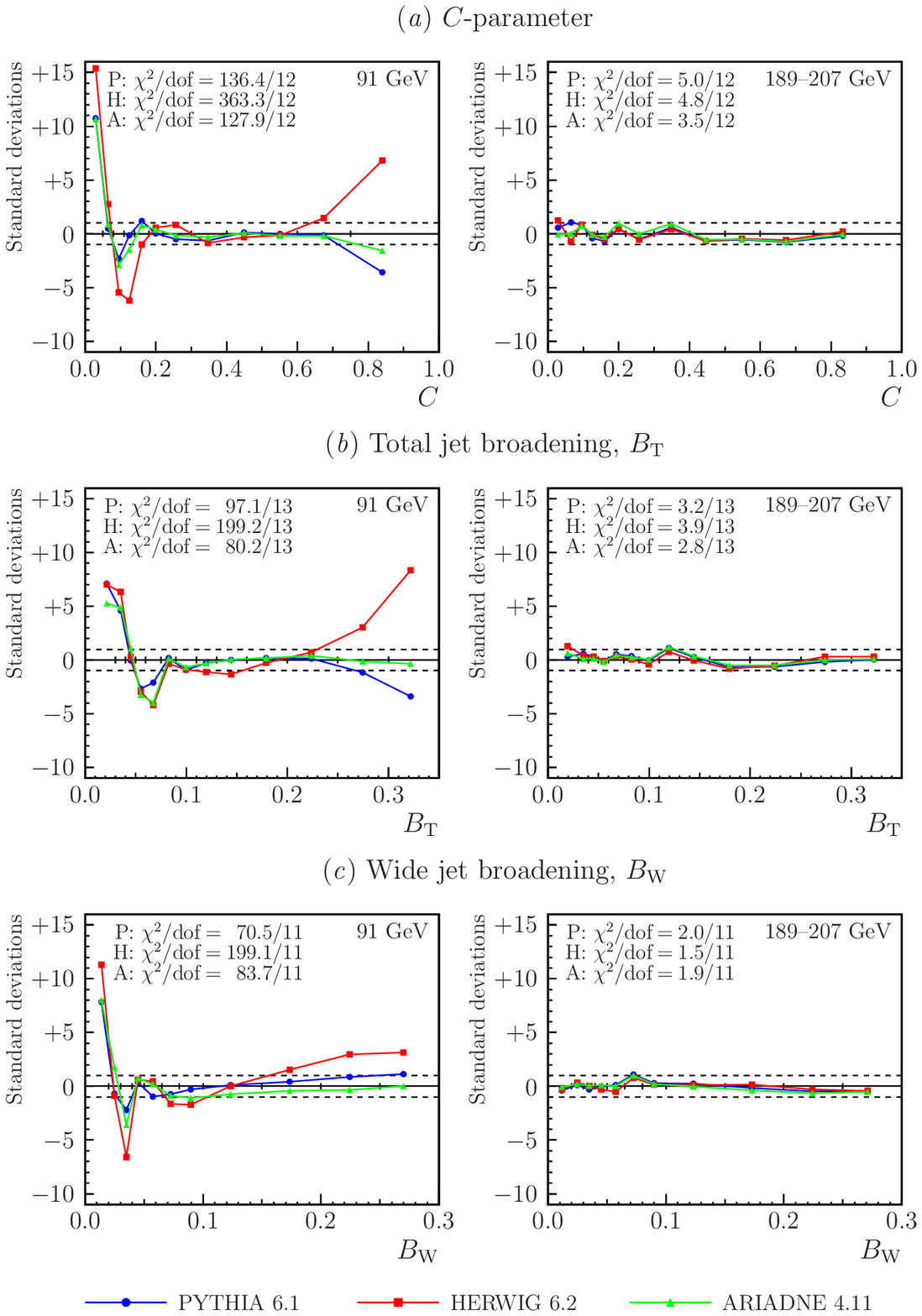}
\end{center}
\caption [Deviations of Monte Carlo predictions from OPAL data, for the
\mbox{$C$-parameter}, total jet broadening and wide jet broadening]
{Deviations of Monte Carlo predictions from OPAL data, for the
\mbox{$C$-parameter}, total jet broadening and wide jet broadening. See 
Figure~\ref{evsh_mcratio1} for full details.}
\label{evsh_mcratio2}
\end{figure}

\begin{figure}
\begin{center}
\includegraphics[width=\linewidth]{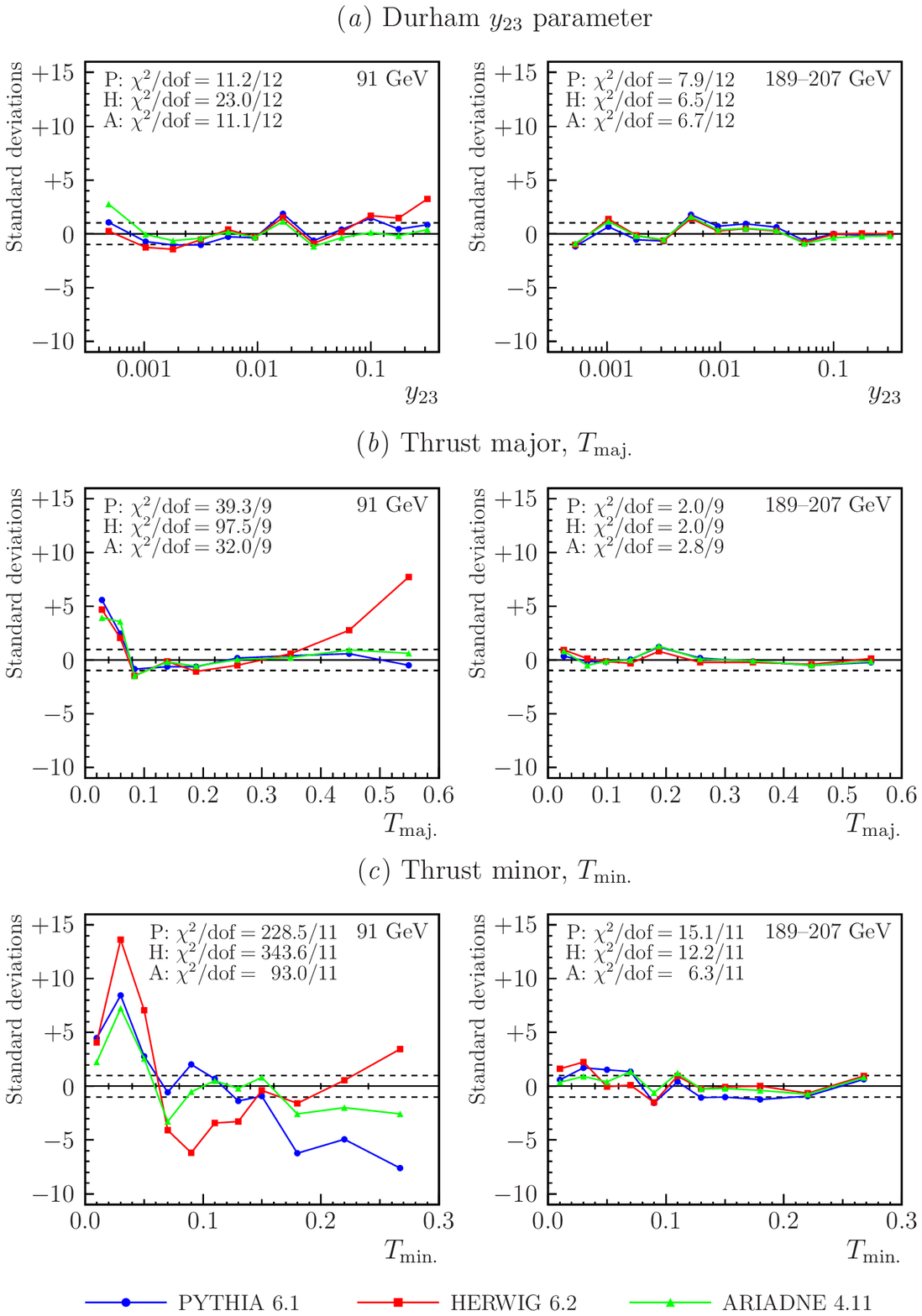}
\end{center}
\caption [Deviations of Monte Carlo predictions from OPAL data, for the
Durham $y_{23}$ parameter, thrust major and thrust minor]
{Deviations of Monte Carlo predictions from OPAL data, for the
Durham $y_{23}$ parameter, thrust major and thrust minor. See 
Figure~\ref{evsh_mcratio1} for full details.}
\label{evsh_mcratio3}
\end{figure}

\begin{figure}
\begin{center}
\includegraphics[width=\linewidth]{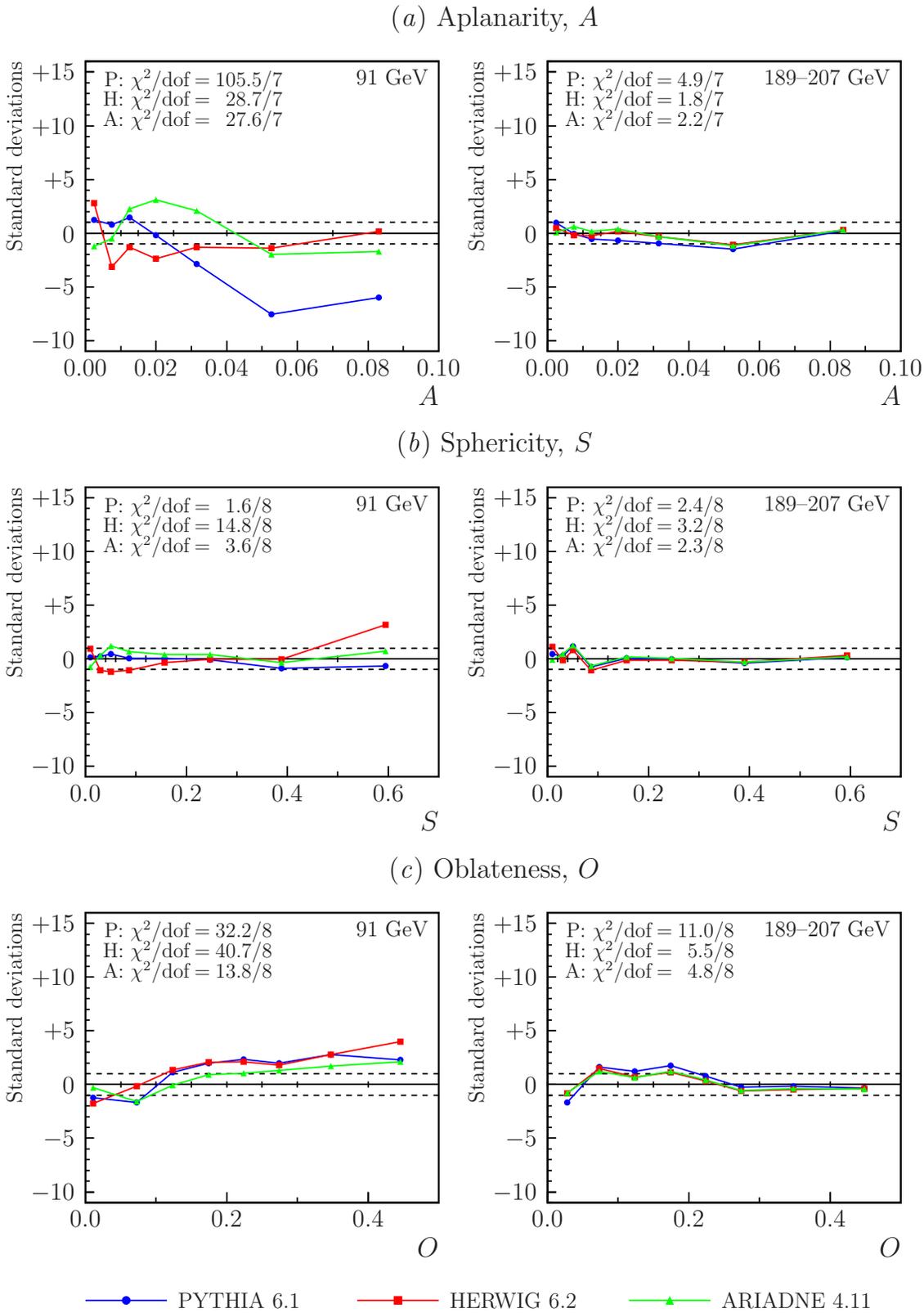}
\end{center}
\caption [Deviations of Monte Carlo predictions from OPAL data, for the
aplanarity, sphericity and oblateness]
{Deviations of Monte Carlo predictions from OPAL data, for the
aplanarity, sphericity and oblateness. See 
Figure~\ref{evsh_mcratio1} for full details.}
\label{evsh_mcratio4}
\end{figure}

\begin{figure}
\begin{center}
\includegraphics[width=\linewidth]{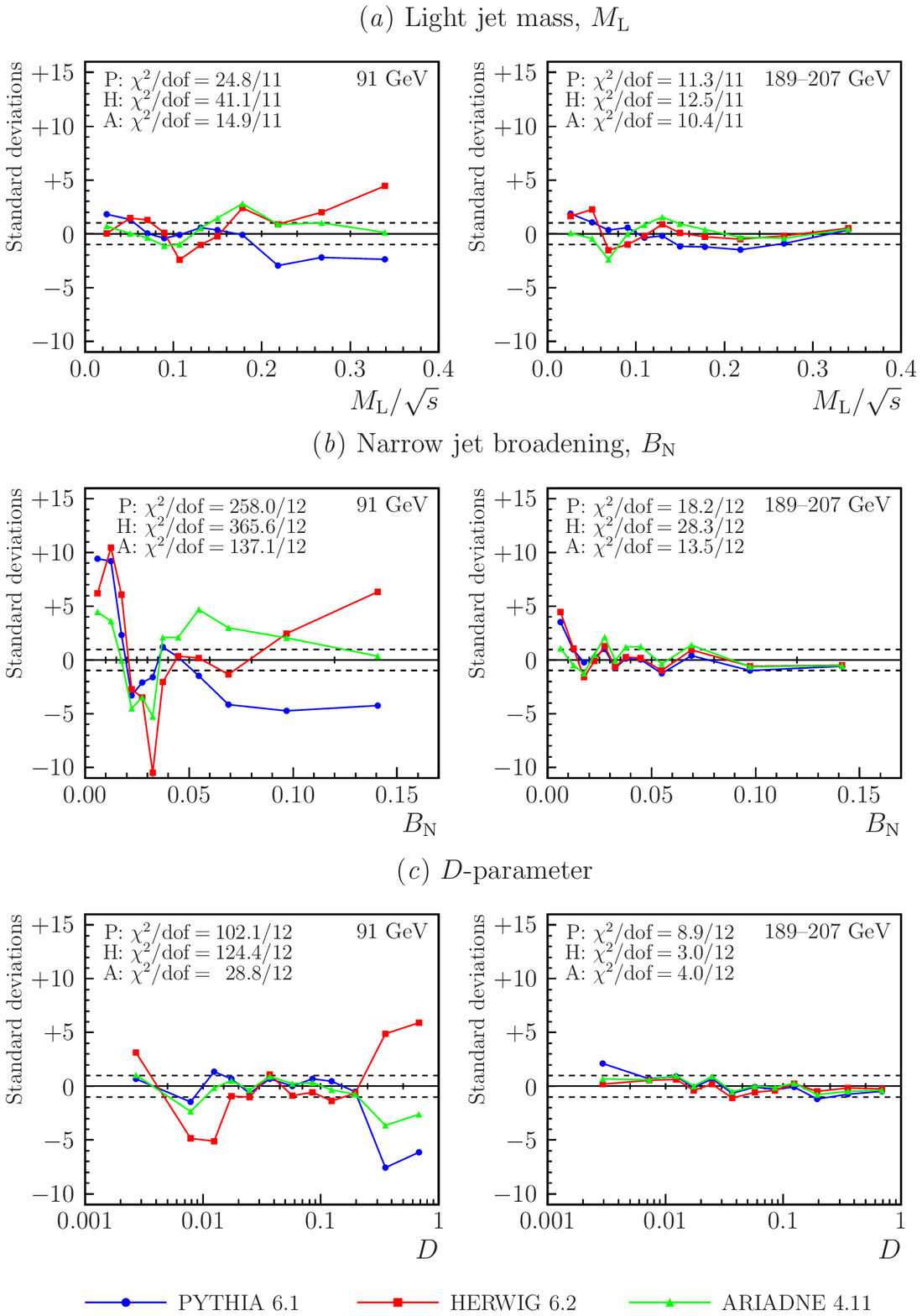}
\end{center}
\caption
[Deviations of Monte Carlo predictions from OPAL data,
for the light jet mass, narrow jet broadening and $D$-parameter.]
{Deviations of Monte Carlo predictions from OPAL data,
for the light jet mass, narrow jet broadening and $D$-parameter. See
Figure~\ref{evsh_mcratio1} for full details.}
\label{evsh_mcratio5}
\end{figure}

\enlargethispage{\baselineskip}To assess the level of agreement
between Monte Carlo models and OPAL data, we have performed a set of
crude $\chi^2$ calculations, which are illustrated in
Figures~\ref{evsh_mcratio1}--\ref{evsh_mcratio5}.  For each bin of the
measured distributions, we have calculated a corresponding
hadron-level prediction using the large PYTHIA, HERWIG and ARIADNE
samples. The differences between Monte Carlo distributions and OPAL
data are displayed as multiples of the total uncertainty in the OPAL
measurement. A $\chi^2$ value is then calculated by summing the
squared deviations across the distribution, assuming no correlation of
uncertainties between bins; this approximation is not accurate, but
provides a rudimentary statistic with which to compare the performance
of different generators.

At $\sqrt{s}=91$~GeV, where the uncertainties of the OPAL measurements
are smallest, we observe significant discrepancies in the modelling of
some distributions: all three generators appear to overestimate the
cross section in the extreme two-jet region of certain
observables.\footnote{These large discrepancies between data and Monte
Carlo, for example in the first bin of thrust, could also indicate
that our systematic uncertainties have been underestimated. We aim to
measure the model-dependence of our detector corrections by taking the
differences between results obtained using PYTHIA- and HERWIG-based
correction factors; however, the uncertainty obtained by this method
is subject to fluctuations, and may happen to give an unrealistically
low value in certain bins. This effect will need to be investigated
further, prior to publication of the final OPAL results.} HERWIG
sometimes underestimates the distributions in the peak region, as does
ARIADNE to a lesser extent. Since the distributions are normalised,
however, it is to be expected that deviations of opposite sign will
appear in different regions of the observables. In the three-jet
regions of the observables used for our \as\ fits ($T$, \MH, $C$, \BT,
\BW\ and \ytwothree), PYTHIA and ARIADNE both tend to give slightly
better predictions than HERWIG. We do not find significant evidence to
discriminate between the quality of PYTHIA and ARIADNE, for these
observables.

All three models generally give less accurate descriptions for
four-jet observables ($T_\text{min}$, $A$, $M_\text{L}$, $B_\text{N}$
and $D$) than for three-jet quantities; the models also do not agree
well with one another, which implies that the fault cannot not lie
entirely in the data. For all five of these observables, ARIADNE gives
the lowest~$\chi^2$.

At LEP2, we have combined the OPAL data at all centre-of-mass energies
\mbox{$\sqrt{s}\geq 189$~GeV}, to reduce statistical uncertainties;
this should also have the effect of reducing fluctuations in our
estimates of the systematic uncertainties. Nonetheless, our
measurements do not have sufficient precision to detect significant
deviations between Monte Carlo and data. This may in part be due to
the scaling properties of non-perturbative QCD; at higher energies,
predictions are less sensitive to the accurate modelling of
non-perturbative effects, since the effects themselves are smaller.

\subsection{A note on the horizontal positioning of data points}
All of our event shape histograms include some wide bins, which span a
substantial range of the observable. The predicted value of the
distribution $\mathrm{d}R/\mathrm{d}y$ may vary significantly across
the width of such bins, and its gradient may not be constant. When
fitting perturbative QCD predictions to the data, we integrate the
predicted distribution across the width of the bin, thereby avoiding
the need to choose a specific value for the observable~$y$. However,
when presenting the distributions graphically, we must take care in the
horizontal positioning of data points, to ensure a valid comparison
with the predicted functions.

Instead of plotting points at the centre of each bin, it is common
practice to plot them at the barycentre; this is the ``centre of
mass,'' or mean value of the observable~$y$ within the bin. However,
it has been shown~\cite{lafferty} that there is a better solution.
We can write our measured distribution in a given bin~$y_i<y<y_{i+1}$
as
\begin{equation}
R'_{i}=\,
\frac{N_i^\mathrm{corr.} / \sum_j N_j^\mathrm{corr.}}
{y_{i+1}-y_i}
\;\;\;,
\end{equation}
where $N_i^\mathrm{corr.}$ is the number of events observed in the
bin, after the background subtraction and detector
correction. The summation over bins $j$ in the normalisation
must include the `underflow' and `overflow' bins at the
extremities of the histogram. The expected value of~$R'_i$ is
\begin{equation}
\left\langle R'_i\,\right\rangle\,=\,
\frac{\int_{y_i}^{y_{i+1}}R'(y)\;
\mathrm{d}y}{y_{i+1}-y_i}\,\equiv\,
\frac{R(y_{i+1})-R(y_i)}{y_{i+1}-y_i} \;\;\;,
\end{equation}
where $R'(y)$ is the true distribution. We wish to make a graphical
comparison between the measurements~$R'_i$ and the
distributions~$R'(y)$ predicted by Monte Carlo models and by
perturbative QCD. If our measurements are in agreement with the
predictions, we would like their expected values to lie on the
curve~$R'(y)$. Adopting the notation of Ref.~\cite{lafferty} for bins
of ``large~width,'' we therefore choose to plot our data points at the
horizontal position~$y_i^\mathrm{lw}$, such that
\begin{equation}
\frac{\int_{y_i}^{y_{i+1}}R'(y)\;
\mathrm{d}y}{y_{i+1}-y_i}=\,R'(y_i^\mathrm{lw})\;\;\;.
\label{eqn_ylw}
\end{equation}
We solve Equation (\ref{eqn_ylw}) numerically at each centre-of-mass
energy, using our PYTHIA hadron-level Monte Carlo samples. The five
million events are placed in histograms comprising 100 bins of equal
width.\footnote{In the case of $y_{23}$ and the $D$-parameter, the
bins were evenly distributed in $\log y_{23}$ and $\log D$
respectively.} Where a bin of the measured distribution spans at least
five bins of the Monte Carlo distribution, we compute the horizontal
position of the data point using Equation~(\ref{eqn_ylw}); the
function $R'(y)$ is estimated by linear interpolation between the
centres of the Monte Carlo bins. For narrower bins, we plot our data
points at the centre.

In Table \ref{fig:ylw}, we show the
horizontal positions of data points for the thrust and $D$-parameter
distributions. We find that there is usually very little difference
between the centres of the bins and the calculated
positions~$y_i^\mathrm{lw}$.

\begin{table}
\begin{center}
\begin{tabular}{|r@{$\,-\,$}l|c|c|}
\hline
\multicolumn{2}{|c|}{$1-T$} & Bin centre & $y_i^\mathrm{lw}$ \bigstrut \\
\hline
\hline
\bigstrut[t]
0.0000 & 0.0100 & 0.0050 & -- \\
0.0100 & 0.0200 & 0.0150 & -- \\
0.0200 & 0.0300 & 0.0250 & -- \\
0.0300 & 0.0400 & 0.0350 & -- \\
0.0400 & 0.0500 & 0.0450 & -- \\
0.0500 & 0.0700 & 0.0600 & -- \\
0.0700 & 0.0900 & 0.0800 & -- \\
0.0900 & 0.1200 & 0.1050 & 0.1039 \\
0.1200 & 0.1500 & 0.1350 & 0.1343 \\
0.1500 & 0.2200 & 0.1850 & 0.1823 \\
0.2200 & 0.3000 & 0.2600 & 0.2565 \bigstrut[b]\\
\hline
\end{tabular}
\hspace{0.5cm}
\begin{tabular}{|r@{$\,-\,$}l|c|c|}
\hline
\multicolumn{2}{|c|}{$D$} & Bin centre & $y_i^\mathrm{lw}$ \bigstrut \\
\hline
\hline
\bigstrut[t]
0.0010 & 0.0050 & 0.0030 & 0.0030 \\
0.0050 & 0.0100 & 0.0075 & 0.0073 \\
0.0100 & 0.0150 & 0.0125 & 0.0124 \\
0.0150 & 0.0200 & 0.0175 & -- \\
0.0200 & 0.0300 & 0.0250 & -- \\
0.0300 & 0.0450 & 0.0375 & 0.0370 \\
0.0450 & 0.0700 & 0.0575 & -- \\
0.0700 & 0.1000 & 0.0850 & -- \\
0.1000 & 0.1500 & 0.1250 & 0.1230 \\
0.1500 & 0.2500 & 0.2000 & 0.1943 \\
0.2500 & 0.5000 & 0.3750 & 0.3523 \\
0.5000 & 1.0000 & 0.7500 & 0.6859 \bigstrut[b] \\
\hline
\end{tabular}
\end{center}
\caption{Horizontal positions $y_i^\text{lw}$ for data points,
calculated from Equation~(\ref{eqn_ylw}) for the thrust and
$D$-parameter distributions at $\sqrt{s}=189$~GeV. Where less than
five bins of the Monte Carlo distribution are contained within a
single bin of the measured distribution (indicated by `--'), our data
points are plotted at the centre.}
\label{fig:ylw}
\end{table}

\section[Measurements of \as]{Measurements of \boldmath\as}
\label{asfits}

We discussed in
Section~\ref{pert_predictions} the prediction of event shape
distributions by perturbative QCD. The distribution of each observable
is dependent on the dimensionless coupling strength \as, which in turn
is predicted to depend on the energy scale~$Q$ of the hard
interaction. By fitting perturbative predictions to each of our
measured distributions, we can determine~\as\ and test its energy
dependence.

Throughout this section, we use the $\mathcal{O}(\as^2)+\mathrm{NLLA}$
prediction for the cumulative distributions $R(y)$, with the Log$(R)$
matching scheme; a kinematic constraint is applied as discussed in
Section~\ref{kinconstraints}. The renormalisation scale~$\mu$ is
assumed equal to the centre-of-mass energy~$\sqrt{s}$, and the
parameters \xl\ and $p$ defined in
Section~\ref{distribution_theory_errors} are set to unity. Variation
of these assumptions will form the basis for our estimate of the
theoretical uncertainty.

\subsection{Predictions at the hadron level}
\label{hadronlevelprediction}

Perturbative QCD does not describe the complete process by which an
electron-positron system annihilates to form hadrons; instead it gives
us an intermediate ``parton~level'' prediction, for a state comprised
of quarks and gluons. In Section~\ref{mcmodels}, we discussed various
models for the non-perturbative fragmentation of quarks into bound
colourless objects. Some of these hadrons may then decay to produce
more stable particles, which are finally observed by our
detector. Before theory and data can be compared, one of these models
must be used to simulate the effects of hadronisation and decays on
the event shape distributions. By default we will use PYTHIA~6.1 with
the OPAL parameter set, as we have done for the detector
correction. Since no simulation of the OPAL detector is required in
this case, it is possible to generate larger samples: we use five
million events at each centre-of-mass energy. Also, since we do not
include initial-state radiation at the hadron and parton levels, there
is no need to simulate interference between initial- and final-state
photons emissions; we therefore generate the q$\bar{\text{q}}$ pairs using
PYTHIA in place of $\mathcal{KK}$2f.

Conventionally, the OPAL Collaboration performs the hadronisation
correction by applying a multiplicative factor to the integrated
parton-level theory prediction~$R_\mathrm{part.}(y)$:
\begin{equation}
R_\mathrm{hadr.}(y)\;=\;
\frac{R_\mathrm{hadr.}^\mathrm{MC}(y)\rule[-0.15cm]{0pt}{0pt}}
{R_\mathrm{part.}^\mathrm{MC}(y)}\;
R_\mathrm{part.}(y)\;\;\;,
\label{opalhadcor}
\end{equation}
where $R_\mathrm{part.}^\mathrm{MC}(y)$ and
$R_\mathrm{hadr.}^\mathrm{MC}(y)$ are results of the same Monte Carlo
simulation at the parton and hadron levels. Many other authors apply a
correction factor to the differential distribution $R'(y)$, or use a
full matrix-based unfolding method:
\begin{eqnarray}
R'_\mathrm{hadr.}(y) & = &
\frac{{R_\mathrm{hadr.}'\!\!\!\!\!\!\!}^\mathrm{MC}(y)\rule[-0.15cm]{0pt}{0pt}}
{{R_\mathrm{part.}'\!\!\!\!\!\!\!}^\mathrm{MC}(y)}\;
R'_\mathrm{part.}(y) \rule[-0.7cm]{0pt}{0pt}
\label{diffhadcor} \\
\mathrm{or} \;\;\;\;\;
R'_\mathrm{hadr.}(y_i) & = & \sum_j A_{ij} R'_\mathrm{part.}(y_j) \;\;\;.
\end{eqnarray}
It can be shown that Equations~(\ref{opalhadcor})
and~(\ref{diffhadcor}) are almost equivalent, provided the simulation
gives a good description of the data, and the correction factor does
not vary rapidly across the distribution. Studies by the LEP QCD
Working Group have shown that hadronisation corrections do differ
between the experiments, but that this difference is largely due to
the tuning of Monte Carlo models rather than the method by which the
correction is applied: details of this comparison will be given in
Section~\ref{lep_consistency_hadr}. It is open to question which
method is most accurate and reliable; however, for consistency with
other OPAL work, we will continue to use the correction defined in
Equation~(\ref{opalhadcor}).

\subsection{Fitting the distributions}
\label{opalfitmethod}

\enlargethispage{-1\baselineskip}Our fits to the data are based on the
method of least squares, with the coupling \as\ being the only free
parameter of the theory. We use a~$\chi^2$ variable which includes
only the statistical uncertainties of the data; experimental
systematic uncertainties in the fitted value of \as\ will be evaluated
separately, by fitting several variants of the measured
distributions.\footnote{Inclusion of systematic uncertainties in the
$\chi^2$ variable would in principle give a more reliable estimate of
\as, because the bins with smallest total uncertainties would be given
the greatest weight. However, the construction of such a $\chi^2$
would require detailed understanding of the covariance matrix relating
uncertainties in different bins. As we shall see in
Chapter~\ref{lepcombinationchapter}, an inappropriate choice of
correlation coefficients can result in a fit which bears little
connection to the data; to avoid such dangers, we will accept a
possible small bias by neglecting systematic uncertainties in our
central fit.}

As we have seen from Figures~\ref{dethad1}--\ref{dethad5}, certain
bins of the event shape distributions have a low efficiency or purity,
or a high probability of bin migration. These are usually the bins at
largest~$y$, corresponding to multiple gluon emissions. Furthermore,
the bins at low~$y$ tend to offer a poor sensitivity to \as, because
the form of the peak in this region is determined primarily by
non-perturbative physics. We therefore restrict our fit to an
intermediate range of bins, for which both our experimental
measurements and the theoretical predictions are sufficiently precise.
The choice of fit range is a trade-off between statistical and
systematic uncertainties, and the four LEP Collaborations have made
different decisions. The OPAL philosophy has been to adopt the same
fit ranges at all energy scales, regardless of variation in the
selection purity; this ensures a high correlation between the
theoretical systematic uncertainties at different
energies.\footnote{Slightly different ranges were used in the original
analysis of 91~GeV data~\cite{OPAL_as_91}, because four-fermion
background events were not considered. In our new 91~GeV measurement,
we will use the standard LEP2 fit ranges.} We therefore use the ranges
chosen for previous LEP2 analyses in
Refs.~\cite{OPAL_as_133,OPAL_as_161,OPAL_as_189,OPAL_PN512}, which are
as follows: \nolinebreak[4]
\begin{center}
\begin{tabular}{r@{$\;\;<\;$}c@{$\;<\;\;$}l}
0.05   & $1-T$       & 0.30   \\
0.17   & $M_\mathrm{H}/\sqrt{s}$ & 0.45   \\
0.18   & $C$         & 0.60   \\
0.075  & \BT         & 0.250  \\
0.05   & \BW         & 0.20   \\
0.0023 & \ytwothree  & 0.1300 \hspace{1.5cm}.
\end{tabular}
\end{center}

\enlargethispage{-\baselineskip}The fit includes statistical
correlations between bins of the normalised distribution, by using the
full covariance matrix~$V$ derived in Appendix~\ref{multinomial}. The
best fit is given by the value of \as\ which minimises the following
$\chi^2$ expression:
\begin{equation}
\chi^2\;=\;\sum_{i,j \,\in\, \mathrm{fit}}
\left[\widetilde{N}_i-\widetilde{N}_i^\mathrm{theo}(\as)\right]\left({\widehat{V}}^{-1}\right)_{ij}\left[\widetilde{N}_j-\widetilde{N}_j^\mathrm{theo}(\as)\right]
\;\;\;,
\end{equation}
where $\widetilde{N}_i$ and $\widetilde{N}_i^\mathrm{theo}(\as)$ are
the measured and predicted numbers of signal events in bin~$i$, and
${\widehat{V}}^{-1}$ is the inverse of the sub-covariance matrix within
the fit range, defined by
\begin{equation}
\sum_{j \,\in\, \mathrm{fit}} \,V_{ij}\,\left({\widehat{V}}^{-1}\right)_{jk} \;=\;\delta_{ik} \hspace{1cm}(i,k \in \mathrm{fit})\;\;\;.
\end{equation}

\subsection{Statistical uncertainties}
\label{as_stat_error}

We consider here two possible methods for the estimation of
statistical uncertainties in our \as\ measurements.

\subsubsection[Width of the $\chi^2$ minimum]{Width of the \boldmath $\chi^2$ minimum}
\label{chisquarewidth}

In a least-squares fit, the uncertainty of the fitted parameter \as\
can be estimated from the width of the minimum in $\chi^2$:
\begin{equation}
\chi^2(\as \pm \sigma_\mathrm{stat.})\;=\;\chi^2_\mathrm{min.}+1 \;\;\;\mathrm{,\hspace{1cm}where}\;\;\;\;\;
\chi^2_\mathrm{min.}=\chi^2(\as) \;\;\;.
\end{equation}
This formula makes a justified assumption that the predicted contents
of each bin varies linearly with \as, over the range of the
corresponding statistical uncertainty.

To obtain a reliable estimate using this method, $\chi^2$ must be
calculated using the full covariance matrix given in
Equation~\ref{cov_unfoldedhistogram}. If only the diagonal elements
were used, we would fail to take into account the anti-correlation
between bins, which is introduced by the normalisation; this would
result in an overestimate of the statistical uncertainty, as shown in
Table~\ref{staterror_subs_fits}.

\begin{table}
\begin{center}
\resizebox{\textwidth}{!}{
\begin{tabular}{|c||c|c|c|c||c|c|c|}
\hline
&
\multicolumn{4}{|c||}{Diagonal covariance matrix} &
\multicolumn{3}{|c|}{Full covariance matrix} \bigstrut \\
\cline{2-8} &
\parbox{1.5cm}{\centering Fitted\vspace{-0.15cm} \as} &
\parbox{2.25cm}{\centering \rule[0.4cm]{0pt}{0pt}$\sigma_\mathrm{stat.}$ from $\chi^2=\chi_\mathrm{min}^2\!+1$}\rule[-0.5cm]{0pt}{0pt} &
\parbox{2.2cm}{\centering \rule[0.4cm]{0pt}{0pt}$\sigma_\mathrm{stat.}$ from \vspace{-0.05cm}\\subsamples\rule[-0.2cm]{0pt}{0pt}} &
\parbox{1.9cm}{\centering \rule[0.4cm]{0pt}{0pt}$\sigma_\mathrm{stat.}$ from Ref.~\cite{OPAL_as_189}\rule[-0.2cm]{0pt}{0pt}} &
\parbox{1.5cm}{\centering Fitted\vspace{-0.15cm} \as} &
\parbox{2.25cm}{\centering \rule[0.4cm]{0pt}{0pt}$\sigma_\mathrm{stat.}$ from $\chi^2=\chi_\mathrm{min}^2\!+1$}\rule[-0.5cm]{0pt}{0pt} &
\parbox{2.2cm}{\centering \rule[0.4cm]{0pt}{0pt}$\sigma_\mathrm{stat.}$ from \vspace{-0.05cm}\\subsamples\rule[-0.2cm]{0pt}{0pt}}
\bigstrut \\ \hline \hline
$T$         & 0.11396 & $\pm\,0.00261$  & $\pm\,0.00222$ & $\pm\,0.0016$ & 0.11469 & $\pm\,0.00212$ & $\pm\,0.00224$ \bigstrut[t] \\
\MH         & 0.10678 & $\pm\,0.00251$  & $\pm\,0.00194$ & $\pm\,0.0017$ & 0.10690 & $\pm\,0.00191$ & $\pm\,0.00198$ \\
$C$         & 0.10840 & $\pm\,0.00241$  & $\pm\,0.00195$ & $\pm\,0.0018$ & 0.10844 & $\pm\,0.00207$ & $\pm\,0.00195$ \\
\BT         & 0.11271 & $\pm\,0.00238$  & $\pm\,0.00195$ & $\pm\,0.0014$ & 0.11284 & $\pm\,0.00198$ & $\pm\,0.00197$ \\
\BW         & 0.10311 & $\pm\,0.00213$  & $\pm\,0.00170$ & $\pm\,0.0013$ & 0.10301 & $\pm\,0.00162$ & $\pm\,0.00171$ \\
\ytwothree  & 0.10672 & $\pm\,0.00228$  & $\pm\,0.00171$ & $\pm\,0.0017$ & 0.10671 & $\pm\,0.00159$ & $\pm\,0.00169$ \bigstrut[b] \\
\hline
\end{tabular}
}
\end{center}
\caption{A comparison of the fitted \as\ values and statistical
uncertainties $\sigma_\mathrm{stat.}$ obtained by different methods at
$\sqrt{s}=189$~GeV. The covariance matrix is calculated using
Equation~(\ref{cov_unfoldedhistogram}) in Appendix~\ref{multinomial};
in the `diagonal' case, we substitute $V_{ij}=0$ for all elements $i
\neq j$. We use two different methods to calculate
$\sigma_\mathrm{stat.}$: in the first case we measure the width of the
minimum in the $\chi^2$ fit, and in the second we find the standard
deviation of 100~fits to simulated data samples.}
\label{staterror_subs_fits}
\end{table}

\subsubsection{Monte Carlo subsample method}
\label{subsamples}

In previous OPAL measurements, an exact statistical covariance matrix
has not been calculated for the measured event shape histograms.
Instead, fits have been performed using a diagonal covariance matrix, 
and a Monte Carlo technique has been used to estimate the statistical
uncertainties.

In this method, we create 100 samples of simulated data at each
centre-of-mass energy. We first calculate the quantity of each event
type expected to occur at each energy; a Poisson random number
generator is then used to determine the numbers of signal and
background events in each sample. The simulated events are drawn
randomly from the appropriate OPAL Monte Carlo sample. Selection is
performed ``with~replacement,'' so the same event may appear more than
once in any given sample.

The Monte Carlo subsamples are then processed in exactly the same way
as the real data; our standard selection criteria are applied to the
simulated events, based on the reconstructed output of the OPAL
detector simulation. Finally, we obtain a set of \as\ measurements
from each sample. The standard deviation of the 100~simulated
measurements for each event shape observable gives an estimate of the
statistical uncertainty~$\sigma_\mathrm{stat.}$. Assuming a Gaussian
distribution of \as\ values, the fractional uncertainty in our
determination of~$\sigma_\mathrm{stat.}$ from $N$~subsamples is
\mbox{$(2N+2)^{-1/2} \approx\; 7\%$}.

Given $\mathcal{N}$ Monte Carlo events from which to construct
subsamples of size $n$, one might na\"{\i}vely expect that the maximum
number of useful subsamples would be $N=\mathcal{N}/n$. If one were
using the subsamples to estimate the \emph{mean} of the statistical
distribution of \as\ values, this statement would be true; in that
case, we would not allow re-sampling of simulated events. However, it
has been shown~\cite{barlow_bootstrap} that the uncertainty in the
\emph{standard deviation} estimated from $N$~subsamples is
proportional to
\[
\sqrt{\frac{1}{N}+\rho^2} \;\;\;,
\]
where $\rho$ is the correlation coefficient between measurements from
different samples. Assuming that this correlation is roughly equal to
the fractional overlap in the event samples, $\rho =
n/\mathcal{N}$, our precision in $\sigma_\mathrm{stat.}$ will
therefore saturate when
\begin{equation}
N \; \sim \; \frac{1}{\rho^2} \; = \; {\left(\frac{\mathcal{N}}{n}\right)}^2 \;\;\;.
\end{equation}
Therefore, provided we have at least ten times more simulated events
than data events, we are justified in using 100 subsamples. This
condition is satisfied at all centre-of-mass energies except 91~GeV,
for which the integrated luminosity of the Monte Carlo sample is
roughly twice that of the data. At $\sqrt{s}=91$~GeV we therefore
generate each of the 100~subsamples with one tenth of the data
luminosity, and multiply the resulting statistical uncertainties by
$1/\sqrt{10}$.

\subsubsection[Comparing the $\chi^2$ and subsample methods]
{Comparing the \boldmath $\chi^2$ and subsample methods}

We conclude from Table~\ref{staterror_subs_fits} that the two methods for
estimating $\sigma_\mathrm{stat.}$ give compatible results when the
full covariance matrix is used. If we include only the diagonal elements of
$V_{ij}$, so that
\begin{equation}
\chi^2\;=\;\sum_{i \,\in\, \mathrm{fit}}
{\left[\frac{\widetilde{N}_i-\widetilde{N}_i^\mathrm{theo}(\as)}{\sigma^\mathrm{stat.}_i}\right]}^2
\;\;\;,
\end{equation}
then the width of the $\chi^2$ minimum gives an overestimate of the
statistical uncertainty in \as, as expected; this is because we are
ignoring the effect of the normalisation, which tends to
``average~out'' statistical fluctuations across the fit range. The
Monte Carlo subsample method, however, yields similar results
regardless of whether correlations are included in the covariance
matrix. It was claimed in Ref.~\cite{OPAL_as_189} that the statistical
uncertainties in the OPAL \as\ measurements at
\mbox{$\sqrt{s}=172$--189~GeV} had been estimated using the Monte
Carlo subsample method with a diagonal covariance matrix. Our own
results in Table~\ref{staterror_subs_fits} suggest that the published
uncertainties were usually underestimated; we believe that this
problem was caused by ignoring fluctuations in the number of
background events in the subsamples.

Since the full statistical covariance matrix is now known, we will use
it in the calculation of $\chi^2$ for our \as\ fits. A choice still
remains, however, as to whether one should use the subsample method or
the width of the $\chi^2$ minimum to estimate the uncertainties. The
subsamples have the advantage that they are based entirely on Monte
Carlo simulations, so the uncertainty will not vary between one data
sample and another.\footnote{A Bayesian statistician might argue that
it is \emph{right} for an uncertainty to depend on the observed data;
a lucky observation will provide more information than an unlucky
one. However, it is conventional in High Energy Physics to use a
purely frequentist interpretation of probability, in which an
uncertainty is defined by the standard deviation of the predicted
distribution of measurements. In an OPAL measurement of the W~boson
mass~\cite{pn279_wmass}, a similar study based on Monte Carlo
subsamples concluded that the statistical uncertainty returned by a
$\chi^2$ fit to data can sometimes be misleading; among the simulated
data samples returning the lowest uncertainties, the spread of fitted
W~mass values was not in fact smaller than average.} However, when we
calculate $\sigma_\mathrm{stat.}$ using the width of the $\chi^2$
minimum for each of the 100 subsamples, we find that the relative
variations in our estimates are of order $\pm1$--2\%, whereas the
uncertainty in our determination of $\sigma_\mathrm{stat.}$ from the
standard deviation of the 100 \as\ fits is around~7\%. We will
therefore take our statistical uncertainties from the width of the
$\chi^2$ minimum observed in the data. The subsamples will be used
later, in Section~\ref{six_observable_mean}, when we combine \as\
measurements obtained from the six observables.

\subsection{Systematic uncertainties}

The systematic uncertainties in our measurements of \as\ are
attributed to three independent sources:
\begin{itemize}
\item An \textbf{experimental systematic uncertainty}, corresponding to the
systematic uncertainty of the measured distribution
\item A \textbf{hadronisation uncertainty}, due to imperfect
simulation of non-perturbative effects by the Monte Carlo model
(PYTHIA, in this case)
\item A \textbf{theoretical uncertainty}, relating to missing
higher-order contributions to the event shape predictions in
perturbative QCD
\end{itemize}

\subsubsection{Experimental systematic uncertainties}

In Section~\ref{evsh_errors} we described several variants of our
standard analysis procedure, which were used to assess the systematic
uncertainties of our event shape measurements. We would now like to
propagate these uncertainties through to our \as\ fits. An analytical
``propagation of errors'' would be difficult in these circumstances,
as it would require detailed knowledge of the correlation between bins
of the distributions. We instead compute a set of alternative
distributions using each of the variant analyses described in
Section~\ref{evsh_errors}, and repeat the fit to determine \as\ in
each case; the difference between the \as\ values obtained from the
variant and standard distributions gives a contribution to the
systematic uncertainty. We then combine these uncertainties in exactly
the same way as for the distributions themselves, taking for example
the larger of the two deviations in \as\ when the subtracted background
is varied by $\pm 5\%$.

\subsubsection{Hadronisation uncertainties}
\label{opal_as_had_error}

To estimate the uncertainties due to modelling of non-perturbative
physics, we repeat our \as\ fits using two alternative hadronisation
corrections calculated using HERWIG~6.2 and ARIADNE~4.11 in place of
PYTHIA~6.1; in each case, we use the parameter sets tuned to OPAL
data. We generate independent samples of five million events using
each Monte Carlo model at each centre-of-mass energy. In accordance
with the current convention within the OPAL Collaboration, we define
the uncertainty in \as\ to be the larger of the two absolute deviations
with respect to the PYTHIA result.

In previous OPAL
analyses~\cite{OPAL_as_91,OPAL_as_133,OPAL_as_161,OPAL_as_189}, some
additional contributions to the hadronisation uncertainty have been
estimated by variation of the PYTHIA parameter set. However, it was
proposed recently by the LEP QCD Working Group that such uncertainties
in the parameters are already included, to a large extent, by
considering differences between independently tuned models. Although
this point may be debatable, we have adopted the LEP convention by
removing these uncertainties, which were anyhow much smaller in most
cases than the differences between models.

\subsubsection{Theoretical uncertainties}

Each of the six event shape distributions has been predicted in
perturbative QCD using $\mathcal{O}\left(\as^2\right)+\text{NLLA}$
calculations. As we have discussed in
Section~\ref{evsh_prediction_errors}, a variety of possible methods
exist to estimate the uncertainty due to missing higher orders in the
calculation. In previous OPAL measurements, the renormalisation
scale~$\mu$ has been varied over the range from $\frac{1}{2}\sqrt{s}$
to $2\sqrt{s}$; we now use the ``uncertainty band'' method, developed
in collaboration with the LEP QCD Working
Group~\cite{uncertaintyband}, which aims to incorporate a wider range
of possible higher-order contributions.

\subsection[Results of the \as\ fits]
{Results of the \boldmath \as\ fits}
\label{opal_as_fit_results}

\begin{figure}
\begin{center}
\includegraphics[width=\textwidth]{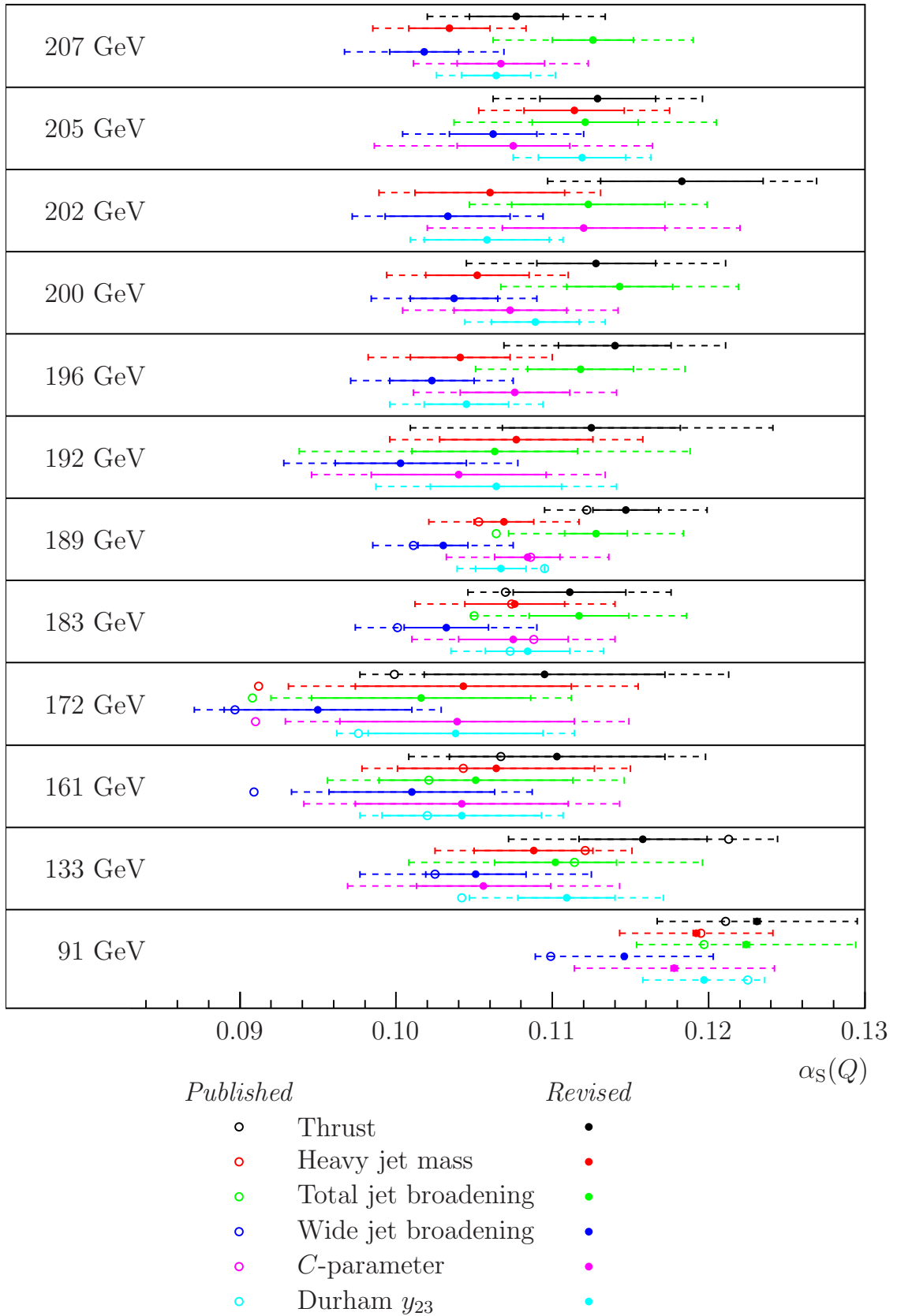}
\end{center}
\vspace{-0.3cm}
\caption{Measurements of \as\ using fits to six event shape
observables. The inner (solid) error bars represent statistical
uncertainties. The open circles indicate previously published OPAL
measurements, which are superseded by our new results.}
\label{asqfits}
\end{figure}

In Figure~\ref{asqfits}, we present our measurements of \as\ using
each of the six observables at each centre-of-mass energy; previous
OPAL measurements are indicated by open circles, where they exist. Our
results are also tabulated with a full breakdown of uncertainties in
Appendix~\ref{asfitappendix}.

Given that a high degree of correlation is expected between the
systematic uncertainties at different energy scales, our fits provide
strong evidence for the evolution of \as\ between $\sqrt{s}=91$~GeV
and the highest LEP2 energies.\footnote{To quantify this evidence in
the form of a $\chi^2$ statistic would be difficult, due to
complicated nature of the covariance matrix; this issue will be
discussed in greater detail in Chapter~\ref{lepcombinationchapter}.}
We find that the \as\ values obtained from the thrust and total jet
broadening distributions tend to be higher than the average at each
energy, while those from the wide jet broadening are consistently
lower: this effect is due to the missing higher orders in the
perturbative predictions, which differ between observables, and are
highly correlated between energy scales.

\enlargethispage{-1\baselineskip}We also see that the statistical
uncertainties are correlated between observables at each energy; at
$\sqrt{s}=161$~GeV, for example, the scatter of the six \as\
measurements less than one might expect from the size of the
statistical error-bars. This should not be surprising, when one
recalls the physical interpretations of the event shape variables:
each observable measures the amount of `branching' in an event, on
some positive scale where zero is defined to represent a perfectly
linear two-jet event. The event shape distributions are all sensitive
to the number and character of the three- and four-jet signal events
present in a given sample. If we have a significant upward fluctuation
in the number of such events (or in the number of background events
passing our selection), we expect an upward fluctuation in each of the
event shape distributions within our fit ranges, and hence an increase
in the fitted values of \as\ from all six observables. Correlations
between the event shapes will be discussed further in
Chapter~\ref{lepcombinationchapter}.

Our new \as\ fits differ in some cases by up to about $\pm 0.01$ with
respect to those previously published by the OPAL Collaboration,
though in most cases the differences are less than 0.004. In the
introduction to this chapter, we listed the main improvements that
have been applied to the analysis since these original results were
published; the observed differences are due to a combination of these.
For comparison, the intermediate results presented at
\mbox{$\sqrt{s}\leq 189$~GeV} in Refs.~\cite{OPAL_PN512,montpellier}
were obtained by fitting the new theory predictions to previously
published experimental distributions.\footnote{These are the
measurements to be used for the LEP \as\ combination presented in
Chapter~\ref{lepcombinationchapter}, and are tabulated in
Appendix~\ref{lepinputappendix}.}

The large differences seen at $\sqrt{s}=172$~GeV are mainly due to a
statistical effect. In the past, the statistical uncertainties of the
distributions were computed from the data, and not from the Monte
Carlo predictions; a downward fluctuation in the distribution could
therefore lead to an underestimation of the statistical
uncertainty. This effect is especially significant when the expected
number of events is small. The bin \mbox{$0.85<T<0.87$} at 172~GeV,
for example, is predicted by Monte Carlo to contain 10.9~events, but
only 4~events are observed in the data. When the statistical
uncertainty is measured from the Monte Carlo, this represents a
difference of 2.1~standard deviations, but using the data it is
3.5~standard deviations. The \as\ fits were consequently ``dragged
down'' in the previous analysis, by excessive $\chi^2$ contributions
from bins with downward fluctuations.\footnote{One can imagine an
extreme case in which \emph{no} events are observed in certain bin. In
this case the fit would fail altogether, due to an infinite~$\chi^2$,
if the uncertainty is estimated from the data.}

At some energies we have cross-checked our distributions with other
members of the OPAL Collaboration~\cite{donkers_kluth_pahl}, and we
have also verified our theory predictions and fitting algorithms with
other members the LEP QCD Working Group, as we will discuss in
Section~\ref{lepconsistency}. We therefore have good faith in our new
\as\ measurements.

\enlargethispage{-1\baselineskip}The figures in
Appendix~\ref{asfitappendix} present a comparison between the
measured and fitted event shape distributions at each centre-of-mass
energy. Each data point represents the ratio of the measured
hadron-level bin contents to the integral of the fitted prediction
over the width of the bin. The statistical and experimental
uncertainties are indicated by the error bars of the data points,
while the hadronisation and theory uncertainties are shown by the
horizontal yellow bands. The fit ranges are represented by arrows. At
$\sqrt{s}=91$~GeV, there is clear evidence of deviations between
theory and data outside the fit ranges, except perhaps in the case of
thrust and the Durham $y_{23}$ parameter; in some cases there is also
some evidence of a slope within the fit range, which is likely to be
caused by the missing higher orders of the perturbative
prediction. Our sensitivity to these higher orders suggests that we
have potential not only to improve our measurements of \as, but also
to test the validity of the $\mathcal{O}\left(\as^3\right)$
predictions when they become available. At the higher centre-of-mass
energies, where our statistical uncertainties are much larger, no
conclusive deviations are seen between data and theory, except in the
extreme two-jet regions of some distributions.

\section[Combined \as\ measurements]
{Combined \boldmath \as\ measurements}
\label{opal_as_combinations}

By combining some of the \as\ fits shown in Figure~\ref{asqfits}, it
should be possible to obtain a smaller number of results, each with
smaller uncertainties. In Section~\ref{six_observable_mean} we will
combine results from the six observables at each energy point, and in
Section~\ref{ascomb_allopal} we will attempt to combine all of
our OPAL results to give a single measurement of \asmz. In
Chapter~\ref{lepcombinationchapter}, we proceed to combine our OPAL
measurements with those from the other LEP experiments.

\subsection{Combining the six observables}
\label{six_observable_mean}

We would like to calculate a single measurement of \as\ at each energy
point, using all six of the event shape observables for which
perturbative predictions exist. There are two possible ways to proceed:
\begin{itemize}
\item Perform a simultaneous fit to the six measured event shape
distributions, with \as\ as a single free parameter.
\item Combine the six existing fits by forming a weighted mean.
\end{itemize}
Ideally, the simultaneous fit would be the more accurate method: it
uses a single $\chi^2$ statistic, depending explicitly on each
individual bin within the fit ranges of all six distributions. As we
have discussed, however, our fits to the individual observables are
based only on statistical uncertainties; we do not have reliable
estimates for the bin-to-bin correlations of the systematic
uncertainties. When combining results from the six event shapes, we
would like to give larger weights to those observables which have
smaller systematic uncertainties. This is not possible using a purely
statistical covariance matrix. We therefore use the second of the two
methods proposed above: we calculate a weighted mean of the six
fits to individual observables, of the form
\begin{equation}
\asq_\text{comb.} \,=\, \sum_{i=1}^6 \,w_i \,\asq_i \;\;\;.
\end{equation}

The calculation of the weights $w_i$ will be discussed in detail in
the next chapter. In summary, they are given by the expression
\begin{equation}
w_i\;=\;\frac{\sum_{j}\;\big(V^{-1}\big)_{ij}}{\sum_{j,k}\;\big(V^{-1}\big)_{jk}}\;\;\;,
\end{equation}
where $V_{ij}$ is the $6\times 6$ covariance matrix relating the
uncertainties of the six measurements. The diagonal elements of this
matrix include all four contributions to the total variance:
statistical, experimental, hadronisation and theory. The off-diagonal
elements, however, which represent correlations between the
uncertainties of different measurements, include only statistical
components.\footnote{For the LEP combination discussed in
Chapter~\ref{lepcombinationchapter}, and for the combination of all
OPAL measurements in the next section, we include both statistical and
experimental uncertainties in the off-diagonal elements of the
covariance matrix. In the present six-variable combination, however,
we find that the weights~$w_i$ are liable to fluctuate excessively
between energies, or become negative, if correlations between
experimental uncertainties are included. This problem is reduced in
the LEP combination by averaging the experimental uncertainties
between different experiments.} Although we know that the
systematic uncertainties \emph{are} correlated between
measurements, we do not include them in the covariance matrix, for
reasons to be discussed in the next chapter. The correlation
coefficients~$\rho$ between the six statistical uncertainties are
estimated using the Monte Carlo subsamples described in
Section~\ref{subsamples}; most are found to be about
0.7. Table~\ref{statcorrcompare} in the next chapter lists the
correlations determined for the 207~GeV measurements.

The calculation of uncertainties in the weighted mean will also be
discussed in the next chapter. Even though we ignore the correlation
of systematic uncertainties when calculating the weights~$w_i$, we
will attempt to re-introduce them when determining the combined
uncertainty.

Our combined results at each energy, and the weights given to each
observable, are tabulated in Appendix~\ref{asfitappendix}. Generally
the Durham \ytwothree\ parameter has the largest weight, due to the
small theoretical uncertainties in its distribution.

\subsection{Combining OPAL measurements at all energies}
\label{ascomb_allopal}

Finally, we present a global combination of all our 72 measurements of
the strong coupling. Since they have been determined at a range of
different energy scales, the first step is to convert each measurement
to a value of \as\ at the Z$^0$ mass scale. We then calculate a
weighted mean of all the results, to obtain a combined value
of~\asmz. Once again, details of the procedure will be left for the
next chapter.

Unlike our weighted means at the individual energies, the global
combination will use standardised LEP definitions for the
uncertainties: the hadronisation uncertainties used in the combined
LEP analysis are slightly different from those used in OPAL, for
example. We will also calculate our OPAL distributions in standard LEP
energy bins, instead of the twelve OPAL bins used elsewhere in this
chapter. For energies \mbox{$\sqrt{s}\leq 183$~GeV}, the OPAL and LEP
bins coincide. At the highest energies, however, seven of the OPAL
bins are combined to give three LEP energy ranges; these are shown in
Table~\ref{opal_lep_bins}. Background subtractions and detector
corrections are calculated using a weighted average of the OPAL Monte
Carlo samples within each LEP energy bin. The \as\ measurements
obtained at $\langle\sqrt{s}\rangle=198.5$~GeV are converted to the
nominal LEP energy point of 200.0~GeV (a~difference of~$\Delta\as\approx-0.0001$).

\begin{table}
\begin{center}
\begin{tabular}{|c|c||c|c|}
\hline
\multicolumn{2}{|c||}{OPAL energy bin} & \multicolumn{2}{|c|}{LEP energy bin} \bigstrut \\
\hline
\parbox{2.5cm}{\centering Mean $\sqrt{s}$\\(GeV)} &
\parbox{2.1cm}{\centering \rule{0pt}{0.4cm}Integrated\\luminosity (pb$^{-1}$)\rule[-0.2cm]{0pt}{0pt}} &
\parbox{2.5cm}{\centering Mean $\sqrt{s}$\\(GeV)} &
\parbox{2.1cm}{\centering \rule{0pt}{0.4cm}Integrated\\luminosity (pb$^{-1}$)\rule[-0.2cm]{0pt}{0pt}} \bigstrut \\
\hline
\hline
188.6 & 185.2 & \multirow{2}[2]{*}{189.0} & \multirow{2}[2]{*}{214.7} \bigstrut[t] \\
191.6 & 29.53 & & \bigstrut[b] \\
\hline
195.5 & 76.67 & \multirow{3}[2]{*}{\parbox{2.7cm}{\centering 198.5\\{\small(200.0~nominal)}}} & \multirow{3}[2]{*}{193.7} \bigstrut[t] \\
199.5 & 79.27 & & \\
201.6 & 37.75 & & \bigstrut[b] \\
\hline
204.9 & 82.01 & \multirow{2}[2]{*}{206.0} & \multirow{2}[2]{*}{220.8} \bigstrut[t] \\
206.6 & 138.8 & & \bigstrut[b] \\
\hline
\end{tabular}
\end{center}
\caption{Centre-of-mass energy bins used for the OPAL analysis and for
the LEP \as\ combination}
\label{opal_lep_bins}
\end{table}

Our result is as follows:\nopagebreak
{\small
\begin{displaymath}
\fbox{$\begin{array}{rcl}
\asmz & = & 0.1189
\;\pm\;0.0005~\mathrm{(stat.)}
\;\pm\;0.0010~\mathrm{(exp.)}
\;\pm\;0.0006~\mathrm{(hadr.)}
~\begin{array}{r}
+0.0039 \vspace{-0.3cm}\\
-0.0040
\end{array} \mathrm{(theo.)} \vspace{-0.2cm} \\
& = & 0.1189
\;\pm\;0.0005~\mathrm{(stat.)}
\;\pm\;0.0041~\mathrm{(syst.)} \\
& = & 0.1189
\;\pm\;0.0042~\mathrm{(total)}
\end{array}$}
\end{displaymath}}

\begin{figure}[tbp!]
\begin{center}
\includegraphics[width=0.8\textwidth]{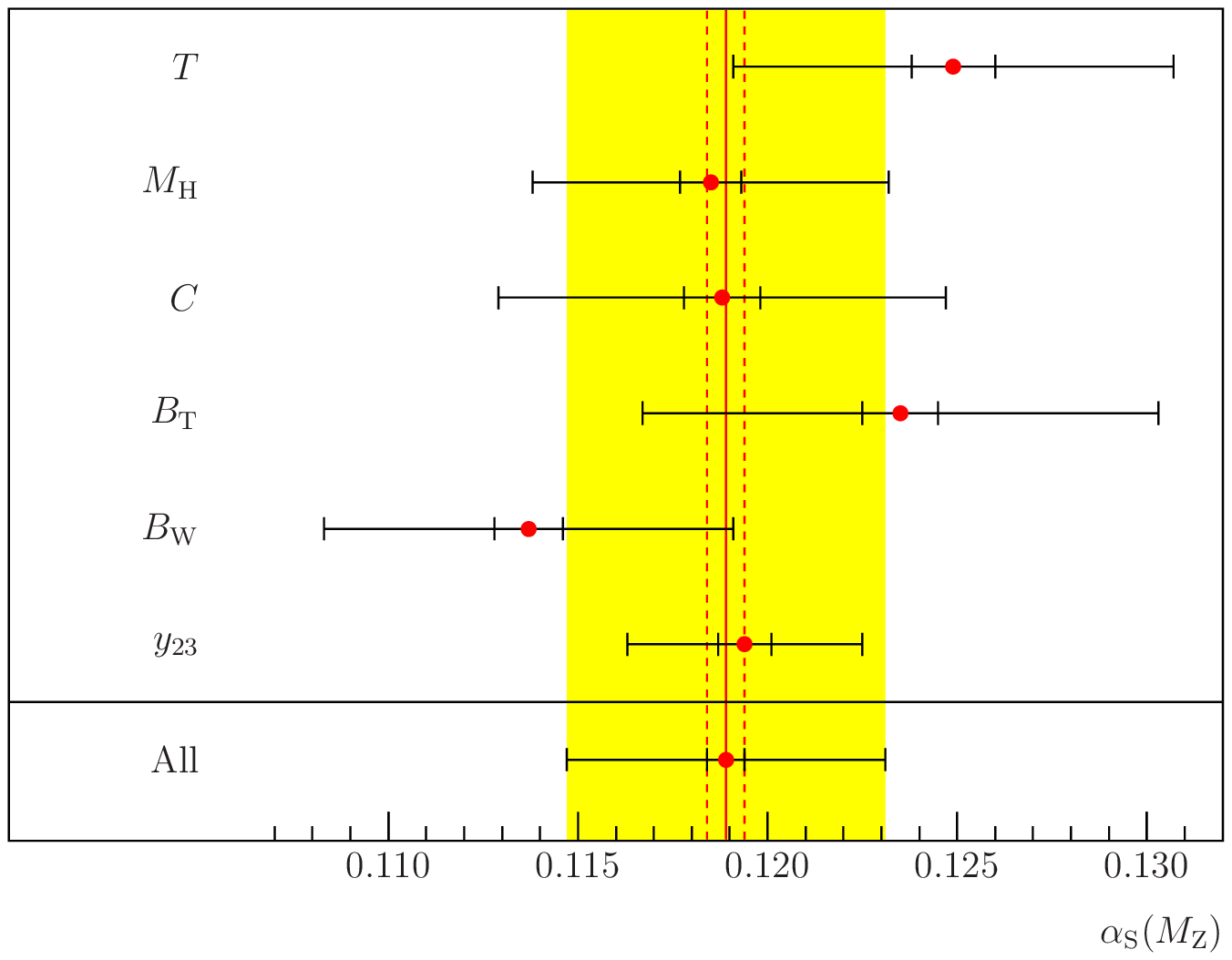}
\caption{OPAL \asmz\ combinations for individual event shape
observables. The inner error bars are statistical, while the outer
bars represent total uncertainties.}
\label{fitsbyvariableplot_opal}
\end{center}
\end{figure}

\begin{figure}[tbp!]
\begin{center}
\includegraphics[width=0.8\textwidth]{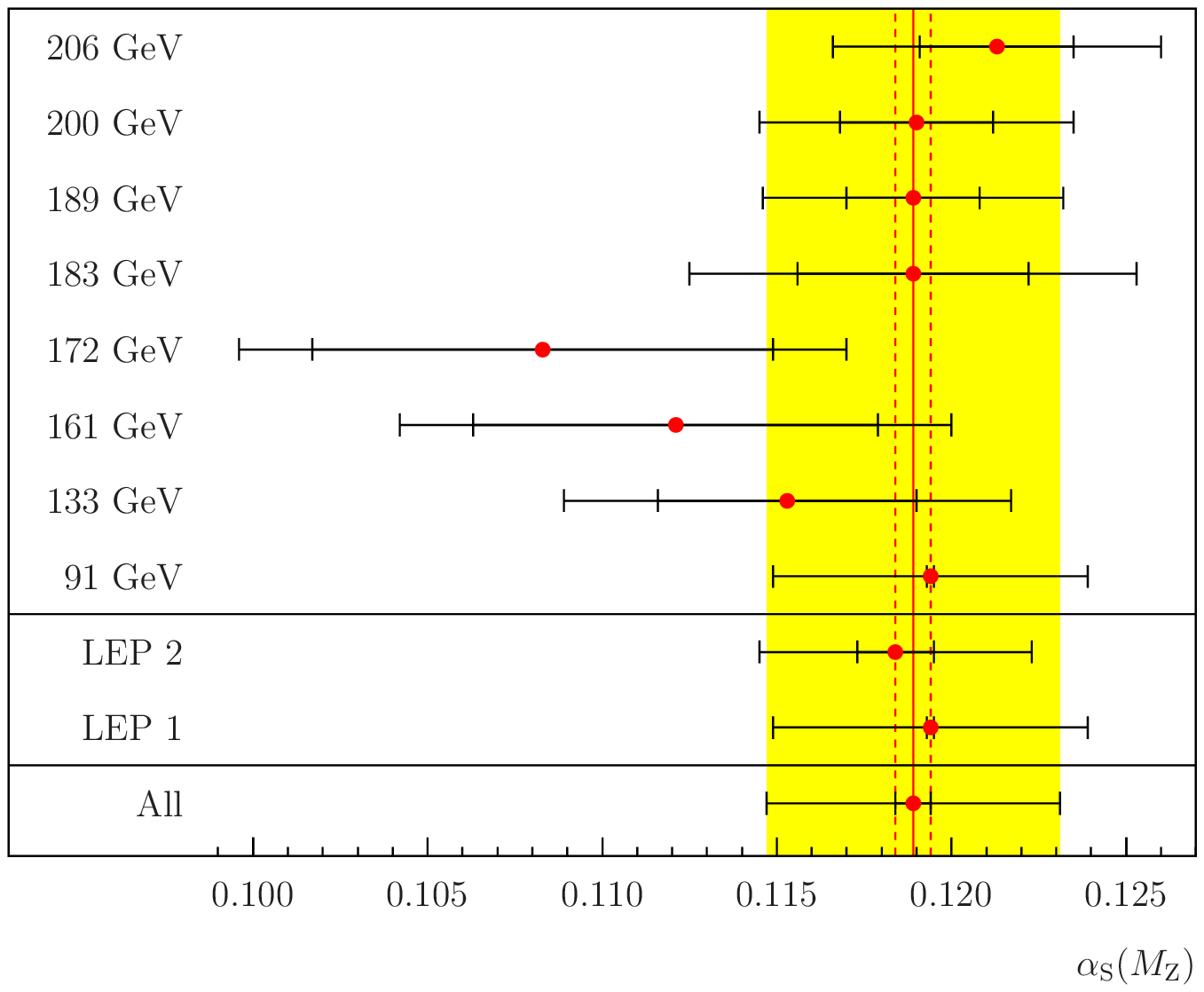}
\caption{OPAL \asmz\ combinations for individual centre-of-mass
  energies. The inner error bars are statistical, while the outer bars
  represent total uncertainties.}
\label{fitsbyenergyplot_opal}
\end{center}
\end{figure}

\begin{figure}[tbp!]
\begin{center}
\includegraphics[width=\textwidth]{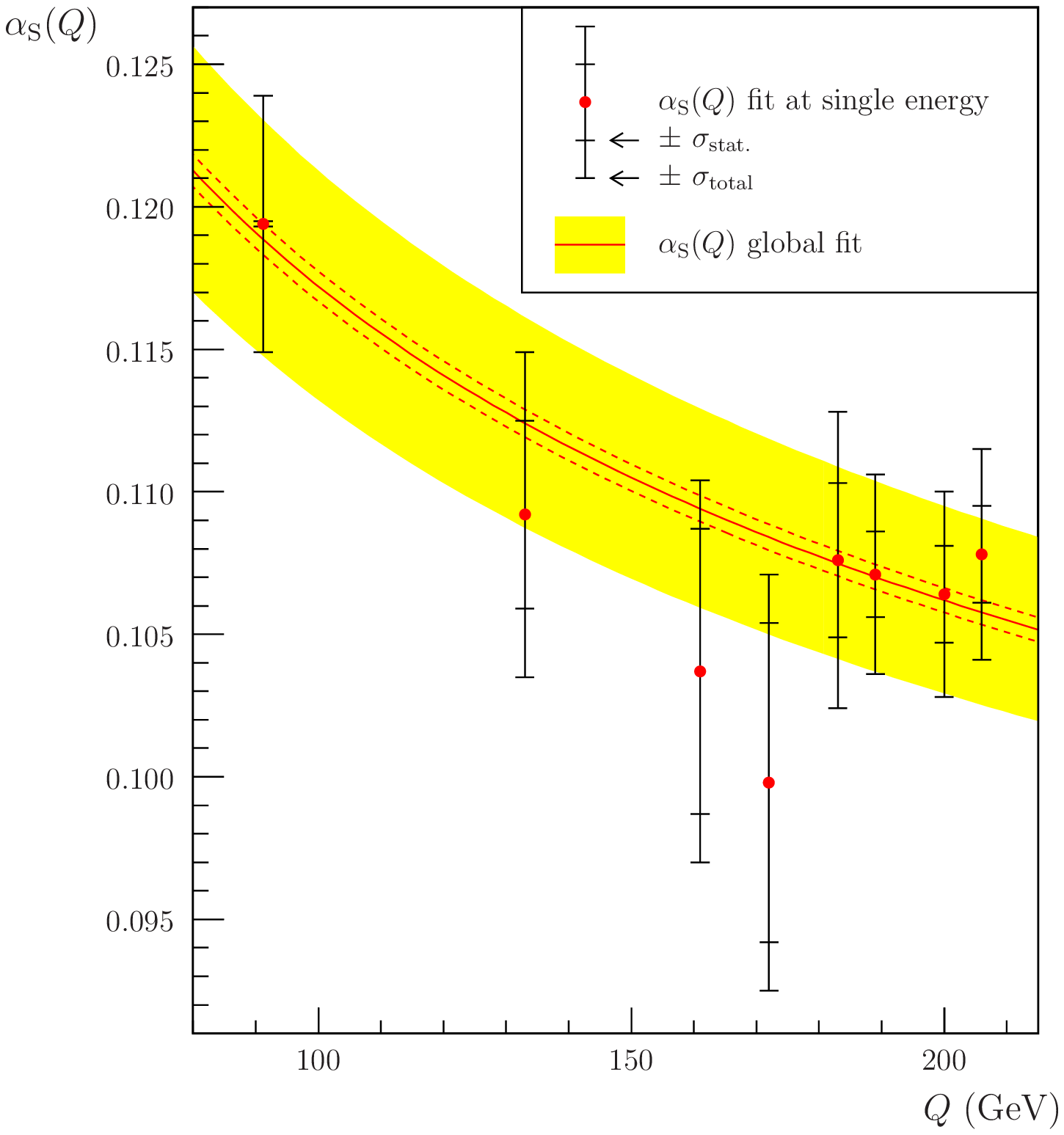}
\caption{A global QCD running fit to the OPAL \as\ measurements. Each
  point represents a fit to the six measurements at an individual
  centre-of-mass energy, while the curve represents a global fit to
  all measurements. The form of the curve is determined by the
  Renormalisation Group Equation of QCD, with \asmz\ as a free
  parameter. The yellow band corresponds to the total uncertainty of
  the fitted \asmz\ value, and the dotted curves indicate the
  statistical uncertainty.}
\label{runningplot_opal}
\end{center}
\end{figure}

We have also calculated combinations for other subsets of the OPAL
measurements. Tables~\ref{tab:fitsbyvariable_opal} gives \as\ values
for individual observables, and Table~\ref{tab:fitsbyenergy_opal}
gives separate results for LEP1\footnote{`LEP1' here refers to
measurements at $\sqrt{s}=91$~GeV, which in this case were performed
during calibration runs with the LEP2 detector.} and LEP2 data. In
both cases, a breakdown of the weights~$w_i$ is
shown. Figures~\ref{fitsbyvariableplot_opal}
and~\ref{fitsbyenergyplot_opal} present the same results, and also
show the \asmz\ values obtained from each of the eight LEP energy
bins. Figure~\ref{runningplot_opal} shows the running of \as\
predicted by the Renormalisation Group Equation for our measured value
of \asmz. The OPAL results are found to be in good agreement with the
QCD prediction, although we do not calculate a~$\chi^2$ value, due to
the complications of correlated uncertainties.

The results given here may be compared with
Tables~\ref{tab:fitsbyvariable} and~\ref{tab:fitsbyenergy}, and
Figures~\ref{runningplot}--\ref{fitsbyenergyplot} in the next chapter,
which show the corresponding results for all four LEP experiments
combined. Counter-intuitively, the total uncertainty of the overall
LEP combination is slightly larger than that of our OPAL
combination. The least-squares method should ordinarily lead to a
combination with the lowest possible uncertainty, but we shall see in
the next chapter that such a combination would not be reliable in this
case. The combined OPAL \as\ measurement is more precise than the
corresponding combinations of DELPHI and L3 results, because the
latter do not include fits to the \ytwothree\ distribution, which has
the smallest theoretical uncertainty of the six observables.

The LEP combination does not currently use the newest set of OPAL
measurements presented here. Instead, we use the most recent set of
preliminary measurements approved by the OPAL Collaboration. These can
be found in Refs.~\cite{OPAL_PN512,montpellier}, and are reproduced in
Appendix~\ref{lepinputappendix}, together with measurements from the
other experiments. The final LEP result will be published after final
measurements have become available from all four Collaborations.

\begin{table}[p]
\begin{leftfullpage}
\begin{center}
{\small
\begin{tabular}{r|r r r r r r r|}
 & \multicolumn{1}{|c}{\boldmath $T$}
 & \multicolumn{1}{c}{\boldmath $M_\mathrm{H}$}
 & \multicolumn{1}{c}{\boldmath $C$}
 & \multicolumn{1}{c}{\boldmath $B_\mathrm{T}$}
 & \multicolumn{1}{c}{\boldmath $B_\mathrm{W}$}
 & \multicolumn{1}{c}{\boldmath $y_{23}$}
 & \multicolumn{1}{c|}{\multirow{2}[1]{*}{\textbf{All}}} \\
 & & & & & & & \vspace{-0.55cm} \\
 & \multicolumn{1}{|c}{\textbf{only}}
 & \multicolumn{1}{c}{\textbf{only}}
 & \multicolumn{1}{c}{\textbf{only}}
 & \multicolumn{1}{c}{\textbf{only}}
 & \multicolumn{1}{c}{\textbf{only}}
 & \multicolumn{1}{c}{\textbf{only}}
 &

\bigstrut[b]
\\
\hline
\multicolumn{1}{l|}{\textbf{\scriptsize Fit results}}
 & & & & & & &  \bigstrut
\\
\bigstrut[t]

\bigstrut[b] $\alpha_\mathrm{s}(M_\mathrm{Z})$
 & $0.1249$
 & $0.1185$
 & $0.1188$
 & $0.1235$
 & $0.1137$
 & $0.1194$
 & $0.1189$
 \\ 
\hline \bigstrut Stat. error
 & $\pm 0.0011$
 & $\pm 0.0008$
 & $\pm 0.0010$
 & $\pm 0.0010$
 & $\pm 0.0009$
 & $\pm 0.0007$
 & $\pm 0.0005$
 \\ 
\hline \vspace{-0.15cm}{\scriptsize Exp. error}
 & {\scriptsize $\pm 0.0017$}
 & {\scriptsize $\pm 0.0017$}
 & {\scriptsize $\pm 0.0019$}
 & {\scriptsize $\pm 0.0020$}
 & {\scriptsize $\pm 0.0015$}
 & {\scriptsize $\pm 0.0012$}
 & {\scriptsize $\pm 0.0010$}
 \\ 
\vspace{-0.15cm}{\scriptsize Hadr. error}
 & {\scriptsize $\pm 0.0014$}
 & {\scriptsize $\pm 0.0009$}
 & {\scriptsize $\pm 0.0016$}
 & {\scriptsize $\pm 0.0014$}
 & {\scriptsize $\pm 0.0005$}
 & {\scriptsize $\pm 0.0004$}
 & {\scriptsize $\pm 0.0006$}
 \\
\hline
\end{tabular}
}
\end{center}
\caption{Combined \asmz\ fit results for different observables, using
OPAL data. These results may be compared with
Table~\ref{tab:fitsbyvariable} in the next chapter, which includes
data from all four LEP experiments.}
\label{tab:fitsbyvariable_opal}
\end{leftfullpage}
\end{table}

\begin{table}[p]
\begin{fullpage}
\begin{center}
{\small
\begin{tabular}{r|r r r|}
 & \multicolumn{1}{|c}{\textbf{LEP1}}
 & \multicolumn{1}{c}{\textbf{LEP2}}
 & \multicolumn{1}{c|}{\multirow{2}[1]{*}{\textbf{All}}} \\
 & & & \vspace{-0.55cm} \\
 & \multicolumn{1}{|c}{\scriptsize \boldmath $(Q=M_\mathrm{Z})$}
 & \multicolumn{1}{c}{\scriptsize \boldmath $(Q>M_\mathrm{Z})$}
 &

\bigstrut[b]
\\
\hline
\multicolumn{1}{l|}{\textbf{\scriptsize Fit results}}
 & & &  \bigstrut
\\
\bigstrut[t]

\bigstrut[b] $\alpha_\mathrm{s}(M_\mathrm{Z})$
 & $0.1194$
 & $0.1184$
 & $0.1189$
 \\ 
\hline \bigstrut Stat. error
 & $\pm 0.0001$
 & $\pm 0.0011$
 & $\pm 0.0005$
 \\ 
\hline \vspace{-0.15cm}{\scriptsize Exp. error}
 & {\scriptsize $\pm 0.0008$}
 & {\scriptsize $\pm 0.0014$}
 & {\scriptsize $\pm 0.0010$}
 \\ 
\vspace{-0.15cm}{\scriptsize Hadr. error}
 & {\scriptsize $\pm 0.0010$}
 & {\scriptsize $\pm 0.0003$}
 & {\scriptsize $\pm 0.0006$}
 \\
\hline
\end{tabular}
}
\end{center}
\caption{Combined \asmz\ fit results at LEP1 and LEP2 centre-of-mass
energies, using OPAL data. These results may be compared with
Table~\ref{tab:fitsbyenergy} in the next chapter, which includes
data from all four LEP experiments.}
\label{tab:fitsbyenergy_opal}
\end{fullpage}
\end{table}

\clearpage

\chapter[A combined LEP measurement of \as]{A combined
measurement of \boldmath{\as} by the four LEP Collaborations}

\label{lepcombinationchapter}

All four of the LEP Collaborations have presented measurements of the
strong coupling derived from \epem\ event shape observables, over the
full range of LEP centre-of-mass
energies~\cite{ALEPH_as_2,ALEPH_as_ALL,DELPHI_as_1,DELPHI_as_2,DELPHI_as_3,DELPHI_as_4,L3_as_1,L3_as_6,L3_as_7,OPAL_as_91,OPAL_as_133,OPAL_as_161,OPAL_as_189,OPAL_PN512}.
The L3 Collaboration has also published measurements at energy scales
lower than the \Zzero\ mass,\footnote{Preliminary measurements at
$Q<M_\text{Z}$ have also been performed using radiative events from
OPAL~\cite{OPAL_PN519}. These are not currently included in the LEP
\as\ combination.} using events with final-state photon radiation at
LEP1~\cite{L3_as_radiative}. In this chapter, we describe a method
developed with the \LEP~QCD Working Group to combine all of these
results into a single measurement. Such a combination should allow a
more precise measurement of \asmz, and a more sensitive test for the
energy-evolution of \asq, than is possible with one experiment alone.

\section{Preliminary consistency tests}
\label{lepconsistency}

When combining results from more than one experiment, it is essential
to ensure that the same quantity is being measured in each
case. Furthermore, when estimating uncertainties in the combined
result, it is desirable that the same analysis methods should be
followed by all Collaborations. Several preliminary tests and
comparisons were therefore performed, before the combination itself.
These consistency checks served not only to improve the reliability of
the combined LEP \as\ measurement, but also to eliminate problems in
the internal procedures of the Collaborations.

It was agreed that some minor differences could remain between the
methods used by different experiments. For example, the ALEPH and L3
Collaborations currently use a matrix unfolding method to apply
hadronisation corrections to the perturbative
predictions~\cite{ALEPH_as_ALL,L3_as_7}; \OPAL\ instead uses a
bin-by-bin correction based on the ratios of integrated distributions,
as described in Section~\ref{hadronlevelprediction}, while DELPHI uses
ratios of differential distributions~\cite{DELPHI_as_4}. Furthermore,
each Collaboration has chosen its own ranges in which to fit the event
shape distributions. However, since significant differences exist
between possible `variants' of the perturbative theory predictions, it
was decided that all Collaborations should use $\mathcal{O}(\as^2)$
matrix elements matched with the most recent NLLA resummations, using
the Log($R$) matching scheme, as described in
Section~\ref{pert_predictions}. The kinematic constraints defined in
Section~\ref{kinconstraints} were also adopted, so that the NLLA
prediction is forced to vanish in unphysical regions.

In the following sections, we outline the investigations performed by
the LEP QCD Working Group to test the consistency of our theory
predictions, fitting procedures and hadronisation corrections.

\subsection{Theoretical predictions}
\label{lepconsistency_theory}

\begin{figure}
\begin{leftfullpage}
\begin{center}
\vspace{0.5cm}
\includegraphics[width=0.85\textwidth]{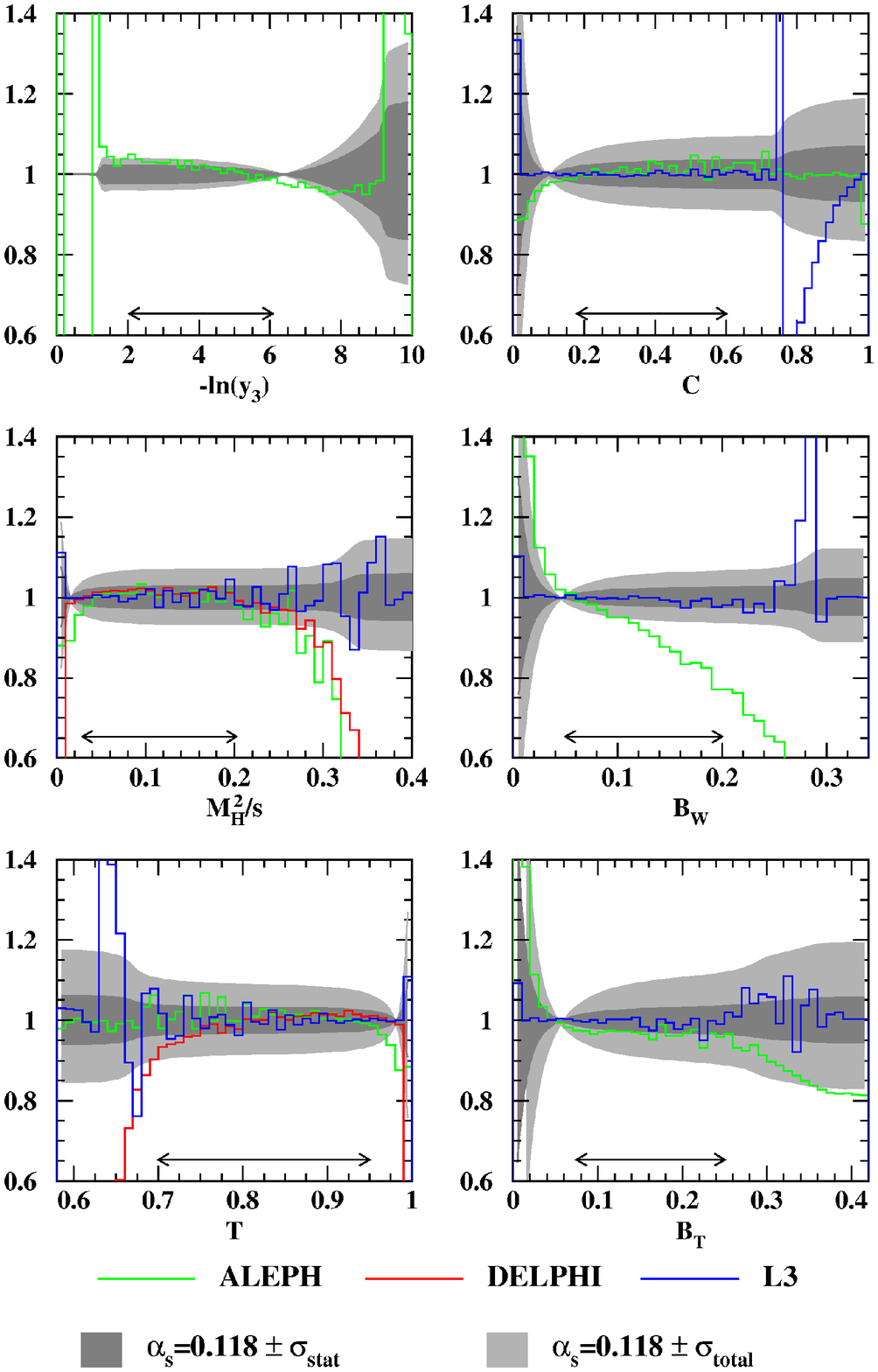}
\end{center}
\caption{A comparison between the theory predictions used for \as\
fits by the LEP Collaborations, in February~2001. The differential
cross sections d$R$/d$y$ used by ALEPH, DELPHI and L3 are shown as
multiples of that used by OPAL. The grey bands indicate fractional
variations of the OPAL prediction with respect to changes in \as: the
dark and light bands correspond to the statistical and total
uncertainties of the \as\ measurements using all OPAL data at
$\sqrt{s}\geq 203$~GeV. The fit ranges used by OPAL are indicated by
arrows.}
\label{leptheorycomp_old}
\end{leftfullpage}
\end{figure}

\enlargethispage{\baselineskip}Figure~\ref{leptheorycomp_old} illustrates a comparison between the
perturbative theory predictions used by the four Collaborations at the
start of our investigation. Differences are shown relative to the
differential cross sections used by OPAL at the time. As a measure of
the effect these deviations had on our \as\ fits, we also show bands
indicating fractional variations of the OPAL prediction with respect
to changes in \as; the widths of the bands correspond to the
statistical and total uncertainties of the OPAL \as\ measurements at
$\sqrt{s}\geq 203$~GeV. Some of the differences between our
distributions were found to be substantial. Although all four
Collaborations were using $\mathcal{O}(\as^2)+\text{NLLA}$
calculations with the Log($R$) matching scheme, the OPAL and L3
Collaborations had not yet adopted the latest NLLA calculations for
the total and wide jet broadenings~\cite{newbroadenings}, and were
also not using the kinematic constraints described in
Section~\ref{kinconstraints}.

\begin{figure}
\begin{fullpage}
\begin{center}
\vspace{0.5cm}
\includegraphics[width=0.85\textwidth]{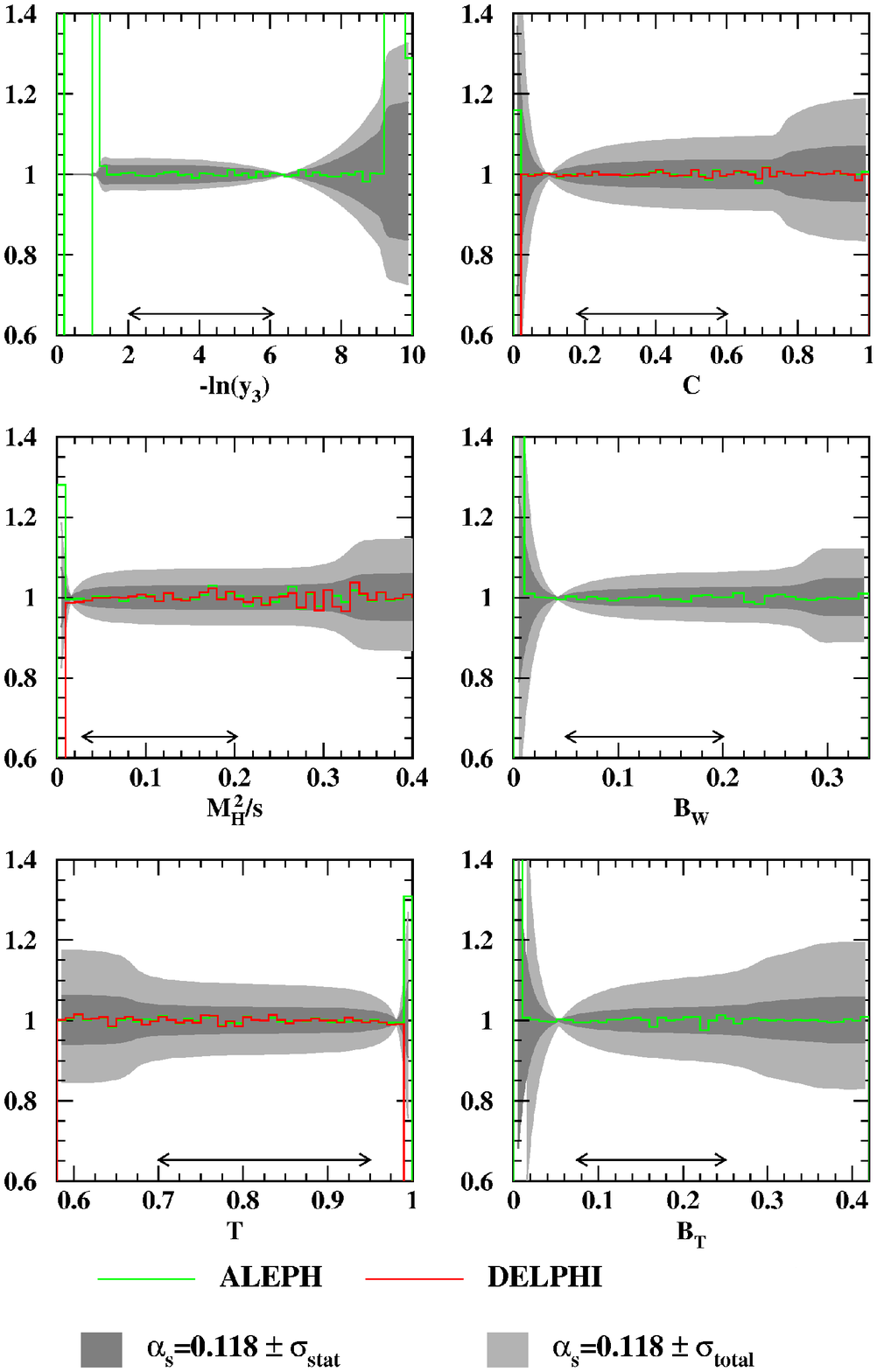}
\end{center}
\caption{A updated comparison between the theory predictions used for
\as\ fits by ALEPH, DELPHI and OPAL, in June~2001. See the caption of
Figure~\ref{leptheorycomp_old} for further details. \textcolor{white}{
FILL FILL FILL FILL FILL FILL FILL FILL FILL FILL FILL FILL FILL FILL
FILL FILL FILL FILL FILL FILL FILL FILL FILL FILL FILL FILL FILL FILL
FILL FILL FILL FILL FILL FILL FILL FILL FILL FILL FILL FILL FILL FILL
FILL FILL FILL FILL FILL FILL FILL FILL FILL FILL FILL FILL FILL FILL}}
\label{leptheorycomp_new}
\end{fullpage}
\end{figure}

After detailed discussion and documentation of the theoretical
predictions and matching schemes, satisfactory agreement was
reached. Figure~\ref{leptheorycomp_new} compares the Log($R$)-matched
predictions now implemented in the ALEPH, DELPHI and OPAL
software. The L3 predictions were not included in this figure, but were
also found to be in good agreement. The small statistical fluctuations
between bins are due to the finite number of $\mathcal{O}(\as^2)$
`events' generated in EVENT2, when estimating the $\mathcal{A}(y)$ and
$\mathcal{B}(y)$ coefficient functions.

\subsection{Fitting procedures}

\begin{figure}
\begin{center}
\includegraphics*[width=\textwidth]{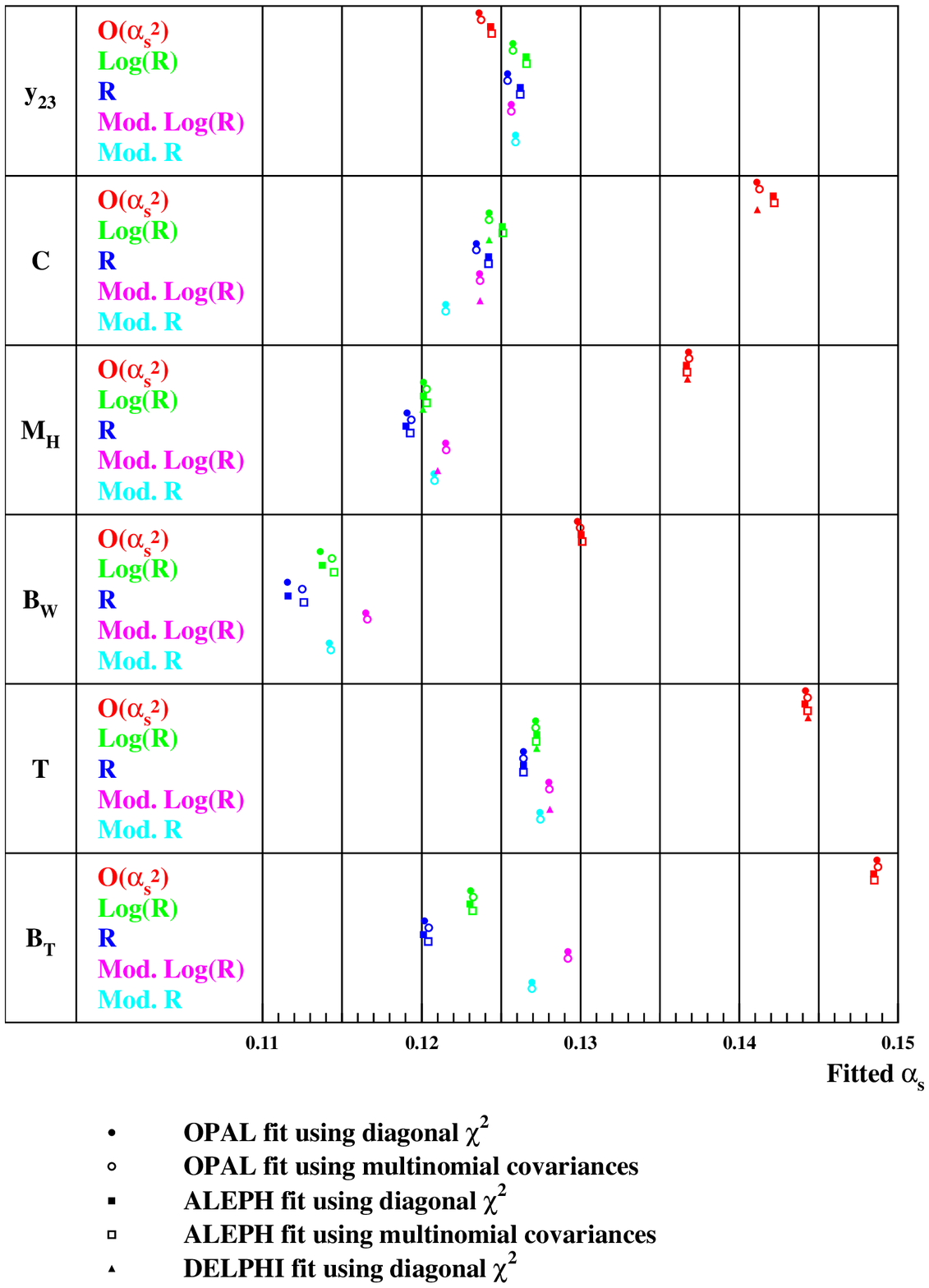}
\end{center}
\caption{Fitted \as\ values calculated by the OPAL, ALEPH and DELPHI
software, from a single set of simulated parton-level events. The five
colours indicate different theoretical predictions used in the fit
(`Mod.'~indicates that kinematic constraints were imposed), and the
different symbols represent fits by different Collaborations or using
different covariance matrices.}
\label{fitcompareplot}
\end{figure}

As a further cross-check on fits to LEP data, a set of event shape
distributions from a parton shower Monte Carlo program was distributed
to the Collaborations. Fits were then performed to determine \as\ from
the simulated parton-level `data', using two different
$\mathcal{O}(\as^2)+\text{NLLA}$ matching schemes, and also using pure
$\mathcal{O}(\as^2)$ matrix element predictions. The same fit ranges
were used by ALEPH, DELPHI and OPAL for the purposes of this test (L3
did not take part), and no hadronisation effects were included. The
results, shown in Figure~\ref{fitcompareplot}, show that the fitted
\as\ values do not differ by more than about~$\pm 0.001$. The residual
discrepancies are much smaller than the theoretical uncertainties of
the \as\ measurements, which are partially reflected in the
differences between matching schemes. The level of agreement between
experiments in this test is again limited in principle by the size of
the EVENT2 samples.

\subsection{Hadronisation corrections}
\label{lep_consistency_hadr}

\begin{figure}
\begin{center}
\includegraphics*[width=0.95\textwidth]{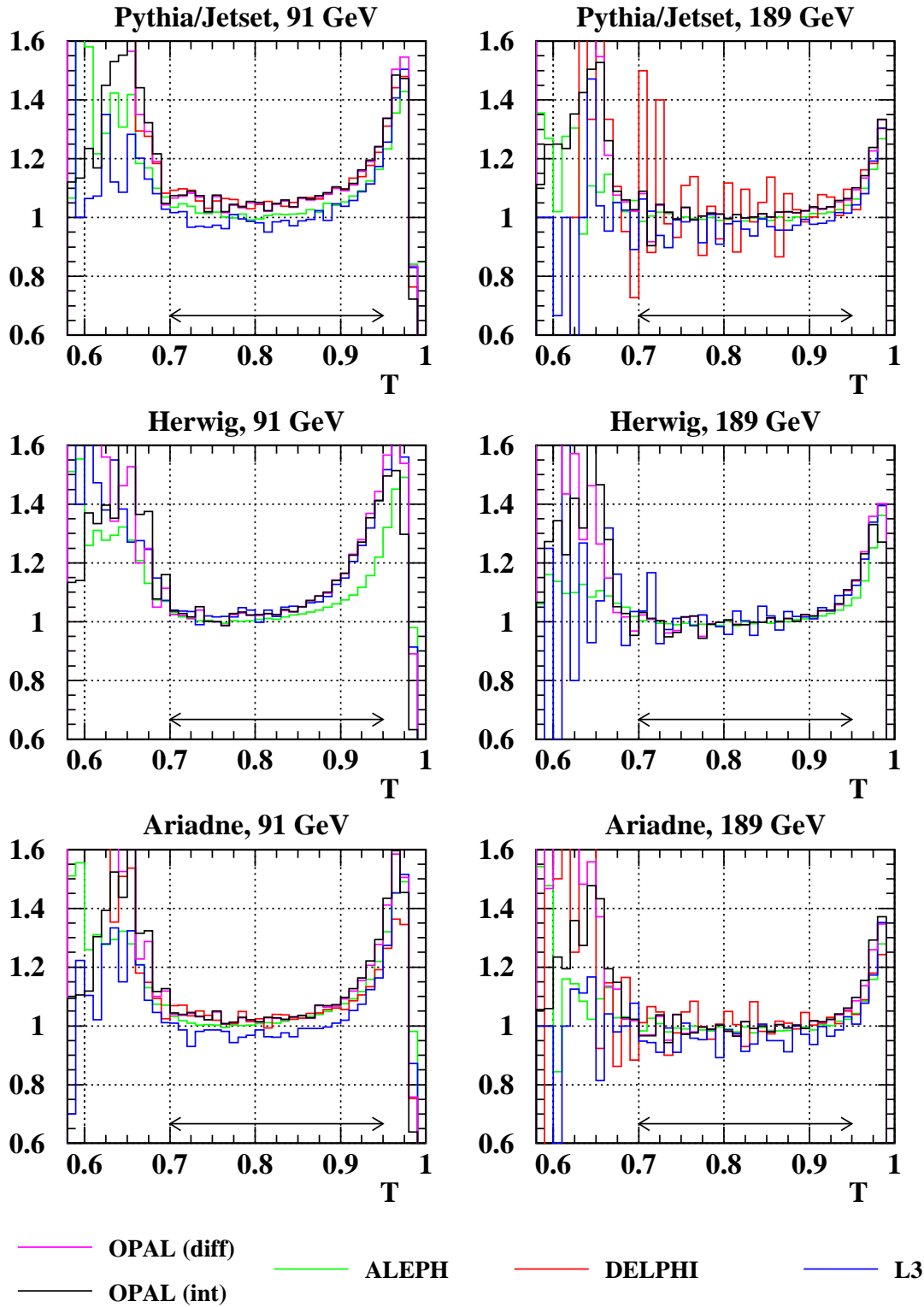}
\end{center}
\caption{Ratios of the hadron- and parton-level differential thrust
distributions,
$(\text{d}R_\text{had}/\text{d}T)/(\text{d}R_\text{part}/\text{d}T)$,
predicted by the tuned PYTHIA, HERWIG and ARIADNE Monte Carlo programs
at $\sqrt{s}=91$~GeV and 189~GeV. In the case of OPAL, we also show
the effective bin-by-bin correction factors ``OPAL~(int)'' based on
the \emph{integrated} distribution of $(1-T)$, as described in
Section~\ref{hadronlevelprediction}. The arrows indicate OPAL fit ranges for
the \as\ measurements.}
\label{lephadronratios}
\end{figure}

\enlargethispage{\baselineskip}When fitting theoretical predictions to
real data, one of the Monte Carlo models described in
Section~\ref{mcmodels} must be used to propagate the predicted
parton-level distributions to the hadron level. Although the models
themselves are universal, the Collaborations must choose values for
various parameters. The string fragmentation model of PYTHIA, for
example, is dependent on many quantities which can be `tuned' to give
the best possible fit to experimental data. As part of our programme
of consistency checks, we have therefore compared the hadronisation
corrections calculated with each tuned version of PYTHIA, HERWIG and
ARIADNE. Our results for the thrust distribution are shown in
Figure~\ref{lephadronratios}. For OPAL, we also compare hadronisation
corrections based on the differential and integrated distributions
$R(y)$ and $R'(y)$, as we described in
Section~\ref{hadronlevelprediction}.\footnote{In notation of
Section~\ref{hadronlevelprediction}, the functions ``OPAL~(diff)'' and
``OPAL~(int)'' displayed in Figure~\ref{lephadronratios} are given
respectively by:
\[ 
\text{``OPAL~(diff)''}\;=\;\frac{\text{d}R_\text{hadr.}^\text{MC}/\text{d}y}
     {\text{d}R_\text{part.}^\text{MC}/\text{d}y}
\;\;\;,\;\;\;\;\;\;\;\;
\text{``OPAL~(int)''}\;=\;
\frac{1}{\text{d}R_\mathrm{part.}/\text{d}y}\Bigg[\frac{\text{d}}{\text{d}y}\Big(\frac{R_\mathrm{hadr.}^\mathrm{MC}(y)}
{R_\mathrm{part.}^\mathrm{MC}(y)}\;R_\mathrm{part.}(y)\Big)\Bigg]
\]}
The discrepancies between hadronisation corrections used by the four
experiments are far more significant at 91~GeV than at 189~GeV. In the
three-jet region at 91~GeV, the differences are typically of order
3\%, although L3 shows slightly larger deviations relative to the
other three experiments. The two OPAL correction methods agree to
better than 1\% within the OPAL fit range; the methods become formally
equivalent if the parton shower model gives a perfect description of
the perturbative theory. Our OPAL analysis described in
Chapter~\ref{opalchapter} uses the `integral' form of the
correction, described in~\ref{hadronlevelprediction}.

{\linespread{1}\section[The input measurements and their
uncertainties]{The input measurements and their\\uncertainties}}

\begin{table}
\begin{center}
\begin{tabular}{|r|c c c c c c|}
\hline
$Q$/GeV & \TR & \MH & \BW & \BT & \CP & \ytwothree \bigstrut\\
\hline \bigstrut[t]
41.4  & \phantom{IAD}L\phantom{OI}
      & \phantom{IAD}L\phantom{OI}
      & \phantom{IAD}L\phantom{OI}
      & \phantom{IAD}L\phantom{OI}
      & \phantom{IAD}L\phantom{OI}
      & \phantom{IADLOI} \\
55.3  & \phantom{AD}L\phantom{O}
      & \phantom{AD}L\phantom{O}
      & \phantom{AD}L\phantom{O}
      & \phantom{AD}L\phantom{O}
      & \phantom{AD}L\phantom{O}
      & \phantom{ADLO} \\
65.4  & \phantom{AD}L\phantom{O}
      & \phantom{AD}L\phantom{O}
      & \phantom{AD}L\phantom{O}
      & \phantom{AD}L\phantom{O}
      & \phantom{AD}L\phantom{O}
      & \phantom{ADLO} \\
75.7  & \phantom{AD}L\phantom{O}
      & \phantom{AD}L\phantom{O}
      & \phantom{AD}L\phantom{O}
      & \phantom{AD}L\phantom{O}
      & \phantom{AD}L\phantom{O}
      & \phantom{ADLO} \\
82.3  & \phantom{AD}L\phantom{O}
      & \phantom{AD}L\phantom{O}
      & \phantom{AD}L\phantom{O}
      & \phantom{AD}L\phantom{O}
      & \phantom{AD}L\phantom{O}
      & \phantom{ADLO} \\
85.1  & \phantom{AD}L\phantom{O}
      & \phantom{AD}L\phantom{O}
      & \phantom{AD}L\phantom{O}
      & \phantom{AD}L\phantom{O}
      & \phantom{AD}L\phantom{O}
      & \phantom{ADLO} \\
91.2  & ADLO
      & ADLO
      & ADLO
      & ADLO
      & ADL\phantom{O}
      & A\phantom{DL}O \\
\hline
\end{tabular}
\end{center}
\caption{Measurements of \asq\ contributing to the \LEP\ combination.
Each letter in the table indicates a fit to the event shape
distribution indicated in the column heading. The letters themselves
indicate the experimental Collaborations (`A'=\ALEPH, `D'=\DELPHI,
`L'=\Lthree, `O'=\OPAL).}
\label{inputtable}
\end{table}

Table \ref{inputtable} lists the 194 available \as\ measurements
contributing to the LEP combination. The measurements themselves are
tabulated for reference in Appendix~\ref{lepinputappendix}. Not all of
the results are published, but all have been approved by the
individual Collaborations for use in a preliminary LEP average. In the
case of OPAL, we will use the preliminary results listed in
Refs.~\cite{OPAL_PN512,montpellier}, and not the most recent
measurements presented in Chapter~\ref{opalchapter}.

The uncertainty of each measurement has four uncorrelated
contributions: a statistical uncertainty, an experimental systematic
uncertainty, a hadronisation uncertainty, and a theoretical
uncertainty:
\begin{equation}
\sigma^2_\mathrm{total}\;=\;
\sigma^2_\mathrm{stat.}\,+\;\sigma^2_\mathrm{exp.}\,+\;
\sigma^2_\mathrm{hadr.}\,+\;\sigma^2_\mathrm{theo.}\;\;\;.
\end{equation}
All four Collaborations have decomposed their quoted uncertainties in
this way.

One can define a \mbox{$194\times 194$} covariance matrix $V_{ij}$,
relating the uncertainties of all input measurements. The diagonal
elements of this matrix are the variances $\sigma^2$ of the individual
measurements, while the off-diagonal terms relate to correlations
between measurements from different experiments, energy scales, or
observables. For example, one of the single elements $V_{ij}$ would be
\begin{displaymath}
\mathrm{Cov}\big(\;\as^{\scriptscriptstyle \left[189\;\mathrm{GeV}, \TR, \OPAL\right]},\;\as^{\scriptscriptstyle \left[161\;\mathrm{GeV}, \MH, \ALEPH\right]} \;\big)\;\;\;.
\end{displaymath}
Since the four uncertainty contributions are independent, the entire covariance
matrix can be expressed as a sum of four parts,
\begin{eqnarray}
V_{ij}^\mathrm{total} & \!=\! &
V_{ij}^\mathrm{stat.}\;+\;V_{ij}^\mathrm{exp.}\;+\;
V_{ij}^\mathrm{hadr.}\;+\;V_{ij}^\mathrm{theo.} \nonumber \\
\label{covsum}
 & \!=\! & \left\{\begin{array}{ll}
(\sigma_i^2)_\mathrm{stat.}\;+\;(\sigma_i^2)_\mathrm{exp.}\;+\;
(\sigma_i^2)_\mathrm{hadr.}\;+\;(\sigma_i^2)_\mathrm{theo.}
& i=j \\
(\rho_{ij}\sigma_i\sigma_j)_\mathrm{stat.}+
(\rho_{ij}\sigma_i\sigma_j)_\mathrm{exp.}+
(\rho_{ij}\sigma_i\sigma_j)_\mathrm{hadr.}+
(\rho_{ij}\sigma_i\sigma_j)_\mathrm{theo.} \phantom{O}
& i\neq j \phantom{OO}\end{array}\right. \;\;\;,
\end{eqnarray}
where the correlation coefficients $\rho_{ij}$ must be estimated for
each class of uncertainty.

\section[Methods for combining \as\ measurements]{Methods for combining \boldmath{\as} measurements}
\label{asmzfitsection}

Before further discussion of the uncertainties $\sigma_i$, and their
correlations, $\rho_{ij}$, we first outline our basic combination
method.

\subsection[The least-squares method for \asq]{The least-squares method for \boldmath{\asq}}

\label{simplerunning}
Given the value of the strong coupling $\as (Q_0)$, at one arbitrary
energy scale $Q_0$, the Renormalisation Group Equation~(RGE) will
predict the running coupling \asq\ at all other scales, as described
in Section~\ref{renormalisation}. Our task, therefore, is to find a
value for the parameter $\lambda = \as (Q_0)$, such that the function
$\as (Q;\lambda)$ makes the best possible fit to the data $y_i =
\left[\as (Q)\right]_i$.  We perform this fit by applying the
principle of least squares, which implies minimisation of the
following expression with respect to $\lambda$: \footnote{This is a
standard formula for $\chi^2$, to be used in the case where the input
measurements have correlated uncertainties. A brief explanation for
this formula is given in
Appendix~\ref{appsec:chisquare}.}\label{sec:chisquare}
\begin{equation}
\label{chisquare}
\chi^2\;=\;\sum_{i,j}\;\big(y_i-\as(Q_i;\lambda)\big)\;{\big(V^{-1}\big)}_{ij}\;
\big(y_j-\as(Q_j;\lambda)\big)\;\;\;.
\end{equation}
Our statistical estimator, $\hat\lambda=\hat\alpha_\text{S}(Q_0)$, is
the value of $\lambda$ corresponding to the minimum of $\chi^2$. By
convention, we set the arbitrary energy scale~$Q_0$ equal to the
\Zzero\ mass, so our parameter $\lambda$ becomes~\asmz. This is a
sensible choice, since the most statistically precise measurements of
\as\ were performed at this scale, where the cross section for
multihadron production is highest. After estimating the covariance
matrix $V$, we can use a numerical minimisation program such as
\texttt{MINUIT} to perform the fit. A numerical solution of the RGE
provides the running coupling $\as(Q_i;\lambda)$ to three-loop
accuracy, and a routine from the CERN program library inverts the
$194\times 194$ matrix~$V$. The uncertainty on \ashatmz\ can be
estimated using variations of $\chi^2$ around its minimum:
\begin{equation}
\label{chisquaredplusone}
\chi^2\big(\hat\lambda \pm \sigma_\lambda\big)=\chi^2\big(\hat\lambda\big)+1\;\;\;.
\end{equation}
This numerical implementation of the least squares method, however,
has several disadvantages:
\begin{itemize}
\item It would be useful to know the `weight' of each input
measurement contributing to the combination. We expect the more precise
measurements to carry a higher weight, and we do not expect any
measurements to carry an unduly large (or negative) weight, which
could destabilise the combination. However, the numerical
minimisation technique does not allow such weights to be calculated easily.
\item The uncertainty determined from
Equation~(\ref{chisquaredplusone}) cannot be decomposed into distinct
sources; only the total uncertainty can be calculated by this method.
\item Equation (\ref{chisquaredplusone}) assumes that the function
$\as(Q_i;\lambda)$ varies linearly with $\lambda$, so that the minimum
of $\chi^2$ is perfectly parabolic. Although true to leading order,
this condition is not strictly satisfied.
\end{itemize}

\subsection{The weighted mean method}

\label{weightssection}

Fortunately, there are conditions under which the minimum of~$\chi^2$
may be found analytically. Instead of fitting the set of measurements
$y_i \equiv \left[\as (Q)\right]_i$ to the predicted running function,
we can convert each of them into a measurement $y'_i$
of~\asmz. Letting $f_i(\as)$ be the function which maps a value of $\as
(Q_i)$ on to a corresponding value of~\asmz, we have\vspace{-0.2cm}
\begin{eqnarray}
\label{runcentral}
y'_i & \!=\! & f_i\,(y_i) \rule[-0.7cm]{0pt}{0pt} \\
\text{and}\;\;\;\;\;\sigma_{y'_i} & \!\approx\! & \frac{ f_i\,(y_i+\sigma_{y_i}) -  f_i\,(y_i-\sigma_{y_i}) }{2}\;\;\;,
\label{runerror}
\end{eqnarray}
where Equation~(\ref{runerror}) applies individually to the
statistical, experimental and hadronisation uncertainties of
$y'_i$.\footnote{Treating each $\sigma_i$ formally as the standard
deviation of some distribution, Equation~(\ref{runerror}) will be
precisely valid only in the case when $f_i\,(y_i)$ is a linear
function of $y_i$. That is to say, the shape of the probability
distribution for $\as (Q_i)$ must be preserved (up to a scale factor)
when running measurements to the scale~$Q_0$. Although this condition
is only approximately satisfied, the uncertainties $\sigma_{y_i}$
are not estimated with sufficient precision for the effects of
non-linearity to be significant.} Equation~(\ref{chisquare}) then simplifies to
\begin{equation}
\label{chisquareprime}
\chi^2\;=\;\sum_{i,j}\;\big(y'_i-\lambda\big)\;{\big({V'}^{\phantom{.}-1}\big)}_{ij}\;
\big(y'_j-\lambda\big)\;\;\;,
\end{equation}
where $V'$ is the full covariance matrix relating the measurements
$y'_i$. To find the minimum of $\chi^2$, we simply differentiate the
above expression with respect the parameter $\lambda$, and exploit the
symmetry of the covariance matrix $V'$:
\begin{equation}
\frac{\mathrm{d}\chi^2}{\mathrm{d}\lambda}\;=\;-2\sum_{i,j}\;{\big({V'}^{\phantom{.}-1}\big)}_{ij}\;\big(y'_j-\lambda\big) \;\;\;.
\end{equation}
Setting this derivative to zero at the minimum $\lambda=\hat\lambda$
gives
\begin{equation}
\sum_{i,j}\;{\big({V'}^{\phantom{.}-1}\big)}_{ij}\,y'_j\;=\;\hat\lambda\sum_{i,j}\;{\big({V'}^{\phantom{.}-1}\big)}_{ij}\;\;\;.
\end{equation}
Hence our estimator for \asmz\ becomes a linear combination of the measurements~$y'_i$,
\begin{equation}
\label{weightedmean}
\ashatmz\;=\;\hat\lambda \;=\; \sum_i\;w_i\,y'_i 
\;\;\;,
\end{equation}
where the weights $w_i$ are given by
\begin{equation}
\label{weightformula}
w_i\;=\;\frac{\sum_{j}\;{\big({V'}^{\phantom{.}-1}\big)}_{ij}}{\sum_{j,k}\;{\big({V'}^{\phantom{.}-1}\big)}_{jk}}\;\;\;.
\end{equation}

\subsection{The uncertainties of the combined measurement}

\label{errorcomb}
Since we now have an explicit expression for our combined measurement
of \asmz\ in terms of the `converted' inputs $y'_i$, we can now find
corresponding expressions for the uncertainties. The variance of
$\lambda$ is given by
\begin{equation}
\sigma^2
\;=\;\Big\langle \; \big(\lambda-\langle\lambda\rangle \big)^2 \; \Big\rangle
\;=\;\bigg\langle \; \Big[\sum_i w_i\big(y'_i-\left\langle y'_i \right\rangle \big) \Big]^2 \; \bigg\rangle \;\;\;.
\end{equation}
Expanding the square of the sum yields
\begin{eqnarray}
\sigma^2
& \!=\! & \sum_{i,j}w_i \, w_j \, \Big\langle \big(y'_i-\left\langle y'_i \right\rangle \big) 
\big(y'_j-\left\langle y'_j \right\rangle \big) \Big\rangle \nonumber \\
& \!=\! & \sum_{i,j}\;w_i \, V'_{ij} \, w_j \nonumber \\
& \!\equiv\! & w^\mathrm{T}\,V'\,w \;\;\;,
\label{totalerror}
\end{eqnarray}
where $w$ represents a vector of weights in the last line.  We have
already noted in Equation~(\ref{covsum}), however, that the covariance
matrix $V$ (and hence~$V'$), can be expressed as a sum of four
independent contributions. Hence we can now deduce the statistical,
experimental, hadronisation and theory uncertainties of our combined
measurement $\lambda$: \footnote{We will return to this point in
Section~\ref{newcombinederror}. For our final combined \asmz\
measurement, we will use Equation~\ref{errorbreakdown} to estimate
only our statistical and experimental uncertainties.}
\begin{equation}
\label{errorbreakdown}
\sigma^2_{\left\{\textrm{\parbox{0.6cm}{\tiny
stat.\vspace{-0.15cm}\\
exp.\vspace{-0.15cm}\\
hadr.\vspace{-0.15cm}\\
theo.}}
\right\}}
\;=\; w^\mathrm{T} \, V'_{
\left\{\textrm{\parbox{0.6cm}{\tiny
stat.\vspace{-0.15cm}\\
exp.\vspace{-0.15cm}\\
hadr.\vspace{-0.15cm}\\
theo.}}
\right\}
} \, w \;\;\;\;.
\end{equation}

\subsection{Minimisation of the total uncertainty}

\label{errormin}
An interesting property of our weighted mean formula,
Equation~(\ref{weightedmean}), is that it automatically minimises the
total uncertainty given in Equation~(\ref{totalerror}) for our
combined result, in addition to minimising $\chi^2$. A proof of this
result is given in Appendix~\ref{appsec:errormin}. We will return to
this property later, in Section~\ref{fitresults}, when considering our
numerical results.

\section{The covariance matrix}

\label{fullcovmatrix}

Given the formulae presented in the previous section, all that remains
for a measurement of \asmz\ is to specify the covariance matrix $V$
(or $V'$). We will first discuss the uncertainties, which form the
leading diagonal of the matrix, and then their correlations.

\subsection{The uncertainties}

\label{errorsubsect}

In previous \LEP\ combinations of \as\ measurements, the four
uncertainties $\sigma_\mathrm{stat.}$, $\sigma_\mathrm{exp.}$,
$\sigma_\mathrm{had.}$ and $\sigma_\mathrm{theo.}$ were taken directly
from values quoted by the Collaborations. However, there are
differences in the methods used to estimate these uncertainties; undue
weight could therefore be given to one Collaboration's results, simply
because they used a less conservative error estimate. We therefore
attempt to re-evaluate the uncertainties independently wherever
possible, or to smooth differences between the experiments. Our
treatments of the four error types are explained in the sections that
follow.

\subsubsection[Statistical uncertainties, $\sigma_\mathrm{stat.}$]{Statistical uncertainties, \boldmath{$\sigma_\mathrm{stat.}$}}

Unlike the three systematic uncertainties, $\sigma_\mathrm{stat.}$ is
a well-defined and calculable quantity: it is the standard deviation
of the ensemble of results obtained, if the complete experiment were
repeated indefinitely.\footnote{There is, however, an ambiguity as to
whether the uncertainty due to the finite sizes of Monte Carlo samples
should be included in $\sigma_\mathrm{stat.}$ or in
$\sigma_\mathrm{exp.}$. With the exception of LEP1 results, this
contribution is generally small in comparison to both
$\sigma_\mathrm{stat.}$ and $\sigma_\mathrm{exp.}$.} As explained in
Section~\ref{as_stat_error}, we have investigated two possible methods
to determine the statistical uncertainty for OPAL measurements, and
found them to be in good agreement. The methods used by other
Collaborations vary, but should in principle be equivalent.  We
therefore insert the quoted statistical uncertainties into the
covariance matrix without modification.

\subsubsection[Experimental systematic uncertainties, $\sigma_\mathrm{exp.}$]{Experimental systematic uncertainties,
\boldmath{$\sigma_\mathrm{exp.}$}}

\begin{table}
\begin{center}
\begin{tabular}{|r|c c c c c c|}
\cline{2-7}
\multicolumn{1}{c|}{ } & \TR    & \MH    & \BW    & \BT    & \CP    & \ytwothree
\bigstrut \\
\hline
\bigstrut[t]
ALEPH  & 0.0008 & 0.0009 & 0.0006 & 0.0007 & 0.0007 & 0.0010 \\
DELPHI & 0.0012 & 0.0019 & 0.0021 & 0.0019 & 0.0022 & --     \\
L3     & 0.0023 & 0.0011 & 0.0018 & 0.0019 & 0.0016 & --     \\
OPAL   & 0.0024 & 0.0017 & 0.0027 & 0.0034 & --     & 0.0042
\bigstrut[b] \\
\hline
\end{tabular}
\linespread{1}
\caption{Experimental systematic uncertainties in the \as\ measurements
quoted by the four Collaborations at \LEP1}
\label{t:experrors}
\end{center}
\end{table}

The experimental systematic uncertainties account for unknown biases
which are largely unique to a specific experiment or analysis. It is
therefore impossible for the \LEP~QCD Working Group to estimate them
independently. However, the problem remains that certain experiments
consistently make more conservative estimates than others. Consider,
for example, the LEP1 measurements shown in Table~\ref{t:experrors},
which suggest that the experimental systematic uncertainties quoted by
\ALEPH\ are consistently lower than those for the other
Collaborations. Experiments do, of course, have their own unique
strengths and weaknesses, which should be manifest in the systematic
uncertainties; however, it was felt that such extreme differences as
seen in Table~\ref{t:experrors} were not justified, and would lead to
unfair weighting in the combined measurement. We have therefore
averaged the experimental systematic uncertainties between all
contributing experiments, for each variable and centre-of-mass energy.

\subsubsection[Hadronisation uncertainties, $\sigma_\mathrm{had.}$]{Hadronisation uncertainties,
\boldmath{$\sigma_\mathrm{had.}$}}
\label{haderrorsubsubsect}

As described in Section~\ref{opal_as_had_error}, the uncertainties due
to non-perturbative modelling are estimated using three different
Monte Carlo programs (PYTHIA\footnote{For some of the earlier LEP
results, the hadronisation corrections were performed using \mbox{JETSET},
which has since merged with PYTHIA.}, HERWIG and ARIADNE) to calculate
hadronisation corrections to the event shape distributions. The
differences between the \as\ values obtained in the three cases should
give an indication of the uncertainty. However, the exact definitions
vary between Collaborations: some have taken the largest deviation
from the PYTHIA result, while others have taken the standard deviation
of the three results, for example.  In the past, \OPAL\ has also
considered variations of certain parameters within the individual
models~\cite{OPAL_as_91,OPAL_as_133,OPAL_as_161,OPAL_as_189}.

For the combined \LEP\ measurement, we have agreed a universal
definition for the hadronisation uncertainty. The Collaborations each
provided three sets of \as\ measurements, corresponding to the three
Monte Carlo event generators; the PYTHIA result is taken as the
central value, and the standard deviation of the three results gives
the uncertainty. The hadronisation uncertainties published
individually by the Collaborations were not used.

\subsubsection[Perturbative theory uncertainties, $\sigma_\mathrm{theo.}$]{Perturbative theory uncertainties,
\boldmath{$\sigma_\mathrm{theo.}$}}

Until recently, all four Collaborations have estimated the uncertainty
associated with the perturbative QCD predictions by varying the
renormalisation scale $\mu$ from $\frac{1}{2}\sqrt{s}$ to
$2\sqrt{s}$. As we described in Section~\ref{evsh_prediction_errors}
and in Ref.~\cite{uncertaintyband}, however, we have now developed a
more sophisticated ``uncertainty~band'' method, involving several
parameter variations. The perturbative theory uncertainties are now
estimated independently by the LEP QCD Working Group, and the values
quoted by the Collaborations are ignored.

\subsection{Correlation terms in the covariance matrix}

\label{corrsubsect}

To determine the off-diagonal parts of the covariance matrix $V$,
certain assumptions must be made about the correlation of
uncertainties between different measurements. The results provided by
the Collaborations do not explicitly include information on these
correlations, so we must make our own estimates. We shall see later,
in Section~\ref{naivecomb}, that our initial na\"{\i}ve choice of
correlations will lead to somewhat unexpected results. However, we
will proceed in this section with a description of our \emph{a~priori}
expectations for the off-diagonal part of the covariance matrix.

As with the uncertainties themselves, the covariances $V_{ij}$ can be
written as a sum of four contributions, and these will be treated
separately.

\subsubsection{Correlation of statistical uncertainties}

\label{subsubsect_statcorr}

The statistical uncertainties for measurements at different
centre-of-mass energies or different experiments are completely
uncorrelated. However, fits to different event shape distributions
measured by the same experiment at the same centre-of-mass
energy will be statistically correlated. For example, an upward
fluctuation in the number of three- or four-jet events may increase
the measured values of \as\ from all six event shape variables.

\begin{table}[tbh!]
\begin{center}
\begin{tabular}{|c|r r r r r r@{}c|}
\cline{2-8}
\multicolumn{1}{c|}{ } & \TR$\phantom{0}$ & \MH$\phantom{.}$ & \BW$\phantom{.}$ & \BT$\phantom{.}$ & \CP$\phantom{0}$ & \ytwothree$\phantom{.}$ & \hspace{0.3cm} \bigstrut \\
\hline
\multirow{2}[1]{*}[3pt]{$T$} \bigstrut[t] & $\phantom{-}1.00$ & $\phantom{-}0.79$ & $\phantom{-}0.77$ & $\phantom{-}0.80$ & $\phantom{-}0.86$ & $\phantom{-}0.75$ & \\[-3pt]
 & $1.00$ & $0.77$ & $0.62$ & $0.64$ & $0.66$ & $0.53$ & \\
 &        &        &        &        &        &        & \\[-12pt]
\multirow{2}{*}[3pt]{\MH}
 &        & $1.00$ & $0.86$ & $0.81$ & $0.78$ & $0.79$ & \\[-3pt]
 &        & $1.00$ & $0.76$ & $0.76$ & $0.75$ & $0.61$ & \\
 &        &        &        &        &        &        & \\[-12pt]
\multirow{2}{*}[3pt]{\BW}
 &        &        & $1.00$ & $0.84$ & $0.80$ & $0.83$ & \\[-3pt]
 &        &        & $1.00$ & $0.81$ & $0.67$ & $0.81$ & \\
 &        &        &        &        &        &        & \\[-12pt]
\multirow{2}{*}[3pt]{\BT}
 &        &        &        & $1.00$ & $0.82$ & $0.79$ & \\[-3pt]
 &        &        &        & $1.00$ & $0.80$ & $0.68$ & \\
 &        &        &        &        &        &        & \\[-12pt]
\multirow{2}{*}[3pt]{$C$}
 &        &        &        &        & $1.00$ & $0.70$ & \\[-3pt]
 &        &        &        &        & $1.00$ & $0.60$ & \\
 &        &        &        &        &        &        & \\[-12pt]
\multirow{2}[1]{*}[3pt]{$y_{23}$}
 &        &        &        &        &        & $1.00$ & \\[-3pt]
\bigstrut[b]
 &        &        &        &        &        & $1.00$ & \\
\hline
\end{tabular}
\linespread{1}
\caption{Statistical correlation coefficients between \as\
measurements using different event shape observables at 206~GeV. The
top coefficients of each pair were estimated using \ALEPH\ Monte Carlo
samples, while the lower coefficients were estimated by \OPAL. Both
sets of fits were performed using the \OPAL\ fit ranges.}
\label{statcorrcompare}
\end{center}
\end{table}

The correlation coefficients between pairs of fits to different
variables may be estimated using simulated data samples. A large
number of detector-level Monte Carlo subsamples were generated at each
centre-of-mass energy, with the same integrated luminosity as the
data, as explained in Section~\ref{subsamples}. The six event shape
distributions were calculated for each subsample, and `unfolded' back
to the hadron level in the same way as the data. A set of six \as\
measurements was hence obtained from each simulated data sample.
Finally, the fifteen correlation coefficients were calculated between
pairs of \as\ values from different observables.

The statistical correlation coefficients used in our \LEP\ \as\
combination were evaluated by the \ALEPH\ Collaboration, using
simulated ALEPH data~\cite{stenzel_private}. Separate correlation
coefficients were estimated for each experiment, however, due to
differing choices of fit range. We have independently cross-checked
the correlations calculated for OPAL measurements at
$\sqrt{s}=206$~GeV, using \OPAL\ Monte Carlo samples: the results of
this comparison are shown in Table~\ref{statcorrcompare}. The
uncertainties in the ALEPH and OPAL values for the correlation
coefficients are estimated to be approximately~$\pm 0.03$ and~$\pm
0.04$ respectively. There is some evidence of a slight discrepancy
(the ALEPH coefficients are all larger than our OPAL estimates, by an
average of~0.10), but tests have shown that a difference at this level
will not affect our results significantly.

\subsubsection{Correlation of experimental systematic uncertainties}

We do not expect the experimental systematic uncertainties to be
correlated between measurements from different Collaborations.
However, there should be a partial correlation between the
experimental uncertainties for measurements from the same experiment,
using different observables or different centre-of-mass energies. It
is impossible to calculate precise correlation coefficients for these
uncertainties, since we do not have a well-defined statistical
ensemble; we would need to simulate an infinite set of
\emph{detectors} and \emph{analyses}, with calibration constants and
cut values varied according according to some arbitrary
distribution. This is in contrast to the correlation of statistical
uncertainties, for which our ensemble is an infinite set of
\emph{events} occurring with well-defined probabilities.

A notable feature of the experimental uncertainty is that it contains
several independent contributions, some of which apply only at certain
centre-of-mass energies. For example, the estimation and subtraction
of four-fermion background does not contribute an uncertainty at
energies below the $\mathrm{W}^+\mathrm{W}^-$ pair
threshold. Similarly, biases related to the response of the detector
may change from year to year, and will affect different observables in
different ways. With this in mind, we will adopt a ``minimum~overlap''
convention for the correlation of experimental systematic
uncertainties; the covariance for a pair of measurements from the same
experiment is defined as the smaller of the two corresponding
variances:
\begin{equation}
\label{min_overlap}
V_{ij}^\mathrm{exp.} = \mathrm{min}\big\{\;(\sigma_i^\mathrm{exp.})^2,\; (\sigma_j^\mathrm{exp.})^2 \;\big\} \;\;\;.
\end{equation}
This is probably a slight overestimate, since it assumes that \emph{all} of
the contributions to the smaller of the two variances are fully
correlated with corresponding equal-sized contributions to the larger
variance. However, it is felt that an overestimate of the covariance
will be `safer' than an underestimate, since it should lead to a
more conservative total uncertainty in the weighted mean.

\subsubsection{Correlation of hadronisation uncertainties}
\label{subsubsect_hadcorr}

Without a complete understanding of non-perturbative QCD, we cannot
precisely calculate the correlations between our hadronisation
uncertainties. Such a calculation would require us to define a large
ensemble of alternative hadronisation models, \emph{and} to know their
relationship to true non-perturbative physics.

However, we can make very rough estimates of these correlations, using
the three Monte Carlo programs available to us. If the fitted \as\ 
results from the three generators always differ by the same amounts,
and in the same directions, we can deduce that the hadronisation
uncertainties are highly correlated; if the pattern of the three
generators' results is different for each measurement, however, the
uncertainties will be uncorrelated. This is based on the (perhaps
na\"{\i}ve) assumption that true non-perturbative QCD can be regarded
as a fourth independent `generator', so that the mutual differences
between PYTHIA, HERWIG and ARIADNE can be used to model the difference
between PYTHIA and true physics.

\subsubsection*{$\mathit{Correlations~between~observables}$}

\begin{figure}[tbp!]
\begin{center}
\includegraphics[width=\textwidth]{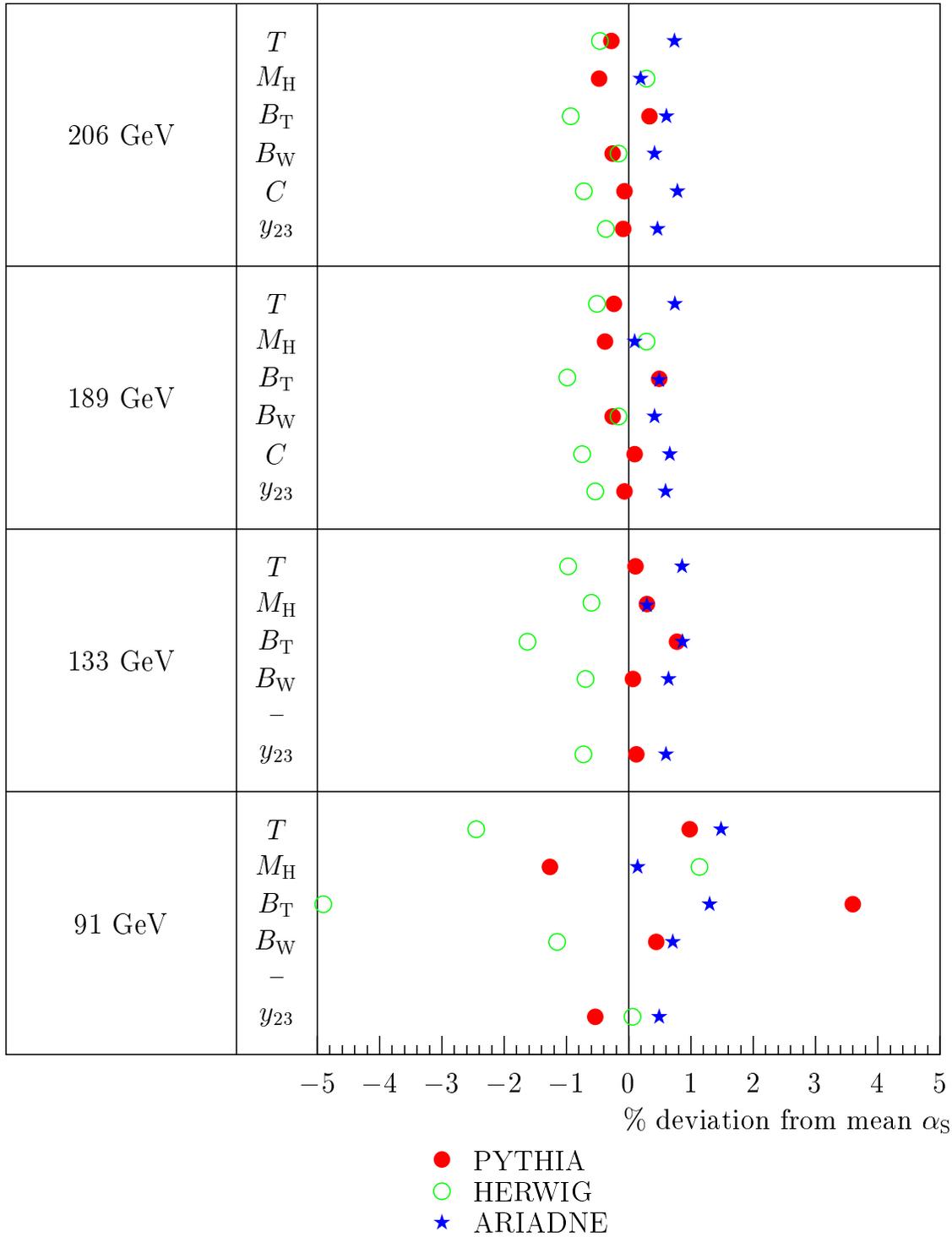}
\caption[Spread of \as\ fit results using PYTHIA, HERWIG and ARIADNE
hadronisation corrections (grouped by observable)]{The spread of \as\
fit results using PYTHIA, HERWIG and ARIADNE hadronisation
corrections, for fits to \TR, \MH, \BT, \BW, \CP\ and \ytwothree. This
comparison uses \OPAL\ data, at centre-of-mass energies of 91, 133,
189 and 206~GeV. In each case, the three results are plotted as
percentage deviations from their mean value. \emph{(Note, however,
that the ``central~value'' used in our analyses is the PYTHIA result,
not the mean result.)}}
\label{hadcorr_vars}
\end{center}
\end{figure}

Figure~\ref{hadcorr_vars} compares OPAL \as\ measurements using the
three generators, for each of the six event shape observables. For
\LEP2 data, we observe a similar pattern for all observables: in most
cases we have
\begin{displaymath}
\as^\mathrm{HERWIG}\;<\;\as^\mathrm{PYTHIA}\;<\;\as^\mathrm{ARIADNE}\;\;\;.
\end{displaymath}
The distribution of 91~GeV fits is somewhat more random. However, for
the majority of our results, we deduce that the hadronisation
uncertainties are highly, but not fully, correlated between fits to
different observables. As a crude estimate of these correlations, we
will make the same ``minimum~overlap'' assumption as defined in
Equation~\ref{min_overlap} for the experimental systematic
uncertainties.
\footnote{The minimum overlap assumption is not so easily justified for
hadronisation uncertainties, because they are not derived from several
independent contributions added in quadrature. However, we will see later
(Section~\ref{newcombinederror}) that precise estimates of these
correlations are not needed for our final combination. The crude
simplifications made in this discussion will not, therefore,
compromise our results.}

\subsubsection*{$\mathit{Correlations~between~energy~scales}$}

\begin{figure}[tbp!]
\begin{center}
\includegraphics[width=\textwidth]{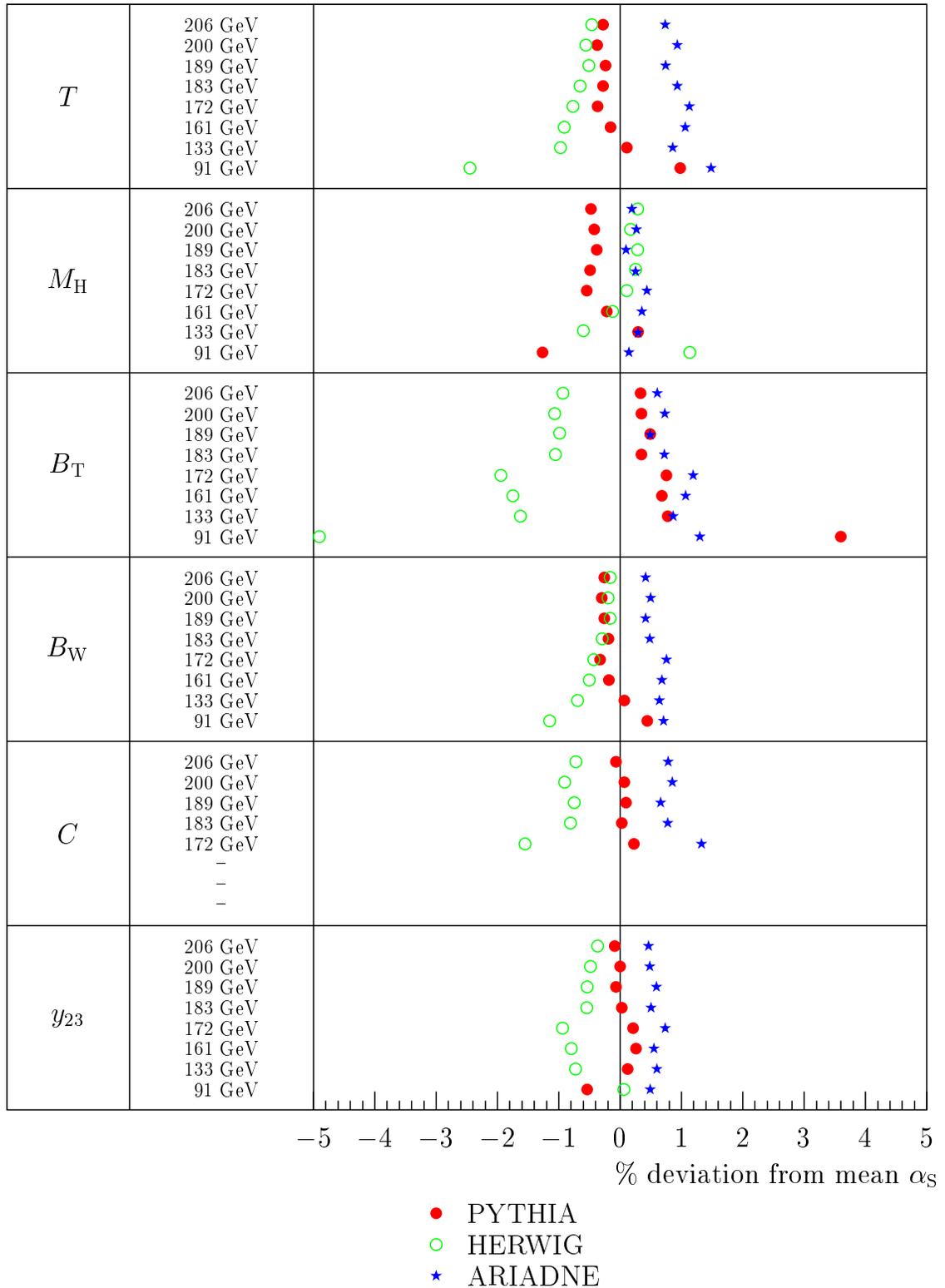}
\caption[Spread of \as\ fit results using PYTHIA, HERWIG and ARIADNE
hadronisation corrections (grouped by energy)]{The spread of \as\
fit results using PYTHIA, HERWIG and ARIADNE hadronisation
corrections, at centre-of-mass energies in the range 91--206~GeV. This
comparison uses the six observables \TR, \MH, \BT, \BW, \CP\
and~\ytwothree, measured by \OPAL. In each case, the three results are
plotted as percentage deviations from their mean value.}
\label{hadcorr_energy}
\end{center}
\end{figure}

In Figure~\ref{hadcorr_energy}, we compare differences between the
three generators at eight centre-of-mass energies. We observe a high
degree of correlation between the hadronisation uncertainties at
neighbouring energy scales. As the energy gap increases, the pattern
of the three generators is seen to change more significantly; however,
between most pairs of \LEP\ energy scales, we can safely assume that
the uncertainties are highly correlated. Once again, we apply the minimum
overlap assumption.

\subsubsection*{$\mathit{Correlations~between~experiments}$}

\begin{figure}[tbp!]
\begin{center}
\includegraphics[width=\textwidth]{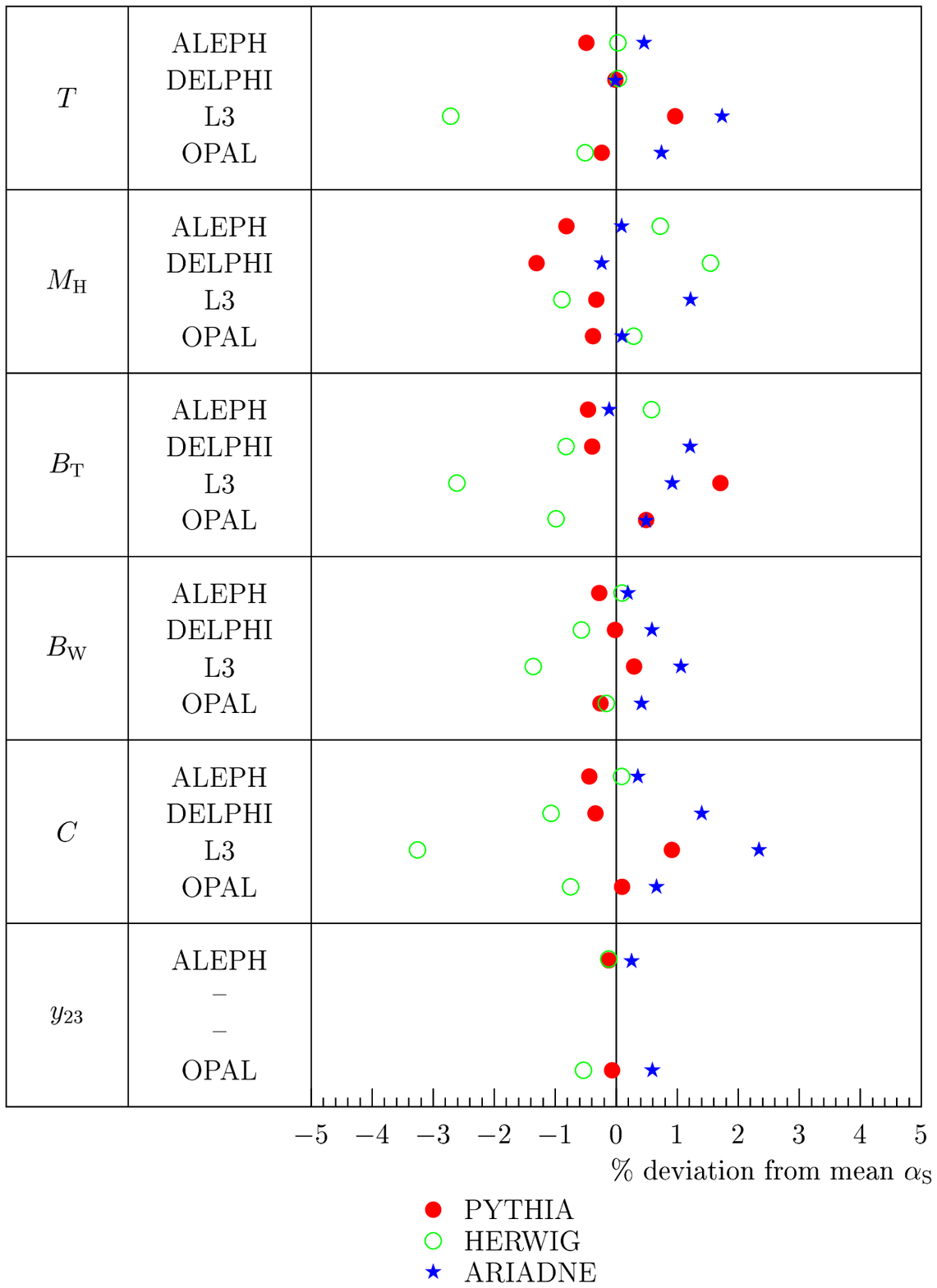}
\caption[Spread of \as\ fit results using PYTHIA, HERWIG and ARIADNE
hadronisation corrections (grouped by experiment)]{The spread of \as\
fit results using PYTHIA, HERWIG and ARIADNE hadronisation
corrections, from \ALEPH, \DELPHI, \Lthree\ and \OPAL. This comparison
uses the six observables \TR, \MH, \BT, \BW, \CP\ and~\ytwothree,
measured at 189~GeV centre-of-mass energy.  In each case, the three
results are plotted as percentage deviations from their mean value.}
\label{hadcorr_expt}
\end{center}
\end{figure}

Finally, in Figure~\ref{hadcorr_expt}, we make the same comparison
between the four \LEP\ experiments. In this case, we see very little
correlation between the patterns of the three generators in each
Collaboration. As we demonstrated in
Section~\ref{lep_consistency_hadr}, the independent tuning of Monte
Carlo event generators by the four Collaborations has led to
significant differences between the predicted hadronisation
corrections. This suggests that the correlation between hadronisation
uncertainties for different experiments can be neglected.

\subsubsection{Correlation of perturbative theory uncertainties}

\label{subsubsect_theocorr}

\subsubsection*{$\mathit{Correlations~between~experiments}$}

As we discussed in Section~\ref{lepconsistency_theory}, we have taken
care to ensure that all four Collaborations are using the same
perturbative theory predictions for the event shape
distributions. Thus the uncertainties due to uncomputed higher-order
terms should be strongly correlated between measurements from
different experiments. They will not be \emph{fully} correlated in
reality, because each Collaboration has chosen its own fit ranges;
however, the extent of the overlap in these ranges suggests that we
can treat the correlations as 100\% in practice.

\subsubsection*{$\mathit{Correlations~between~energy~scales}$}
Between different energy scales, we also expect the theory
uncertainties to be highly correlated. Setting the renormalisation
scale $\mu=Q$, our order-by-order theory prediction for the cumulative
distribution of the event shape $y$ is given by
\begin{eqnarray}
R(y,Q) & = & 1\;+\;\mathcal{A}(y)\,\asq\;
+\;\mathcal{B}(y)\,\alpha_\mathrm{S}^2(Q)\;
+\;\underbrace{\mathcal{C}(y)\,\alpha_\mathrm{S}^3(Q)\;+\;
\mathcal{D}(y)\,\alpha_\mathrm{S}^4(Q)\;+\;\ldots}_\mathrm{unknown~
higher~orders}
\nonumber \\
\end{eqnarray}
If the unknown coefficient functions
$\mathcal{C}(y),\mathcal{D}(y),\ldots$ in this expansion are of
similar magnitudes, which we intuitively expect, our theoretical error
will be dominated by the first unknown term:
\begin{equation}
\mathcal{C}(y)\,\alpha^3_\mathrm{S}(Q)
\;\gg\;\mathcal{D}(y)\,\alpha^4_\mathrm{S}(Q)
\;\gg\;\mathcal{E}(y)\,\alpha^5_\mathrm{S}(Q)\;\ldots\;\;\;.
\end{equation}
Since $\mathcal{C}(y)$ does not depend on the energy scale $Q$, our
uncertainty on this first term will be fully correlated between energy
scales.  To see the effect of the second unknown term, let us consider
the functions $\mathcal{C}(y)$ and $\mathcal{D}(y)$ to be independent
random variables. We can calculate the covariance between the
cumulative distributions $R(y,Q_i)$ at two different energy scales
$Q_1$ and $Q_2$, as follows:
\begin{eqnarray}
 & & \!\!\!\!\!\!\!\!\!\!\mathrm{Cov}\big[ R(y,Q_1), R(y,Q_2)\big] \nonumber \\
 & & \!\!\!\!\!\!\!\!\!\!\phantom{AAAA} = \;
 \mathrm{Cov}\left[\,\mathcal{C}(y)\,\alpha^3_\mathrm{S}(Q_1)
 +\mathcal{D}(y)\,\alpha^4_\mathrm{S}(Q_1)\;,\;
 \mathcal{C}(y)\,\alpha^3_\mathrm{S}(Q_2)
 +\mathcal{D}(y)\,\alpha^4_\mathrm{S}(Q_2) \,\right] \nonumber \\
 & & \!\!\!\!\!\!\!\!\!\!\phantom{AAAA} = \;
 \left(\,\alpha_\mathrm{S}(Q_1)\,\alpha_\mathrm{S}(Q_2)\,\right)^3\mathrm{Var}\left[\,\mathcal{C}(y)\,\right]\;+\;\left(\,\alpha_\mathrm{S}(Q_1)\,\alpha_\mathrm{S}(Q_2)\,\right)^4\mathrm{Var}\left[\,\mathcal{D}(y)\,\right] \nonumber \\
 & & \!\!\!\!\!\!\!\!\!\!\phantom{AAAA} \equiv \;
 \left(\,\alpha_1\alpha_2\,\right)^3\sigma^2_\mathcal{C}
 \;+\;\left(\,\alpha_1\alpha_2\,\right)^4\sigma^2_\mathcal{D} \;\;\;,
\end{eqnarray}
where we have defined
$\sigma^2_\mathcal{C}\equiv \mathrm{Var}\left[\,\mathcal{C}(y)\,\right]$,
$\sigma^2_\mathcal{D}\equiv \mathrm{Var}\left[\,\mathcal{D}(y)\,\right]$
and $\alpha_i \equiv \alpha_\mathrm{S}(Q_i)$.
The corresponding correlation coefficient $\rho$, between $R(y,Q_1)$ and
$R(y,Q_2)$, is then given by
\begin{eqnarray}
\rho & \!=\! & \frac{\left(\,\alpha_1\alpha_2\,\right)^3\sigma^2_\mathcal{C}
\;+\; \left(\,\alpha_1\alpha_2\,\right)^4\sigma^2_\mathcal{D}}
{\sqrt{\left(
\left(\,\alpha_1^3\,\sigma_\mathcal{C}\,\right)^2
\;+\;\left(\,\alpha_1^4\,\sigma_\mathcal{D}\,\right)^2
\right)\,\left(
\left(\,\alpha_2^3\,\sigma_\mathcal{C}\,\right)^2
\;+\;\left(\,\alpha_2^4\,\sigma_\mathcal{D}\,\right)^2
\right)
}} \nonumber \\
& \!=\! & \frac{\sigma^2_\mathcal{C}\;+\;\alpha_1\alpha_2\sigma^2_\mathcal{D}\,}
{\sqrt{
\big(\,\sigma^2_\mathcal{C}\;+\;\alpha_1^2\,\sigma^2_\mathcal{D}\,\big)
\big(\,\sigma^2_\mathcal{C}\;+\;\alpha_2^2\,\sigma^2_\mathcal{D}\,\big)
}} \nonumber \\
& \!\approx\! & 1\;-\;\frac{\sigma^2_\mathcal{D}}{2\,\sigma^2_\mathcal{C}}\,
\left(\alpha_1-\alpha_2\right)^2 \;\;\;.
\end{eqnarray}
Over the full range of \LEP\ energy scales, including those of
radiative events used by the \Lthree\ Collaboration, we have
\begin{displaymath}
\left|\alpha_1-\alpha_2\right|\;\leq\;\as(\mathrm{206~GeV})-\as(\mathrm{41~GeV})
\;\approx\;0.03 \;\;\;,
\end{displaymath}
so that
\begin{displaymath}
\rho \; \gtrsim \; 1 \; - \; 0.0005 \, \frac{\sigma^2_\mathcal{D}}{\sigma^2_\mathcal{C}}\;\;\;.
\end{displaymath}
Therefore, provided the uncertainties $\sigma_\mathcal{C}$ and
$\sigma_\mathcal{D}$ on the two coefficient functions are of similar
magnitude, the correlation coefficient $\rho$ will be very close to
unity. Since we have no reason to suspect \textit{a priori} that
$\mathcal{C}(y)$ and $\mathcal{D}(y)$ should be of vastly differing
magnitude, we thus assert that the uncertainties of our perturbative
theory predictions $R(y,Q)$ are fully correlated between energy
scales. Assuming that changes in our fit-ranges are negligible, this
also implies that the corresponding theoretical uncertainties in our
measurements of \as\ are fully correlated.

\subsubsection*{$\mathit{Correlations~between~observables}$}
We must also consider the correlation of theory uncertainties between
measurements using different observables. We know that the observables
themselves are statistically correlated, and that their perturbative
QCD calculations have much in common. Therefore it should be expected
that the uncalculated higher-order corrections to these calculations
will also be correlated. However, the extent of this
correlation is difficult to estimate.

The LEP QCD Working Group has attempted to study the correlation of
theory uncertainties between observables in the following way. We
define an ``ensemble of theories,'' by varying the parameters $x_\mu$,
\xl\ and $p$.\footnote{These are the same parameters used in the
``uncertainty~band'' method to estimate our theoretical uncertainties,
as described in Section~\ref{evsh_prediction_errors}.} Our ensemble
comprises an $N_1\times N_2\times N_3$ grid, with each point
representing a particular set of values for these three parameters. At
every point on the grid, we fit the distributions of the six event
shape distributions, using the same (arbitrarily chosen) sample of
events in each case. We can then measure the correlation coefficient
between fits to each pair of observables. This method has two
drawbacks, however:
\begin{itemize}
\item The results will depend on the size, density and metric of the
grid, along its three axes. For example, a grid spaced evenly with
respect to $x_\mu$ may give different results from a grid spaced
evenly with respect to $\ln x_\mu$.
\item Without justification, we are varying the parameters $x_\mu$,
\xl\ and $p$ \emph{simultaneously} for all six event shape
observables. In reality, the true higher-order corrections to these
six distributions may be best represented by \emph{different} sets of
values for these parameters.
\end{itemize}
Nonetheless, all the results \cite{stenzel_salam_private} obtained by
this method indicated high degrees of correlation, of the order $\rho
\gtrsim 0.9$.

Based on the conclusions of this section, it is tempting to define a
100\% correlation between all pairs of theoretical uncertainties in
our covariance matrix. However, this would imply that every
theoretical error\footnote{The word `error' here refers strictly to
the unknown deviation between the true and measured values of~\as; the
`uncertainty' is the RMS of the error distribution.} can be expressed
in terms of a single unknown random variable~$\delta$, with zero
expectation and unit variance:
\begin{equation}
\left[ \; \alpha_i^\mathrm{measured} \,-\, \alpha_i^\mathrm{true} \; \right]^\mathrm{theo.}
\;=\; \sigma^\mathrm{theo.}_i\,\delta \;\;\;.
\end{equation}
If the measurements $\alpha_i$ are combined in a weighted mean
$\as=\sum w_i \alpha_i$~, such that the weights $w_i$ satisfy
\begin{equation}
\sum_i w_i \, \sigma_i^\mathrm{theo.} \;=\; 0 \;\;,
\end{equation}
then the theoretical uncertainty of our combined \as\ will be
zero. This phenomenon has been observed numerically in our weighted
means, and is clearly unrealistic.

We therefore modify our ansatz, for the moment, by defining the
correlation coefficient $\rho$ between theoretical uncertainties for
different observables to be 0.9 instead of 1. Later we will see that
even this assumption is unsatisfactory; however, it does avoid the
immediate difficulty described above, and the resulting \as\ is
reasonably stable with respect to small changes in $\rho$.

\subsubsection{Summary}

The anticipated off-diagonal elements of the covariance matrix, as
discussed in
Sections~\ref{subsubsect_statcorr}--\ref{subsubsect_theocorr}, are
summarised in Table~\ref{table:naivecorr}.

\begin{table}
\begin{center}
\begin{tabular}{|l|ccc|}
\cline{2-4}
\multicolumn{1}{c|}{ } & Different & Different & Different \bigstrut \\[-0.15cm]
\multicolumn{1}{c|}{ } & experiments & energies & observables \\
\hline
\bigstrut
$V_{ij}^\mathrm{stat.}$ & 0 & 0 & $\rho^\mathrm{stat.}_{ij}\sigma_i\,\sigma_j$ \\
\bigstrut
$V_{ij}^\mathrm{exp.} $ & 0 & $\mathrm{min}(\sigma^2_i,\sigma^2_j)$ & $\mathrm{min}(\sigma^2_i,\sigma^2_j)$ \\
\bigstrut
$V_{ij}^\mathrm{hadr.}$ & 0 & $\mathrm{min}(\sigma^2_i,\sigma^2_j)$ & $\mathrm{min}(\sigma^2_i,\sigma^2_j)$ \\
\bigstrut
$V_{ij}^\mathrm{theo.}$ & $\sigma_i\,\sigma_j$ & $\sigma_i\,\sigma_j$ & $0.90\,\sigma_i\,\sigma_j$
\bigstrut[b] \\ \hline \end{tabular}
\end{center} \caption{The approximate values expected for off-diagonal
elements of the covariance matrix $V$}
\label{table:naivecorr} \end{table}

Where more than one of the three column-headings applies in this
table, the covariance is taken to be the smallest of the entries in
the corresponding columns. For example in the case
\[
i=\left[\,\mathrm{OPAL,~189~GeV,~}\TR\,\right] \;\;,\;\;\;
j=\left[\,\mathrm{OPAL,~133~GeV,~}\BW\,\right] \;\;\;,
\]
we take the smaller of the two covariances listed in the ``different
energies'' and ``different observables'' columns:
\[
V_{ij}^\mathrm{stat.} = 0 \;,\;\;
V_{ij}^\mathrm{exp.} = \mathrm{min}(\sigma^2_i,\sigma^2_j) \;,\;\;
V_{ij}^\mathrm{hadr.} = \mathrm{min}(\sigma^2_i,\sigma^2_j) \;,\;\;
V_{ij}^\mathrm{theo.} = 0.90\,\sigma_i\,\sigma_j \;\;\;.
\]
The uncertainties $\sigma_i$ here refer to the corresponding
contribution $\sigma_i^\mathrm{exp.}$, $\sigma_i^\mathrm{hadr.}$ or
$\sigma_i^\mathrm{theo.}$.

\section[A na\"{\i}ve measurement of \asmz]
{A na\"{\i}ve measurement of \boldmath{\asmz}}
\label{naivecomb}

Using the input measurements tabulated in
Appendix~\ref{lepinputappendix}, and the covariance matrix defined in
the previous section, we can apply the ``weighted~mean'' method
described in Section~\ref{weightssection} to obtain the following
measurement: {\small
\begin{eqnarray*}
\asmz & = & 0.1162
\;\pm\;0.0008~\mathrm{(stat.)}
\;\pm\;0.0008~\mathrm{(exp.)}
\;\pm\;0.0003~\mathrm{(hadr.)}
\;\pm\;0.0017~\mathrm{(theo.)} \\[-0.3cm]
& = & 0.1162
\;\pm\;0.0008~\mathrm{(stat.)}
\;\pm\;0.0019~\mathrm{(syst.)} \\[-0.3cm]
& = & 0.1162
\;\pm\;0.0021~\mathrm{(total)}
\end{eqnarray*}
}with $\chi^2/\mathrm{d.o.f.}=432/193$. As a cross-check, we also
performed a numerical minimisation of $\chi^2$ as described in
Section~\ref{simplerunning}; the result was identical, except that no
breakdown of the total uncertainty was possible.

Our measurement suffers three obvious problems:
\begin{itemize}
\item The $\chi^2$ value indicates unreasonably poor agreement between
the weighted average and its contributing measurements.
\item The statistical uncertainty of the combined value is larger than
  that of several contributing measurements.
\item The current world average quoted by the Particle Data Group
\cite{PDbook} is $\asmz=0.1172\pm0.0020$. Although our own central
result is in good numerical agreement with this
value,\footnote{$\Delta\asmz=-0.0010\pm0.0029$, assuming no
correlation between the respective uncertainties.} we would expect our
result to have a much larger total uncertainty than that of the PDG;
the PDG average includes \as\ measurements from many different
processes, including $\tau$~decays and deep inelastic scattering,
which are more precise than those from \epem\ event shapes.
\end{itemize}
These problems stem from our poor knowledge of certain parts of the
covariance matrix. Various attempts have been made in the past by the
LEP QCD Working Group, to reduce our sensitivity to fluctuations in
the covariance matrix. These are described in a previous report by the
group~\cite{lepfest}, and have led to results with more reasonable
uncertainties and $\chi^2$ values.

\begin{figure}
  \begin{center}
     \includegraphics[width=\textwidth]{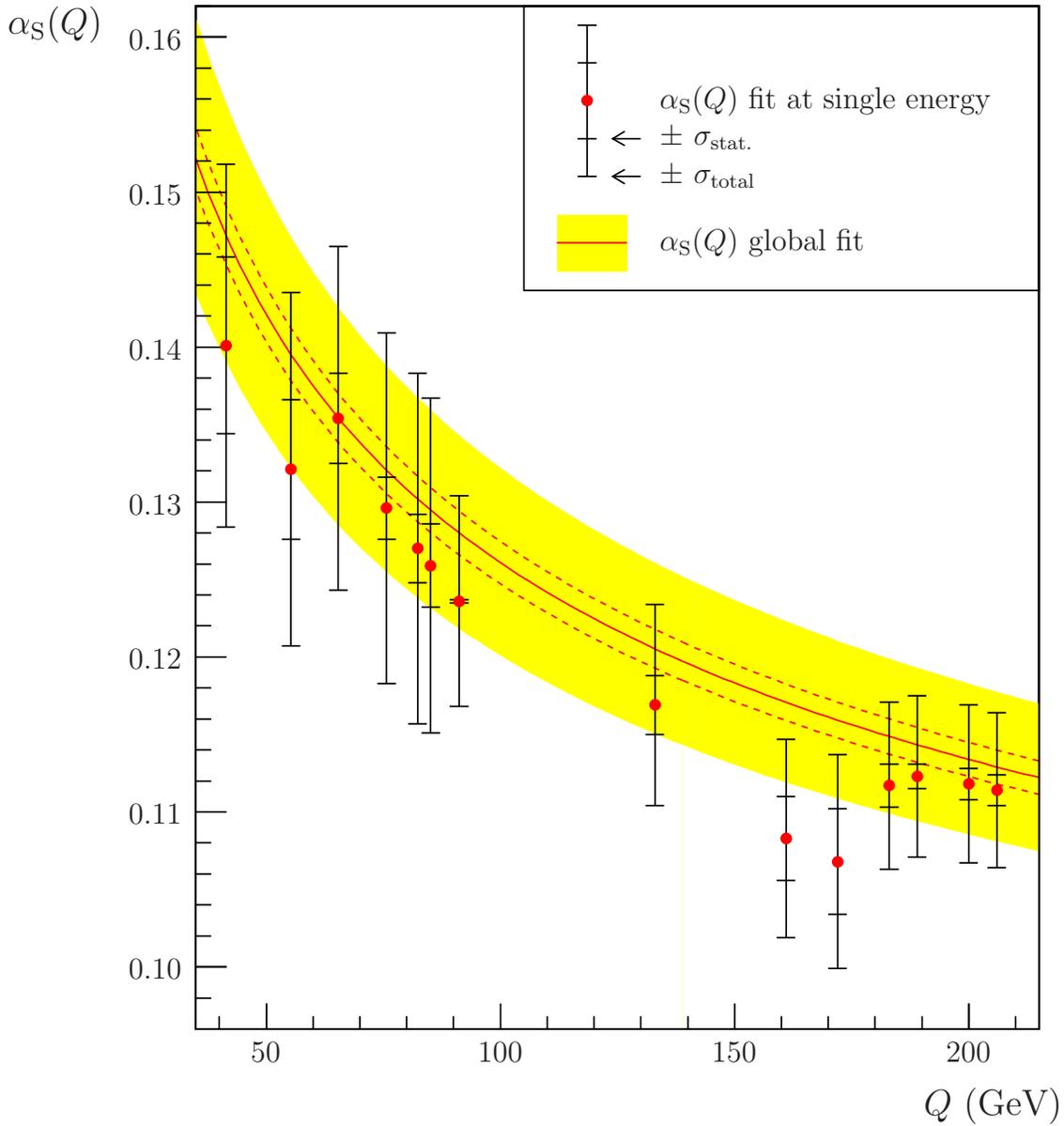}
  \end{center}
  \caption{A pathological combination of 35 \asq\ measurements derived
  from the total jet broadening, \BT, using the `na\"{\i}ve'
  covariance matrix described in Section~\ref{fullcovmatrix}. Each
  point represents a combination of all available measurements at a
  single energy, while the curve represents a simultaneous fit at all
  energies.}
  \label{badbtfit}
\end{figure}

However, further investigation revealed severe internal
inconsistencies in these results. Figure~\ref{badbtfit} shows a fit to
all \LEP\ \as\ measurements derived from the total jet broadening,
\BT, using the covariance matrix defined in
Section~\ref{fullcovmatrix}. The global fit in this case predicts
\asq\ to be higher than \emph{all} of the fits at individual
centre-of-mass energies. Such pathological fits have been confirmed
independently by another member of the LEP QCD Working
Group~\cite{wicke_private}.  In the sections that follow, we discuss
the symptoms, cause and remedy for these unacceptable fits.

\section{Dependence on the correlation coefficients}
\label{depcorrsect}

\begin{figure}[p]
  \begin{leftfullpage}
    \vspace{1.0cm}
    \begin{center}
      \resizebox{!}{0.85\textheight}{
      \includegraphics[width=\textwidth]{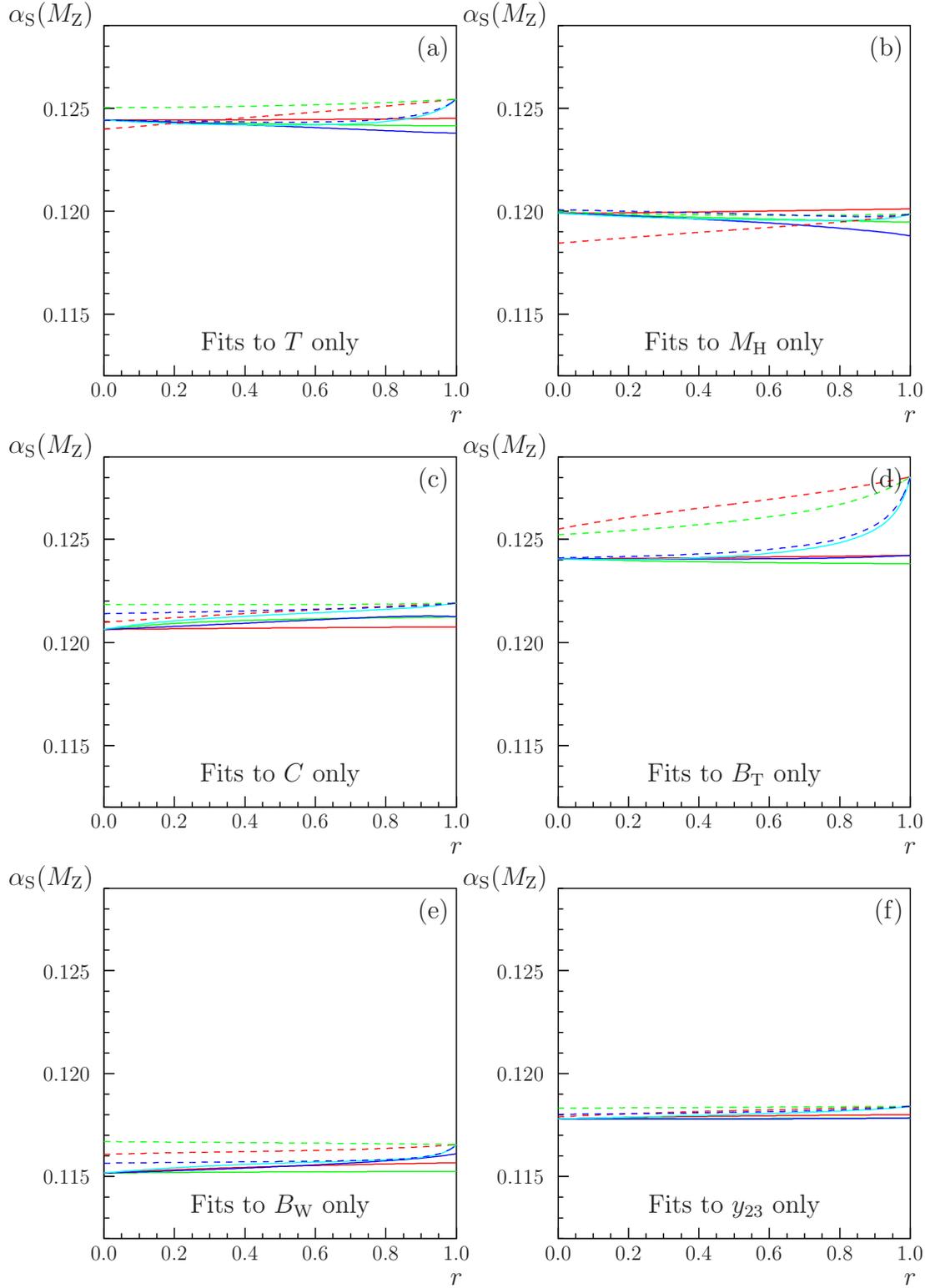}}
      \caption{Variation of the fitted \asmz\ value with respect to
      certain subsets of terms in the covariance matrix $V$.  Each
      colour and line-style in the legend (shown on the next page)
      represents a different variation of the covariance matrix with
      respect to the parameter $r$; only the off-diagonal terms
      ($V_{ij}:\,i\neq j$) are varied.  For clarity, the covariance
      terms which depend on $r$ are highlighted in red. The same
      legend applies to all seven plots.}
      \label{rhovary}
    \end{center}
  \end{leftfullpage}
\end{figure}

When defining the correlations between uncertainties in different
measurements, many imprecise and subjective assumptions were made.  We
should therefore ensure that our combined \asmz\ measurement is not
excessively sensitive to any reasonable variation of these
estimates. In Figure~\ref{rhovary}, we show the dependence of our
fitted \asmz\ values on various subsets of the correlation
coefficients between systematic uncertainties. For the purposes of
this plot, we define five independent groups of coefficients:
\begin{enumerate}
\item Correlations of statistical uncertainties between measurements
  from different observables, using the same experiments at the same
  centre-of-mass energy: these correlations are not varied, as they
  are known to about 10\% precision, as described in
  Section~\ref{subsubsect_statcorr}.
\item Correlations of experimental systematic uncertainties, between
pairs of measurements from the same experiment
\item Correlations of hadronisation uncertainties, between pairs of
measurements from the same experiment
\item Correlations of theory uncertainties, between measurements using
  the same observable
\item Correlations of theory uncertainties, between measurements using
  different observables
\end{enumerate}
To study the effects of these correlations on our \as\ fits, we vary them
in three different ways:
\begin{itemize}
\item Switch the correlations `on', one at a time, leaving the other
types of correlation switched `off'. This mode of variation is
indicated in Figure~\ref{rhovary} by the solid red, green, blue and
magenta curves.
\item Switch the correlations `off', one at a time, leaving the
  other types of correlation fixed at their default values. This mode
  of variation is indicated by the dashed red, green, blue and magenta
  curves.
\item Switch all correlations `on' simultaneously. This mode of
  variation is indicated by the solid cyan curve.
\end{itemize}
These variations have been applied to fits to each individual event
shape observable, in Figures~\ref{rhovary}(\emph{a})--(\emph{f}), and
also for global fits to all observables, in
Figure~\ref{rhovary}(\emph{g}). In all cases, the variable $r$ is used
as a ``switching~parameter'' for the correlations which are to be
varied: $r=0$ corresponds to zero correlation, while $r=1$ corresponds
to the correlation assumed in Section~\ref{corrsubsect}.

\begin{figure}[p]
  \begin{fullpage}
    \vspace{-1.6cm}
    \begin{center}
      \scalebox{0.89}{
      \parbox{\linewidth}{\centering
      \includegraphics[width=0.7\linewidth]{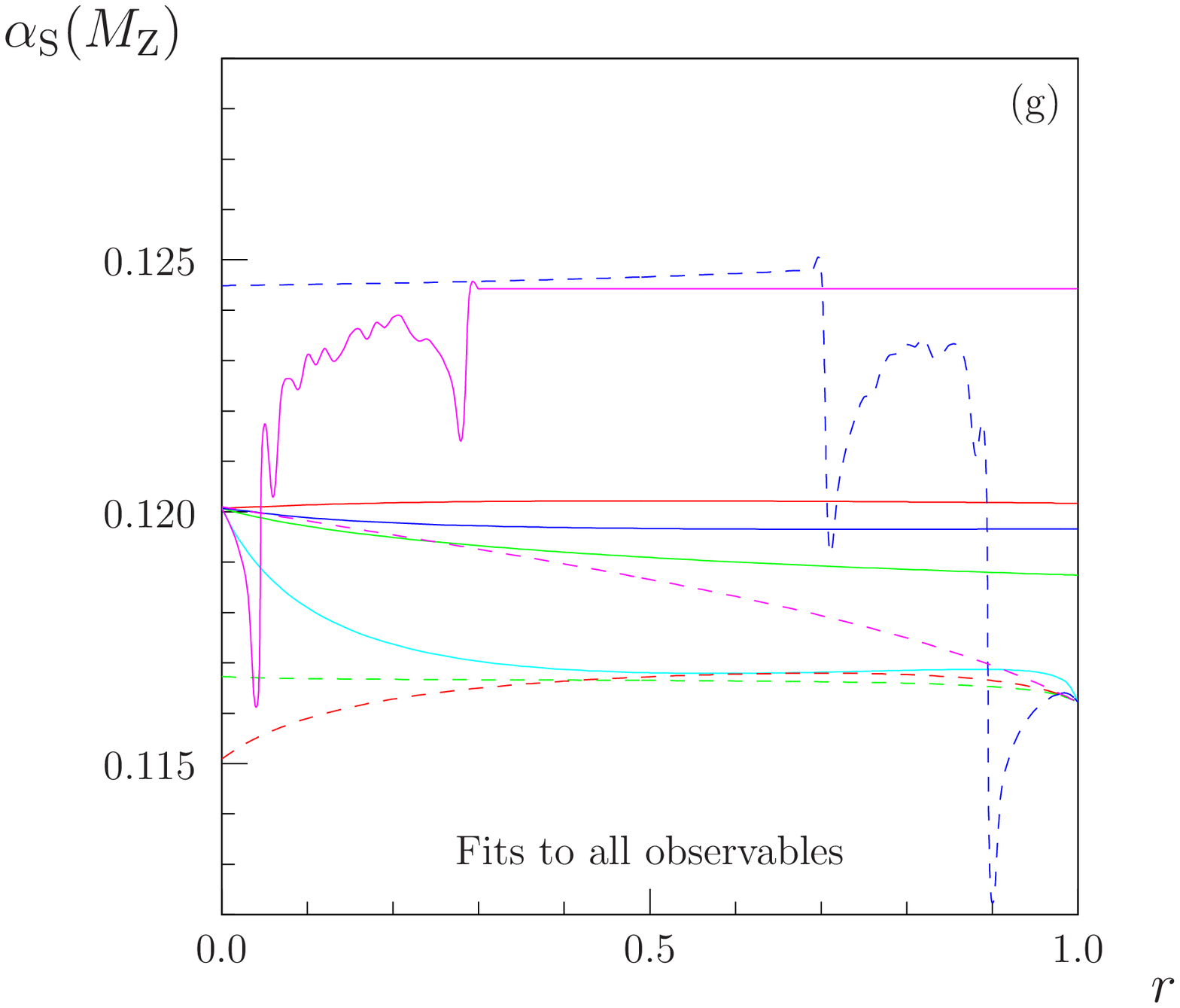}
      \vspace{1.0cm}\\
      \includegraphics[width=1.0\linewidth]{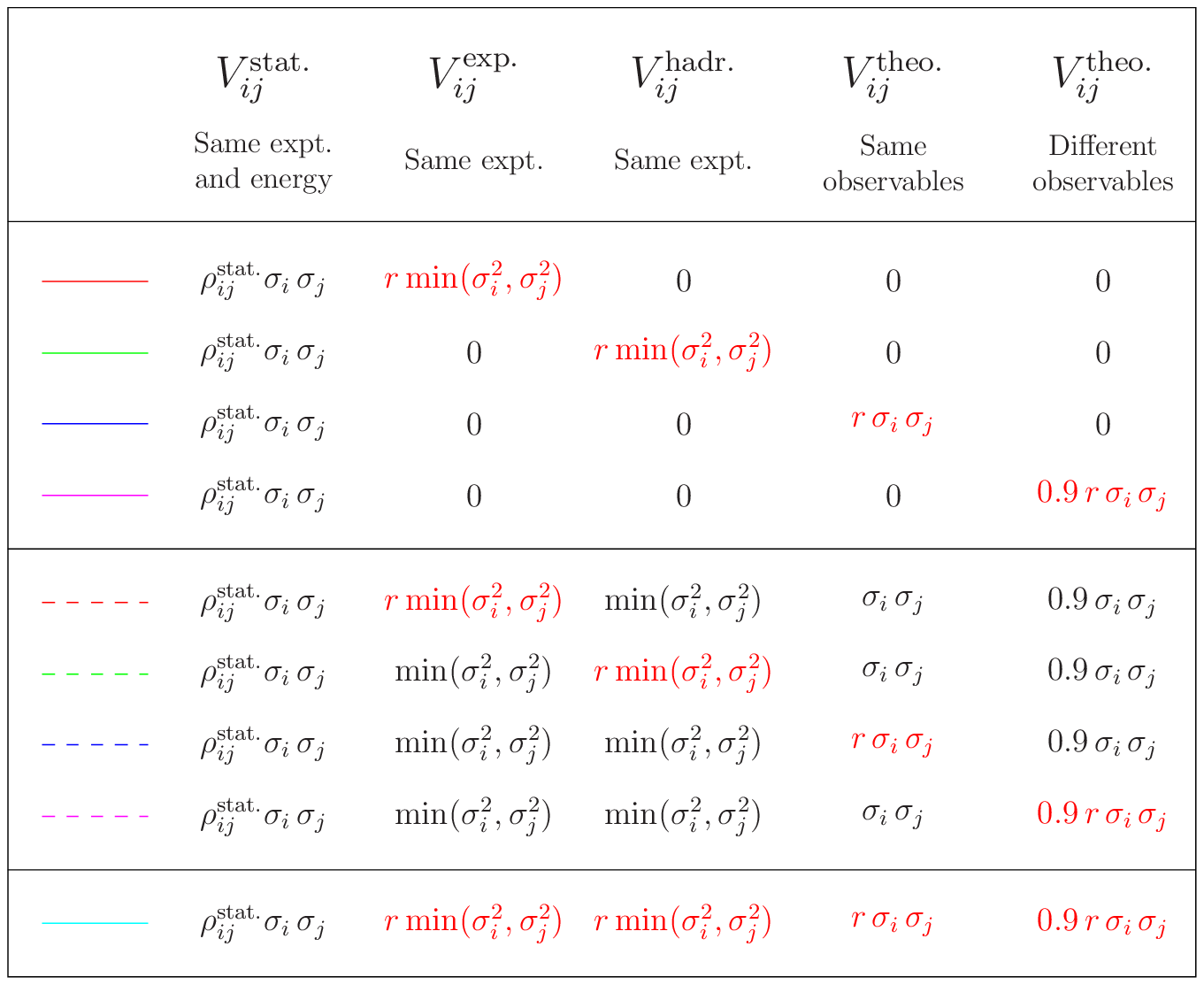}
      }}
      \vspace{1.25cm} \\
      Figure~\ref{rhovary} (contd.)
    \end{center}
  \end{fullpage}
\end{figure}

With the exception of \BT, shown in Figure~\ref{rhovary}(\emph{d}), we
find that the fits using single observables are relatively insensitive
to variation of the correlation coefficients. Generally the range of
variation in the combined \asmz\ value is much smaller than the
systematic uncertainties of the input measurements. When all
observables are used simultaneously in a global fit, however, as shown
in Figure~\ref{rhovary}(\emph{g}), we find a more pronounced
dependence on the covariance matrix. Several pieces of evidence in
Figure~\ref{rhovary} suggest that this dependence can be attributed to
the correlation of theory uncertainties between measurements using
different observables (`type~5'):
\begin{itemize}
\item Correlation type 5 is only present in the global fit, where more
  than one observable is involved. This is also the only case in which
  a substantial variation is seen in the \asmz\ fit.
\item When correlation type 5 remains switched off, the \asmz\ fit
  shows very little dependence on the other correlation types; this is
  indicated by the solid red, green and blue curves in
  Figure~\ref{rhovary}(\emph{g}).
\item When correlation type 5 is switched on \emph{alone}, we find a
  large amplitude of variation in the \asmz\ fit, with some apparently
  chaotic behaviour. For $r\gtrsim 0.3$, the weighted mean contains
  only measurements using thrust ($T$), and the weights $w_i$ for all
  other measurements vanish; therefore no further dependence on $r$ is
  observed. This is indicated by the solid magenta curve in
  Figure~\ref{rhovary}(\emph{g}).
\item A large amplitude of variation is also seen when correlations of
  theory uncertainties between measurements using the \emph{same}
  observable (`type~4') are switched off, while leaving type-5
  correlations switched on. Switching off the type-4 correlations in
  this case allows the type-5 correlations to dominate the
  off-diagonal part of the covariance matrix. This is indicated by the
  dashed blue curve in Figure~\ref{rhovary}(\emph{g}).
\end{itemize}
We must therefore seek to explain why the fits are so sensitive to this
type of correlation, and attempt to stabilise them.

\section{The problem of negative weights}

In our formulation of the weighted mean method, in
Section~\ref{weightssection}, we did not place any restriction on
the signs of the weights $w_i$; the only constraint was that their sum
should be unity. However, one intuitively expects that a weighted mean
of $N$ measurements should move in the ``right direction'' if one of
the contributing measurements changes:\footnote{As in
Section~\ref{weightssection}, each input measurement has been
converted to the $\mathrm{Z}^0$ mass scale, and denoted by $y'_i
\equiv \left[\asmz\right]_i$.}
\begin{equation}
\frac{\partial\hat\lambda}{\partial y'_i}\;\geq \;0\;\;\;\forall i\;\;\;.
\end{equation}
Recalling that $\hat\lambda$, our measurement of \asmz, is given by
$\sum_i w_i y'_i$, this implies
\begin{equation}
w_i\;\geq \;0\;\;\;\forall i\;\;\;.
\end{equation}
Requiring the weights $w_i$ to be non-negative also implies that the
weighted mean will be \emph{bounded} by the range of the input
measurements, as one expects.

The weights contributing to the combination in Section~\ref{naivecomb}
do not satisfy the above condition. The 194 weights vary in this case
between $-30.5\%$ and $+53.6\%$,\footnote{The minimum and maximum
weights are those for the \OPAL\ \BW\ and \ALEPH\ $y_{23}$ measurements
respectively, both at 91~GeV.} with a total positive weight of $+400.8\%$
being balanced by a total negative weight of $-300.8\%$. This explains
why the global fit presented in Figure~\ref{badbtfit} appears
inconsistent with the fits at individual energy points; it also
explains the large statistical uncertainty of the combined value.

We therefore wish to reformulate our combination algorithm, such that
the weights $w_i$ are constrained to be positive. One approach to this
problem would be to vary the 194 weights, subject to the constraints
$0 \leq w_i \leq 1$, such that $\chi^2$ is minimised.  This would lead
to two difficulties:
\begin{itemize}
\item The practical implementation of this constrained minimisation
would not be straightforward. We would need to minimise over the
volume of a 194-dimensional hypercube, which has an enormous number of
`edges' and `surfaces'.
\item More importantly, our $\chi^2$ would still be defined by
inverting the same poorly-known covariance matrix which led to our unstable
`na\"{\i}ve' combination.  A small change in our correlation
assumptions could therefore lead to a dramatic change in our combined
result, even if the weights are constrained to be positive.
\end{itemize}
Since the weights are functions of the covariance matrix $V$, as
defined by Equation~(\ref{weightformula}), we will instead attempt to
eliminate the negative weights by modifying~$V$. As we shall see
later, this does inevitably lead to some compromises; however it will
not prevent us from taking full account of the uncertainties in
our final results.

\section{Criteria for the avoidance of negative weights}

As we derived in Section~\ref{weightssection}, the weights $w_i$
can be expressed as a function of the covariance matrix $V$:
\footnote{Here $V$ is the covariance matrix relating measurements of
\as\ at the $\mathrm{Z}^0$ mass scale. We have dropped the notation
$V'$, used in Section~\ref{weightssection}.}
\begin{equation}
\label{weightformula1}
w_i\;=\;\frac{\sum_{j}\;\big(V^{-1}\big)_{ij}}{\sum_{j,k}\;\big(V^{-1}\big)_{jk}}\;\;\;.
\end{equation}
We wish to find some conditions on the matrix $V$, to ensure that none
of the weights are negative.  In the general case, it is difficult to
define any necessary or sufficient conditions under which this
requirement is satisfied. However, there are some special cases in
which the covariance matrix can be inverted analytically to obtain
simple expressions for the weights $w_i$.

\subsection{The trivial case: uncorrelated measurements}

When all correlations are removed from the covariance matrix,
including those between statistical uncertainties, $V$ and its inverse
are both diagonal:
\begin{eqnarray}
V_{ij}&\!\!=\!\!&\sigma_i^2\,\delta_{ij} \rule[-0.5cm]{0pt}{0pt}\\
\big(V^{-1}\big)_{ij}&\!\!=\!\!&\frac{1}{\sigma_i^2}\,\delta_{ij}\;\;\;.
\end{eqnarray}
The weights in this case are guaranteed be positive, and are
inversely proportional to the squares of the total uncertainties:
\begin{equation}
w_i=\frac{1/\sigma_i^2}{\sum_j 1/\sigma_j^2}\;\;\;.
\end{equation}

\subsection{The weighted mean of two correlated measurements}

Given two measurements $y_1$ and $y_2$, with variances $\sigma_1^2$ and 
$\sigma_2^2$ respectively, the covariance matrix and its 
inverse are given by
\begin{eqnarray}
V&\!\!=\!\!&\left(\begin{array}{cc}
\sigma_1^2 & \mathrm{Cov}(y_1,y_2) \\
\mathrm{Cov}(y_1,y_2) & \sigma_2^2
\end{array}\right) \rule[-1cm]{0pt}{0pt}\\
V^{-1}&\!\!=\!\!&
\frac{1}{\sigma_1^2\sigma_2^2\;-\;
\left[\mathrm{Cov}(y_1,y_2)\right]^2}
\left(\begin{array}{cc}
\sigma_2^2 & -\mathrm{Cov}(y_1,y_2) \\
-\mathrm{Cov}(y_1,y_2) & \sigma_1^2   
\end{array}\right)\;\;\;,
\end{eqnarray}
and hence the ratio of weights contributing to the least-squares fit is
\begin{equation}
\frac{w_1}{w_2}=\frac{\sigma_2^2-\mathrm{Cov}(y_1,y_2)}
{\sigma_1^2-\mathrm{Cov}(y_1,y_2)}\;\;\;.
\end{equation}
Both weights will therefore be positive, if and only if the covariance of 
$y_1$ and $y_2$ is less than both of the two variances:
\footnote{It is impossible for both weights to be negative, as their
sum is constrained to be unity.}
\begin{equation}
\label{2x2cond}
\mathrm{Cov}(y_1,y_2)\;<\;\sigma_\mathrm{min}^2
\end{equation}
When Condition (\ref{2x2cond}) is violated, one of the two weights
will become negative, and the weighted mean $w_1y_1+w_2y_2$ may not
lie between the measurements $y_1$ and $y_2$. This does not
\emph{necessarily} imply that the weighted mean is unreliable; it
simply means that the uncertainties are dominated by a correlated
systematic contribution. For example, if we have two measurements with
fully correlated uncertainties, \mbox{$y_1\pm \sigma_1$} and
\mbox{$y_2\pm \sigma_2$}, we could write
\begin{eqnarray} 
y_1 & \!=\! & y_\mathrm{true} + \sigma_1 \delta \nonumber \\
y_2 & \!=\! & y_\mathrm{true} + \sigma_2 \delta \;\;\;,
\end{eqnarray}
for some single unknown quantity $\delta$ with zero expectation and unit
variance. As we described in Section~\ref{subsubsect_theocorr}, we
could then deduce
\begin{equation}
y_\mathrm{true}\;=\;\frac{\sigma_1 y_2 - \sigma_2 y_1 }{\sigma_1 -
  \sigma_2} \;\;\;,
\end{equation}
with \emph{no uncertainty}. A negative weight has been used in this
example to cancel out completely the systematic uncertainty. There is
nothing improper about this, provided the ratio of uncertainties
$\sigma_1/\sigma_2$, and their correlation coefficient ($\rho=1$ in
this example), are known with sufficiently high precision. A
legitimate physical example of this is discussed in Ref.~\cite{cowanbook}.
In the case of our \asmz\ combination, however, we do not have
sufficient knowledge of the covariance matrix to ``cancel~out''
systematic uncertainties in this way. 

For cases with more than two measurements, it is difficult to find
criteria analogous to Condition~(\ref{2x2cond}) to prevent negative
weights. We can, of course, still write out the inverse covariance
matrix explicitly, and determine relationships between the variances
and covariances to ensure positive weights. However, even in the
$3\times 3$ case, it is difficult to interpret the constraints in
meaningful terms. We will therefore consider another special case in
which the covariance matrix may be inverted analytically.

\subsection
[A single fully-correlated source of systematic 
error, and uncorrelated statistical errors]
{\setstretch{1}A single fully-correlated source of systematic 
error, and uncorrelated statistical errors}

Suppose each measurement $y_i$ is subject to a statistical uncertainty
$\sigma_i$ and a \emph{single} systematic error $s_i$. Suppose further
that the systematic errors are fully correlated between all
measurements. The covariance matrix is therefore
\begin{eqnarray}
V_{ij}&\!\!=\!\!&\sigma_i\sigma_j\,\delta_{ij}\,+\,s_is_j \nonumber\\
&\!\!=\!\!&\sigma_i\sigma_j\;(\delta_{ij}\,+\,\beta_i\beta_j)
\hspace{1.0cm}\mathrm{[no~summation]}\;\;,
\end{eqnarray}
where $\beta_i\equiv s_i/\sigma_i$. For example in the case of two 
measurements, we have
\begin{equation}
V\;=\;\left(\begin{array}{cc}\sigma_1^2+s_1^2 & s_1s_2 \\
s_1s_2 & \sigma_2^2+s_2^2 \end{array}\right)\;\;\;.
\end{equation}
As shown in Ref.~\cite{alekhin}, this covariance matrix may 
be inverted analytically:\footnote{It is straightforward to verify
  that this expression satisfies $V^{-1}V=I$.}
\begin{equation}
\big(V^{-1}\big)_{ij}\;=\;\frac{1}{\sigma_i\sigma_j}\left(\delta_{ij}\;-\;
\frac{\beta_i\beta_j}{1+\sum_k \beta_k^2}\right)\;\;\;.
\end{equation}
Substituting this into Equation~(\ref{weightformula1}), we obtain the
weights.
\begin{eqnarray}
w_i & \!=\! & \frac{1}{\sum_{j,k}{\big(V^{-1}\big)_{jk}}}\;\left[
\sum_j{\frac{1}{\sigma_i\sigma_j}\left(\delta_{ij}\;-\;
\frac{\left(s_i/\sigma_i\right)\left(s_j/\sigma_j\right)}
     {1+\sum_k\left(s_k^2/\sigma_k^2\right)}\right)}\right] \rule[-1cm]{0pt}{0pt}
\nonumber \\
&\!=\!&\frac{1}{\sigma_i^2\sum_{j,k}{\big(V^{-1}\big)_{jk}}}\;\left(
1\;-\;\frac{s_i\sum_j (s_j/\sigma_j^2)}{1+\sum_j 
(s_j^2/\sigma_j^2)}\right)
\end{eqnarray}
To avoid negative weights, we therefore require
\begin{equation}
1\;-\;\frac{s_i\sum_j (s_j/\sigma_j^2)}{1+\sum_j 
(s_j^2/\sigma_j^2)} \;>\;0 \;\;\;\;\forall i \;\;\;,
\end{equation}
so
\begin{equation}
\sum_j \frac{s_j(s_i-s_j)}{\sigma_j^2}\;<\;1 \;\;\;\;\forall i\;\;\;.
\end{equation}
Choosing $s_i$ to be the largest of the systematic uncertainties
$s_\mathrm{max}$, we arrive at the following necessary and sufficient
condition for all weights~$w_i$ to be positive:
\begin{equation}
\label{gencond}
\fbox{
$\displaystyle{
\rule{0pt}{0.7cm}
\sum_j \frac{s_j(s_\mathrm{max}-s_j)}{\sigma_j^2}\;<\;1}$
}\;\;\;.
\end{equation}
In the case of only two measurements, this inequality simplifies to
\begin{equation}
\label{2x2case}
s_1s_2<\sigma_1^2+s_1^2\;\;\;,\;\;\;\;\;\mathrm{where}\;\;\;s_1<s_2\;\;\;,
\end{equation}
in agreement with our earlier result,
Condition~(\ref{2x2cond}).\footnote{The smallest total uncertainty
$\sigma_\mathrm{min}$ in Condition~(\ref{2x2cond}) corresponds to the
expression $\sqrt{\sigma_1^2+s_1^2}$ in Condition~(\ref{2x2case}).}
One obvious situation in which Condition~(\ref{gencond}) will be
satisfied is the case in which all the systematic uncertainties are
equal. In this case, the left-hand side of the inequality is zero, and
the weights are identical to those in the uncorrelated case:
\begin{equation}
w_i=\frac{1/\sigma_i^2}{\sum_j 1/\sigma_j^2} \;\;\;.
\end{equation}

\subsection{Examples}

Table \ref{gencondexamples} illustrates four hypothetical weighted
means, with dominant but variable systematic uncertainties. In each case,
the middle column shows the terms contributing to the sum in Equation~(\ref{gencond}).

\begin{table}[p]
\begin{center}
\scalebox{0.9}{
\begin{minipage}{\textwidth}
\begin{center}
\small
{\normalsize Example (\emph{a})}
\vspace{0.3cm}

\begin{tabular}{| c c | r | r |} 
\hline
\multicolumn{2}{|c|}{Uncertainties} &
\multicolumn{1}{|c|}{\multirow{2}[2]{*}[-0.07cm]{\scalebox{1.15}{$\textstyle \frac{s_i(s_\mathrm{max}-s_i)}{\sigma_i^2}$}}} &
\multicolumn{1}{|c|}{\multirow{2}[2]{*}[-0.07cm]{$\begin{array}{c}\mathrm{Weight}\vspace{-0.1cm}\\w_i\end{array}$}}\\
$\sigma_i$~(stat.) & $s_i$~(syst.) & & \bigstrut \\
\hline
\hline
2.0 & \textbf{7.9} & 0.20 & $0.202$ \bigstrut[t]\\
2.0 & \textbf{8.0} & 0.00 & $0.012$ \\
3.0 & 7.7 & 0.26 & $0.258$ \\
3.0 & 7.9 & 0.09 & $0.090$ \\
3.5 & 7.3 & 0.42 & $0.438$ 
\bigstrut[b] \\
\hline 
\multicolumn{2}{|c|}{} & 
Total: 0.96 & Total: 1.000 
\bigstrut[tb] \\
\hline
\end{tabular}

\vspace{0.7cm}
{\normalsize Example (\emph{b})}
\vspace{0.3cm}

\begin{tabular}{| c c | r | r |} 
\hline
\multicolumn{2}{|c|}{Uncertainties} &
\multicolumn{1}{|c|}{\multirow{2}[2]{*}[-0.07cm]{\scalebox{1.15}{$\textstyle \frac{s_i(s_\mathrm{max}-s_i)}{\sigma_i^2}$}}} &
\multicolumn{1}{|c|}{\multirow{2}[2]{*}[-0.07cm]{$\begin{array}{c}\mathrm{Weight}\vspace{-0.1cm}\\w_i\end{array}$}}\\
$\sigma_i$~(stat.) & $s_i$~(syst.) & & \bigstrut \\
\hline
\hline
2.0 & \textbf{7.8} & 0.39 & $0.333$ \bigstrut[t]\\
2.0 & \textbf{8.0} & 0.00 & $-0.046$ \\
3.0 & 7.7 & 0.26 & $0.232$ \\
3.0 & 7.9 & 0.09 & $0.064$ \\
3.5 & 7.3 & 0.42 & $0.417$ 
\bigstrut[b] \\
\hline 
\multicolumn{2}{|c|}{} & 
Total: 1.15 & Total: 1.000 
\bigstrut[tb] \\
\hline
\end{tabular}

\vspace{0.7cm}
{\normalsize Example (\emph{c})}
\vspace{0.3cm}

\begin{tabular}{| c c | r | r |} 
\hline
\multicolumn{2}{|c|}{Uncertainties} &
\multicolumn{1}{|c|}{\multirow{2}[2]{*}[-0.07cm]{\scalebox{1.15}{$\textstyle \frac{s_i(s_\mathrm{max}-s_i)}{\sigma_i^2}$}}} &
\multicolumn{1}{|c|}{\multirow{2}[2]{*}[-0.07cm]{$\begin{array}{c}\mathrm{Weight}\vspace{-0.1cm}\\w_i\end{array}$}}\\
$\sigma_i$~(stat.) & $s_i$~(syst.) & & \bigstrut \\
\hline
\hline
2.0 & \textbf{7.9} & 0.39 & $0.261$ \bigstrut[t]\\
2.0 & \textbf{8.1} & 0.00 & $-0.116$ \\
3.0 & 7.7 & 0.34 & $0.284$ \\
3.0 & 7.9 & 0.18 & $0.116$ \\
3.5 & 7.3 & 0.48 & $0.455$ 
\bigstrut[b] \\
\hline 
\multicolumn{2}{|c|}{} & 
Total: 1.39 & Total: 1.000 
\bigstrut[tb] \\
\hline
\end{tabular}

\vspace{0.7cm}
{\normalsize Example (\emph{d})}
\vspace{0.3cm}

\begin{tabular}{| c c | r | r |} 
\hline
\multicolumn{2}{|c|}{Uncertainties} &
\multicolumn{1}{|c|}{\multirow{2}[2]{*}[-0.07cm]{\scalebox{1.15}{$\textstyle \frac{s_i(s_\mathrm{max}-s_i)}{\sigma_i^2}$}}} &
\multicolumn{1}{|c|}{\multirow{2}[2]{*}[-0.07cm]{$\begin{array}{c}\mathrm{Weight}\vspace{-0.1cm}\\w_i\end{array}$}}\\
$\sigma_i$~(stat.) & $s_i$~(syst.) & & \bigstrut \\
\hline
\hline
2.0 & \textbf{7.9} & 0.20 & $0.050$ \bigstrut[t]\\
2.0 & \textbf{7.9} & 0.20 & $0.050$ \\
2.0 & \textbf{8.0} & 0.00 & $-0.134$ \\
2.0 & \textbf{8.0} & 0.00 & $-0.134$ \\
3.0 & 7.7 & 0.26 & $0.186$ \\
3.0 & 7.7 & 0.26 & $0.186$ \\
3.0 & 7.9 & 0.09 & $0.022$ \\
3.0 & 7.9 & 0.09 & $0.022$ \\
3.5 & 7.3 & 0.42 & $0.376$ \\
3.5 & 7.3 & 0.42 & $0.376$ 
\bigstrut[b] \\
\hline 
\multicolumn{2}{|c|}{} & 
Total: 1.92 & Total: 1.000 
\bigstrut[tb] \\
\hline
\end{tabular}
\end{center}
\end{minipage}
}
\end{center}
\caption{Examples illustrating the application of
  Condition~(\ref{gencond}), in combinations of hypothetical
  measurements.}
\label{gencondexamples}
\end{table}

In Example (\emph{a}), we have
\begin{displaymath}
\sum_{i=1}^5\;\frac{s_i(s_\mathrm{max}-s_i)}{\sigma_i^2}\;=\;0.96\;\;\;
,\;\;\;\;\mathrm{where}\;\;s_\mathrm{max}=8.0 \;\;\;,
\end{displaymath}
so Condition~(\ref{gencond}) is satisfied; therefore, all five weights
$w_i$ are positive.

In Example (\emph{b}), one of the systematic uncertainties (printed in
boldface) has been modified from $s=7.9$ to $s=7.8$. Since this
measurement has a small statistical uncertainty ($\sigma=2.0$), the
effect of this change is sufficient to violate
Condition~(\ref{gencond}):
\begin{displaymath}
\sum_{i=1}^5\;\frac{s_i(s_\mathrm{max}-s_i)}{\sigma_i^2}\;=\;1.15\;\;\;
,\;\;\;\;\mathrm{where}\;\;s_\mathrm{max}=8.0 \;\;\;.
\end{displaymath}
We now find that one of the weights $w_i$ (the weight corresponding to
the measurement with the largest systematic uncertainty
$s_\mathrm{max}$) has become negative.

In Example (\emph{c}), the largest systematic uncertainty $s_\mathrm{max}$
(also printed in boldface) has been modified from $s=8.0$ to $s=8.1$;
all other uncertainties are identical to those in Example~(\emph{a}). In this
case we have violated Condition~(\ref{gencond}) more severely than in
Example~(\emph{b}), because the largest systematic uncertainty contributes to
every term in the summation:
\begin{displaymath}
\sum_{i=1}^5\;\frac{s_i(s_\mathrm{max}-s_i)}{\sigma_i^2}\;=\;1.39\;\;\;
,\;\;\;\;\mathrm{where}\;\;s_\mathrm{max}=8.1 \;\;\;.
\end{displaymath}
We now have a negative weight of $-11.6\%$ in our hypothetical
combination, compared with $-4.6\%$ in Example~(\emph{b}).

In Example (\emph{d}), we have repeated each of the five measurements used in
Example~(\emph{a}). The new measurements have uncorrelated statistical
uncertainties, but their systematic uncertainties remain fully
correlated with the original measurements. Condition~(\ref{gencond})
is now violated, since each term of the summation has been duplicated:
\begin{displaymath}
\sum_{i=1}^{10}\;\frac{s_i(s_\mathrm{max}-s_i)}{\sigma_i^2}\;=\;2\times
0.96\;=\;1.92\;\;\; ,\;\;\;\;\mathrm{where}\;\;s_\mathrm{max}=8.0
\;\;\;.
\end{displaymath}
Once again, we have a negative weight. It is worth noting, in
particular, that the weights in Example~(\emph{d}) are \emph{not}
equal to half of the corresponding weights in Example~(\emph{a}).

\subsection[Application to the LEP \asmz\ combination]{Application to the LEP \boldmath{\asmz} combination}
\label{weightslepapp}

\enlargethispage{-1\baselineskip}Our calculations discussed in the
preceding sections are not strictly applicable to the combination of
\LEP\ \asmz\ measurements. Condition~(\ref{gencond}) is only valid in
the case where \emph{all} systematic uncertainties are \emph{fully}
correlated with all others; it does not allow for four independent
sources of systematic uncertainty, nor for the detailed correlation
assumptions discussed in Section~\ref{corrsubsect}. However, we can
expect some of the same principles to apply. We have seen that the
following factors influence the tendency toward negative weights:
\begin{itemize}
\item \textbf{Relative magnitudes of systematic and statistical
  uncertainties:}\\Measurements with systematically dominated
  uncertainties are most prone to negative weights. Previous
  combinations of statistically limited \LEP\ measurements, such as
  the W$^\pm$~boson mass~\cite{wmasscomb}, have not suffered from
  negative weights.
\item \textbf{Spread in the magnitudes of systematic
  uncertainties:}\\When systematic uncertainties have the same
  magnitude for all measurements, they do not influence the weights;
  they simply ``carry through'' as an uncertainty in the weighted mean.
  If the magnitudes vary, however, their spread must be small in comparison to
  the statistical uncertainties if negative weights are to be
  avoided.
\item \textbf{Largest systematic uncertainty:}\\In
  Condition~(\ref{gencond}), the `spread' of systematic uncertainties
  is measured in relation to the largest of them,~$s_\mathrm{max}$. As
  we saw in Example~(\emph{c}) of the preceding section, an increase in the
  largest uncertainty has a much greater effect than a decrease in one
  of the others.\enlargethispage{-1\baselineskip}
\item \textbf{Number of measurements:}\\The left-hand side of
  Condition~(\ref{gencond}) scales in proportion to the number of
  measurements. As we increase the number of terms contributing to the
  sum, as in Example~(\emph{d}), we must decrease the average size of
  each term to prevent the inequality being violated. For $N$
  measurements, the condition can be rewritten as
\begin{equation}
\left\langle\frac{s_i(s_\mathrm{max}-s_i)}{\sigma^2_i}\right\rangle\;<\;\frac{1}{N}\;\;\;.
\end{equation}
  When the uncertainties of all measurements are roughly equal, and
  are dominated by systematics,
\begin{equation}
\left\langle s_\mathrm{max}-s_i\right\rangle
\;\lesssim\;\frac{\langle\sigma\rangle^2}{N\langle s\rangle}\;<\;\frac{\langle\sigma\rangle}{N}\;\;\;.
\label{minsysrange}
\end{equation}
  For our 194 measurements of \asmz, which have
  $\sigma_\mathrm{stat.}\sim 0.002$ and $s=\sigma_\mathrm{syst.}\sim
  0.005$, the fractional range of our systematic uncertainties would need to
  satisfy
\begin{displaymath}
\left\langle \frac{s_\mathrm{max}-s_i}{s_i}\right\rangle \;\lesssim\;
\frac{0.002^2}{194\times 0.005^2}\;=\;0.08\,\% \;\;\;.
\end{displaymath}
  This condition is obviously not satisfied, so our na\"{\i}ve
  combination presented in Section~\ref{naivecomb} inevitably had
  large negative weights.
\end{itemize}

To ensure that our \asmz\ combination has only positive weights, we
therefore have three options.
\begin{enumerate}
\item Remove the systematic uncertainties from the covariance matrix
  altogether, and instead apply an `averaged' systematic uncertainty
  to the final weighted mean.  The weights will then be proportional
  to the inverse squares of the statistical uncertainties.
\item Artificially set the systematic uncertainties for different
  measurements equal to one another.
\item Remove some, or all, of the \emph{correlations} between
  systematic uncertainties, but leave the diagonal elements of the
  covariance matrix unchanged.
\end{enumerate}
The first option is undesirable, as it does not give extra weight
to measurements with the smallest systematic uncertainties: for
example, measurements at \LEP2 are less sensitive to the effects of
non-perturbative QCD than those at \LEP1. The second option is
mathematically equivalent to the first, and will not be considered
further. The third option is preferable, as it allows the weights to
be influenced by systematic uncertainties, without permitting them to
become negative. In the next section, we determine which correlations
need to be removed from the covariance matrix to avoid negative
weights.

\subsection[Effects of correlations on the \asmz\ weights]{Effects of
  correlations on the \boldmath{\asmz}\ weights}

\enlargethispage{-1\baselineskip}In Section~\ref{depcorrsect}, we
studied the dependence of our fitted \asmz\ values on the correlation
coefficients between systematic uncertainties. We now perform a
similar investigation, but focusing on the distribution of weights,
$w_i$.

\begin{figure}[p]
\begin{leftfullpage}
\begin{multicols}{2}
\includegraphics[width=\columnwidth]{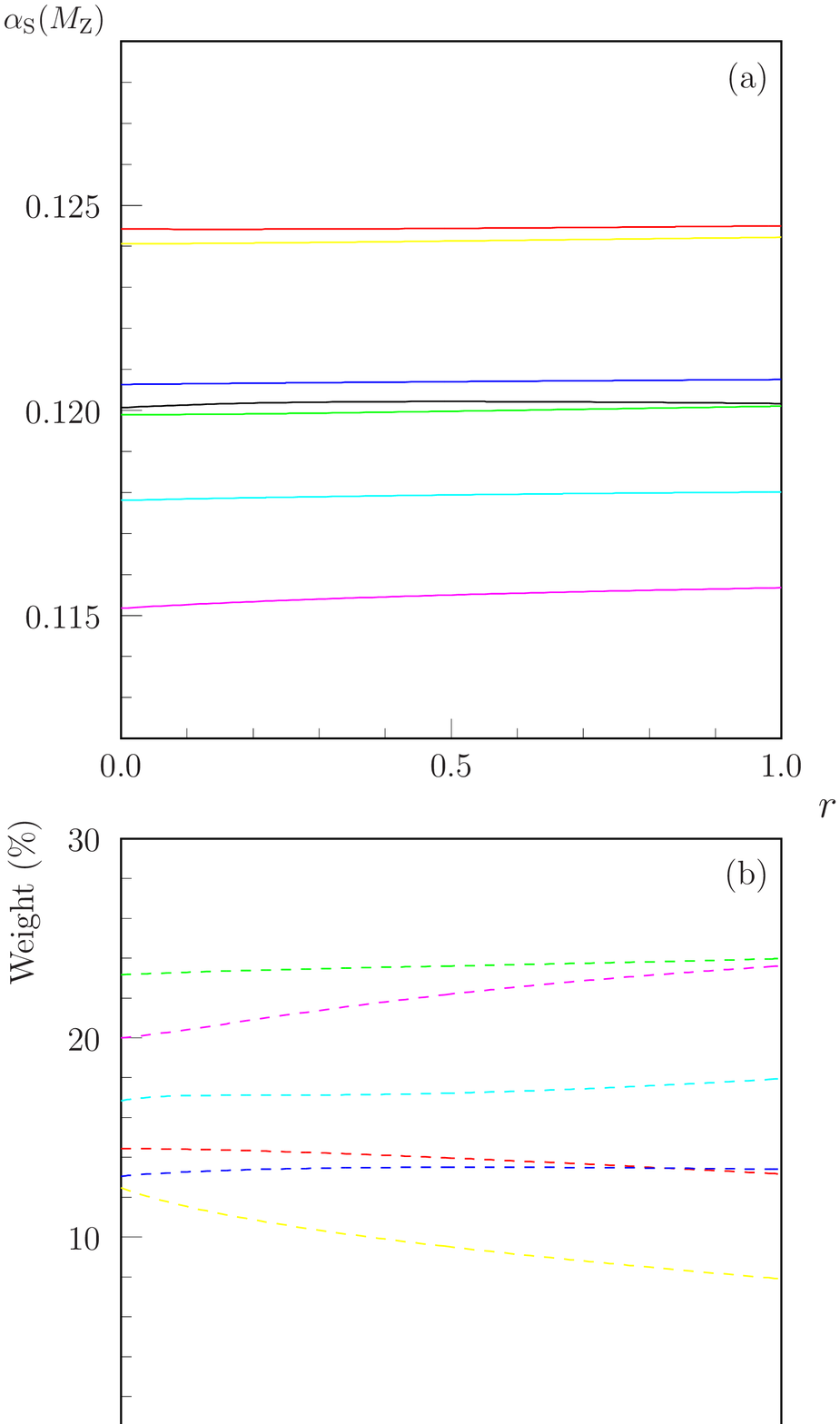}
\vspace{1.0cm}

\begin{center}\scalebox{0.8}{\framebox{\parbox[t]{5.5cm}{\linespread{0}
{\centering Off-diagonal elements of $V$:\vspace{0.2cm}
\begin{displaymath}
V_{ij}^\mathrm{stat.} = \rho_{ij}^\mathrm{stat.}\,\sigma_i\,\sigma_j
\end{displaymath}
{\footnotesize (same energy and experiment)}\vspace{0.2cm}
\begin{displaymath}
V_{ij}^\mathrm{exp.} = \textcolor{red}{r\,\mathrm{min}(\sigma^2_i,\sigma^2_j)}
\end{displaymath}
{\footnotesize (same experiment)}\vspace{0.2cm}
\begin{displaymath}
V_{ij}^\mathrm{hadr.} = 0
\end{displaymath}
{\footnotesize (same experiment)}\vspace{0.2cm}
\begin{displaymath}
V_{ij}^\mathrm{theo.} = 0
\end{displaymath}
{\footnotesize (same observable)}\vspace{0.2cm}
\begin{displaymath}
V_{ij}^\mathrm{theo.} = 0
\end{displaymath}
{\footnotesize (different observables)\\}}}}}

\end{center}
\caption{\small Variation of \asmz\ fits (Fig.~(\emph{a})),
and weights (Fig.~(\emph{b})), for the six observables, as the
correlations of experimental systematic uncertainties are ``switched
on''. The hadronisation and theory uncertainties remain uncorrelated.
See Fig.~\ref{rhovary_weightskey} for the legend.
\textcolor{white}{FILL FILL FILL FILL FILL FILL FILL FILL}}
\label{rhovary_weights1}
\pagebreak
\includegraphics[width=\columnwidth]{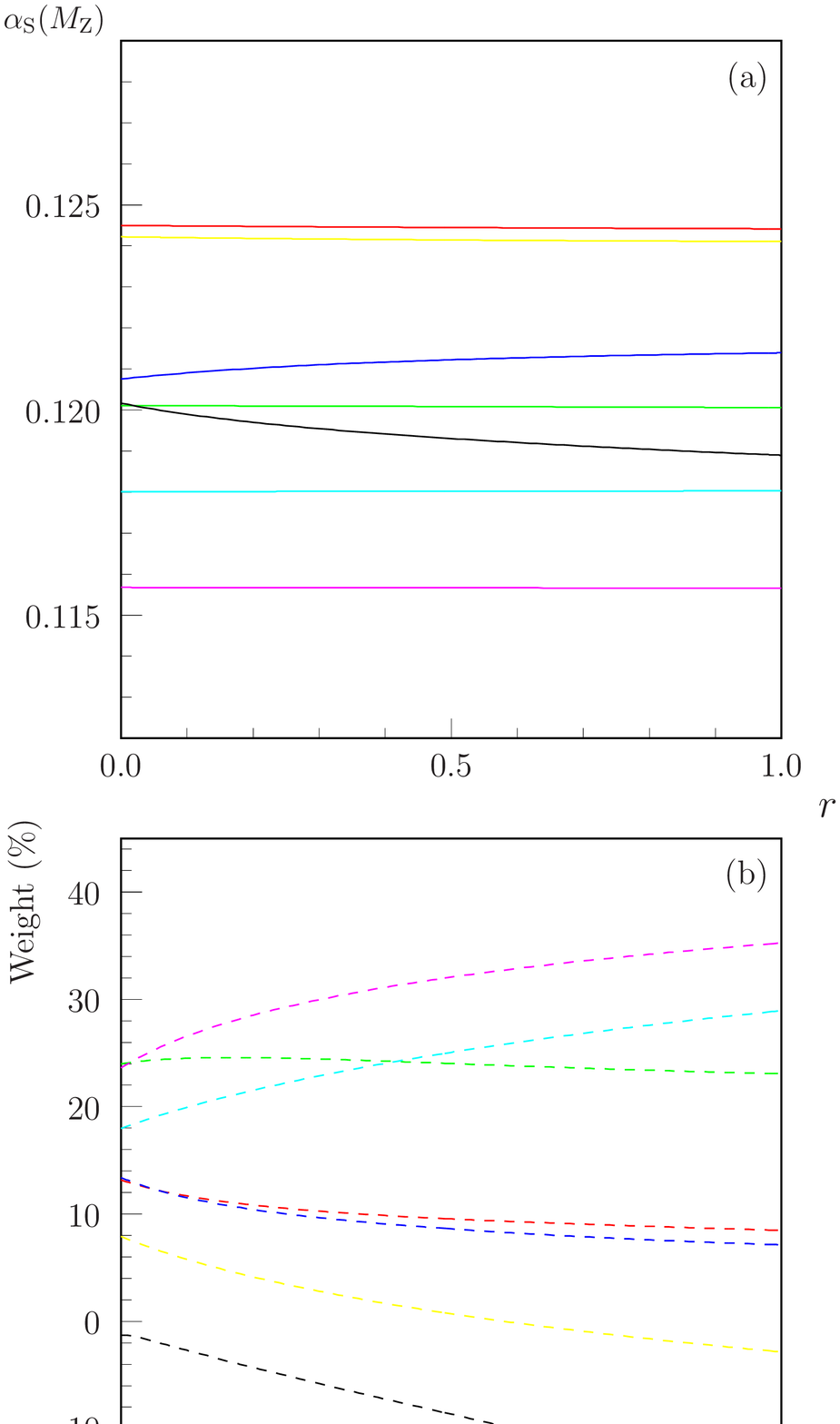}
\vspace{1.0cm}

\begin{center}\scalebox{0.8}{\framebox{\parbox[t]{5.5cm}{\linespread{0}
{\centering Off-diagonal elements of $V$:\vspace{0.2cm}
\begin{displaymath}
V_{ij}^\mathrm{stat.} = \rho_{ij}^\mathrm{stat.}\,\sigma_i\,\sigma_j
\end{displaymath}
{\footnotesize (same energy and experiment)}\vspace{0.2cm}
\begin{displaymath}
V_{ij}^\mathrm{exp.} = \mathrm{min}(\sigma^2_i,\sigma^2_j)
\end{displaymath}
{\footnotesize (same experiment)}\vspace{0.2cm}
\begin{displaymath}
V_{ij}^\mathrm{hadr.} =  \textcolor{red}{r\,\mathrm{min}(\sigma^2_i,\sigma^2_j)}
\end{displaymath}
{\footnotesize (same experiment)}\vspace{0.2cm}
\begin{displaymath}
V_{ij}^\mathrm{theo.} = 0
\end{displaymath}
{\footnotesize (same observable)}\vspace{0.2cm}
\begin{displaymath}
V_{ij}^\mathrm{theo.} = 0
\end{displaymath}
{\footnotesize (different observables)\\}}}}}

\end{center}

\caption{\small Variation of \asmz\ fits (Fig.~(\emph{a})), and
weights (Fig.~(\emph{b})), for the six observables, as the
correlations of hadronisation uncertainties are ``switched on''.  The
experimental systematic uncertainties have a ``minimum overlap''
correlation, while the theory uncertainties remain uncorrelated.  See
Fig.~\ref{rhovary_weightskey} for the legend.}
\label{rhovary_weights2}
\end{multicols}
\end{leftfullpage}
\end{figure}

\begin{figure}[p]
\begin{fullpage}
\vspace{0.5cm}
\begin{multicols}{2}
\includegraphics[width=\columnwidth]{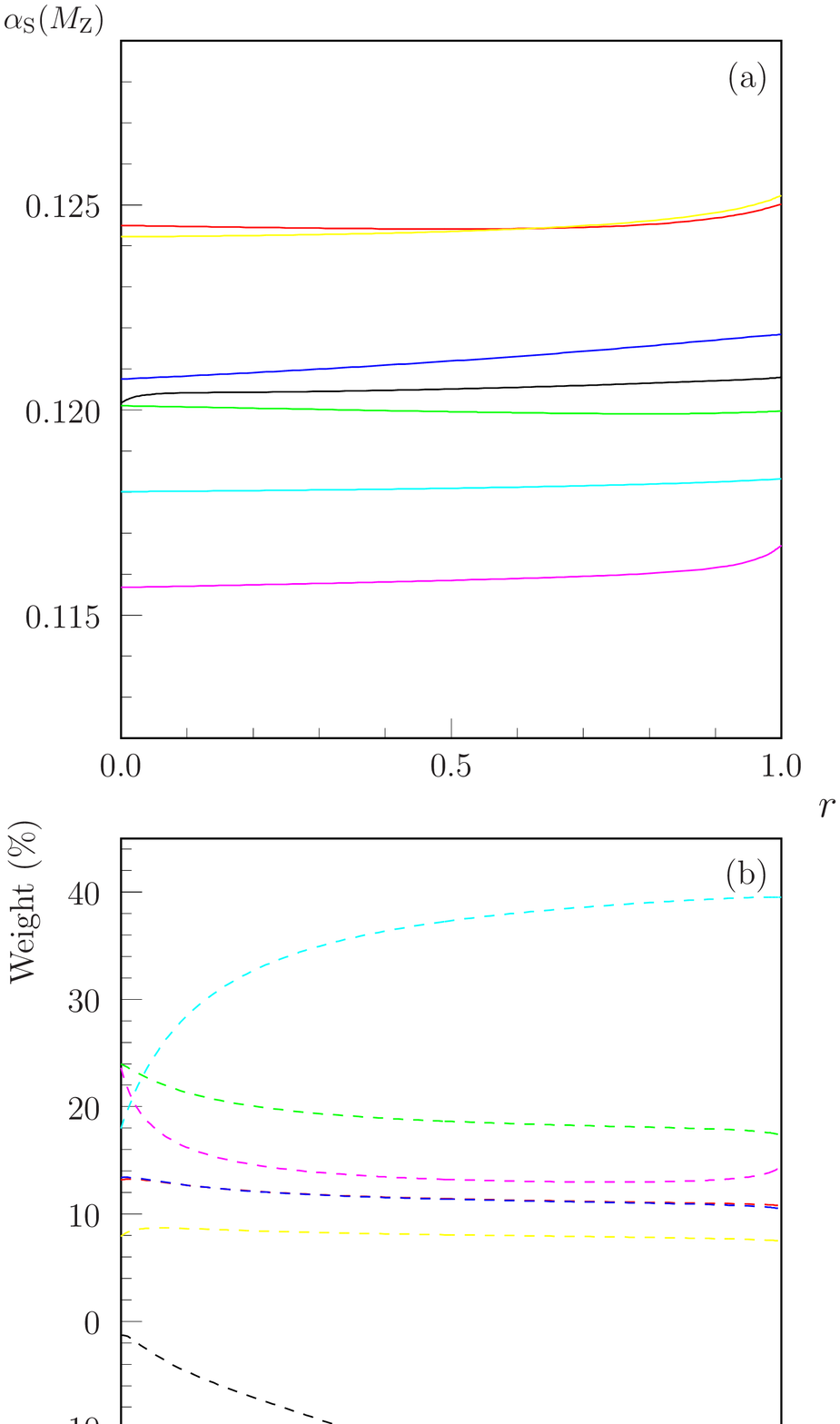}
\vspace{1.0cm}

\begin{center}\scalebox{0.8}{\framebox{\parbox[t]{5.5cm}{\linespread{0}
{\centering Off-diagonal elements of $V$:\vspace{0.2cm}
\begin{displaymath}
V_{ij}^\mathrm{stat.} = \rho_{ij}^\mathrm{stat.}\,\sigma_i\,\sigma_j
\end{displaymath}
{\footnotesize (same energy and experiment)}\vspace{0.2cm}
\begin{displaymath}
V_{ij}^\mathrm{exp.} = \mathrm{min}(\sigma^2_i,\sigma^2_j)
\end{displaymath}
{\footnotesize (same experiment)}\vspace{0.2cm}
\begin{displaymath}
V_{ij}^\mathrm{hadr.} = 0
\end{displaymath}
{\footnotesize (same experiment)}\vspace{0.2cm}
\begin{displaymath}
V_{ij}^\mathrm{theo.} = \textcolor{red}{r\,\sigma_i\,\sigma_j}
\end{displaymath}
{\footnotesize (same observable)}\vspace{0.2cm}
\begin{displaymath}
V_{ij}^\mathrm{theo.} = 0
\end{displaymath}
{\footnotesize (different observables)\\}}}}}

\end{center}
\caption{\small Variation of
\asmz\ fits (Fig.~(\emph{a})) and weights (Fig.~(\emph{b})) for
the six observables, as the correlations of theory uncertainties are
``switched on'' between measurements using the same observables.  The
experimental systematic uncertainties have a ``minimum overlap''
correlation, while the hadronisation uncertainties remain
uncorrelated.  See Fig.~\ref{rhovary_weightskey} for the legend.}
\label{rhovary_weights3}
\pagebreak
\includegraphics[width=\columnwidth]{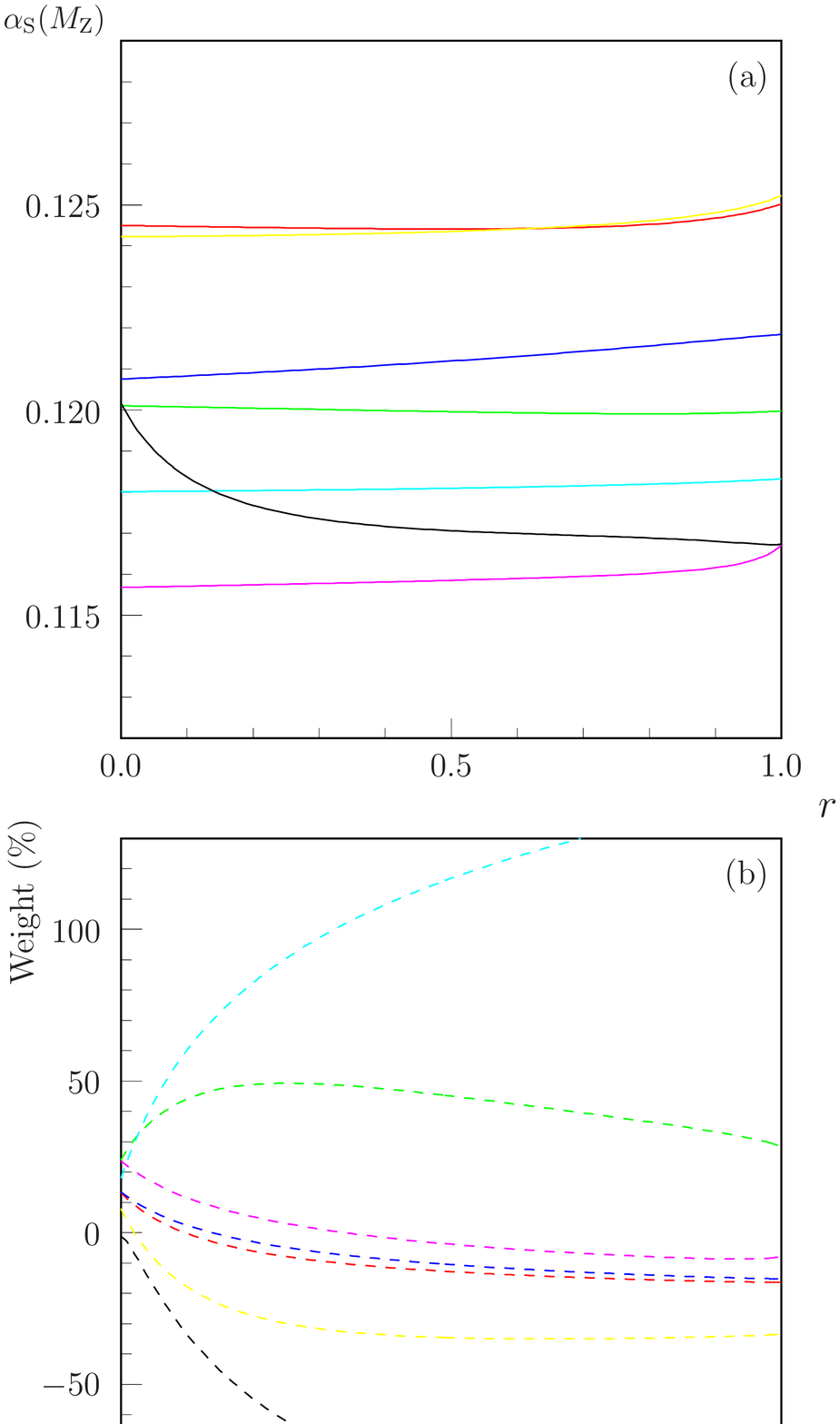}
\vspace{1.0cm}

\begin{center}\scalebox{0.8}{\framebox{\parbox[t]{5.5cm}{\linespread{0}
{\centering Off-diagonal elements of $V$:\vspace{0.2cm}
\begin{displaymath}
V_{ij}^\mathrm{stat.} = \rho_{ij}^\mathrm{stat.}\,\sigma_i\,\sigma_j
\end{displaymath}
{\footnotesize (same energy and experiment)}\vspace{0.2cm}
\begin{displaymath}
V_{ij}^\mathrm{exp.} = \mathrm{min}(\sigma^2_i,\sigma^2_j)
\end{displaymath}
{\footnotesize (same experiment)}\vspace{0.2cm}
\begin{displaymath}
V_{ij}^\mathrm{hadr.} = 0
\end{displaymath}
{\footnotesize (same experiment)}\vspace{0.2cm}
\begin{displaymath}
V_{ij}^\mathrm{theo.} = \textcolor{red}{r\,\sigma_i\,\sigma_j}
\end{displaymath}
{\footnotesize (same observable)}\vspace{0.2cm}
\begin{displaymath}
V_{ij}^\mathrm{theo.} = \textcolor{red}{0.9\,r\,\sigma_i\,\sigma_j}
\end{displaymath}
{\footnotesize (different observables)\\}}}}}

\end{center}

\caption{\small Variation of \asmz\ fits (Fig.~(\emph{a})) and weights
(Fig.~(\emph{b})) for the six observables, as the correlations between all
theory uncertainties are ``switched on,'' including those between
measurements derived from different observables.  The experimental
systematic uncertainties have a ``minimum overlap'' correlation, while
the hadronisation uncertainties remain uncorrelated.  See
Fig.~\ref{rhovary_weightskey} for the legend.}
\label{rhovary_weights4}
\end{multicols}
\end{fullpage}
\end{figure}

\begin{figure}[tbp!]
\begin{center}
\includegraphics[width=\textwidth]{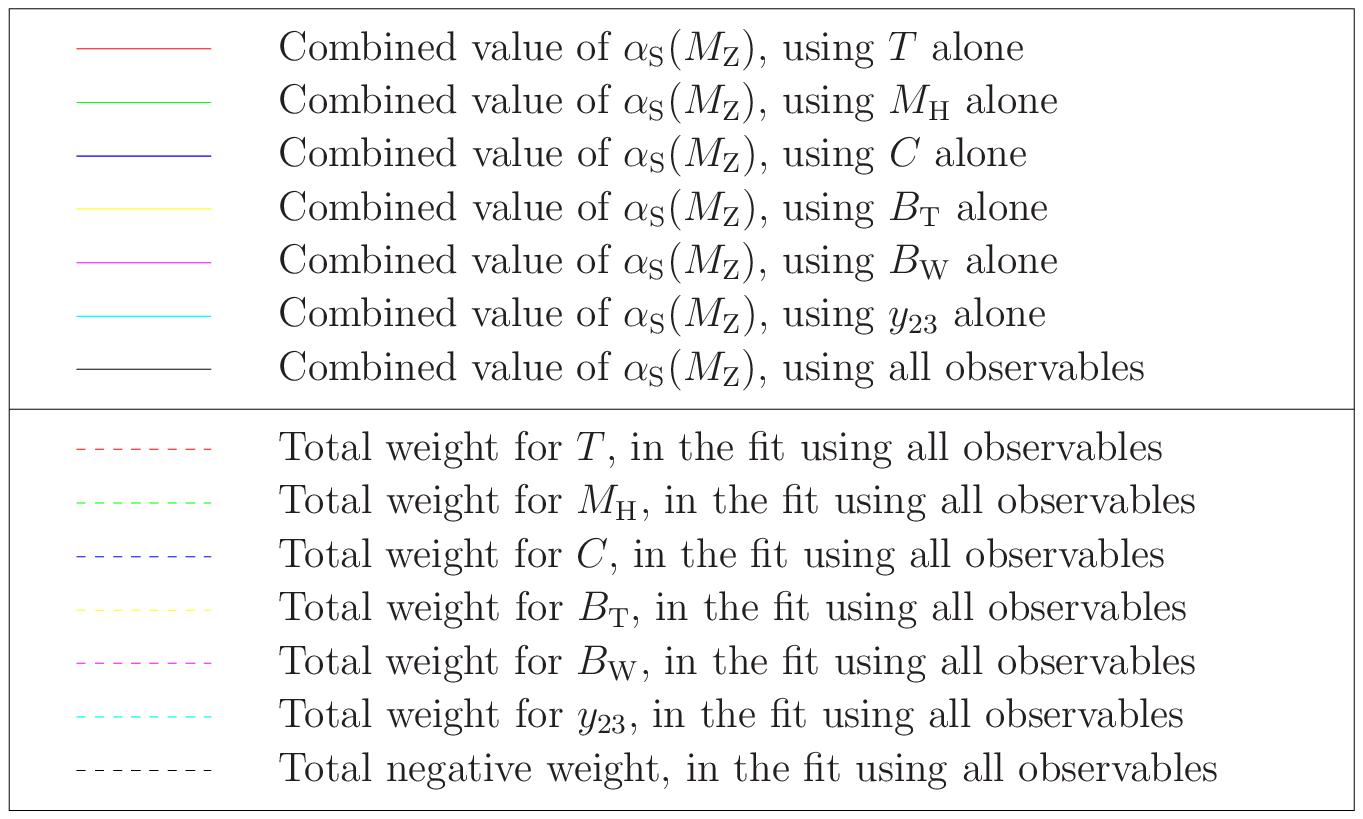}
\end{center}
\caption{Legend for Figures \ref{rhovary_weights1}--\ref{rhovary_weights4}.}
\label{rhovary_weightskey}
\end{figure}

In each of Figures \ref{rhovary_weights1}--\ref{rhovary_weights4}, we
show the variation of the following quantities with respect to certain
correlation coefficients in the covariance matrix:
\begin{itemize}
\item[(\emph{a})]the fitted \asmz\ values derived from individual event shape
  observables, and from a global fit to all measurements.
\item[(\emph{b})]the total weights assigned to each event shape observable,
  and the sum of the negative weights, in the global fit.
\end{itemize}
The legend for these plots is given in
Figure~\ref{rhovary_weightskey}; note that the vertical scales for the
weight variations differ between the four plots. We will discuss
separately the effects of correlations in the four types of
uncertainty.

\begin{description}
\item[Statistical uncertainties:]As in Section~\ref{depcorrsect}, we
  do not vary the correlations of statistical uncertainties, as
  they are well understood.
\item[Experimental systematic uncertainties:]In
  Figure~\ref{rhovary_weights1}, we smoothly switch on the correlation
  of experimental systematic uncertainties, starting from a covariance
  matrix with only statistical correlations. The right-hand side of
  the plot corresponds to a ``minimum overlap'' correlation between
  measurements within the same experiment, as described in
  Section~\ref{corrsubsect}. We observe very little variation in the
  fitted \asmz\ values, and also no significant negative weights. With
  correlations fully switched on, the total negative weight is
  only~$-1.3\%$. Furthermore, the distribution of weights among the
  six observables is rather insensitive to small variations of the
  correlation from its maximum value. We will therefore continue to
  apply the minimum overlap assumption in the covariance
  matrix.\enlargethispage{\baselineskip}
\item[Hadronisation uncertainties:]In Figure~\ref{rhovary_weights2},
  we switch on the correlation of hadronisation uncertainties, while
  leaving the statistical and experimental uncertainties correlated as
  described above. The right-hand side of the plot corresponds to a
  ``minimum overlap'' correlation between measurements within the same
  experiment. When the correlation is fully switched on, we find a
  significant total negative weight ($-15.6\%$). This also leads to a
  stronger variation of the weight distributions than was seen in the
  case of experimental uncertainties. These observations can be
  attributed to the energy-dependence of the hadronisation
  uncertainties, which is far greater than that of the experimental
  systematics. As we saw in Section~\ref{weightslepapp}, a wider
  spread in the magnitudes of systematic uncertainties leads to an
  increased tendency towards negative weights. In order to prevent an
  unreasonable increase in negative weights, we will therefore
  consider hadronisation uncertainties to be
  uncorrelated.\footnote{This assumption will necessitate a
  modification in our definition of the combined uncertainty, which
  will be discussed in Section~\ref{newcombinederror}.}
\item[Theory uncertainties:]\enlargethispage{\baselineskip}In
  Section~\ref{depcorrsect}, we distinguished between two types of
  correlations between theoretical uncertainties: those between fits
  to the same observable, and those between measurements using
  different observables. We will maintain this distinction here. In
  Figure~\ref{rhovary_weights3}, we switch on the correlation of
  theory uncertainties between fits to the same observable, while
  leaving the statistical and experimental uncertainties correlated as
  described above. In this case, the right-hand side of the plot
  corresponds to 100\%~correlation.  Although the \asmz\ fit results
  do not change significantly, there is a rapid increase in negative
  weights as the correlation is switched on: when the correlation
  reaches 100\%, the total negative weight is~$-79\%$. In
  Figure~\ref{rhovary_weights4}, we simultaneously switch on
  \emph{all} correlations between theory uncertainties, including
  those between measurements derived from different observables; as
  before, the maximum correlation coefficient between fits to
  different observables is taken to be~0.9. We now see a dramatic
  increase in the negative weights, reaching $-179\%$ when the
  correlations are fully switched on. The total weight assigned to the
  observable \ytwothree\ increases to $+144\%$, leaving overall
  negative weights assigned to $T$, $C$, \BT\ and \BW. This result
  supports the conclusions of Section~\ref{depcorrsect}, that the
  instability of our na\"{\i}ve combination can be attributed
  primarily to the correlation of theoretical uncertainties between
  different observables. These uncertainties are large, and differ
  more significantly between observables than between experiments or
  centre-of-mass energies, so such correlations are highly prone to
  introduce negative weights.  As with hadronisation uncertainties, we
  therefore choose to remove all correlations between theory
  uncertainties from the covariance matrix.
\end{description}

In summary, our covariance matrix will contain all uncertainties in
the diagonal terms, but only statistical and experimental
uncertainties in the off-diagonal terms.

\section{The combined uncertainty revisited}
\label{newcombinederror}

In the previous section, we have discussed the need to modify the
covariance matrix, thereby eliminating large negative weights from the
combination. This will not bias our estimator, $\ashatmz=\sum w_i
\alpha_i$, but will certainly change the variance of the combination,
$\sigma^2=\sum w_i V_{ij} w_j$, defined in
Section~\ref{errorcomb}. Regarding the hadronisation and theory
uncertainties as uncorrelated between measurements would reduce the
combined uncertainty. To take account of correlations, we instead use
the following method:
\begin{description}
\item[Statistical and experimental uncertainties:] The
  covariance matrices associated with these uncertainties,
  $V_\mathrm{stat.}$ and $V_\mathrm{exp.}$, are unaltered by our new
  combination algorithm. We can therefore still apply
  Equation~(\ref{errorbreakdown}) to estimate the corresponding
  uncertainties $\sigma_\mathrm{stat.}$ and $\sigma_\mathrm{exp.}$.
\item[Combined hadronisation uncertainties:] We repeat the entire
  combination procedure three times, using different sets of input
  measurements $y_i$. Each set uses a different Monte Carlo event
  generator to estimate hadronisation corrections:
  \begin{itemize}
   \item $y_{i,\mathrm{P}} \; \equiv \; \left[\; \alpha_\mathrm{S}(Q)^\mathrm{PYTHIA}\;\right]_i$ 
   \item $y_{i,\mathrm{H}} \; \equiv \; \left[\; \alpha_\mathrm{S}(Q)^\mathrm{HERWIG}\;\right]_i$ 
   \item $y_{i,\mathrm{A}} \; \equiv \; \left[\; \alpha_\mathrm{S}(Q)^\mathrm{ARIADNE}\;\right]_i$ 
  \end{itemize}
  The covariance matrix $V$, and hence the weights $w_i$, are the same
  in all three cases. Our hadronisation uncertainty for the combined
  measurement is then defined as the standard deviation of the three
  resulting estimators: $\ashatmz_\mathrm{P}$, $\ashatmz_\mathrm{H}$
  and $\ashatmz_\mathrm{A}$. This definition is analogous to that of
  the hadronisation uncertainties for the individual measurements,
  discussed in Section~\ref{haderrorsubsubsect}.
\item[Combined theory uncertainties:] We again repeat the combination
  with three sets of input values:
  \begin{itemize}
    \item `Central' values, $y_i^0 \; \equiv \; \left[\;
    \alpha_\mathrm{S}(Q) \;\right]_i$ 
    \item `Raised' values, at the upper extremity of the theory
    error-bar,
    \footnote{The theory uncertainties predicted by the ``uncertainty
    band'' method are generally asymmetric. We denote the upper and
    lower uncertainties by $\sigma^+_\mathrm{theo.}$ and
    $\sigma^-_\mathrm{theo.}$ respectively.}
    \\$y_i^+\;\equiv\;\left[\;\alpha_\mathrm{S}(Q)\;\right]_i+\sigma^+_{i,\,\mathrm{theo.}}$
    \item `Lowered' values, at the lower extremity of the theory
    error-bar,
    \\$y_i^-\;\equiv\;\left[\;\alpha_\mathrm{S}(Q)\;\right]_i+\sigma^-_{i,\,\mathrm{theo.}}$
  \end{itemize}
The upper and lower uncertainties for the combined result are then
given by the differences $\ashatmz^+-\ashatmz^0$ and
$\ashatmz^0-\ashatmz^-$.
\end{description}

\subsection{A minor caveat}
It is sometimes claimed that this
treatment described above takes full account of correlations between
the systematic uncertainties. This is not strictly true, unless the
uncertainties are 100\% correlated.  For example, our combined upper
theory uncertainty is given by
\begin{equation}
\sigma_\mathrm{theo.}\;=\;\sum_i \;w_i\,(y_i'\,^+-y_i'\,^0) \;\;\;,
\end{equation}
where $y_i'$ are the individual input measurements after `running' to
the Z$^0$~mass. The corresponding variance is therefore
\begin{equation}
\sigma^2_\mathrm{theo.}\;=\;\sum_{i,j} 
\;w_i\,\left[(y_i'\,^+-y_i'\,^0)\,(y_j'\,^+-y_j'\,^0)\right]\,w_j\;\;\;,
\end{equation}
where the term in square brackets is equivalent to the covariance
matrix $(V'_{ij})^\mathrm{theo.}$ for a theory uncertainty which is
\emph{fully} correlated between the input measurements.  Our combined
hadronisation uncertainty can similarly be written as
\begin{eqnarray}
\sigma^2_\mathrm{hadr.}&=&\frac{1}{9}\,\Big(\;
(\hat\alpha_\mathrm{S,\,P}-\hat\alpha_\mathrm{S,\,H})^2 +
(\hat\alpha_\mathrm{S,\,H}-\hat\alpha_\mathrm{S,\,A})^2 +
(\hat\alpha_\mathrm{S,\,A}-\hat\alpha_\mathrm{S,\,P})^2\;\Big)
\rule[-0.7cm]{0pt}{0pt}
\nonumber \\
&=&\frac{1}{9}\,\left(\;
\Big[\sum_i w_i\,(y'_{i,\mathrm{P}}-y'_{i,\mathrm{H}}) \Big]^2 \! +
\Big[\sum_i w_i\,(y'_{i,\mathrm{H}}-y'_{i,\mathrm{A}}) \Big]^2 \! +
\Big[\sum_i w_i\,(y'_{i,\mathrm{A}}-y'_{i,\mathrm{P}}) \Big]^2
\;\right)\;.
\nonumber \\
\end{eqnarray}
$\phantom{DUMMY}$\vspace{-1cm}\\ This expression can be regarded as a
quadratic sum of three fully-correlated `uncertainties', provided the
signs of the differences $(y'_{i,\mathrm{P}}-y'_{i,\mathrm{H}})$,
$(y'_{i,\mathrm{H}}-y'_{i,\mathrm{A}})$ and
$(y'_{i,\mathrm{A}}-y'_{i,\mathrm{P}})$ do not vary between
measurements.\footnote{As we saw in Section~\ref{subsubsect_hadcorr},
a high proportion of our measurements satisfy
$y'_{i,\mathrm{H}}<y'_{i,\mathrm{P}}<y'_{i,\mathrm{A}}$.}

Our algorithm therefore includes no information about the \emph{true}
correlations of hadronisation and theory uncertainties between
measurements, except in the signs of the differences between
measurements using different event generators.

\section{Fitting the hadronisation uncertainties}
\label{hadfit}

\begin{figure}[tbp!]
\begin{center}
\includegraphics[width=\textwidth]{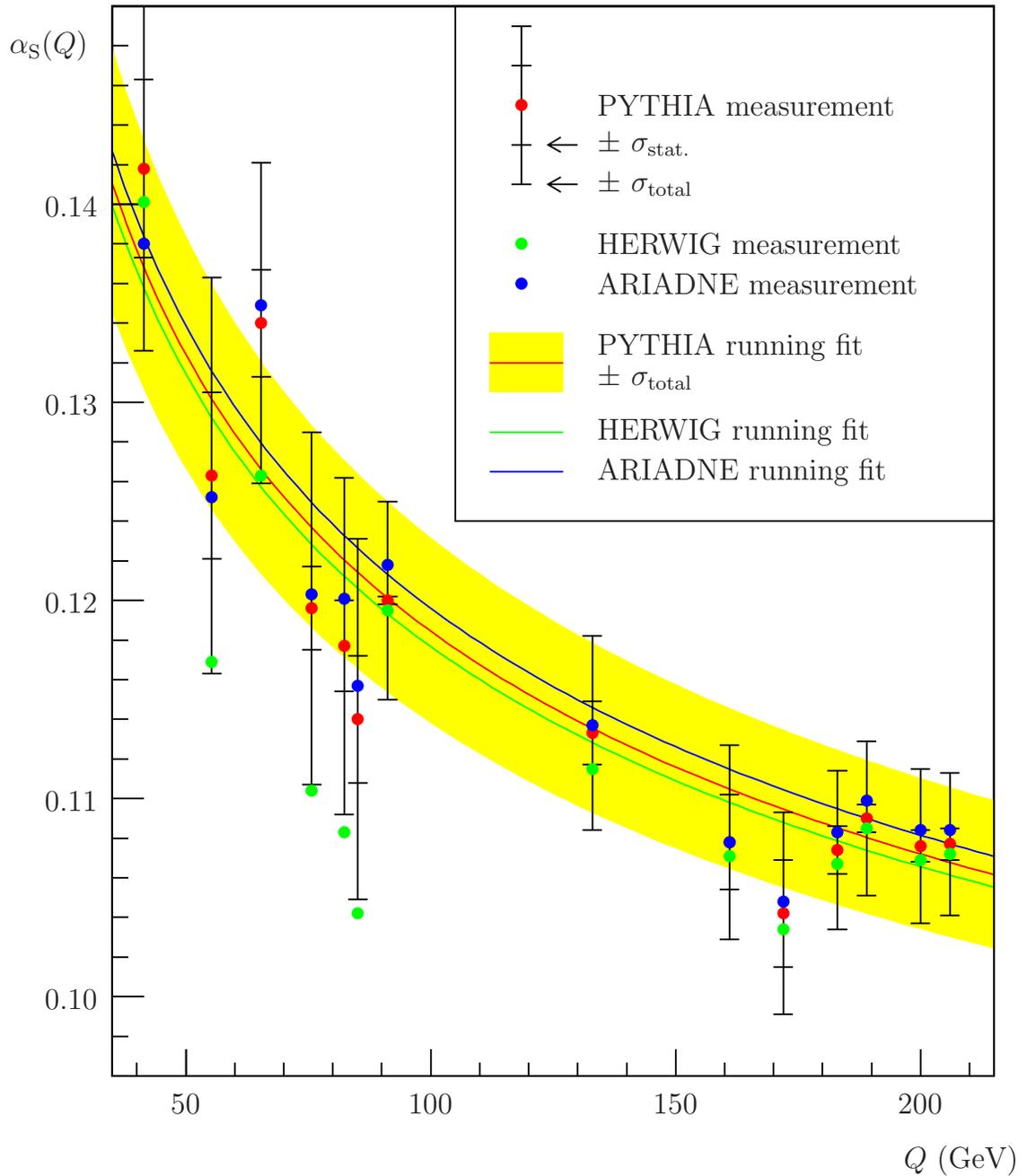}
\caption[Fits to LEP \as\ measurements using PYTHIA, HERWIG and
ARIADNE hadronisation corrections]{Fits to LEP \as\ measurements using
PYTHIA, HERWIG and ARIADNE hadronisation corrections. Each point is a
weighted average of measurements derived from all available
experiments and event-shape observables at the given centre-of-mass
energy. The curves are obtained by fitting all measurements from a
given generator, after conversion to the $M_\mathrm{Z}$ scale.}
\label{runningplot_hadr}
\end{center}
\end{figure}

The uncertainties in our hadronisation corrections should vary
smoothly with the energy scale, provided the same models, parameters
and fit ranges are used. However, as discussed in
Section~\ref{subsubsect_hadcorr}, there are significant differences
between the tuned parameters used by the four Collaborations, which
lead to variations in the estimated hadronisation
uncertainties. Figure~\ref{runningplot_hadr} illustrates the fits
obtained for \asq\ at each centre-of-mass energy, and also the global
running fits, using each of the three generators. The differences
between generators clearly do not vary smoothly between energies,
especially between $Q=91$~GeV and the lower-energy measurements from
L3 data.\footnote{Much of this variation is due to differing
contributions from the four experiments entering into our combination
at different energies.} To prevent these fluctuations from affecting
the weights, it is desirable to smooth them out.

Various theoretical models
\cite{powercorr_Dokshitzer,powercorr_Manohar,powercorr_Akhoury,powercorr_gardi1,powercorr_gardi2,powercorr_Korchemsky}
have predicted that non-perturbative contributions to the moments and
distributions of the event shapes should scale as $1/Q^n$ for some
$n$. For the observables involved in our combination, we expect $n=1$
for $T$, \MH\, $C$, \BT\ and \BW\, and $n=2$ for \ytwothree. A recent
review of these ``power correction'' models, and of their experimental
tests, can be found in Ref.~\cite{dasgupta_salam_review}.

We would intuitively expect that the uncertainties should depend
linearly on the correction itself, which in turn scales as $1/Q^n$. We
therefore adopt the form
\begin{equation}
\sigma'_\mathrm{hadr.}=\frac{A_y}{Q}+B_y \;\;\;,
\end{equation}
for our new hadronisation uncertainty, where the constants $A$ and $B$
are fitted to the data. A separate fit is calculated for each
observable~$y$, since the event shapes differ in their sensitivity to
non-perturbative effects.\footnote{It could be argued that the
uncertainty for \ytwothree\ measurements should include a $1/Q^2$
term; however, it was decided that the input uncertainties are not
sufficiently precise to distinguish between different scaling laws,
and that the two terms of the fit would be sufficient to mimic any
significant $1/Q^2$ behaviour in the 91--206~GeV energy range.}
The fit is performed with weights $w_i$ equal to those used 
in the \asmz\ fit itself. We minimise the following expression, where
$(\sigma_\mathrm{hadr.})_i$ is the original hadronisation uncertainty
defined in Section~\ref{haderrorsubsubsect}, and the summation runs
over all measurements using a single observable $y$:
\begin{equation}
\chi^2\;=\;\sum_i w_i \, \left(\frac{A_y}{Q}+B_y-(\sigma_\mathrm{hadr.})_i\right)^2
\end{equation}
Since the weights vary as functions of covariance matrix, which
includes the hadronisation uncertainty, we must iterate the fit until
the coefficients are stable. The results of our iterated fit are as
follows:\vspace{0.4cm}

\begin{center}
\begin{tabular}{rl}
Thrust, $T$:               & $\sigma_\mathrm{hadr.}=0.29/Q - 0.00067$ \\
Heavy Jet mass, \MH:       & $\sigma_\mathrm{hadr.}=0.25/Q - 0.00063$ \\
C-parameter, $C$:          & $\sigma_\mathrm{hadr.}=0.42/Q - 0.00093$ \\
Total jet broadening, \BT: & $\sigma_\mathrm{hadr.}=0.17/Q + 0.00041$ \\
Wide jet broadening, \BW:  & $\sigma_\mathrm{hadr.}=0.13/Q - 0.00009$ \\
Durham \ytwothree:         & $\sigma_\mathrm{hadr.}=0.05/Q + 0.00009$ \vspace{0.6cm}
\end{tabular}\\
\end{center}
Note that we cannot specify the precision of these coefficients,
because the hadronisation uncertainties entering in the fit do not
have well-defined uncertainties of their own.

\begin{figure}[p]
  \begin{leftfullpage}
  \begin{center}
  \includegraphics[width=0.95\textwidth]{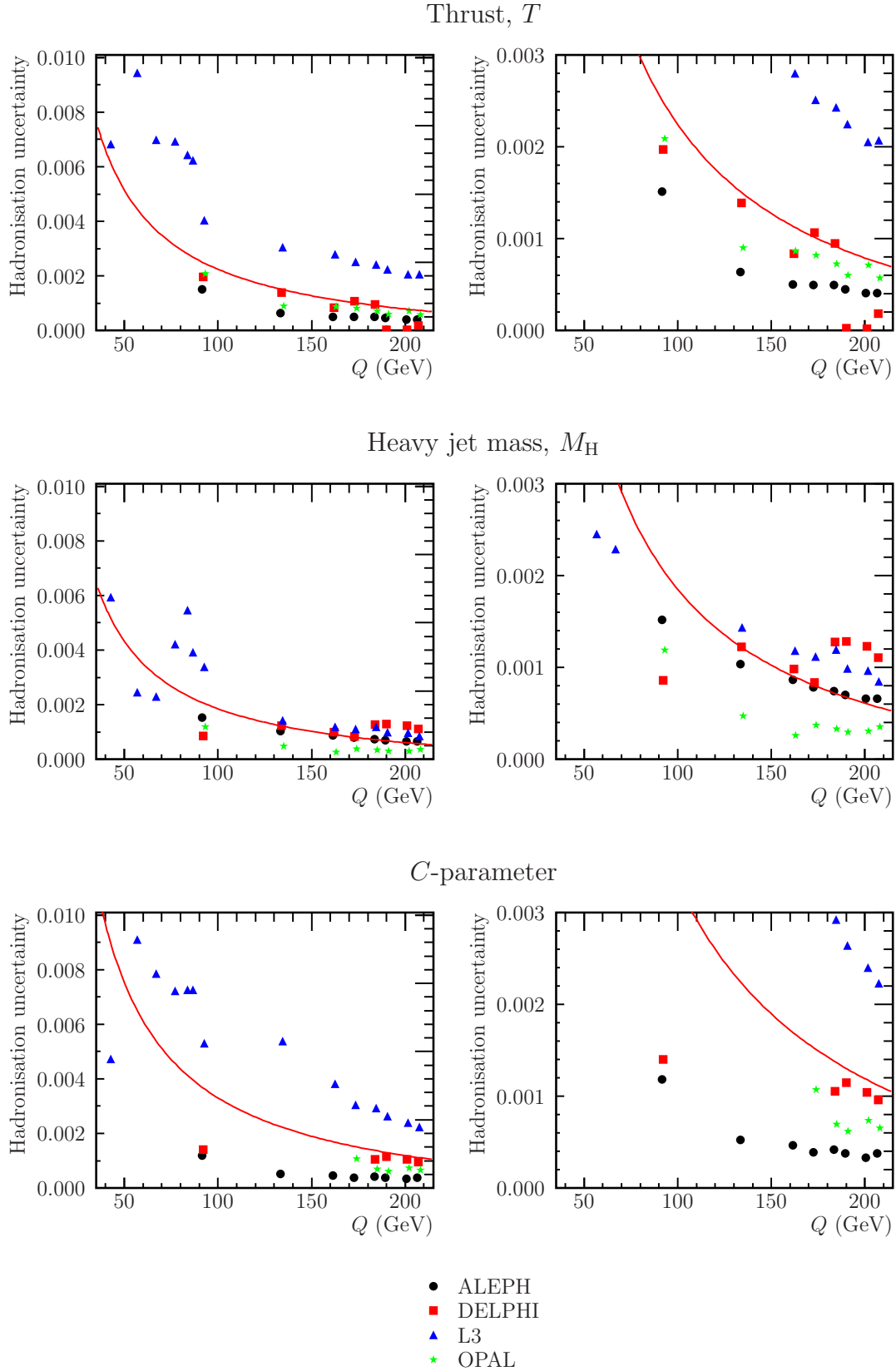}
  \caption{Fitted hadronisation uncertainties for $T$, \MH\ and $C$.
    Each point represents the standard deviation of three results
    using PYTHIA, HERWIG and ARIADNE hadronisation corrections, for an
    individual LEP \as\ measurement.  The left and right plots contain
    the same data on different scales.}
  \label{hadfitplot1}
  \end{center}
  \end{leftfullpage}
\end{figure}

\begin{figure}[p]
  \begin{fullpage}
  \begin{center}
  \includegraphics[width=0.95\textwidth]{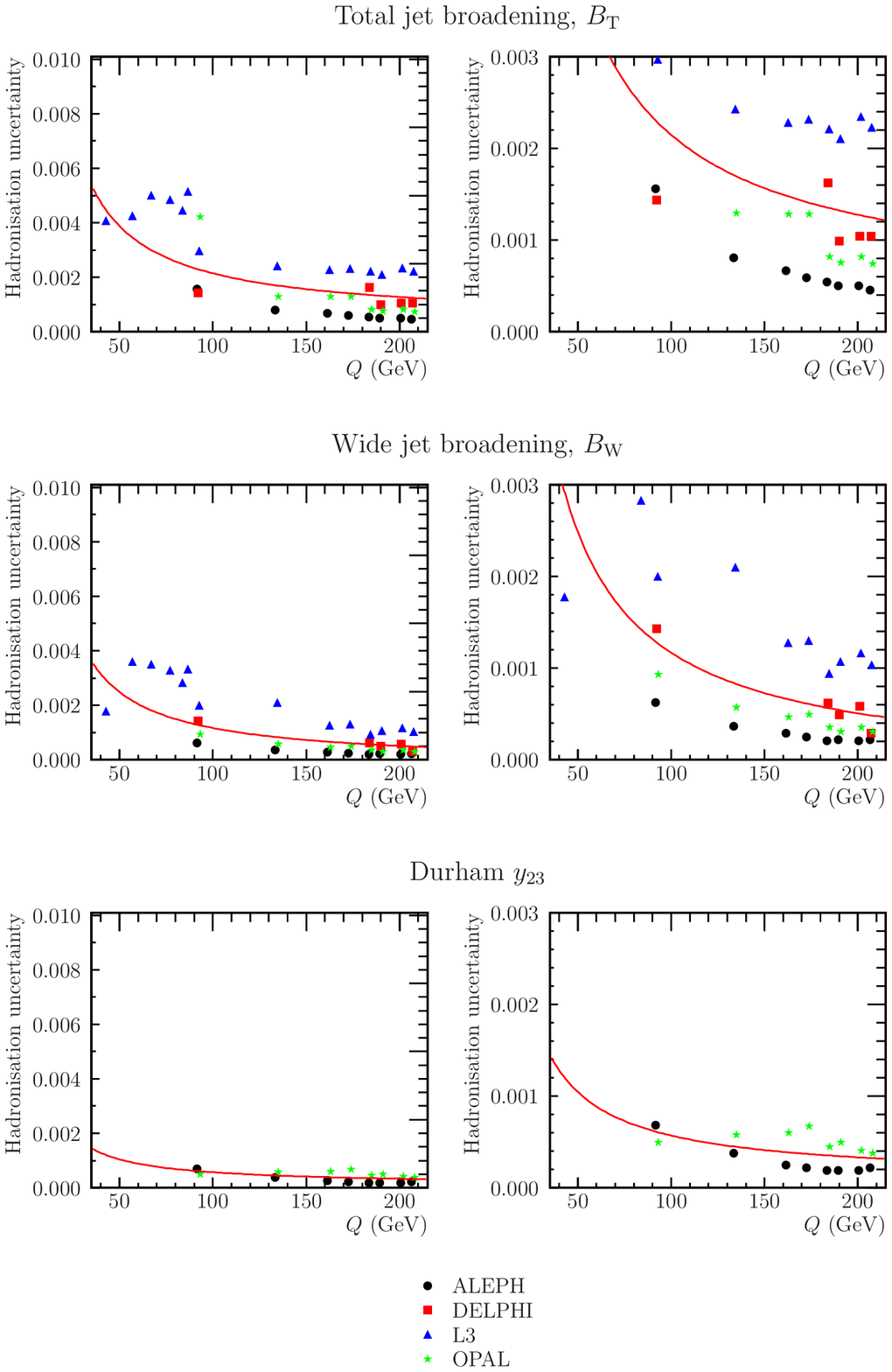}
   \caption{Fitted hadronisation
    uncertainties for \BT, \BW\ and \ytwothree. Each point represents
    the standard deviation of three results using PYTHIA, HERWIG and
    ARIADNE hadronisation corrections, for an individual LEP \as\
    measurement. The left and right plots contain the same data on
    different scales.}
  \label{hadfitplot2}
  \end{center}
  \end{fullpage}
\end{figure}

Our fitted hadronisation uncertainties for each observable are
illustrated in Figures~\ref{hadfitplot1} and~\ref{hadfitplot2}. The
fits do not describe the data well, because there are large
differences between hadronisation uncertainties from different
experiments. In particular, the measurements from L3 tend to suggest a
much steeper energy-dependence than those from other experiments; this
causes an apparent anomaly in the fit for \ytwothree, which is not
measured by L3. However, the fits do provide a satisfactory smoothing
algorithm to remove unwanted fluctuations between energies, so we will
adopt them in our covariance matrix~$V$.

At present, the hadronisation uncertainties on our \emph{combined}
results continue to be evaluated as described in
Section~\ref{newcombinederror}, by repeating the entire combination
for each generator. For the final LEP results, however, it is planned
that the smoothed hadronisation uncertainties should be used
throughout, both for the weights and for the combined uncertainty.


\section{Weight distributions}

\begin{figure}
\begin{center}
\includegraphics[width=0.90\textwidth]{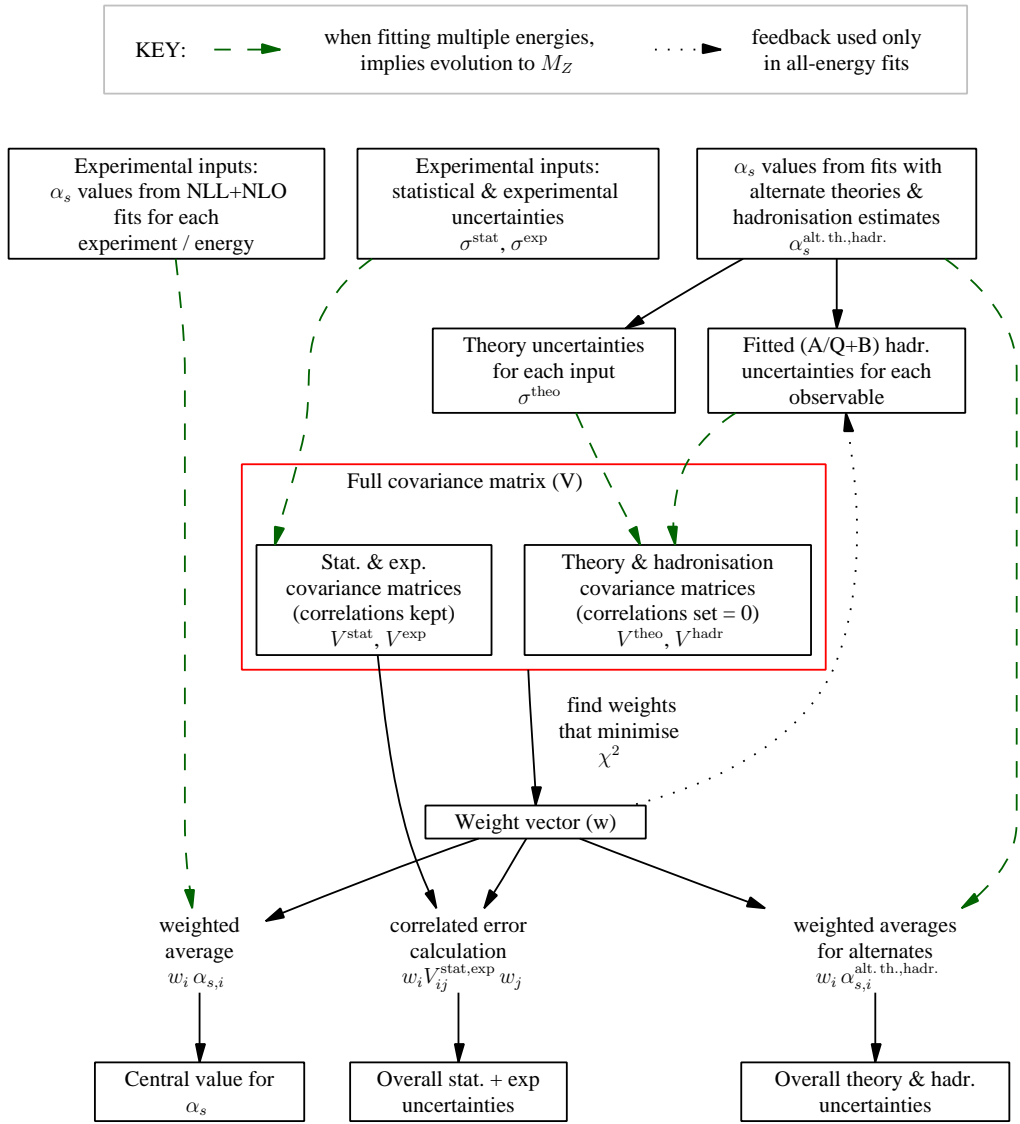}
\vspace{0.5cm}
\caption[Summary of the LEP \asmz\ combination method]{Summary of the
LEP \asmz\ combination method \emph{(Figure prepared by G.P.~Salam)}.}
\label{flow_salam}
\end{center}
\end{figure}

\begin{table}
\begin{center}
\begin{tabular}{|rr|cccc|}
\hline
\multicolumn{2}{|c|}{Covariance element} & Matrix 1 & Matrix 2 & Matrix 3 
& Matrix 4 \bigstrut\\
\hline \hline \bigstrut[t]
\multirow{2}[2]{*}{Statistical}   & {\small Diagonal} & \textbullet & \textbullet & \textbullet & \textbullet \\
                              & {\small Off-diagonal} & \textbullet &             & \textbullet & \textbullet
\bigstrut [b] \\ \hline \bigstrut[t]
\multirow{2}[2]{*}{Experimental}  & {\small Diagonal} & \textbullet & \textbullet & \textbullet & \textbullet \\
                              & {\small Off-diagonal} & \textbullet &             & \textbullet & \textbullet
\bigstrut [b] \\ \hline \bigstrut[t]
\multirow{2}[2]{*}{Hadronisation} & {\small Diagonal} & \textbullet & \textbullet &             & \textbullet \\
                              & {\small Off-diagonal} & \textbullet &             &             &
\bigstrut [b] \\ \hline \bigstrut[t]
\multirow{2}[2]{*}{Theory}        & {\small Diagonal} & \textbullet & \textbullet &             & \textbullet \\
                              & {\small Off-diagonal} & \textbullet &             &             & 
\bigstrut [b] \\ \hline
\end{tabular}
\caption{The diagonal and off-diagonal components of the four
alternative covariance matrices compared in
Table~\ref{tab:fitsbymethod} and in
Figures~\ref{meth1plots}--\ref{meth4plots}. Matrix~4 corresponds to
the covariances used in our final LEP \asmz\ combination.}
\label{weight_dist_covs}
\end{center}
\end{table}

\begin{figure}[tbp!]
\begin{center}
\includegraphics[width=\textwidth]{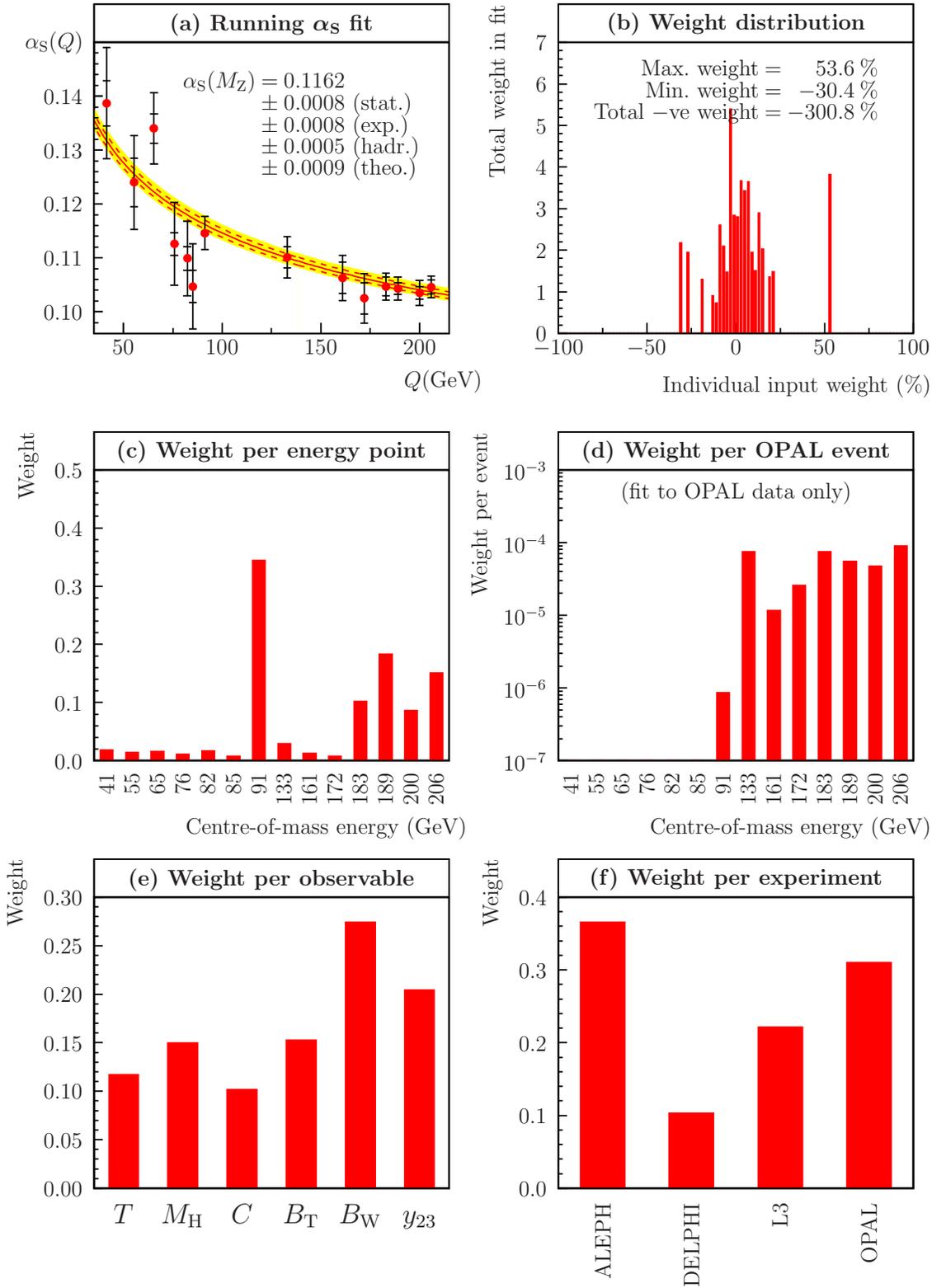}
\caption[Fit results and distributions of weights, using covariance
matrix~1]{Fit results and distributions of weights, using
\textbf{covariance~matrix~1}, the na\"{\i}ve method discussed in
Section~\ref{naivecomb}. In this method, all of the uncertainties and
correlations described in Section~\ref{fullcovmatrix} are included in
the covariance matrix.}
\label{meth1plots}
\end{center}
\end{figure}

\begin{figure}[tbp!]
\begin{center}
\includegraphics[width=\textwidth]{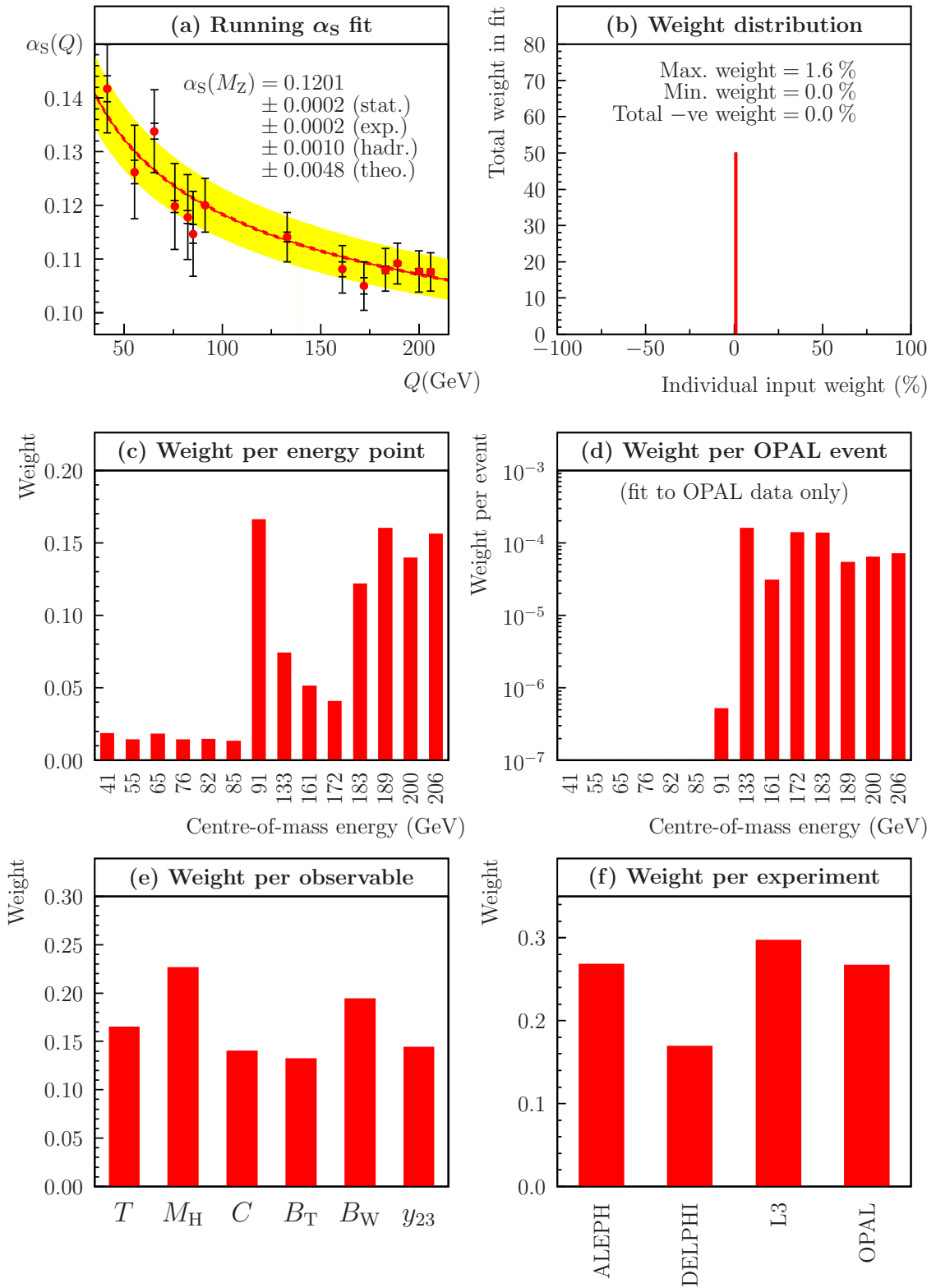}
\caption[Fit results and distributions of weights, using covariance
matrix~2]{Fit results and distributions of weights, using
\textbf{covariance~matrix~2}. In this method, all of the uncertainties
defined in Section~\ref{errorsubsect} are included in the covariance
matrix, but they are treated as uncorrelated.}
\label{meth2plots}
\end{center}
\end{figure}

\begin{figure}[tbp!]
\begin{center}
\includegraphics[width=\textwidth]{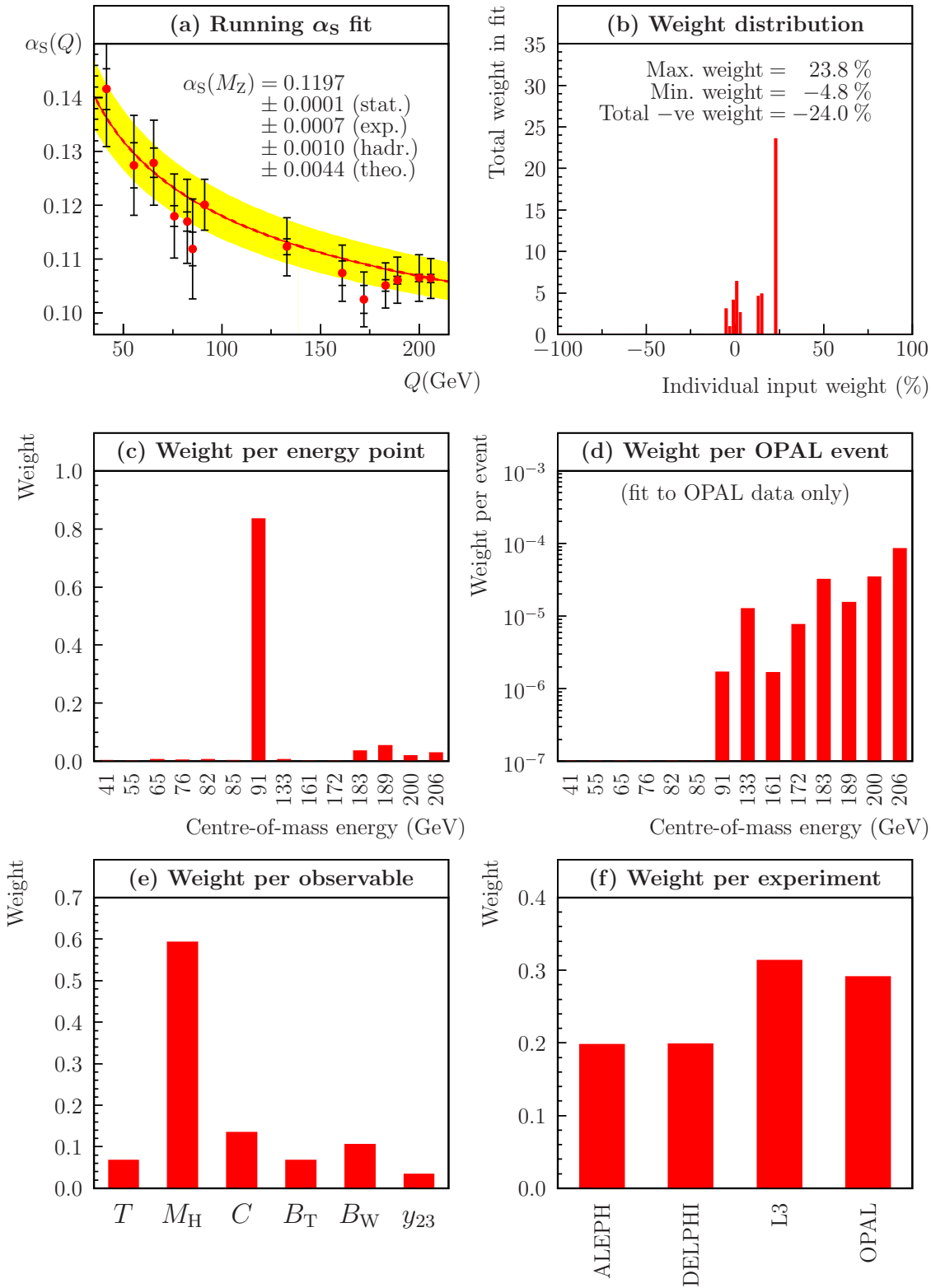}
\caption[Fit results and distributions of weights, using covariance matrix~3]{Fit results and distributions of weights, using
\textbf{covariance~matrix~3}. In this method, only statistical and
experimental uncertainties are included in the covariance matrix.
Correlations for these uncertainties are included, as defined in
Section~\ref{corrsubsect}.}
\label{meth3plots}
\end{center}
\end{figure}

\begin{figure}[tbp!]
\begin{center}
{\small \includegraphics[width=\textwidth]{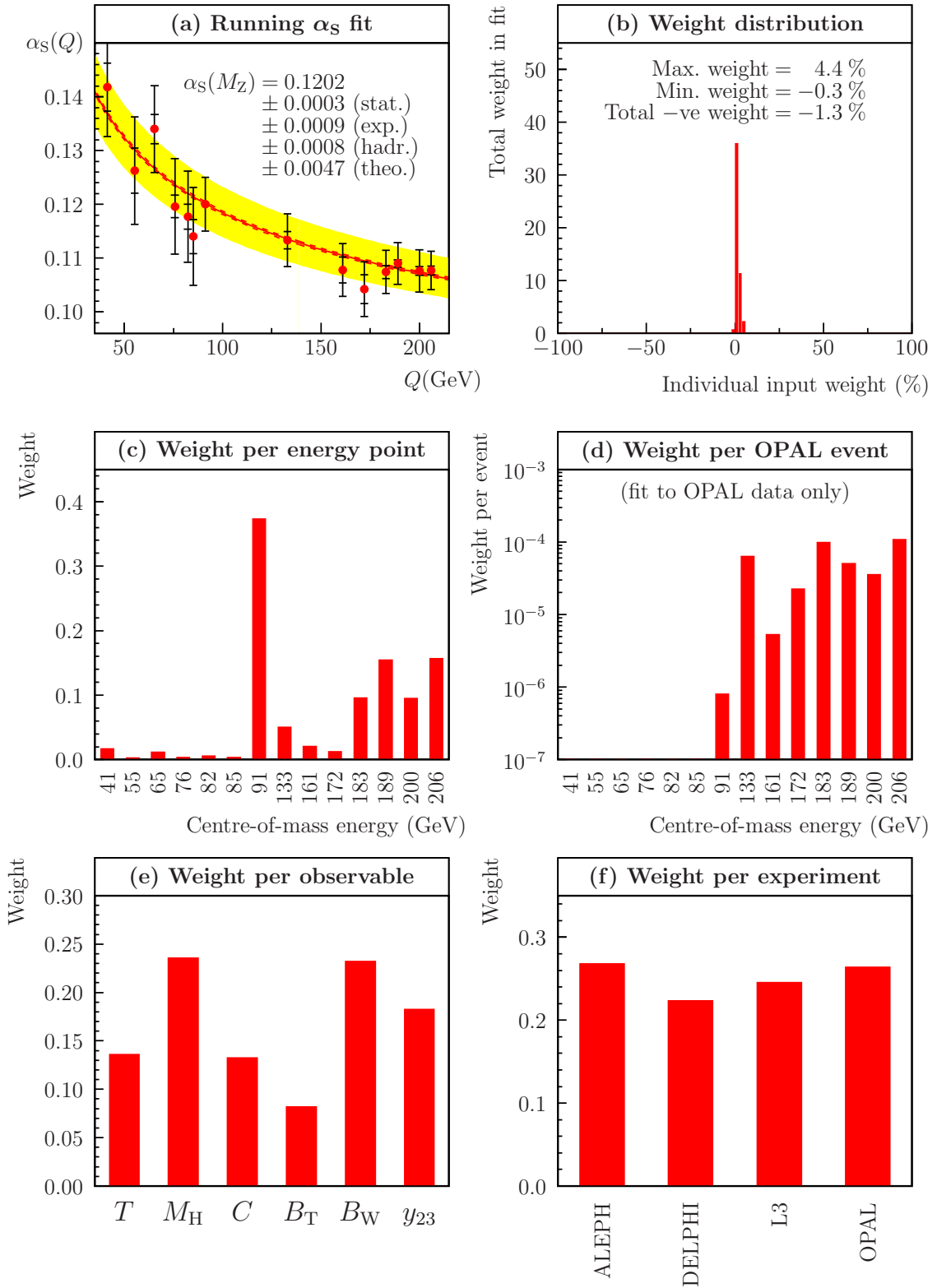}}
\caption[Fit results and distributions of weights, using covariance
matrix~4]{Fit results and distributions of weights, using
\textbf{covariance~matrix~4}, the method adopted for our final
combination.  In this method, all of the uncertainties defined in
Section~\ref{errorsubsect} are included in the diagonal terms of
the covariance matrix, but only the statistical and experimental
uncertainties are regarded as correlated.}
\label{meth4plots}
\end{center}
\end{figure}

\begin{figure}[tbp!]
\begin{center}
{\small \includegraphics[width=\textwidth]{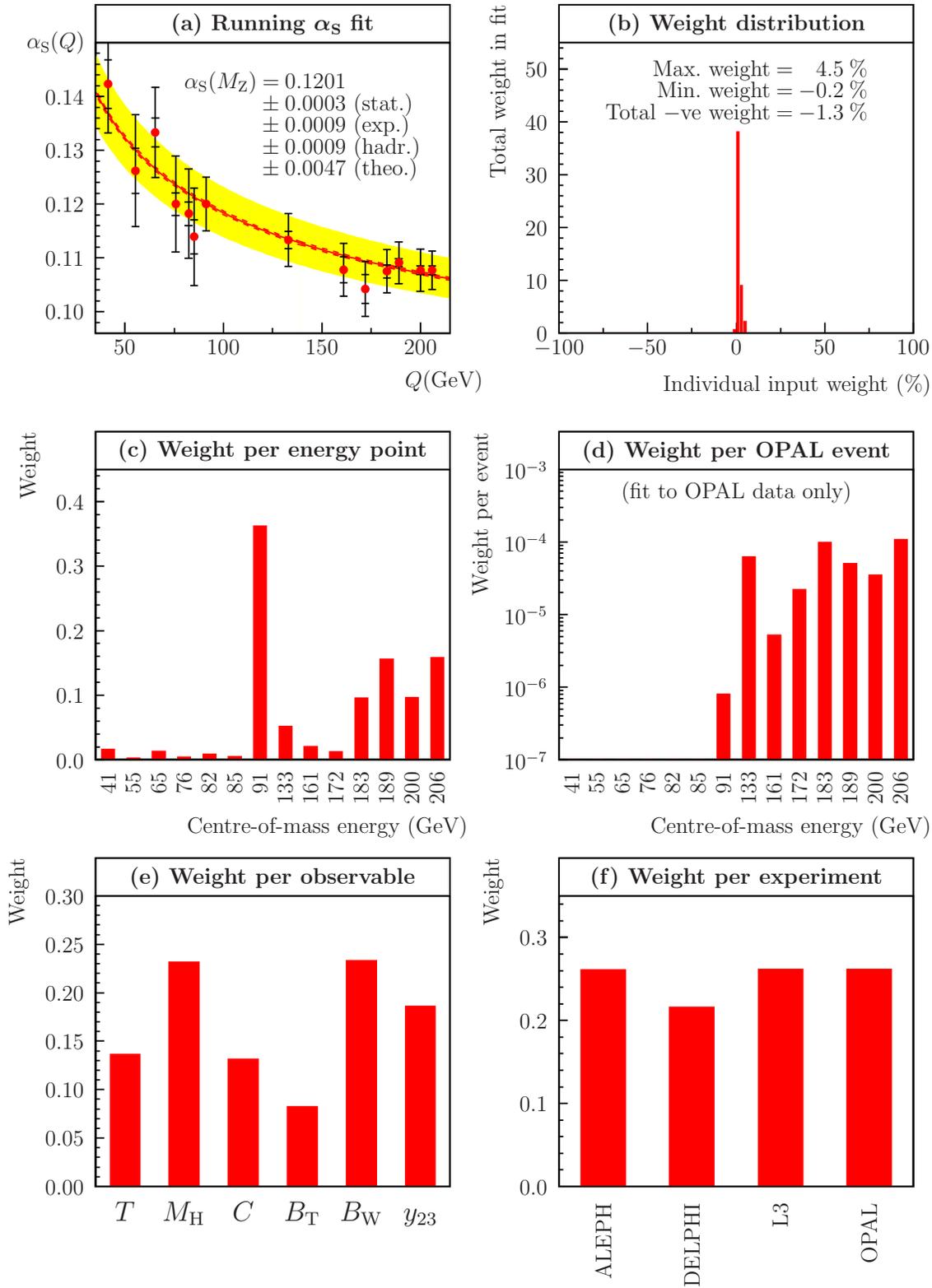}}
\caption[Fit results and distributions of weights, using covariance
matrix~4, after smoothing of hadronisation uncertainties]{Fit results
and distributions of weights, using \textbf{covariance~matrix~4},
after smoothing of the hadronisation uncertainties as described in
Section~\ref{hadfit}.}
\label{meth5plots}
\end{center}
\end{figure}

Our complete combination procedure is summarised in
Figure~\ref{flow_salam}.  Up to this point, our main criterion for
choosing the covariance matrix $V$ has been the prevention of negative
weights. Before presenting our final results, however, we will
investigate the distribution of weights allocated to the various input
measurements. We define four alternative covariance matrices, as follows:
\begin{description}
\item[Matrix 1:] The `na\"{\i}ve' covariance matrix, containing all
the uncertainties and correlations discussed in Section~\ref{fullcovmatrix}.
\item[Matrix 2:] A matrix containing all four types of uncertainty,
but regarding them as uncorrelated.
\item[Matrix 3:] A matrix containing only statistical and experimental
uncertainties, including their correlations.
\item[Matrix 4:] The matrix chosen for our LEP \asmz\ combination. All
uncertainties are included in the diagonal terms of the covariance
matrix, but only statistical and experimental uncertainties are
regarded as correlated.
\end{description}
The diagonal and off-diagonal parts of these matrices are summarised
in Table~\ref{weight_dist_covs}.  The resulting fits and weight
distributions are presented in Table~\ref{tab:fitsbymethod}, and in
Figures~\ref{meth1plots}--\ref{meth4plots}. We will discuss in turn
each of the plots~(\emph{a})--(\emph{f}) within these figures.
\begin{description}
\item[(\emph{a}) Running \boldmath\as\ fit:] This plot shows the
  combined \asq\ measurement at each energy point, and the QCD running
  curve predicted from a global \asmz\ fit at all energies. The inner
  error-bars and dotted red curves show the statistical uncertainty,
  while the outer bars and yellow band show the total uncertainty. The
  running curve describes the \asq\ points satisfactorily for all
  matrices except Matrix~3.
\item[(\emph{b}) Weight distribution:] This histogram shows the
  distribution of weights assigned to individual \as\
  measurements. The number of measurements in each bin is multiplied
  by the weight itself, so the histogram is automatically
  normalised. We require the contribution from negative weights to be
  negligible, and the size of each positive weight to be reasonably
  small. These conditions are satisfied only by matrices~2 and~4
  (negative weights cannot arise from Matrix~2, since there are no
  correlations).
\item[(\emph{c}) Weights per energy point:] This bar-chart shows the
  total weight assigned to measurements at each centre-of-mass
  energy. We expect a large weight for measurements at 91~GeV, which
  have very small statistical uncertainties, but also a significant
  weight for higher-energy LEP2 data, where the theory and hadronisation
  uncertainties are smaller. These conditions are satisfied best by
  Matrices 1 and~4. An uncorrelated fit (Matrix~2) does not give
  sufficient weight to LEP1 measurements, while a fit which omits
  theory and hadronisation uncertainties (Matrix~3) does not give
  sufficient weight to LEP2.
\item[(\emph{d}) Weights per OPAL event:] Here we present the weight
  for each centre-of-mass energy divided by the number of selected
  signal events used. Since this information is not readily available
  for other LEP experiments, the fit is restricted to OPAL data; as
  with all OPAL measurements used in this chapter, the results at
  $\sqrt{s}\leq 200$~GeV are based on previously published or approved
  distributions, and not on those presented in
  Chapter~\ref{opalchapter}. In a fit based entirely on statistical
  uncertainties, we would expect the weight per event to be roughly
  constant,\footnote{This assumes that the same set of observables is
  used at all energies. In fact, the $C$-parameter is omitted at
  91--161~GeV, which should reduce the weight at these energies by
  $\mathord{\sim} 20\%$.} except at 91~GeV, where Monte Carlo
  statistics form a significant contribution to the uncertainty.
\item[(\emph{e}) Weight per observable:] The perturbative predictions
  have different uncertainties, which should be reflected in the total
  weights assigned to each observable. Based on the theoretical
  uncertainty estimates discussed in
  Section~\ref{evsh_prediction_errors}, we would expect \ytwothree\ to
  carry the most weight, followed by~\MH. \BT\ should carry the least
  weight. This is confirmed in the fit using Matrix~4, except that the
  weight for \ytwothree\ is suppressed, because only two experiments
  have measured it.
\item[(\emph{f}) Weight per experiment:] We would expect the total weights for
  each experiment to be roughly equal, although differences will arise
  from the availability of input measurements. Once again, Matrix~4
  provides the most even distribution between the experiments.
\end{description}
In Figure \ref{meth5plots}, we repeat the combination with
Matrix~4, but apply the smoothing of hadronisation uncertainties
described in Section~\ref{hadfit}. This smoothing was not applied in
Figures~\ref{meth1plots}--\ref{meth4plots}. Comparing the fits and
weight distributions in Figures~\ref{meth4plots} and~\ref{meth5plots},
we find the effect of this smoothing to be negligible.

In conclusion, the distribution of weights derived from for our chosen
covariance matrix (Matrix~4) is satisfactory. All weights assigned to
individual input measurements are between $-0.5\%$ and $+4.5\%$, with
a total negative weight of only $-1.3\%$. Although a significant
weight (37\%) is allocated to measurements at 91~GeV, these are
complemented in the combination by measurements at higher energy
scales, which have smaller systematic uncertainties. The balance of
weights among the six observables reflects the relative theoretical
uncertainties and the number of available input measurements; all
observables contribute between 7\% and~24\% to the combination. The
weight distribution among the four experiments is roughly equal,
though the DELPHI Collaboration has a slightly smaller contribution
since they have measured \as\ neither from $y_{23}$ nor from radiative
($Q<91$~GeV) events.

\begin{table}[p]
\begin{center}
\scalebox{0.94}{\small
\begin{tabular}{r|r r r r r|}
 & \multicolumn{1}{|c}{\textbf{Matrix 1}}
 & \multicolumn{1}{c}{\textbf{Matrix 2}}
 & \multicolumn{1}{c}{\textbf{Matrix 3}}
 & \multicolumn{1}{c}{\textbf{Matrix 4}}
 & \multicolumn{1}{c|}{\textbf{Matrix 4}} \\
 & & & & & \vspace{-0.45cm} \\
 & \multicolumn{1}{|c}{}
 & \multicolumn{1}{c}{}
 & \multicolumn{1}{c}{}
 & \multicolumn{1}{c}{\parbox{1.7cm}{\centering\textbf
   {\scriptsize\boldmath without\vspace{-0.2cm}\\$A/Q+B$~fit}}}
 & \multicolumn{1}{c|}{\parbox{1.7cm}{\centering\textbf
   {\scriptsize\boldmath with\vspace{-0.2cm}\\$A/Q+B$~fit}}}

\bigstrut[b]
\\
\hline
\multicolumn{1}{l|}{\textbf{\scriptsize Fit results}}
 & & & & &  \bigstrut
\\
\bigstrut[t]

\bigstrut[b] $\alpha_\mathrm{s}(M_\mathrm{Z})$
 & $0.1162$
 & $0.1201$
 & $0.1197$
 & $0.1202$
 & $0.1201$
 \\ 
\hline \bigstrut Stat. error
 & $\pm 0.0008$
 & $\pm 0.0002$
 & $\pm 0.0001$
 & $\pm 0.0003$
 & $\pm 0.0003$
 \\ 
\hline \vspace{-0.15cm}{\scriptsize Exp. error}
 & {\scriptsize $\pm 0.0008$}
 & {\scriptsize $\pm 0.0002$}
 & {\scriptsize $\pm 0.0007$}
 & {\scriptsize $\pm 0.0009$}
 & {\scriptsize $\pm 0.0009$}
 \\ 
\vspace{-0.15cm}{\scriptsize Hadr. error}
 & {\scriptsize $\pm 0.0005$}
 & {\scriptsize $\pm 0.0010$}
 & {\scriptsize $\pm 0.0010$}
 & {\scriptsize $\pm 0.0008$}
 & {\scriptsize $\pm 0.0009$}
 \\
\hline
\end{tabular}
}
\end{center}

\caption{Combined \as\ fit results and 
weights for the four different correlation assumptions}
\label{tab:fitsbymethod}
\end{table}

\section[\asmz\ fit results]{\boldmath\asmz\ fit results}
\label{fitresults}

\begin{figure}[tbp!]
\begin{center}
\includegraphics[width=\textwidth]{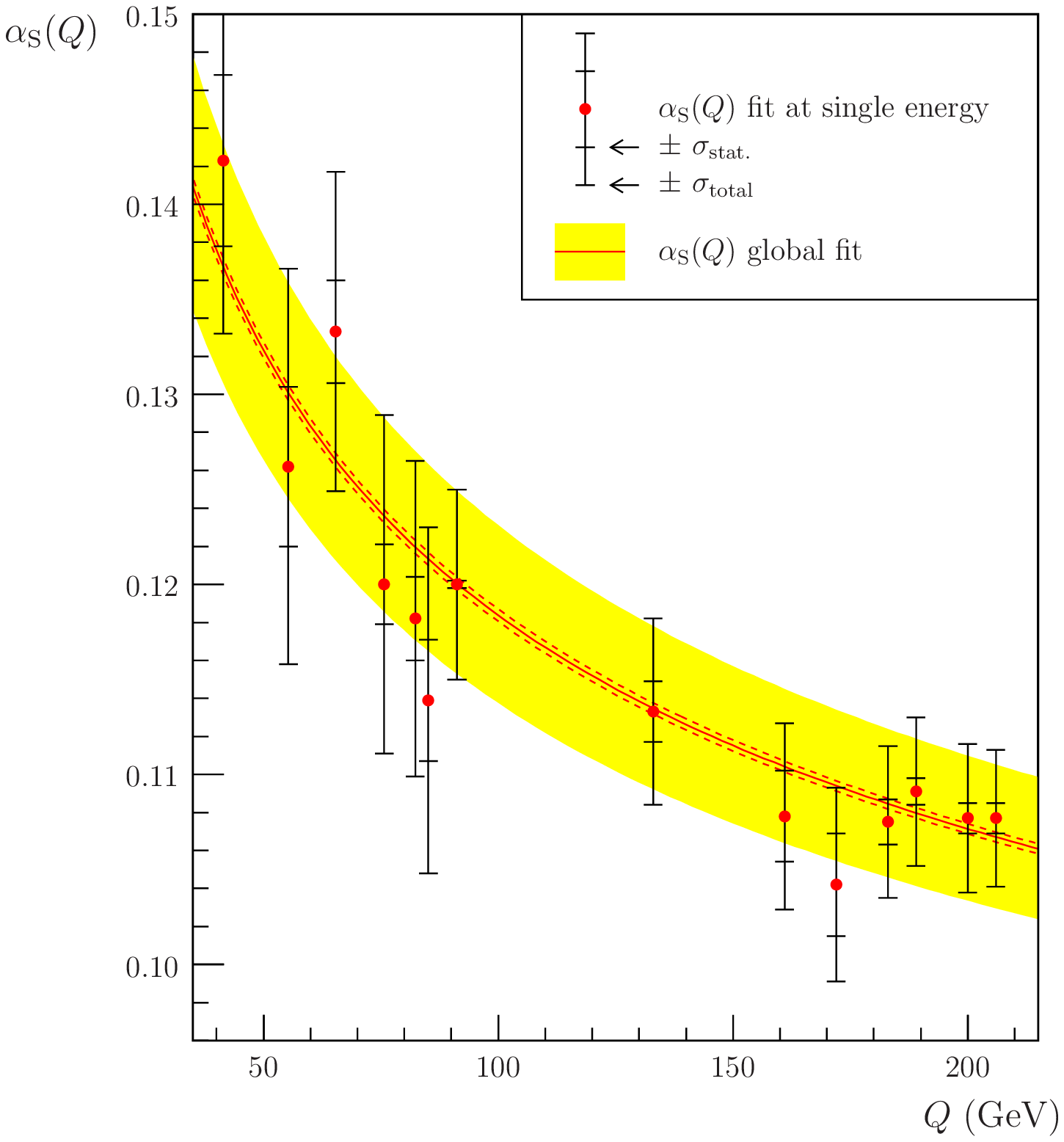}
\caption[A global QCD running fit to the LEP \as\ measurements]{A
  global QCD running fit to the LEP \as\ measurements. Each point
  represents a fit to the available measurements at an individual
  centre-of-mass energy, while the curve represents a global fit to
  all measurements. The form of the curve is determined by the
  Renormalisation Group Equation of QCD, with \asmz\ as a free parameter.
  The yellow band corresponds to the total uncertainty of
  the fitted \asmz\ value, and the dotted curves indicate the
  statistical uncertainty.
}
\label{runningplot}
\end{center}
\end{figure}

Our final combined measurement of \asmz, using all available
measurements, is:\\* {\small
\begin{displaymath}
\fbox{$\begin{array}{rcl}
\asmz & = & 0.1201
\;\pm\;0.0003~\mathrm{(stat.)}
\;\pm\;0.0009~\mathrm{(exp.)}
\;\pm\;0.0009~\mathrm{(hadr.)}
~\begin{array}{r}
+0.0046 \vspace{-0.3cm}\\
-0.0047
\end{array} \mathrm{(theo.)} \vspace{-0.2cm} \\
& = & 0.1201
\;\pm\;0.0003~\mathrm{(stat.)}
\;\pm\;0.0048~\mathrm{(syst.)} \\
& = & 0.1201
\;\pm\;0.0048~\mathrm{(total)}
\end{array}$}
\end{displaymath}}
\textcolor{white}{DUMMY}\\[-0.5cm]
The \asq\ results at individual energy points are presented in
Table~\ref{tab:alphasq}. These have been combined without conversion
to the $M_\mathrm{Z}$ scale. Figure~\ref{runningplot} illustrates these
fits, together with the \asq\ running curve predicted from the global
\asmz\ combination. 

\begin{figure}[tbp!]
\begin{center}
\includegraphics[width=0.8\textwidth]{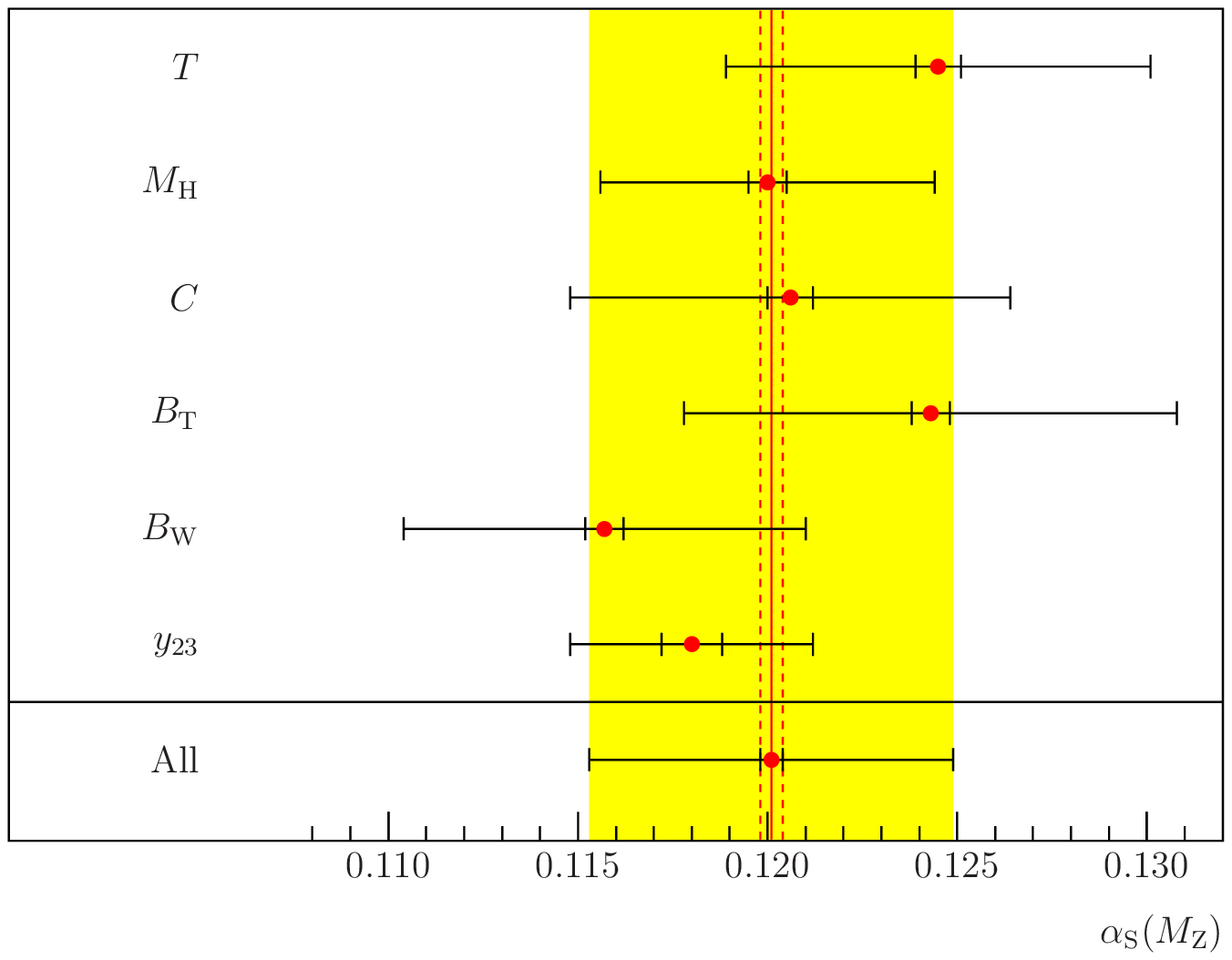}
\caption[LEP \asmz\ combinations for individual event shape
observables]{LEP \asmz\ combinations for individual event shape
observables. The inner error bars are statistical, while the outer
bars represent total uncertainties.}
\label{fitsbyvariableplot}
\end{center}
\end{figure}

\begin{figure}[tbp!]
\begin{center}
\includegraphics[width=0.8\textwidth]{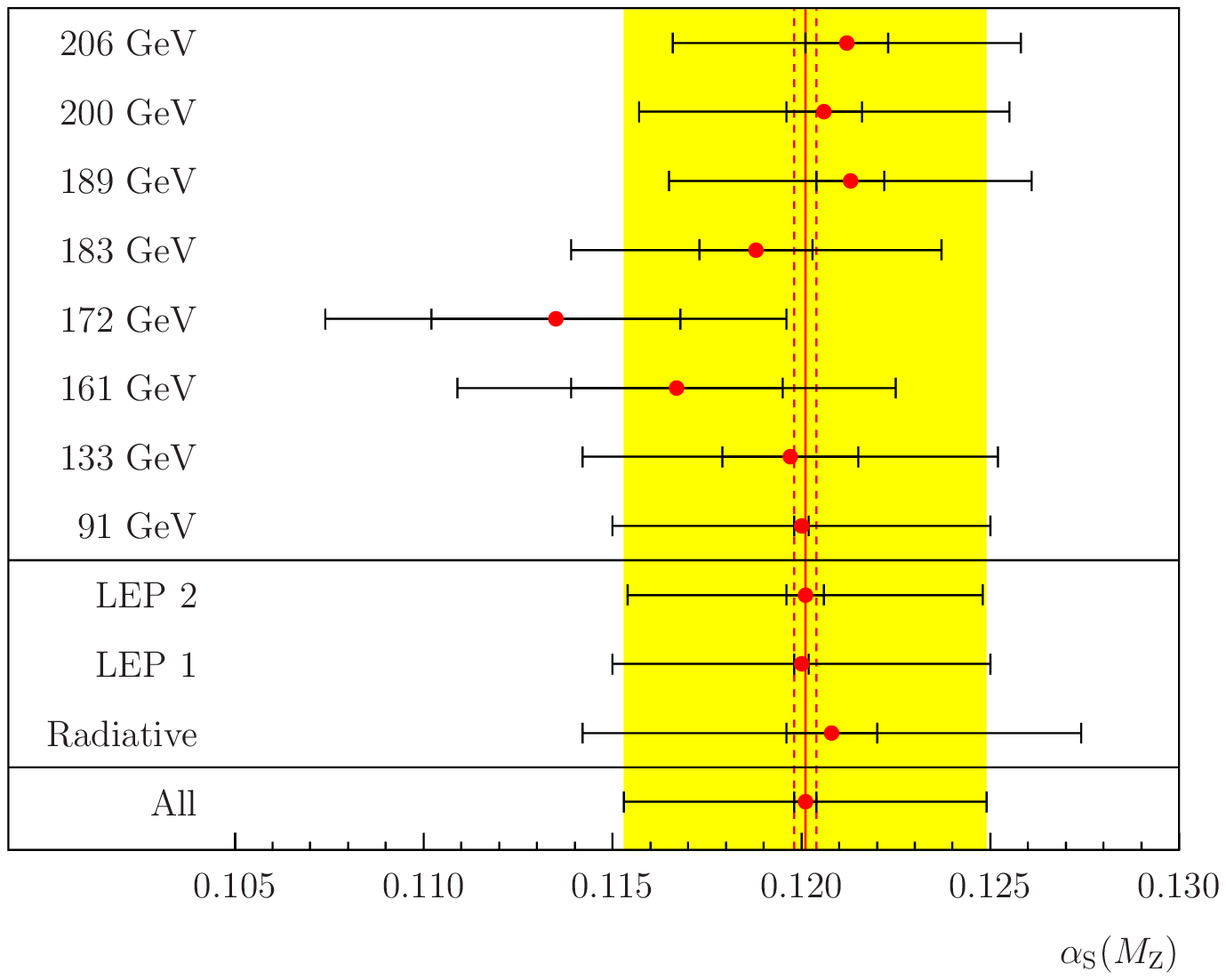}
\caption[LEP \asmz\ combinations for individual centre-of-mass
  energies]{LEP \asmz\ combinations for individual centre-of-mass
  energies. The inner error bars are statistical, while the outer bars
  represent total uncertainties.}
\label{fitsbyenergyplot}
\end{center}
\end{figure}

\begin{figure}[tbp!]
\begin{center}
\includegraphics[width=0.8\textwidth]{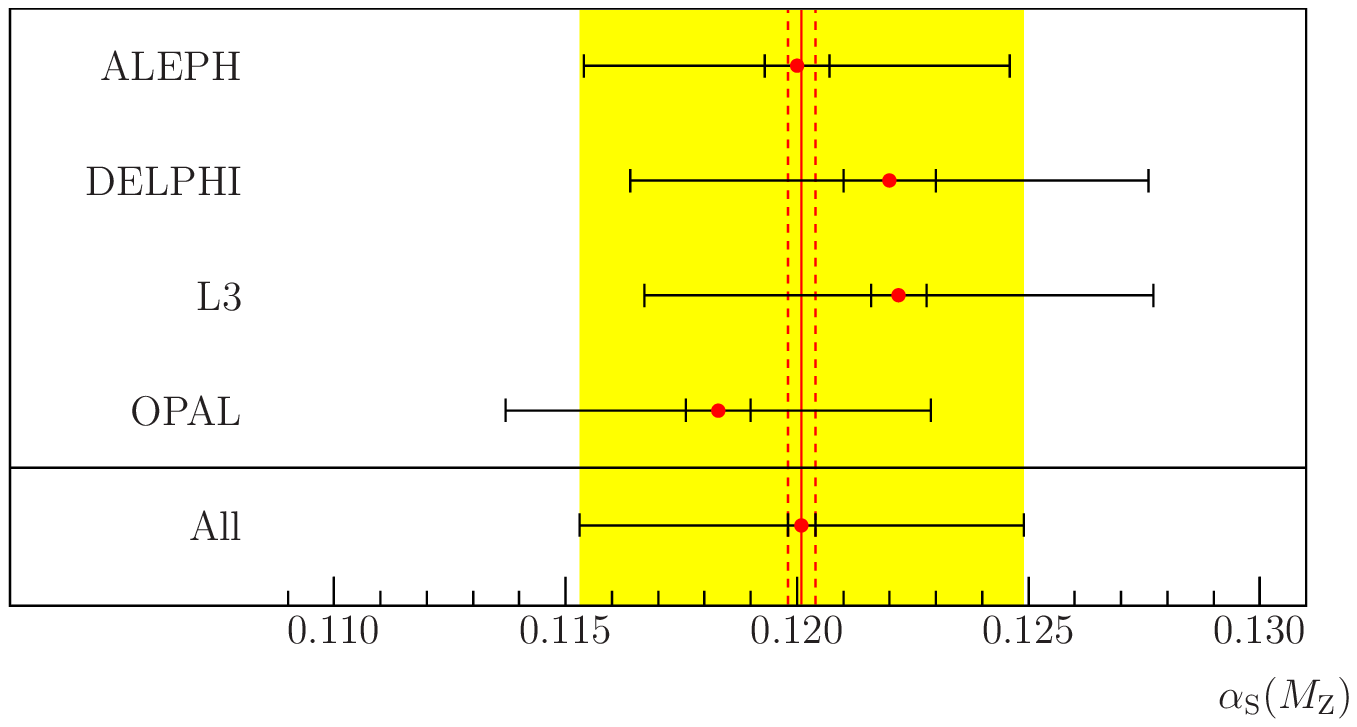}
\caption[LEP \asmz\ combinations for individual experiments]{LEP
  \asmz\ combinations for individual experiments. The inner error bars
  are statistical, while the outer bars represent total
  uncertainties.}
\label{fitsbyexptplot}
\end{center}
\end{figure}

In Tables~\ref{tab:fitsbyvariable}--\ref{tab:fitsbyexperiment} and
Figures~\ref{fitsbyvariableplot}--\ref{fitsbyexptplot}, we present
combinations of \asmz\ measurements based on individual observables,
energy ranges, and experiments. The breakdown of weights by energy, by
observable and by experiment is given for each combination. The value
of the minimised~$\chi^2$ is also quoted for each fit; however, this
is of limited use for judging the quality of the fit, since it is
calculated using an incomplete covariance matrix. In almost all cases,
the $\chi^2$ value is ``too~small'' for the number of degrees of
freedom, indicating that the spread of measurements is inconsistent
with the assumption of uncorrelated theory and hadronisation
uncertainties. This should not concern us, however, since the
correlations have been `re-included' in the uncertainties of our combined
results, as described in Section~\ref{newcombinederror}.

\begin{table}
\begin{center}
\begin{tabular}{|r|r r r r r|}
\hline
   \multicolumn{1}{|c|}{Energy scale\rule[-0.2cm]{0pt}{0pt}}
 & \multirow{2}[1]{*}[0pt]{$\alpha_\mathrm{s}(Q)$}
 & \multirow{2}[1]{*}[0pt]{$\sigma_{\text{stat.}}\phantom{I}$}
 & \multirow{2}[1]{*}[0pt]{$\sigma_{\text{exp.}}\phantom{I}$}
 & \multirow{2}[1]{*}[0pt]{$\sigma_{\text{hadr.}}\phantom{I}$}
 & \multirow{2}[1]{*}[0pt]{$\sigma_{\text{theo.}}\phantom{I}$}
\bigstrut[t] \\
 & & & & & \vspace{-0.6cm} \\
   \multicolumn{1}{|c|}{$Q$~(GeV)}
 &
 & & & & \bigstrut[b] \\
\hline \hline \bigstrut[t]

 & & & & & \vspace{-0.3cm} \\
$ 41.4\phantom{000}$ &
$\phantom{0}0.1423$ &
$\pm 0.0045 $ &
$\pm 0.0025 $ & 
$\pm 0.0015 $ &
${\scriptsize
\begin{array}{r}
+0.0076 \\
-0.0071
\end{array}}$\\ 

 & & & & & \vspace{-0.3cm} \\
$ 55.3\phantom{000}$ &
$\phantom{0}0.1262$ &
$\pm 0.0042 $ &
$\pm 0.0049 $ & 
$\pm 0.0049 $ &
${\scriptsize
\begin{array}{r}
+0.0067 \\
-0.0064
\end{array}}$\\ 

 & & & & & \vspace{-0.3cm} \\
\hline
\end{tabular}
\end{center}

\caption{Combined LEP \asq\ measurements at individual energy scales}
\label{tab:alphasq}
\end{table}

\begin{table}[p]
\begin{center}
\scalebox{0.94}{\small
\begin{tabular}{r|r r r r r r r|}
 & \multicolumn{1}{|c}{\boldmath $T$}
 & \multicolumn{1}{c}{\boldmath $M_\mathrm{H}$}
 & \multicolumn{1}{c}{\boldmath $C$}
 & \multicolumn{1}{c}{\boldmath $B_\mathrm{T}$}
 & \multicolumn{1}{c}{\boldmath $B_\mathrm{W}$}
 & \multicolumn{1}{c}{\boldmath $y_{23}$}
 & \multicolumn{1}{c|}{\multirow{2}[1]{*}{\textbf{All}}} \\
 & & & & & & & \vspace{-0.55cm} \\
 & \multicolumn{1}{|c}{\textbf{only}}
 & \multicolumn{1}{c}{\textbf{only}}
 & \multicolumn{1}{c}{\textbf{only}}
 & \multicolumn{1}{c}{\textbf{only}}
 & \multicolumn{1}{c}{\textbf{only}}
 & \multicolumn{1}{c}{\textbf{only}}
 &

\bigstrut[b]
\\
\hline
\multicolumn{1}{l|}{\textbf{\scriptsize Fit results}}
 & & & & & & &  \bigstrut
\\
\bigstrut[t]

\bigstrut[b] $\alpha_\mathrm{s}(M_\mathrm{Z})$
 & $0.1245$
 & $0.1200$
 & $0.1206$
 & $0.1243$
 & $0.1157$
 & $0.1180$
 & $0.1201$
 \\ 
\hline \bigstrut Stat. error
 & $\pm 0.0006$
 & $\pm 0.0005$
 & $\pm 0.0006$
 & $\pm 0.0005$
 & $\pm 0.0005$
 & $\pm 0.0008$
 & $\pm 0.0003$
 \\ 
\hline \vspace{-0.15cm}{\scriptsize Exp. error}
 & {\scriptsize $\pm 0.0011$}
 & {\scriptsize $\pm 0.0010$}
 & {\scriptsize $\pm 0.0011$}
 & {\scriptsize $\pm 0.0012$}
 & {\scriptsize $\pm 0.0010$}
 & {\scriptsize $\pm 0.0014$}
 & {\scriptsize $\pm 0.0009$}
 \\ 
\vspace{-0.15cm}{\scriptsize Hadr. error}
 & {\scriptsize $\pm 0.0015$}
 & {\scriptsize $\pm 0.0008$}
 & {\scriptsize $\pm 0.0018$}
 & {\scriptsize $\pm 0.0013$}
 & {\scriptsize $\pm 0.0007$}
 & {\scriptsize $\pm 0.0004$}
 & {\scriptsize $\pm 0.0009$}
 \\
\hline
\end{tabular}
}
\end{center}

\caption{Combined \as\ fit results for 
different observables}
\label{tab:fitsbyvariable}
\end{table}

\begin{table}[p]
\begin{center}
\scalebox{0.94}{\small
\begin{tabular}{r|r r r r|}
 & \multicolumn{1}{|c}{\textbf{Radiative}}
 & \multicolumn{1}{c}{\textbf{LEP1}}
 & \multicolumn{1}{c}{\textbf{LEP2}}
 & \multicolumn{1}{c|}{\multirow{2}[1]{*}{\textbf{All}}} \\
 & & & & \vspace{-0.55cm} \\
 & \multicolumn{1}{|c}{\scriptsize \boldmath $(Q<M_\mathrm{Z})$}
 & \multicolumn{1}{c}{\scriptsize \boldmath $(Q=M_\mathrm{Z})$}
 & \multicolumn{1}{c}{\scriptsize \boldmath $(Q>M_\mathrm{Z})$}
 &

\bigstrut[b]
\\
\hline
\multicolumn{1}{l|}{\textbf{\scriptsize Fit results}}
 & & & &  \bigstrut
\\
\bigstrut[t]

\bigstrut[b] $\alpha_\mathrm{s}(M_\mathrm{Z})$
 & $0.1208$
 & $0.1200$
 & $0.1201$
 & $0.1201$
 \\ 
\hline \bigstrut Stat. error
 & $\pm 0.0012$
 & $\pm 0.0002$
 & $\pm 0.0005$
 & $\pm 0.0003$
 \\ 
\hline \vspace{-0.15cm}{\scriptsize Exp. error}
 & {\scriptsize $\pm 0.0023$}
 & {\scriptsize $\pm 0.0008$}
 & {\scriptsize $\pm 0.0010$}
 & {\scriptsize $\pm 0.0009$}
 \\ 
\vspace{-0.15cm}{\scriptsize Hadr. error}
 & {\scriptsize $\pm 0.0032$}
 & {\scriptsize $\pm 0.0010$}
 & {\scriptsize $\pm 0.0007$}
 & {\scriptsize $\pm 0.0009$}
 \\
\hline
\end{tabular}
}
\end{center}

\caption{\as\ fit results for different 
\LEP\ energy ranges}
\label{tab:fitsbyenergy}
\end{table}

\begin{table}[p]
\begin{center}
\scalebox{0.94}{\small
\begin{tabular}{r|r r r r r|}
 & \multicolumn{1}{|c}{\textbf{ALEPH}}
 & \multicolumn{1}{c}{\textbf{DELPHI}}
 & \multicolumn{1}{c}{\textbf{L3}}
 & \multicolumn{1}{c}{\textbf{OPAL}}
 & \multicolumn{1}{c|}{\multirow{2}[1]{*}{\textbf{All}}} \\
 & & & & & \vspace{-0.55cm} \\
 & \multicolumn{1}{|c}{\textbf{only}}
 & \multicolumn{1}{c}{\textbf{only}}
 & \multicolumn{1}{c}{\textbf{only}}
 & \multicolumn{1}{c}{\textbf{only}}
 &

\bigstrut[b]
\\
\hline
\multicolumn{1}{l|}{\textbf{\scriptsize Fit results}}
 & & & & &  \bigstrut
\\
\bigstrut[t]

\bigstrut[b] $\alpha_\mathrm{s}(M_\mathrm{Z})$
 & $0.1200$
 & $0.1220$
 & $0.1222$
 & $0.1183$
 & $0.1201$
 \\ 
\hline \bigstrut Stat. error
 & $\pm 0.0007$
 & $\pm 0.0010$
 & $\pm 0.0006$
 & $\pm 0.0007$
 & $\pm 0.0003$
 \\ 
\hline \vspace{-0.15cm}{\scriptsize Exp. error}
 & {\scriptsize $\pm 0.0008$}
 & {\scriptsize $\pm 0.0016$}
 & {\scriptsize $\pm 0.0014$}
 & {\scriptsize $\pm 0.0020$}
 & {\scriptsize $\pm 0.0009$}
 \\ 
\vspace{-0.15cm}{\scriptsize Hadr. error}
 & {\scriptsize $\pm 0.0005$}
 & {\scriptsize $\pm 0.0008$}
 & {\scriptsize $\pm 0.0018$}
 & {\scriptsize $\pm 0.0005$}
 & {\scriptsize $\pm 0.0009$}
 \\
\hline
\end{tabular}
}
\end{center}

\caption{Combined \as\ fit results for 
different experiments}
\label{tab:fitsbyexperiment}
\end{table}

All of the partial combinations are consistent with the global
combination, within the appropriate uncertainties. However, it should
be noted that some have smaller total uncertainties than that of the
global combination. For example, when we fit LEP1 and
LEP2\footnote{LEP2 includes all centre-of-mass energies $\sqrt{s} \geq 133$~GeV
in this context.} data alone, we obtain the following results:\\*[-1.2cm]
\begin{center}
{\small 
\scalebox{0.95}{$\begin{array}{rcl}
\mathrm{LEP1:~} \asmz &=&0.1200
\pm 0.0002~\mathrm{(stat.)}
\pm 0.0008~\mathrm{(exp.)}
\pm 0.0010~\mathrm{(hadr.)}
~\begin{array}{r}
+0.0048 \vspace{-0.3cm}\\
-0.0048
\end{array} \mathrm{(theo.)} \\
\mathrm{LEP2:~} \asmz &=&0.1201
\pm 0.0005~\mathrm{(stat.)}
\pm 0.0010~\mathrm{(exp.)}
\pm 0.0007~\mathrm{(hadr.)}
~\begin{array}{r}
+0.0044 \vspace{-0.3cm}\\
-0.0045
\end{array} \mathrm{(theo.)} \end{array}$}} \\
\end{center}
\textcolor{white}{DUMMY}\\[-0.5cm] The combined uncertainty for LEP2
measurements alone is less than that for the all-energies
combination. Similarly the combinations of ALEPH and OPAL results, and
also those using the observables \MH\ and \ytwothree\ alone, have
smaller uncertainties than the global LEP combination. Ordinarily this
situation should not occur, since the minimisation of $\chi^2$ should
automatically select the same weights which minimise the total
uncertainty: a proof of this result is given in
Appendix~\ref{appsec:errormin}. As we have seen, however, the choice
of weights which minimises the `true' $\chi^2$ is not acceptable,
since the result depends strongly on our subjective estimates of the
uncertainties and correlation coefficients. Also, many of the weights
in such a combination are negative. We have therefore compromised by
developing an alternative algorithm, which gives more reasonable
weights, but does not minimise the uncertainty.  It is debatable
whether one should regard the global combination as our `best'
measurement of \asmz, when several of the partial combinations (and
indeed several individual input measurements) have smaller
uncertainties. However, we consider the global combination to be
robust, since it takes input from a variety of observables, and does
not rely excessively on any single assumption used in the construction
of the covariance matrix. Although the `uncertainty band' method
presented in Section~\ref{evsh_prediction_errors} represents an
improvement over previous theoretical uncertainty estimates based
exclusively on variation of the renormalisation scale~$\mu$, it cannot
guarantee to reflect the true magnitude of the uncalculated terms in the
perturbative predictions. Similarly, a high degree of arbitrariness
remains in our estimation of the hadronisation uncertainties. It would
therefore be inappropriate to focus \emph{exclusively} on the
measurements or observables which return the smallest uncertainties;
instead we have performed a global combination, giving higher weight to the
measurements which appear most precise.

The fits for individual observables, presented in
Table~\ref{tab:fitsbyvariable} and Figure~\ref{fitsbyvariableplot},
provide an interesting test for our theory uncertainties. The spread
of \asmz\ values among these six fits suggests that the correlation of
theory uncertainties between observables may be lower than our
previous crude estimate in
Section~\ref{subsubsect_theocorr}.\footnote{Alternatively, the
theory uncertainty itself could have been drastically underestimated. This is
unlikely, since all our results are in good agreement with independent
measurements of \asmz\ from other sources.} It also indicates that the
six central values agree with one another, within their total
uncertainties. One could argue that the theory uncertainty for our
global combination is too conservative, as it assumes 100\%
correlation of the theory uncertainty between observables. However, in
the absence of firm evidence to the contrary, we continue to apply the
method described in Section~\ref{newcombinederror}; to do otherwise
would risk underestimating the dominant source of uncertainty in our
combination.

\chapter{Summary and outlook}
\label{conclusionchapter}

In this work, we have presented the culmination of a series of event
shape measurements published by OPAL and the other LEP
Collaborations.

Using a consistent set of theoretical predictions and experimental
methods, we have re-analysed samples of
$\epem\to\text{Z}^0/\gamma\to\text{q}\bar{\text{q}}$ events collected
with the OPAL detector in the energy range $\sqrt{s}=91$--202~GeV, and
have performed new measurements of event shape distributions and of
the strong coupling at the highest LEP energies, up to 209~GeV. We
have presented distributions of fourteen observables, including three
`four-jet' quantities which have not previously been measured from
OPAL data.

Our measurements use a new event selection based on four-fermion
likelihood variables, which have not been applied in previous OPAL
event shape studies; the overall purities of our
$\text{q}\bar{\text{q}}$ multihadronic event samples are predicted to
be 94\% at the highest collision energies. We have studied the
efficiency and purity of this selection as functions of the event
shape observables, and have also investigated the bin-to-bin response
matrix of the detector. We conclude that the unfolding methods and fit
ranges used in previous OPAL measurements are suitable for our own
analysis.

A comparison was made between the measured event shape distributions
and those predicted by Monte Carlo models. Significant discrepancies
were found at $\sqrt{s}=91$~GeV, where the measurements are
statistically most precise; the modelling of observables sensitive to
three-jet events, however, is superior to that for four-jet
quantities. For the three-jet observables used in our \as\
measurements, the modelling is satisfactory within the appropriate fit
ranges: PYTHIA and ARIADNE generally give the best description. We
therefore have reason to trust the hadronisation corrections applied
to perturbative predictions when measuring \as. At energies above
$\sqrt{s}=91$~GeV, we do not have sufficient events to detect a
deviation between the OPAL measurements and Monte Carlo predictions.

Fits have been performed to determine the strong coupling over the
full range of LEP energy scales, based on the same six event shape
observables used in previous OPAL publications. Our results strongly
disfavour a constant \as\ over this energy range, and are compatible
with the `running' predicted by perturbative QCD. Combining OPAL
measurements from all observables and energy scales, we obtain the
following weighted average of \as\ at the Z$^0$ mass scale:
\[
\asmz = 0.1189 \pm 0.0005\;\mathrm{(stat.)}\pm0.0041\;\mathrm{(syst.)} \;\;\;.
\]
A new method was investigated to determine the statistical
uncertainties of the event shape distributions and \as\ measurements,
using an exact analytical form for the covariance matrix; results were
found to be compatible with previous estimates based on simulated data
samples. The estimation of theoretical uncertainties in the fits has
also been improved: an ``uncertainty band'' algorithm, which
considers the variation of several arbitrary parameters in the theory,
has been introduced in collaboration with the LEP QCD working group.

We have developed a successful method to combine \as\ measurements
obtained using different observables, experiments and energy
scales. Detailed comparisons were made between the results and
analysis methods of the four LEP Collaborations, to ensure consistent
implementation of the perturbative theory predictions and fitting
algorithms: the agreement was initially unsatisfactory, but is now
almost perfect. Estimates were then made for the covariance matrix
relating uncertainties of different measurements. The four
contributions to the covariance (statistical, experimental,
hadronisation and theory uncertainties) were treated separately,
including correlations between all possible pairs of measurements.

Initial attempts to form a weighted mean of \as\ measurements were
unreliable, due to the appearance of negative weights in the
combination. We have demonstrated that the negative weights arise from
the presence of large correlated uncertainties with differing
magnitudes in the covariance matrix. By removing the larger correlated
uncertainties from the off-diagonal elements of the matrix, we were
able to form a stable combination with reasonable weights. Our final
combined result, using all available measurements from event shape
observables at LEP, is
\[
\asmz = 0.1201 \pm 0.0003\;\mathrm{(stat.)}\pm0.0048\;\mathrm{(syst.)} \;\;\;.
\]
We have also performed combinations of certain subsets of
measurements, such as those for individual energies, observables or
experiments. These results are all in good agreement with the global
combination, and with the predicted running of \as\ with energy. Some
of the partial combinations, however, such as the OPAL result, have a
smaller uncertainty than that of the complete LEP combination: this is
a consequence of the incomplete correlations used in the covariance
matrix.

The dominant uncertainty in our combined \as\ measurement, and in many
of the original input measurements, arises from the higher-order terms
missing from the perturbative calculations; at $\sqrt{s}=91$~GeV, the
measured event shape distributions are sufficiently precise to observe
significant deviations from the fitted
$\mathcal{O}(\as^2)+\text{NLLA}$ theory predictions. When the next
generation of QCD calculations become available, at
$\mathcal{O}(\as^3)$ and next-to-next-to-leading logarithmic order, a
more precise measurement of \asmz\ will be possible using the existing
LEP data. It is also likely that the data at 91~GeV will permit a
test of the higher-order terms.

As we discussed in Section~\ref{nlla_advances}, several new
resummations have recently become available at NLLA precision,
including the distributions of the light jet mass, narrow jet
broadening and $D$-parameter. Although we have not performed fits to
these `four-jet' observables ourselves, it is planned that OPAL data
will be used to test these predictions, and to extract a corresponding
measurement of \as.

It is expected that the measurements presented in this
dissertation will form the basis for two publications in the near
future, containing the final measurement of \as\ from event shapes at
OPAL, and the final combination of published \as~results from the four
LEP Collaborations. A small number of improvements may be made to the
analysis presented here, prior to publication.

\appendix
\chapter[Explanations of statistical results]{Explanations and proofs of some statistical results}
\label{statappendix}

In this appendix, we shall elaborate on some of the statistical
methods applied in our analysis. With the possible exception of
Section~\ref{multinomial}, the results presented here are standard,
and can be can be found in textbooks such as Ref.~\cite{cowanbook}.

\section[The covariance matrix for a normalised 
histogram]
{\setstretch{1}The statistical covariance matrix for a\\corrected and
normalised histogram}
\label{multinomial}

In Chapter~\ref{opalchapter}, we described measurements of \asq\
obtained by fitting theoretical predictions to several event shape
distributions constructed from OPAL data. The best fit between the
prediction and the data was determined by minimisation of a $\chi^2$
variable, based only on statistical uncertainties.

When we construct a `raw' histogram from the events seen in the
detector, without normalisation, the statistical uncertainties are
straightforward: the number of events, $N_i$, in each bin conforms to a
Poisson distribution, and there is no correlation between
bins.\footnote{There may be `migrations' between bins, as discussed
in Section~\ref{detsimresults}, but these only introduce
\emph{systematic} correlations.} Before making a fit, however, we
apply three modifications to our data:
\begin{enumerate}
\item We subtract the expected background $\beta_i$ from each bin:
\begin{equation}
N_i \;\to\; N_i-\beta_i
\end{equation}
\item We multiply the remaining events by a ``detector correction,''
$\alpha_i$:
\begin{equation}
N_i-\beta_i \;\;\to\;\; \widetilde{N}_i=\alpha_i\left(N_i-\beta_i\right)
\end{equation}
\item Finally we normalise the distribution, to obtain a measure $P_i$
of the probability for a given event to occur in bin $i$:
\begin{equation}
\label{P_idef}
P_i\;=\;\frac{\widetilde{N}_i}{\sum_k \widetilde{N}_k}
\end{equation}
In practice we also divide the contents of each bin by the
width~$\Delta y_i$, to measure the probability density function. This
factor will be omitted here, and can easily be applied to each element
of our final covariance matrix.
\end{enumerate}
In the final normalisation step above, we have introduced correlations
between bins, since the contents of each bin now appears in the
denominator of every other. For a trivial case of only two bins, their
correlation coefficient $\rho$ would be $-1$, because an upward fluctuation
in one bin must be precisely compensated by a downward fluctuation in
the other. Here we derive the general covariance matrix
$V_{ij}=\mathrm{Cov}\left[P_i,P_j\right]$, which is used in
Section~\ref{asfits} to perform a least-squares fit to the
distribution, and to determine the corresponding statistical
uncertainty. The uncertainty can also be measured numerically by a
Monte Carlo method, as we discussed in Section~\ref{subsamples}.

The covariance matrix for the corrected bin contents
$\widetilde{N}_i$ is given by
\begin{equation}
\mathrm{Cov}\left[\widetilde{N}_i,\widetilde{N}_j\right]\;=\;
\alpha_i \alpha_j \; \mathrm{Cov}\left[N_i,N_j\right]\;=\;\left\{
\begin{array}{ll}
\alpha_i^2 N_i \hspace{0.5cm} & i=j \\
0   & i \neq j \;\;\;.
\end{array}
\right.
\end{equation}
We can propagate these uncertainties into a covariance matrix for
$P_i$, as follows:
\begin{eqnarray}
V_{ij}\;=\;\mathrm{Cov}\left[P_i,P_j\right] & = &
\sum_{k,l}\;
\frac{\partial P_i}{\partial \widetilde{N}_k}
\frac{\partial P_j}{\partial \widetilde{N}_l}
\;\mathrm{Cov}\left[\widetilde{N}_k,\widetilde{N}_l\right] \nonumber \\
& = & \sum_k \;
\frac{\partial P_i}{\partial \widetilde{N}_k}
\frac{\partial P_j}{\partial \widetilde{N}_k}
\;\alpha_k^2 N_k \;\;\;.
\end{eqnarray}
The partial derivatives here are obtained directly from
Equation~(\ref{P_idef}):
\begin{equation}
\frac{\partial P_i}{\partial \widetilde{N}_k}
\;=\;\left\{\renewcommand\arraystretch{1.3}
\begin{array}{rcl}\displaystyle
\frac{\sum_{l \neq i} \widetilde{N}_l}{\left[\sum_l \widetilde{N}_l\right]^2}
& \hspace{1cm} & i=k \rule[-1.1cm]{0pt}{0pt} \\
\displaystyle
-\frac{\widetilde{N}_i}{\left[\sum_l \widetilde{N}_l\right]^2}
& & i \neq k \;\;\;,
\end{array}
\right.
\end{equation}
leading to the following expressions for the covariance matrix $V_{ij}$:
\begin{equation}
V_{ij}\;\;=\;\;\left\{
\begin{array}{ll}\displaystyle
\frac{
\alpha_i^2 N_i {\left[\sum_{k \neq i} \widetilde{N}_k \right]}^2
+\;\left[\sum_{k \neq i} \alpha_k^2 N_k \right] {\widetilde{N}_i}^2}
{\left[\sum_k \widetilde{N}_k\right]^4}
& i=j \rule[-1.1cm]{0pt}{0pt}\\
\displaystyle
\frac{
\left[\sum_{k \neq i,j} \alpha_k^2 N_k \right] \widetilde{N}_i \widetilde{N}_j
- \alpha_i^2 N_i \widetilde{N}_j \left[\sum_{k \neq i} \widetilde{N}_k \right]
- \alpha_j^2 N_j \widetilde{N}_i \left[\sum_{k \neq j} \widetilde{N}_k \right]}
{\left[\sum_k \widetilde{N}_k\right]^4}
& i \neq j
\end{array}
\right.
\label{cov_unfoldedhistogram}
\end{equation}
In the case where no background subtraction or detector correction
takes place, namely $\alpha_i=1$ and $\widetilde{N}_i=N_i$, the above
formulae simplify to
\begin{equation}
V_{ij} \;\; = \;\; \left\{
\begin{array}{lcll}\displaystyle
\frac{N_i \left[\sum_{k \neq i} N_k\right]}
{\Big[\sum_k N_k\Big]^3}
& \displaystyle = & \displaystyle \frac{P_i \left(1-P_i\right)}{\sum_k N_k}
\hspace{1.5cm}
& i=j \rule[-1.1cm]{0pt}{0pt} \\
\displaystyle
-\frac{N_i N_j}
{\Big[\sum_k N_k\Big]^3}
& \displaystyle = & \displaystyle -\frac{P_i P_j}{\sum_k N_k}
& i \neq j \;\;\;\;\;.
\end{array}
\right.
\end{equation}
This is the standard covariance matrix for a normalised multinomial
distribution (or binomial distribution, in the case of the
diagonal elements).

In practice, we calculate the covariance matrix using the
\emph{expected} numbers of events $N_i$ and $\widetilde{N}_i$,
estimated from Monte Carlo samples, and not from the actual
data. Using real data would allow a downward fluctuation in the
observed number of events $N_i$ to reduce the corresponding
uncertainty, as described in Section~\ref{opal_as_fit_results}.

It should be emphasised that it is \emph{not} correct, as is sometimes
suggested, to assume a multinomial covariance matrix for the bins
$N_i$ at the detector level and then propagate this matrix to the
normalised hadron level. This approach would give covariances for a
distribution which has been normalised at the detector level, and is
no longer normalised after unfolding. The statistical properties of
the normalisation factor, which is responsible for the bin-to-bin
correlations, can differ substantially between the detector and hadron
levels. At the detector level, the bin contents are Poisson
distributions, so the bins with the largest contents also have the
largest absolute uncertainties; after unfolding to the detector level,
however, the uncertainties in the tail of the distribution may be
amplified by large detector corrections, and can even exceed those in
the most highly populated bins. This effect is especially significant
for `four-jet' observables, such as the aplanarity, which are measured
relatively poorly by the detector. One cannot, therefore, expect a
matrix of correlation coefficients derived for a normalised
\emph{detector-level} distribution to be valid for an unfolded
hadron-level distribution.

Table~\ref{bt_apl_detcor} shows the predicted statistical
uncertainties for the total jet broadening and aplanarity
distributions, at 189~GeV. In the unfolded aplanarity distribution, we
see that the largest contribution to the uncertainty of the
normalisation factor is due to the overflow bin; this bin would be
completely neglected if the uncertainty were derived by propagation
from a normalised detector-level distribution.

\begin{table}[p]
\begin{center}
\resizebox{\textwidth}{!}{
\begin{tabular}{|c|c@{\hspace{1cm}}r@{$\,\pm\,$}l|c|c|c@{\hspace{1cm}}r@{$\,\pm\,$}l|}
\hline
\rule[-0.7cm]{0pt}{0pt}\parbox{2.8cm}{\centering Total jet broadening, \BT} &
\multicolumn{3}{c|}{\parbox{4cm}{\centering Detector level events \\ $N_i\pm\sqrt{N_i}$}} &
 \multicolumn{1}{c|}{\parbox{2.2cm}{\centering Background \\ $\beta_i$}} &
 \multicolumn{1}{c|}{\rule{0pt}{0.95cm}\parbox{2.1cm}{\centering Detector\\[-0.05cm]correction\vspace{-0.1cm}\\ $\alpha_i$}} &
\multicolumn{3}{c|}{\parbox{4cm}{\centering Hadron-level events \\ $\alpha_i(N_i - \beta_i)\pm \alpha_i\sqrt{N_i}$}} \bigstrut \\
\hline
\hline
0.000 -- 0.030          &&  147   & 12   &  0   &  1.27 &&  187 & 15 \bigstrut[t] \\
0.030 -- 0.040          &&  297   & 17   &  1   &  1.06 &&  315 & 18 \\
0.040 -- 0.050          &&  351   & 19   &  1   &  0.91 &&  317 & 17 \\
0.050 -- 0.060          &&  327   & 18   &  1   &  0.84 &&  273 & 15 \\
0.060 -- 0.075          &&  405   & 20   &  2   &  0.83 &&  334 & 17 \\
0.075 -- 0.090          &&  313   & 18   &  3   &  0.84 &&  261 & 15 \\
0.090 -- 0.110          &&  324   & 18   &  7   &  0.87 &&  276 & 16 \\
\hline
\end{tabular}
}
\end{center} \vspace{0.5cm}
\caption{Monte Carlo predictions for the total jet broadening and
aplanarity distributions at 189~GeV, corresponding to the luminosity
of the OPAL measurement. The expected number of events in each bin is
shown without normalisation, at the detector and hadron levels; the
detector level includes background. The predicted statistical
uncertainties are those of the OPAL measurement (not of the Monte
Carlo prediction itself), and are uncorrelated between bins in this
unnormalised distribution.}
\label{bt_apl_detcor}
\end{table}

Fortunately, most bins of our event shape distributions are `narrow',
so they behave approximately like a set of uncorrelated Poisson
variables divided by a constant normalisation factor: this was the
implicit assumption made in previous OPAL analyses. For wide bins such
as the first bin of aplanarity, however, the uncertainty of the
normalisation cannot be neglected. Furthermore, the correlations
between bins can never be neglected when fitting a theoretical
prediction to the distribution, if a significant proportion of events
fall inside the fit range.\footnote{This statement is true even if the
bins themselves are infinitesimally narrow. As the number of bins~$n$
within the fit range increases, the covariance between any pair of
bins will fall as~$1/n^2$ (and~the correlation coefficient as~$1/n$),
but the number of entries in the covariance matrix will scale
with~$n^2$.}  To calculate the uncertainty of the fit, one must either
use a Monte Carlo `subsample' method, as described in
Section~\ref{subsamples}, or use a $\chi^2$ parameter based on the full
analytical covariance matrix.

\newpage
\section[$\chi^2$ in the presence of correlations]{\boldmath{$\chi^2$} in the presence of correlations}

\label{appsec:chisquare}
In Section~\ref{sec:chisquare}, we claimed that the following
expression should be minimised with respect to the parameter $\lambda$ to
determine the best fit between theory and experiment:
\begin{eqnarray}
\chi^2 & = & \sum_{ij}\;\big(y_i-\as(Q_i;\lambda)\big)\;\big(V^{-1}\big)_{ij}\;
\big(y_j-\as(Q_j;\lambda)\big) \nonumber \\
\label{chisquare_appendix}
 & \equiv & 
\sum_{ij}\;\delta y_i\;\big(V^{-1}\big)_{ij}\;\delta y_j
\end{eqnarray}
Here $y_i$ is an experimental measurement at an energy scale $Q_i$,
and $\as(Q_i;\lambda)$ is a theoretical prediction for the measured
quantity at the same energy scale. $V$ is the covariance matrix
relating the uncertainties of the measurements. $\lambda$ is an
unknown parameter of the theory, which we wish to measure (in our
case, \asmz).

\subsubsection*{Explanation:}
It is well-known~\cite{cowanbook} that the following definition of
$\chi^2$ is a measure of the negative log-likelihood for data with
\emph{uncorrelated} Gaussian distributions:
\begin{equation}
\label{chisquare_nocor}
\chi^2\;=\;\sum_i\;\frac{\big(u_i-f(x_i;\lambda)\big)^2}{\sigma_i^2}
\;\equiv \; \frac{(\delta u_i)^2}{\sigma_i^2} \;\;\;.
\end{equation}
In this expression, $f(x_i;\lambda)$ is the predicted mean value for
the measurement $u_i$, and $\sigma_i$ is its standard deviation.
Therefore $\delta u_i$, the difference between the measurement and its
predicted mean, has an expectation value of zero, and a standard
deviation of~$\sigma_i$.

Equation~(\ref{chisquare_nocor}) can be written in matrix form, by
defining the covariance matrix $W$ for the vector of deviations
$\delta u$:
\begin{equation}
W=\left(\begin{array}{c c c c}
\sigma_1^2 & 0 & 0 & \hdots \\
0 & \sigma_2^2 & 0 & \hdots \\
0 & 0 & \sigma_3^2 & \hdots \\
\vdots & \vdots & \vdots & \ddots
\end{array}\right)
\;\;\;\;\;\mathrm{and}\;\;\;\;
W^{-1}=\left(\begin{array}{c c c c}
\frac{1}{\sigma_1^2} & 0 & 0 & \hdots \\
0 & \frac{1}{\sigma_2^2} & 0 & \hdots \\
0 & 0 & \frac{1}{\sigma_3^2} & \hdots \\
\vdots & \vdots & \vdots & \ddots
\end{array}\right)\;\;\;,
\end{equation}
so that 
\begin{equation}
\chi^2\;=\;(\delta u)^\mathrm{T}\;W^{-1}\;(\delta u) \;\;\;.
\end{equation}
Suppose we now apply a linear transformation represented by the matrix~$A$:
\begin{eqnarray}
\label{chisquarerotate}
\chi^2 & \!=\! & \big(A\,\delta u\big)^\mathrm{T}\;\big(A\,WA^\mathrm{T}\big)^{-1}\;\big(A\,\delta u\big) \nonumber \\
 & \equiv & \big(\delta u'\big)^\mathrm{T}\;\big(W'\big)^{-1}\;\big(\delta u'\big)
\end{eqnarray}
Here each component of the vector $\delta u' = A\,\delta u$
represents a linear combination of the original elements $\delta u_i$
defined in Equation~(\ref{chisquare_nocor}). Furthermore, the matrix
$W'= A\,WA^\mathrm{T}$ is the
covariance matrix relating these linear combinations; this can be proven
by the same method used in Section~\ref{errorcomb} to find
the variance of a single linear combination.

We now wish to establish the validity of
Equation~(\ref{chisquare_appendix}), for a set of measurements~$y_i$
with covariance matrix~$V$. This will be proven if we can find a set
of uncorrelated quantities~$\delta u_i$ with covariance matrix~$W$,
and a matrix~$A$, such that
\begin{eqnarray}
\label{vprimeeqv}
\delta y & \!=\! & A\,\delta u \\
\mathrm{and}\;\;\;\;\;\; V & \!=\! & A\,WA^\mathrm{T} \;\;\;.
\end{eqnarray}
The covariance matrix $V$ is real and symmetric, by definition, so~$A$
is simply the orthonormal matrix which diagonalises it. The elements
of the diagonal matrix $W$ are the eigenvalues of $V$. Furthermore,
the matrix $A$ is non-singular (since $A^\mathrm{T}=A^{-1}$), so
Equation~(\ref{vprimeeqv}) may be inverted to find $\delta u$. Each
element $\delta u_i$ is an uncorrelated linear combination of the
correlated deviations $\delta y$.

Hence we have shown that $\chi^2$, as defined in
Equation~(\ref{chisquare_appendix}), is a valid measure of the
negative log-likelihood when the Gaussian uncertainties of the
measurements $y_i$ are correlated. Applying the principle of maximum
likelihood, we should therefore minimise this quantity to estimate the
parameter~$\lambda$.

\newpage
\section[Minimisation of the total uncertainty]
{\setstretch{1}Minimisation of the total 
uncertainty in a weighted mean\footnote{The primed quantities $y'_i$ and $V'$ used in Equations~(\ref{chisquareprime}), (\ref{weightformula}) and~(\ref{totalerror}) will be denoted by $y_i$ and $V$ respectively in this section, for clarity.}}

\label{appsec:errormin}
In Section \ref{weightssection} we chose the weights
\begin{equation}
\label{app_weightsformula}
w_i\;=\;\frac{\sum_{j}\;\big(V^{-1}\big)_{ij}}{\sum_{jk}\;\big(V^{-1}\big)_{jk}}\;\;\;,
\end{equation}
so as to minimise the following $\chi^2$, which measures the likelihood of the data~$y_i$ for a weighted mean $\sum_k w_k y_k$:
\begin{equation}
\chi^2\;=\;\sum_{ij}\;\big(y_i-\left[\,{\textstyle \sum_k} w_k y_k\,\right] \big)\;\big(V^{-1}\big)_{ij}\;
\big(y_j-\left[\,{\textstyle \sum_k} w_k y_k\,\right]\big)\;\;\;.
\end{equation}
But we claimed without proof, in Section \ref{errormin}, that
that these same weights would also minimise the total variance $\sigma^2$
of the weighted mean:
\begin{equation}
\sigma^2 \; = \; \sum_{ij}\;w_i \, V_{ij} \, w_j \; \equiv \; w^\mathrm{T}\,V\,w \;\;\;.
\label{varmean}
\end{equation}

\subsubsection*{Proof:}

We wish to minimise the variance $\sigma^2$, as defined in
Equation~(\ref{varmean}), subject to the constraint that the sum of
the weights $w_i$ should be unity:
\begin{equation}
\label{weightsum}
\sum_k w_k=1 \;\;\;.
\end{equation}
This is achieved by introducing a Lagrange multiplier $\mu$, and requiring
the following set of partial derivatives to vanish:
\begin{equation}
\frac{\partial}{\partial w_i}\left(\sigma^2\,-\,\mu\sum_k w_k\right)=0\;, \;\;\;\;\forall i \;\;\;.
\end{equation}
Substituting $\sigma^2$ from Equation~(\ref{varmean}), and exploiting
the symmetry of the covariance matrix $V$, we find
\begin{equation}
2\left(\sum_{j}\,V_{ij}\,w_j\right)\,-\,\mu\,=\,0\;, \;\;\;\;\forall i \;\;\;.
\end{equation}
In matrix notation, we have
\begin{equation}
2\,Vw=\mu \, a \;\;\;,\;\;\;\mathrm{where\ we\ have\ defined}\;\;\;{\textstyle a\,=\,
\left(\begin{array}{c}
1 \\[-7pt]
1 \\[-7pt]
1 \\[-7pt]
\vdots
\end{array}\right)}
\end{equation}
Therefore, provided $V$ is non-singular\footnote{The statement that
$V$ is non-singular is equivalent to the requirement that no
measurement can be predicted with absolute precision from the other
measurements. If this condition were violated, a linear combination of
measurements would exist with a variance of zero, and so one of the
eigenvalues of $V$ would vanish.}, we may invert this equation to find
the vector of weights $w$:
\begin{equation}
w\,=\,\frac{\mu}{2}\,V^{-1}a \;\;\;.
\end{equation}
Re-writing Equation~(\ref{weightsum}) in vector notation, we have
\begin{equation}
a^\mathrm{T}w=1 \;\;\;,
\end{equation}
so we can now eliminate the Lagrange multiplier $\mu$ to find
\begin{equation}
w\,=\,\frac{\displaystyle V^{-1}a}{\displaystyle a^\mathrm{T}V^{-1}a} 
\;\;\;.
\end{equation}
This is the same vector of weights given in
Equation~(\ref{app_weightsformula}). We have therefore shown that our
choice of weights simultaneously maximises the likelihood of the data,
and minimises the total variance of the weighted mean.

\chapter{OPAL event shape distributions}
\label{evshdistappendix}

In this appendix, we present the distributions of the following event shape 
observables, measured by OPAL at centre-of-mass energies 
$\sqrt{s}=91-207$~GeV:

{\setstretch{1.0}
\begin{enumerate}
\item Thrust, $T$
\item Heavy jet mass, $M_\mathrm{H}$
\item $C$-parameter
\item Total jet broadening, $B_\mathrm{T}$
\item Wide jet broadening, $B_\mathrm{W}$
\item Durham $y_{23}$ parameter
\item Thrust major, $T_\mathrm{maj.}$
\item Thrust minor, $T_\mathrm{min.}$
\item Aplanarity, $A$
\item Sphericity, $S$
\item Oblateness, $O$
\item Light jet mass, $M_\mathrm{L}$
\item Narrow jet broadening, $B_\mathrm{N}$
\item $D$-parameter
\end{enumerate}}

\noindent Definitions of these observables can be found in
Section~\ref{evsh_defs}, and a full discussion of the analysis
procedure is given in Chapter~\ref{opalchapter}.

\newpage
\section[Thrust, $T$]{Thrust, \boldmath{$T$}}

\begin{table}[hb!]
\begin{center}
\scalebox{0.90}{
\begin{minipage}{\linewidth}
\begin{center}
\begin{tabular}{| r @{ -- } l | r @{.} l @{ $\pm$ } r @{.} l @{ $\pm$ } r @{.} l @{} l | r @{.} l @{ $\pm$ } r @{.} l @{ $\pm$ } r @{.} l @{} l |}
\hline \multicolumn{2}{|c|}{$1-T$} & \multicolumn{7}{|c|}{$R'\,(1-T)$ at 91 GeV} & \multicolumn{7}{|c|}{$R'\,(1-T)$ at 133 GeV} \bigstrut \\ 
\hline \bigstrut[t]
$0.00$ & $0.01$  & 1 & 273 & 0 & 019 & 0 & 043 &  & 4 & 37 & 0 & 82 & 0 & 55 & \\ 
$0.01$ & $0.02$  & 12 & 26 & 0 & 05 & 0 & 40 &  & 20 & 4 & 1 & 6 & 3 & 8 & \\ 
$0.02$ & $0.03$  & 18 & 38 & 0 & 07 & 0 & 28 &  & 20 & 4 & 1 & 5 & 1 & 7 & \\ 
$0.03$ & $0.04$  & 13 & 86 & 0 & 06 & 0 & 15 &  & 10 & 5 & 1 & 2 & 0 & 9 & \\ 
$0.04$ & $0.05$  & 9 & 80 & 0 & 05 & 0 & 18 &  & 6 & 7 & 1 & 0 & 0 & 4 & \\ 
$0.05$ & $0.07$  & 6 & 502 & 0 & 027 & 0 & 086 &  & 4 & 70 & 0 & 58 & 0 & 35 & \\ 
$0.07$ & $0.09$  & 4 & 133 & 0 & 022 & 0 & 037 &  & 3 & 64 & 0 & 51 & 0 & 29 & \\
\hline 

\end{tabular}
\end{center}
\end{minipage}}
\end{center}
\caption{Distributions for the thrust, $T$, measured by
OPAL at centre-of-mass energies $\sqrt{s}=91$--189~GeV.
The first uncertainty is statistical, while the second is systematic.}
\label{tabdist10}
\end{table}
\clearpage

\begin{figure}[p]
\begin{center}
\includegraphics[width=\textwidth]{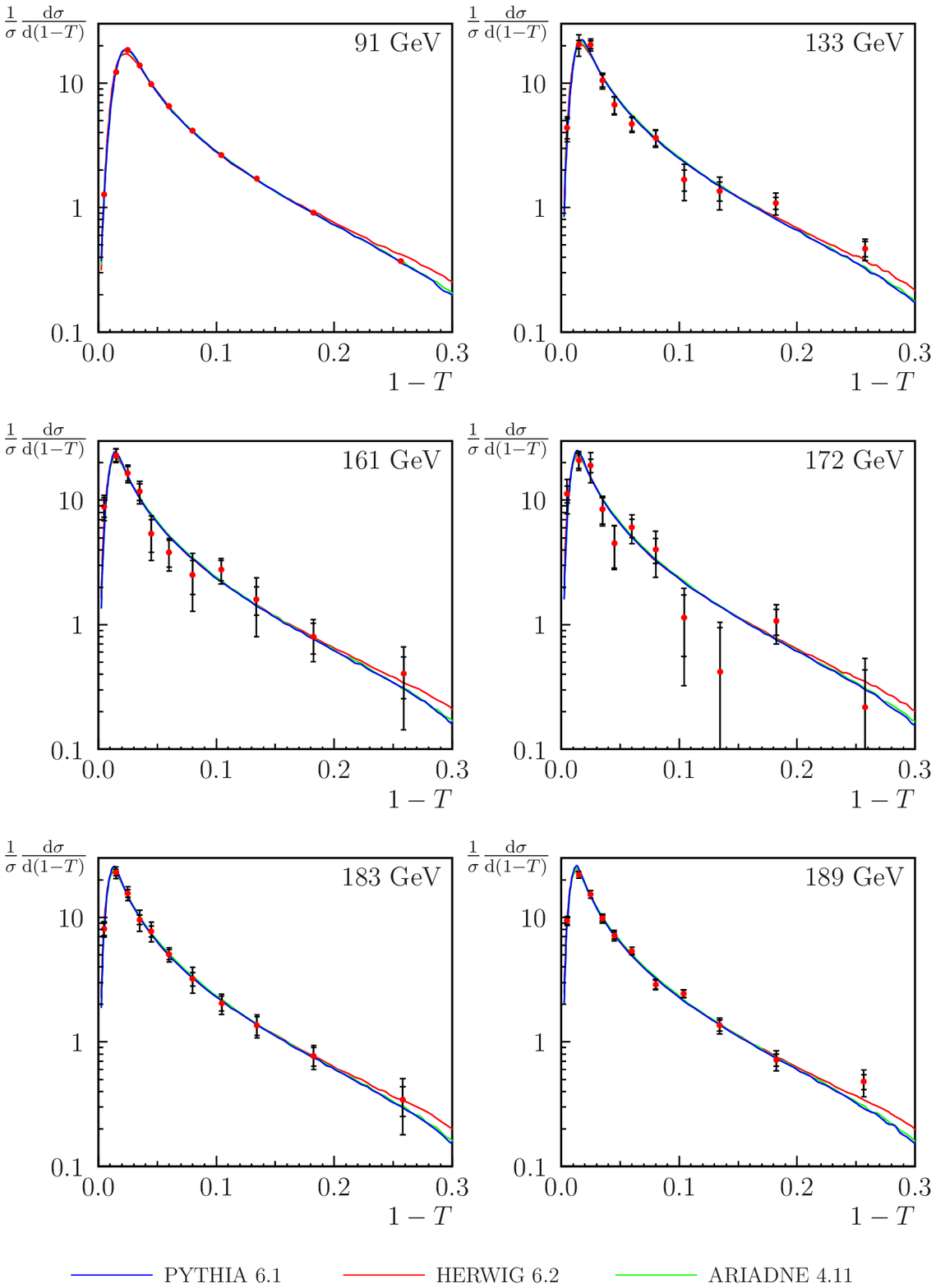}
\caption{Distributions for the thrust, $T$, measured by OPAL
at centre-of-mass energies $\sqrt{s}=91$--189~GeV. The inner error
bars indicate statistical uncertainties. Each curve is generated using
five million non-radiative Monte Carlo events, after hadronisation.}
\label{figdist10}
\end{center}
\end{figure}
\clearpage
\section*{Thrust, \boldmath{$T$}~(contd.)}

\begin{table}[hb!]
\begin{center}
\scalebox{0.90}{
\begin{minipage}{\linewidth}
\begin{center}
\begin{tabular}{| r @{ -- } l | r @{.} l @{ $\pm$ } r @{.} l @{ $\pm$ } r @{.} l @{} l | r @{.} l @{ $\pm$ } r @{.} l @{ $\pm$ } r @{.} l @{} l |}
\hline \multicolumn{2}{|c|}{$1-T$} & \multicolumn{7}{|c|}{$R'\,(1-T)$ at 192 GeV} & \multicolumn{7}{|c|}{$R'\,(1-T)$ at 196 GeV} \bigstrut \\ 
\hline \bigstrut[t]
$0.00$ & $0.01$  & 9 & 0 & 1 & 5 & 1 & 6 &  & 8 & 70 & 0 & 95 & 0 & 53 & \\ 
$0.01$ & $0.02$  & 22 & 6 & 2 & 1 & 3 & 0 &  & 23 & 6 & 1 & 4 & 1 & 4 & \\ 
$0.02$ & $0.03$  & 13 & 2 & 1 & 7 & 1 & 4 &  & 15 & 3 & 1 & 1 & 1 & 2 & \\ 
$0.03$ & $0.04$  & 9 & 3 & 1 & 4 & 1 & 8 &  & 9 & 86 & 0 & 85 & 0 & 94 & \\ 
$0.04$ & $0.05$  & 7 & 0 & 1 & 2 & 0 & 9 &  & 7 & 13 & 0 & 78 & 0 & 95 & \\ 
$0.05$ & $0.07$  & 5 & 67 & 0 & 73 & 0 & 85 &  & 5 & 21 & 0 & 46 & 0 & 22 & \\ 
$0.07$ & $0.09$  & 3 & 16 & 0 & 60 & 0 & 83 &  & 2 & 59 & 0 & 39 & 0 & 48 & \\
\hline 

\end{tabular}
\end{center}
\end{minipage}}
\end{center}
\caption{Distributions for the thrust, $T$, measured by
OPAL at centre-of-mass energies $\sqrt{s}=192$--207~GeV.
The first uncertainty is statistical, while the second is systematic.}
\label{tabdist20}
\end{table}
\clearpage

\begin{figure}[p]
\begin{center}
\includegraphics[width=\textwidth]{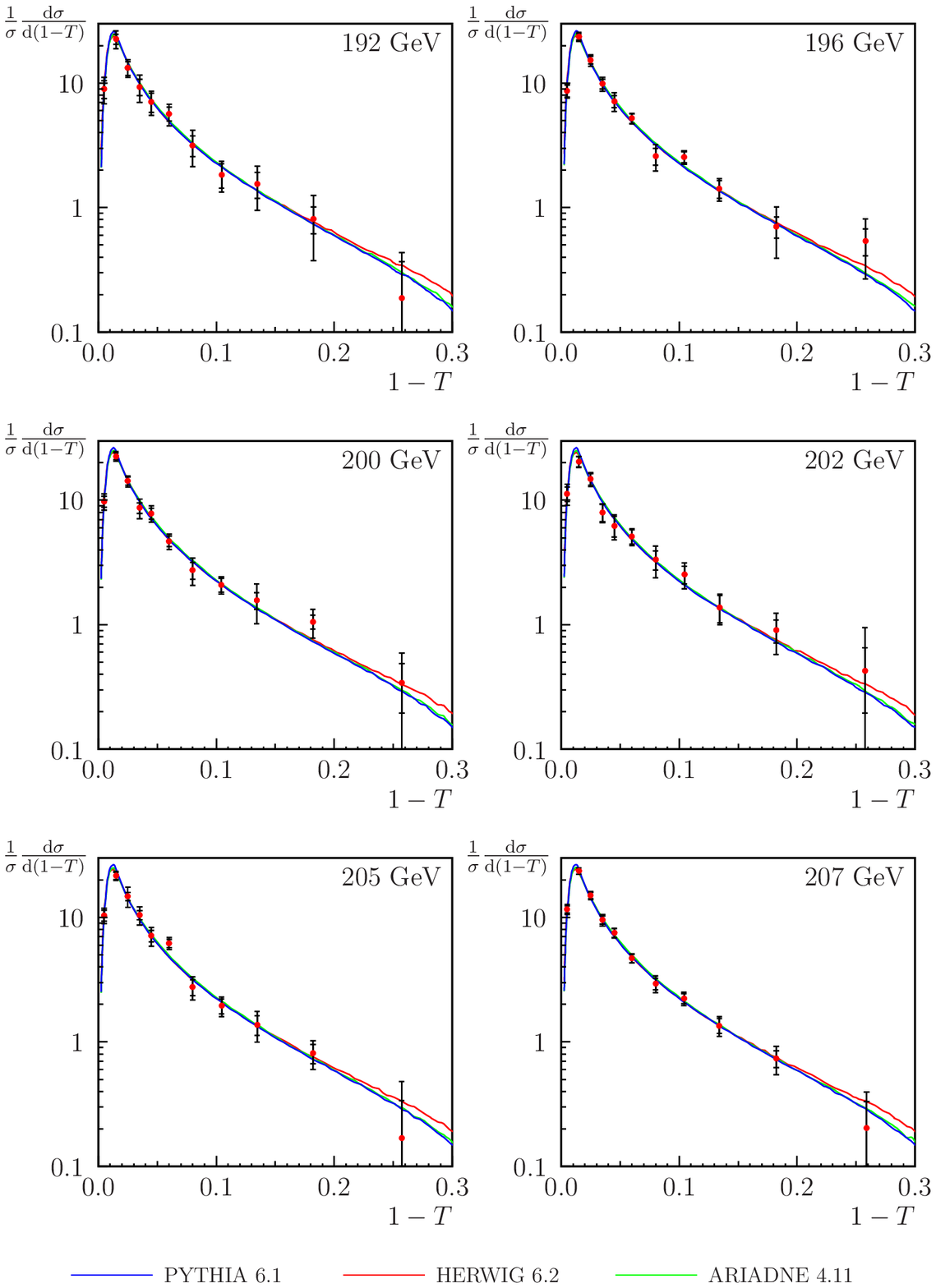}
\caption{Distributions for the thrust, $T$, measured by OPAL
at centre-of-mass energies $\sqrt{s}=192$--207~GeV. The inner error
bars indicate statistical uncertainties. Each curve is generated using
five million non-radiative Monte Carlo events, after hadronisation.}
\label{figdist20}
\end{center}
\end{figure}
\clearpage
\section[Heavy jet mass, $M_\mathrm{H}$]{Heavy jet mass, \boldmath{$M_\mathrm{H}$}}

\begin{table}[hb!]
\begin{center}
\scalebox{0.90}{
\begin{minipage}{\linewidth}
\begin{center}
\begin{tabular}{| r @{ -- } l | r @{.} l @{ $\pm$ } r @{.} l @{ $\pm$ } r @{.} l @{} l | r @{.} l @{ $\pm$ } r @{.} l @{ $\pm$ } r @{.} l @{} l |}
\hline \multicolumn{2}{|c|}{$M_\mathrm{H}/\sqrt{s}$} & \multicolumn{7}{|c|}{$R'\,(M_\mathrm{H}/\sqrt{s})$ at 91 GeV} & \multicolumn{7}{|c|}{$R'\,(M_\mathrm{H}/\sqrt{s})$ at 133 GeV} \bigstrut \\ 
\hline \bigstrut[t]
$0.060$ & $0.075$  & 0 & 1185 & 0 & 0030 & 0 & 0098 &  & 0 & 38 & 0 & 16 & 0 & 13 & \\ 
$0.075$ & $0.090$  & 0 & 55 & 0 & 01 & 0 & 13 &  & 1 & 56 & 0 & 35 & 0 & 36 & \\ 
$0.090$ & $0.110$  & 2 & 17 & 0 & 01 & 0 & 38 &  & 4 & 7 & 0 & 5 & 1 & 3 & \\ 
$0.110$ & $0.140$  & 5 & 729 & 0 & 020 & 0 & 057 &  & 8 & 2 & 0 & 5 & 1 & 0 & \\ 
$0.140$ & $0.170$  & 6 & 65 & 0 & 02 & 0 & 35 &  & 4 & 99 & 0 & 48 & 0 & 82 & \\ 
$0.170$ & $0.200$  & 4 & 88 & 0 & 02 & 0 & 15 &  & 3 & 92 & 0 & 43 & 0 & 28 & \\ 
$0.200$ & $0.250$  & 3 & 30 & 0 & 01 & 0 & 11 &  & 2 & 47 & 0 & 27 & 0 & 30 & \\
\hline 

\end{tabular}
\end{center}
\end{minipage}}
\end{center}
\caption{Distributions for the heavy jet mass, $M_\mathrm{H}$, measured by
OPAL at centre-of-mass energies $\sqrt{s}=91$--189~GeV.
The first uncertainty is statistical, while the second is systematic.}
\label{tabdist11}
\end{table}
\clearpage

\begin{figure}[p]
\begin{center}
\includegraphics[width=\textwidth]{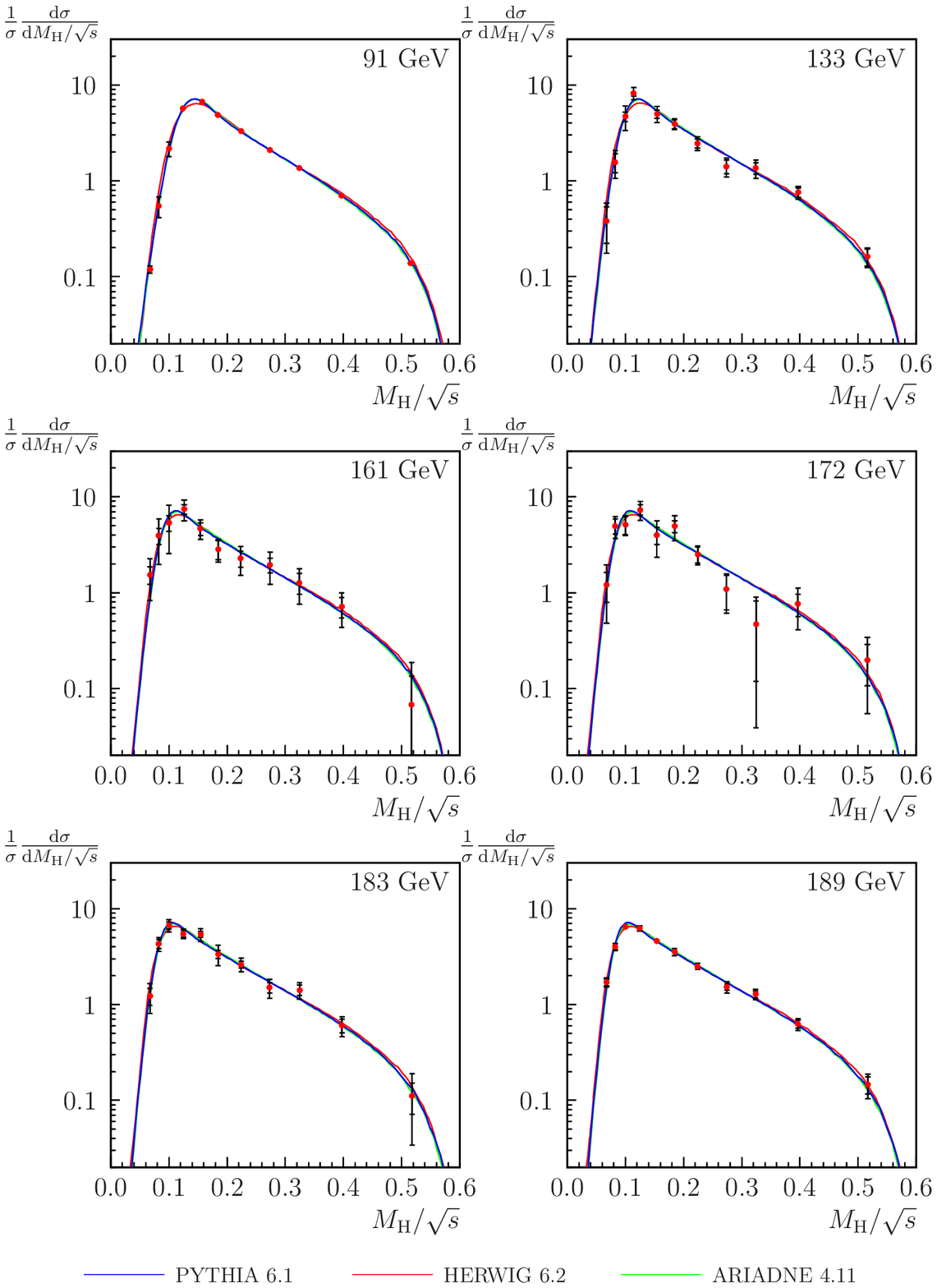}
\caption{Distributions for the heavy jet mass, $M_\mathrm{H}$, measured by OPAL
at centre-of-mass energies $\sqrt{s}=91$--189~GeV. The inner error
bars indicate statistical uncertainties. Each curve is generated using
five million non-radiative Monte Carlo events, after hadronisation.}
\label{figdist11}
\end{center}
\end{figure}
\clearpage
\section*{Heavy jet mass, \boldmath{$M_\mathrm{H}$}~(contd.)}

\begin{table}[hb!]
\begin{center}
\scalebox{0.90}{
\begin{minipage}{\linewidth}
\begin{center}
\begin{tabular}{| r @{ -- } l | r @{.} l @{ $\pm$ } r @{.} l @{ $\pm$ } r @{.} l @{} l | r @{.} l @{ $\pm$ } r @{.} l @{ $\pm$ } r @{.} l @{} l |}
\hline \multicolumn{2}{|c|}{$M_\mathrm{H}/\sqrt{s}$} & \multicolumn{7}{|c|}{$R'\,(M_\mathrm{H}/\sqrt{s})$ at 192 GeV} & \multicolumn{7}{|c|}{$R'\,(M_\mathrm{H}/\sqrt{s})$ at 196 GeV} \bigstrut \\ 
\hline \bigstrut[t]
$0.060$ & $0.075$  & 1 & 44 & 0 & 37 & 0 & 54 &  & 1 & 55 & 0 & 25 & 0 & 22 & \\ 
$0.075$ & $0.090$  & 4 & 96 & 0 & 72 & 0 & 84 &  & 4 & 30 & 0 & 48 & 0 & 46 & \\ 
$0.090$ & $0.110$  & 6 & 85 & 0 & 83 & 0 & 83 &  & 7 & 11 & 0 & 53 & 0 & 63 & \\ 
$0.110$ & $0.140$  & 5 & 31 & 0 & 61 & 0 & 88 &  & 6 & 08 & 0 & 39 & 0 & 54 & \\ 
$0.140$ & $0.170$  & 4 & 47 & 0 & 55 & 0 & 52 &  & 4 & 38 & 0 & 35 & 0 & 59 & \\ 
$0.170$ & $0.200$  & 4 & 36 & 0 & 49 & 0 & 54 &  & 3 & 23 & 0 & 30 & 0 & 27 & \\ 
$0.200$ & $0.250$  & 2 & 23 & 0 & 34 & 0 & 34 &  & 2 & 30 & 0 & 21 & 0 & 18 & \\
\hline 

\end{tabular}
\end{center}
\end{minipage}}
\end{center}
\caption{Distributions for the heavy jet mass, $M_\mathrm{H}$, measured by
OPAL at centre-of-mass energies $\sqrt{s}=192$--207~GeV.
The first uncertainty is statistical, while the second is systematic.}
\label{tabdist21}
\end{table}
\clearpage

\begin{figure}[p]
\begin{center}
\includegraphics[width=\textwidth]{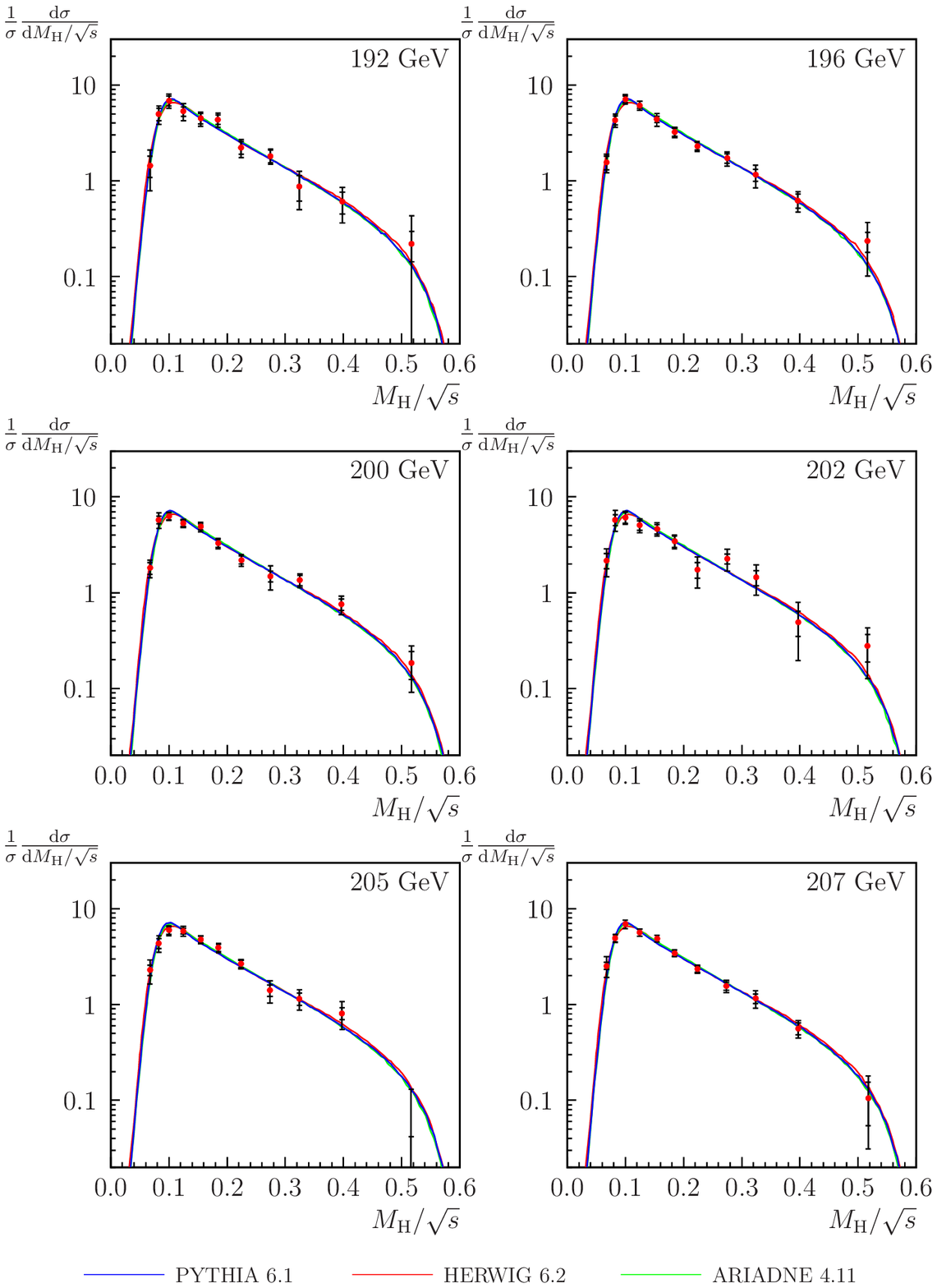}
\caption{Distributions for the heavy jet mass, $M_\mathrm{H}$, measured by OPAL
at centre-of-mass energies $\sqrt{s}=192$--207~GeV. The inner error
bars indicate statistical uncertainties. Each curve is generated using
five million non-radiative Monte Carlo events, after hadronisation.}
\label{figdist21}
\end{center}
\end{figure}
\clearpage
\section[$C$-parameter]{\boldmath{$C$}-parameter}

\begin{table}[hb!]
\begin{center}
\scalebox{0.90}{
\begin{minipage}{\linewidth}
\begin{center}
\begin{tabular}{| r @{ -- } l | r @{.} l @{ $\pm$ } r @{.} l @{ $\pm$ } r @{.} l @{} l | r @{.} l @{ $\pm$ } r @{.} l @{ $\pm$ } r @{.} l @{} l |}
\hline \multicolumn{2}{|c|}{$C$} & \multicolumn{7}{|c|}{$R'\,(C)$ at 91 GeV} & \multicolumn{7}{|c|}{$R'\,(C)$ at 133 GeV} \bigstrut \\ 
\hline \bigstrut[t]
$0.00$ & $0.05$  & 0 & 2177 & 0 & 0038 & 0 & 0068 &  & 0 & 83 & 0 & 17 & 0 & 11 & \\ 
$0.05$ & $0.08$  & 2 & 07 & 0 & 01 & 0 & 10 &  & 4 & 1 & 0 & 5 & 1 & 2 & \\ 
$0.08$ & $0.11$  & 4 & 033 & 0 & 018 & 0 & 049 &  & 5 & 70 & 0 & 46 & 0 & 51 & \\ 
$0.11$ & $0.14$  & 4 & 154 & 0 & 019 & 0 & 053 &  & 3 & 86 & 0 & 39 & 0 & 41 & \\ 
$0.14$ & $0.18$  & 3 & 224 & 0 & 014 & 0 & 046 &  & 2 & 55 & 0 & 29 & 0 & 14 & \\ 
$0.18$ & $0.22$  & 2 & 418 & 0 & 012 & 0 & 060 &  & 1 & 65 & 0 & 25 & 0 & 19 & \\ 
$0.22$ & $0.30$  & 1 & 705 & 0 & 007 & 0 & 020 &  & 1 & 22 & 0 & 15 & 0 & 07 & \\
\hline 

\end{tabular}
\end{center}
\end{minipage}}
\end{center}
\caption{Distributions for the $C$-parameter, measured by
OPAL at centre-of-mass energies $\sqrt{s}=91$--189~GeV.
The first uncertainty is statistical, while the second is systematic.}
\label{tabdist12}
\end{table}
\clearpage

\begin{figure}[p]
\begin{center}
\includegraphics[width=\textwidth]{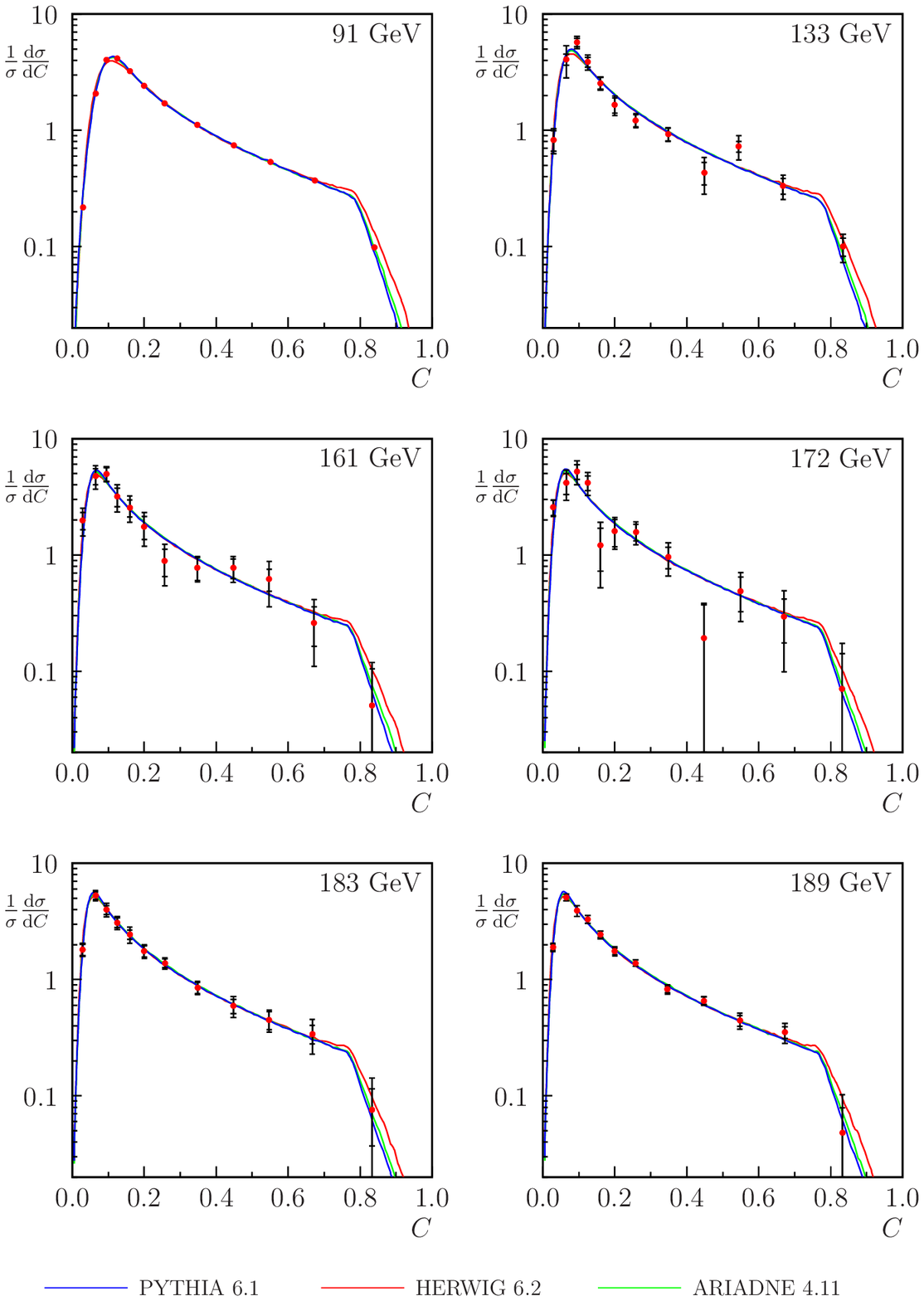}
\caption{Distributions for the $C$-parameter, measured by OPAL
at centre-of-mass energies $\sqrt{s}=91$--189~GeV. The inner error
bars indicate statistical uncertainties. Each curve is generated using
five million non-radiative Monte Carlo events, after hadronisation.}
\label{figdist12}
\end{center}
\end{figure}
\clearpage
\section*{\boldmath{$C$}-parameter~(contd.)}

\begin{table}[hb!]
\begin{center}
\scalebox{0.90}{
\begin{minipage}{\linewidth}
\begin{center}
\begin{tabular}{| r @{ -- } l | r @{.} l @{ $\pm$ } r @{.} l @{ $\pm$ } r @{.} l @{} l | r @{.} l @{ $\pm$ } r @{.} l @{ $\pm$ } r @{.} l @{} l |}
\hline \multicolumn{2}{|c|}{$C$} & \multicolumn{7}{|c|}{$R'\,(C)$ at 192 GeV} & \multicolumn{7}{|c|}{$R'\,(C)$ at 196 GeV} \bigstrut \\ 
\hline \bigstrut[t]
$0.00$ & $0.05$  & 1 & 79 & 0 & 31 & 0 & 31 &  & 1 & 78 & 0 & 20 & 0 & 13 & \\ 
$0.05$ & $0.08$  & 5 & 1 & 0 & 6 & 1 & 0 &  & 5 & 39 & 0 & 39 & 0 & 45 & \\ 
$0.08$ & $0.11$  & 3 & 92 & 0 & 51 & 0 & 70 &  & 4 & 12 & 0 & 32 & 0 & 48 & \\ 
$0.11$ & $0.14$  & 2 & 99 & 0 & 43 & 0 & 64 &  & 3 & 08 & 0 & 29 & 0 & 24 & \\ 
$0.14$ & $0.18$  & 2 & 20 & 0 & 34 & 0 & 24 &  & 2 & 28 & 0 & 21 & 0 & 46 & \\ 
$0.18$ & $0.22$  & 1 & 72 & 0 & 30 & 0 & 21 &  & 1 & 88 & 0 & 19 & 0 & 20 & \\ 
$0.22$ & $0.30$  & 1 & 32 & 0 & 19 & 0 & 26 &  & 1 & 23 & 0 & 12 & 0 & 12 & \\
\hline 

\end{tabular}
\end{center}
\end{minipage}}
\end{center}
\caption{Distributions for the $C$-parameter, measured by
OPAL at centre-of-mass energies $\sqrt{s}=192$--207~GeV.
The first uncertainty is statistical, while the second is systematic.}
\label{tabdist22}
\end{table}
\clearpage

\begin{figure}[p]
\begin{center}
\includegraphics[width=\textwidth]{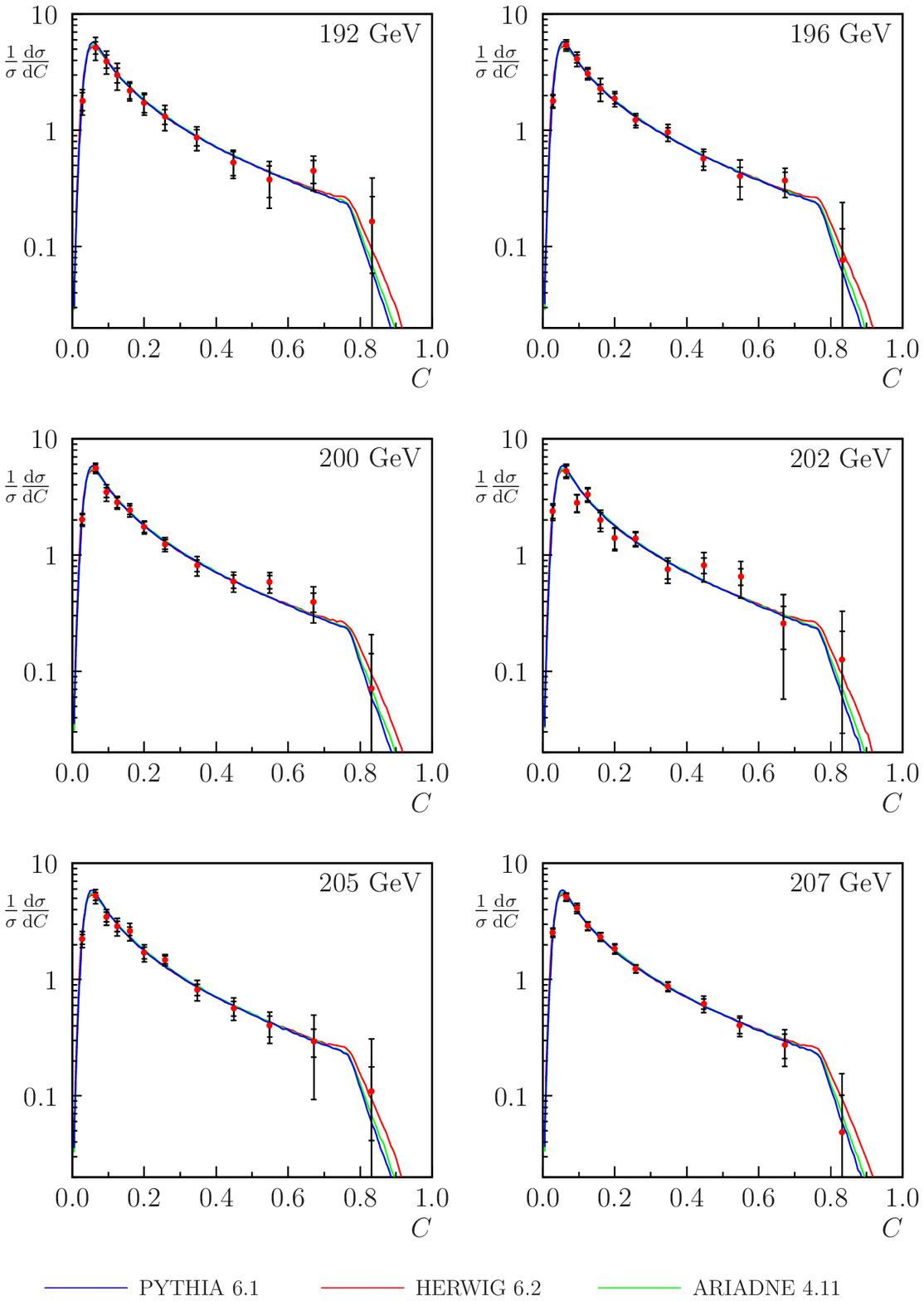}
\caption{Distributions for the $C$-parameter, measured by OPAL
at centre-of-mass energies $\sqrt{s}=192$--207~GeV. The inner error
bars indicate statistical uncertainties. Each curve is generated using
five million non-radiative Monte Carlo events, after hadronisation.}
\label{figdist22}
\end{center}
\end{figure}
\clearpage
\section[Total jet broadening, $B_\mathrm{T}$]{Total jet broadening, \boldmath{$B_\mathrm{T}$}}

\begin{table}[hb!]
\begin{center}
\scalebox{0.90}{
\begin{minipage}{\linewidth}
\begin{center}
\begin{tabular}{| r @{ -- } l | r @{.} l @{ $\pm$ } r @{.} l @{ $\pm$ } r @{.} l @{} l | r @{.} l @{ $\pm$ } r @{.} l @{ $\pm$ } r @{.} l @{} l |}
\hline \multicolumn{2}{|c|}{$B_\mathrm{T}$} & \multicolumn{7}{|c|}{$R'\,(B_\mathrm{T})$ at 91 GeV} & \multicolumn{7}{|c|}{$R'\,(B_\mathrm{T})$ at 133 GeV} \bigstrut \\ 
\hline \bigstrut[t]
$0.000$ & $0.030$  & 0 & 116 & 0 & 005 & 0 & 013 &  & 0 & 58 & 0 & 26 & 0 & 16 & \\ 
$0.030$ & $0.040$  & 2 & 49 & 0 & 03 & 0 & 12 &  & 10 & 2 & 1 & 1 & 1 & 3 & \\ 
$0.040$ & $0.050$  & 6 & 98 & 0 & 05 & 0 & 16 &  & 9 & 6 & 1 & 3 & 1 & 9 & \\ 
$0.050$ & $0.060$  & 9 & 780 & 0 & 051 & 0 & 083 &  & 11 & 9 & 1 & 2 & 0 & 9 & \\ 
$0.060$ & $0.075$  & 10 & 755 & 0 & 041 & 0 & 098 &  & 10 & 06 & 0 & 86 & 0 & 52 & \\ 
$0.075$ & $0.090$  & 8 & 77 & 0 & 04 & 0 & 24 &  & 6 & 31 & 0 & 76 & 0 & 36 & \\ 
$0.090$ & $0.110$  & 6 & 595 & 0 & 026 & 0 & 072 &  & 4 & 35 & 0 & 59 & 0 & 82 & \\
\hline 

\end{tabular}
\end{center}
\end{minipage}}
\end{center}
\caption{Distributions for the total jet broadening, $B_\mathrm{T}$, measured by
OPAL at centre-of-mass energies $\sqrt{s}=91$--189~GeV.
The first uncertainty is statistical, while the second is systematic.}
\label{tabdist13}
\end{table}
\clearpage

\begin{figure}[p]
\begin{center}
\includegraphics[width=\textwidth]{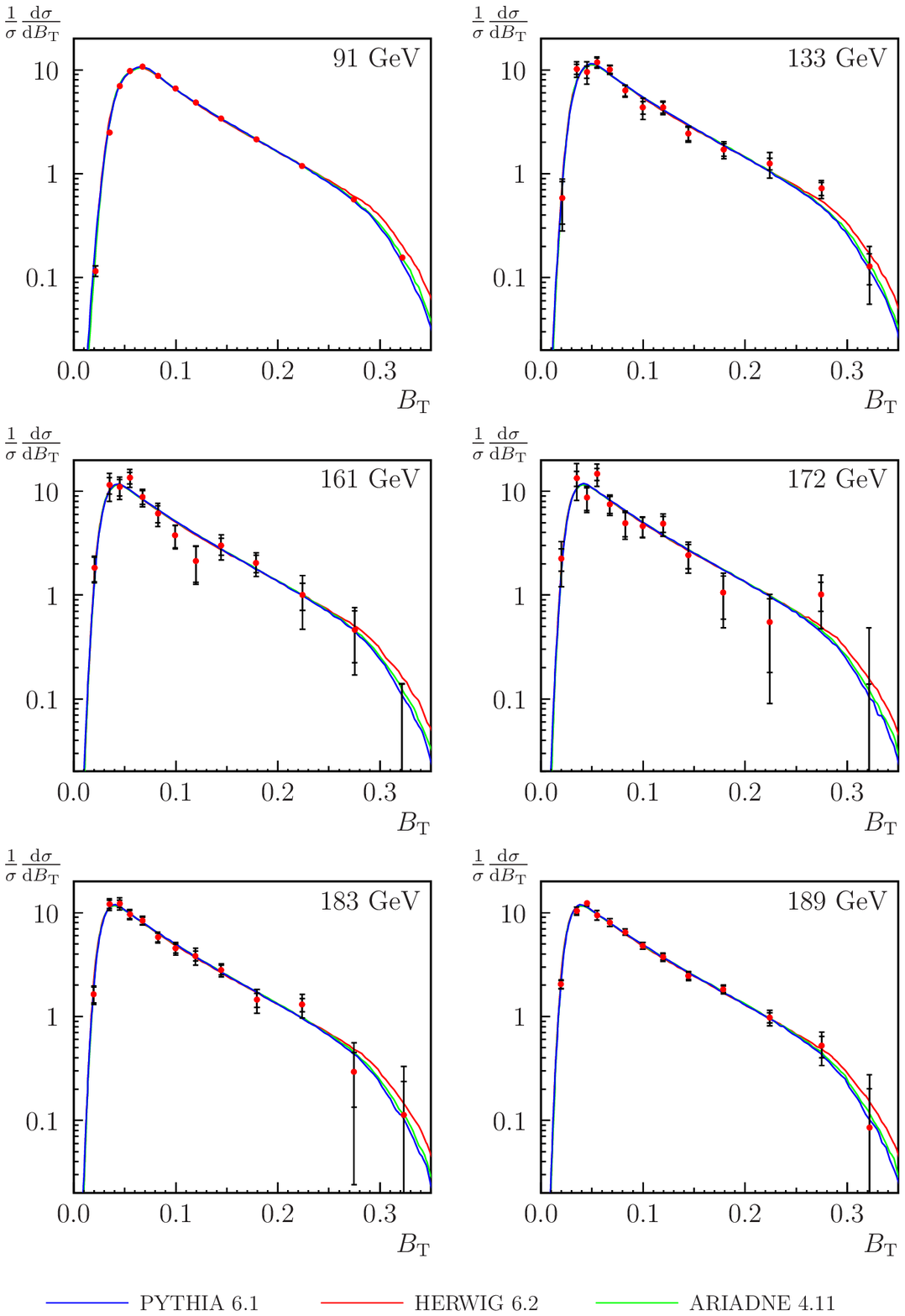}
\caption{Distributions for the total jet broadening, $B_\mathrm{T}$, measured by OPAL
at centre-of-mass energies $\sqrt{s}=91$--189~GeV. The inner error
bars indicate statistical uncertainties. Each curve is generated using
five million non-radiative Monte Carlo events, after hadronisation.}
\label{figdist13}
\end{center}
\end{figure}
\clearpage
\section*{Total jet broadening, \boldmath{$B_\mathrm{T}$}~(contd.)}

\begin{table}[hb!]
\begin{center}
\scalebox{0.90}{
\begin{minipage}{\linewidth}
\begin{center}
\begin{tabular}{| r @{ -- } l | r @{.} l @{ $\pm$ } r @{.} l @{ $\pm$ } r @{.} l @{} l | r @{.} l @{ $\pm$ } r @{.} l @{ $\pm$ } r @{.} l @{} l |}
\hline \multicolumn{2}{|c|}{$B_\mathrm{T}$} & \multicolumn{7}{|c|}{$R'\,(B_\mathrm{T})$ at 192 GeV} & \multicolumn{7}{|c|}{$R'\,(B_\mathrm{T})$ at 196 GeV} \bigstrut \\ 
\hline \bigstrut[t]
$0.000$ & $0.030$  & 2 & 40 & 0 & 47 & 0 & 26 &  & 2 & 12 & 0 & 31 & 0 & 45 & \\ 
$0.030$ & $0.040$  & 10 & 5 & 1 & 6 & 2 & 7 &  & 10 & 9 & 1 & 0 & 1 & 7 & \\ 
$0.040$ & $0.050$  & 10 & 3 & 1 & 5 & 1 & 0 &  & 11 & 9 & 0 & 9 & 1 & 3 & \\ 
$0.050$ & $0.060$  & 10 & 0 & 1 & 4 & 1 & 3 &  & 12 & 1 & 0 & 9 & 1 & 2 & \\ 
$0.060$ & $0.075$  & 8 & 6 & 1 & 0 & 1 & 2 &  & 7 & 62 & 0 & 64 & 0 & 97 & \\ 
$0.075$ & $0.090$  & 4 & 58 & 0 & 87 & 0 & 62 &  & 6 & 10 & 0 & 57 & 0 & 57 & \\ 
$0.090$ & $0.110$  & 5 & 3 & 0 & 7 & 1 & 4 &  & 4 & 70 & 0 & 44 & 0 & 32 & \\
\hline 

\end{tabular}
\end{center}
\end{minipage}}
\end{center}
\caption{Distributions for the total jet broadening, $B_\mathrm{T}$, measured by
OPAL at centre-of-mass energies $\sqrt{s}=192$--207~GeV.
The first uncertainty is statistical, while the second is systematic.}
\label{tabdist23}
\end{table}
\clearpage

\begin{figure}[p]
\begin{center}
\includegraphics[width=\textwidth]{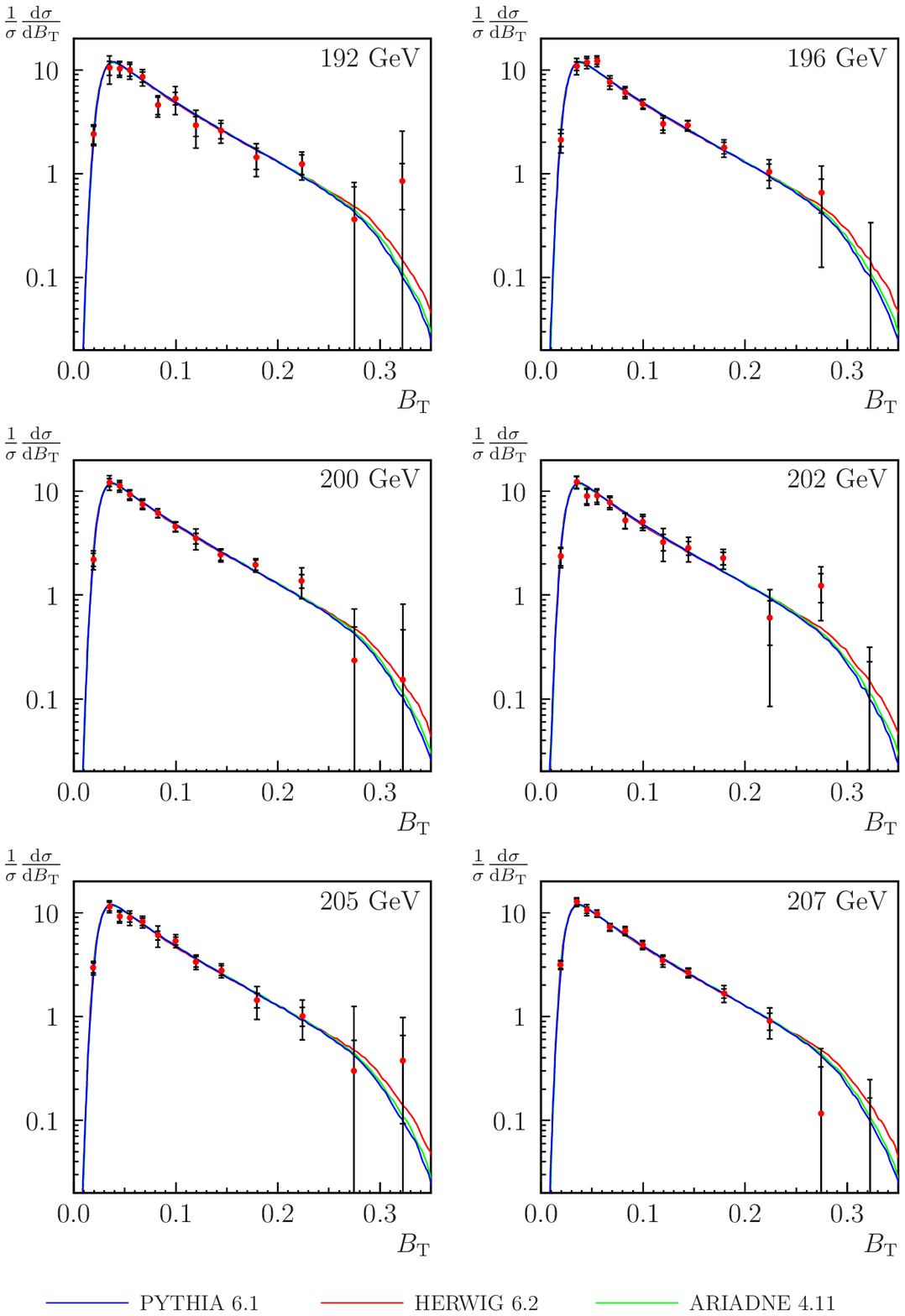}
\caption{Distributions for the total jet broadening, $B_\mathrm{T}$, measured by OPAL
at centre-of-mass energies $\sqrt{s}=192$--207~GeV. The inner error
bars indicate statistical uncertainties. Each curve is generated using
five million non-radiative Monte Carlo events, after hadronisation.}
\label{figdist23}
\end{center}
\end{figure}
\clearpage
\section[Wide jet broadening, $B_\mathrm{W}$]{Wide jet broadening, \boldmath{$B_\mathrm{W}$}}

\begin{table}[hb!]
\begin{center}
\scalebox{0.90}{
\begin{minipage}{\linewidth}
\begin{center}
\begin{tabular}{| r @{ -- } l | r @{.} l @{ $\pm$ } r @{.} l @{ $\pm$ } r @{.} l @{} l | r @{.} l @{ $\pm$ } r @{.} l @{ $\pm$ } r @{.} l @{} l |}
\hline \multicolumn{2}{|c|}{$B_\mathrm{W}$} & \multicolumn{7}{|c|}{$R'\,(B_\mathrm{W})$ at 91 GeV} & \multicolumn{7}{|c|}{$R'\,(B_\mathrm{W})$ at 133 GeV} \bigstrut \\ 
\hline \bigstrut[t]
$0.000$ & $0.020$  & 0 & 557 & 0 & 011 & 0 & 021 &  & 3 & 15 & 0 & 53 & 0 & 83 & \\ 
$0.020$ & $0.030$  & 10 & 39 & 0 & 06 & 0 & 17 &  & 16 & 4 & 1 & 5 & 2 & 5 & \\ 
$0.030$ & $0.040$  & 16 & 50 & 0 & 06 & 0 & 12 &  & 14 & 6 & 1 & 3 & 0 & 6 & \\ 
$0.040$ & $0.050$  & 13 & 34 & 0 & 05 & 0 & 63 &  & 10 & 7 & 1 & 1 & 0 & 7 & \\ 
$0.050$ & $0.065$  & 9 & 82 & 0 & 04 & 0 & 13 &  & 9 & 47 & 0 & 84 & 0 & 84 & \\ 
$0.065$ & $0.080$  & 7 & 16 & 0 & 03 & 0 & 15 &  & 5 & 44 & 0 & 72 & 0 & 40 & \\ 
$0.080$ & $0.100$  & 5 & 067 & 0 & 024 & 0 & 062 &  & 3 & 96 & 0 & 54 & 0 & 45 & \\
\hline 

\end{tabular}
\end{center}
\end{minipage}}
\end{center}
\caption{Distributions for the wide jet broadening, $B_\mathrm{W}$, measured by
OPAL at centre-of-mass energies $\sqrt{s}=91$--189~GeV.
The first uncertainty is statistical, while the second is systematic.}
\label{tabdist14}
\end{table}
\clearpage

\begin{figure}[p]
\begin{center}
\includegraphics[width=\textwidth]{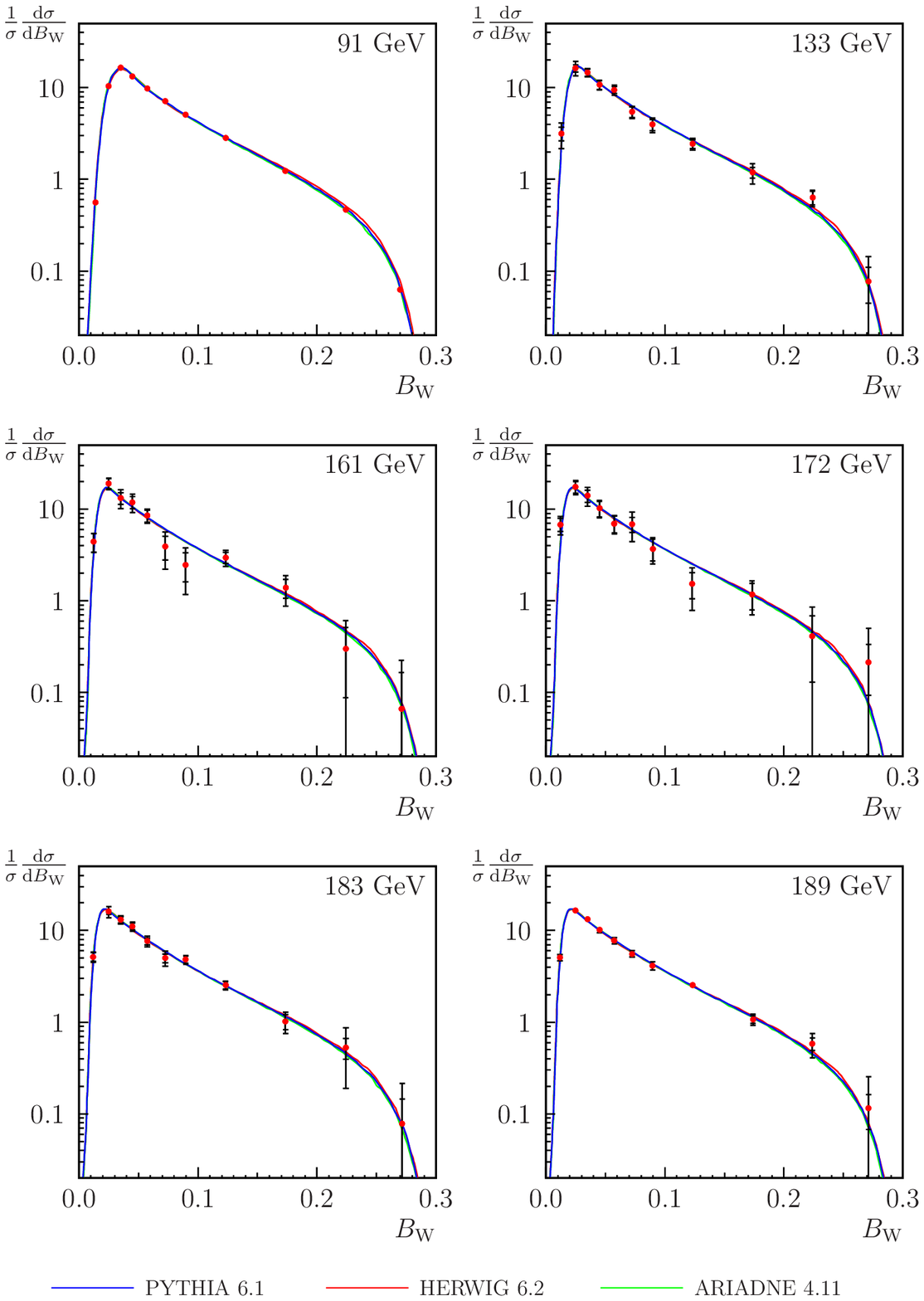}
\caption{Distributions for the wide jet broadening, $B_\mathrm{W}$, measured by OPAL
at centre-of-mass energies $\sqrt{s}=91$--189~GeV. The inner error
bars indicate statistical uncertainties. Each curve is generated using
five million non-radiative Monte Carlo events, after hadronisation.}
\label{figdist14}
\end{center}
\end{figure}
\clearpage
\section*{Wide jet broadening, \boldmath{$B_\mathrm{W}$}~(contd.)}

\begin{table}[hb!]
\begin{center}
\scalebox{0.90}{
\begin{minipage}{\linewidth}
\begin{center}
\begin{tabular}{| r @{ -- } l | r @{.} l @{ $\pm$ } r @{.} l @{ $\pm$ } r @{.} l @{} l | r @{.} l @{ $\pm$ } r @{.} l @{ $\pm$ } r @{.} l @{} l |}
\hline \multicolumn{2}{|c|}{$B_\mathrm{W}$} & \multicolumn{7}{|c|}{$R'\,(B_\mathrm{W})$ at 192 GeV} & \multicolumn{7}{|c|}{$R'\,(B_\mathrm{W})$ at 196 GeV} \bigstrut \\ 
\hline \bigstrut[t]
$0.000$ & $0.020$  & 6 & 42 & 0 & 85 & 0 & 59 &  & 5 & 45 & 0 & 55 & 0 & 40 & \\ 
$0.020$ & $0.030$  & 14 & 8 & 1 & 7 & 2 & 6 &  & 15 & 1 & 1 & 1 & 1 & 7 & \\ 
$0.030$ & $0.040$  & 14 & 2 & 1 & 5 & 2 & 1 &  & 14 & 1 & 1 & 0 & 1 & 2 & \\ 
$0.040$ & $0.050$  & 10 & 0 & 1 & 4 & 1 & 5 &  & 10 & 2 & 0 & 9 & 1 & 1 & \\ 
$0.050$ & $0.065$  & 7 & 52 & 0 & 97 & 0 & 45 &  & 7 & 22 & 0 & 63 & 0 & 69 & \\ 
$0.065$ & $0.080$  & 6 & 15 & 0 & 93 & 0 & 82 &  & 5 & 57 & 0 & 56 & 0 & 39 & \\ 
$0.080$ & $0.100$  & 3 & 68 & 0 & 69 & 0 & 75 &  & 4 & 91 & 0 & 46 & 0 & 74 & \\
\hline 

\end{tabular}
\end{center}
\end{minipage}}
\end{center}
\caption{Distributions for the wide jet broadening, $B_\mathrm{W}$, measured by
OPAL at centre-of-mass energies $\sqrt{s}=192$--207~GeV.
The first uncertainty is statistical, while the second is systematic.}
\label{tabdist24}
\end{table}
\clearpage

\begin{figure}[p]
\begin{center}
\includegraphics[width=\textwidth]{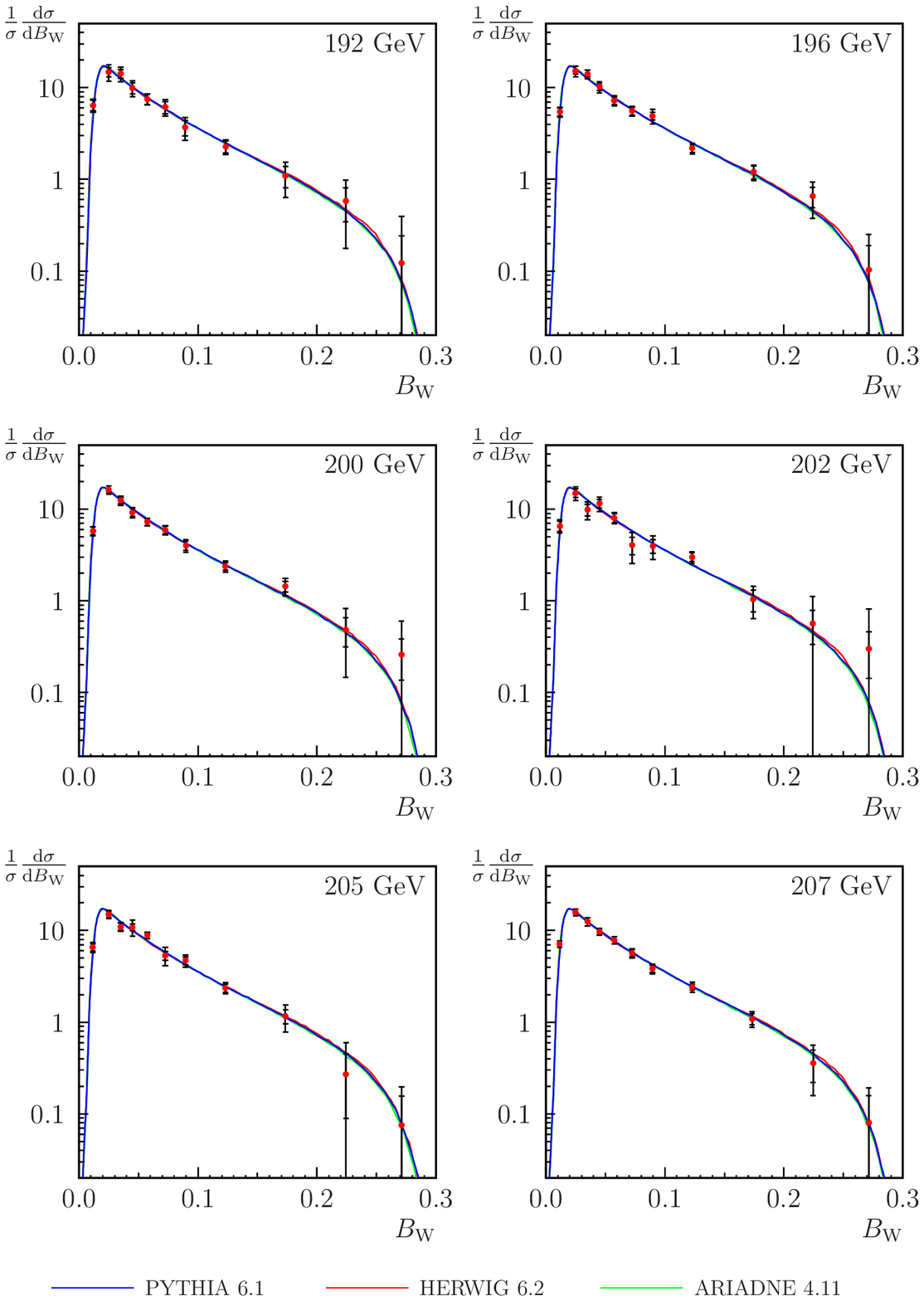}
\caption{Distributions for the wide jet broadening, $B_\mathrm{W}$, measured by OPAL
at centre-of-mass energies $\sqrt{s}=192$--207~GeV. The inner error
bars indicate statistical uncertainties. Each curve is generated using
five million non-radiative Monte Carlo events, after hadronisation.}
\label{figdist24}
\end{center}
\end{figure}
\clearpage
\section[Durham \MakeLowercase{$y_{23}$} parameter]{Durham \boldmath{$y_{23}$} parameter}

\begin{table}[hb!]
\begin{center}
\scalebox{0.90}{
\begin{minipage}{\linewidth}
\begin{center}
\begin{tabular}{| r @{ -- } l | r @{.} l @{ $\pm$ } r @{.} l @{ $\pm$ } r @{.} l @{} l | r @{.} l @{ $\pm$ } r @{.} l @{ $\pm$ } r @{.} l @{} l |}
\hline \multicolumn{2}{|c|}{$y_{23}$} & \multicolumn{7}{|c|}{$R'\,(y_{23})$ at 91 GeV} & \multicolumn{7}{|c|}{$R'\,(y_{23})$ at 133 GeV} \bigstrut \\ 
\hline \bigstrut[t]
$0.00030$ & $0.00075$  & 146 & 5 & 1 & 0 & 2 & 8 &  & 299 & & 30 & & 41 & & \\ 
$0.00075$ & $0.00130$  & 183 & 1 & 0 & 9 & 5 & 5 &  & 201 & & 22 & & 21 & & \\ 
$0.00130$ & $0.00230$  & 141 & 5 & 0 & 6 & 2 & 8 &  & 116 & & 12 & & 22 & & \\ 
$0.00230$ & $0.00400$  & 81 & 7 & 0 & 3 & 1 & 5 &  & 62 & 6 & 6 & 7 & 8 & 4 & \\ 
$0.00400$ & $0.00700$  & 39 & 5 & 0 & 2 & 1 & 2 &  & 33 & 0 & 3 & 9 & 5 & 1 & \\ 
$0.00700$ & $0.01200$  & 19 & 92 & 0 & 09 & 0 & 22 &  & 19 & 9 & 2 & 2 & 3 & 1 & \\ 
$0.01200$ & $0.02250$  & 9 & 73 & 0 & 05 & 0 & 20 &  & 6 & 9 & 1 & 1 & 0 & 4 & \\
\hline 

\end{tabular}
\end{center}
\end{minipage}}
\end{center}
\caption{Distributions for the Durham $y_{23}$ parameter, measured by
OPAL at centre-of-mass energies $\sqrt{s}=91$--189~GeV.
The first uncertainty is statistical, while the second is systematic.}
\label{tabdist15}
\end{table}
\clearpage

\begin{figure}[p]
\begin{center}
\includegraphics[width=\textwidth]{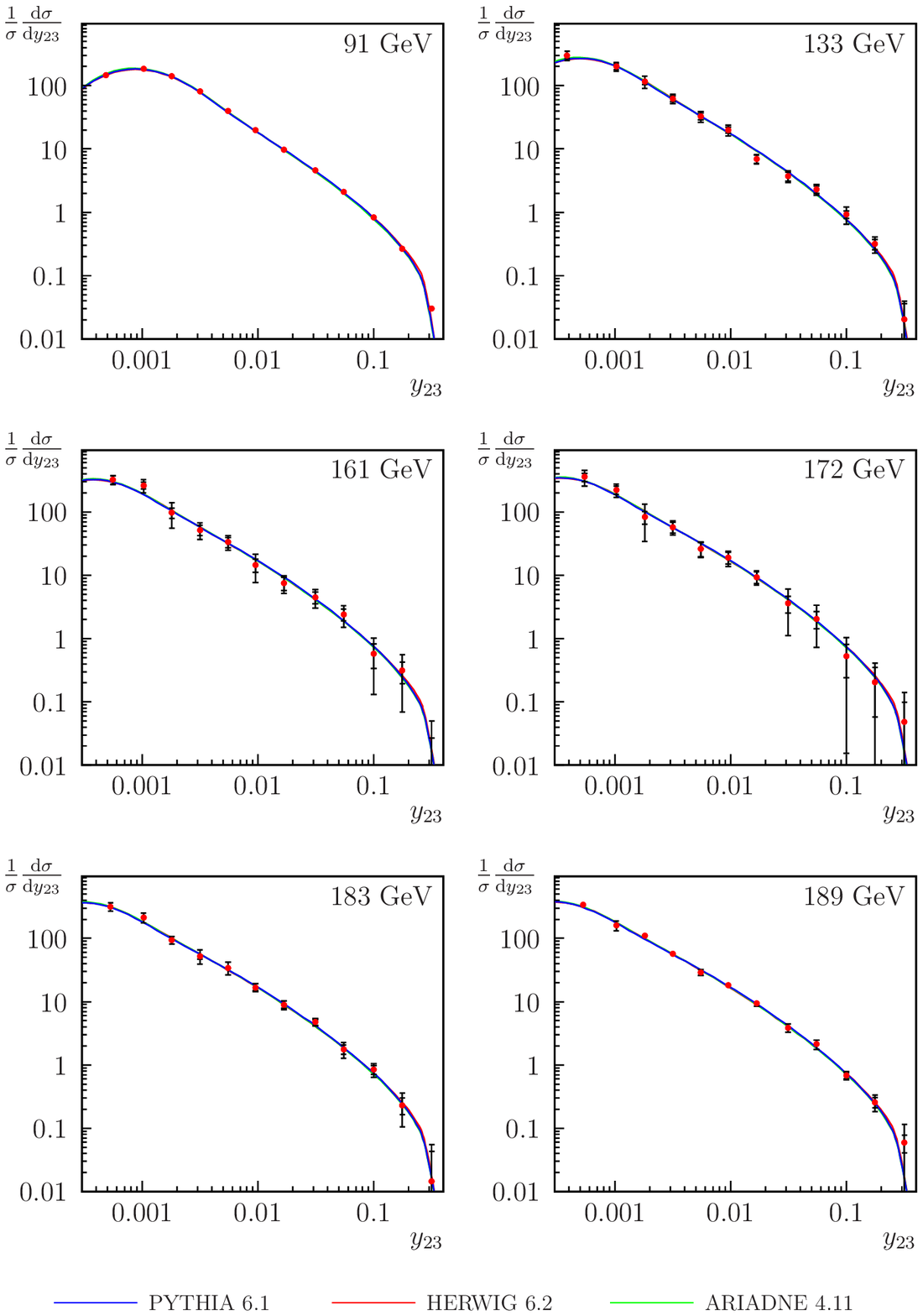}
\caption{Distributions for the Durham $y_{23}$ parameter, measured by OPAL
at centre-of-mass energies $\sqrt{s}=91$--189~GeV. The inner error
bars indicate statistical uncertainties. Each curve is generated using
five million non-radiative Monte Carlo events, after hadronisation.}
\label{figdist15}
\end{center}
\end{figure}
\clearpage
\section*{Durham \boldmath{$y_{23}$} parameter~(contd.)}

\begin{table}[hb!]
\begin{center}
\scalebox{0.90}{
\begin{minipage}{\linewidth}
\begin{center}
\begin{tabular}{| r @{ -- } l | r @{.} l @{ $\pm$ } r @{.} l @{ $\pm$ } r @{.} l @{} l | r @{.} l @{ $\pm$ } r @{.} l @{ $\pm$ } r @{.} l @{} l |}
\hline \multicolumn{2}{|c|}{$y_{23}$} & \multicolumn{7}{|c|}{$R'\,(y_{23})$ at 192 GeV} & \multicolumn{7}{|c|}{$R'\,(y_{23})$ at 196 GeV} \bigstrut \\ 
\hline \bigstrut[t]
$0.00030$ & $0.00075$  & 283 & & 37 & & 88 & &  & 325 & & 24 & & 30 & & \\ 
$0.00075$ & $0.00130$  & 194 & & 24 & & 37 & &  & 195 & & 15 & & 14 & & \\ 
$0.00130$ & $0.00230$  & 106 & & 13 & & 11 & &  & 102 & 5 & 8 & 6 & 5 & 0 & \\ 
$0.00230$ & $0.00400$  & 66 & 0 & 7 & 9 & 8 & 4 &  & 54 & 5 & 5 & 2 & 2 & 5 & \\ 
$0.00400$ & $0.00700$  & 27 & 8 & 4 & 6 & 5 & 2 &  & 31 & 1 & 2 & 8 & 5 & 5 & \\ 
$0.00700$ & $0.01200$  & 17 & 8 & 2 & 8 & 2 & 5 &  & 19 & 3 & 1 & 8 & 1 & 9 & \\ 
$0.01200$ & $0.02250$  & 8 & 8 & 1 & 4 & 1 & 6 &  & 7 & 70 & 0 & 91 & 0 & 99 & \\
\hline 

\end{tabular}
\end{center}
\end{minipage}}
\end{center}
\caption{Distributions for the Durham $y_{23}$ parameter, measured by
OPAL at centre-of-mass energies $\sqrt{s}=192$--207~GeV.
The first uncertainty is statistical, while the second is systematic.}
\label{tabdist25}
\end{table}
\clearpage

\begin{figure}[p]
\begin{center}
\includegraphics[width=\textwidth]{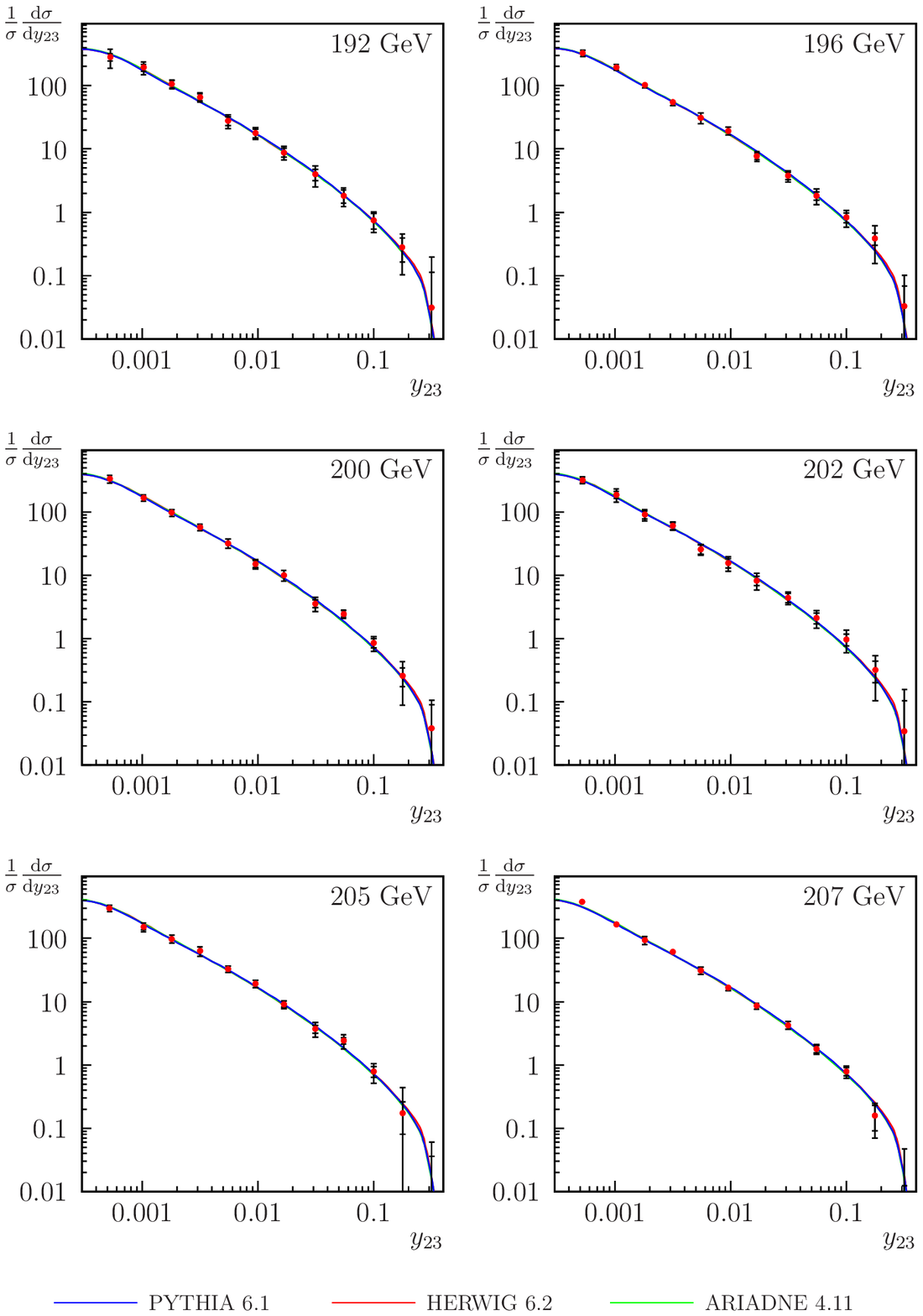}
\caption{Distributions for the Durham $y_{23}$ parameter, measured by OPAL
at centre-of-mass energies $\sqrt{s}=192$--207~GeV. The inner error
bars indicate statistical uncertainties. Each curve is generated using
five million non-radiative Monte Carlo events, after hadronisation.}
\label{figdist25}
\end{center}
\end{figure}
\clearpage
\section[Thrust major, $T_\mathrm{\MakeLowercase{maj.}}$]{Thrust major, \boldmath{$T_\mathrm{maj.}$}}

\begin{table}[hb!]
\begin{center}
\scalebox{0.90}{
\begin{minipage}{\linewidth}
\begin{center}
\begin{tabular}{| r @{ -- } l | r @{.} l @{ $\pm$ } r @{.} l @{ $\pm$ } r @{.} l @{} l | r @{.} l @{ $\pm$ } r @{.} l @{ $\pm$ } r @{.} l @{} l |}
\hline \multicolumn{2}{|c|}{$T_\mathrm{maj.}$} & \multicolumn{7}{|c|}{$R'\,(T_\mathrm{maj.})$ at 91 GeV} & \multicolumn{7}{|c|}{$R'\,(T_\mathrm{maj.})$ at 133 GeV} \bigstrut \\ 
\hline \bigstrut[t]
$0.00$ & $0.04$  & 0 & 0356 & 0 & 0028 & 0 & 0067 &  & 0 & 27 & 0 & 15 & 0 & 18 & \\ 
$0.04$ & $0.08$  & 3 & 374 & 0 & 016 & 0 & 046 &  & 6 & 00 & 0 & 44 & 0 & 60 & \\ 
$0.08$ & $0.12$  & 6 & 50 & 0 & 02 & 0 & 12 &  & 5 & 93 & 0 & 39 & 0 & 16 & \\ 
$0.12$ & $0.16$  & 4 & 417 & 0 & 015 & 0 & 069 &  & 3 & 32 & 0 & 33 & 0 & 40 & \\ 
$0.16$ & $0.22$  & 2 & 755 & 0 & 010 & 0 & 048 &  & 2 & 29 & 0 & 23 & 0 & 18 & \\ 
$0.22$ & $0.30$  & 1 & 578 & 0 & 007 & 0 & 025 &  & 1 & 12 & 0 & 15 & 0 & 16 & \\ 
$0.30$ & $0.40$  & 0 & 825 & 0 & 004 & 0 & 021 &  & 0 & 94 & 0 & 10 & 0 & 11 & \\
\hline 

\end{tabular}
\end{center}
\end{minipage}}
\end{center}
\caption{Distributions for the thrust major, $T_\mathrm{maj.}$, measured by
OPAL at centre-of-mass energies $\sqrt{s}=91$--189~GeV.
The first uncertainty is statistical, while the second is systematic.}
\label{tabdist16}
\end{table}
\clearpage

\begin{figure}[p]
\begin{center}
\includegraphics[width=\textwidth]{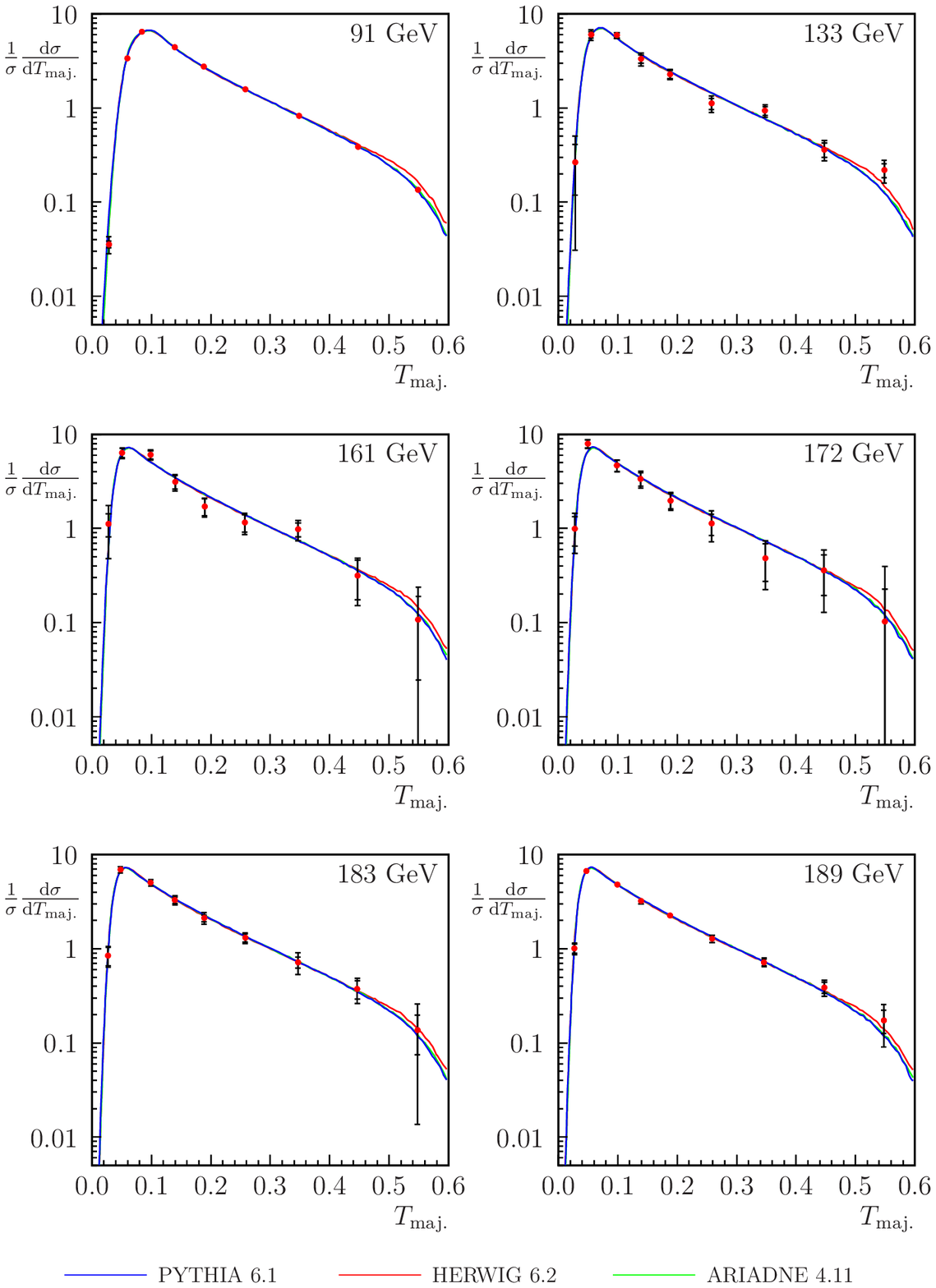}
\caption{Distributions for the thrust major, $T_\mathrm{maj.}$, measured by OPAL
at centre-of-mass energies $\sqrt{s}=91$--189~GeV. The inner error
bars indicate statistical uncertainties. Each curve is generated using
five million non-radiative Monte Carlo events, after hadronisation.}
\label{figdist16}
\end{center}
\end{figure}
\clearpage
\section*{Thrust major, \boldmath{$T_\mathrm{maj.}$}~(contd.)}

\begin{table}[hb!]
\begin{center}
\scalebox{0.90}{
\begin{minipage}{\linewidth}
\begin{center}
\begin{tabular}{| r @{ -- } l | r @{.} l @{ $\pm$ } r @{.} l @{ $\pm$ } r @{.} l @{} l | r @{.} l @{ $\pm$ } r @{.} l @{ $\pm$ } r @{.} l @{} l |}
\hline \multicolumn{2}{|c|}{$T_\mathrm{maj.}$} & \multicolumn{7}{|c|}{$R'\,(T_\mathrm{maj.})$ at 192 GeV} & \multicolumn{7}{|c|}{$R'\,(T_\mathrm{maj.})$ at 196 GeV} \bigstrut \\ 
\hline \bigstrut[t]
$0.00$ & $0.04$  & 1 & 07 & 0 & 30 & 0 & 16 &  & 1 & 06 & 0 & 19 & 0 & 31 & \\ 
$0.04$ & $0.08$  & 6 & 62 & 0 & 54 & 0 & 73 &  & 6 & 78 & 0 & 35 & 0 & 34 & \\ 
$0.08$ & $0.12$  & 5 & 30 & 0 & 45 & 0 & 54 &  & 4 & 86 & 0 & 29 & 0 & 28 & \\ 
$0.12$ & $0.16$  & 3 & 05 & 0 & 39 & 0 & 49 &  & 3 & 37 & 0 & 25 & 0 & 27 & \\ 
$0.16$ & $0.22$  & 2 & 06 & 0 & 28 & 0 & 22 &  & 2 & 11 & 0 & 18 & 0 & 14 & \\ 
$0.22$ & $0.30$  & 1 & 48 & 0 & 20 & 0 & 10 &  & 1 & 21 & 0 & 13 & 0 & 13 & \\ 
$0.30$ & $0.40$  & 0 & 42 & 0 & 15 & 0 & 20 &  & 0 & 79 & 0 & 10 & 0 & 16 & \\
\hline 

\end{tabular}
\end{center}
\end{minipage}}
\end{center}
\caption{Distributions for the thrust major, $T_\mathrm{maj.}$, measured by
OPAL at centre-of-mass energies $\sqrt{s}=192$--207~GeV.
The first uncertainty is statistical, while the second is systematic.}
\label{tabdist26}
\end{table}
\clearpage

\begin{figure}[p]
\begin{center}
\includegraphics[width=\textwidth]{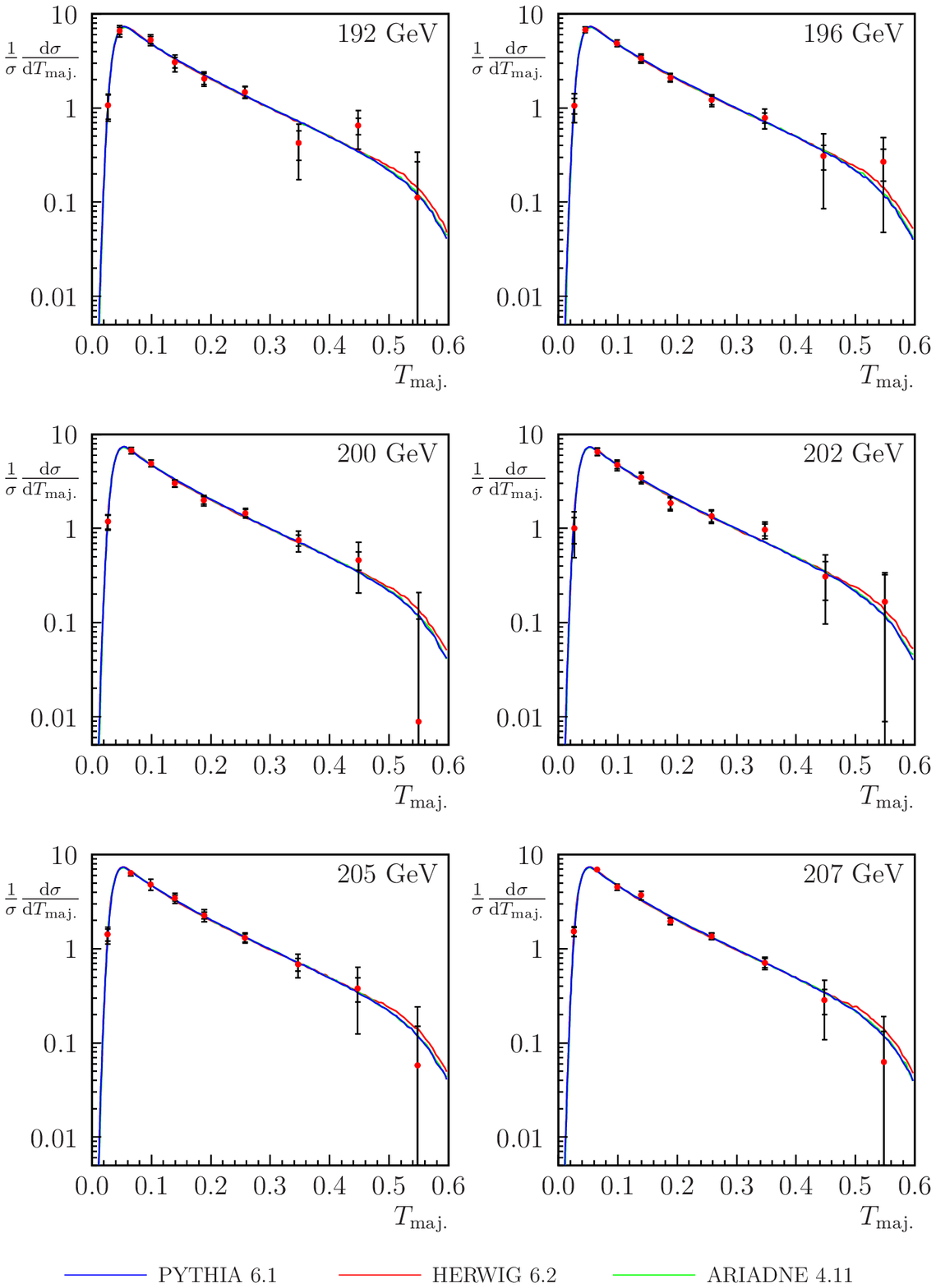}
\caption{Distributions for the thrust major, $T_\mathrm{maj.}$, measured by OPAL
at centre-of-mass energies $\sqrt{s}=192$--207~GeV. The inner error
bars indicate statistical uncertainties. Each curve is generated using
five million non-radiative Monte Carlo events, after hadronisation.}
\label{figdist26}
\end{center}
\end{figure}
\clearpage
\section[Thrust minor, $T_\mathrm{\MakeLowercase{min.}}$]{Thrust minor, \boldmath{$T_\mathrm{min.}$}}

\begin{table}[hb!]
\begin{center}
\scalebox{0.90}{
\begin{minipage}{\linewidth}
\begin{center}
\begin{tabular}{| r @{ -- } l | r @{.} l @{ $\pm$ } r @{.} l @{ $\pm$ } r @{.} l @{} l | r @{.} l @{ $\pm$ } r @{.} l @{ $\pm$ } r @{.} l @{} l |}
\hline \multicolumn{2}{|c|}{$T_\mathrm{min.}$} & \multicolumn{7}{|c|}{$R'\,(T_\mathrm{min.})$ at 91 GeV} & \multicolumn{7}{|c|}{$R'\,(T_\mathrm{min.})$ at 133 GeV} \bigstrut \\ 
\hline \bigstrut[t]
$0.00$ & $0.02$  & 0 & 0229 & 0 & 0029 & 0 & 0038 &  & 0 & 07 & 0 & 08 & 0 & 19 & \\ 
$0.02$ & $0.04$  & 1 & 386 & 0 & 016 & 0 & 045 &  & 5 & 13 & 0 & 62 & 0 & 44 & \\ 
$0.04$ & $0.06$  & 8 & 147 & 0 & 032 & 0 & 070 &  & 13 & 9 & 0 & 9 & 1 & 8 & \\ 
$0.06$ & $0.08$  & 12 & 413 & 0 & 036 & 0 & 097 &  & 12 & 1 & 0 & 8 & 1 & 7 & \\ 
$0.08$ & $0.10$  & 10 & 335 & 0 & 033 & 0 & 056 &  & 6 & 9 & 0 & 7 & 1 & 1 & \\ 
$0.10$ & $0.12$  & 6 & 850 & 0 & 027 & 0 & 065 &  & 4 & 69 & 0 & 49 & 0 & 62 & \\ 
$0.12$ & $0.14$  & 4 & 195 & 0 & 021 & 0 & 046 &  & 2 & 75 & 0 & 38 & 0 & 69 & \\
\hline 

\end{tabular}
\end{center}
\end{minipage}}
\end{center}
\caption{Distributions for the thrust minor, $T_\mathrm{min.}$, measured by
OPAL at centre-of-mass energies $\sqrt{s}=91$--189~GeV.
The first uncertainty is statistical, while the second is systematic.}
\label{tabdist17}
\end{table}
\clearpage

\begin{figure}[p]
\begin{center}
\includegraphics[width=\textwidth]{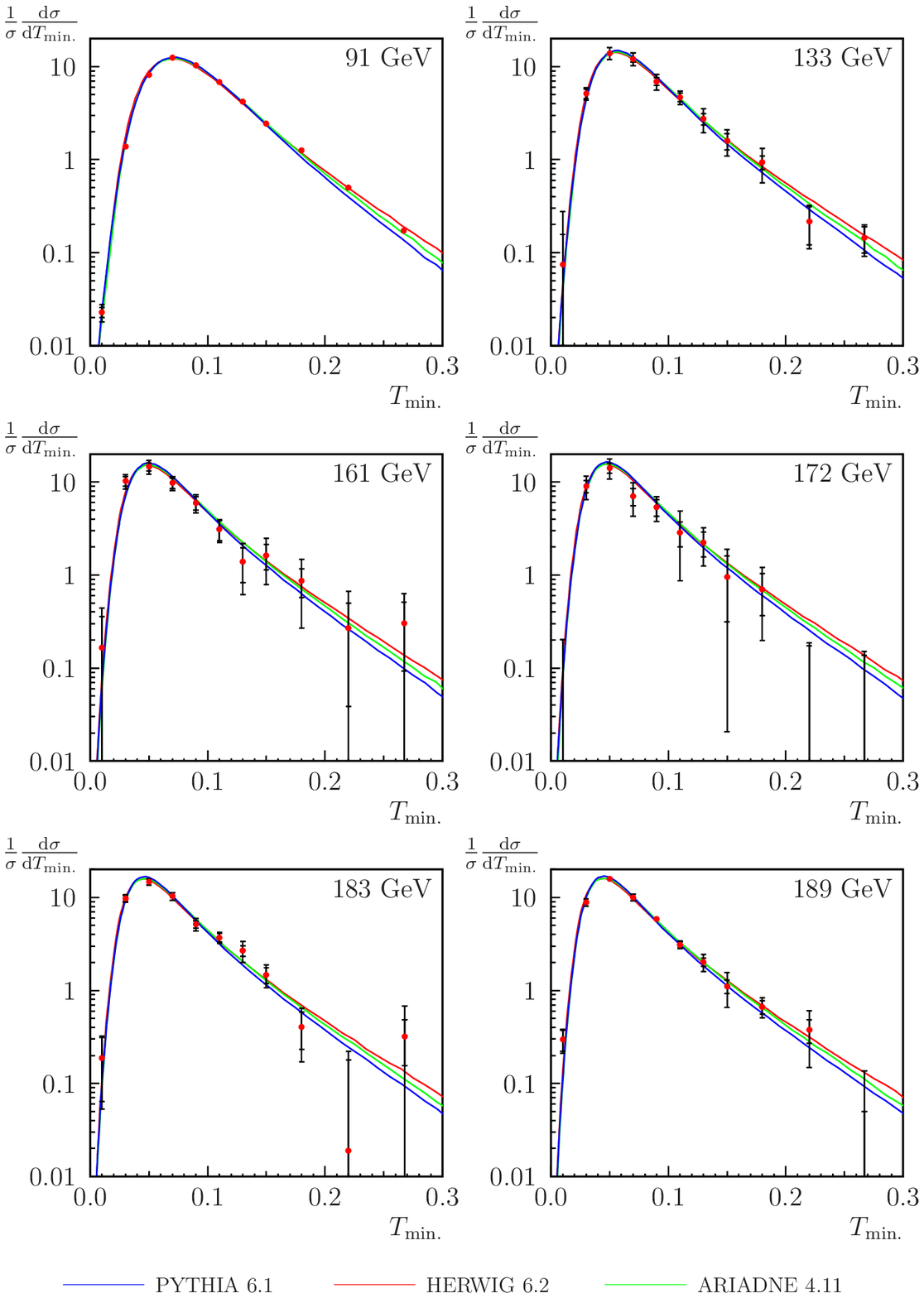}
\caption{Distributions for the thrust minor, $T_\mathrm{min.}$, measured by OPAL
at centre-of-mass energies $\sqrt{s}=91$--189~GeV. The inner error
bars indicate statistical uncertainties. Each curve is generated using
five million non-radiative Monte Carlo events, after hadronisation.}
\label{figdist17}
\end{center}
\end{figure}
\clearpage
\section*{Thrust minor, \boldmath{$T_\mathrm{min.}$}~(contd.)}

\begin{table}[hb!]
\begin{center}
\scalebox{0.90}{
\begin{minipage}{\linewidth}
\begin{center}
\begin{tabular}{| r @{ -- } l | r @{.} l @{ $\pm$ } r @{.} l @{ $\pm$ } r @{.} l @{} l | r @{.} l @{ $\pm$ } r @{.} l @{ $\pm$ } r @{.} l @{} l |}
\hline \multicolumn{2}{|c|}{$T_\mathrm{min.}$} & \multicolumn{7}{|c|}{$R'\,(T_\mathrm{min.})$ at 192 GeV} & \multicolumn{7}{|c|}{$R'\,(T_\mathrm{min.})$ at 196 GeV} \bigstrut \\ 
\hline \bigstrut[t]
$0.00$ & $0.02$  & 0 & 14 & 0 & 18 & 0 & 17 &  & 0 & 28 & 0 & 15 & 0 & 08 & \\ 
$0.02$ & $0.04$  & 8 & 4 & 1 & 2 & 2 & 9 &  & 9 & 1 & 0 & 7 & 1 & 1 & \\ 
$0.04$ & $0.06$  & 12 & 8 & 1 & 5 & 3 & 5 &  & 16 & 4 & 0 & 8 & 2 & 2 & \\ 
$0.06$ & $0.08$  & 7 & 6 & 1 & 1 & 1 & 5 &  & 8 & 93 & 0 & 61 & 0 & 62 & \\ 
$0.08$ & $0.10$  & 5 & 2 & 0 & 8 & 1 & 4 &  & 5 & 87 & 0 & 47 & 0 & 58 & \\ 
$0.10$ & $0.12$  & 2 & 89 & 0 & 61 & 0 & 66 &  & 3 & 14 & 0 & 37 & 0 & 51 & \\ 
$0.12$ & $0.14$  & 1 & 43 & 0 & 52 & 0 & 31 &  & 2 & 63 & 0 & 35 & 0 & 53 & \\
\hline 

\end{tabular}
\end{center}
\end{minipage}}
\end{center}
\caption{Distributions for the thrust minor, $T_\mathrm{min.}$, measured by
OPAL at centre-of-mass energies $\sqrt{s}=192$--207~GeV.
The first uncertainty is statistical, while the second is systematic.}
\label{tabdist27}
\end{table}
\clearpage

\begin{figure}[p]
\begin{center}
\includegraphics[width=\textwidth]{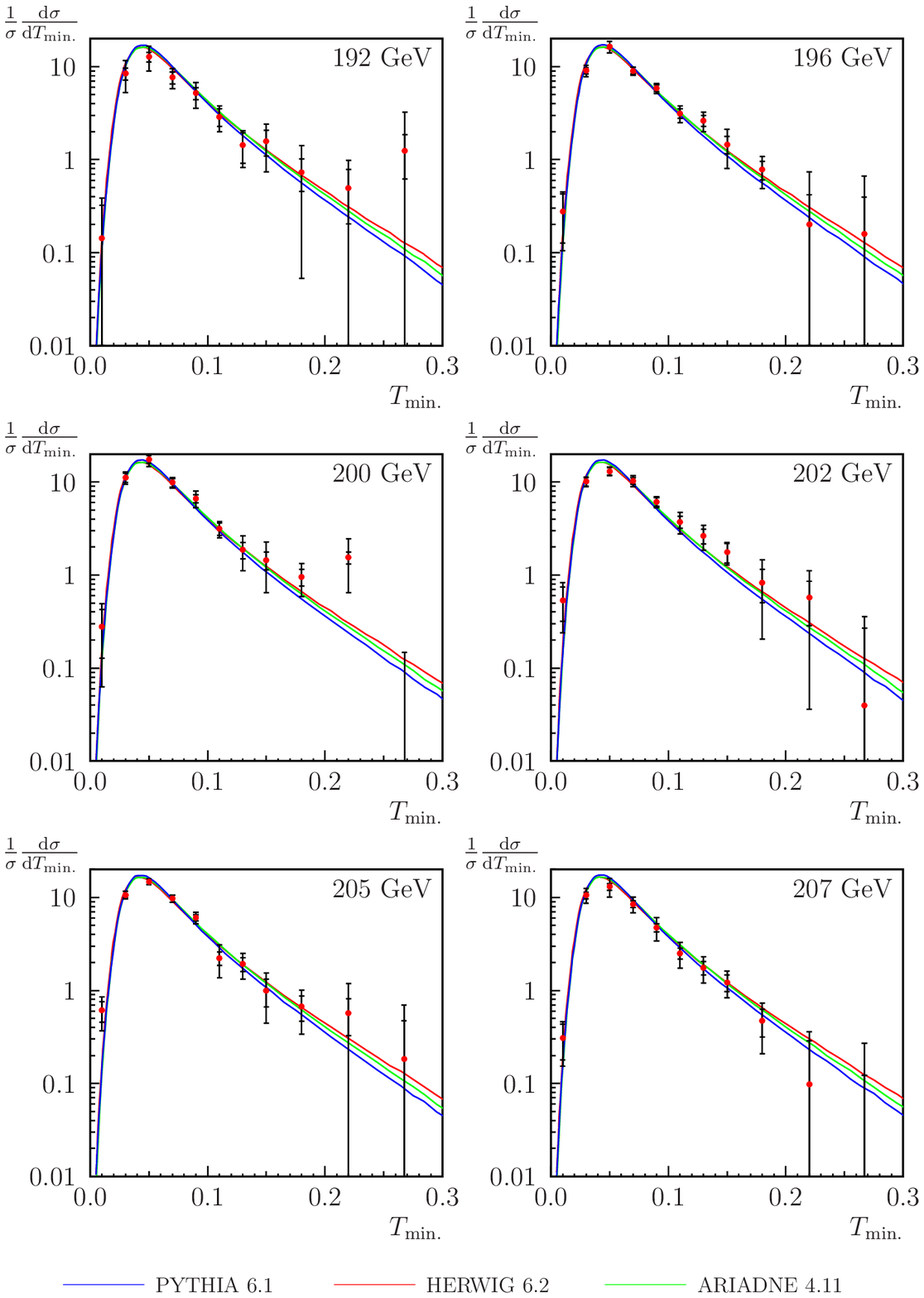}
\caption{Distributions for the thrust minor, $T_\mathrm{min.}$, measured by OPAL
at centre-of-mass energies $\sqrt{s}=192$--207~GeV. The inner error
bars indicate statistical uncertainties. Each curve is generated using
five million non-radiative Monte Carlo events, after hadronisation.}
\label{figdist27}
\end{center}
\end{figure}
\clearpage
\section[Aplanarity, $A$]{Aplanarity, \boldmath{$A$}}

\begin{table}[hb!]
\begin{center}
\scalebox{0.90}{
\begin{minipage}{\linewidth}
\begin{center}
\begin{tabular}{| r @{ -- } l | r @{.} l @{ $\pm$ } r @{.} l @{ $\pm$ } r @{.} l @{} l | r @{.} l @{ $\pm$ } r @{.} l @{ $\pm$ } r @{.} l @{} l |}
\hline \multicolumn{2}{|c|}{$A$} & \multicolumn{7}{|c|}{$R'\,(A)$ at 91 GeV} & \multicolumn{7}{|c|}{$R'\,(A)$ at 133 GeV} \bigstrut \\ 
\hline \bigstrut[t]
$0.000$ & $0.005$  & 76 & 9 & 0 & 2 & 1 & 2 &  & 115 & 0 & 3 & 8 & 7 & 4 & \\ 
$0.005$ & $0.010$  & 54 & 66 & 0 & 15 & 0 & 55 &  & 38 & 3 & 3 & 1 & 2 & 7 & \\ 
$0.010$ & $0.015$  & 25 & 50 & 0 & 11 & 0 & 41 &  & 16 & 8 & 2 & 0 & 1 & 9 & \\ 
$0.015$ & $0.025$  & 10 & 90 & 0 & 05 & 0 & 14 &  & 7 & 7 & 0 & 9 & 1 & 7 & \\ 
$0.025$ & $0.040$  & 3 & 708 & 0 & 023 & 0 & 064 &  & 2 & 60 & 0 & 47 & 0 & 24 & \\ 
$0.040$ & $0.070$  & 1 & 108 & 0 & 008 & 0 & 022 &  & 0 & 76 & 0 & 17 & 0 & 16 & \\ 
$0.070$ & $0.100$  & 0 & 320 & 0 & 005 & 0 & 011 &  & 0 & 204 & 0 & 076 & 0 & 072 &  \bigstrut[b] \\
\hline 

\end{tabular}
\end{center}
\end{minipage}}
\end{center}
\caption{Distributions for the aplanarity, $A$, measured by
OPAL at centre-of-mass energies $\sqrt{s}=91$--189~GeV.
The first uncertainty is statistical, while the second is systematic.}
\label{tabdist18}
\end{table}
\clearpage

\begin{figure}[p]
\begin{center}
\includegraphics[width=\textwidth]{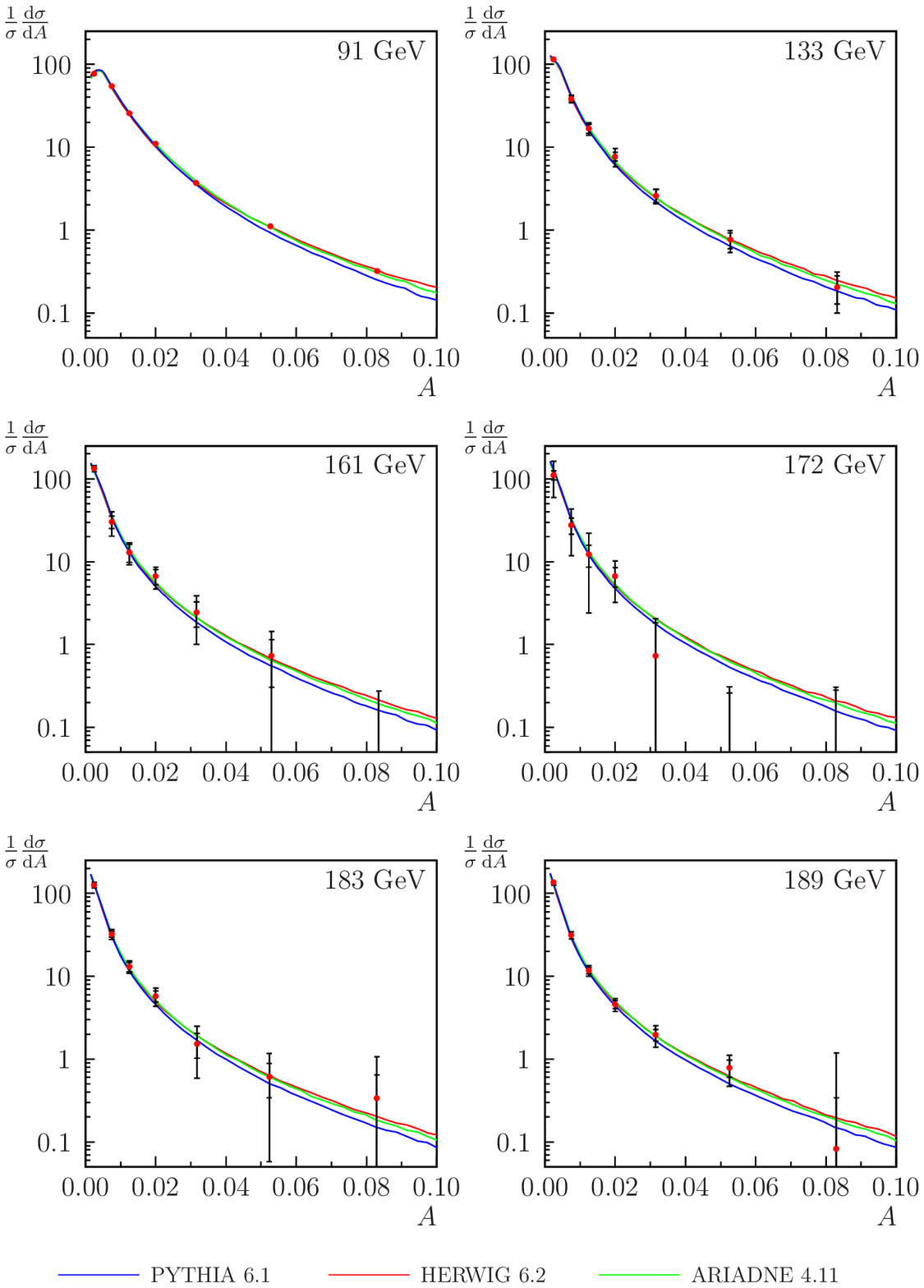}
\caption{Distributions for the aplanarity, $A$, measured by OPAL
at centre-of-mass energies $\sqrt{s}=91$--189~GeV. The inner error
bars indicate statistical uncertainties. Each curve is generated using
five million non-radiative Monte Carlo events, after hadronisation.}
\label{figdist18}
\end{center}
\end{figure}
\clearpage
\section*{Aplanarity, \boldmath{$A$}~(contd.)}

\begin{table}[hb!]
\begin{center}
\scalebox{0.90}{
\begin{minipage}{\linewidth}
\begin{center}
\begin{tabular}{| r @{ -- } l | r @{.} l @{ $\pm$ } r @{.} l @{ $\pm$ } r @{.} l @{} l | r @{.} l @{ $\pm$ } r @{.} l @{ $\pm$ } r @{.} l @{} l |}
\hline \multicolumn{2}{|c|}{$A$} & \multicolumn{7}{|c|}{$R'\,(A)$ at 192 GeV} & \multicolumn{7}{|c|}{$R'\,(A)$ at 196 GeV} \bigstrut \\ 
\hline \bigstrut[t]
$0.000$ & $0.005$  & 123 & & 12 & & 13 & &  & 136 & 1 & 4 & 8 & 9 & 7 & \\ 
$0.005$ & $0.010$  & 29 & 2 & 4 & 2 & 6 & 6 &  & 32 & 6 & 2 & 4 & 3 & 0 & \\ 
$0.010$ & $0.015$  & 12 & 0 & 2 & 7 & 7 & 0 &  & 12 & 1 & 1 & 6 & 1 & 9 & \\ 
$0.015$ & $0.025$  & 5 & 0 & 1 & 4 & 1 & 7 &  & 5 & 98 & 0 & 87 & 0 & 96 & \\ 
$0.025$ & $0.040$  & 3 & 3 & 0 & 8 & 1 & 4 &  & 1 & 59 & 0 & 53 & 0 & 50 & \\ 
$0.040$ & $0.070$  & 0 & 4 & 0 & 5 & 1 & 1 &  & 0 & 68 & 0 & 36 & 0 & 73 & \\ 
$0.070$ & $0.100$  & 0 & 92 & 0 & 49 & 0 & 87 &  & $-$0 & 03 & 0 & 46 & 0 & 67 &  \bigstrut[b] \\
\hline 

\end{tabular}
\end{center}
\end{minipage}}
\end{center}
\caption{Distributions for the aplanarity, $A$, measured by
OPAL at centre-of-mass energies $\sqrt{s}=192$--207~GeV.
The first uncertainty is statistical, while the second is systematic.}
\label{tabdist28}
\end{table}
\clearpage

\begin{figure}[p]
\begin{center}
\includegraphics[width=\textwidth]{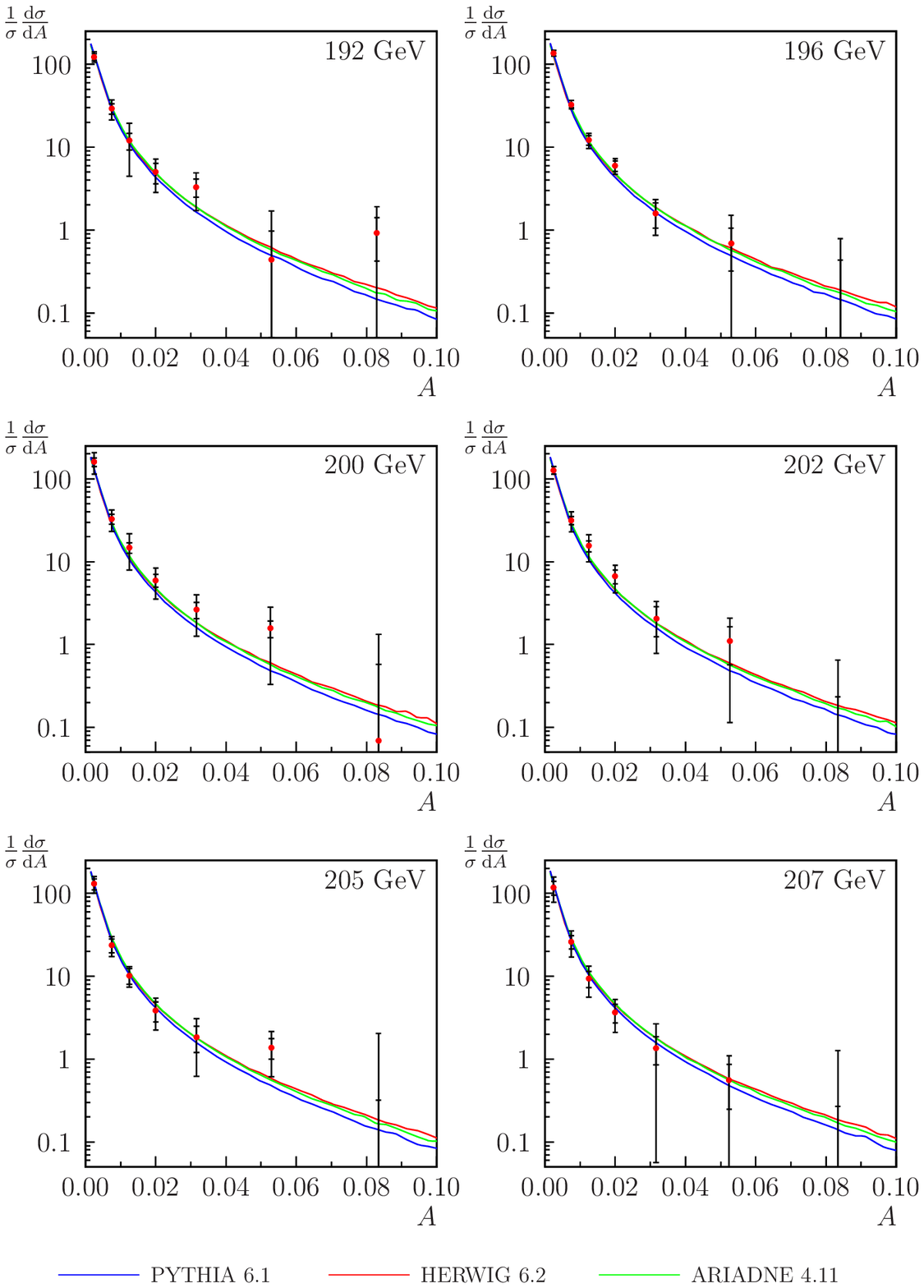}
\caption{Distributions for the aplanarity, $A$, measured by OPAL
at centre-of-mass energies $\sqrt{s}=192$--207~GeV. The inner error
bars indicate statistical uncertainties. Each curve is generated using
five million non-radiative Monte Carlo events, after hadronisation.}
\label{figdist28}
\end{center}
\end{figure}
\clearpage
\section[Sphericity, $S$]{Sphericity, \boldmath{$S$}}

\begin{table}[hb!]
\begin{center}
\scalebox{0.90}{
\begin{minipage}{\linewidth}
\begin{center}
\begin{tabular}{| r @{ -- } l | r @{.} l @{ $\pm$ } r @{.} l @{ $\pm$ } r @{.} l @{} l | r @{.} l @{ $\pm$ } r @{.} l @{ $\pm$ } r @{.} l @{} l |}
\hline \multicolumn{2}{|c|}{$S$} & \multicolumn{7}{|c|}{$R'\,(S)$ at 91 GeV} & \multicolumn{7}{|c|}{$R'\,(S)$ at 133 GeV} \bigstrut \\ 
\hline \bigstrut[t]
$0.00$ & $0.02$  & 18 & 06 & 0 & 04 & 0 & 41 &  & 23 & 6 & 1 & 0 & 1 & 8 & \\ 
$0.02$ & $0.04$  & 10 & 56 & 0 & 03 & 0 & 16 &  & 8 & 15 & 0 & 72 & 0 & 77 & \\ 
$0.04$ & $0.06$  & 5 & 185 & 0 & 024 & 0 & 068 &  & 4 & 07 & 0 & 53 & 0 & 28 & \\ 
$0.06$ & $0.12$  & 2 & 408 & 0 & 009 & 0 & 051 &  & 1 & 65 & 0 & 22 & 0 & 13 & \\ 
$0.12$ & $0.20$  & 0 & 989 & 0 & 005 & 0 & 033 &  & 0 & 93 & 0 & 12 & 0 & 22 & \\ 
$0.20$ & $0.30$  & 0 & 467 & 0 & 003 & 0 & 015 &  & 0 & 377 & 0 & 068 & 0 & 048 & \\ 
$0.30$ & $0.50$  & 0 & 200 & 0 & 002 & 0 & 011 &  & 0 & 293 & 0 & 034 & 0 & 033 & \\
\hline 

\end{tabular}
\end{center}
\end{minipage}}
\end{center}
\caption{Distributions for the sphericity, $S$, measured by
OPAL at centre-of-mass energies $\sqrt{s}=91$--189~GeV.
The first uncertainty is statistical, while the second is systematic.}
\label{tabdist19}
\end{table}
\clearpage

\begin{figure}[p]
\begin{center}
\includegraphics[width=\textwidth]{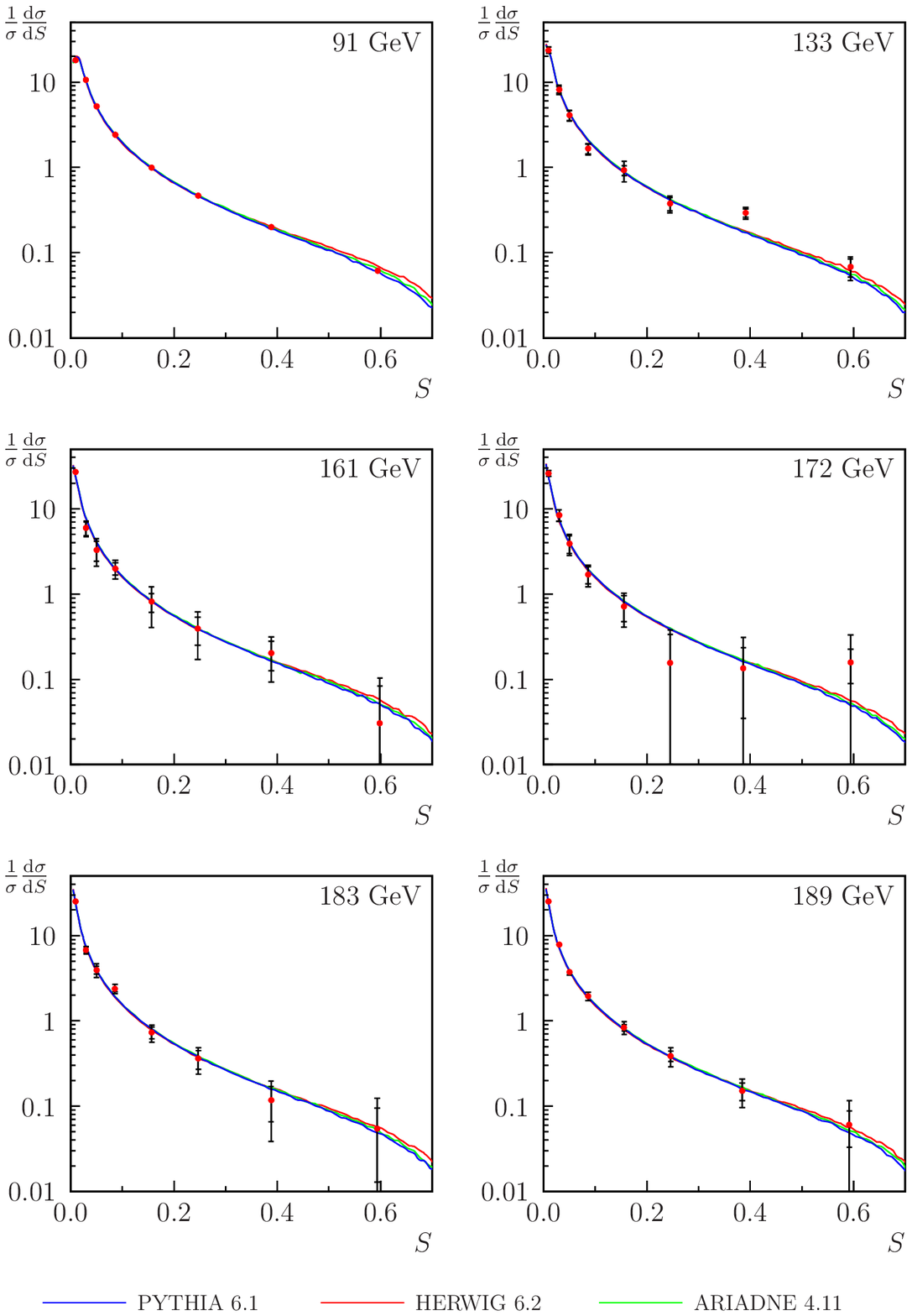}
\caption{Distributions for the sphericity, $S$, measured by OPAL
at centre-of-mass energies $\sqrt{s}=91$--189~GeV. The inner error
bars indicate statistical uncertainties. Each curve is generated using
five million non-radiative Monte Carlo events, after hadronisation.}
\label{figdist19}
\end{center}
\end{figure}
\clearpage
\section*{Sphericity, \boldmath{$S$}~(contd.)}

\begin{table}[hb!]
\begin{center}
\scalebox{0.90}{
\begin{minipage}{\linewidth}
\begin{center}
\begin{tabular}{| r @{ -- } l | r @{.} l @{ $\pm$ } r @{.} l @{ $\pm$ } r @{.} l @{} l | r @{.} l @{ $\pm$ } r @{.} l @{ $\pm$ } r @{.} l @{} l |}
\hline \multicolumn{2}{|c|}{$S$} & \multicolumn{7}{|c|}{$R'\,(S)$ at 192 GeV} & \multicolumn{7}{|c|}{$R'\,(S)$ at 196 GeV} \bigstrut \\ 
\hline \bigstrut[t]
$0.00$ & $0.02$  & 25 & 1 & 1 & 4 & 1 & 9 &  & 26 & 3 & 1 & 0 & 1 & 1 & \\ 
$0.02$ & $0.04$  & 6 & 42 & 0 & 85 & 0 & 49 &  & 6 & 44 & 0 & 56 & 0 & 41 & \\ 
$0.04$ & $0.06$  & 3 & 93 & 0 & 70 & 0 & 76 &  & 3 & 42 & 0 & 43 & 0 & 72 & \\ 
$0.06$ & $0.12$  & 1 & 76 & 0 & 27 & 0 & 24 &  & 2 & 07 & 0 & 18 & 0 & 27 & \\ 
$0.12$ & $0.20$  & 0 & 98 & 0 & 19 & 0 & 28 &  & 0 & 83 & 0 & 12 & 0 & 18 & \\ 
$0.20$ & $0.30$  & 0 & 25 & 0 & 13 & 0 & 17 &  & 0 & 373 & 0 & 092 & 0 & 070 & \\ 
$0.30$ & $0.50$  & 0 & 31 & 0 & 09 & 0 & 14 &  & 0 & 155 & 0 & 061 & 0 & 087 & \\
\hline 

\end{tabular}
\end{center}
\end{minipage}}
\end{center}
\caption{Distributions for the sphericity, $S$, measured by
OPAL at centre-of-mass energies $\sqrt{s}=192$--207~GeV.
The first uncertainty is statistical, while the second is systematic.}
\label{tabdist29}
\end{table}
\clearpage

\begin{figure}[p]
\begin{center}
\includegraphics[width=\textwidth]{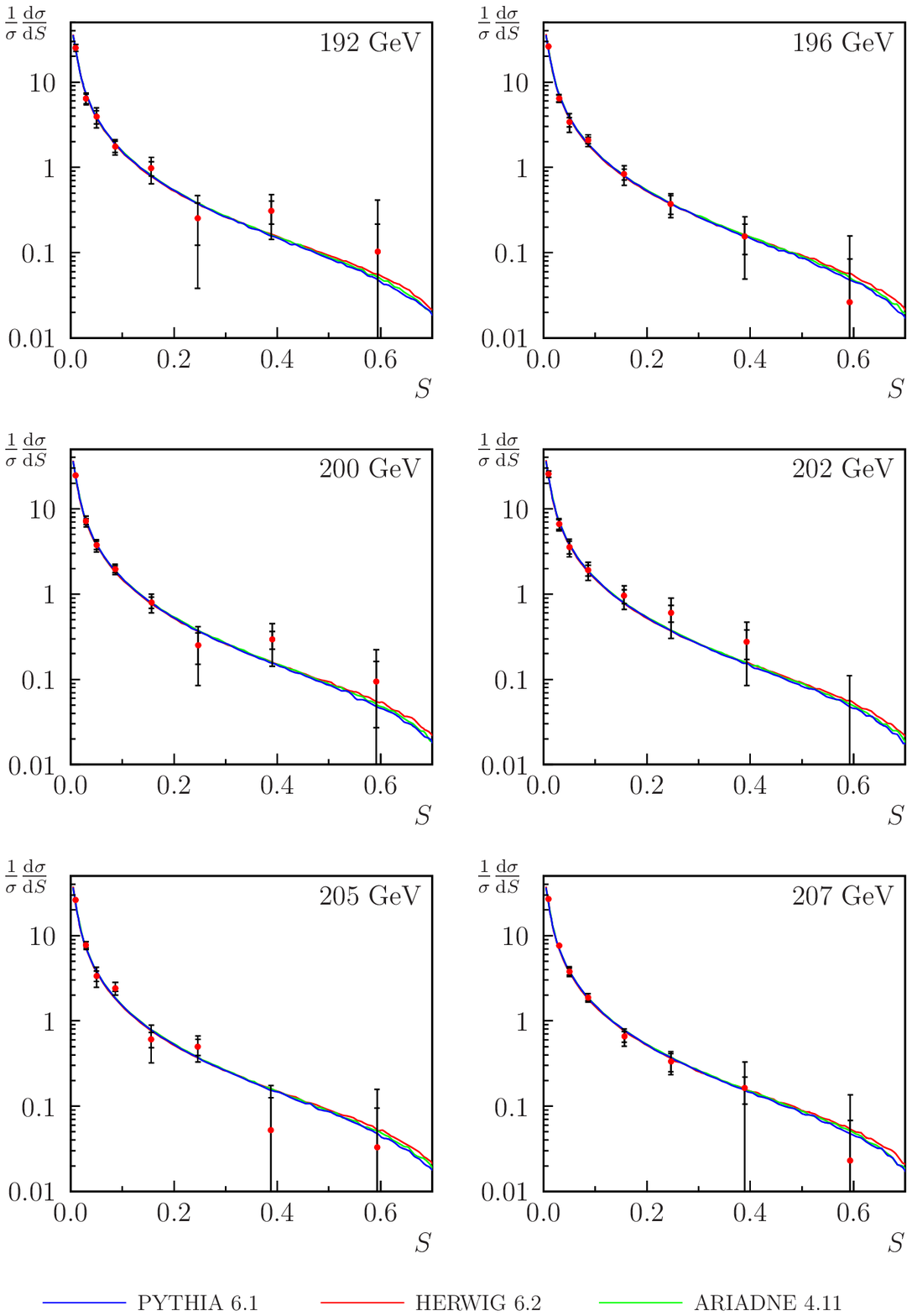}
\caption{Distributions for the sphericity, $S$, measured by OPAL
at centre-of-mass energies $\sqrt{s}=192$--207~GeV. The inner error
bars indicate statistical uncertainties. Each curve is generated using
five million non-radiative Monte Carlo events, after hadronisation.}
\label{figdist29}
\end{center}
\end{figure}
\clearpage
\section[Oblateness, $O$]{Oblateness, \boldmath{$O$}}

\begin{table}[hb!]
\begin{center}
\scalebox{0.90}{
\begin{minipage}{\linewidth}
\begin{center}
\begin{tabular}{| r @{ -- } l | r @{.} l @{ $\pm$ } r @{.} l @{ $\pm$ } r @{.} l @{} l | r @{.} l @{ $\pm$ } r @{.} l @{ $\pm$ } r @{.} l @{} l |}
\hline \multicolumn{2}{|c|}{$O$} & \multicolumn{7}{|c|}{$R'\,(O)$ at 91 GeV} & \multicolumn{7}{|c|}{$R'\,(O)$ at 133 GeV} \bigstrut \\ 
\hline \bigstrut[t]
$0.00$ & $0.05$  & 9 & 93 & 0 & 02 & 0 & 14 &  & 9 & 71 & 0 & 38 & 0 & 20 & \\ 
$0.05$ & $0.10$  & 4 & 579 & 0 & 013 & 0 & 041 &  & 4 & 52 & 0 & 31 & 0 & 24 & \\ 
$0.10$ & $0.15$  & 2 & 273 & 0 & 010 & 0 & 037 &  & 2 & 17 & 0 & 25 & 0 & 13 & \\ 
$0.15$ & $0.20$  & 1 & 307 & 0 & 008 & 0 & 031 &  & 1 & 20 & 0 & 19 & 0 & 13 & \\ 
$0.20$ & $0.25$  & 0 & 811 & 0 & 007 & 0 & 016 &  & 0 & 82 & 0 & 15 & 0 & 07 & \\ 
$0.25$ & $0.30$  & 0 & 489 & 0 & 005 & 0 & 022 &  & 0 & 65 & 0 & 11 & 0 & 13 & \\ 
$0.30$ & $0.40$  & 0 & 2364 & 0 & 0025 & 0 & 0078 &  & 0 & 318 & 0 & 062 & 0 & 045 & \\
\hline 

\end{tabular}
\end{center}
\end{minipage}}
\end{center}
\caption{Distributions for the oblateness, $O$, measured by
OPAL at centre-of-mass energies $\sqrt{s}=91$--189~GeV.
The first uncertainty is statistical, while the second is systematic.}
\label{tabdist110}
\end{table}
\clearpage

\begin{figure}[p]
\begin{center}
\includegraphics[width=\textwidth]{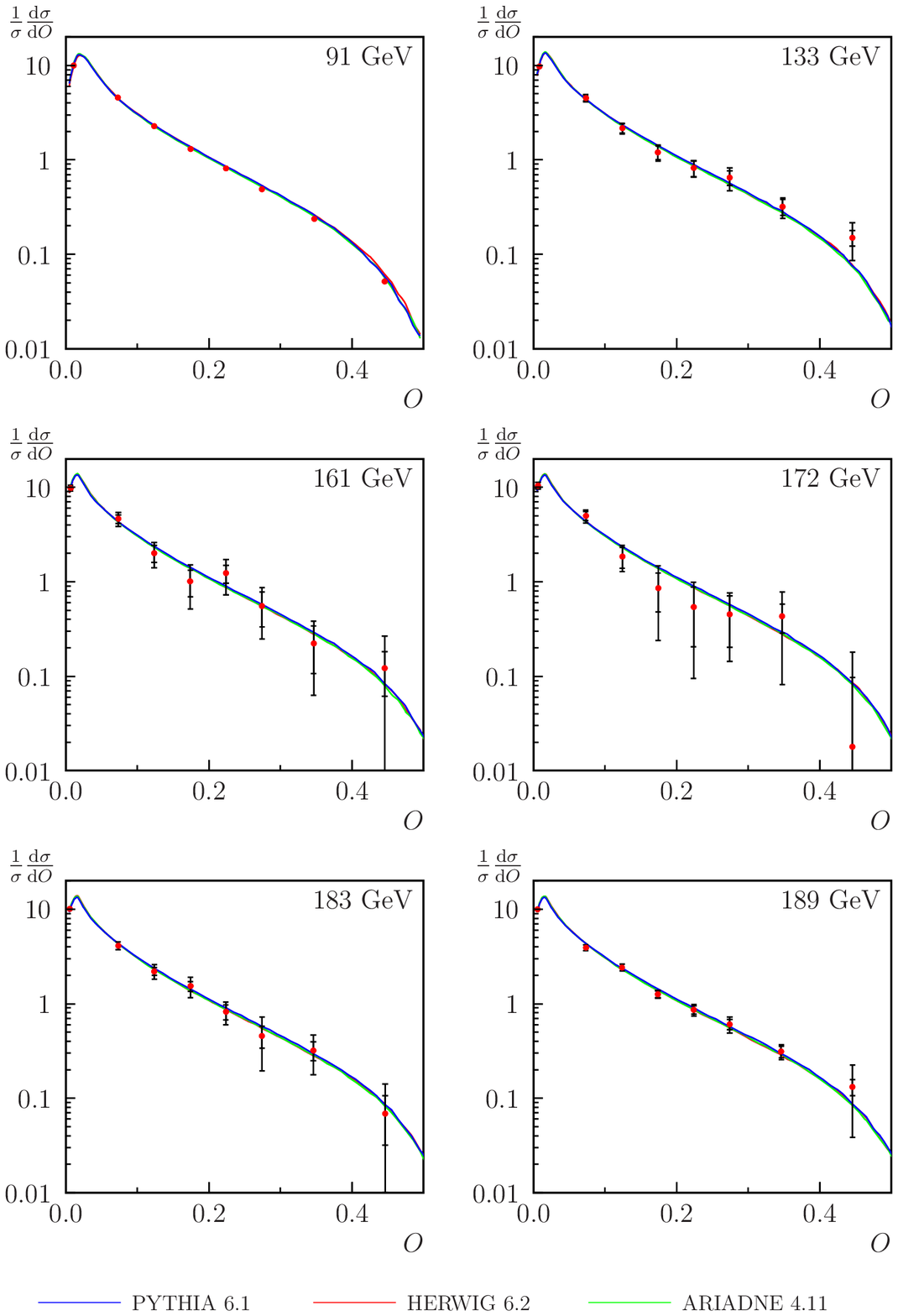}
\caption{Distributions for the oblateness, $O$, measured by OPAL
at centre-of-mass energies $\sqrt{s}=91$--189~GeV. The inner error
bars indicate statistical uncertainties. Each curve is generated using
five million non-radiative Monte Carlo events, after hadronisation.}
\label{figdist110}
\end{center}
\end{figure}
\clearpage
\section*{Oblateness, \boldmath{$O$}~(contd.)}

\begin{table}[hb!]
\begin{center}
\scalebox{0.90}{
\begin{minipage}{\linewidth}
\begin{center}
\begin{tabular}{| r @{ -- } l | r @{.} l @{ $\pm$ } r @{.} l @{ $\pm$ } r @{.} l @{} l | r @{.} l @{ $\pm$ } r @{.} l @{ $\pm$ } r @{.} l @{} l |}
\hline \multicolumn{2}{|c|}{$O$} & \multicolumn{7}{|c|}{$R'\,(O)$ at 192 GeV} & \multicolumn{7}{|c|}{$R'\,(O)$ at 196 GeV} \bigstrut \\ 
\hline \bigstrut[t]
$0.00$ & $0.05$  & 10 & 37 & 0 & 49 & 0 & 65 &  & 10 & 00 & 0 & 31 & 0 & 40 & \\ 
$0.05$ & $0.10$  & 4 & 24 & 0 & 40 & 0 & 48 &  & 4 & 41 & 0 & 25 & 0 & 39 & \\ 
$0.10$ & $0.15$  & 2 & 36 & 0 & 32 & 0 & 19 &  & 2 & 06 & 0 & 21 & 0 & 07 & \\ 
$0.15$ & $0.20$  & 1 & 15 & 0 & 27 & 0 & 41 &  & 1 & 21 & 0 & 17 & 0 & 26 & \\ 
$0.20$ & $0.25$  & 0 & 64 & 0 & 24 & 0 & 21 &  & 0 & 72 & 0 & 15 & 0 & 23 & \\ 
$0.25$ & $0.30$  & 0 & 52 & 0 & 19 & 0 & 13 &  & 0 & 71 & 0 & 13 & 0 & 22 & \\ 
$0.30$ & $0.40$  & 0 & 26 & 0 & 12 & 0 & 14 &  & 0 & 26 & 0 & 07 & 0 & 12 & \\
\hline 

\end{tabular}
\end{center}
\end{minipage}}
\end{center}
\caption{Distributions for the oblateness, $O$, measured by
OPAL at centre-of-mass energies $\sqrt{s}=192$--207~GeV.
The first uncertainty is statistical, while the second is systematic.}
\label{tabdist210}
\end{table}
\clearpage

\begin{figure}[p]
\begin{center}
\includegraphics[width=\textwidth]{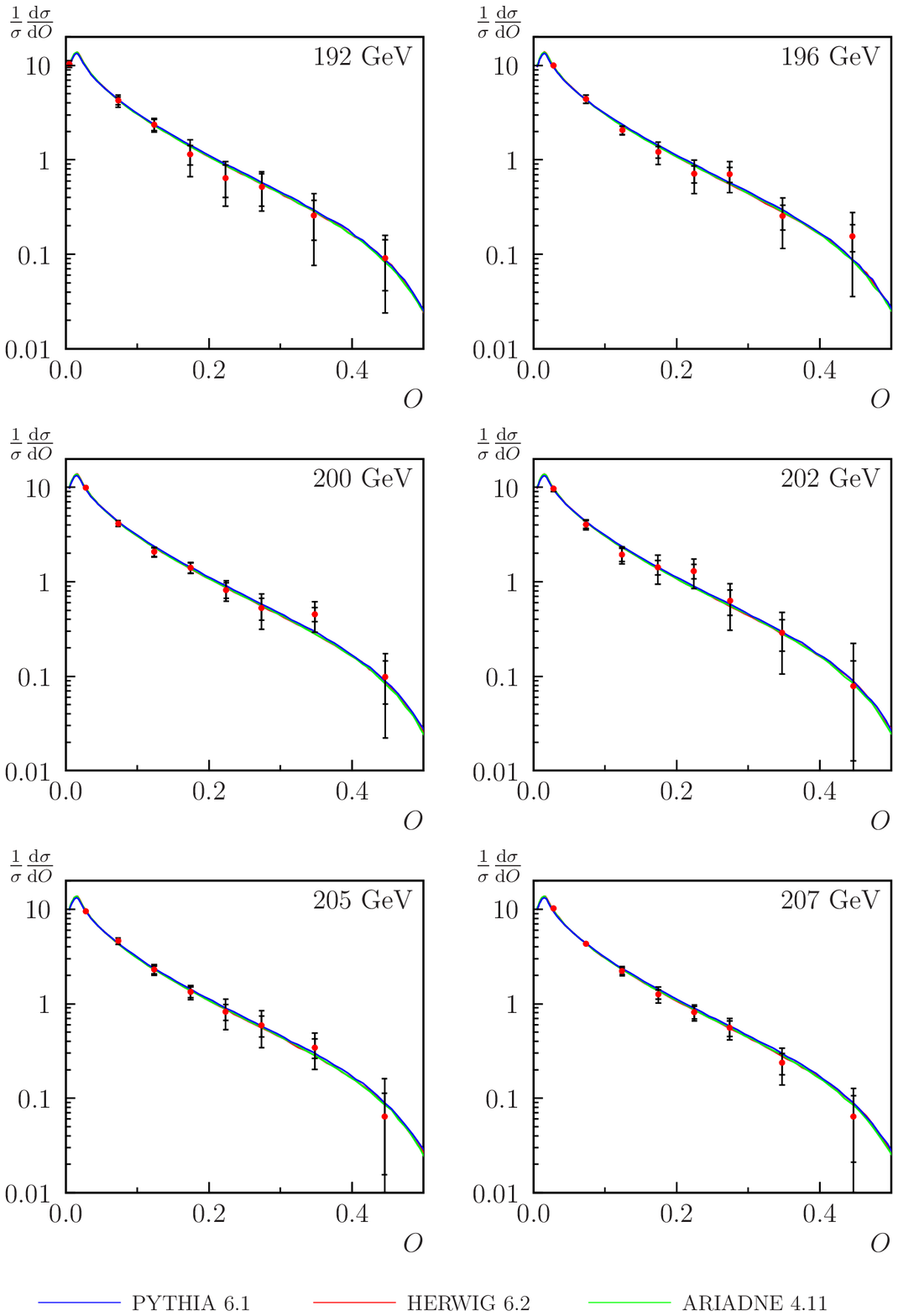}
\caption{Distributions for the oblateness, $O$, measured by OPAL
at centre-of-mass energies $\sqrt{s}=192$--207~GeV. The inner error
bars indicate statistical uncertainties. Each curve is generated using
five million non-radiative Monte Carlo events, after hadronisation.}
\label{figdist210}
\end{center}
\end{figure}
\clearpage
\section[Light jet mass, $M_\mathrm{L}$]{Light jet mass, \boldmath{$M_\mathrm{L}$}}

\begin{table}[hb!]
\begin{center}
\scalebox{0.90}{
\begin{minipage}{\linewidth}
\begin{center}
\begin{tabular}{| r @{ -- } l | r @{.} l @{ $\pm$ } r @{.} l @{ $\pm$ } r @{.} l @{} l | r @{.} l @{ $\pm$ } r @{.} l @{ $\pm$ } r @{.} l @{} l |}
\hline \multicolumn{2}{|c|}{$M_\mathrm{L}/\sqrt{s}$} & \multicolumn{7}{|c|}{$R'\,(M_\mathrm{L}/\sqrt{s})$ at 91 GeV} & \multicolumn{7}{|c|}{$R'\,(M_\mathrm{L}/\sqrt{s})$ at 133 GeV} \bigstrut \\ 
\hline \bigstrut[t]
$0.00$ & $0.04$  & 0 & 124 & 0 & 002 & 0 & 038 &  & 0 & 292 & 0 & 067 & 0 & 080 & \\ 
$0.04$ & $0.06$  & 1 & 108 & 0 & 008 & 0 & 089 &  & 2 & 29 & 0 & 29 & 0 & 57 & \\ 
$0.06$ & $0.08$  & 3 & 72 & 0 & 02 & 0 & 55 &  & 7 & 5 & 0 & 6 & 1 & 4 & \\ 
$0.08$ & $0.10$  & 8 & 22 & 0 & 03 & 0 & 56 &  & 11 & 1 & 0 & 8 & 1 & 7 & \\ 
$0.10$ & $0.12$  & 10 & 73 & 0 & 04 & 0 & 37 &  & 10 & 3 & 0 & 8 & 1 & 1 & \\ 
$0.12$ & $0.14$  & 9 & 11 & 0 & 04 & 0 & 54 &  & 6 & 2 & 0 & 7 & 1 & 0 & \\ 
$0.14$ & $0.16$  & 6 & 22 & 0 & 03 & 0 & 22 &  & 4 & 7 & 0 & 6 & 1 & 1 & \\
\hline 

\end{tabular}
\end{center}
\end{minipage}}
\end{center}
\caption{Distributions for the light jet mass, $M_\mathrm{L}$, measured by
OPAL at centre-of-mass energies $\sqrt{s}=91$--189~GeV.
The first uncertainty is statistical, while the second is systematic.}
\label{tabdist111}
\end{table}
\clearpage

\begin{figure}[p]
\begin{center}
\includegraphics[width=\textwidth]{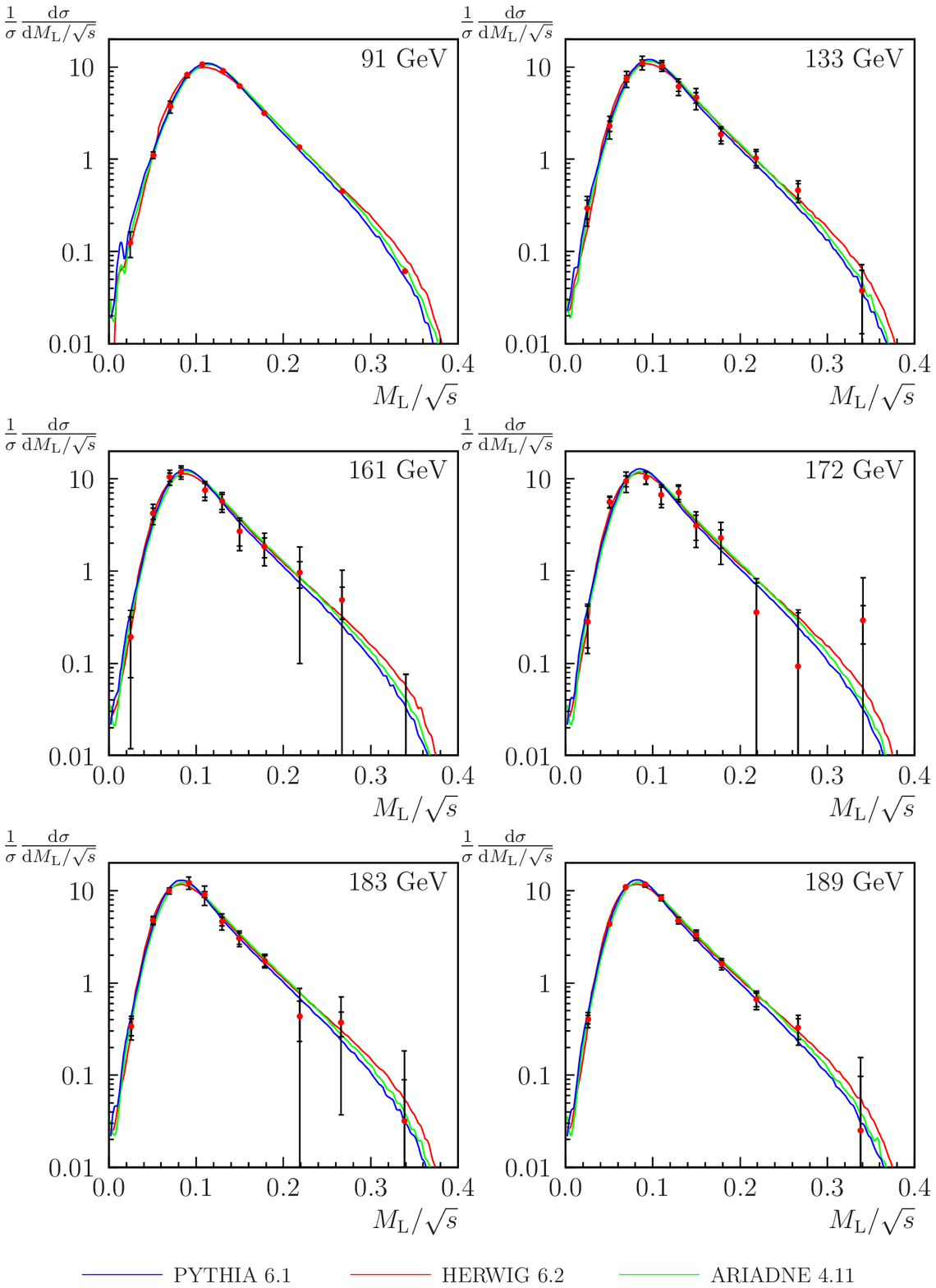}
\caption{Distributions for the light jet mass, $M_\mathrm{L}$, measured by OPAL
at centre-of-mass energies $\sqrt{s}=91$--189~GeV. The inner error
bars indicate statistical uncertainties. Each curve is generated using
five million non-radiative Monte Carlo events, after hadronisation.}
\label{figdist111}
\end{center}
\end{figure}
\clearpage
\section*{Light jet mass, \boldmath{$M_\mathrm{L}$}~(contd.)}

\begin{table}[hb!]
\begin{center}
\scalebox{0.90}{
\begin{minipage}{\linewidth}
\begin{center}
\begin{tabular}{| r @{ -- } l | r @{.} l @{ $\pm$ } r @{.} l @{ $\pm$ } r @{.} l @{} l | r @{.} l @{ $\pm$ } r @{.} l @{ $\pm$ } r @{.} l @{} l |}
\hline \multicolumn{2}{|c|}{$M_\mathrm{L}/\sqrt{s}$} & \multicolumn{7}{|c|}{$R'\,(M_\mathrm{L}/\sqrt{s})$ at 192 GeV} & \multicolumn{7}{|c|}{$R'\,(M_\mathrm{L}/\sqrt{s})$ at 196 GeV} \bigstrut \\ 
\hline \bigstrut[t]
$0.00$ & $0.04$  & 0 & 26 & 0 & 12 & 0 & 06 &  & 0 & 48 & 0 & 07 & 0 & 11 & \\ 
$0.04$ & $0.06$  & 3 & 94 & 0 & 57 & 0 & 55 &  & 4 & 59 & 0 & 40 & 0 & 64 & \\ 
$0.06$ & $0.08$  & 9 & 27 & 0 & 97 & 0 & 75 &  & 11 & 10 & 0 & 73 & 0 & 65 & \\ 
$0.08$ & $0.10$  & 12 & 2 & 1 & 1 & 1 & 4 &  & 12 & 65 & 0 & 79 & 0 & 86 & \\ 
$0.10$ & $0.12$  & 8 & 4 & 0 & 9 & 1 & 4 &  & 7 & 96 & 0 & 65 & 0 & 42 & \\ 
$0.12$ & $0.14$  & 4 & 7 & 0 & 8 & 1 & 4 &  & 4 & 51 & 0 & 51 & 0 & 57 & \\ 
$0.14$ & $0.16$  & 4 & 44 & 0 & 67 & 0 & 55 &  & 2 & 89 & 0 & 42 & 0 & 28 & \\
\hline 

\end{tabular}
\end{center}
\end{minipage}}
\end{center}
\caption{Distributions for the light jet mass, $M_\mathrm{L}$, measured by
OPAL at centre-of-mass energies $\sqrt{s}=192$--207~GeV.
The first uncertainty is statistical, while the second is systematic.}
\label{tabdist211}
\end{table}
\clearpage

\begin{figure}[p]
\begin{center}
\includegraphics[width=\textwidth]{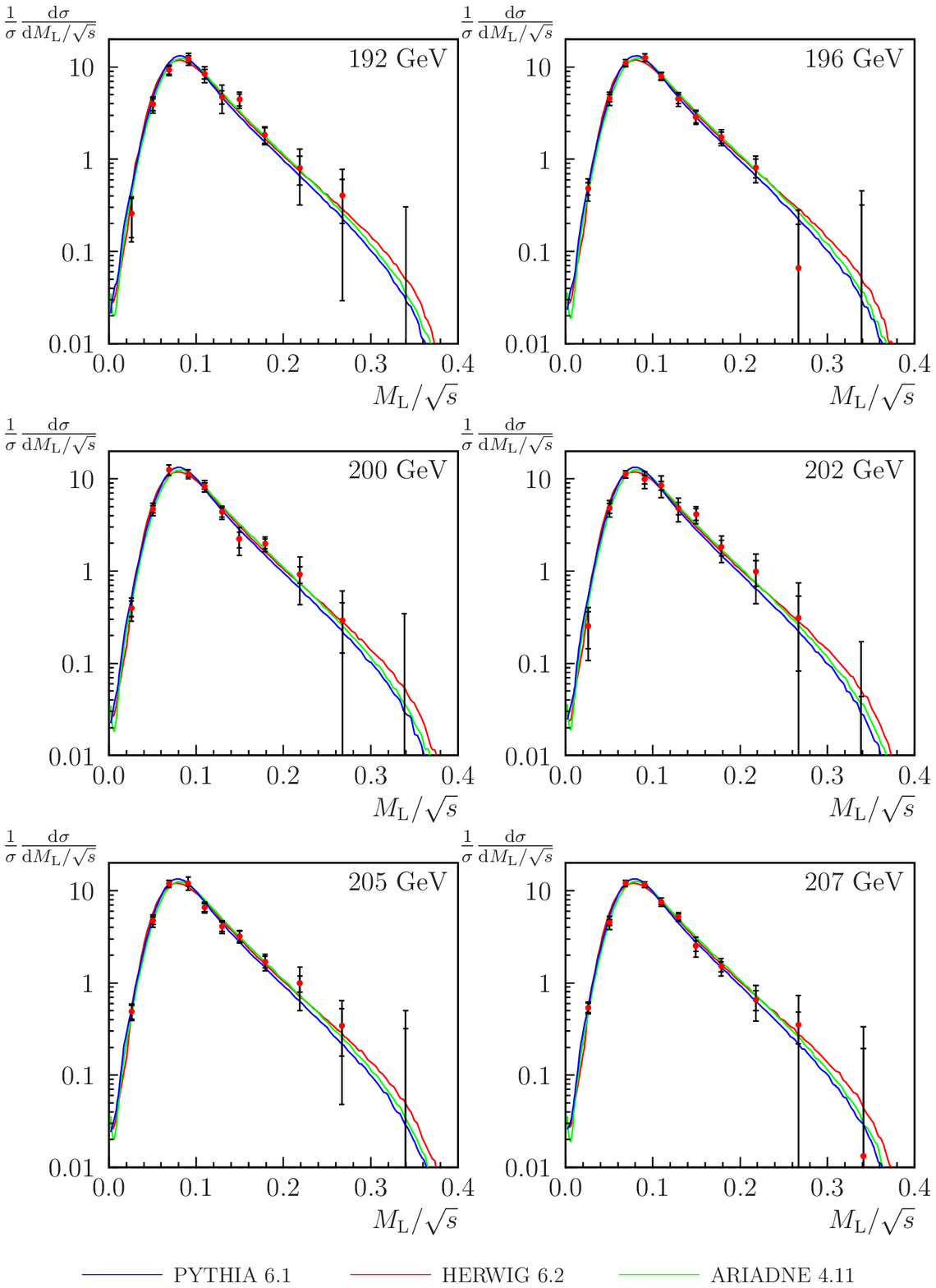}
\caption{Distributions for the light jet mass, $M_\mathrm{L}$, measured by OPAL
at centre-of-mass energies $\sqrt{s}=192$--207~GeV. The inner error
bars indicate statistical uncertainties. Each curve is generated using
five million non-radiative Monte Carlo events, after hadronisation.}
\label{figdist211}
\end{center}
\end{figure}
\clearpage
\section[Narrow jet broadening, $B_\mathrm{N}$]{Narrow jet broadening, \boldmath{$B_\mathrm{N}$}}

\begin{table}[hb!]
\begin{center}
\scalebox{0.90}{
\begin{minipage}{\linewidth}
\begin{center}
\begin{tabular}{| r @{ -- } l | r @{.} l @{ $\pm$ } r @{.} l @{ $\pm$ } r @{.} l @{} l | r @{.} l @{ $\pm$ } r @{.} l @{ $\pm$ } r @{.} l @{} l |}
\hline \multicolumn{2}{|c|}{$B_\mathrm{N}$} & \multicolumn{7}{|c|}{$R'\,(B_\mathrm{N})$ at 91 GeV} & \multicolumn{7}{|c|}{$R'\,(B_\mathrm{N})$ at 133 GeV} \bigstrut \\ 
\hline \bigstrut[t]
$0.000$ & $0.010$  & 0 & 693 & 0 & 020 & 0 & 081 &  & 1 & 83 & 0 & 71 & 1 & 00 & \\ 
$0.010$ & $0.015$  & 8 & 00 & 0 & 08 & 0 & 23 &  & 30 & 0 & 2 & 8 & 1 & 6 & \\ 
$0.015$ & $0.020$  & 21 & 56 & 0 & 11 & 0 & 35 &  & 37 & 7 & 3 & 2 & 4 & 9 & \\ 
$0.020$ & $0.025$  & 32 & 43 & 0 & 12 & 0 & 46 &  & 34 & 1 & 3 & 2 & 6 & 9 & \\ 
$0.025$ & $0.030$  & 33 & 64 & 0 & 13 & 0 & 70 &  & 26 & 3 & 2 & 5 & 3 & 7 & \\ 
$0.030$ & $0.035$  & 28 & 37 & 0 & 11 & 0 & 17 &  & 15 & 6 & 2 & 0 & 1 & 0 & \\ 
$0.035$ & $0.040$  & 19 & 99 & 0 & 09 & 0 & 29 &  & 10 & 3 & 1 & 7 & 1 & 2 & \\
\hline 

\end{tabular}
\end{center}
\end{minipage}}
\end{center}
\caption{Distributions for the narrow jet broadening, $B_\mathrm{N}$, measured by
OPAL at centre-of-mass energies $\sqrt{s}=91$--189~GeV.
The first uncertainty is statistical, while the second is systematic.}
\label{tabdist112}
\end{table}
\clearpage

\begin{figure}[p]
\begin{center}
\includegraphics[width=\textwidth]{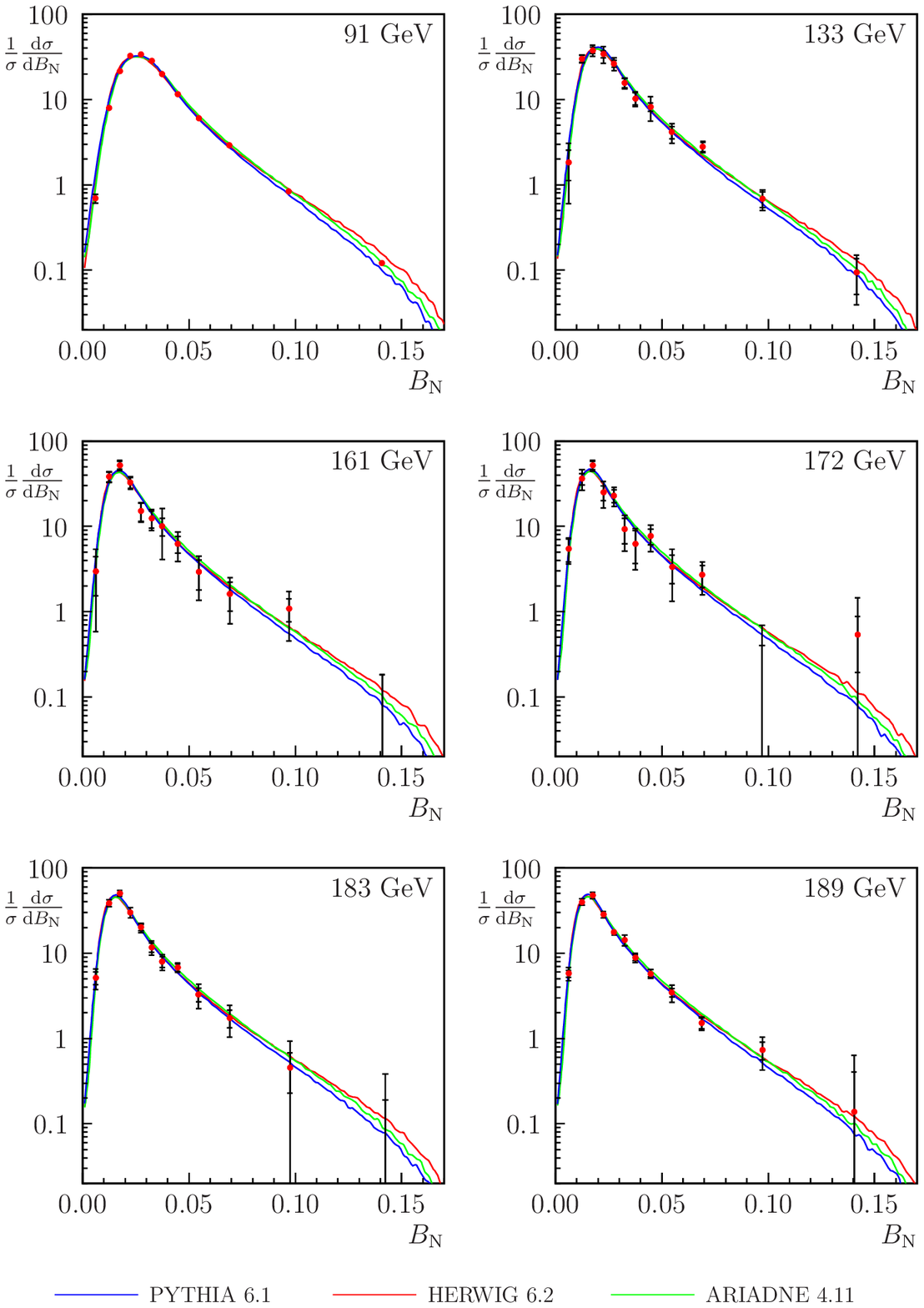}
\caption{Distributions for the narrow jet broadening, $B_\mathrm{N}$, measured by OPAL
at centre-of-mass energies $\sqrt{s}=91$--189~GeV. The inner error
bars indicate statistical uncertainties. Each curve is generated using
five million non-radiative Monte Carlo events, after hadronisation.}
\label{figdist112}
\end{center}
\end{figure}
\clearpage
\section*{Narrow jet broadening, \boldmath{$B_\mathrm{N}$}~(contd.)}

\begin{table}[hb!]
\begin{center}
\scalebox{0.90}{
\begin{minipage}{\linewidth}
\begin{center}
\begin{tabular}{| r @{ -- } l | r @{.} l @{ $\pm$ } r @{.} l @{ $\pm$ } r @{.} l @{} l | r @{.} l @{ $\pm$ } r @{.} l @{ $\pm$ } r @{.} l @{} l |}
\hline \multicolumn{2}{|c|}{$B_\mathrm{N}$} & \multicolumn{7}{|c|}{$R'\,(B_\mathrm{N})$ at 192 GeV} & \multicolumn{7}{|c|}{$R'\,(B_\mathrm{N})$ at 196 GeV} \bigstrut \\ 
\hline \bigstrut[t]
$0.000$ & $0.010$  & 4 & 6 & 1 & 5 & 0 & 9 &  & 6 & 3 & 1 & 0 & 1 & 8 & \\ 
$0.010$ & $0.015$  & 35 & 2 & 4 & 8 & 4 & 8 &  & 41 & 6 & 3 & 7 & 8 & 7 & \\ 
$0.015$ & $0.020$  & 36 & 5 & 4 & 6 & 6 & 4 &  & 45 & 8 & 3 & 6 & 9 & 8 & \\ 
$0.020$ & $0.025$  & 26 & 5 & 3 & 4 & 3 & 8 &  & 29 & 6 & 2 & 5 & 7 & 6 & \\ 
$0.025$ & $0.030$  & 20 & 2 & 2 & 7 & 2 & 8 &  & 16 & 1 & 1 & 8 & 3 & 9 & \\ 
$0.030$ & $0.035$  & 15 & 1 & 2 & 0 & 2 & 7 &  & 11 & 5 & 1 & 5 & 2 & 3 & \\ 
$0.035$ & $0.040$  & 9 & 9 & 1 & 8 & 2 & 2 &  & 6 & 8 & 1 & 2 & 1 & 2 & \\
\hline 

\end{tabular}
\end{center}
\end{minipage}}
\end{center}
\caption{Distributions for the narrow jet broadening, $B_\mathrm{N}$, measured by
OPAL at centre-of-mass energies $\sqrt{s}=192$--207~GeV.
The first uncertainty is statistical, while the second is systematic.}
\label{tabdist212}
\end{table}
\clearpage

\begin{figure}[p]
\begin{center}
\includegraphics[width=\textwidth]{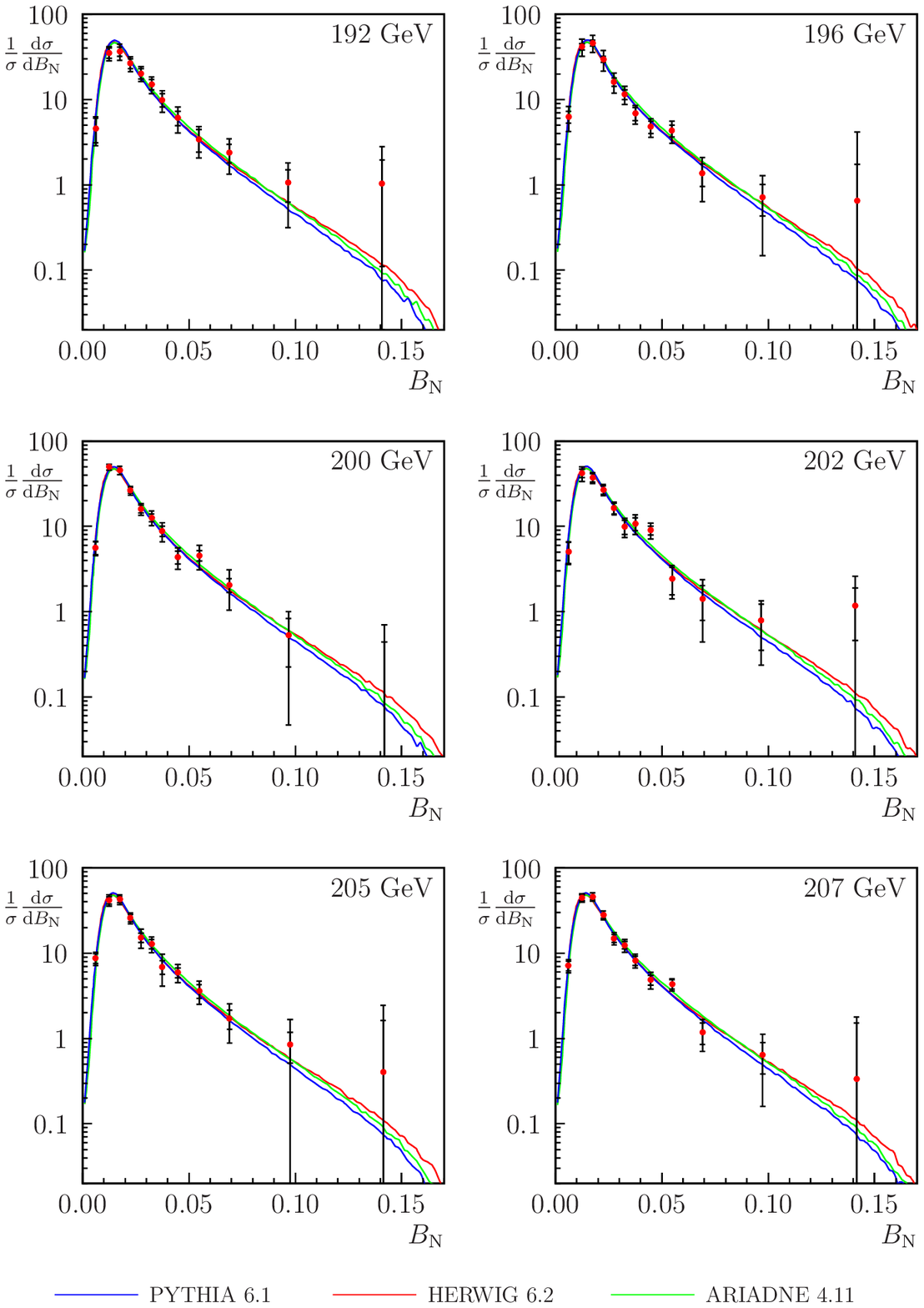}
\caption{Distributions for the narrow jet broadening, $B_\mathrm{N}$, measured by OPAL
at centre-of-mass energies $\sqrt{s}=192$--207~GeV. The inner error
bars indicate statistical uncertainties. Each curve is generated using
five million non-radiative Monte Carlo events, after hadronisation.}
\label{figdist212}
\end{center}
\end{figure}
\clearpage
\section[$D$-parameter]{\boldmath{$D$-parameter}}

\begin{table}[hb!]
\begin{center}
\scalebox{0.90}{
\begin{minipage}{\linewidth}
\begin{center}
\begin{tabular}{| r @{ -- } l | r @{.} l @{ $\pm$ } r @{.} l @{ $\pm$ } r @{.} l @{} l | r @{.} l @{ $\pm$ } r @{.} l @{ $\pm$ } r @{.} l @{} l |}
\hline \multicolumn{2}{|c|}{$D$} & \multicolumn{7}{|c|}{$R'\,(D)$ at 91 GeV} & \multicolumn{7}{|c|}{$R'\,(D)$ at 133 GeV} \bigstrut \\ 
\hline \bigstrut[t]
$0.001$ & $0.005$  & 23 & 1 & 0 & 1 & 1 & 1 &  & 43 & 9 & 3 & 9 & 7 & 9 & \\ 
$0.005$ & $0.010$  & 29 & 70 & 0 & 12 & 0 & 38 &  & 40 & 9 & 2 & 9 & 3 & 0 & \\ 
$0.010$ & $0.015$  & 21 & 72 & 0 & 10 & 0 & 20 &  & 20 & 2 & 2 & 2 & 1 & 3 & \\ 
$0.015$ & $0.020$  & 15 & 64 & 0 & 09 & 0 & 38 &  & 10 & 5 & 1 & 9 & 2 & 4 & \\ 
$0.020$ & $0.030$  & 10 & 83 & 0 & 05 & 0 & 12 &  & 8 & 4 & 1 & 1 & 1 & 1 & \\ 
$0.030$ & $0.045$  & 6 & 64 & 0 & 03 & 0 & 11 &  & 4 & 37 & 0 & 72 & 0 & 93 & \\ 
$0.045$ & $0.070$  & 4 & 036 & 0 & 019 & 0 & 030 &  & 3 & 31 & 0 & 43 & 0 & 26 & \\
\hline 

\end{tabular}
\end{center}
\end{minipage}}
\end{center}
\caption{Distributions for the $D$-parameter, measured by
OPAL at centre-of-mass energies $\sqrt{s}=91$--189~GeV.
The first uncertainty is statistical, while the second is systematic.}
\label{tabdist113}
\end{table}
\clearpage

\begin{figure}[p]
\begin{center}
\includegraphics[width=\textwidth]{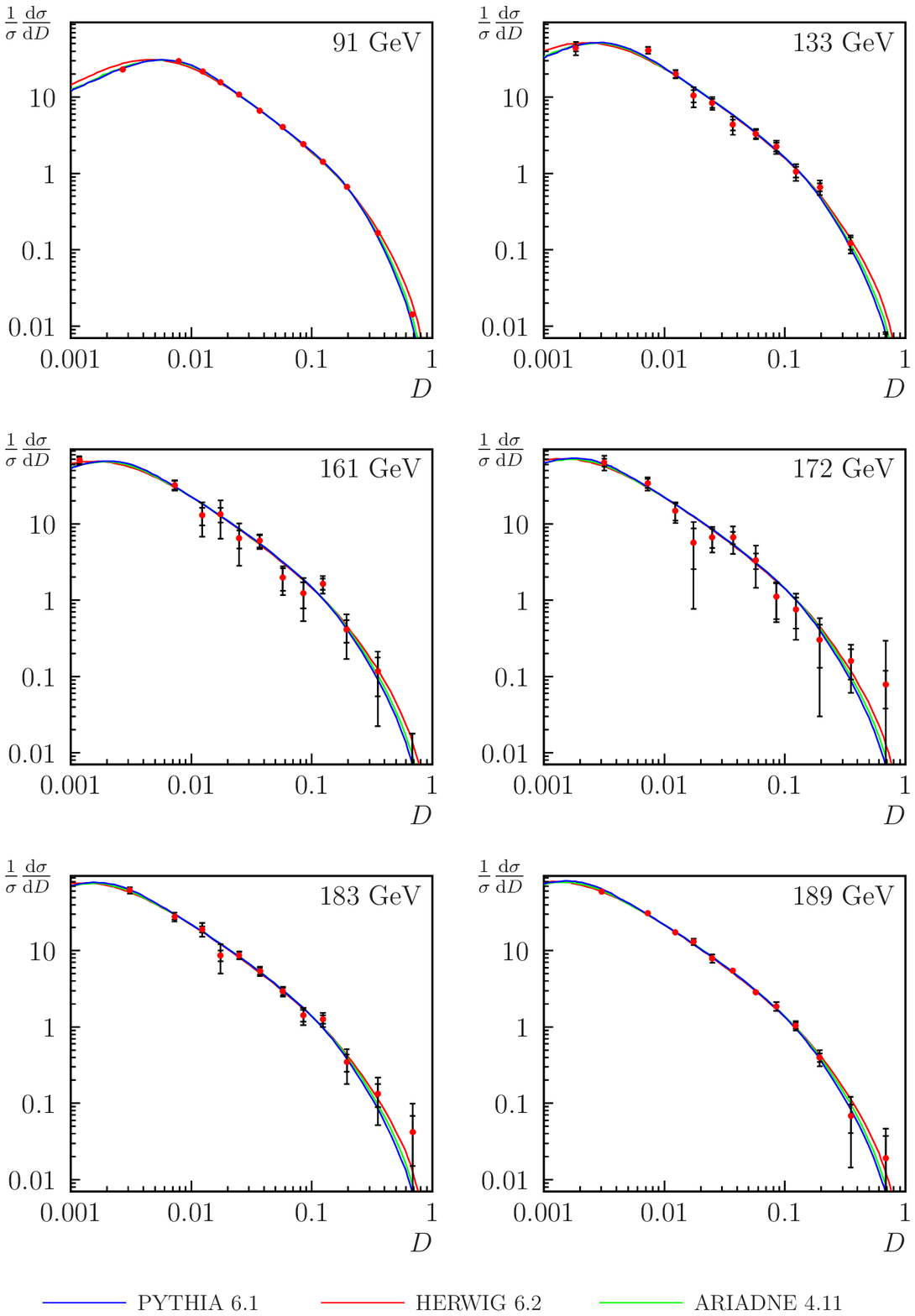}
\caption{Distributions for the $D$-parameter, measured by OPAL
at centre-of-mass energies $\sqrt{s}=91$--189~GeV. The inner error
bars indicate statistical uncertainties. Each curve is generated using
five million non-radiative Monte Carlo events, after hadronisation.}
\label{figdist113}
\end{center}
\end{figure}
\clearpage
\section*{\boldmath{$D$-parameter}~(contd.)}

\begin{table}[hb!]
\begin{center}
\scalebox{0.90}{
\begin{minipage}{\linewidth}
\begin{center}
\begin{tabular}{| r @{ -- } l | r @{.} l @{ $\pm$ } r @{.} l @{ $\pm$ } r @{.} l @{} l | r @{.} l @{ $\pm$ } r @{.} l @{ $\pm$ } r @{.} l @{} l |}
\hline \multicolumn{2}{|c|}{$D$} & \multicolumn{7}{|c|}{$R'\,(D)$ at 192 GeV} & \multicolumn{7}{|c|}{$R'\,(D)$ at 196 GeV} \bigstrut \\ 
\hline \bigstrut[t]
$0.001$ & $0.005$  & 55 & 5 & 5 & 7 & 9 & 2 &  & 62 & 1 & 3 & 6 & 2 & 5 & \\ 
$0.005$ & $0.010$  & 25 & 4 & 3 & 4 & 3 & 8 &  & 29 & 5 & 2 & 1 & 4 & 9 & \\ 
$0.010$ & $0.015$  & 15 & 1 & 2 & 5 & 4 & 0 &  & 17 & 3 & 1 & 7 & 1 & 8 & \\ 
$0.015$ & $0.020$  & 9 & 1 & 2 & 3 & 3 & 7 &  & 11 & 6 & 1 & 4 & 1 & 8 & \\ 
$0.020$ & $0.030$  & 8 & 3 & 1 & 3 & 1 & 2 &  & 7 & 59 & 0 & 83 & 0 & 33 & \\ 
$0.030$ & $0.045$  & 5 & 00 & 0 & 86 & 0 & 77 &  & 4 & 98 & 0 & 54 & 0 & 61 & \\ 
$0.045$ & $0.070$  & 2 & 90 & 0 & 52 & 0 & 51 &  & 3 & 11 & 0 & 33 & 0 & 42 & \\
\hline 

\end{tabular}
\end{center}
\end{minipage}}
\end{center}
\caption{Distributions for the $D$-parameter, measured by
OPAL at centre-of-mass energies $\sqrt{s}=192$--207~GeV.
The first uncertainty is statistical, while the second is systematic.}
\label{tabdist213}
\end{table}
\clearpage

\begin{figure}[p]
\begin{center}
\includegraphics[width=\textwidth]{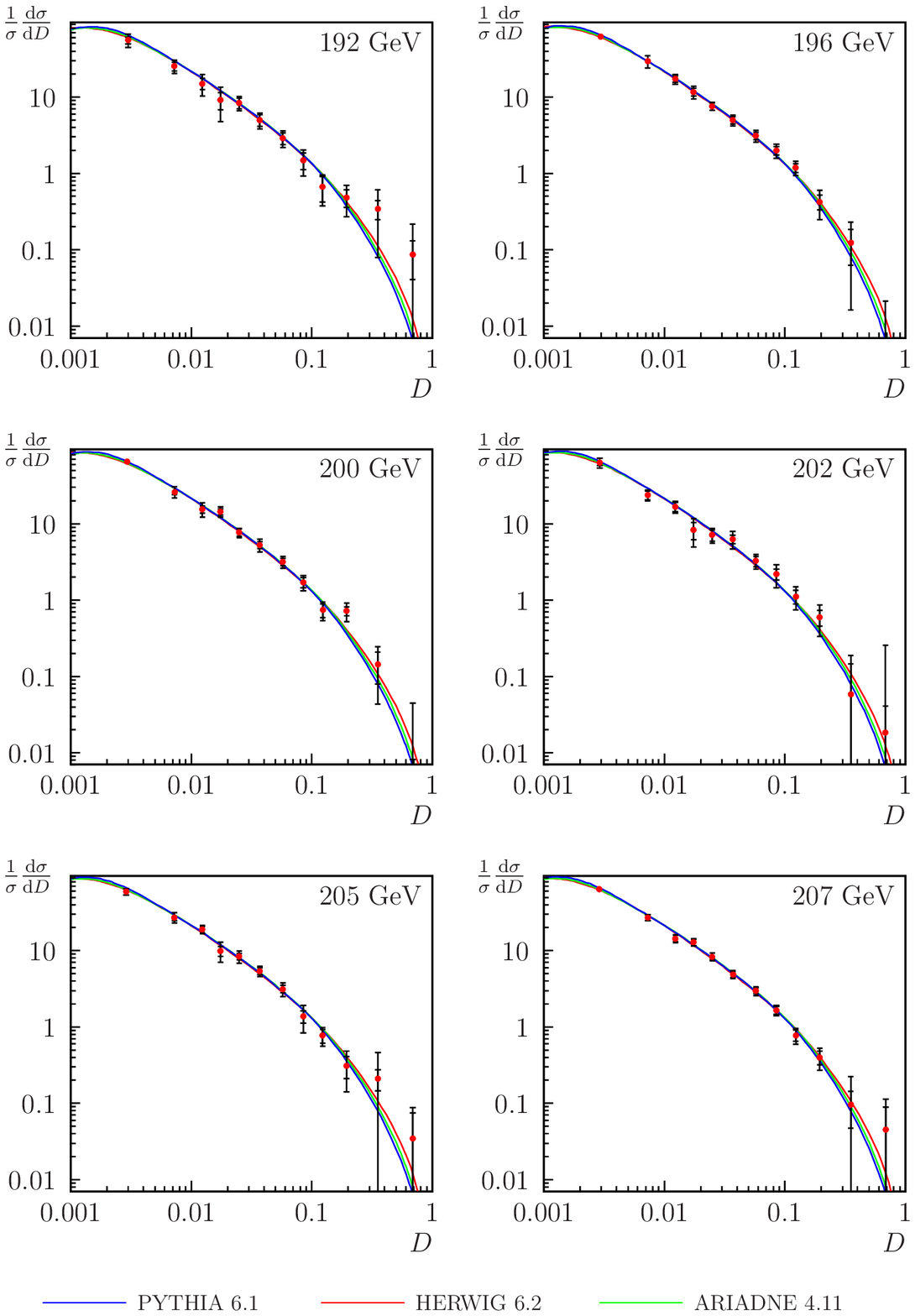}
\caption{Distributions for the $D$-parameter, measured by OPAL
at centre-of-mass energies $\sqrt{s}=192$--207~GeV. The inner error
bars indicate statistical uncertainties. Each curve is generated using
five million non-radiative Monte Carlo events, after hadronisation.}
\label{figdist213}
\end{center}
\end{figure}
\clearpage

\chapter{Fits to the OPAL event shape measurements}
\label{asfitappendix}

We present here the measurements of \asq~derived from OPAL event shape
measurements at each energy scale, as discussed in
Chapter~\ref{opalchapter}.

In each of Tables~\ref{tab:alphasfits91 }--\ref{tab:alphasfits207}, we
list the \as\ values obtained from the thrust~($T$), the heavy jet
mass~($M_\mathrm{H}$), the $C$-parameter, the total jet
broadening~($B_\mathrm{T}$), the wide jet broadening~($B_\mathrm{W}$)
and the Durham $y_{23}$ parameter. A weighted mean of the six \as\
values at each energy is also given; the weights, which are listed
under each observable, are determined as outlined in
Section~\ref{six_observable_mean}.

For each observable, and for the weighted mean, we give a full
breakdown of the uncertainties in \as. The experimental systematic
uncertainty is given by the quadratic sum of the contributions
labelled with asterisks~($^*$); these sources are described in
Section~\ref{evsh_errors}. The hadronisation uncertainty is the larger
of the two absolute deviations in our fitted \as\ when HERWIG and
ARIADNE are used in place of PYTHIA, to apply hadronisation
corrections to the predicted distribution; this deviation is also
marked with an asterisk. We take our theory uncertainty as the mean of
the upper and lower absolute deviations given by the ``uncertainty
band method,'' which is outlined in Section~\ref{evsh_prediction_errors} 
and described fully in Ref.~\cite{uncertaintyband}.

\begin{figure}[p]
 \begin{center}
 \includegraphics[width=\textwidth]{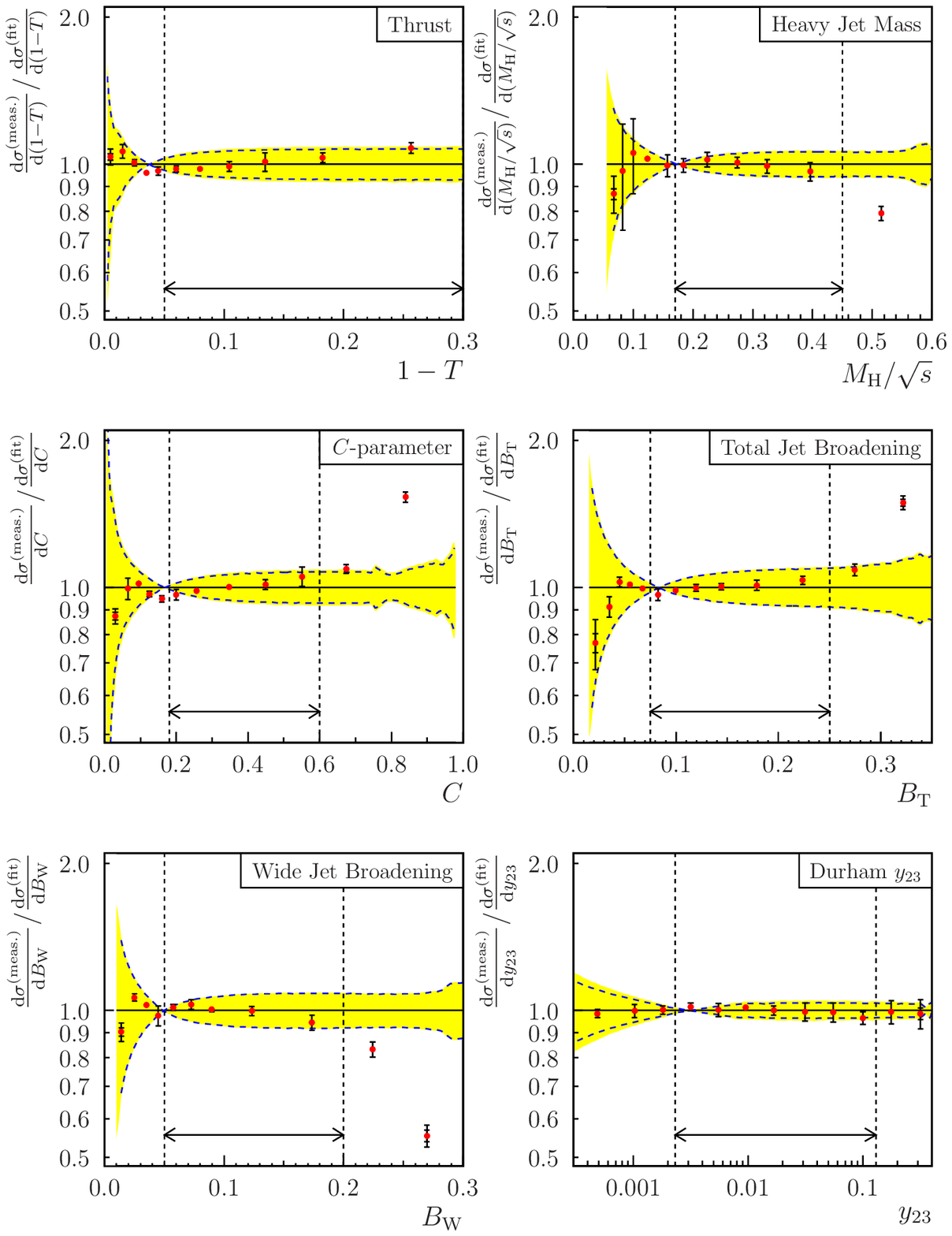}
 \end{center}
 \caption{Ratios of measured event shape distributions to hadron-level predictions, at $\sqrt{s}=91 $~GeV. An explanation is given in the introduction to this       appendix, on page~\pageref{asfitappendix}.}
 \label{fig:alphasfits91 }
 \end{figure}

\begin{figure}[p]
 \begin{center}
 \includegraphics[width=\textwidth]{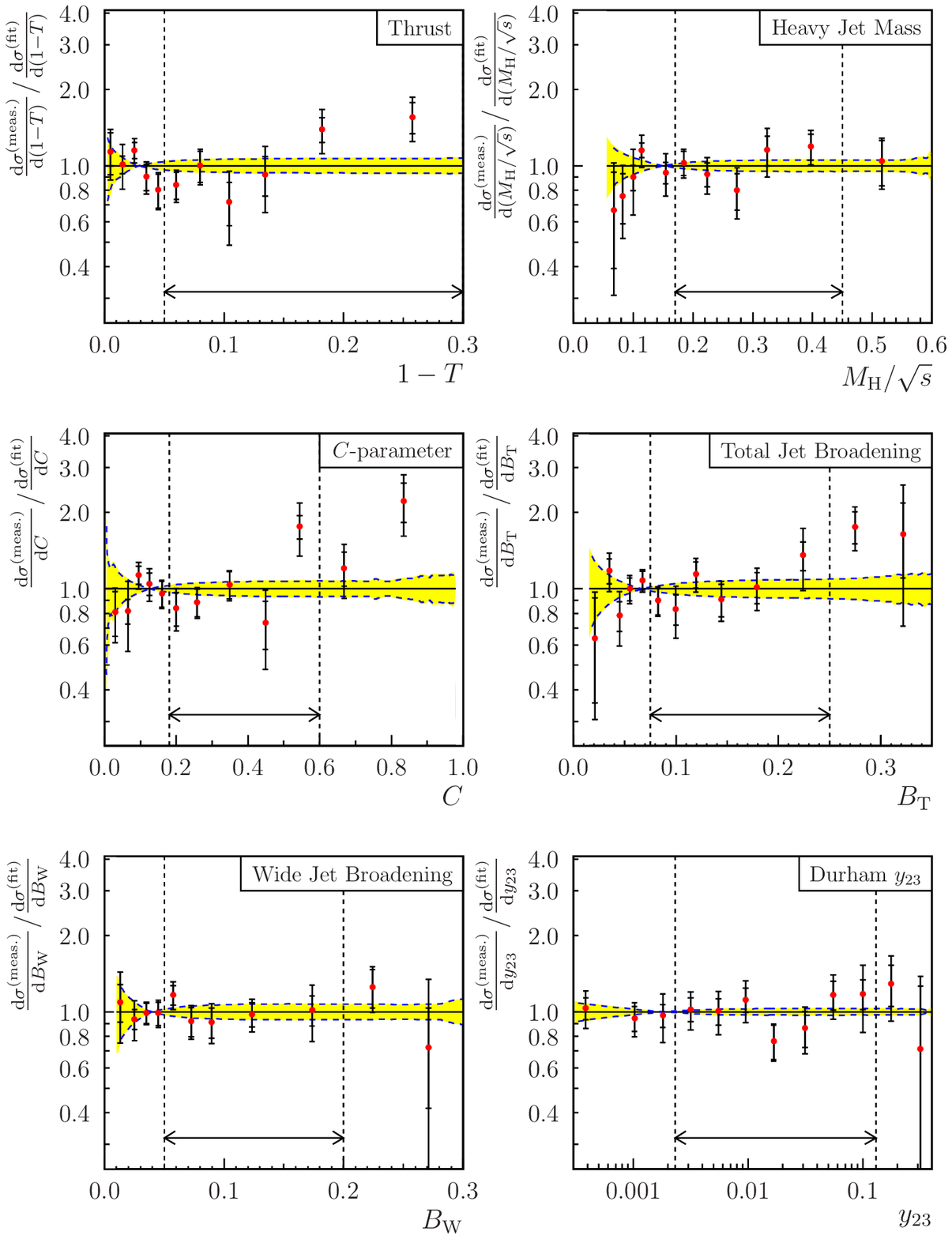}
 \end{center}
 \caption{Ratios of measured event shape distributions to hadron-level predictions, at $\sqrt{s}=133$~GeV. An explanation is given in the introduction to this       appendix, on page~\pageref{asfitappendix}.}
 \label{fig:alphasfits133}
 \end{figure}

\begin{figure}[p]
 \begin{center}
 \includegraphics[width=\textwidth]{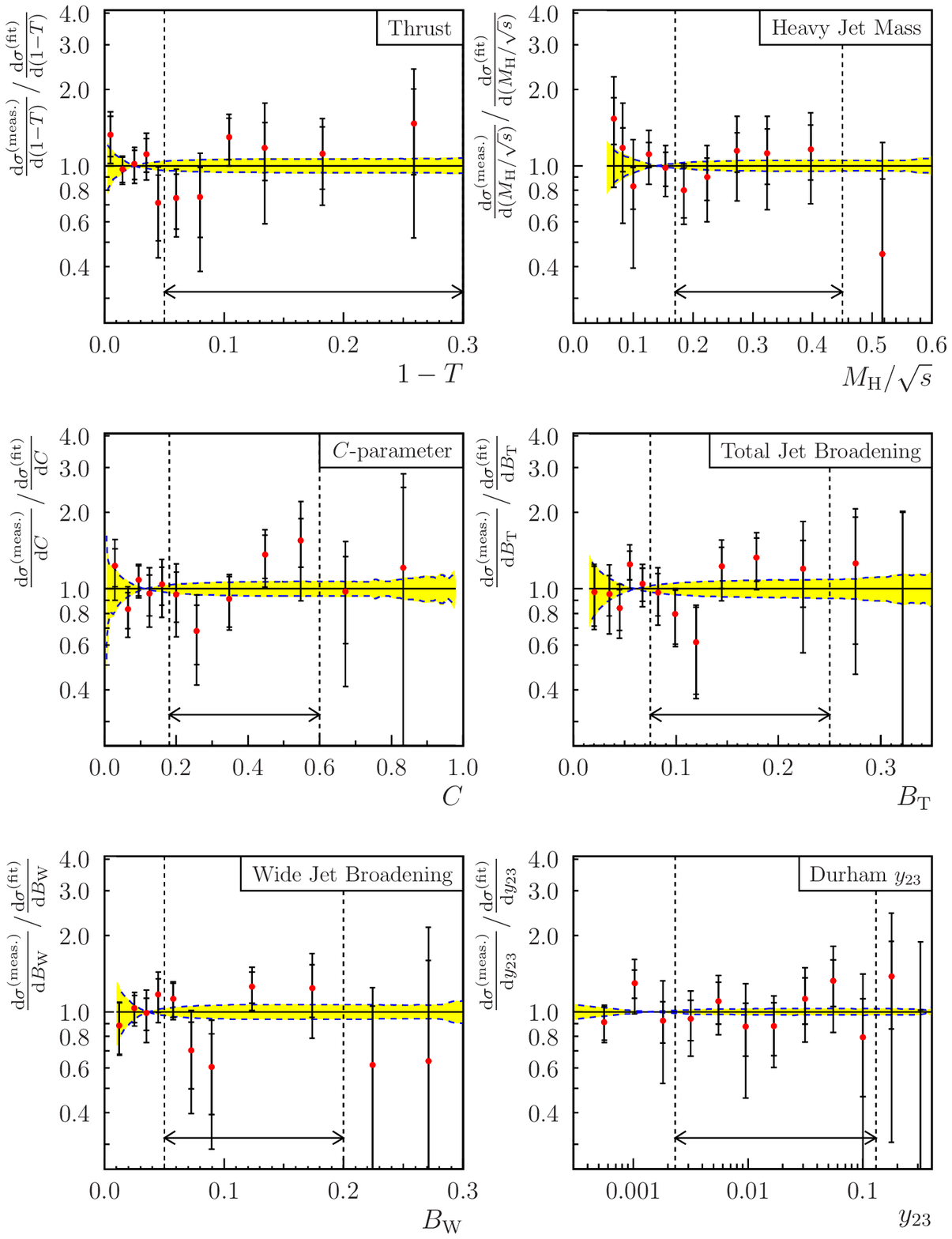}
 \end{center}
 \caption{Ratios of measured event shape distributions to hadron-level predictions, at $\sqrt{s}=161$~GeV. An explanation is given in the introduction to this       appendix, on page~\pageref{asfitappendix}.}
 \label{fig:alphasfits161}
 \end{figure}

\begin{figure}[p]
 \begin{center}
 \includegraphics[width=\textwidth]{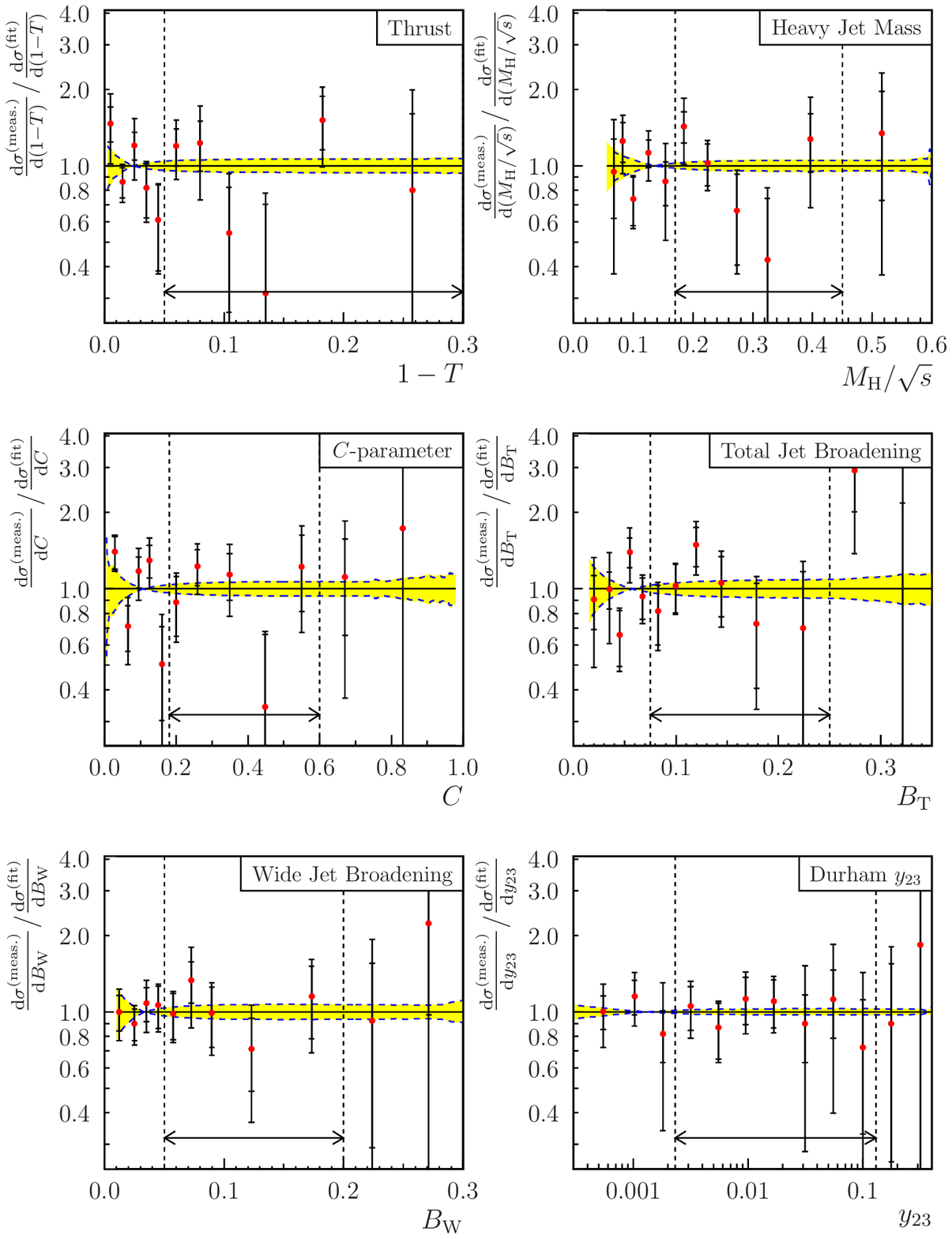}
 \end{center}
 \caption{Ratios of measured event shape distributions to hadron-level predictions, at $\sqrt{s}=172$~GeV. An explanation is given in the introduction to this       appendix, on page~\pageref{asfitappendix}.}
 \label{fig:alphasfits172}
 \end{figure}

\begin{figure}[p]
 \begin{center}
 \includegraphics[width=\textwidth]{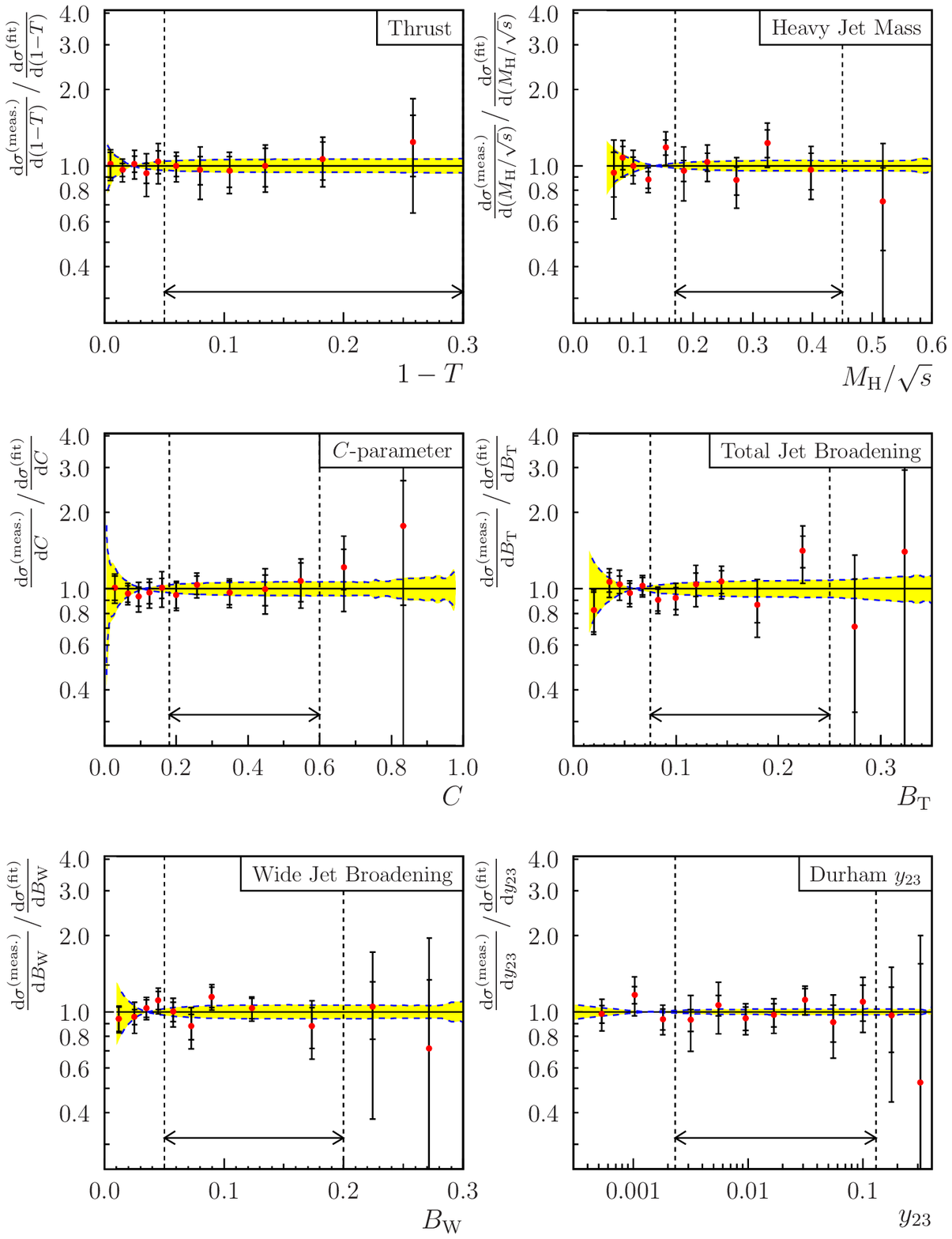}
 \end{center}
 \caption{Ratios of measured event shape distributions to hadron-level predictions, at $\sqrt{s}=183$~GeV. An explanation is given in the introduction to this       appendix, on page~\pageref{asfitappendix}.}
 \label{fig:alphasfits183}
 \end{figure}

\begin{figure}[p]
 \begin{center}
 \includegraphics[width=\textwidth]{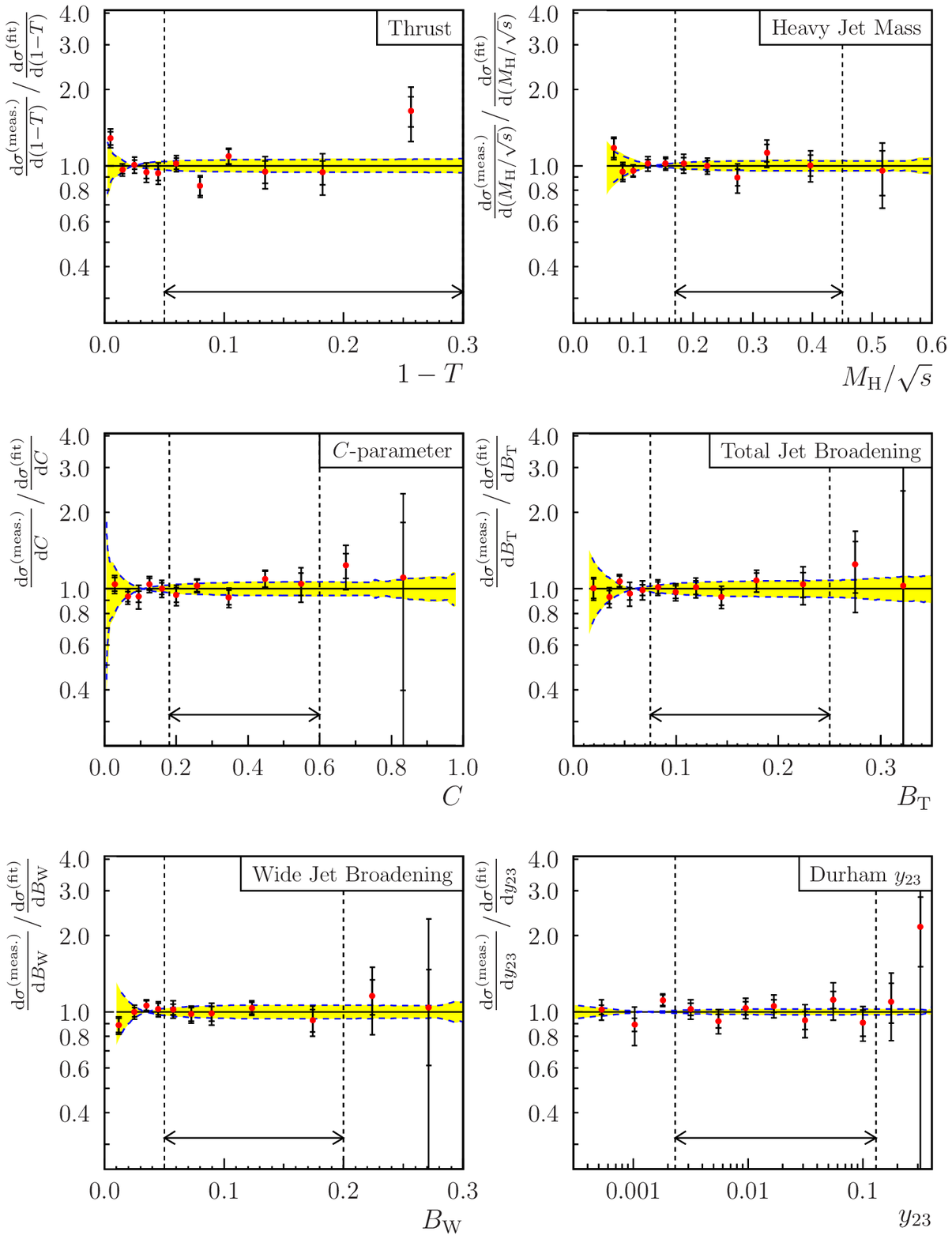}
 \end{center}
 \caption{Ratios of measured event shape distributions to hadron-level predictions, at $\sqrt{s}=189$~GeV. An explanation is given in the introduction to this       appendix, on page~\pageref{asfitappendix}.}
 \label{fig:alphasfits189}
 \end{figure}

\begin{figure}[p]
 \begin{center}
 \includegraphics[width=\textwidth]{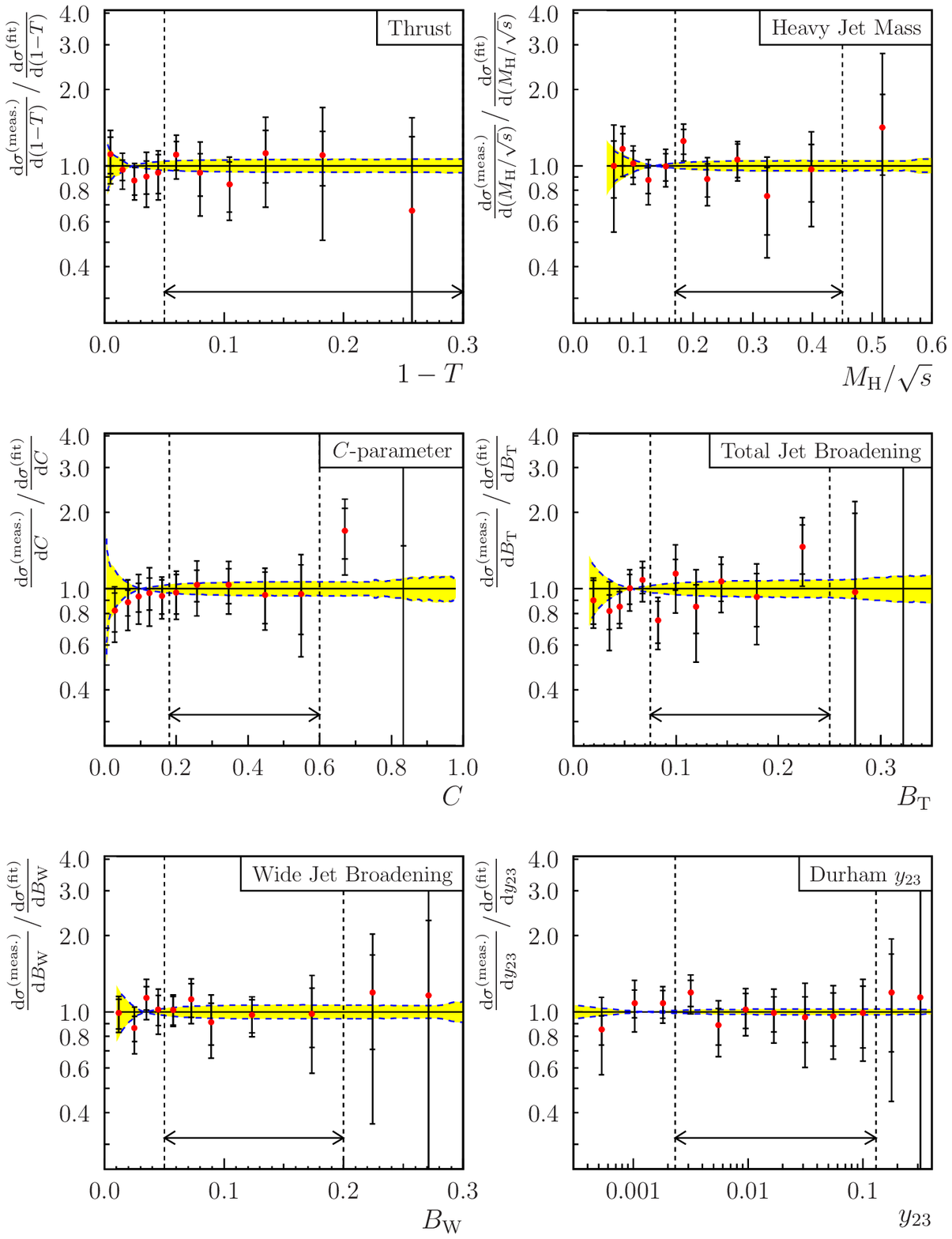}
 \end{center}
 \caption{Ratios of measured event shape distributions to hadron-level predictions, at $\sqrt{s}=192$~GeV. An explanation is given in the introduction to this       appendix, on page~\pageref{asfitappendix}.}
 \label{fig:alphasfits192}
 \end{figure}

\begin{figure}[p]
 \begin{center}
 \includegraphics[width=\textwidth]{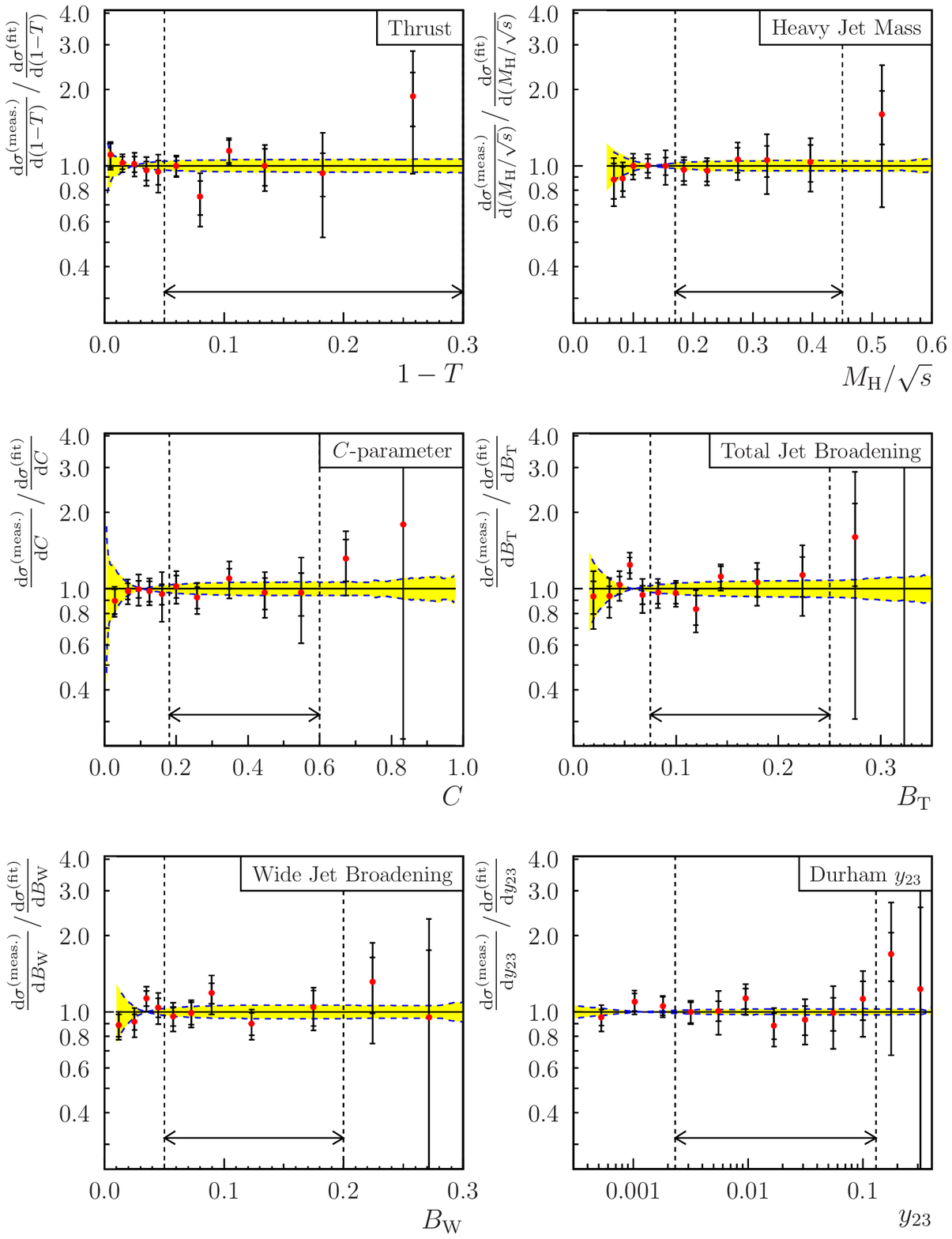}
 \end{center}
 \caption{Ratios of measured event shape distributions to hadron-level predictions, at $\sqrt{s}=196$~GeV. An explanation is given in the introduction to this       appendix, on page~\pageref{asfitappendix}.}
 \label{fig:alphasfits196}
 \end{figure}

\begin{figure}[p]
 \begin{center}
 \includegraphics[width=\textwidth]{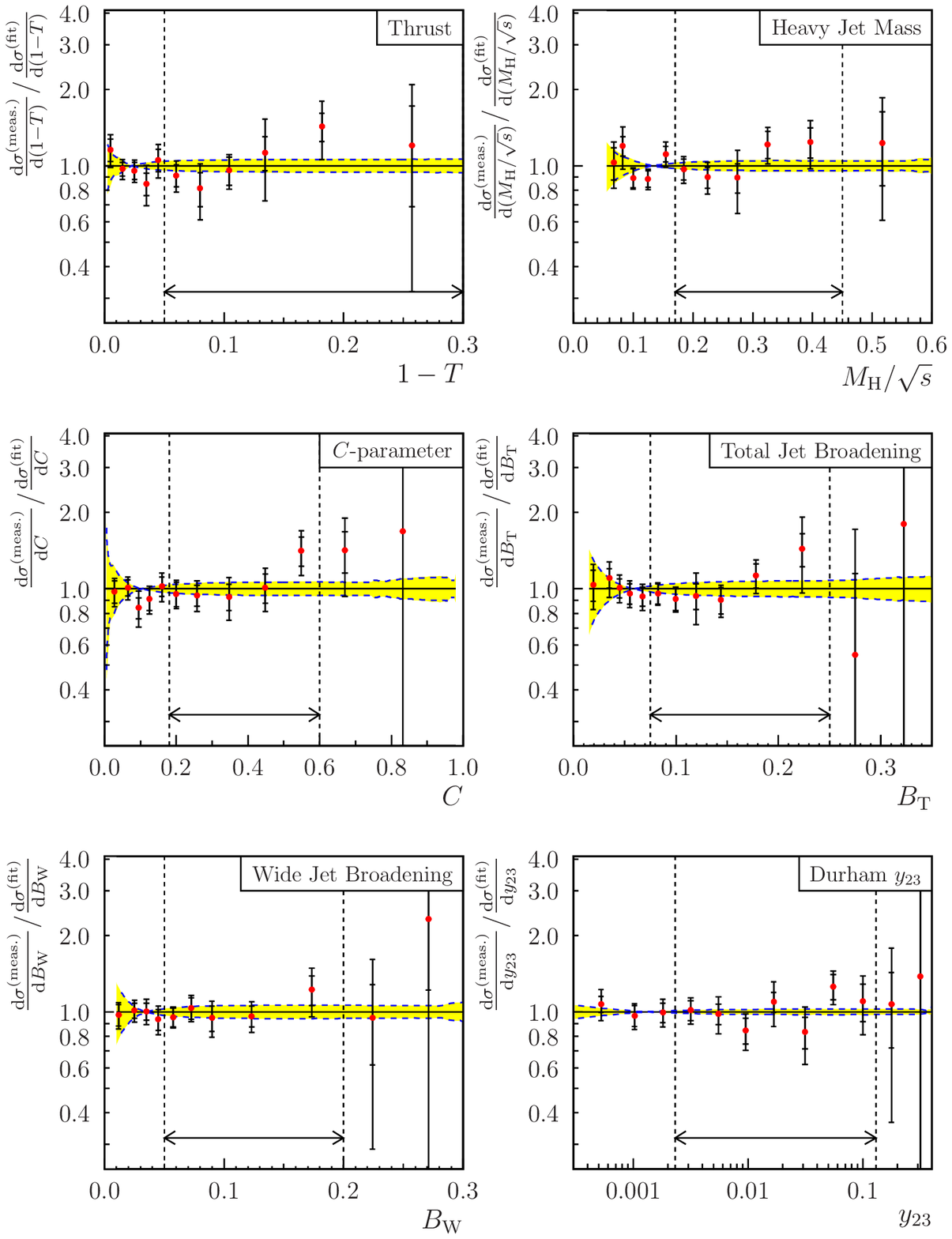}
 \end{center}
 \caption{Ratios of measured event shape distributions to hadron-level predictions, at $\sqrt{s}=200$~GeV. An explanation is given in the introduction to this       appendix, on page~\pageref{asfitappendix}.}
 \label{fig:alphasfits200}
 \end{figure}

\begin{figure}[p]
 \begin{center}
 \includegraphics[width=\textwidth]{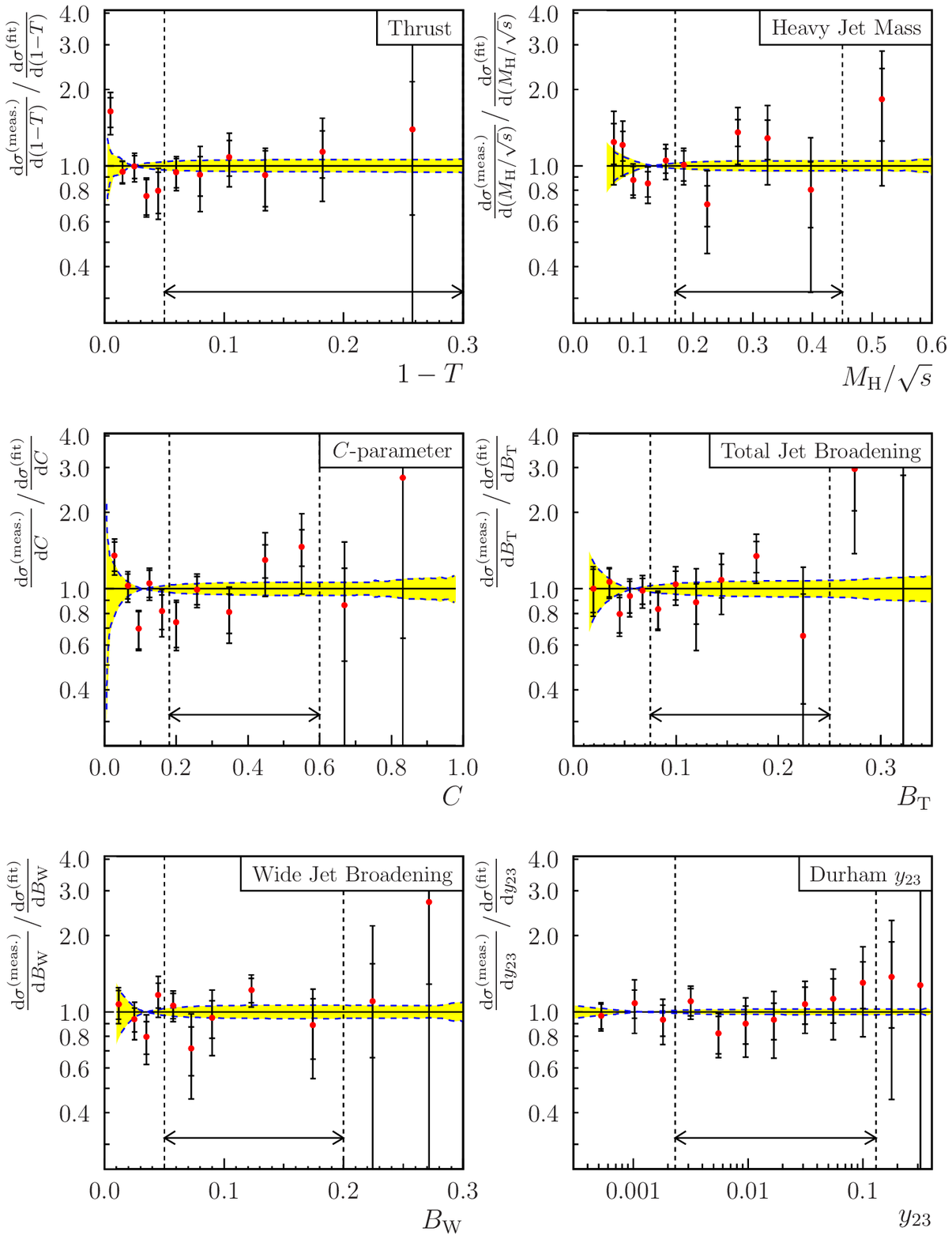}
 \end{center}
 \caption{Ratios of measured event shape distributions to hadron-level predictions, at $\sqrt{s}=202$~GeV. An explanation is given in the introduction to this       appendix, on page~\pageref{asfitappendix}.}
 \label{fig:alphasfits202}
 \end{figure}

\begin{figure}[p]
 \begin{center}
 \includegraphics[width=\textwidth]{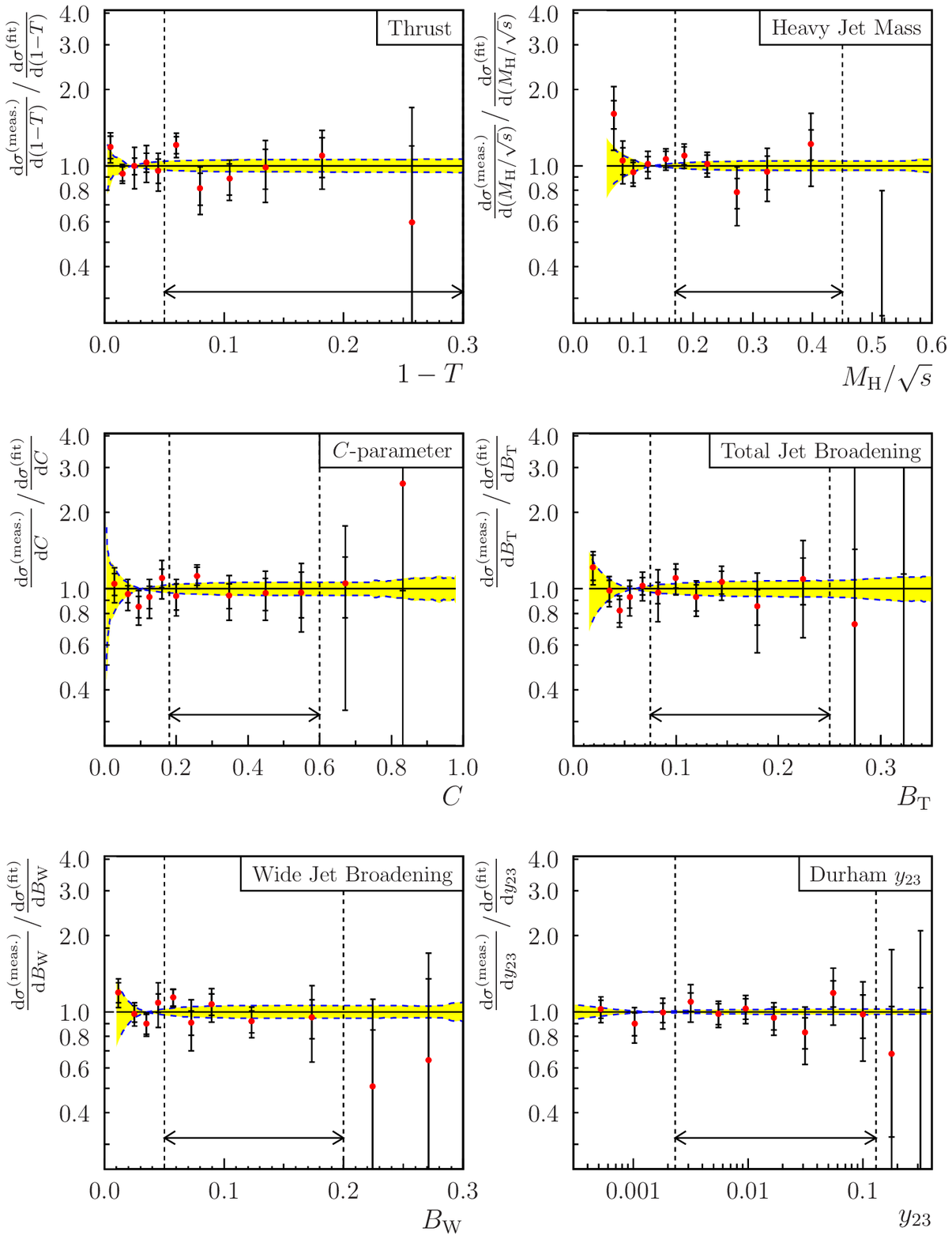}
 \end{center}
 \caption{Ratios of measured event shape distributions to hadron-level predictions, at $\sqrt{s}=205$~GeV. An explanation is given in the introduction to this       appendix, on page~\pageref{asfitappendix}.}
 \label{fig:alphasfits205}
 \end{figure}

\begin{figure}[p]
 \begin{center}
 \includegraphics[width=\textwidth]{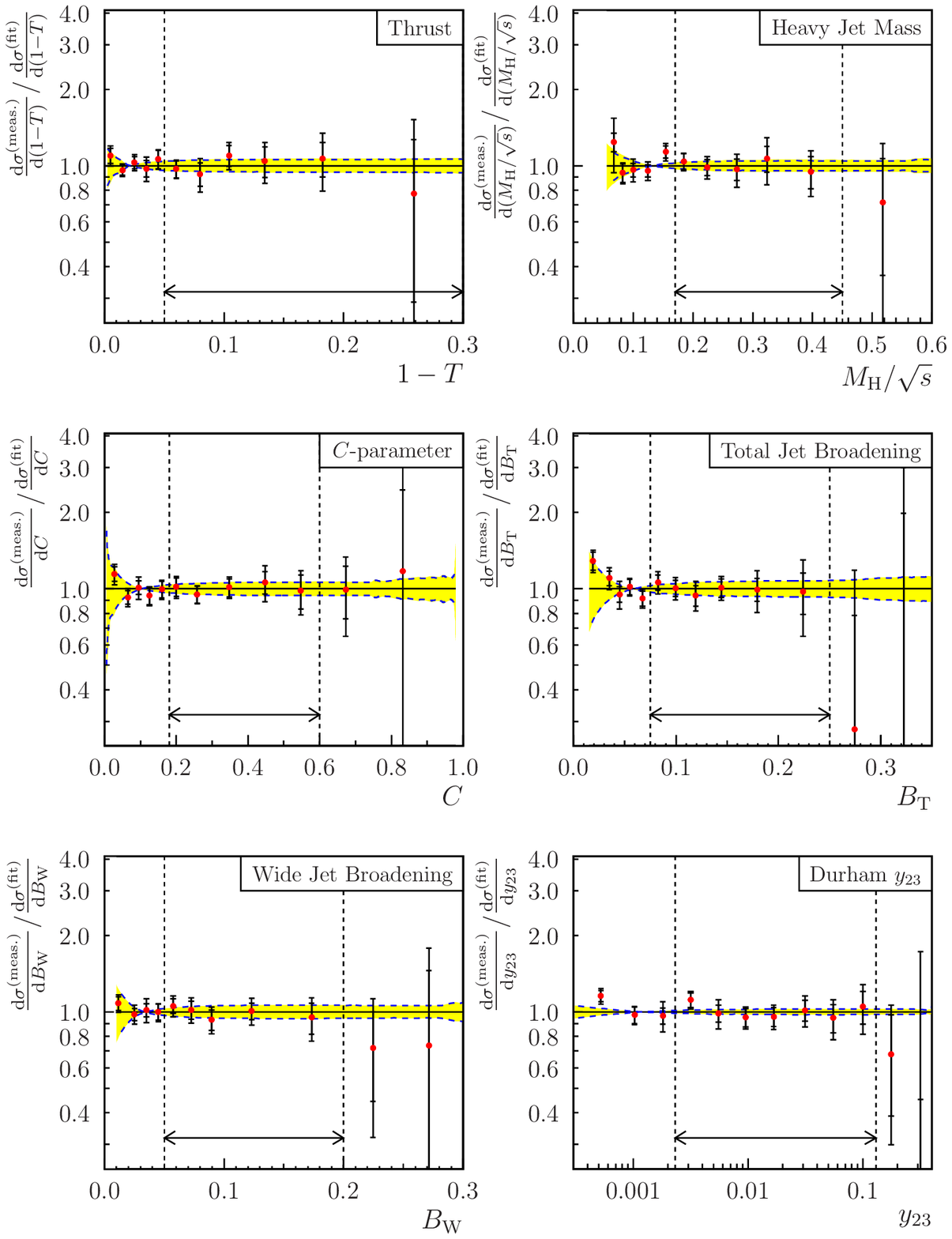}
 \end{center}
 \caption{Ratios of measured event shape distributions to hadron-level predictions, at $\sqrt{s}=207$~GeV. An explanation is given in the introduction to this       appendix, on page~\pageref{asfitappendix}.}
 \label{fig:alphasfits207}
 \end{figure}

Figures~\ref{fig:alphasfits91 }--\ref{fig:alphasfits207} compare our
measured distributions with the fitted hadron-level predictions. Each
data point shows the measured bin contents divided by the integral of
the predicted distribution across the bin; the inner error bars
indicate statistical uncertainties, and the outer bars show the
combined statistical and experimental contributions. The blue dashed
curves represent fractional variations in the predicted distributions,
corresponding to our perturbative theory uncertainties in~\as. The
slightly wider yellow bands indicate the combined theory and
hadronisation uncertainties. The ranges used for fitting each
distribution are shown by horizontal arrows. Note that the range of the
vertical scale in the 91~GeV plots (Figure~\ref{fig:alphasfits91 }) is 
different from that used at higher energies.

 \section[OPAL measurements of $\alpha_\mathrm{S}$ at $\sqrt{\lowercase{s}}=91 $~G\lowercase{e}V]{\boldmath OPAL measurements of $\alpha_\mathrm{S}$ at $\sqrt{s}=91 $~GeV}
 \begin{table}[hb!]
 \begin{center}
 \scalebox{0.80}{
 \begin{minipage}{\linewidth}
 \begin{center}
\begin{tabular}{ | r || r@{}l r@{}l r@{}l || r@{}l |}
 \cline{1-7}
 {\parbox{1.7cm}{\centering \rule{0cm}{0.45cm}\textcolor{white}{Weighted}\vspace{-0.15cm}\\\textcolor{white}{mean}\vspace{-0.3cm}\\$\phantom{A}$}} & \multicolumn{2}{|c}{$T$} & \multicolumn{2}{c}{$M_\mathrm{H}$} & \multicolumn{2}{c||}{$C$} & \multicolumn{2}{c}{ }
 \bigstrut \\
 \cline{1-7} \multicolumn{9}{c}{ } \vspace{-0.45cm} \\         \cline{1-7}  \bigstrut
 $\alpha_\mathrm{S}$(91  GeV)
  & $    0.1231$
  &
  & $    0.1192$
  &
  & $    0.1178$
  &
 & \multicolumn{2}{c}{ }
  \\
 \cline{1-7} \multicolumn{9}{c}{ } \vspace{-0.45cm} \\         \cline{1-7}  \bigstrut
 Statistical error
  & $\pm 0.0002$
  &
  & $\pm 0.0002$
  &
  & $\pm 0.0002$
  &
 & \multicolumn{2}{c}{ }
  \\
 \cline{1-7} \multicolumn{9}{c}{ } \vspace{-0.45cm} \\         \cline{1-7}  \bigstrut[t]
 No \texttt{MT} algorithm
  & $+   0.0002$
  & $^*$
  & $+   0.0003$
  & $^*$
  & $+   0.0003$
  & $^*$
 & \multicolumn{2}{c}{ }
  \\
 HERWIG det.~corr.
  & $   -0.0010$
  & $^*$
  & $   -0.0001$
  & $^*$
  & $   -0.0007$
  & $^*$
 & \multicolumn{2}{c}{ }
  \\
\end{tabular}
 \end{center}
 \end{minipage}
 }
 \end{center}
 \caption{OPAL measurements of $\alpha_\mathrm{S}$ at $\sqrt{s}=91 $~GeV}
 \label{tab:alphasfits91 }
 \end{table}
 \clearpage

\clearpage
 \section[OPAL measurements of $\alpha_\mathrm{S}$ at $\sqrt{\lowercase{s}}=133$~G\lowercase{e}V]{\boldmath OPAL measurements of $\alpha_\mathrm{S}$ at $\sqrt{s}=133$~GeV}
 \begin{table}[hb!]
 \begin{center}
 \scalebox{0.80}{
 \begin{minipage}{\linewidth}
 \begin{center}

 \end{center}
 \end{minipage}
 }
 \end{center}
 \caption{OPAL measurements of $\alpha_\mathrm{S}$ at $\sqrt{s}=133$~GeV}
 \label{tab:alphasfits133}
 \end{table}
 \clearpage

\clearpage
 \section[OPAL measurements of $\alpha_\mathrm{S}$ at $\sqrt{\lowercase{s}}=161$~G\lowercase{e}V]{\boldmath OPAL measurements of $\alpha_\mathrm{S}$ at $\sqrt{s}=161$~GeV}
 \begin{table}[hb!]
 \begin{center}
 \scalebox{0.80}{
 \begin{minipage}{\linewidth}
 \begin{center}

 \end{center}
 \end{minipage}
 }
 \end{center}
 \caption{OPAL measurements of $\alpha_\mathrm{S}$ at $\sqrt{s}=161$~GeV}
 \label{tab:alphasfits161}
 \end{table}
 \clearpage

\clearpage
 \section[OPAL measurements of $\alpha_\mathrm{S}$ at $\sqrt{\lowercase{s}}=172$~G\lowercase{e}V]{\boldmath OPAL measurements of $\alpha_\mathrm{S}$ at $\sqrt{s}=172$~GeV}
 \begin{table}[hb!]
 \begin{center}
 \scalebox{0.80}{
 \begin{minipage}{\linewidth}
 \begin{center}

 \end{center}
 \end{minipage}
 }
 \end{center}
 \caption{OPAL measurements of $\alpha_\mathrm{S}$ at $\sqrt{s}=172$~GeV}
 \label{tab:alphasfits172}
 \end{table}
 \clearpage

\clearpage
 \section[OPAL measurements of $\alpha_\mathrm{S}$ at $\sqrt{\lowercase{s}}=183$~G\lowercase{e}V]{\boldmath OPAL measurements of $\alpha_\mathrm{S}$ at $\sqrt{s}=183$~GeV}
 \begin{table}[hb!]
 \begin{center}
 \scalebox{0.80}{
 \begin{minipage}{\linewidth}
 \begin{center}

 \end{center}
 \end{minipage}
 }
 \end{center}
 \caption{OPAL measurements of $\alpha_\mathrm{S}$ at $\sqrt{s}=183$~GeV}
 \label{tab:alphasfits183}
 \end{table}
 \clearpage

\clearpage
 \section[OPAL measurements of $\alpha_\mathrm{S}$ at $\sqrt{\lowercase{s}}=189$~G\lowercase{e}V]{\boldmath OPAL measurements of $\alpha_\mathrm{S}$ at $\sqrt{s}=189$~GeV}
 \begin{table}[hb!]
 \begin{center}
 \scalebox{0.80}{
 \begin{minipage}{\linewidth}
 \begin{center}

 \end{center}
 \end{minipage}
 }
 \end{center}
 \caption{OPAL measurements of $\alpha_\mathrm{S}$ at $\sqrt{s}=189$~GeV}
 \label{tab:alphasfits189}
 \end{table}
 \clearpage

\clearpage
 \section[OPAL measurements of $\alpha_\mathrm{S}$ at $\sqrt{\lowercase{s}}=192$~G\lowercase{e}V]{\boldmath OPAL measurements of $\alpha_\mathrm{S}$ at $\sqrt{s}=192$~GeV}
 \begin{table}[hb!]
 \begin{center}
 \scalebox{0.80}{
 \begin{minipage}{\linewidth}
 \begin{center}

 \end{center}
 \end{minipage}
 }
 \end{center}
 \caption{OPAL measurements of $\alpha_\mathrm{S}$ at $\sqrt{s}=192$~GeV}
 \label{tab:alphasfits192}
 \end{table}
 \clearpage

\clearpage
 \section[OPAL measurements of $\alpha_\mathrm{S}$ at $\sqrt{\lowercase{s}}=196$~G\lowercase{e}V]{\boldmath OPAL measurements of $\alpha_\mathrm{S}$ at $\sqrt{s}=196$~GeV}
 \begin{table}[hb!]
 \begin{center}
 \scalebox{0.80}{
 \begin{minipage}{\linewidth}
 \begin{center}

 \end{center}
 \end{minipage}
 }
 \end{center}
 \caption{OPAL measurements of $\alpha_\mathrm{S}$ at $\sqrt{s}=196$~GeV}
 \label{tab:alphasfits196}
 \end{table}
 \clearpage

\clearpage
 \section[OPAL measurements of $\alpha_\mathrm{S}$ at $\sqrt{\lowercase{s}}=200$~G\lowercase{e}V]{\boldmath OPAL measurements of $\alpha_\mathrm{S}$ at $\sqrt{s}=200$~GeV}
 \begin{table}[hb!]
 \begin{center}
 \scalebox{0.80}{
 \begin{minipage}{\linewidth}
 \begin{center}

 \end{center}
 \end{minipage}
 }
 \end{center}
 \caption{OPAL measurements of $\alpha_\mathrm{S}$ at $\sqrt{s}=200$~GeV}
 \label{tab:alphasfits200}
 \end{table}
 \clearpage

\clearpage
 \section[OPAL measurements of $\alpha_\mathrm{S}$ at $\sqrt{\lowercase{s}}=202$~G\lowercase{e}V]{\boldmath OPAL measurements of $\alpha_\mathrm{S}$ at $\sqrt{s}=202$~GeV}
 \begin{table}[hb!]
 \begin{center}
 \scalebox{0.80}{
 \begin{minipage}{\linewidth}
 \begin{center}

 \end{center}
 \end{minipage}
 }
 \end{center}
 \caption{OPAL measurements of $\alpha_\mathrm{S}$ at $\sqrt{s}=202$~GeV}
 \label{tab:alphasfits202}
 \end{table}
 \clearpage

\clearpage
 \section[OPAL measurements of $\alpha_\mathrm{S}$ at $\sqrt{\lowercase{s}}=205$~G\lowercase{e}V]{\boldmath OPAL measurements of $\alpha_\mathrm{S}$ at $\sqrt{s}=205$~GeV}
 \begin{table}[hb!]
 \begin{center}
 \scalebox{0.80}{
 \begin{minipage}{\linewidth}
 \begin{center}

 \end{center}
 \end{minipage}
 }
 \end{center}
 \caption{OPAL measurements of $\alpha_\mathrm{S}$ at $\sqrt{s}=205$~GeV}
 \label{tab:alphasfits205}
 \end{table}
 \clearpage

\clearpage
 \section[OPAL measurements of $\alpha_\mathrm{S}$ at $\sqrt{\lowercase{s}}=207$~G\lowercase{e}V]{\boldmath OPAL measurements of $\alpha_\mathrm{S}$ at $\sqrt{s}=207$~GeV}
 \begin{table}[hb!]
 \begin{center}
 \scalebox{0.80}{
 \begin{minipage}{\linewidth}
 \begin{center}

 \end{center}
 \end{minipage}
 }
 \end{center}
 \caption{OPAL measurements of $\alpha_\mathrm{S}$ at $\sqrt{s}=207$~GeV}
 \label{tab:alphasfits207}
 \end{table}
 \clearpage

\clearpage

\chapter[Inputs to the combined LEP \as\ measurement]{Inputs to the 
combined LEP \boldmath{\as} measurement}

\label{lepinputappendix}

This appendix contains the \as\ measurements from the four LEP
Collaborations, which contribute to the combination discussed in
Chapter~\ref{lepcombinationchapter}.

The statistical uncertainties~$\sigma_\mathrm{stat.}$, and
experimental systematic uncertainties~$\sigma_\mathrm{exp.}$, listed
in these tables are the values quoted by the Collaborations. The
experimental uncertainties will be averaged between Collaborations in
the construction of the covariance matrix. The hadronisation and
theory uncertainties given in the tables are the values calculated
independently for the \LEP\ combination, as described in
Section~\ref{errorsubsect}. The total systematic uncertainties
$\sigma_\mathrm{syst.}$, and total uncertainties
$\sigma_\mathrm{total.}$, which do not enter directly into the
combination algorithm, have been evaluated after symmetrisation of the
theory uncertainties.

Where an asterisk (\textasteriskcentered) appears next to a
measurement, the values have not been derived from a single fit to an
event shape distribution.  Instead, a number of published or
preliminary results from neighbouring energy scales have been
combined, in order to standardise the energy scales of the input
measurements. These `pre-combinations' were carried out using
weights based only on statistical uncertainties, and assuming full
correlation between all systematic uncertainties.

The OPAL measurements given in Section~\ref{opal_inputs_appendix} are
not the latest results presented in
Chapter~\ref{opalchapter}. The LEP \as\ combination uses the most
recent preliminary results approved by the Collaborations, which in
the case of OPAL were presented in
Refs.~\cite{OPAL_PN512,montpellier}.

%
\newpage\section{Fits to ALEPH  data}
$\phantom{A}$\\
\noindent{\small

}
 
\newpage\section{Fits to DELPHI data}
$\phantom{A}$\\
\noindent{\small
%
}
 
\newpage\section{Fits to L3     data}
$\phantom{A}$\\
\scalebox{0.85}{
\parbox{\linewidth}{
\noindent{\small
%
}}}
 
\newpage\section{Fits to OPAL   data}
\label{opal_inputs_appendix}
$\phantom{A}$\\
\noindent{\small
%
}

\bibliography{thesis}

\providecommand{\href}[2]{#2}\begingroup\raggedright\begin{thebibliography}{10%
0}

\bibitem{greenberg1964}
O.~W. Greenberg, {\em Spin and unitary spin independence in a paraquark model
  of baryons and mesons}, Phys.~Rev.~Lett.~{\bf 13}~(1964)~598.

\bibitem{fritzsch_gellmann}
H.~Fritzsch and M.~Gell-Mann, {\em Current algebra: Quarks and what else?},
  Proceedings of the 16th International Conference on High Energy Physics,
  Chicago, IL, USA, 6--13~Sept~1972, Vol.~2, 135.
  \href{http://www.arXiv.org/abs/hep-ph/0208010}{E-print~number:~{\tt
  hep-ph/0208010}}.

\bibitem{asymptotic_freedom}
D.~J. Gross and F.~Wilczek, {\em Ultraviolet behaviour of non-Abelian gauge
  theories}, Phys.~Rev.~Lett.~{\bf 30}~(1973)~1343.

\bibitem{webber_qcdbook}
R.~K. Ellis, W.~J. Stirling, and B.~R. Webber, {\em QCD and Collider Physics}.
\newblock Cambridge,~UK: CUP~(1998).

\bibitem{pi0decay}
G.~von Dardel et al., {\em Mean life of the neutral pion}, Phys.~Lett.~{\bf
  4}~(1963)~51.

\bibitem{lep_electroweak_comb}
ALEPH, DELPHI, L3 and OPAL Collaborations, {\em A combination of preliminary
  electroweak measurements and constraints on the standard model}.
  Report~number:~ALEPH-2003-017-PHYSICS-2003-005~/ DELPHI-2003-072-PHYS-937~/
  L3-NOTE-2825~/ OPAL-PR-392,
  \href{http://www.arXiv.org/abs/hep-ex/0312023}{E-print~number:~{\tt
  hep-ex/0312023}}.

\bibitem{gluon_discovery}
D.~P. Barber et al., {\em Discovery of three jet events and a test of quantum
  chromodynamics at PETRA energies}, Phys.~Rev.~Lett.~{\bf 43}~(1979)~830.

\bibitem{OPAL_as_91}
OPAL Collaboration (P.~D. Acton et al.), {\em A Determination of
  {$\as\left(M_{\text{Z}^0}\right)$} at {LEP} using resummed {QCD}
  calculations}, Z.~Phys.~{\bf C59}~(1993)~1.

\bibitem{OPAL_as_133}
OPAL Collaboration (G.~Alexander et al.), {\em {QCD} studies with {\epem}
  annihilation data at 130~GeV and 136~GeV}, Z.~Phys.~{\bf C72}~(1996)~191.

\bibitem{OPAL_as_161}
OPAL Collaboration (K.~Ackerstaff et al.), {\em {QCD} studies with {\epem}
  annihilation data at 161~GeV}, Z.~Phys.~{\bf C75}~(1997)~193.

\bibitem{OPAL_as_189}
OPAL Collaboration (G.~Abbiendi et al.), {\em {QCD} studies with {\epem}
  annihilation data at 172--189~GeV}, Eur.~Phys.~J.~{\bf C16}~(2000)~185.
  \href{http://www.arXiv.org/abs/hep-ex/0002012}{E-print~number:~{\tt
  hep-ex/0002012}}.

\bibitem{uncertaintyband}
R.~W.~L. Jones, M.~Ford, G.~P. Salam, H.~Stenzel, and D.~Wicke, {\em
  Theoretical uncertainties on {\as} from event-shape variables in {\epem}
  annihilations}, JHEP~{\bf 12}~(2003)~007.
  \href{http://www.arXiv.org/abs/hep-ph/0312016}{E-print~number:~{\tt
  hep-ph/0312016}}.

\bibitem{PDbook}
{Particle Data Group (K.~Hagiwara et al.)}, {\em Review of particle physics},
  Phys.~Rev.~{\bf D66}~(2002)~010001.

\bibitem{martin_shaw}
B.~R. Martin and G.~Shaw, {\em Particle Physics}.
\newblock Chichester,~UK: Wiley~(1997).

\bibitem{aitchison_hey}
I.~J.~R. Aitchison and A.~J.~G. Hey, {\em Gauge Theories in Particle Physics}.
\newblock Bristol,~UK: IOP~(1989).

\bibitem{yang_mills}
C.-N. Yang and R.~L. Mills, {\em Conservation of isotopic spin and isotopic
  gauge invariance}, Phys.~Rev.~{\bf 96}~(1954)~191.

\bibitem{cpar_ellis}
R.~K. Ellis, D.~A. Ross, and A.~E. Terrano, {\em The perturbative calculation
  of jet structure in {\epem} annihilation}, Nucl.~Phys.~{\bf B178}~(1981)~421.

\bibitem{thrust1964}
S.~Brandt, C.~Peyrou, R.~Sosnowski, and A.~Wroblewski, {\em The principal axis
  of jets---an attempt to analyze high-energy collisions as two-body
  processes}, Phys.~Lett.~{\bf 12}~(1964)~57.

\bibitem{thrust1977}
E.~Farhi, {\em A {QCD} test for jets}, Phys.~Rev.~Lett.~{\bf 39}~(1977)~1587.

\bibitem{sphericity}
G.~Parisi, {\em Superinclusive cross sections}, Phys.~Lett.~{\bf
  B74}~(1978)~65.

\bibitem{pythiamanual}
T.~Sj{\"o}strand, L.~L{\"o}nnblad, and S.~Mrenna, {\em PYTHIA~6.2: Physics and
  manual}~(2001).
  \href{http://www.arXiv.org/abs/hep-ph/0108264}{E-print~number:~{\tt
  hep-ph/0108264}}.

\bibitem{cpar_donoghue}
J.~F. Donoghue, F.~E. Low, and S.-Y. Pi, {\em Tensor analysis of hadronic jets
  in quantum chromodynamics}, Phys.~Rev.~{\bf D20}~(1979)~2759.

\bibitem{foxwolfram1979}
G.~C. Fox and S.~Wolfram, {\em Event shapes in {\epem} annihilation},
  Nucl.~Phys.~{\bf B149}~(1979)~413.

\bibitem{broadenings_catani}
S.~Catani, G.~Turnock, and B.~R. Webber, {\em Jet broadening measures in
  {\epem} annihilation}, Phys.~Lett.~{\bf B295}~(1992)~269.

\bibitem{durham_catani}
S.~Catani, Y.~L. Dokshitzer, M.~Olsson, G.~Turnock, and B.~R. Webber, {\em New
  clustering algorithm for multijet cross sections in {\epem} annihilation},
  Phys.~Lett.~{\bf B269}~(1991)~432.

\bibitem{pr054}
OPAL Collaboration (P.~D. Acton et al.), {\em A global determination of
  {$\as\left(M_{\text{Z}^0}\right)$} at {LEP}}, Z.~Phys.~{\bf C55}~(1992)~1.

\bibitem{jade}
JADE Collaboration (S.~Bethke et al.), {\em Experimental investigation of the
  energy dependence of the strong coupling strength}, Phys.~Lett.~{\bf
  B213}~(1988)~235.

\bibitem{sigmatot_gorishnii}
S.~G. Gorishnii, A.~L. Kataev, and S.~A. Larin, {\em The {$\mathcal{O}(\as^3)$}
  corrections to {$\sigma_\text{tot}(\epem\to \text{hadrons})$} and
  {$\Gamma(\tau^- \to \nu_\tau + \text{hadrons})$} in {QCD}}, Phys.~Lett.~{\bf
  B259}~(1991)~144.

\bibitem{sigmatot_surguladze}
L.~R. Surguladze and M.~A. Samuel, {\em Total hadronic cross-section in {\epem}
  annihilation at the four-loop level of perturbative {QCD}}, Presented at
  Beyond the Standard Model~II Conf., Norman, OK, USA, 1--3~Nov~1990.
  Phys.~Rev.~Lett.~{\bf 66}~(1991)~560.

\bibitem{event2}
S.~Catani and M.~H. Seymour, {\em The dipole formalism for the calculation of
  {QCD} jet cross sections at next-to-leading order}, Phys.~Lett.~{\bf
  B378}~(1996)~287.
  \href{http://www.arXiv.org/abs/hep-ph/9602277}{E-print~number:~{\tt
  hep-ph/9602277}}.

\bibitem{dglap}
G.~Altarelli and G.~Parisi, {\em Asymptotic freedom in parton language},
  Nucl.~Phys.~{\bf B126}~(1977)~298.

\bibitem{resum_catani}
S.~Catani, L.~Trentadue, G.~Turnock, and B.~R. Webber, {\em Resummation of
  large logarithms in {\epem} event shape distributions}, Nucl.~Phys.~{\bf
  B407}~(1993)~3.

\bibitem{newbroadenings}
{\textrm Yu}.~L. Dokshitzer et al., {\em On the {QCD} analysis of jet
  broadening}, JHEP~{\bf 01}~(1998)~011.
  \href{http://www.arXiv.org/abs/hep-ph/9801324}{E-print~number:~{\tt
  hep-ph/9801324}}.

\bibitem{cpar_resum}
S.~Catani and B.~R. Webber, {\em Resummed {$C$}-parameter distribution in
  {\epem} annihilation}, Phys.~Lett.~{\bf B427}~(1998)~377.
  \href{http://www.arXiv.org/abs/hep-ph/9801350}{E-print~number:~{\tt
  hep-ph/9801350}}.

\bibitem{durham_dissertori}
G.~Dissertori and M.~Schmelling, {\em An Improved theoretical prediction for
  the two-jet rate in {\epem} annihilation}, Phys.~Lett.~{\bf B361}~(1995)~167.

\bibitem{numres}
A.~Banfi, G.~P. Salam, and G.~Zanderighi, {\em Semi-numerical resummation of
  event shapes}, JHEP~{\bf 01}~(2002)~018.
  \href{http://www.arXiv.org/abs/hep-ph/0112156}{E-print~number:~{\tt
  hep-ph/0112156}}.

\bibitem{quarkmass_oasq}
P.~Nason and C.~Oleari, {\em Next-to-leading-order corrections to the
  production of heavy-flavour jets in {\epem} collisions}, Nucl.~Phys.~{\bf
  B521}~(1998)~237.
  \href{http://www.arXiv.org/abs/hep-ph/9709360}{E-print~number:~{\tt
  hep-ph/9709360}}.

\bibitem{quarkmass_resum}
F.~Krauss and G.~Rodrigo, {\em Resummed jet rates for {\epem} annihilation into
  massive quarks}, Phys.~Lett.~{\bf B576}~(2003)~135.
  \href{http://www.arXiv.org/abs/hep-ph/0303038}{E-print~number:~{\tt
  hep-ph/0303038}}.

\bibitem{fourjetresum}
S.~J. Burby and E.~W.~N. Glover, {\em Resumming the light hemisphere mass and
  narrow jet broadening distributions in {\epem} annihilation}, JHEP~{\bf
  04}~(2001)~029.
  \href{http://www.arXiv.org/abs/hep-ph/0101226}{E-print~number:~{\tt
  hep-ph/0101226}}.

\bibitem{dparresum}
A.~Banfi, Y.~L. Dokshitzer, G.~Marchesini, and G.~Zanderighi, {\em {QCD}
  analysis of {$D$}-parameter in near-to-planar three-jet events}, JHEP~{\bf
  05}~(2001)~040.
  \href{http://www.arXiv.org/abs/hep-ph/0104162}{E-print~number:~{\tt
  hep-ph/0104162}}.

\bibitem{menlo_parc_mc}
L.~J. Dixon and A.~Signer, {\em Complete {$\mathcal{O}(\as^3)$} results for
  {$\epem\to(\gamma,\text{Z})\to\text{four~jets}$}}, Phys.~Rev.~{\bf
  D56}~(1997)~4031.
  \href{http://www.arXiv.org/abs/hep-ph/9706285}{E-print~number:~{\tt
  hep-ph/9706285}}.

\bibitem{debrecen_mc}
Z.~Nagy and Z.~Tr{\'o}cs{\'a}nyi, {\em Next-to-leading order calculation of
  four-jet observables in electron positron annihilation}, Phys.~Rev.~{\bf
  D59}~(1999)~014020.
  \href{http://www.arXiv.org/abs/hep-ph/9806317}{E-print~number:~{\tt
  hep-ph/9806317}}.

\bibitem{mercutio_mc}
S.~Weinzierl and D.~A. Kosower, {\em {QCD} corrections to four-jet production
  and three-jet structure in {\epem} annihilation}, Phys.~Rev.~{\bf
  D60}~(1999)~054028.
  \href{http://www.arXiv.org/abs/hep-ph/9901277}{E-print~number:~{\tt
  hep-ph/9901277}}.

\bibitem{eerad2_mc}
J.~M. Campbell, M.~A. Cullen, and E.~W.~N. Glover, {\em Four jet event shapes
  in electron positron annihilation}, Eur.~Phys.~J.~{\bf C9}~(1999)~245.
  \href{http://www.arXiv.org/abs/hep-ph/9809429}{E-print~number:~{\tt
  hep-ph/9809429}}.

\bibitem{xscale}
M.~Dasgupta and G.~P. Salam, {\em Resummed event-shape variables in DIS},
  JHEP~{\bf 08}~(2002)~032.
  \href{http://www.arXiv.org/abs/hep-ph/0208073}{E-print~number:~{\tt
  hep-ph/0208073}}.

\bibitem{kk2f}
S.~Jadach, B.~F.~L. Ward, and Z.~Was, {\em The precision Monte Carlo event
  generator {$\mathcal{KK}$} for two-fermion final states in {\epem}
  collisions}, Comput.~Phys.~Commun.~{\bf 130}~(2000)~260.
  \href{http://www.arXiv.org/abs/hep-ph/9912214}{E-print~number:~{\tt
  hep-ph/9912214}}.

\bibitem{pythia6.1}
T.~Sj{\"o}strand et al., {\em High-energy-physics event generation with
  PYTHIA~6.1}, Comput.~Phys.~Commun.~{\bf 135}~(2001)~238.
  \href{http://www.arXiv.org/abs/hep-ph/0010017}{E-print~number:~{\tt
  hep-ph/0010017}}.

\bibitem{herwig6}
G.~Corcella et al., {\em HERWIG 6: An event generator for hadron emission
  reactions with interfering gluons (including supersymmetric processes)},
  JHEP~{\bf 01}~(2001)~010.
  \href{http://www.arXiv.org/abs/hep-ph/0011363}{E-print~number:~{\tt
  hep-ph/0011363}}.

\bibitem{ariadne4}
L.~L{\"o}nnblad, {\em ARIADNE version 4: A Program for simulation of QCD
  cascades implementing the color dipole model}, Comput.~Phys.~Commun.~{\bf
  71}~(1992)~15.

\bibitem{colour_dipole}
G.~Gustafson, {\em Dual description of a confined color field},
  Phys.~Lett.~{\bf B175}~(1986)~453.

\bibitem{colour_dipole2}
G.~Gustafson and U.~Pettersson, {\em Dipole formulation of QCD cascades},
  Nucl.~Phys.~{\bf B306}~(1988)~746.

\bibitem{stringfrag}
B.~Andersson, G.~Gustafson, and B.~S{\"o}derberg, {\em A general model for jet
  fragmentation}, Z. Phys.~{\bf C20}~(1983)~317.

\bibitem{clustermodel}
B.~R. Webber, {\em A QCD model for jet fragmentation including soft gluon
  interference}, Nucl.~Phys.~{\bf B238}~(1984)~492.

\bibitem{preconfinement}
D.~Amati and G.~Veneziano, {\em Preconfinement as a property of perturbative
  QCD}, Phys.~Lett.~{\bf B83}~(1979)~87.

\bibitem{alphas_bethke}
S.~Bethke, {\em {\as} 2002}, Proceedings of the 9th High Energy Physics
  International Conference on Quantum ChromoDynamics (`QCD~02'), Montpellier,
  France, 2--9~July~2002, Nucl.~Phys.~Proc.~Suppl.~{\bf 121}~(2003)~74.
  \href{http://www.arXiv.org/abs/hep-ex/0211012}{E-print~number:~{\tt
  hep-ex/0211012}}.

\bibitem{alphas_taudecay}
OPAL Collaboration (K.~Ackerstaff et al.), {\em Measurement of the strong
  coupling constant {\as} and the vector and axial-vector spectral functions in
  hadronic {$\tau$}~decays}, Eur.~Phys.~J.~{\bf C7}~(1999)~571.
  \href{http://www.arXiv.org/abs/hep-ex/9808019}{E-print~number:~{\tt
  hep-ex/9808019}}.

\bibitem{alphas_jets}
JADE and OPAL Collaborations (P.~Pfeifenschneider et al.), {\em QCD analyses
  and determinations of {\as} in {\epem} annihilation at energies between
  35~GeV and 189~GeV}, Eur.~Phys.~J.~{\bf C17}~(2000)~19.
  \href{http://www.arXiv.org/abs/hep-ex/0001055}{E-print~number:~{\tt
  hep-ex/0001055}}.

\bibitem{alphas_f2gamma}
S.~Albino, M.~Klasen, and S.~S{\"o}ldner-Rembold, {\em Strong coupling constant
  from the photon structure function}, Phys.~Rev.~Lett.~{\bf 89}~(2002)~122004.
  \href{http://www.arXiv.org/abs/hep-ph/0205069}{E-print~number:~{\tt
  hep-ph/0205069}}.

\bibitem{alphas_fragfun}
B.~A. Kniehl, G.~Kramer, and B.~Potter, {\em Strong coupling constant from
  scaling violations in fragmentation functions}, Phys.~Rev.~Lett.~{\bf
  85}~(2000)~5288.
  \href{http://www.arXiv.org/abs/hep-ph/0003297}{E-print~number:~{\tt
  hep-ph/0003297}}.

\bibitem{powercorr_Dokshitzer}
Y.~L. Dokshitzer and B.~R. Webber, {\em Calculation of power corrections to
  hadronic event shapes}, Phys.~Lett.~{\bf B352}~(1995)~451.
  \href{http://www.arXiv.org/abs/hep-ph/9504219}{E-print~number:~{\tt
  hep-ph/9504219}}.

\bibitem{powercorr_Manohar}
A.~V. Manohar and M.~B. Wise, {\em Power suppressed corrections to hadronic
  event shapes}, Phys.~Lett.~{\bf B344}~(1995)~407.
  \href{http://www.arXiv.org/abs/hep-ph/9406392}{E-print~number:~{\tt
  hep-ph/9406392}}.

\bibitem{powercorr_Akhoury}
R.~Akhoury and V.~I. Zakharov, {\em On the universality of the leading {$1/Q$}
  power corrections in {QCD}}, Phys.~Lett.~{\bf B357}~(1995)~646.
  \href{http://www.arXiv.org/abs/hep-ph/9504248}{E-print~number:~{\tt
  hep-ph/9504248}}.

\bibitem{powercorr_gardi1}
E.~Gardi and G.~Grunberg, {\em Power corrections in the single dressed gluon
  approximation: The average thrust as a case study}, JHEP~{\bf 11}~(1999)~016.
  \href{http://www.arXiv.org/abs/hep-ph/9908458}{E-print~number:~{\tt
  hep-ph/9908458}}.

\bibitem{powercorr_gardi2}
E.~Gardi, {\em Suppressed power corrections for moments of event-shape
  variables in {\epem} annihilation}, JHEP~{\bf 04}~(2000)~030.
  \href{http://www.arXiv.org/abs/hep-ph/0003179}{E-print~number:~{\tt
  hep-ph/0003179}}.

\bibitem{powercorr_Korchemsky}
G.~P. Korchemsky and G.~Sterman, {\em Nonperturbative corrections in resummed
  cross-sections}, Nucl.~Phys.~{\bf B437}~(1995)~415.
  \href{http://www.arXiv.org/abs/hep-ph/9411211}{E-print~number:~{\tt
  hep-ph/9411211}}.

\bibitem{DELPHI_as_1}
DELPHI Collaboration (P.~Abreu et al.), {\em Energy dependence of event shapes
  and of {\as} at {LEP2}}, Phys.~Lett.~{\bf B456}~(1999)~322.

\bibitem{alphas_mrst}
A.~D. Martin, R.~G. Roberts, W.~J. Stirling, and R.~S. Thorne, {\em MRST2001:
  Partons and {\as} from precise deep inelastic scattering and Tevatron jet
  data}, Eur.~Phys.~J.~{\bf C23}~(2002)~73.
  \href{http://www.arXiv.org/abs/hep-ph/0110215}{E-print~number:~{\tt
  hep-ph/0110215}}.

\bibitem{opal_colourfactors1}
OPAL Collaboration (G.~Abbiendi et al.), {\em A simultaneous measurement of the
  QCD colour factors and the strong coupling}, Eur.~Phys.~J.~{\bf
  C20}~(2001)~601.
  \href{http://www.arXiv.org/abs/hep-ex/0101044}{E-print~number:~{\tt
  hep-ex/0101044}}.

\bibitem{opal_colourfactors2}
OPAL Collaboration (R.~Akers et al.), {\em A Study of QCD structure constants
  and a measurement of {$\as(M_{{\text{Z}}^0})$} at LEP using event shape
  observables}, Z.~Phys.~{\bf C68}~(1995)~519.

\bibitem{alphas_flavour1}
OPAL Collaboration (R.~Akers et al.), {\em A test of the flavor independence of
  the strong interaction for five flavors}, Z.~Phys.~{\bf C60}~(1993)~397.

\bibitem{alphas_flavour2}
OPAL Collaboration (G.~Abbiendi et al.), {\em Test of the flavour independence
  of {\as} using next-to-leading order calculations for heavy quarks},
  Eur.~Phys.~J.~{\bf C11}~(1999)~643.
  \href{http://www.arXiv.org/abs/hep-ex/9904013}{E-print~number:~{\tt
  hep-ex/9904013}}.

\bibitem{multiplicity1}
OPAL Collaboration (G.~Abbiendi et al.), {\em Charged multiplicities in Z
  decays into u, d, and s quarks}, Eur.~Phys.~J.~{\bf C19}~(2001)~257.
  \href{http://www.arXiv.org/abs/hep-ex/0011022}{E-print~number:~{\tt
  hep-ex/0011022}}.

\bibitem{multiplicity2}
OPAL Collaboration (G.~Abbiendi et al.), {\em Charged particle multiplicities
  in heavy and light quark initiated events above the {Z$^0$} peak},
  Phys.~Lett.~{\bf B550}~(2002)~33.
  \href{http://www.arXiv.org/abs/hep-ex/0211007}{E-print~number:~{\tt
  hep-ex/0211007}}.

\bibitem{alephdetector}
ALEPH Collaboration (D.~Decamp et al.), {\em {ALEPH:} a detector for
  electron-positron annihilations at {LEP}}, Nucl.~Instrum.~Meth.~{\bf
  A294}~(1990)~121.

\bibitem{delphidetector}
DELPHI Collaboration (P.~A. Aarnio et al.), {\em The {DELPHI} detector at
  {LEP}}, Nucl.~Instrum.~Meth.~{\bf A303}~(1991)~233.

\bibitem{l3detector}
L3 Collaboration (B.~Adeva et al.), {\em The construction of the {L3}
  experiment}, Nucl.~Instrum.~Meth.~{\bf A289}~(1990)~35.

\bibitem{nimpaper}
OPAL Collaboration (K.~Ahmet et al.), {\em The {OPAL} detector at {LEP}},
  Nucl.~Instrum.~Meth.~{\bf A305}~(1991)~275.

\bibitem{gargamelle}
F.~J. Hasert et al., {\em Observation of neutrino-like interactions without
  muon or electron in the Gargamelle neutrino experiment}, Phys.~Lett.~{\bf
  B46}~(1973)~138.

\bibitem{ua1_zdiscovery}
UA1 Collaboration (G.~Arnison et al.), {\em Experimental observation of lepton
  pairs of invariant mass around 95~{$\text{GeV}/c^2$} at the CERN SPS
  collider}, Phys.~Lett.~{\bf B126}~(1983)~398.

\bibitem{ua2_zdiscovery}
UA2 Collaboration (P.~Bagnaia et al.), {\em Evidence for {$\text{Z}^0 \to
  \epem$} at the CERN {$\bar{\text{p}}\text{p}$} collider}, Phys.~Lett.~{\bf
  B129}~(1983)~130.

\bibitem{ua1_wdiscovery}
UA1 Collaboration (G.~Arnison et al.), {\em Experimental observation of
  isolated large transverse energy electrons with associated missing energy at
  {$\sqrt{s}=\text{540}$}~GeV}, Phys.~Lett.~{\bf B122}~(1983)~103.

\bibitem{ua2_wdiscovery}
UA2 Collaboration (M.~Banner et al.), {\em Observation of single isolated
  electrons of high transverse momentum in events with missing transverse
  energy at the CERN {$\bar{\text{p}}\text{p}$} collider}, Phys.~Lett.~{\bf
  B122}~(1983)~476.

\bibitem{firstlepproposal}
J.~R.~J. Bennett et al., {\em Design concept for a 100~GeV {\epem} storage ring
  (LEP)}, CERN report number CERN-77-14~(1977).

\bibitem{lepdesign1}
{\em LEP Design Report: Vol. 1. The LEP Injector Chain}, CERN report number
  CERN-LEP/TH/83-29~(1983).

\bibitem{lepdesign2}
{\em LEP Design Report: Vol. 2. The LEP Main Ring}, CERN report number
  CERN-LEP/84-01~(1984).

\bibitem{lepmachine}
E.~Picasso and G.~Plass, {\em The machine design: {LEP}}, Europhys.~News~{\bf
  20}~(1989)~80.

\bibitem{lepdesign3}
{\em LEP Design Report: Vol. 3. LEP2}, CERN report number
  CERN-AC-96-01-LEP-2~(1996).

\bibitem{opalsilicon1}
P.~P. Allport et al., {\em The {OPAL} silicon microvertex detector},
  Nucl.~Instrum.~Meth.~{\bf A324}~(1993)~34.

\bibitem{opalsilicon2}
P.~P. Allport et al., {\em The OPAL silicon strip microvertex detector with two
  coordinate readout}, Nucl.~Instrum.~Meth.~{\bf A346}~(1994)~476.

\bibitem{opalsilicon3}
S.~Anderson et al., {\em The extended {OPAL} silicon strip microvertex
  detector}, Nucl.~Instrum.~Meth.~{\bf A403}~(1998)~326.

\bibitem{opal_te}
G.~Aguillion et al., {\em Thin scintillating tiles with high light yield for
  the OPAL endcaps}, Nucl.~Instrum.~Meth.~{\bf A417}~(1998)~266.

\bibitem{opal_sw}
B.~E. Anderson et al., {\em The OPAL silicon-tungsten calorimeter front end
  electronics}, IEEE~Trans.~Nucl.~Sci.~{\bf 41}~(1994)~845.

\bibitem{opal_pretrigger}
M.~Arignon et al., {\em The pretrigger system of the OPAL experiment at LEP},
  Nucl.~Instrum.~Meth.~{\bf A333}~(1993)~330.

\bibitem{opalcv}
J.~R. Carter et al., {\em The {OPAL} vertex drift chamber},
  Nucl.~Instrum.~Meth.~{\bf A286}~(1990)~99.

\bibitem{opalcj}
H.~M. Fischer et al., {\em The OPAL jet chamber}, Nucl.~Instrum.~Meth.~{\bf
  A283}~(1989)~492.

\bibitem{opalcz}
H.~Mes et al., {\em Design and tests of the {$z$} coordinate drift chamber
  system for the {OPAL} central detector at {LEP}}, Nucl.~Instrum.~Meth.~{\bf
  A265}~(1988)~445.

\bibitem{opal_ee}
P.~W. Jeffreys et al., {\em Development studies for the OPAL endcap
  electromagnetic calorimeter using vacuum photo triode instrumented lead
  glass}, Nucl.~Instrum.~Meth.~{\bf A290}~(1990)~76.

\bibitem{opal_ecal_presamplers}
C.~Beard et al., {\em Thin, high gain wire chambers for electromagnetic
  presampling in OPAL}, Nucl.~Instr.~Meth.~{\bf A286}~(1990)~117.

\bibitem{opal_hcal_streamers}
G.~Artusi et al., {\em Limited streamer tubes for the OPAL hadron calorimeter},
  Nucl.~Instr.~Meth.~{\bf A279}~(1989)~523.

\bibitem{opal_mb}
R.~J. Akers et al., {\em The OPAL muon barrel detector},
  Nucl.~Instrum.~Meth.~{\bf A357}~(1995)~253.

\bibitem{opal_me}
G.~T.~J. Arnison et al., {\em Production and testing of limited streamer tubes
  for the endcap muon subdetector of OPAL}, Nucl.~Instrum.~Meth.~{\bf
  A294}~(1990)~431.

\bibitem{sw_lumi_pr289}
OPAL Collaboration (G.~Abbiendi et al.), {\em Precision luminosity for Z{$^0$}
  lineshape measurements with a silicon-tungsten calorimeter},
  Eur.~Phys.~J.~{\bf C14}~(2000)~373.
  \href{http://www.arXiv.org/abs/hep-ex/9910066}{E-print~number:~{\tt
  hep-ex/9910066}}.

\bibitem{opal_trigger}
M.~Arignon et al., {\em The Trigger system of the OPAL experiment at LEP},
  Nucl.~Instrum.~Meth.~{\bf A313}~(1992)~103.

\bibitem{opal_tt}
A.~A. Carter et al., {\em A fast track trigger processor for the OPAL detector
  at LEP}, Nucl.~Instrum.~Meth.~{\bf A250}~(1986)~503.

\bibitem{opal_daq}
J.~T.~M. Baines et al., {\em The data acquisition system of the OPAL detector
  at LEP}, Nucl.~Instrum.~Meth.~{\bf A325}~(1993)~271.

\bibitem{opal_filter}
D.~G. Charlton, F.~Meijers, T.~J. Smith, and P.~S. Wells, {\em The online event
  filter of the OPAL experiment at LEP}, Nucl.~Instrum.~Meth.~{\bf
  A325}~(1993)~129.

\bibitem{zphyslep1}
Z.~Kunszt and P.~Nason, {\em QCD}, in ``Z~Physics at LEP 1'', CERN report
  number CERN-89-08, vol.~1, p.~373~(1989).

\bibitem{mt}
T.~Omori, {\em A Matching Algorithm: MT package}, OPAL Technical
  Note~381~(1996). [{\tt
  http://opal.web.cern.ch/opal/doc/technote/html/tn381.html}].

\bibitem{lep_higgs}
ALEPH, DELPHI, L3 and OPAL Collaborations (R.~Barate et al.), {\em Search for
  the standard model Higgs boson at LEP}, Phys.~Lett.~{\bf B565}~(2003)~61.
  \href{http://www.arXiv.org/abs/hep-ex/0306033}{E-print~number:~{\tt
  hep-ex/0306033}}.

\bibitem{gopal}
J.~Allison et al., {\em The detector simulation program for the {OPAL}
  experiment at {LEP}}, Nucl.~Instrum.~Meth.~{\bf A317}~(1992)~47.

\bibitem{pr141}
OPAL Collaboration (G.~Alexander et al.), {\em A comparison of b and uds quark
  jets to gluon jets}, Z.~Phys.~{\bf C69}~(1996)~543.

\bibitem{pr379_tunings}
OPAL Collaboration (G.~Abbiendi et al.), {\em Tests of models of color
  reconnection and a search for glueballs using gluon jets with a rapidity
  gap}, Submitted to Eur.~Phys.~J.~(2003).
  \href{http://www.arXiv.org/abs/hep-ex/0306021}{E-print~number:~{\tt
  hep-ex/0306021}}.

\bibitem{koralw}
M.~Skrzypek, S.~Jadach, W.~Placzek, and Z.~Was, {\em Monte Carlo program
  KORALW~1.02 for W~pair production at LEP2/NLC energies with
  Yennie-Frautschi-Suura exponentiation}, Comput.~Phys.~Commun.~{\bf
  94}~(1996)~216.

\bibitem{koralw1.42}
S.~Jadach, W.~Placzek, M.~Skrzypek, B.~F.~L. Ward, and Z.~Was, {\em Monte Carlo
  program KORALW~1.42 for all four-fermion final states in {\epem} collisions},
  Comput.~Phys.~Commun.~{\bf 119}~(1999)~272.
  \href{http://www.arXiv.org/abs/hep-ph/9906277}{E-print~number:~{\tt
  hep-ph/9906277}}.

\bibitem{grc4f}
J.~Fujimoto et al., {\em {\lowercase{G}}rc4f v1.1: a four-fermion event
  generator for {\epem} collisions}, Comput.~Phys.~Commun.~{\bf
  100}~(1997)~128.
  \href{http://www.arXiv.org/abs/hep-ph/9605312}{E-print~number:~{\tt
  hep-ph/9605312}}.

\bibitem{sprime}
D.~Ward, {\em Separation of non-radiative
  $\text{Z}^0/\gamma\rightarrow\text{q}\bar{\text{q}}$ events at {LEP2}},
  OPAL~Technical~Note~394~(1996). [{\tt
  http://opal.web.cern.ch/opal/doc/technote/html/tn394.html}].

\bibitem{ainsley}
C.~Ainsley, {\em Studies of {$\text{Z}/\gamma\to\text{q}\bar{\text{q}}$} events
  with the {OPAL} detector at {LEP~II}}, PhD thesis, University of
  Cambridge~(2003).

\bibitem{pr321}
OPAL Collaboration (G.~Abbiendi et al.), {\em W{$^+$}W{$^-$} production cross
  section and W~branching fractions in {\epem} collisions at 189~GeV},
  Phys.~Lett.~{\bf B493}~(2000)~249.
  \href{http://www.arXiv.org/abs/hep-ex/0009019}{E-print~number:~{\tt
  hep-ex/0009019}}.

\bibitem{pr362}
OPAL Collaboration (G.~Abbiendi et al.), {\em Charged particle momentum spectra
  in {\epem} annihilation at {$\sqrt{s}=\text{192}$}--209~GeV},
  Eur.~Phys.~J.~{\bf C27}~(2003)~467.
  \href{http://www.arXiv.org/abs/hep-ex/0209048}{E-print~number:~{\tt
  hep-ex/0209048}}.

\bibitem{lqqln}
M.~Thomson, {\em The OPAL qq{$\ell\nu$} event selection},
  OPAL~Technical~Note~635~(2000). [{\tt
  http://opal.web.cern.ch/opal/doc/technote/html/tn635.html}].

\bibitem{cowanbook}
G.~Cowan, {\em Statistical Data Analysis}.
\newblock Oxford,~UK: Clarendon~(1998).

\bibitem{ALEPH_as_1}
ALEPH Collaboration (D.~Buskulic et al.), {\em Studies of {QCD} in
  {$\epem\to$}~hadrons at {$E_\text{cm}=\text{130}$~GeV} and 136~GeV},
  Z.~Phys.~{\bf C73}~(1997)~409.

\bibitem{lafferty}
G.~D. Lafferty and T.~R. Wyatt, {\em Where to stick your data points: the
  treatment of measurements within wide bins}, Nucl.~Instrum.~Meth.~{\bf
  A355}~(1995)~541.

\bibitem{OPAL_PN512}
OPAL Collaboration, {\em Updated measurements of {\as} using event shape
  observables}, OPAL Physics Note 512~(2002). [{\tt
  http://opal.web.cern.ch/opal/doc/physnote/html/pn512.html}].

\bibitem{barlow_bootstrap}
R.~Barlow, {\em Application of the Bootstrap resampling technique to particle
  physics experiments}, Manchester University HEP preprint number
  MAN/HEP/99/4~(1999). [{\tt
  http://www.hep.man.ac.uk/preprints/manhep99-4.ps}].

\bibitem{pn279_wmass}
OPAL Collaboration, {\em A measurement of the W boson mass by direct
  reconstruction}, OPAL Physics Note 279~(1997). [{\tt
  http://opal.web.cern.ch/opal/doc/physnote/html/pn279.html}].

\bibitem{montpellier}
M.~T. Ford for the OPAL Collaboration, {\em Measurements of {\as} from QCD
  event shapes at OPAL and LEP}, Proceedings of the 9th High Energy Physics
  International Conference on Quantum ChromoDynamics (`QCD~02'), Montpellier,
  France, 2--9~July~2002, Nucl.~Phys.~Proc.~Suppl.~{\bf 121}~(2003)~65.

\bibitem{donkers_kluth_pahl}
M.~Donkers, S.~Kluth, and C.~Pahl, Private communications~(2003). [{\tt
  http://opalinfo.cern.ch/opal/group/qcd/lep2/lep2tables.html}].

\bibitem{ALEPH_as_2}
ALEPH Collaboration (R.~Barate et al.), {\em Studies of quantum chromodynamics
  with the {ALEPH} detector}, Phys.~Rept.~{\bf 294}~(1998)~1.

\bibitem{ALEPH_as_ALL}
ALEPH Collaboration (A.~Heister et al.), {\em Studies of QCD at {\epem}
  centre-of-mass energies between 91~GeV and 209~GeV}, Submitted to
  Eur.~Phys.~J.~(2003). Report~number:~CERN-EP-2003-084.

\bibitem{DELPHI_as_2}
DELPHI Collaboration (P.~Abreu et al.), {\em Consistent measurements of {\as}
  from precise oriented event shape distributions}, Eur.~Phys.~J.~{\bf
  C14}~(2000)~557.
  \href{http://www.arXiv.org/abs/hep-ex/0002026}{E-print~number:~{\tt
  hep-ex/0002026}}.

\bibitem{DELPHI_as_3}
J.~Drees et al. for the DELPHI Collaboration, {\em QCD results from the
  measurements of event shape distributions between 41~GeV and 189~GeV},
  Proceedings of the International Europhysics Conference on High-Energy
  Physics (`EPS-HEP~99'), Tampere, Finland, 15--21~July~1999
  (Bristol,~UK:~IOP). Report~number:~CERN-OPEN-99-387~/ DELPHI-99-114-CONF-301.

\bibitem{DELPHI_as_4}
J.~Drees et al. for the DELPHI Collaboration, {\em The running of the strong
  coupling and a study of power corrections to hadronic event shapes with the
  DELPHI detector at LEP}, Proceedings of the 30th International Conference on
  High Energy Physics, Osaka, Japan, 27~July--2~Aug~2000
  (Singapore:~World~Sci.). Report~number:~DELPHI-2000-116-CONF-415.

\bibitem{L3_as_1}
L3 Collaboration (O.~Adriani et al.), {\em Determination of {\as} from hadronic
  event shapes measured on the Z{$^0$} resonance}, Phys.~Lett.~{\bf
  B284}~(1992)~471.

\bibitem{L3_as_6}
L3 Collaboration (M.~Acciarri et al.), {\em {QCD} studies in {\epem}
  annihilation from 30~GeV to 189~GeV}, Phys.~Lett.~{\bf B489}~(2000)~65.
  \href{http://www.arXiv.org/abs/hep-ex/0005045}{E-print~number:~{\tt
  hep-ex/0005045}}.

\bibitem{L3_as_7}
L3 Collaboration (P.~Achard et al.), {\em Determination of {\as} from hadronic
  event shapes in {\epem} annihilation at {$192\leq\sqrt{s}\leq 208$~GeV}},
  Phys. Lett.~{\bf B536}~(2002)~217.
  \href{http://www.arXiv.org/abs/hep-ex/0206052}{E-print~number:~{\tt
  hep-ex/0206052}}.

\bibitem{OPAL_PN519}
OPAL Collaboration, {\em Measurement of {\as} in radiative hadronic events},
  OPAL Physics Note 519~(2003). [{\tt
  http://opal.web.cern.ch/opal/doc/physnote/html/pn519.html}].

\bibitem{L3_as_radiative}
L3 Collaboration (M.~Acciarri et al.), {\em Study of hadronic events and
  measurements of {\as} between 30~GeV and 91~GeV}, Phys.~Lett.~{\bf
  B411}~(1997)~339.

\bibitem{stenzel_private}
H.~Stenzel, Private communication~(July 2002).

\bibitem{stenzel_salam_private}
H.~Stenzel and G.~P. Salam, Private communication~(August 2002).

\bibitem{lepfest}
{The LEP QCD Working Group}, {\em Preliminary combination of {\as} values
  derived from event shape variables at {LEP}}.
  Report~number:~ALEPH~01-038~Physic~01-012~/ DELPHI~2001-043~PHYS~893~/
  L3~note~2661~/ OPAL~Technical Note 689. [{\tt
  http://delphiwww.cern.ch/\~{}pubxx/www/delsec/delnote/public/
  2001\_043\_phys\_893.ps.gz}].

\bibitem{wicke_private}
D.~Wicke, Private communication~(2002).

\bibitem{alekhin}
S.~I. Alekhin, {\em Statistical properties of the estimator using covariance
  matrix}~(2000).
  \href{http://www.arXiv.org/abs/hep-ex/0005042}{E-print~number:~{\tt
  hep-ex/0005042}}.

\bibitem{wmasscomb}
ALEPH, DELPHI, L3 and OPAL Collaborations, {\em A combination of preliminary
  electroweak measurements and constraints on the standard model}, Prepared
  from contributions of the LEP and SLD experiments to the 2002 summer
  conferences.
  \href{http://www.arXiv.org/abs/hep-ex/0212036}{E-print~number:~{\tt
  hep-ex/0212036}}, ALEPH-2002-042-PHYSIC-2002-018, DELPHI-2002-098-PHYS-927,
  CERN-L3-NOTE-2788, OPAL-PR-370.

\bibitem{dasgupta_salam_review}
M.~Dasgupta and G.~P. Salam, {\em Event shapes in {\epem} annihilation and deep
  inelastic scattering}, Submitted to J.~Phys.~G~(2003).
  \href{http://www.arXiv.org/abs/hep-ph/0312283}{E-print~number:~{\tt
  hep-ph/0312283}}.

\end{thebibliography}\endgroup
\bibliographystyle{fordthesis}

\end{fmffile}
\end{document}